\pdfoutput=1
\documentclass[graphicx,amssymb,amsmath,enumerate,11pt]{report}
\usepackage[a4paper,inner=3.5cm,outer=2.5cm,top=2.5cm,bottom=2.5cm,pdftex]{geometry}
\usepackage{graphicx}
\usepackage{epsfig}
\usepackage{amsmath}
\usepackage{amssymb}
\usepackage{enumerate}
\usepackage{axodraw}
\usepackage{url}
\newcommand\mchapter[2]{\chapter*{#1}
\vskip -0.5cm
\noindent
{\it #2}
\addcontentsline{toc}{chapter}{#1\\{\normalsize\it #2}}}

\begin{document}
\vskip 3cm
\begin{center}
{\Large\bf\sc
Proceedings of the Workshop \\
on Monte Carlo's, Physics \\
and Simulations at the LHC \\
\vskip 4cm
PART I
}
\end{center}
\newpage
{\it
\begin{center}
Authors
\end{center}
\vskip 0.5cm
\noindent
F. Ambroglini~$^{\ref{UniPG}}$,
R. Armillis~$^{\ref{Lecce}}$,
P. Azzi~$^{\ref{INFNPD}}$,
G. Bagliesi~$^{\ref{INFNPI}}$,
A. Ballestrero~$^{\ref{INFNTO}}$,
G. Balossini~$^{\ref{UniPV}}$,
A. Banfi~$^{\ref{UniMIB}}$,
P. Bartalini~$^{\ref{Florida}}$,
D. Benedetti~$^{\ref{Northeastern}}$,
G. Bevilacqua~$^{\ref{Demokritos}}$,
S. Bolognesi~$^{\ref{UniTO}}$,
A. Cafarella~$^{{\ref{Demokritos},\ref{Lecce}}}$,
C.M. Carloni Calame~$^{\ref{Southampton}}$,
L. Carminati~$^{\ref{UniMI}}$,
M. Cobal~$^{\ref{Udine}}$,
G. Corcella~$^{{\ref{FermiZ},\ref{Normale}}}$,
C. Corian\`{o}~$^{\ref{Lecce}}$,
A. Dainese~$^{\ref{Legnaro}}$,
V. Del Duca~$^{\ref{LNF}}$,
F. Fabbri~$^{\ref{INFNBO}}$,
M. Fabbrichesi~$^{\ref{INFNTS}}$,
L. Fan\`{o}~$^{\ref{INFNPG}}$,
Alon E. Faraggi~$^{\ref{Liverpool}}$,
S. Frixione~$^{{\ref{CERNTH},\ref{EPFL}}}$,
L. Garbini~$^{\ref{UniPG}}$,
A. Giammanco~$^{\ref{Louvain}}$,
M. Guzzi~$^{\ref{Lecce}}$,
N. Irges~$^{\ref{Crete}}$,
E. Maina~$^{\ref{UniTO}}$,
C. Mariotti~$^{\ref{INFNTO}}$,
G. Masetti~$^{\ref{UniBO}}$,
B. Mele~$^{\ref{INFNRM1}}$,
E. Migliore~$^{\ref{UniTO}}$,
G. Montagna~$^{\ref{UniPV}}$,
M. Monteno~$^{\ref{INFNTO}}$,
M. Moretti~$^{\ref{Ferrara}}$,
P. Nason~$^{\ref{INFNMIB}}$,
O. Nicrosini~$^{\ref{INFNPV}}$,
A. Nisati~$^{\ref{INFNRM1}}$,
A. Perrotta~$^{\ref{INFNBO}}$,
F. Piccinini~$^{\ref{INFNPV}}$,
G. Polesello~$^{\ref{INFNPV}}$,
D. Rebuzzi~$^{\ref{INFNPV}}$,
A. Rizzi~$^{\ref{ETH}}$,
S. Rolli~$^{\ref{Tufts}}$,
C. Roda~$^{\ref{INFNPI}}$,
S. Rosati~$^{\ref{INFNRM1}}$,
A. Santocchia~$^{\ref{UniPG}}$,
D. Stocco~$^{{\ref{Subatech},\ref{UniTO}}}$,
F. Tartarelli~$^{\ref{UniMI}}$,
R. Tenchini~$^{\ref{INFNPI}}$,
A. Tonero~$^{\ref{SISSA}}$,
M. Treccani~$^{{\ref{granada},\ref{Ferrara}}}$,
D. Treleani~$^{\ref{UniTS}}$,
A. Tricoli~$^{\ref{Rutherford}}$,
D. Trocino~$^{\ref{UniTO}}$,
L. Vecchi~$^{\ref{SISSA}}$,
A. Vicini~$^{\ref{UniMI}}$,
I. Vivarelli~$^{\ref{INFNPI}}$.

{\scriptsize\it
\begin{enumerate}\addtolength{\itemsep}{-0.7\baselineskip}

\item \label{Lecce}       University of Salento and INFN, Lecce, Italy

\item \label{INFN} INFN, Frascati, Italy

\item \label{UniTO}      University of Torino and INFN, Torino, Italy

\item \label{INFNTO}      INFN, Sezione di Torino, Torino, Italy

\item \label{UniMIB}     University of Milano Bicocca and Sezione INFN, Milano, Italy

\item \label{UniMI}      University of Milano and Sezione INFN, Milano, Italy

\item \label{UniTS}      University of Trieste and Sezione INFN, Trieste, Italy

\item \label{UniBO}      University of Bologna and Sezione INFN, Bologna, Italy

\item \label{INFNBO}     INFN, Sezione di Bologna, Bologna, Italy

\item \label{INFNMIB}    INFN, Sezione di Milano Bicocca, Milano, Italy

\item \label{INFNMI}     INFN, Sezione di Milano, Milano, Italy

\item \label{INFNRM1}     INFN, Sezione di Roma, Roma, Italy

\item \label{UniPV}      University of Pavia and INFN, Pavia, Italy

\item \label{INFNPV}     INFN, Sezione di Pavia, Pavia, Italy

\item \label{INFNTS}     INFN, Sezione di Trieste, Trieste, Italy

\item \label{INFNPG}     INFN, Sezione di Perugia, Perugia, Italy

\item \label{SISSA}      SISSA/ISAS, Trieste, Italy

\item \label{Ferrara}     University of Ferrara and INFN, Ferrara, Italy

\item \label{UniPG}       University of Perugia and INFN, Perugia, Italy

\item \label{INFNPD}      INFN, Sezione di Padova, Padova, Italy

\item \label{UniPA}       University of Padova and INFN, Padova, Italy

\item \label{UniPI}       University of Pisa and INFN, Pisa, Italy

\item \label{INFNPI}      INFN, Sezione di Pisa, Pisa, Italy

\item \label{Normale} Scuola Normale Superiore and INFN, Pisa, Italy

\item \label{Crete}       Department of Physics and Institute of Plasma Physics,
                         University of Crete, Heraklion, Greece

\item \label{Florida}     University of Florida, Gainesville, Florida, USA

\item \label{Udine}       INFN Gruppo Collegato di Udine, Udine, Italy

\item \label{FermiZ}      Museo Storico della Fisica e Centro Studi e Ricerche E. Fermi, Roma, Italy

\item \label{Legnaro}     INFN, Laboratori Nazionali di Legnaro, Padova, Italy

\item \label{Demokritos}  Institute of Nuclear Physics, NCSR ``Demokritos'', Athens, Greece

\item \label{Liverpool}   Department of Mathematical Sciences,
             University of Liverpool, Liverpool, United Kingdom

\item \label{granada} Departamento de F\'{i}sica Te\'orica y del Cosmos, University of Granada, Granada, Spain

\item \label{Southampton} INFN and School of Physics and Astronomy, University of Southampton, 
      Highfield, Southampton, UK

\item \label{Subatech} Subatech (Universit\'e de Nantes, Ecole des Mines and CNRS/IN2P3),
Nantes, France

\item \label{Northeastern} Northeastern University, Department of Physics, Boston, MA, USA

\item \label{Louvain}     Universit\'e Catholique de Louvain, Louvain-la-Neuve, Belgium

\item \label{ETH}   Institute for Particle Physics, ETH Zurich, Zurich, Switzerland

\item \label{Rutherford}
    Rutherford Appleton Laboratory, Science and Technology Facilities
Council, Harwell Science and Innovation Campus, Didcot OX11 0QX,
United Kingdom

\item \label{CERNTH} PH Department, TH Unit, CERN, Geneva, Switzerland

\item \label{EPFL} ITPP, EPFL, Lausanne, Switzerland

\item \label{LNF} INFN, Laboratori Nazionali di Frascati, Frascati, Italy

\item\label{Tufts} Tufts University, Medford, Massachusetts, USA

\end{enumerate}
}
}

\newpage
\vskip 2cm
\begin{center}
{\bf \Large Preface}
\end{center}
\vskip 1cm
These proceedings collect the presentations given at the
first three meetings of the
``Workshop on Monte Carlo's, Physics and Simulations
at the LHC'', held on February 27-28, May 22-24 and October 23-25 2006
in Frascati (Italy). The purpose of the workshop, sponsored by the INFN,
was to bring together all the complementary Italian
scientific communities interested into high $p_T$ physics at the LHC.
The workshop was thus attended by LHC experimental physicists,
theoretical physicists dedicated to the calculation
of matrix elements for collider processes and to the implementation
of Monte Carlo programs, and theoretical physicists interested into model
building and physics beyond the Standard Model. Theoretical Standard
Model prediction, as well as physics signals from new models, are made available
to the experimental community as Monte Carlo generators, that thus constitute
the meeting points of the three communities mentioned above.
The aim of the workshop was essentially to start to talk to each other,
and to begin to understand the methods, the problems, and the language
of the complementary communities.

Many of the presentations held at the first three workshop meetings were
basic introductions to important theoretical and experimental topics
relevant to LHC physics, and the speakers were requested to use a
language suitable for people with no expertise in their field.  The
collection of these presentations constitutes thus an introduction to
a few basic aspects of high $p_T$ LHC physics. It was decided to put
them in the form of proceedings, maintaining the requirements of a language
suitable for the complementary physics communities.
In order to achieve this goal, the contributions were refereed internally,
and have gone through several revisions.
The second part of these proceedings collects more
specialised presentations held at the workshop.

Although the very ambitious plan for these proceedings was not totally
fulfilled (for instance, a few chapters were never completed),
we feel that, at least for some of the chapters,
we have met our goal. In particular, the first chapter constitute a very
condensed presentation of the basics about LHC high $p_T$ physics, that
can be used as a first introductory reading for the subject.  The last
chapter summarizes the basic features of the most important component of the
ATLAS and CMS experiments, written in a way that should be easily
understandable also by theorists.
Many chapters of these
proceedings\footnote{together with the original slides of the presentations,
available at \url{http://moby.mib.infn.it/~nason/mcws/scientific_programme.htm}.}
can be used for an introductory class on LHC high $p_T$ physics for
graduate students in experimental and theoretical physics.

Although LHC physics is evolving rapidly, we believe that the basic argument
treated in this volume will remain valid for an introduction, and that
this effort will remain useful for the years to come.

\vskip 0.6cm
{\flushright {\it Paolo Nason}}

\vskip 1cm
\noindent
Workshop's Organizing Committee:\\
{\it V. Del Duca,
B. Mele,
P. Nason,
G. Polesello (ATLAS),
R. Tenchini (CMS).}\\
 \noindent
Workshop's Conveners: \\ \noindent
Shower Monte Carlo: {\it S. Frixione, L. Fan\'o (CMS) S. Rolli (ATLAS);}\\ \noindent
Exact calculations at fixed order: {\it F. Piccinini, P. Azzi (CMS);}\\ \noindent
SM and BSM Physics at LHC: {\it B. Mele, M. Cobal (ATLAS), F. Fabbri (CMS);}\\ \noindent
Experimental Studies: {\it F. Tartarelli (ATLAS), C. Mariotti, E. Migliore (CMS).}

\newpage
\tableofcontents
\newpage
\addtocounter{chapter}{1}
\newcommand{\cca}{$\sim$}
\newcommand{\ra}{$\rightarrow$}

\newcommand\pt{p_{\scriptscriptstyle T}}
\newcommand\mt{m_{\scriptscriptstyle T}}
\newcommand\lt{l_{\scriptscriptstyle T}}
\newcommand\Ecm{E_{\scriptscriptstyle CM}}
\newcommand\muF{\mu_{\scriptscriptstyle F}}
\newcommand\muR{\mu_{\scriptscriptstyle R}}
\newcommand\alphas{\alpha_{\scriptscriptstyle S}}
\newcommand\LambdaQCD{\Lambda_{\rm \scriptscriptstyle QCD}}
\newcommand\Cf{C_{\scriptscriptstyle F}}
\newcommand\Ca{C_{\scriptscriptstyle A}}
\newcommand\nf{n_f}


\mchapter{Introduction}
{Authors: Chiara Mariotti,
Ernesto Migliore and Paolo Nason}\label{chap:Intro}\label{cap:intro}
\vskip 0.3cm\noindent
{\it Revisors: Sara Bolognesi}
\vskip 1cm
\section{Physics at the Large Hadron Collider (LHC)}
\subsection{Introduction and basic references}
In this chapter we give a very condensed summary of LHC
physics. We assume that the reader has a basic familiarity
with the Standard Model of electroweak and strong interactions.
We summarize here some easily accessible basic references
to introductory material. In ref.~\cite{Altarelli:2000ma}
an introduction to the Standard Electroweak theory can be found,
together with a summary of precision tests, and hints on physics
bejond the Standard Model. A very basic introduction to the
theory of strong interaction can be found in
refs.~\cite{Nason:1997zu}. Summaries on the Electroweak and strong
interaction, as well as on experimental methods, can be found in the
reviews of the Particle Data Group, available at the URL
{\tt http://pdg.lbl.gov}.
\subsection{Why LHC?}
The Standard Model (SM) of electroweak interactions has been
established experimentally by the observation of neutral current
interactions in 1973 \cite{Hasert:1973ff} and of the W and Z bosons in 1983
\cite{Arnison:1983rp,Arnison:1983mk}.
From 1989 to 2000, the LEP and SLC experiments measured with a better than
per-mill precision the properties of the W and Z bosons: their masses,
their widths, their couplings with fermions and among themselves. 
These measurements were complemented by the Tevatron observation of the top
quark.
The piece of the SM which is still missing is the Higgs boson, which is
a remnant of the scalars that
provide masses to the particles.
Actually, the same precision measurements of the electroweak
observables hint to a light Higgs boson.
In fact the accuracy reached requires that, when relating them among
each other, genuine electroweak quantum corrections $\Delta r$
should be included, namely:
\begin{equation}
    m^2_W = \frac{\pi \alpha_{em}}{G_F\sqrt{2}} \frac{1}{sin^2\theta_W (1-\Delta r)}
\end{equation}
(see ref.~\cite{Altarelli:2000ma})
where the quantum corrections have a quadratic dependence on the mass of the top
quark $m_{top}$ and a logarithmic dependence on the mass of the Higgs:
$\Delta r=f(m_{top}^2, \ln m_H)$. 
With $m_W$, $m_{top}$ and $\sin^2\theta_W$ being measured, $m_H$ can be
extracted from a global fit of the electroweak observables.\\
On the other hand the lower limit on the Higgs mass from direct searches is
currently 114.4 GeV at 95\%{} confidence level. An upper limit on $m_H$ around 1.2 TeV is derived
within the SM requiring that the amplitude for the scattering of
longitudinally polarized vector bosons $V_L V_L\rightarrow V_L V_L$
does not violate unitarity.\\
The discovery of the mechanism which gives origin to the
masses requires the investigation of the energy range from 100 GeV to
1 TeV, and actually LHC has been designed as a discovery machine for processes with
cross-sections down to some tens of fb and in the energy range from 100
GeV to 1-2 TeV.
This physics goal influenced the main design parameters of the
machine:
\begin{itemize}
\item It is a hadron collider: the fundamental constituents entering
  in the scattering are the partons which carry a variable fraction $x$ of the beam 
  four-momentum. Therefore the centre-of-mass
  energy of the hard scattering process $\sqrt{\hat{s}}$ can span different orders of magnitude.
\item The centre-of-mass energy will be $\sqrt{s}$=14 TeV. In this way, incoming
  partons carrying momentum fractions $x_1,x_2 \approx 0.15-0.20$ of the incoming
  hadrons momenta, yield
  a partonic CM energy $\hat{s}=x_{1}x_{2}s \approx 1-2\;$TeV, the energy range
  one wants to explore.
\item It is a proton-proton collider since it is difficult to accumulate high intensity beam 
  of anti-protons. Furthermore,
  the Higgs production process is dominated by gluon fusion, and therefore its cross section
  is nearly the same in proton-antiproton and proton-collision.
\item The time interval between consecutive bunch crossing will be 25
  ns, as the luminosity depends linearly on the bunch crossing
  frequency. The very short bunch crossing interval and the high
  number of bunches accelerated by the machine (2808 per beam) will
  allow to reach the peak luminosity of 10$^{34}$ cm$^{-2}$s$^{-1}$.
\end{itemize}
With respect to an electron-positron machine, it is easier to accelerate protons
to high energy since the energy lost for synchrotron radiation, proportional to $\gamma^4$ (where $\gamma=E/m$)
is much lower than for the electrons.
On the contrary precision measurements are more difficult. Since the
kinematics of the initial state of the hard process can change from
event to event, it is possible to have more than one fundamental
interaction per bunch crossing, with the fragments of the protons mixing
with the products of the hard process in the final state.

The luminosity delivered by LHC during the first 3 years will be of $L_{int}=20$ fb$^{-1}$ per year or 
$L=2 \times 10^{33}$ cm$^{-2}$s$^{-1}$, and later
$L_{int}=100$ fb$^{-1}$ per year or $L=10^{34}$ cm$^{-2}$s$^{-1}$.\\
Two general-purpose experiments are in construction: ATLAS (A Toroidal Lhc ApparatuS)
and CMS (Compact Muon Solenoid). A B-physics dedicated experiment (LHCb) is also 
in preparation, and is dedicated to the study of b-hadrons produced at small angle
in p-p collisions at low luminosity.  Finally, the LHC will also provide Pb-Pb collision
with $\sqrt{s}=1312$ TeV and a luminosity of $L=10^{29}$ cm$^{-2}$s$^{-1}$. The ALICE 
experiment will be devoted to study these collisions.

\subsection{The ATLAS and CMS physics program}
The main goals of the two general purpose experiments are:
\begin{itemize}
\item study the mechanism that breaks the symmetry of the SM
  Lagrangian giving rise to the particle masses. \\
  Whitin the SM this means to search for
  the SM Higgs boson from $m_H$=100 GeV to $m_H$=1 TeV. If the Higgs is found, understand if it is a SM Higgs or a
  SUSY Higgs; if the Higgs is not found, look for alternative models.
\item search for new physics, especially if the Higgs is not found. \\
  Concerning supersimmetry, all the s-particles with mass $m_{\tilde s} \le$ 3 TeV will be accessible.
  For exotic models (lepto-quark, technicolor, new strong interaction, new lepton families, additional
  bosons, extra-dimensions \ldots) the mass reach is 5 TeV.
\item perform precision measurements in the electroweak sector ($m_W$,
  $m_{top}$, triple gauge couplings, $\sin^2\theta_W$), in QCD, and in the CP violation
  and B physics sector.\\
  Concerning the precision electroweak measurements, it should be
  noted that, in order to have a comparable impact in the determination
  of the Higgs mass from the fit of the electroweak observables, 
  the top mass and the W mass should be measured with a relative precision given by:
  \[
  \Delta m_W = 0.7 \times 10^{-2} \Delta m_{top}
  \]
  Therefore the target precision on these quantities will be $\Delta
  m_W \le$ 15 MeV and $\Delta m_{top} \le$ 2 GeV. These precisions will not be trivial to achieve,
  since at a hadron collider the initial state of the parton-parton collision is not
  well known, and the final state is complicated by the presence of many other produced particles.

\end{itemize}

\begin{figure}[hbtp]
  \begin{center}
    \includegraphics[width=\textwidth]{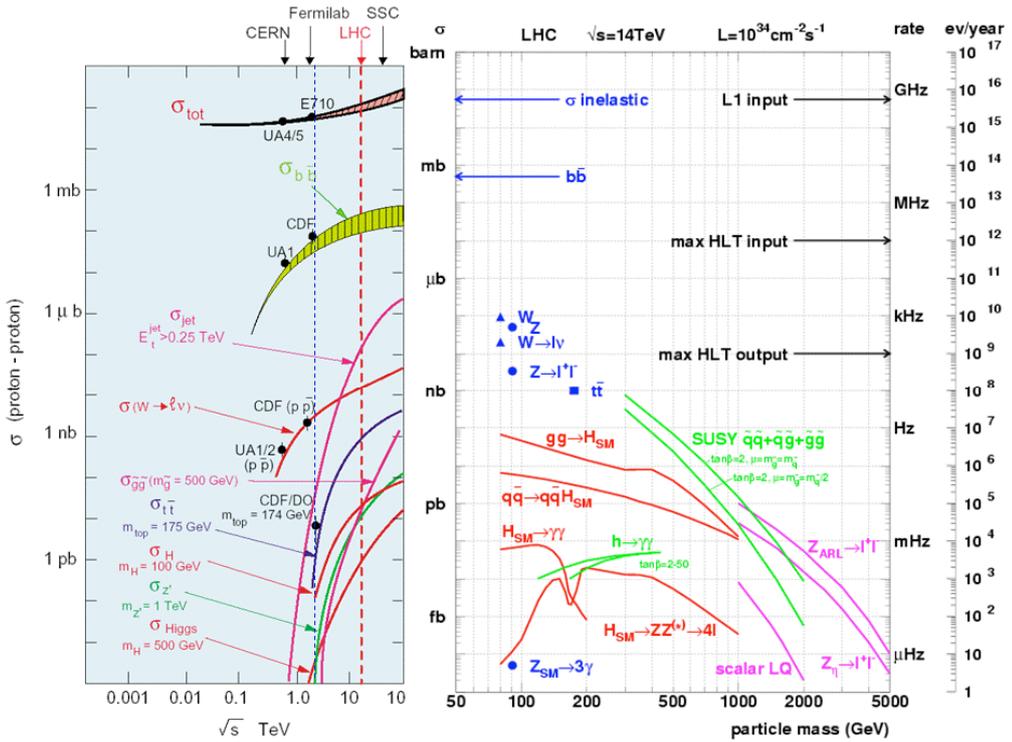}
    \caption{Cross-section as a function of the centre-of-mass energy for
   interesting processes, and the rate of events at LHC.}
    \label{fig:sigma_pp}
  \end{center}
\end{figure}

In figure \ref{fig:sigma_pp} are shown some of the cross-sections (as a function of the
centre-of-mass energy and the rate of production) of interesting processes.
In table \ref{tab:sigma_pp} we report the cross-section and the number of events produced
per experiment for a given process, for low luminosity ($L=2 \times 10^{33}$ cm$^{-2}$s$^{-1}$).

\begin{table}
\begin{center}
\begin{tabular} {|c|c|c|c|c|}
\hline \hline
Process & $\sigma$ & Events/sec & Events/year& Other machine \\
\hline \hline
$W \rightarrow e \nu$ & 20 nb & 15 & 10$^8$ & 10$^4$ LEP / 10$^7$ Tevatron \\
$Z \rightarrow e e$ & 2 nb & 1.5 & 10$^7$ & 10$^7$ LEP  \\
$t \bar t$ & 1 nb & 0.8 & 10$^7$ & 10$^5$ Tevatron \\
$b \bar b$ & 0.8 mb & 10$^5$ & 10$^{12}$ & 10$^8$ Belle/BaBar \\
$\tilde g \tilde g$ (m = 1 TeV)& 1 pb & 0.001 & 10$^4$ & \\
H (m = 0.8 TeV) & 1 pb & 0.001 & 10$^4$ & \\
H (m = 0.2 TeV)& 20 pb & 0.01 & 10$^5$ & \\ 
\hline 
\end{tabular}
\caption{Expected cross-sections and number of events per second and 1 year for one of the
experiments at LHC}
\label{tab:sigma_pp}
\end{center}
\end{table}


\section{The theory of Hadronic collisions}
\subsection{Hadron collider kinematics}
A convenient set of kinematic variables for
particles produced in hadronic collisions is the transverse momentum $\pt$, the rapidity $y$
and the azymuthal angle $\phi$. Assuming that the collision axis is the third axis, in the
CM frame of the collision, for a particle with energy $E$ and three momentum $\vec{p}=\{p_1,p_2,p_3\}$
we write
\begin{equation}
\pt=\sqrt{p_1^2+p_2^2}\,,\quad y=\frac{1}{2}\ln\frac{E+p_3}{E-p_3}\,,\quad
p_1=\pt\cos \phi\,,\quad p_2=\pt\sin \phi\;.
\end{equation}
These variables have simple transformation properties under longitudinal boosts
(i.e. boosts along the beam line direction), $\pt$ and $\phi$ being invariant, and
\begin{equation}
y\Longrightarrow y+\frac{1}{2}\ln\frac{1+\beta}{1-\beta}\,,
\end{equation}
where $\beta$ is the boost velocity along the third direction. The energy and the longitudinal
component of the momentum of a particle have the expression
\begin{equation}\label{eq:Ep3mty}
E=\mt\cosh y\,,\quad\quad p_3=\mt\sinh y\,,
\end{equation}
where $m$ is the mass of the particle, and $\mt=\sqrt{m^2+\pt^2}$ is called the \emph{transverse mass}.\footnote{
In $W$ mass measurements at hadronic colliders the term ``$W$ transverse mass'' is used to denote
the observable $m_W^T=\sqrt{2 p_T^\nu p_T^l (1-\cos\Delta\phi)}$, and has a totally different meaning
from the one introduced here.}
One usually refers to the regions
$y\gg 0$, $y\ll 0$ and $y\approx 0$ as to the forward, backward and central region.

Observe that the single particle invariant phase space is written in terms of rapidity and transverse
momentum as
\begin{equation}
\frac{d^3 p}{2 E (2 \pi)^3}=\frac{1}{2 (2 \pi)^3} d^2\pt\, dy\,,
\end{equation}
and is thus flat in rapidity. Furthermore, the cross section for the
production of an object of mass $M$, for not too extreme
values of the mass and rapidity, is typically flat in rapidity. This can be seen as follows.
The production cross section is proportional to
\begin{equation}\label{eq:diffsig}
 d x_1 d x_2 f(x_1) f(x_2) \delta(s x_1 x_2 - M^2),
\end{equation}
where $f$ are the parton densities and $x_1,x_2$ are the momentum fraction of the
incoming partons.
The rapidity of the produced particle is given by $y=\frac{1}{2} \log(x_1/x_2)$.
Defining $\tau = x_1 x_2$ one can easily show that
\begin{equation}
dx_1 dx_2 = d \tau\, d y\,.
\end{equation}
Assuming\footnote{This is the typical small-$x$ behaviour of parton densities,
with $\delta \approx 0.1 \div 0.5$.}
that $f(x)\approx 1/x^{1+\delta}$, we see that formula~(\ref{eq:diffsig})
yields a flat rapidity distribution.
This of course holds as long as the rapidity is not close to its maximum value
$\ln \frac{\Ecm}{M}$. Thus the typical rapidity distribution has a bell shape with
a central plateau, the plateau becoming wider as $M$ becomes small.

The use of boost invariant variables facilitates the description of particle production in hadronic
collisions, since these phenomena are approximately boost invariant for not too extreme values of
rapidity. This fact is particularly simple to understand for high energy scattering phenomena,
where the incoming hadrons behave as beams of quark and gluons, with a given distribution in longitudinal momenta and limited transverse momentum. It is clear that, depending upon the energy of the incoming constituents,
the same hard scattering phenomenon can take place with an effective center of mass (i.e. with a center of mass
for the incoming constituents) that is moving along the collision direction.

Experimentally, it is more convenient to use the
pseudorapidity, rather than the rapidity. It is defined as
\begin{equation}
\eta=\frac{1}{2}\ln\frac{|\vec{p}|+p_3}{|\vec{p}|-p_3}=-\ln\tan\frac{\theta}{2}\,,
\end{equation}
where $\theta$ is the angle of $\vec{p}$ with respect to the positive 3 direction. Being only a function
of the angle, pseudorapidity is much easier to measure than rapidity, and for ultrarelativistic
particles it coincides with rapidity. The analogue of eqs.~(\ref{eq:Ep3mty}) for pseudorapidity
are
\begin{equation}\label{eq:pp3mteta}
|\vec{p}|=\pt\cosh \eta\,,\quad\quad p_3=\pt\sinh \eta\,.
\end{equation}

The particles distribution in hadronic collisions are often represented as a 3-dimensional histogram
on a rectangle in the $\eta$, $\phi$ plane
(the range in $\eta$ being dependent upon the detector's capability) obtained by cutting the $\eta$,
$\phi$ cylinder surrounding the colliding beams, and centered at the collision point, along the $\phi=\pi$
line. On this plane, one also defines a distance $\Delta R=\sqrt{\Delta \eta^2+\Delta \phi^2}$.

\subsection{Orders of magnitude}
\begin{figure}[htb]
\begin{center}
\epsfig{file=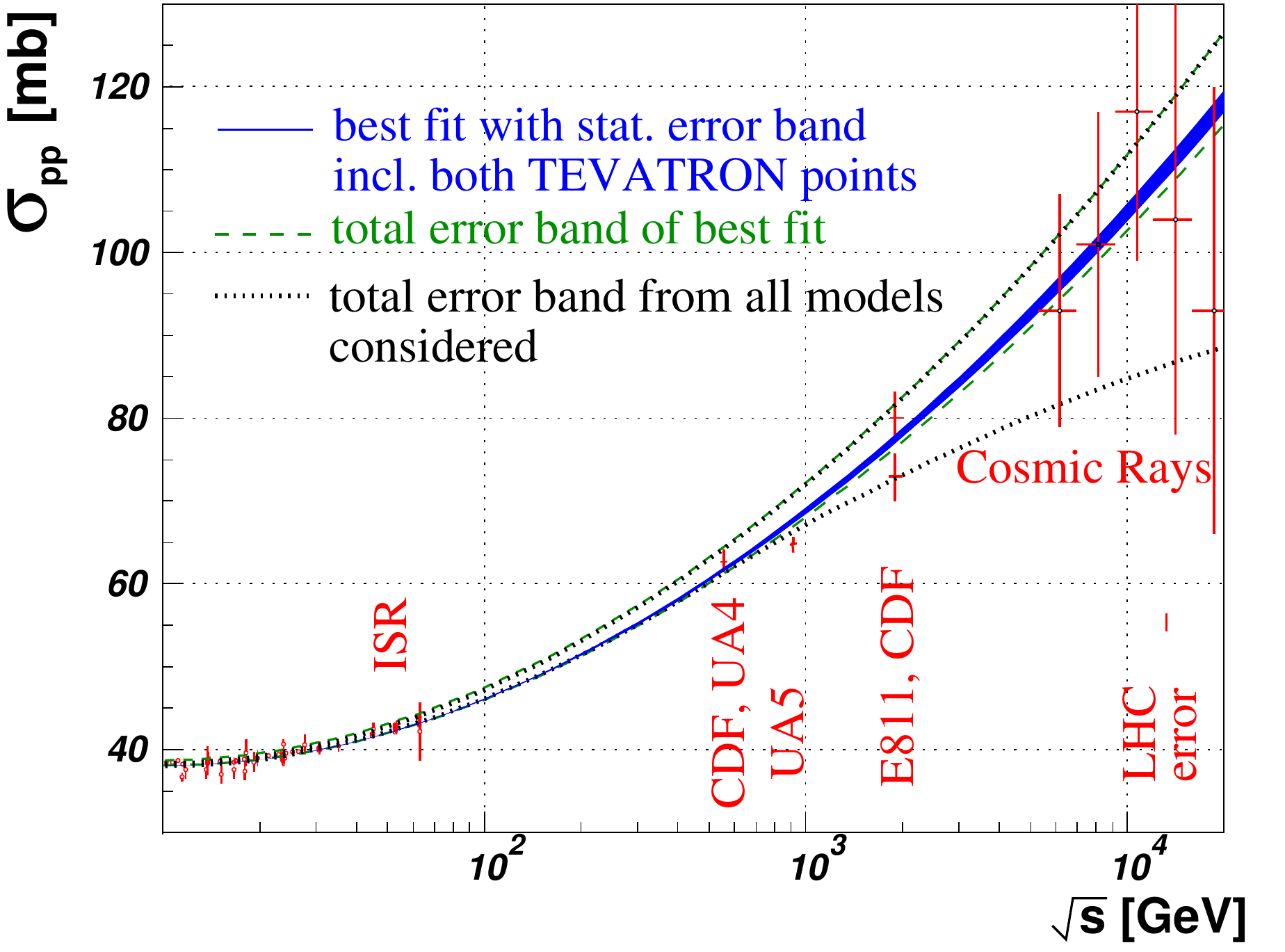,width=0.7\textwidth}
\end{center}
\caption{\label{fig:sigtot}
Total cross section for $pp$ and $p\bar{p}$ collisions, from ref.~\cite{Deile:2006tt}.}
\end{figure}

The total cross section for proton-proton collisions is in the range of few tens
of millibarns ($1\,{\rm mb}=10^{-31}\,{\rm m}^{2}$). This is consistent with a transverse size
of a hadron of the order of a femtometer ($1\, {\rm fm}=10^{-15}\,{\rm m}$), a ${\rm fm}^2$ being equal to
10~mb. The cross section grows slowly with energy, so that projected values at LHC
energies are around 100 mb, as one can see from fig.~\ref{fig:sigtot}.

\begin{figure}[htb]
\begin{center}
\epsfig{file=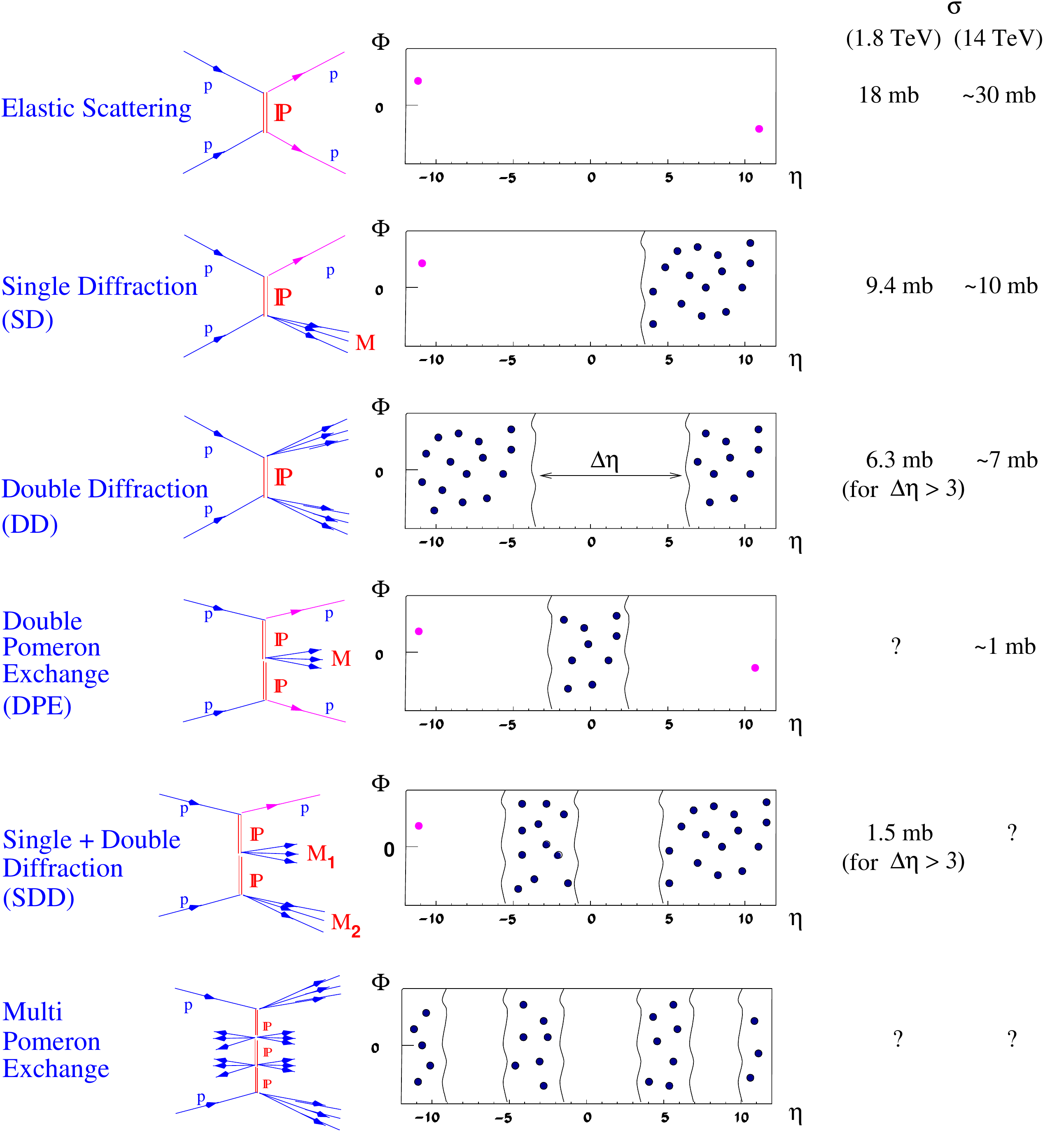,width=0.7\textwidth}
\end{center}
\caption{\label{fig:eventclasses}
Elastic and diffractive processes classes, their cross section at Tevatron energy, and their expected
cross section at LHC energy, from ref.~\cite{Deile:2006tt}.}
\end{figure}

The total inelastic cross section is mainly made up of events with low $\pt$ hadrons in the final state.\footnote{
These events are often called \emph{minimum bias events}. The term originates from the so called minimum bias
trigger selection, that was the trigger using the less stringent criteria, thus yielding the least biased
event sample. In older experiments, these events had typically some particles in the central region.
Today, more detailed definitions are used, and we refer the reader to sec. \ref{chap:MBandUE} for further discussion.}
These events are sometimes classified according to the distributions of the
produced particles in the available rapidity (or pseudorapidity) interval.
If there are empty gaps, inelastic events are classified as diffractive events, as shown
in fig.~\ref{fig:eventclasses}. The value of their measured cross
section at Tevatron energy, and their extrapolated cross section al LHC energy is also shown in the figure.
Most of the total cross section is composed by events with no evident gaps.
They are characterized by transverse
momenta of the produced particles of the order of few hundred MeV, and particle multiplicity (that is to say,
number of produced particles) of order of a few per unit of
rapidity, with fluctuations of the order of 100\%. A more detailed description of typical
inelastic events will be given in chapter~\ref{chap:MBandUE}.

\begin{figure}[htb]
\begin{center}
\epsfig{file=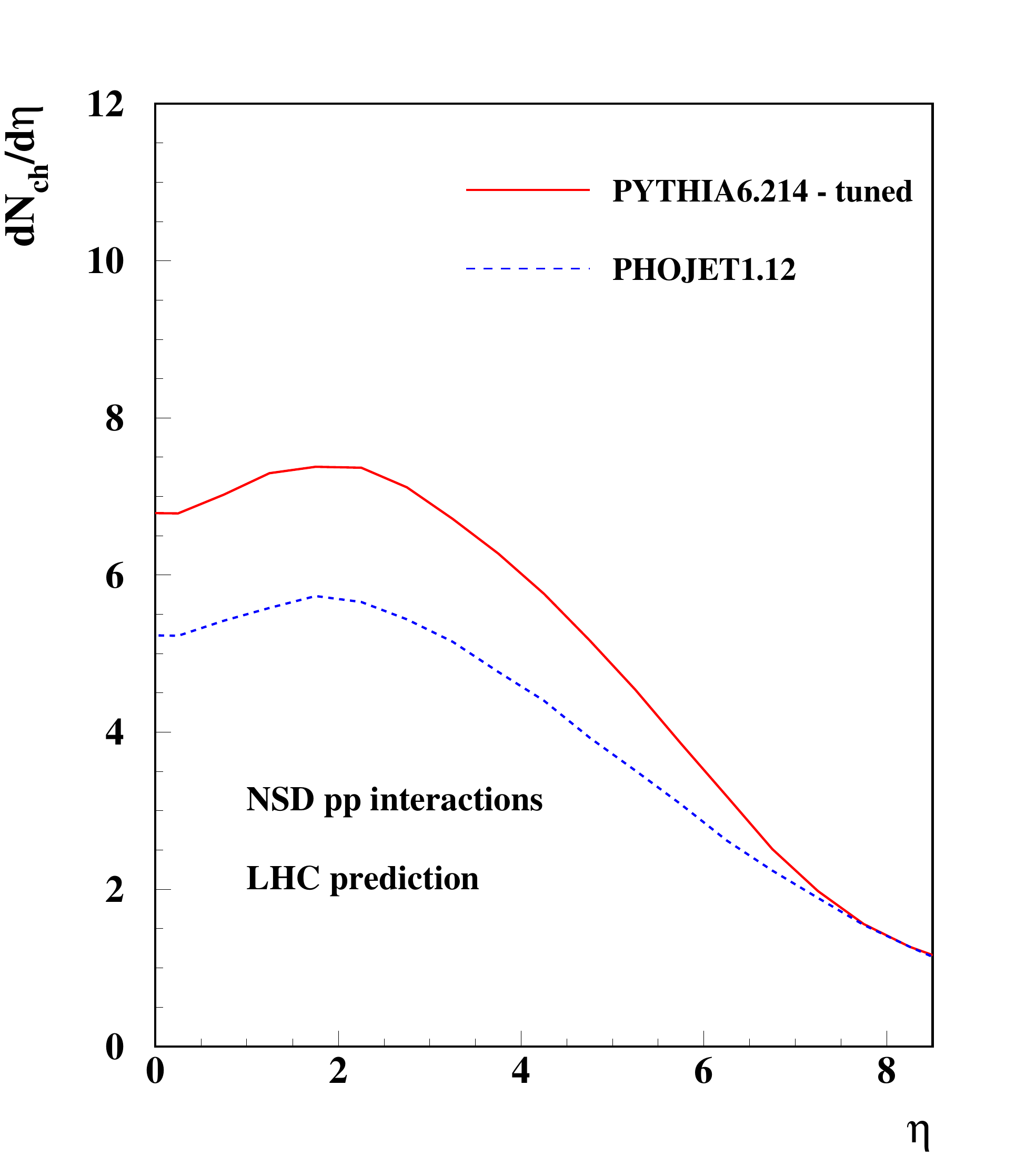,width=0.7\textwidth}
\end{center}
\caption{\label{fig:nch14tev}
The average number of charged particles per unit of rapidity at the LHC from ref.~\cite{Dobbs:2004bu}.}
\end{figure}

An extrapolation of the average number of charged particles per unit of pseudorapidity to LHC energy is shown in
fig.~\ref{fig:nch14tev}.
Observe that the distributions in fig.~\ref{fig:nch14tev} have a dip at $\eta=0$. This is a kinematical
effect, due to the use of pseudorapidity instead of rapidity. Rapidity distributions would instead
be flat in the small rapidity region, thus showing the boost invariant feature of particle
production in hadronic collisions.\footnote{From eqs.~(\ref{eq:Ep3mty},\ref{eq:pp3mteta}),
we have $\sinh\eta = \sinh y \,\mt/\pt$, showing that, when mass effects are not negligible,
for small $\eta,y$ pseudorapidity intervals are wider than the corresponding rapidity intervals.}

The main purpose of high energy colliders is to study scattering
phenomena with very large momentum transfer, or production processes
of very massive, pointlike particles. These processes are collectively
called ``hard scattering'' phenomena. In these cases, incoming hadrons
can be regarded as beams of pointlike constituents. The cross sections
for constituents hard scattering are much smaller than the total
hadronic cross section. They are of the order of the squared inverse of
the typical energy scale entering the process, sometimes called the
hardness of the process, times a dimensionless coupling
constant. Thus, for example, the cross section for $t\bar{t}$
production is of order\footnote{The notation $\alpha_S(t)$ stands for the
running QCD coupling constant at the scale $t$.}
$\alpha_S^2(M_t^2)/M_t^2$, corresponding
to about $10^{-7}$~GeV$^{-2}$ or $10^{-7}$~mb. This is to be
compared with the 100~mb cross section for a typical low transverse
momentum reaction, the ratio being $10^9$.  We notice that an estimate
of the cross section for jet production is given by
$\alpha_S^2(p_T^2)/p_T^2$, where $p_T$ is the transverse momentum of
the jet. For $p_T$ of the order of few hundreds MeV, the strong
coupling becomes of order one, and thus this cross section becomes of
the same order of the typical low transverse momentum reactions.

\section{QCD description of hadronic collisions}
\subsection{Hard production processes}
Inclusive cross sections for hard production processes are calculable in terms of the
so called QCD-improved parton model formula
\begin{equation}\label{eq:QCDPMformula}
\sigma_{H_1 H_2}(p_1,p_2)=\sum_{ij}\int dx_1 dx_2\, f^{H_1}_i (x_1,\muF)\,f^{H_2}_j (x_1,\muF)
\,\hat{\sigma}_{ij}(\muF,\muR)(x_1 p_1, x_2 p_2)\,.
\end{equation}
Here $\sigma$ represents the cross section for some hard phenomenon to take place (for example,
the production of some heavy particle, eventually with some kinematics constraints, the production
of jets with a large transverse momentum, etc.).
The labels $i$ and $j$ run over all quarks, antiquarks, and the gluon.
The parton densities $f^H_i(x,\muF)$ represent the probability to find the constituent $i$ inside
the hadron $H$, carrying a fraction $x$ of its momentum. The scale $\muF$ must be chosen to be of the
order of the typical hard scale entering the process (like, for example, the mass of the $W$ in $W$ production,
the transverse momentum of the jet in jet production, etc.).
The so called ``short distance'' or ``partonic'' cross section $\hat{\sigma}_{ij}$ is calculable
in perturbation theory as an espansion in powers of the strong coupling constant $\alphas$, evaluated at a scale
$\muR$ also of the order of the typical hard scale entering the process.

Formula (\ref{eq:QCDPMformula}) cannot be used to predict the
detailed distribution of all final state particles. It must instead refer to a sum of final states,
that is to say it must be \emph{inclusive} to some degree. First of all, the formula does not say anything about
the fate of the remnants of the incoming hadrons after the collisions.
Beside this, certain characteristics of infrared insensitivity
should be requested to the cross section in order for it to be calculable in QCD. In other words,
our cross section should not depend upon the infrared cutoffs that we use in our calculation.
In general,
the differential production cross sections for heavy particles satisfy these requirements. Not so for massless
coloured particle production. High transverse momenta coloured partons in the final state materialize as
jets, that is to say as sprays of relatively collimated particles. Rather than the kinematical distribution
of final state particles, it is the jet kinematics that is calculable. Thus, the cross section must be inclusive
also in the composition of a jet, in that it must include all final states that have the same jet structure.
Finally, one cannot forbid QCD radiation in any angular region
of phase space, in order for the cross section to be calculable, that is to say, as in QED, one must
require a minimal energy resolution of the measuring apparatus in order to obtain sensible results. This
last point is basically related to the requirement of cancellation of soft divergences.

\begin{figure}[htb]
\begin{center}
\epsfig{file=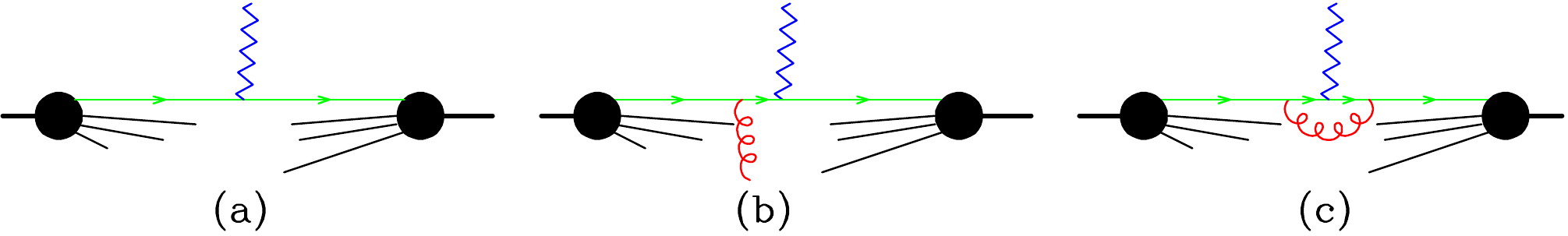,width=0.9\textwidth}
\end{center}
\caption{\label{fig:wprodexa}
$W$ production in hadronic collisions.}
\end{figure}

To better clarify the meaning of
formula (\ref{eq:QCDPMformula}) and the origin of its limitations, we will now
consider the example of $W$ production in hadronic collisions, depicted in fig.~\ref{fig:wprodexa}.
The diagram (a) represents the Born level cross sections, while (b) and (c) are next-to-leading order
(NLO from now on) contributions.
The Born result represents the cross section for the production of a $W$ with given kinematic
properties, plus anything else. The next-to-leading contributions (b) and (c) represent corrections
to (a) that are formally of order $\alphas$, but receive large contributions in particular regions of phase
space. In particular, if the outgoing gluon momentum is parallel to one of the incoming quarks.
\subsection{Collinear singularities}
\begin{figure}[htb]
\begin{center}
\epsfig{file=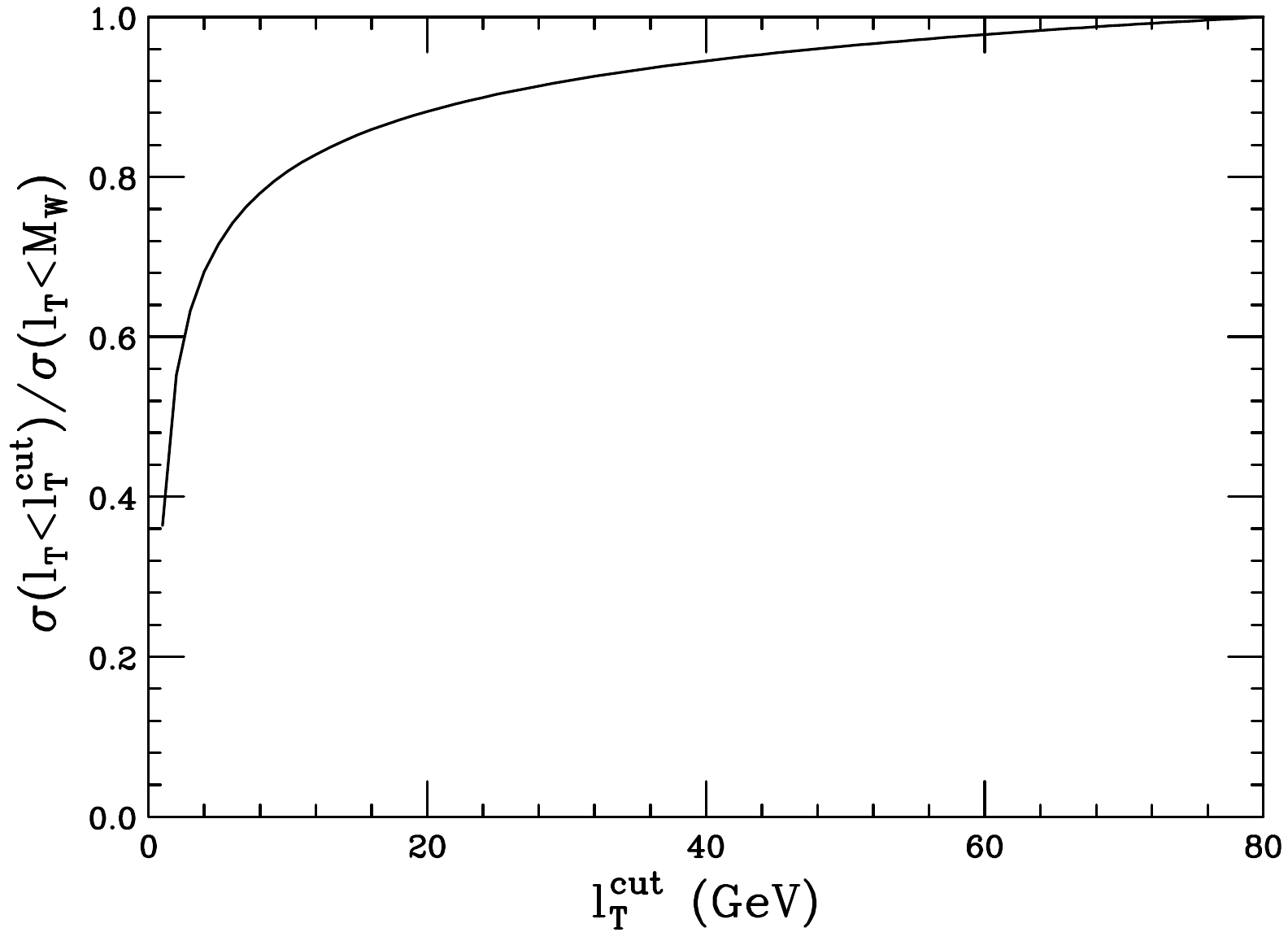,width=0.5\textwidth}
\end{center}
\caption{\label{fig:ptgluon}
Relative probability for the emission of a gluon with transverse momentum below a given cut,
according to eq.~(\ref{eq:ptprob}).}
\end{figure}

Let us briefly review what happens when a gluon is emitted collinearly by an initial state
parton (like the incoming quark in our example).
In lowest order perturbation theory, the emission cross section factorizes as
\begin{equation}\label{eq:isrcoll}
d\sigma_{q\bar{q}\to W+g}(p_q,p_{\bar{q}},p_W,l)
\stackrel{\lt\to 0}{\longrightarrow} \frac{\alphas(\lt^2)}{2\pi}\,P_{gq}(z)\,dz\frac{d\lt^2}{\lt^2}
d\sigma_{q\bar{q}\to W}((1-z)p_q,p_{\bar{q}},p_W)\,,
\end{equation}
where $l$ is the gluon momentum, and we have assumed $\vec{l}=z\vec{p}_q+\vec{l}_T$, i.e. we have decomposed
the gluon momentum into a component parallel to the initial state parton $z\vec{p}_q$,
and a component orthogonal to it, $\vec{l}_T$.
The collinear limit
is reached when $\lt\to 0$, so that the direction of the incoming quark and of the emitted gluon coincide.
The cross section becomes the product of the $W$ production cross section with a reduced momentum for the
incoming quark, times a splitting probability, proportional to the strong coupling constant
evaluated at the characteristic transverse momentum of the splitting process.
$P_{gq}(z)$ is the Altarelli-Parisi splitting function for finding a gluon in a quark
with a fraction $z$ of its momentum
\begin{equation}
P_{gq}(z)=\Cf\frac{1+(1-z)^2}{z}\,,
\end{equation}
and it describes the distribution of the splitting probability as a function of the fraction of longitudinal
momentum.
Of course, formula~(\ref{eq:isrcoll}) makes no sense for $\lt$ too
near $\LambdaQCD$ ($\alphas$ becomes too large and perturbation theory
can no longer be applied), and for $\lt$ of order $M_W$ (factorization only works for small $\lt$).
We can however use it to estimate the
order of magnitude of the probability for the emission of an extra
gluon. Using the one loop formula for the strong coupling constant
\begin{equation}
\alphas(\lt^2)=\frac{1}{b_0\,\ln\frac{\lt^2}{\LambdaQCD^2}}\,,\quad\quad b_0=\frac{33-2\nf}{12\pi}\,,
\quad\quad \nf=\mbox{number of light flavours,}
\end{equation}
we can estimate the size of the contribution of one extra collinear emission in formula (\ref{eq:isrcoll})
by performing the $dl_T^2$ integration
\begin{equation}\label{eq:ptprob}
\int_{l_T^{\rm min}}^{l_T^{\rm max}}\alphas(\lt^2) \frac{d\lt^2}{\lt^2} = \frac{1}{b_0}\,
\ln\frac{\ln\frac{\lt^{\rm max}}{\LambdaQCD}}{{\ln\frac{\lt^{\rm min}}{\LambdaQCD}}}
\end{equation}
Assuming that $\lt^{\rm min} \gtrsim \LambdaQCD $ and $\lt^{\rm max}$ of the order of
the transverse mass of the $W$, 
we see that the right hand side of eq.~(\ref{eq:ptprob}) is not small, i.e. is not suppressed
by a power of the strong coupling at the hard scale.
However, most of the contribution to the integral take place at low $\lt$, as can be seen from fig.~\ref{fig:ptgluon},
so that the probability to emit a hard gluon is indeed of order $\alphas(\lt^{\rm max})$.
This reasoning can be extended for the emission for any number of gluons, with
$\lt^{\rm max}\gg {\lt}_1 \gg {\lt}_2 \ldots\gg \lt^{\rm min} $. We would find that also multiple emissions have
probability of order 1. The presence of these large contributions due to collinear parton
emissions are among the reasons why the detailed final state structure is not described
by formula (\ref{eq:QCDPMformula}). In fact, the sum of all
these large corrections (and of the associated large corrections present in the
virtual contribution of fig~\ref{fig:wprodexa} $(c)$) is already included in eq.~(\ref{eq:QCDPMformula}), since
the scale dependent parton densities $f_i(x,\muF)$ include the effect of all emissions with $l_T$ up to
$\muF$. Thus, if $\muF \approx M_W$, the only left over corrections in eq.~(\ref{eq:QCDPMformula})
are genuine strong radiative corrections, suppressed by powers of $\alphas(M_W)$. We must however assume that
we sum over all final state configurations that have any number of collinear partons in the
direction of the incoming quark or antiquark.

In processes with a massless parton with large transverse momentum in the final state, one also expects large
corrections due to radiation collinear to it. The factorization formula is similar to eq.~(\ref{eq:isrcoll}).
Consider the example of diagram (b) in fig~\ref{fig:wprodexa}, under the assumption that the final state
gluon has large transverse momentum. The gluon can split into a pair of collinear gluons, with cross section
\begin{equation}\label{eq:fsrcoll}
d\sigma_{q\bar{q}\to W+g+g}(p_q,p_{\bar{q}},p_W,p_g,l)
\stackrel{\lt\to 0}{\longrightarrow} \frac{\alphas(\lt^2)}{2\pi}\,P_{gg}(z)\,dz\frac{d\lt^2}{\lt^2}
d\sigma_{q\bar{q}\to W+g}(p_q,p_{\bar{q}},p_W,p_g)\,,
\end{equation}
where now $\vec{l}=z\vec{p}_g+\vec{l}_T$, and $\vec{l}_T$ is transverse with respect to the direction of the outgoing
gluon. One noticeable difference between eq.~(\ref{eq:fsrcoll}) and eq.~(\ref{eq:isrcoll}) is that
in the former case gluon radiation reduces the momentum entering in the hard cross section.
For final state splitting, instead, the momenta flowing through the hard scattering process
are unchanged. These large final state radiation corrections, in fact, cancel out when virtual
corrections are included. Virtual corrections do not change the structure of the final state,
since they correspond to the emission and reabsorption of a virtual parton.
Thus, the cancellation works only if one is allowed to sum contributions with different final state structure.
In other words, rather than requiring a given
light parton in the final state, we should require the presence of a jet, that is to say of a
bunch of collimated particles.
\subsection{Soft singularities}
Soft gluon emission also gives rise to divergencies of the form $dl^0/l^0$, where $l^0$ is the gluon energy.
These divergences are present whenever coloured particles are involved in a reaction, and are analogous to
the soft divergences in electrodynamics. In electrodynamics, soft singularities are a basic consequence
of quantization. In fact, the brehmmstralung frequency spectrum $dE/d\nu$ is known to approach a constant
for small frequencies. Because of quantization, $dE= h\nu\,dn$, so that, if $dE/d\nu$ is a constant,
$dn/d\nu\propto 1/\nu$, that is to say it diverges at low frequencies. Because of this fact, QED radiative
corrections have always infrared divergences related to (both real and virtual) emission of soft photons.
It turns out, however, that if one assumes that there is a finite energy resolution for soft
photons (i.e., that one cannot detect photons with energy below a given cut $E_{\rm soft}$), soft
divergences between real and virtual corrections cancel out, yielding a finite result.
The same thing happens in QCD. Because of this virtual-real infrared cancellation, it also turns out
that we cannot totally veto radiation in any angular region of the phase space. This veto would affect
the soft gluon emission contribution to the cross section, but would leave the corresponding virtual
correction unchanged, thus spoiling the cancellation.

If massless coloured particles are present in a reaction,
collinear and soft singularities are present at the same time. In particular, in formula (\ref{eq:isrcoll}),
since $P_{gq}(z)=\Cf (1+(1-z)^2)/z$, we see that the $z$ integration yields a $dz/z$ integral, that corresponds
(when $\lt$ is already small) to the gluon energy becoming small. The region where a gluon
becomes both collinear and soft is sometimes referred to as the double logarithmic region, because
of the presence of two factorized logarithmic singularities.
\subsection{Jet definitions}
We have seen that we cannot predict the distribution of individual massless partons in the final
state, but we should instead talk about jets. There is much freedom in the way jets are defined.
In order for a jet cross section to be calculable in perturbation theory, it should be insensitive
to the splitting of a massless particle into two collinear partons, in such a way that the cancellation
between real and virtual contributions to the cross section can take place. A typical jet definition
may deal with the total amount of energy flowing from the interaction vertex into a cone of fixed
aperture. A jet definition must be collinear insensitive in order to be calculable in perturbation
theory.
This is however not enough. We cannot characterize the final state by requiring that
all final state particles lie inside cones of given aperture. In fact, because of soft divergences,
we cannot require that there are regions in solid angle where there are no particles at all.
This requirement would spoil the cancellation of infrared divergences. We should thus allow a limited
fraction of the total energy to be present in arbitrary directions outside the cones (the so called
Sterman-Weinberg jet definition), or we should instead deal with inclusive jet cross section,
meaning of a given number of jets above a given energy plus an arbitrary number of jets of smaller energy.
\subsection{Fixed order calculations}
The QCD improved parton model formula, eq~(\ref{eq:QCDPMformula}), applied at any given finite order in the
strong coupling constant, can be used to compute certain cross sections, i.e. inclusive cross
section for the production of heavy particles and/or jets. For these quantities, soft singularities cancel
(and so do the collinear final state singularities) between real and virtual contributions. Initial state
collinear singularities survive, but are subtracted and absorbed into the scale dependent parton densities.
At the Born level, one only considers the cross section for the production of a given number of heavy
particles, and of well separated massless partons. At this level, one associates a jet to each
massless parton. At the NLO level, corrections due to the virtual exchange of a massless parton, or to its
emission, are included. At this level, it is possible to have jets containing more than one massless parton,
since the radiated particle can be collinear to any other massless parton in the process.
The emitted parton can be soft, and their emission cross section diverges in the soft limit.
These divergences cancel among real and virtual graphs. Fixed order calculations are the backbone of
our theoretical ability to predict hard scattering cross sections. Automatic algorithms exist that
allow one to compute arbitrary Born level cross sections, the only limitation being the computer
time for very complex processes. For many processes of interest, the NLO corrections have also been
computed, within a framework to deal with collinear and soft singularities in a consistent way.
Since a few years, some results at NNLO level have also become available.

\subsection{Shower Monte Carlo}
From the discussion carried out so far, it is clear that the QCD parton model formula alone does not
fully describe the final state in hadronic collisions. First of all, we lack a practical way of
computing QCD at low energy, and thus we are incapable to describe the formation of a final state
made up of hadrons. But, even if we put aside this problem,
we have seen that the formation of a specific final state involves
an arbitrary number of collinear (and soft) emissions, all of them contributing corrections of order one.
These corrections cancel for inclusive observables. Sadly, a specific final state is not inclusive by
definition, and so fixed order calculations cannot predict its probability.
There are, however, algorithms that are capable to resum to all order in perturbation theory all most
important real and virtual emission corrections (namely the corrections that are collinear divergent),
the so called \emph{shower} algorithms. 
These algorithms are thus capable to associate with a given hard event an arbitrary number of accompaning
partons. In order to yield a finite result, they must contain an explicit cut-off on the transverse momentum
of emitted partons, and on their energy. The final state they generate is still unphysical, since it is
made up by quarks and gluons, rather than hadrons. In order to generate a physical final state,
phenomenological models of hadron formation are used. These models are not soundly motivated from a
theoretical viewpoint. However, it is also true that the same models should be applicable to any hard collision,
since the part of the collision that involves distances below the typical hadronic scale is well described by
the perturbative QCD formalism embodied in the computation of the short distance cross section and of the shower
development.
Shower Monte Carlo programs are thus event generators that can model the full final state. They have been
successfully employed in several experimental frameworks, like electron-positron, lepton-hadron, and
hadron-hadron collisions, at several different energies. We expect that they should work with
some tuning also for LHC.
\subsubsection{Underlying event}
The QCD picture of an event, as represented by a shower Monte Carlo and a hadronization model, still misses one
ingredient for the description of the full final state. In fact, at the end of the shower, the system
is represented by two incoming partons and a number of final state partons. Given that the fragmentation
process turns final state partons into real hadrons, we still have to specify how to treat the remnant of the
two initial hadrons that have provided the initial state partons. These remnants carry energy, momentum,
flavour and colour, so their treatment is indispensable for a realistic description of the event.
We are unable, at present, to describe this aspect of the process within QCD, and thus
we have to resort to a model, the so called ``Underlying event model''. One very crude model would  be to
include the hadron remnant in the final state (for example, if the incoming parton is a quark, the hadron
remnant is a di-quark), and provide a mechanism for the hadronization of the remnant (that will be typically
correlated in colour to some final state partons) in the fragmentation stage of the shower.

The name ``Underlying event'' is somewhat unfortunate, since it gives the false impression that there
is some uncorrelated activity accompaning any hard scattering process.
It is also used with some ambiguity in the literature. In fact, there is no precise way to separate
the underlying event from the hard process.
In the framework of the Event Generators, the underlying event model
(in general made up of several components) describes the physics of the hadronic
remnants. When fitting the parameters
of these components, one typically looks at regions of phase space where the influence of the hard process
is as small as possible, like for example angular regions in the azimuth $\phi$ that are as far as
possible from the jets in jet pair production. In case of production of colour neutral particles,
like the $W$ or the $Z$, after the removal of the $W$ or $Z$ decay products from the final state, the distribution
of all remaining particles should be strongly dependent upon the ``Underlying event'' model.

\subsubsection{Multiple parton interactions}
The physics of the hadron remnants may influence
in several ways the formation of the final state, and not only in the very forward direction.
First of all, the remnants are coloured, and thus can radiate
soft gluons. They can also give rise to secondary parton interactions. Thus, for example, in our $W$ production
process, another pair of constituents from the incoming hadron can collide, and generate two balanced
jets that accompany the $W$, or they can even produce another $W$.
\begin{figure}[htb]
\begin{center}
\epsfig{file=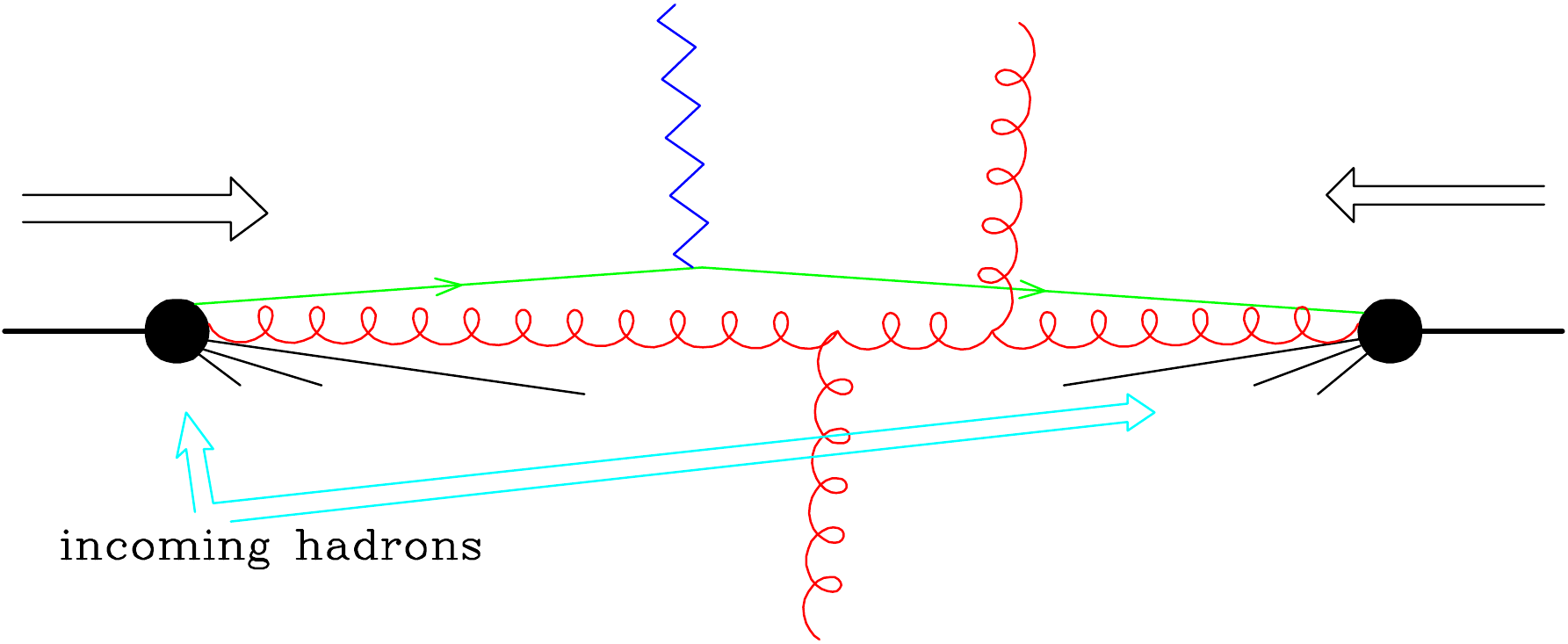,width=0.8\textwidth}
\end{center}
\caption{\label{fig:secondaryint}
Pictorial representation of multiple interactions.}
\end{figure}
 The probability for a secondary interaction
can be easily estimated. Since a hard scattering has taken place, this means that the two hadrons
have overlapped in the transverse plane, with an overlapping area of the order of $1\,{\rm fm}^2$.
An estimate of the probability for another hard cross section is given by the ratio of the
hard cross section divided by the transverse area of the overlapping region.
In the case of the production of an extra pair of jets with transverse momentum $\pt$, the probability is
\begin{equation}
\frac{\alphas^2(\pt^2)}{\pt^2 \times (1\,{\rm fm})^2}\approx 
\frac{\LambdaQCD^2 \alphas^2(\pt^2)}{\pt^2}\;.
\end{equation}
It is thus a power suppressed effect. The perturbative description of the collision we have introduced so far
deals with effects that are at most suppressed by powers of the coupling constant, not by powers of the strong
interaction scale. Nevertheless, for relatively small $\pt$, multiparton interactions can be important for a
full description of the event.
Modern underlying event models do include secondary interactions, as discussed in \ref{chap:MBandUE}.

\section{The detectors}
This section is meant to be an introduction to the experimental
aspects of the investigation of high energy proton-proton events. The
outline of the section is the following: 
\begin{itemize}
\item review the basic physics processes involved in the detection of
  particles in the energy regime typical of LHC;
\item describe how different detection systems can be gathered into
  {\it general purpose} apparatus to provide the most exhaustive
  picture of each proton-proton collision; 
\item a short description of how these different components have been implemented
  in the ATLAS and in the CMS experiments.
\end{itemize}
Deeper discussion of physics objects, trigger, simulation will
be done to Chapter/Session 4.
\subsection{Basics of particle detection}\label{sec:pardet}
Experimentally the measurement of a particle is the
determination of its four-momentum and the identification of
its type, namely mass and charge.
Modern particle detectors are based on the conversion of an absorbed
energy into an electrical signal.
The processes leading to the formation of the signal depend on
the particle type and energy. 
Considering only the energy range typical of particles produced in
high energy collisions, i.e. from several hundreds MeV to several
hundreds GeV, and the particles whose lifetimes are long 
enough to reach the detectors, i.e. $c\tau \gtrsim 2.5$ cm, the main
processes are:
\begin{itemize}
\item electrons/positrons: the energy loss is determined by ionization and
  bremsstrahlung processes the latter being dominant from the
  energy at which they are produced down to a {\it critical energy}
  $E_c$ whose typical value is of the order of ten(s) MeV.   
\item muons: the energy loss is determined by ionization and
  bremsstrahlung processes. As the dependence of the bremsstrahlung
  energy loss per unit length on the particle energy $E$ and mass $m$ goes as
  $E/m^2$, the bremsstrahlung process for the muons starts to be
  relevant at hundreds GeV. As they do not interact strongly and as
  the probability of showering~\footnote{See below for a description of
  a shower process.} is small, they can penetrate deeply in the
  material, as Minimum Ionizing Particle ({\it MIP}).
\item charged hadrons (essentially $\pi^{\pm}$ and protons): the
  energy loss is determined by ionization and strong
  interactions with the nuclei of the material. 
\item neutral hadrons (essentially $\pi^0$ and neutrons): the energy
  loss is determined by strong interactions
  with the nuclei of the material. It should be noted that neutral
  pions quickly decay electromagnetically $\pi^0\rightarrow
  \gamma\gamma$ ($c\tau$=25 nm) before having a chance of
  re-interacting hadronically.
\item photons: the energy loss is determined essentially by pair production
  for energies above some MeV. The electron and positron produced
  behave as described above. 
\end{itemize}
As the main interaction mechanisms of high energy electrons
(i.e. Coulomb scattering and bremsstrahlung) and of high energy photons
(i.e. pair production) are described by closely related diagrams they all
can be characterized by the same parameter: the radiation length $X_0$, which
expresses the mean distance over which a high-energy electron loses
all but $1/e$ of its energy by bremsstrahlung or equivalently
$\frac{7}{9}$ of the mean free path for pair production by a
high-energy photon. 

All the charged particles interacts by elastic Coulomb scattering at
low momentum transfer off
the nuclei of the medium thus resulting in a change of direction which
affects the direction of $\vec{p}$ ({\it Multiple Scattering}).  
The distribution of initially collimated unit charge particles
emerging from a slab of material of thickness $X$ and radiation length
$X_0$ has a Gaussian core with a spread projected into a plane $\sigma_\theta =\frac{13.6
  {\mathrm MeV}}{\beta p c} \sqrt{\frac{X}{X_0}}$ with non-gaussian
tails generated by the collisions at large momentum transfer.%
\footnote{A very rough explanation of this formula is as follows.
Coulomb scattering
and bremsstrahlung effects are related, since Coulomb scattering
is the cause of bremsstrahlung. Thus the number and intensity of random transverse
kicks is proportional to $X/X_0$, and the variance in the total transverse kick
goes like $\sqrt{X/X_0}$. The $1/\beta$ accounts
for the fact that slower particles spend more time in traversing the
atoms they collide with,
and the kick in momentum is proportional to the product of force and time. 
The $1/p$ factor translates a transverse momentum kick into an angle.}

\subsection{Measurement of the four-momentum: spectrometers and calorimeters}

\begin{figure}[hbtp]
  \begin{center}
    \includegraphics[width=0.3\linewidth]{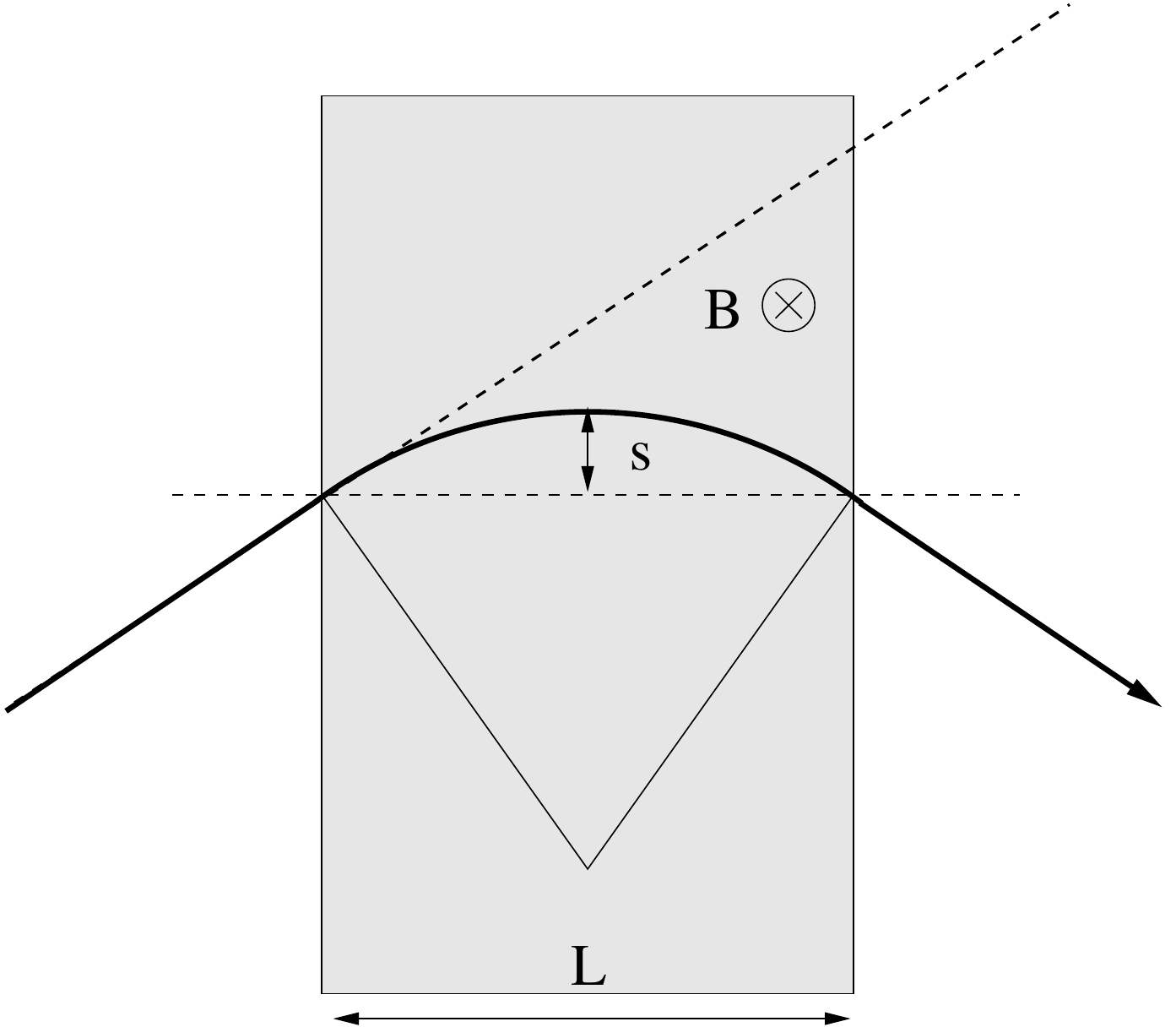}
    \includegraphics[width=0.3\linewidth]{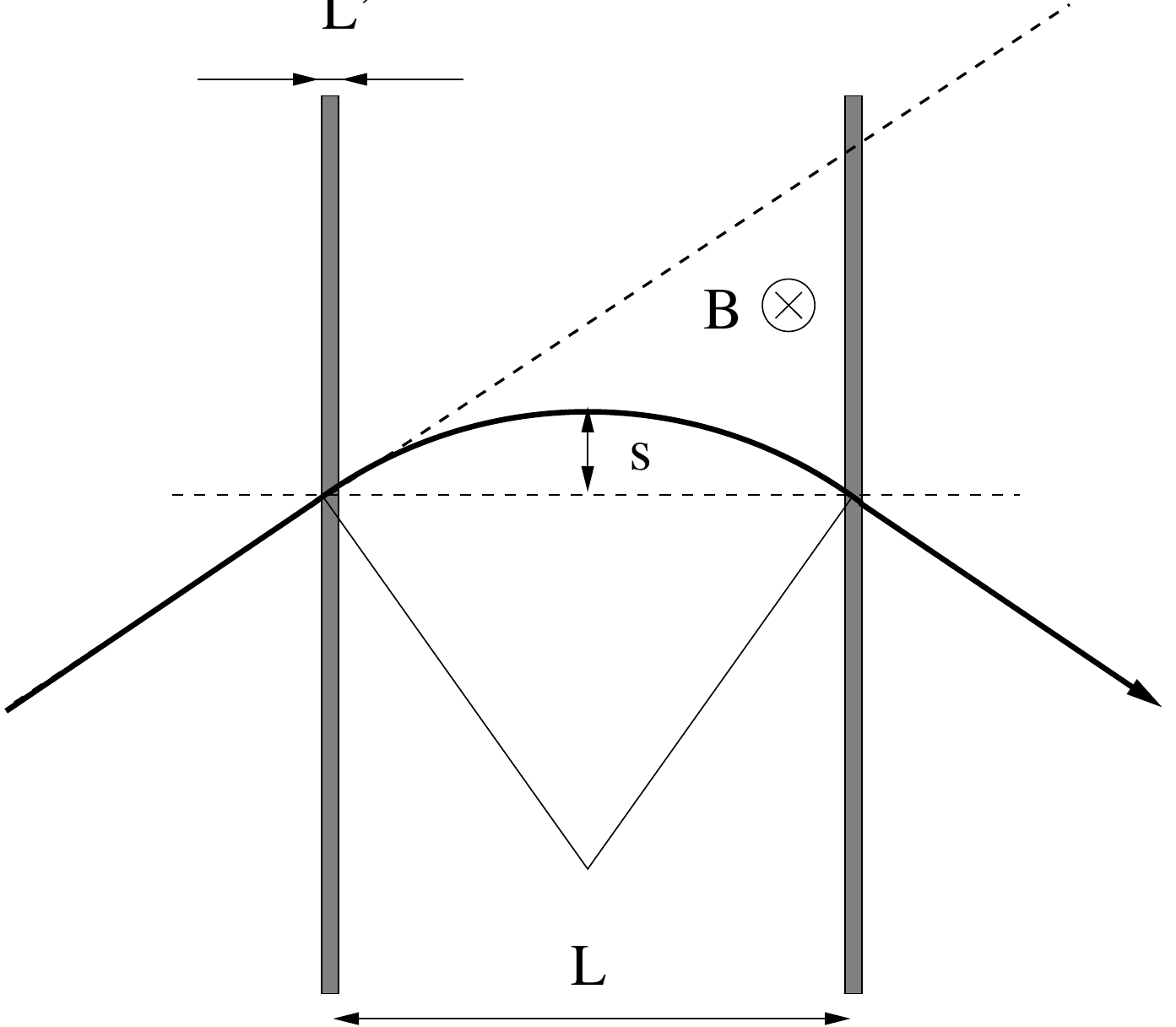}
    \caption{Deflection of a charged particle in a magnetic field:
    measurament of the trajectory performed in a homogeneous medium (left)
    or in an air-gap spectrometer (right).}
    \label{fig:sagitta}
  \end{center}
\end{figure}

\paragraph{Spectrometers.} Spectrometers determine the three-momentum
of a charged particle from observing how it bends in a magnetic field. 
A unit charge particle crossing a uniform magnetic field $\vec{B}$
with momentum $p_T$ in the plane perpendicular to the $\vec{B}$ direction
for a region of length $L$ deflects from the straight line of a
sagitta $s$ (see figure~\ref{fig:sagitta}): 
\begin{equation}
s\approx\frac{0.3}{8}\frac{BL^2}{p_T} 
\end{equation}
with $B$ in T, $L$ in m and $p_T$ in GeV.\\
From $p_T$ and from the knowledge of the trajectory 
(i.e. the angle $\theta$ between the track and the field directions) 
the three-momentum is determined as $p=|\vec{p}|=p_T/\sin\theta$.
If no particle identification is available, the energy of the particle
is then computed assuming the mass of the charged pion which is the most
abundant charged particle in the event. \\
The resolution on $p_T$ is related to the geometric accuracy on the
measurement of the sagitta $\sigma_s$ by
\begin{equation}
\left.\frac{\sigma_{p_T}}{p_T}\right|_{geom} =\frac{\sigma_s}{s}=\frac{8}{0.3BL^2}\:p_T\:\sigma_s
\end{equation}
To achieve observable bendings, in collider experiments magnetic
fields of several Teslas are used (2 T in the ATLAS and 4 T in the CMS
solenoids respectively). 
For example the sagitta of a particle with $p_T$=1 TeV  crossing a 1 m long region with a 4 T
$B$ field is about 150 $\mu$m. Thus a 10\% measurement of the momentum
requires an accuracy on the sagitta of 15 $\mu$m.
As the geometrical accuracy on the sagitta $\sigma_s$ does not depend
on $p_T$, $\left.\frac{\sigma_{p_T}}{p_T}\right|_{geom}$ decreases for
increasing $BL^2$ values.\\

Multiple scattering mimics the magnetic field deflection thus
affecting the resolution by a factor which does not depend on $p$. 
The actual effect on the resolution depends on whether the measurament
of the trajectory is performed in a homogeneous volume (figure
\ref{fig:sagitta} left), as in the case of Time Projection Chambers, 
or in an air-gap spectrometer with the track measured by discrete
tracking planes (figure \ref{fig:sagitta} right), as in the case of
silicon strip tracking detectors.\\  
In the case of the homogeneous volume the determination of the sagitta
is mainly affected by the lateral dispersion, $\sigma_s|_{ms} =
\frac{1}{\sqrt{3}}\sigma_{\theta} L$, leading to: 
\begin{equation}
\left.\frac{\sigma_{p_T}}{p_T}\right|_{ms} \approx\frac{0.21}{\beta B\sqrt{L
    X_0 \sin\theta}}
\end{equation}
In case of air-gap spectrometer the dominant contribution is the kink
$\sigma_{\theta}$ experienced by the particle when crossing a measurement
plane of thickness $L'$: 
\begin{equation}
\left.\frac{\sigma_{p_T}}{p_T}\right|_{ms} \approx\frac{0.045}{\beta B L}\sqrt{\frac{L'}{X_0\sin\theta}}
\end{equation}

The trajectory of the charged particle is measured by tracking
detectors (cf. Sect.~\ref{sec:tkvtx}) which determine the points where
the particle crossed their sensitive volume. Because of the multiple
scattering, tracking detectors ideally should be massless and with no
passive material in front. The multiple scattering dominates at low values of
$p_T$, while at high values the geometrical term dominates.

\paragraph{Calorimeters} Calorimeters are detectors which measure the
energy of a particle by means of total absorption of its energy. 
The process at the basis of the degradation of the energy of the
incoming particle is the development of a shower. Two kinds of showers
can be distinguished according to the nature of the incoming particle:
\begin{itemize}
\item electromagnetic showers. \\
  The interaction of photons and electrons above 10
  MeV is dominated by the processes:
  \begin{itemize}
  \item pair-production $\gamma \rightarrow e^+ e^-$
  \item bremsstrahlung $e^{\pm} \rightarrow e^{\pm}\gamma$
  \end{itemize}
  both characterized by the radiation length $X_0$.
  On average after 1 $X_0$ a $\gamma$ produces an $e^+e^-$ pair while an
  $e^{\pm}$  radiates a bremsstrahlung photon, with the produced
  particles sharing the energy of the initial one. 
  The cascade of these processes produces a so-called electromagnetic shower
  which evolves with the number of particles increasing at each step
  while their energy is decreasing. Below a certain threshold
  $E_{thr}$ the process stops with a number of particles in the
  shower related to the energy of the initial particle.    
  The depth at which the shower stops grows logarithmically with the
  initial energy $E_0$  of the incoming particle.~\footnote{
  Assuming that at each branching the energy equally splits among the
  decay products, the length $L$ for a full containment of a shower
  developed by an incoming particle of energy $E_0$ is 
  \[
  L = X_0\frac{\ln{E_0/E_{thr}}}{\ln{2}}
  \]
  }
  The typical depth of an electromagnetic calorimeters in a high
  energy physics experiment is about 27 $X_0$.
  The energy of an electron/photon is then obtained from counting the
  charged particles which are in the shower, actually measuring the
  energy deposited by them, and applying some
  calibration function which has been determined previously in
  conditions where the energy of the incoming particle was known, as in
  test-beam or {\it in-situ}, i.e. in the experiment itself using events
  where the energy is well known for example by kinematic constraints of
  the event.
  The determination of the energy with a calorimeter is a stochastic process 
  (the measured quantity is the signal released by $N$ particles with
  $N\approx E_0/E_{thr}$) and therefore its resolution improves with
  increasing energy~\footnote{The notation $\oplus$ indicates the sum in quadrature.} 
  \begin{equation}
  \frac{\sigma_E}{E}= \frac{a}{E} \oplus \frac{b}{\sqrt{E}} \oplus c 
 \end{equation}
  The first term is referred to as the 
  {\it noise term} (signal fluctuations independent from the shower
  energy, typically electronics read-out noise), the second as the
  {\it stochastic term} (statistics fluctuations in the various
  processes from the shower development up to the signal formation),
  the last as the {\it constant term} (typically due to detector
  inhomogeneities).\\ 
  Multiple scattering of $e^{\pm}$  produces a broadening of the shower
  also in the transverse direction which is characterized by the Moliere
  radius $R_M$: 95\% of the energy of the shower is contained within a
  cylinder of radius $2R_M$.
\item Hadronic showers: if the incoming particle is a hadron, the 
  showering process is dominated by a sequence of inelastic hadronic
  interactions. At high energy, these are characterized by multi-particle
  production and particle emission originating from nuclear decay of
  excited nuclei. The principle of the energy measurement is the same
  as for the electromagnetic case: counting the number of charged
  particles in the shower and converting it into an energy value by
  means of a known calibration function.
  Similar considerations for the energy resolution holds as for the
  case of electromagnetic calorimeters, but the energy resolution is
  worse than that of electromagnetic calorimeters essentially because
  of three effects:    
  \begin{enumerate}
  \item part of the energy goes into excitation of break-up nuclei
    without being detected;
  \item on average 1/3 of the pions produced (i.e. the lightest
    hadron) are $\pi^0$ which immediately decays
    electromagnetically giving a larger signal than a charged pion of
    the same energy;
  \item hadrons can decay in final state with a neutrino
    (undetected) or a muon (small signal).
  \end{enumerate}
  The typical scale of the process is the nuclear interaction
  length $\lambda_I$ which for materials heavier than iron is one order
  of magnitude larger than the electromagnetic interaction length
  $X_0$. For this reason hadronic calorimeters are longer and placed
  besides electromagnetic calorimeters.
\end{itemize}
Technically two types of calorimeters exists: homogeneous calorimeters
where the same material acts as the medium where shower develops ({\it
  absorber}) and signal is produced ({\it detector}) and sampling
calorimeters (absorber material different from detector material).
Typical examples of homogeneous calorimeters are high-Z material
crystals while for sampling calorimeters are sandwiches of high-Z
materials and gas or liquid detectors or plastic scintillators.
In the {\it detector} material the produced ionization or the scintillation light
emitted in the de-excitation of the crystal lattice for inorganic scintillators
(or of vibrational modes for organic ones) is collected.
Hadronic calorimeters are usually sampling calorimeters.

The determination of the direction of the incoming particle with a
calorimeter is achieved by means of a segmentation of calorimeter into
cells read-out separately. In case of an electromagnetic shower the
three-momentum is then just computed assuming a massless kinematics, i.e. the
measured energy as the absolute value of the momentum and the same
direction for $\vec{p}$ and the shower.

\subsection{Measurement of the topology of the event: vertex and tracking detectors}\label{sec:tkvtx}
Because of their fast response, the LHC tracking systems are mainly
based on finely segmented solid-state detectors: two or three shells
of 2D pixel detectors close to the interaction point at $r\approx$ 10
cm with a typical pixel size of 150x150 $\mu$m$^2$,
followed by 4 shells of 1D strip detectors at 20$<r<$50 cm with a
typical pitch distance between neighboring strips of about
100 $\mu$m.
This layout is dictated by the requirement of keeping the fraction of
hit channels of the detectors per LHC bunch crossing, called occupancy,  
at a level of 1-2\% thus keeping at an
affordable level the combinatorics that the track finding algorithms
have to deal with.\footnote{At $\sqrt{s}$=14 TeV and at $L=10^{34}$
  cm$^{-2}$s$^{-1}$ there are about 1000 tracks per bunch crossing
  which, in absence of magnetic field, would be distributed as $1/r^2$.
  The magnetic field actually confines charged particle and the actual
  distribution is more (less) pronounced than the $1/r^2$ at lower
  (higher) radii.}
The position of the measured point is provided by the channel fired by
the passing charged particle. Therefore the resolution (for a single
$\eta,\phi$ coordinate) is basically
the width of the read-out cell divided by $\sqrt{12}$ that for the
geometry described above is typically of about 30 $\mu$m 
(a 3D point is reconstructed by the pixel detectors while a 2D point is
typically reconstructed by the strip detectors: the extra coordinate is
given by the $r$ position of the detector itself.)
Accurate measurements of the particle tracks close to their production
point can be exploited for:
\begin{itemize}
\item reconstruct the primary vertex of the event. At high luminosity
  there will be on average 20 inelastic proton-proton collisions at each
  bunch-crossing: grouping all particles coming from the same primary
  vertex is an essential simplification for the analysis of the LHC events;
\item identify long-lived particles (typically hadrons containing a heavy
  quark) which travel up to few mm before decaying. They can be
  identified as their decay (secondary) vertex is displaced by their production
  (primary) vertex. This identification ({\it tagging}) is based on the fact
  that the impact parameter of the daughter particles produced at the
  decay vertex, i.e. the distance of minimum approach of daughter
  tracks to the primary vertex, is significantly different from
  zero. The impact parameter is larger than the one of particles
  coming from the interaction vertex, because of the large mass of the
  long living hadron (thus the relatively high $p_T$ of the decay products
  with respect to the hadron flight direction)
  and because the tracks come from a secondary vertex that is 
  produced few millimeters away from the primary one.
\end{itemize}

The resolution on the impact parameter depends on the geometrical
resolution of the detector and on the multiple scattering:
\begin{equation}
  \sigma_{IP}= \sigma_{geom}\oplus\sigma_{ms}
\end{equation}
The geometrical resolution depends on the layout of the detector, namely
the intrinsic resolution $\sigma_{int}$ of the sensing element
(i.e. the pitch of the strips or the dimension of the pixels, typically
$\sigma_{int} \approx 30 \mu$m), the distance $r$ from the primary
vertex of the layer of the detector that gives the first measurement
of the track, and on the total lever arm $l$ of the vertex-detector:
\begin{equation}
  \sigma_{geom}= \sigma_{int} \times \left( \frac{r}{l}  \oplus \frac{r+l}{l}\right)=
\sigma_{int}\times \sqrt{1+2 \frac{r}{l} +2 \frac{r^2}{l^2}}
\end{equation}
The multiple scattering term is due to the presence of material along
the particle trajectory and it depends on the momentum $p$ of the particle, 
on the amount of material crossed (which in turn,
assuming a detector with cylindrical geometry,
depends  on the angle $\theta$ at which the particle is emitted,
$\theta$ being the angle with respect to the beam direction):
\begin{equation}
  \sigma_{ms}= a\oplus\frac{b}{p \sin^{3/2}\theta}
\end{equation}
Thus a performant detector must have the first layer as close as possible to the interaction
region (i.e. $r$ as small as possible), limited by the dimension of the
beam pipe, the lever arm $l$ as large as possible and the material as thin as possible in unit
of $X_0$. Moreover the detector should maximize the  number of layers, in order to minimize the error
and ambiguities in the pattern recognition during the reconstruction of the trajectory.

\subsection{General purpose detectors}
General purpose detectors are designed to measure as many as possible
of the particles produced in each proton-proton collision to get the
most precise picture of each event.
This is achieved basically by:
\begin{itemize}
\item organizing the detector in a ``onion-like'' structure
  (i.e. cylindric shells concentric with the beams direction) where
  each layer/subsystem measures the particles unmeasured by the
  previous layer;
\item embedding the tracking detectors in a magnetic field in order to
  determine the momentum of charged particles from the deflection of
  their tracks.
\end{itemize}

\begin{figure}[hbtp]
  \begin{center}
    \includegraphics[width=1.0\linewidth]{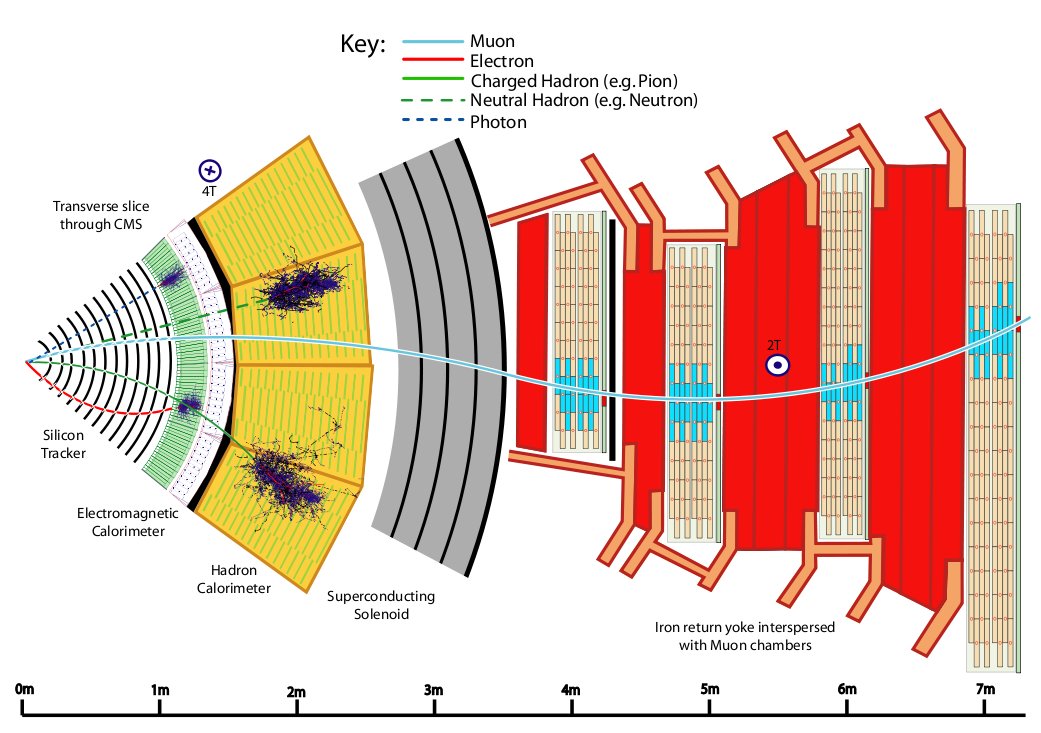}
    \caption{Slice through CMS showing particles incident on the different sub-detectors.}
    \label{cms-sector}
  \end{center}
\end{figure}

Following the drawing of figure \ref{cms-sector}, first the trajectory
of all the stable charged particles are measured by the tracking
detector.\\ 
The electron and the photon are then stopped and their energy measured by the electromagnetic 
calorimeter. When an electromagnetic shower is associated with a track it can be identified
as an electron. If no track corresponds to the shower, then this last one is produced by 
a photon. Electron-photon identification is thus reached.\\
Then all the hadrons are stopped by the hadronic calorimeter and
their energy measured. Their momenta were already measured in the tracker, from the curvature 
of their trajectory due to the magnetic field. For this, it is assumed that all the hadrons
have a pion mass.\\
The only particles that can travel through the full detector as MIP are the muons.
Their tracks are detected in the {\it muon detector} thus also identifying the presence of a
muon.\\
The momentum of the muon is measured by both the tracker and the muon chambers. 
Depending on the strength of the magnetic field only muons above a certain $p_T$ could reach 
the muon chambers (3 GeV for CMS, as an example). 
The number of high $p_T$ muons in a LHC event is not very high, 
since most of the muons coming from the semileptonic decay of B (D) hadrons have a small $p_T$
 (typically of the order of $M_B (M_D) /2$) a large fraction of the muons reaching the muon
chambers are those from W and Z decay.
This means that muons are a very clean and powerful signature of interesting events at LHC.

\subsection{ATLAS and CMS}

ATLAS and CMS 
implement this general purpose structure in a different
way because of the different configuration of the magnetic field.
The chosen magnetic field intend to maximize the $BL^2$ term 
determining the resolution on the
measurement of the momentum of the muon. Good resolution for muons from few 
GeV up to 1 TeV are mandatory to fulfill the physics program.

The size of both the experiments are determined mainly by the fact that they are designed 
to identify and measure the energy and momentum of most of the very energetic particles 
emerging from the proton-proton collision. The interesting particles are produced over a 
wide range of energy (from few hundreds of MeV to a few TeV) and over the full solid angle.
No particle of interest should escape unseen (except neutrinos that are instead identified
by inbalance in the energy-momentum conservation).This means that the two experiments should
avoid any cracks in the acceptance.

The configurations of the magnetic fields are the following: 
\begin{itemize}
\item ATLAS has adopted a toroidal configuration where the relative
  lower magnetic field, $B$=0.6 T, is balanced by a large lever arm
  $L$= 11 m. The toroid is then complemented by an inner solenoid of
  reduced dimensions $R$=1.2 m and relatively high value of the field
  $B$=2 T.
  The calorimeters are placed in the field-free region.
\item CMS has adopted a {\it compact} layout with a solenoid with a very intense field $B$=4 T
  and moderate dimensions  $R$=3 m. The calorimeters are inside the field.
\end{itemize}
The main technological challenge for ATLAS is the mechanical precision
at which the tracking elements should be positioned over such a large
lever arm, while for CMS is to reach this high and uniform value of the $B$ field
over such a large volume.

The ATLAS tracker is made by an inner part of silicon pixels and strips
and an outer part made of TRD (Transition Radiation Detectors)
 in order to identify particles, and in particular
the electrons. It is embedded in the 2~T magnetic field. The resolution on the 
charged particle momentum is $\sigma_{p_T}/p_T \approx 5 \cdot 10^{-4} p_T \oplus 0.01$.
The CMS tracker is inside the 4~T magetic field and it is made entirely of silicon 
sensors (pixels and strips). The  resolution on the 
charged particle momentum is $\sigma_{p_T}/p_T \approx 1.5 \cdot 10^{-4} p_T \oplus 0.005$.
The outer radius of the tracking detectors of the two experiments is similar ($\approx$ 110~cm).
 
The other important difference between ATLAS and CMS concerns the
choice of the electromagnetic calorimeter:
\begin{itemize}
\item ATLAS has a Pb-LAr sampling calorimeter;
\item CMS has an homogeneous calorimeter made of PbWO$_4$ crystals.
\end{itemize}
The sampling structure of the ATLAS electromagnetic calorimeter allows the measurement
of the shower developement at different depths, thus allowing a better
determination of the shower axis and consequently of the
electron/photon direction.\\
Being homogeneous, the CMS electromagnetic calorimeter has instead an intrinsically better energy
resolution for electrons and photons.

The hadronic calorimeters of ATLAS is made by Fe-scintillator (in the barrel) and Cu-liquid argon
(end caps) for a total of 10~$\lambda_I$. It has a relatively good energy resolution
$\sigma_E/E \approx 50\% /\sqrt{E/\rm{GeV}} \oplus 0.03$.
The CMS hadronic calorimeter is made of Cu-scintillator with an energy resolution 
of $\sigma_E / E \approx 100\% / \sqrt {E/\rm{GeV}} \oplus 0.05$. Due to the constraint of beeing inside the magnet
the calorimeter is not long enough to contain the full hadronic shower, being only $\approx 10 \lambda_I$. 
Thus an additional
tail catcher (the HO detector) has been placed after the calorimeter in order to limit
the punch through into the muon system. 

Finally muons are very robust, clean and unambiguous signature of much of the physics
that ATLAS and CMS were designed to study. The ability to trigger  and reconstruct
muons at the highest luminosities of the LHC was incorporated into the design of the two detectors. 
The choice of the magnet, as already said, was motivated by the necessity to 
measure TeV muons.
The ATLAS muon detector is placed in the air and the resolution on the
muon transverse momentum is  $\sigma_{p_T} / p_T \approx 7\% $ at 1 TeV. It provides an independent and
high-accuracy measurement of muons over the full $\eta$ coverage required by physics.
In CMS the muon chambers are placed in the iron of the magnet yoke and the 
muon transverse momentum is  $\sigma_{p_T}/p_T \approx 5\%$ at 1 TeV. 

Both the system can trigger muons from 3-5 GeV of momentum. Moreover they are
able to distinguish bewtween successive beam crossing (spaced 25~ns in time).
%
\subsection{Physics objects}
The objects measured in collider experiments are: muons, electrons,
photons, tau-jets, 
jets of hadrons (as signature of colored partons) and missing
transverse energy (as signature of neutrinos and  particles which have little or no
interaction with ordinary matter).

Out of the elementary particles, only muons,
electrons and photons are directly detected.
From the processes listed in Section~\ref{sec:pardet}, those relevant for
ATLAS and CMS detectors are: the ionization for muons, electrons and
charged hadrons, and
the detection of the low energy charged particles produced in the
electromagnetic cascade for electrons and photons, or in the hadronic
cascades for hadrons. The latter occurs collecting either
the produced ionization or the scintillation light
emitted in the de-excitation of the crystal lattice or of vibrational
modes for inorganic and organic scintillators respectively.

Tau leptons are usually identified by their decay into a lighter
charged lepton plus two neutrinos or in their decay into 1, 3 or 5
charged tracks, thus in collimated jets with a low number of particles.
 
The constituents of the
hadronic matter (quarks and gluons) are revealed only in the form of
jets of hadrons. 
 
The presence of particles which have little or no
interaction with ordinary matter, as neutrinos or neutral SUSY LSP,
can be inferred only by the so-called missing transverse energy. 
In a hadron collider, because of the beam remnants carrying part of the
longitudinal momentum of the incoming beams, kinematics constraints
can be applied in the transverse plane only and the presence of such
particles can be inferred from the component of the missing energy in
the transverse plane.

More details about the physics objects  can be found in session \ref{detectors}.
 
\subsection{Trigger for experiment at hadronic collider}

Compared to electron-positron collider where all the inelastic cross section
can be considered as a signal, in hadronic colliders the cross section
for interesting processes is a very small fraction of the inelastic cross 
section. 

At LHC, the inelastic, non single diffractive, proton-proton 
cross-section $\sigma_{inel}$ is expected to be $\sigma_{inel}\sim$ 80
mb; this corresponds to an interaction rate at the LHC nominal
luminosity of the order of 1 GHz.  
As the raw event has a typical size of \cca  1 MB, the resulting amount of data 
would be way too prohibitive to record and process for 
a later offline analysis. 
Such rate has therefore to be reduced online to the order of 100 Hz, which is 
the upper limit for storing and processing data.

\begin{figure}[hbtp]
  \begin{center}
    \includegraphics[width=0.8\linewidth]{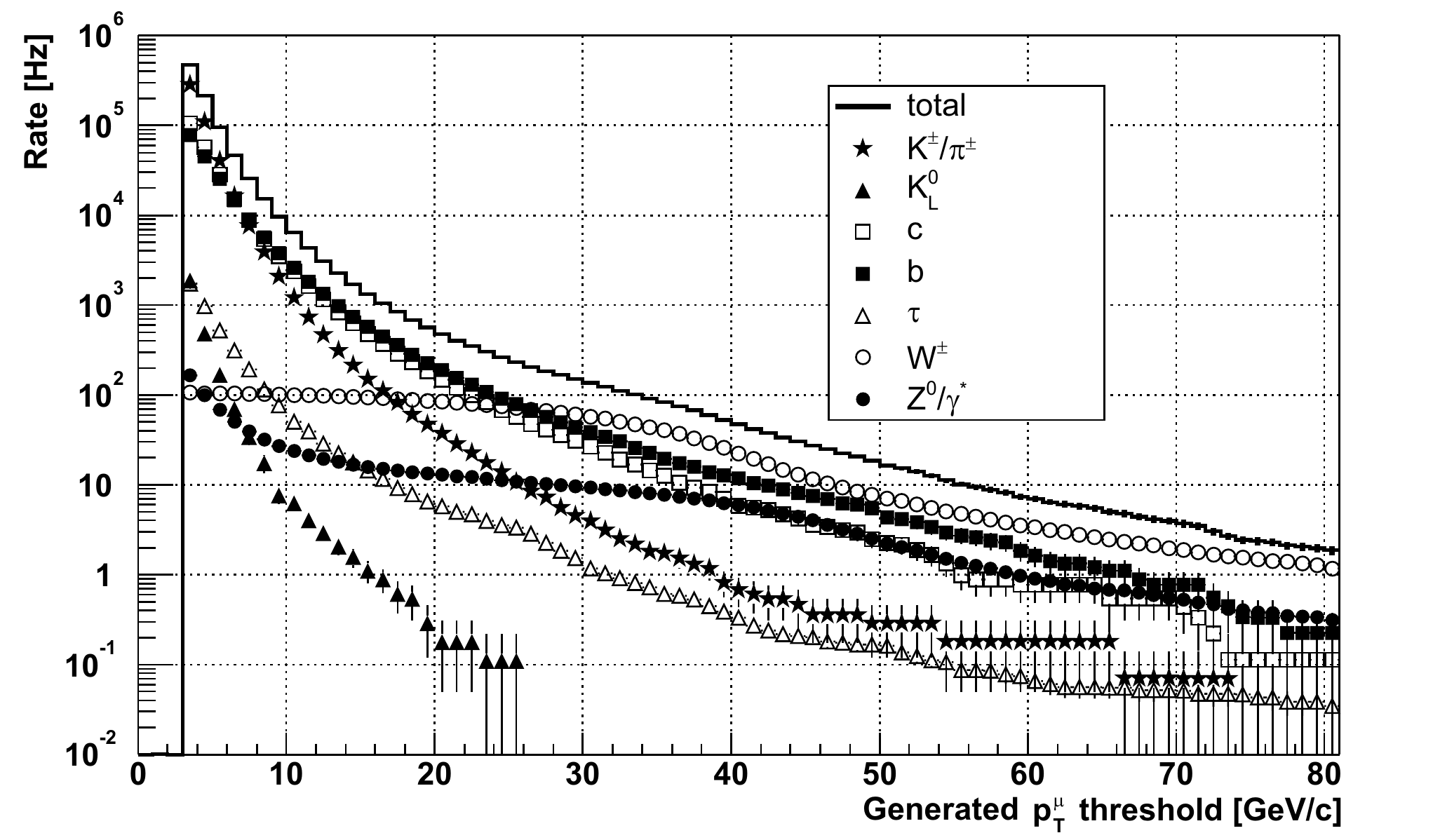}
    \includegraphics[width=0.8\linewidth]{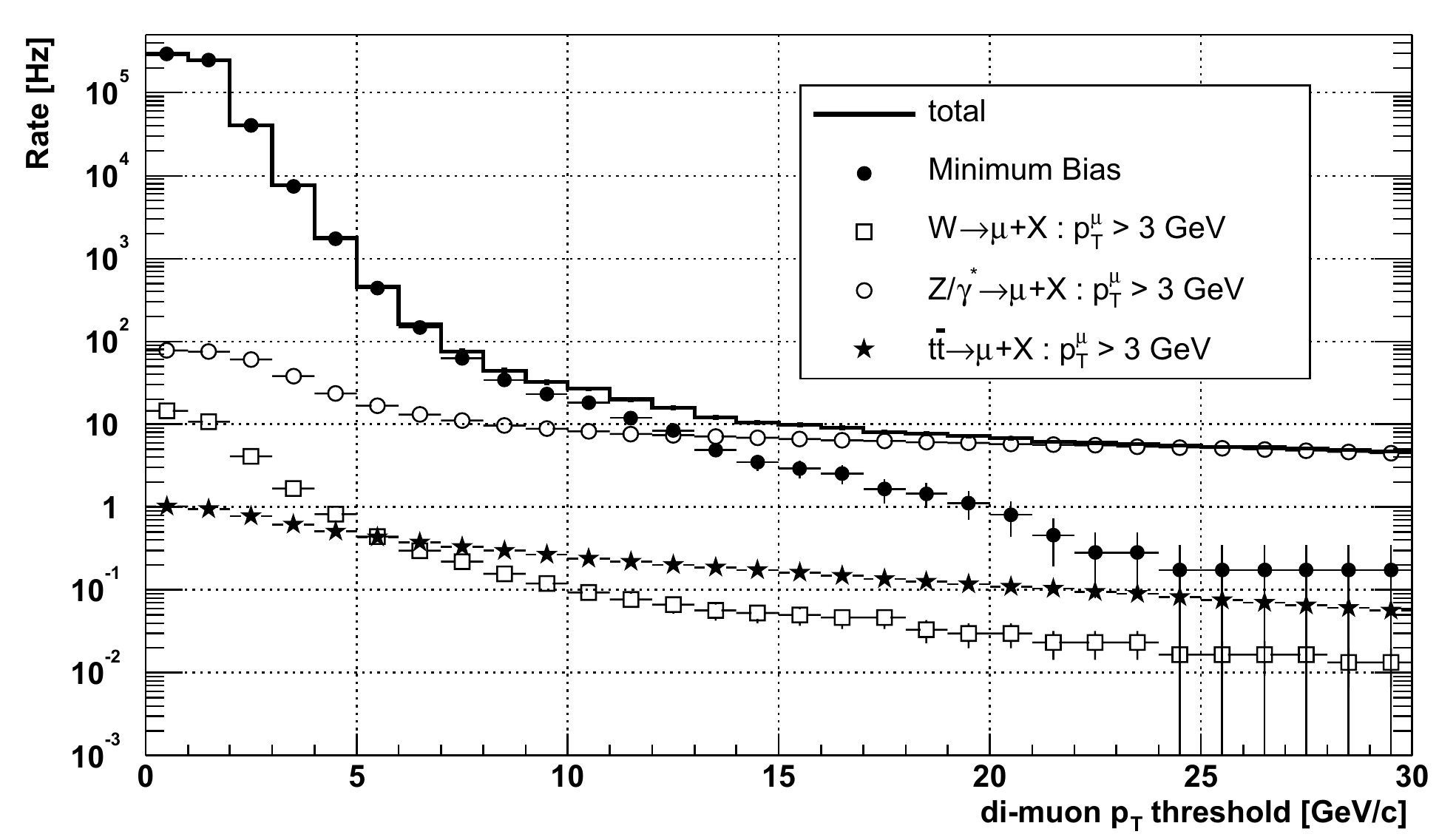}
    \caption{The rates of muons as a function of threshold on their
    transverse momentum: single muons (top), double muons (bottom).}
    \label{hlt-muon}
  \end{center}
\end{figure}

However the rate is dominated by low $p_T$ processes and most of the
events available in this reaction is of no interest. The reduction
corresponds therefore to selecting the events
which have actually some physics interest, events which are a low
fraction of the total (see Fig.~\ref{fig:sigma_pp}).   
As an example, the rate of the SM Higgs boson
decaying to the 2-photon final state is expected to be 1 Hz for m$_H$=100 GeV/$c^2$. 
Fitting the selection of high-$p_T$ processes within the allowed output rate is anyway
difficult because processes like $W$\ra$\ell\nu_{\ell}$ 
and  $Z$\ra$\ell^+\ell^-$ already saturate the output rate if no
selection is applied. 
In figure \ref{hlt-muon} the rates of single muons and two muons generated by 
different processes as a function of the threshold on their transverse momentum 
are shown. 

The online selection of collisions potentially containing physics of interest,
is accomplished by the trigger system.
In order to provide huge rejection factors (\cca$10^7$) and to keep at
the same time the efficiency for interesting events as high as
possible, the online selection has a level of complexity comparable to that of offline reconstruction.
In addition, the trigger algorithms must be fast. In fact, the time available to perform the online 
selection is the interval between two bunch crossings, that at LHC is 25 ns: this time is too small even 
to read out all the raw data from the detector. Hence the final decision must be divided into 
subsequent steps of increasing refinement and length. Each step (\textit{level}) accesses and uses only 
a part of the available data in order to take its \textit{accept}/\textit{reject} decision within the 
required time constraints. The following levels have a lower rate of events to process and more time 
available for the decision, so they can use larger sets of data and more refined algorithms.

The first level (\textit{L1}) is hardware implemented, due to the strict timing constraints. 
It accesses data from the calorimeters and from the muon detectors with coarse granularity and performs low 
level analysis in custom trigger processors. On the basis of this limited information, it has to reduce 
the data rate to \cca 100 kHz, which is the maximum input accepted by the \textit{Data Acquisition system} 
(\textit{DAQ}) at high luminosity. 

All further levels are referred to as \textit{High Level Trigger} (\textit{HLT}). The HLT is fully 
implemented on software running on a farm of commercial processors: this ensures more flexibility 
and the possibility to improve the selection algorithms. The HLT performs the final selection and 
achieves the output rate of the order of 100 Hz. Only data accepted by the HLT are recorded for 
offline physics analysis. 

Finally the selection efficiencies of the different trigger levels should be precisely known in order not
to introduce biases that affect physics results. 

For a detailed description of the ATLAS and CMS trigger see chapter \ref{detectors},
section \ref{sec:trigger}.


\addtocounter{chapter}{1}

\newcommand{\tmem}[1]{{\em #1\/}}
\newcommand{\tmop}[1]{\ensuremath{\operatorname{#1}}}
\newcommand{\tmtextit}[1]{{\itshape{#1}}}
\newcommand{\tmtexttt}[1]{{\ttfamily{#1}}}
\newenvironment{enumeratealpha}{\begin{enumerate}[a{\textup{)}}] }{\end{enumerate}}
\newenvironment{enumeratenumeric}{\begin{enumerate}[1.] }{\end{enumerate}}
\newenvironment{enumerateroman}{\begin{enumerate}[i.] }{\end{enumerate}}
\newenvironment{itemizedot}{\begin{itemize} \renewcommand{\labelitemi}{$\bullet$}\renewcommand{\labelitemii}{$\bullet$}\renewcommand{\labelitemiii}{$\bullet$}\renewcommand{\labelitemiv}{$\bullet$}}{\end{itemize}}


\mchapter{Shower Monte Carlo programs}
{ Author: Paolo Nason}\label{ch:shower}
\vskip 0.3cm\noindent
{\it Revisors: Stefano Frixione and Roberto Tenchini}
\vskip 1cm
\section{Introduction}

In modern experimental particle physics, Shower Monte Carlo programs have
become an indispensable tool for data analysis. From a user perspective, these
programs provide an approximate but extremely detailed description of the
final state in a high energy reaction involving hadrons. They provide an
{\tmem{exclusive}} description of the reaction, as opposite to typical QCD
calculations, that are only suitable to compute {\tmem{inclusive}} quantities.

Shower Monte Carlo programs are a mixture of several heterogeneous
components, that are all needed to give a realistic description of the
formation of the final state:
\begin{enumeratenumeric}
  \item \label{item:hardscatt}A large library of Standard Model and Beyond the
  Standard Model cross sections. The user can choose the hard scattering
  process within this library.
  
  \item \label{item:shower}An algorithm for the generation of dominant
  perturbative QCD effects, called the {\tmem{shower algorithm}}. The shower
  algorithm adds to a given hard scattering a number of enhanced coloured
  parton emission processes. The enhancement is given by collinear and soft
  singularities, that can contribute large logarithms of the hard scale of the
  process over some typical strong interaction scale cutoff. These large log
  are of the order of the inverse of a strong coupling constant, and can thus
  give contributions of order 1 to the hard process.
  
  \item They implement some model of hadron formation, given the state of high
  energy partons that arises from steps \ref{item:hardscatt} and
  \ref{item:shower}.
  
  \item They implement some model for the underlying event.
  
  \item They include libraries for the decay of weakly unstable hadrons. 
\end{enumeratenumeric}
The name ``Shower'' is from item \ref{item:shower}, that can be considered the
kernel of a Shower Monte Carlo program. The shower generation algorithm is in
essence a method for the computation of a potentially infinite number of
Feynman graphs (i.e. all those that are enhanced by infrared logarithms, so
that their contribution to the cross section can be considered of order one).
Besides being useful for simulation of physical processes, the shower
algorithms also provide a remarkably simple mental model of the most important
QCD effects in high energy processes, providing insights into jet structure,
fragmentation functions, structure functions and their Altarelli-Parisi
evolution.

\section{Shower basics}

\subsection{Collinear Factorization}

QCD emission processes are enhanced in the collinear limit, that is to say,
when an emitted parton (gluon or quark) is collinear to an incoming or
outgoing parton in the scattering process. In this limit, the cross section is
dominated by a subprocess in which a parent parton with small virtuality is
produced that decays into the two collinear partons. There are three possible
decay processes: $q \rightarrow q g$, $g \rightarrow g g$ and $g \rightarrow q
\bar{q}$. The cross section factorizes into the product of a cross section for
the production of the parent parton times a splitting factor. This
factorization is depicted schematically in the following graphical formula, for the case
of the $q \rightarrow q g$ splitting process
\begin{equation}
  \raisebox{-1.3cm}[1.3cm][1.3cm]{\epsfig{file=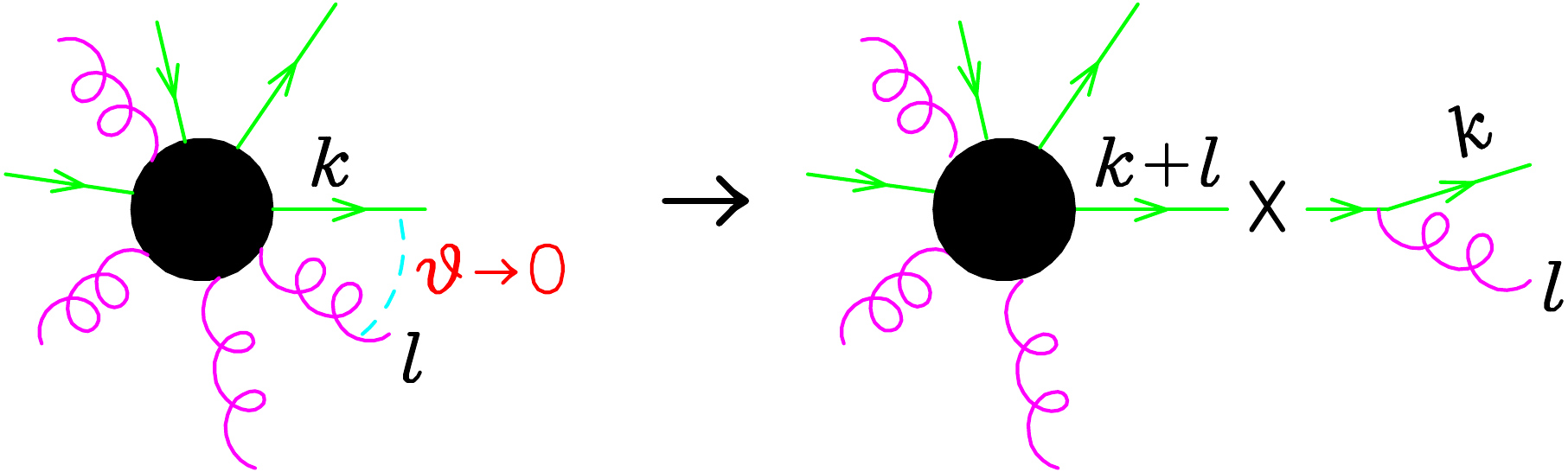,width=0.6\textwidth}},
\label{fig:factorization1}
\end{equation}
that has the following meaning: given a tree level amplitude with $n + 1$
final state particles, assuming that a final state quark becomes collinear to
a final state gluon (i.e. their relative angle goes to zero), we have:
\begin{equation}
  \label{eq:factorization1} |M_{n + 1} |^2 d \Phi_{n + 1} \Rightarrow |M_n |^2
  d \Phi_n \hspace{0.75em} \frac{\alpha_S}{2 \pi} \hspace{0.25em} \frac{dt}{t}
  \hspace{0.25em} P_{q, q g} (z) \hspace{0.25em} dz \hspace{0.25em} \frac{d
  \phi}{2 \pi} .
\end{equation}
where $M_{n + 1}$ and $M_n$ are the amplitudes for the $n + 1$ and $n$ body
processes, represented by the black blobs in fig.~\ref{fig:factorization1}.
The $n$ and particle phase space is defined as usual
\begin{equation}
  d \Phi_n = (2 \pi)^4 \delta^4 \left( \sum^n_{i = 1} k_i - q \right) \prod_{i
  = 1}^n \frac{d^3 k_i}{2 k_i^0 (2 \pi)^3},
\end{equation}
where $q$ is the total incoming momentum. The parameters $t$, $z$ and $\phi$
describe the kinematics of the splitting process: $t$ is a parameter with the
dimension of a mass, vanishing in the collinear limit, $z$ a variable that, in
the collinear limit, yields the momentum fraction of the outgoing quark
relative to the momentum of the quark that has split
\begin{equation}
  \label{eq:zdef} k \rightarrow z (k + l) \tmop{for} t \rightarrow 0,
\end{equation}
and $\phi$ is the azimuth of the $\vec{k}, \vec{l}$ plane around to the
$\overrightarrow{k + l}$ direction. $P_{q, q g} (z)$ is the Altarelli-Parisi
splitting function
\begin{equation}
  \label{eq:pqqap} P_{q, q g} (z) = C_F \hspace{0.25em} \frac{1 + z^2}{1 - z}
  .
\end{equation}
Observe that there is some arbitrariness in the definition of $t$ and $z$,
since $d t / t$ is invariant if we change $t$ by some (possibly $z$ dependent)
scale factor, and for $z$ we only require that eq.~(\ref{eq:zdef}) is
satisfied in the collinear limit. We can, for example, define
\begin{equation}
  \label{eq:zfromk0} z = \frac{k^0}{k^0 + l^0},
\end{equation}
or more generally define
\begin{equation}
  \label{eq:zfrometa} z = \frac{k \cdot \eta}{k \cdot \eta + l \cdot \eta},
\end{equation}
that reduces to the definition (\ref{eq:zfromk0}) for $\eta = (1, \vec{0}$),
and is perfectly acceptable as long as $\eta$ does not coincide with the
collinear direction. For $t$ we can use, for example
\begin{eqnarray}
  \tmop{virtuality} : & t = & (k + l)^2 \approx E^2 \theta^2 z (1 - z), 
  \label{eq:tvirt}\\
  \tmop{transverse} \tmop{momentum} : & t = & k_{\perp}^2 = l_{\perp}^2
  \approx E^2 \theta^2 z^2 (1 - z)^2,  \label{eq:tpt}\\
  \tmop{angular} \tmop{variable} : & t = & E^2 \theta^2,  \label{eq:tangle}
\end{eqnarray}
where the kinematic is illustrated in the following figure
\begin{equation}
\raisebox{-1cm}[1cm][1cm]{\epsfig{file=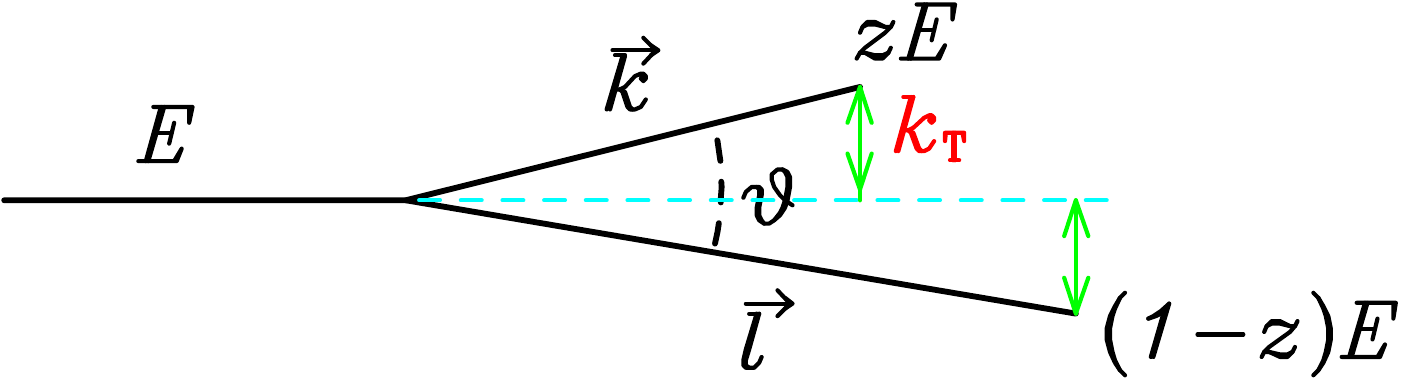,width=0.5\textwidth}}
     \label{fig:splitkin}
\end{equation}
where $E \approx (k + l)^0$, $\theta$ is the angle between $\vec{k}$ and
$\vec{l}$ and the $\approx$ relations hold for small $\theta$. Assuming that
there is nothing special about the $z \rightarrow 0$ and $z \rightarrow 1$
points, alternative choices in the definition of $t$ and $z$ make a difference
in subleading terms in eq.~(\ref{eq:factorization1}), that is to say, for terms
that are non-singular when $t \rightarrow 0$. Unfortunately, the $z
\rightarrow 1$ and $z \rightarrow 0$ points are special: in fact,
eq.~(\ref{eq:pqqap}) yields a divergent integration when $z \rightarrow 1$.
This is an infrared divergence in QCD, since when $z \rightarrow 1$ the energy
of the radiated gluon goes to zero. We will forget for the moment about this
complication, and deal with collinear divergences only. The treatment of the
soft region will be discussed later on.

The factorization of eq.~(\ref{eq:factorization1}) holds as long as the angle
(or, more generally, the $t$ variable) between the collinear partons is the
smallest in the whole amplitude. This is, in some sense, natural:
factorization holds if the intermediate quark with momentum $k + l$ can be
considered, to all effects, as if it was on shell, that is to say, its
virtuality must be negligible compared to all other energy scales entering the
amplitude. It follows then that factorization can be applied recursively to an
amplitude, to obtain its most singular contribution. This is shown pictorially
in the following graphical formula
\begin{equation}
\raisebox{-2.2cm}[1.6cm][2.1cm]{\epsfig{file=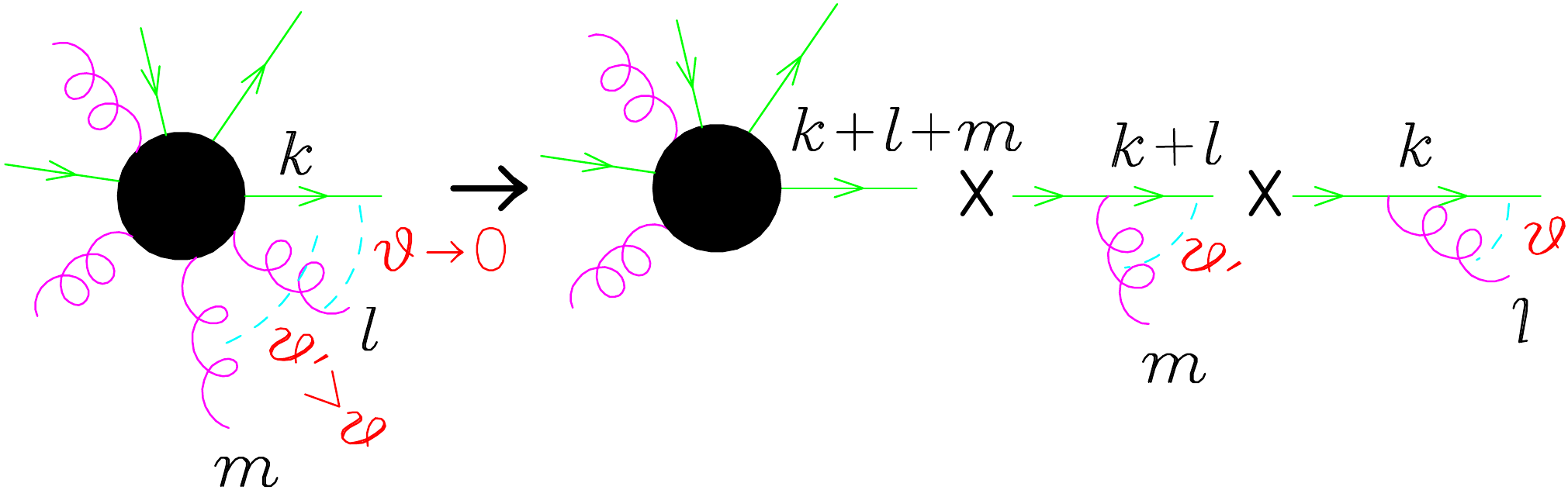,width=0.8\textwidth}},
     \label{fig:factorization2}
\end{equation}
where we have two angles becoming small, maintaining a strong ordering
relation, $\theta' \gg \theta \rightarrow 0$.

Factorization formulae, similar to the one for a $q g$ collinear configuration
(illustrated in eq.~(\ref{fig:factorization1}) and eq.~(\ref{eq:factorization1})),
also hold for the case of a $g g$, and $q \bar{q}$
collinear configuration, the only difference being in the form of the
splitting functions. We thus have three possibilities
\begin{eqnarray}
  P_{q, q g} (z) & = & C_F \hspace{0.25em} \frac{1 + z^2}{1 - z}, \nonumber\\
  P_{g, g g} (z) & = & C_A \left( \frac{z}{1 - z} + \frac{1 - z}{z} + z (1 -
  z) \right) \nonumber\\
  P_{g, q \bar{q}} (z) & = & t_f \left( z^2 + (1 - x)^2 \right)  
  \label{eq:apexclkernels}
\end{eqnarray}
Some of the $P_{i, j l} (z)$ functions are singular for $z \rightarrow 1$ or
$z \rightarrow 0$. These singularities have an infrared origin. In the
following, we tacitly assume that they are \ regularized by a tiny parameter
$\eta$
\begin{equation}
  \label{eq:APregeta} \frac{1}{1 - z} \Longrightarrow \frac{1}{1 - z + \eta},
  \frac{1}{z} \Longrightarrow \frac{1}{z + \eta} .
\end{equation}
Notice that the $P_{i, j l}$ functions in eqs. (\ref{eq:apexclkernels}) are
related to the standard{\footnote{In fact, the unregularized Altarelli-Parisi
splitting function. The difference with the standard, regularized splitting
function will be clarified later.}} Altarelli-Parisi splitting functions
{\cite{Altarelli:1977zs}}, that are given by
\begin{eqnarray}
  P_{g g} (z) & = & 2 P_{g, g g} (z), \nonumber\\
  P_{q q} (z) & = & P_{q, q g} (z), \nonumber\\
  P_{q g} (z) & = & P_{q, q g} (1 - z), \nonumber\\
  P_{g q} (z) & = & P_{g, q \bar{q}} (z) .  \label{eq:APsinc}
\end{eqnarray}
The difference lies in the fact that the Altarelli-Parisi splitting functions
tag one of the final state partons. Thus, in the $g \rightarrow g g$ case
there is an extra factor of 2, because we can tag either gluons. Similarly,
the $q \rightarrow q g$ splitting process is associated to two different
Altarelli-Parisi splitting functions, since one can tag the final quark or the
final gluon.

Strictly speaking, in the case of the $g \rightarrow g g$ and $g \rightarrow
q \bar{q}$ a complication arises: an azimuthal dependent term, that has zero
azimuthal average should be added to eq.~(\ref{eq:factorization1}). This term
is a consequence of the fact that, at fixed helicities of the final state $g
g$ or $q \bar{q}$ partons, the intermediate gluon can have two helicities, and
they can interfere. We will ignore this complication in the following,
reminding the reader that in some shower algorithms this angular correlation
effects are dealt with to some extent.

\subsection{Fixed order calculations}

The factorization formula, eq.~(\ref{eq:factorization1}), reminds us
immediately that real radiative corrections to any inclusive quantity are
divergent. This is better seen in the simple example of $e^+ e^- \rightarrow q
\bar{q}$. The real radiative corrections to this process are given by the $e^+
e^- \rightarrow q \bar{q} g$ emission process. When the gluon becomes
collinear to the quark or to the antiquark, eq.~(\ref{eq:factorization1})
implies that there is a divergent $d t / t$ integration. This divergence is,
of course, limited by some physical cutoff, like the quark masses, or by
confinement effects. But, even if we can reassure ourselves that no real
infinity arises, the divergence implies that the real cross section is
sensitive to low energy phenomena, that we cannot control or understand within
perturbative QCD. Furthermore, the divergence yields a factor $\alpha_S (Q)
\log Q / \lambda$, where $Q$ is the annihilation scale, and $\lambda$ some
typical hadronic scale, that acts as a cut-off. This factor is of order 1,
since $\alpha_S (Q)$ is of order $1 / \log Q / \lambda$. Fortunately, one can
show that, if virtual corrections are included, these divergences cancel,
thanks to a mechanism known as the Kinoshita-Lee-Nauemberg theorem. In the
case at hand, the order $\alpha_S$ virtual correction to the $e^+ e^-
\rightarrow q \bar{q}$ process contains a negative term behaving as \
$\alpha_S (Q) \log Q / \lambda$, that cancels the divergence in the real
emission term. Thus, the inclusive cross section, (that, being inclusive,
requires that we sum over both the $q \bar{q}$ and the $q \bar{q} g$ final
states) does not depend upon the cutoff $\lambda$, and gives rise to the
well-known $1 + \alpha / \pi$ correction factor to the total hadronic cross
section in $e^+ e^-$ annihilation. At the same time, however, it becomes clear
that it is impossible, at fixed order in QCD, to give a realistic description
of the final state.

\subsubsection{\label{sec:simQED}Similarities with QED}

The reader familiar with the infrared problem in QED will find there some
similarities with the problems discussed above. Also in QED, in order to get
finite cross sections at any finite order in perturbation theory, one has to
sum virtual contributions to real photon emission contributions, where photons
with energy below a given resolution must be included. Thus, also in QED, at
fixed order in the coupling constant, we cannot compute fully exclusive cross
sections: we must always sum inclusively over soft photons below the
resolution parameter.

While soft divergences are normally treated in textbooks on QED, collinear
divergences are seldom considered. In fact, in electrodynamics, the mass of
the electron screens the collinear divergences. This is easily understood: a
massive, on shell electron cannot decay into an electron plus a photon, unless
the photon has zero energy. At very high energy, however, the electron mass
becomes negligible, and one should also consider the collinear singularities
in QED. Charge particles, as well as photons, produced at ultra-high energy,
will give rise to true electromagnetic jets. Even at more moderate energies,
when considering, for example, the electron produced in the decay of a heavy
object, for the purpose of mass measurements, it is better to measure the
energy of the associated electromagnetic jet (as measured, for example, by an
electromagnetic calorimeter) rather than that of the electron (as measured by
a tracker), in order not to become sensitive to photon collinear emissions.

\subsection{Exclusive final states}

In order to describe the exclusive, detailed final state, we must thus sum the
perturbative expansion to all orders in $\alpha_S$. This is in fact possible
if we limit ourselves to the most singular terms of the perturbative
expansion, that is to say, all terms that carry the collinear singularities $d
t / t$, in strongly ordered sequences of angles. Sticking to our $e^{+_{}}
e^-$ example, we consider configurations where the final state $q$ and
$\bar{q}$ split into a $q g$ ($\bar{q} g$) pair at small angle. Each final
state parton is allowed to split in turn into a pair of partons with even
smaller angle. Thanks to the factorization properties of the amplitude, one
can easily estimate the corresponding cross section. If one allows for $n$
splitting processes, the cross section goes as
\begin{equation}
  \label{eq:leadinglog} \sigma_0 \alpha_S^n \int \frac{d t_1}{t_1}  \frac{d
  t_2}{t_2} \ldots \frac{d t_n}{t_n} \times \theta (Q^2 > t_1 > t_2 > \ldots >
  t_n > \lambda^2) = \sigma_0 \frac{1}{n!} \alpha_S^n \log^n
  \frac{Q^2}{\lambda^2},
\end{equation}
where $Q$ is the annihilation energy (that provides an upper cut-off to the
virtualities in the splitting processes) and $\lambda$ is an infrared cut-off.
The $\theta$ function here is defined to be equal to 1 if its argument is
true, zero otherwise. It is because of eq.~(\ref{eq:leadinglog}) that the
collinear approximation is sometimes called leading log approximation. As
discussed previously, virtual corrections to all orders in perturbation theory
yield a comparable term. Their leading logarithmic contribution should then be
included in order to get sensible results.

\subsection{Counting logs}

The leading logarithmic approximation requires some more explanation. Let us
look at a simplified factorization formula
\begin{equation}
  M_1 d \Phi_1 \approx M_0 \frac{d t}{t},
\end{equation}
that holds when $t \ll Q^2$, $Q$ being the typical scales in the amplitude
$M_1$. We have
\begin{equation}
  \int M_1 d \Phi_1 = M_0 \int \frac{d t}{t} \theta (Q^2 > t > \lambda^2) +
  \mathcal{O} (1) = \log \frac{Q^2}{\lambda^2} + \mathcal{O} (1) .
\end{equation}
which follows from the fact that in the difference
\begin{equation}
  \int M_1 d \Phi_1 - M_0 \int \frac{d t}{t} 
\end{equation}
the singularity for small $t$ cancels. Thus the difference must be of order 1.
So, even if we have said that the factorization formula holds for $t \ll Q^2$,
in order to get the leading logarithm, we can integrate it for $t$ up to
$Q^2$. And furthermore, if we instead integrate it, for example, up to $Q^2 /
2$ instead of $Q^2$, the difference is $\log 2$, and thus is of order 1, and
the leading logarithm remains the same.

\subsection{Leading log calculation of multiparticle production}

I will now just give the recipe for the calculation of our multiparticle cross
section, with the inclusion of the virtual corrections at the leading log
level. The outcome of the recipe is the cross section associated to each given
final state. We assume that we start from some hard process, like, for
example, the production of a $q \bar{q}$ pair in $e^+ e^-$ annihilation. The
cross section for the hard process is computed by usual means. The recipe
tells us how to compute a weight for the evolution of each coloured parton in
the hard process into an arbitrary number of coloured partons.

We begin by specifying how to construct all possible event structures:
\begin{enumerateroman}
  \item \label{enum:recipe1}We choose a Born kinematics, specifying the hard
  interaction.
  
  \item For each primary coloured parton produced in the hard interaction, we
  consider all possible tree-level graphs that can arise from it, obtained by
  letting the quark split into a $q g$ pair, the gluon split into a $g g$ or
  $q \bar{q}$ pair for any quark flavour, as many times as one wishes.
  
  \item With each splitting vertex in the graph, one associates a $t$, $z$,
  and $\phi$ value.
  
  \item \label{enum:tordering}One imposes that the $t$ are ordered: the $t$
  for splitting near the hard process must be less than the hard process scale
  $Q^2$, and all subsequent $t$'s are in decreasing order as we go toward the
  branches of the tree-graph.
  
  \item \label{enum:momentumreshuffling}Given the initial hard parton momenta,
  and the $t$, $z$ and $\varphi$ variables at each splitting vertex, one
  reconstructs all the momenta in the tree graph.
\end{enumerateroman}
We now specify the weight to be assigned to the given configuration:
\begin{enumeratealpha}
  \item \label{enum:recipe2}The hard process has weight equal to its
  differential (Born level) cross section.
  
  \item \label{enum:vertexweight}Each vertex has the weight
  \begin{equation}
    \label{eq:splitvert} \theta (t - t_0) \hspace{0.75em} \frac{\alpha_S
    (t)}{2 \pi} \hspace{0.25em} \frac{dt}{t} \hspace{0.25em} P_{i, j l} (z)
    \hspace{0.25em} dz \hspace{0.25em} \frac{d \phi}{2 \pi}
  \end{equation}
  where $\alpha_S (t)$ is the QCD running coupling
  \begin{equation}
    \label{eq:alfas1loop} \alpha_S (t) = \frac{1}{b_0 \log
    \frac{t}{\Lambda_{\tmop{QCD}}^2}} .
  \end{equation}
  In order not to reach unphysical values of the running coupling constant, \
  we must introduce an infrared cutoff $t_0 > \Lambda_{\tmop{QCD}}^2$. The
  $\theta$ function in eq.~(\ref{eq:splitvert}) sets the lower bound on $t$.
  The upper bound is determined by the $t$ ordering of point
  (\ref{enum:tordering}).
  
  \item \label{enum:sudakov}Each line in the graph has weight $\Delta_i (t',
  t'')$, where $t'$ is the $t$ value associated with the upstream vertex,
  $t''$ with the downstream vertex, and
  \begin{equation}
    \label{eq:sudadef} \Delta_i (t', t'') = \exp \left[ - \sum_{(j l)}
    \int_{t''}^{t'} \frac{dt}{t} \int_0^1 d z \hspace{0.75em} \frac{\alpha_S
    (t)}{2 \pi} \hspace{0.25em} \hspace{0.25em} P_{i, j l} (z) \right]
  \end{equation}
  In case the line is a final one, $t''$ is replaced by an infrared cutoff
  $t_0$. The weights $\text{$\Delta_i (t', t'')$}$ are called Sudakov form
  factors. They represent all the dominant virtual corrections to our tree
  graph.
\end{enumeratealpha}
At the end of this procedure, some hadronization model will be invoked, in
order to convert the showered final state partons into hadrons. For now, in
order to better clarify the shower mechanisms, we will just neglect the
hadronization stage, and consider the final states (and the initial states) as
made of partons.

The form of the weight at (\ref{enum:vertexweight}) is simply a consequence
of a recursive application of the factorization formula. The prescription for
the argument of $\alpha_S$ and the Sudakov form factors (\ref{enum:sudakov})
are slightly more subtle: they arise from the inclusion of all leading-log
virtual corrections to the process.

\subsubsection{Momentum reshuffling}

The final momentum assignment of step \ref{enum:momentumreshuffling} is
affected by some ambiguities, due to the fact that a parton line, when
followed by a splitting process, acquires a positive virtuality larger than
its mass. Because of these virtualities, the momenta of the parton must be
adjusted, in order to conserve energy and momentum. For example, in the
process $e^+ e^- \rightarrow q \bar{q}$, the initial quarks have energy $Q /
2$, and (neglecting masses) momenta equal to their energy and opposite. If the
quark undergoes a splitting process, it can no longer be considered an
on-shell parton, and thus its momentum must be adjusted according to the
standard formulae for two body decays, including the effect of the masses of
the decay products. This procedure (referred to as \tmtextit{momentum
reshuffling}) does not affect the leading logarithmic structure of the result.

\subsection{Typical structure of a shower}

\begin{figure}[h]
\begin{center}
\epsfig{file=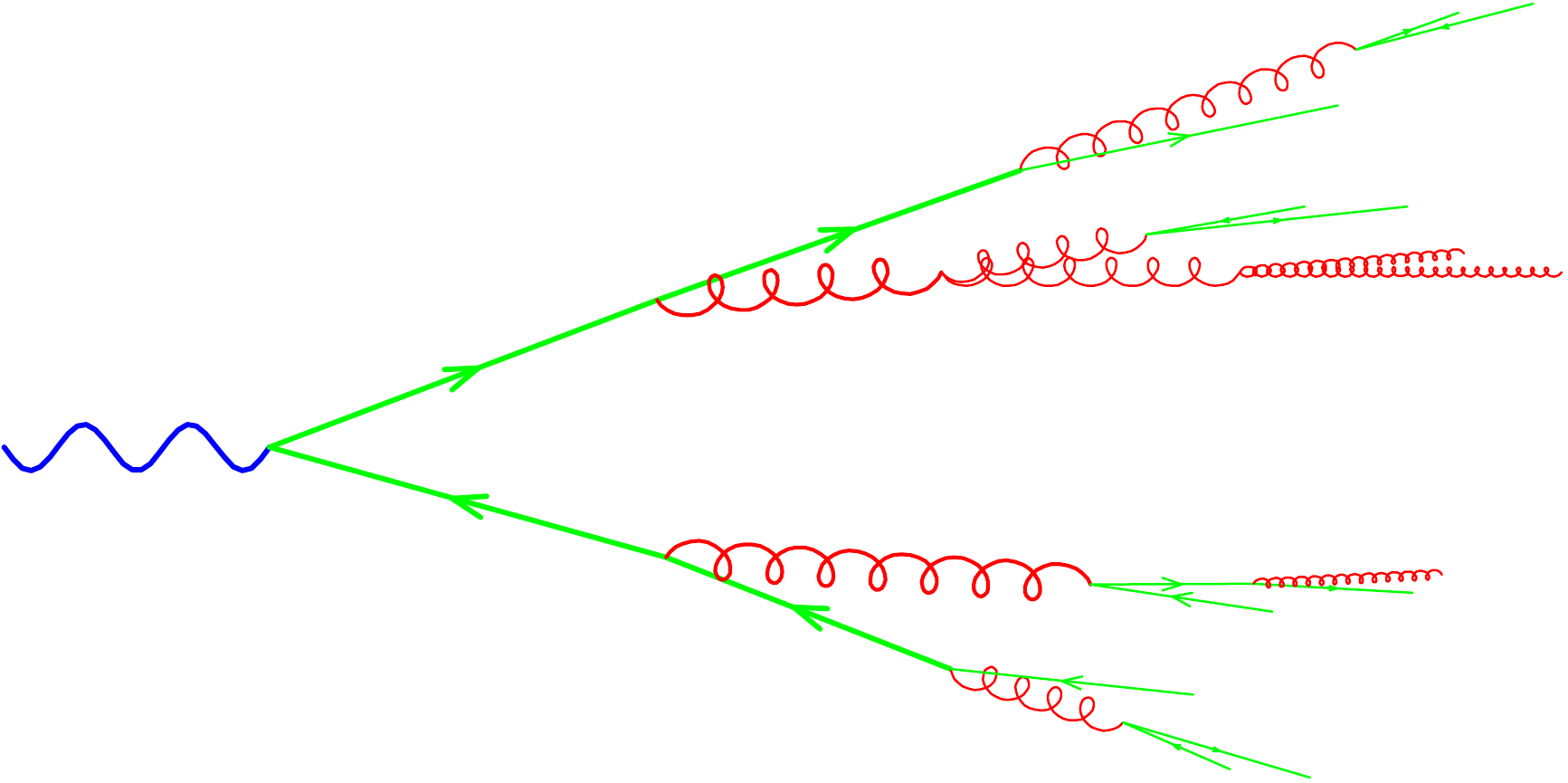,width=0.8\textwidth}
\end{center}
\caption{\label{fig:typicalsh} Typical shower development.}
\end{figure}

\begin{figure}[h]
  \begin{center}
    \epsfig{file=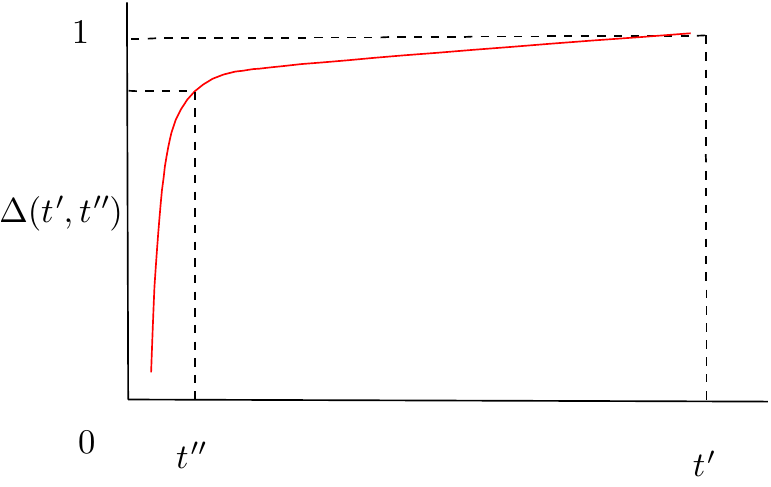}
  \end{center}
  \caption{\label{fig:Sudaplot}Typical behaviour of the Sudakov form
       factor.}
\end{figure}

According to the recipe (\ref{enum:recipe1}-\ref{enum:momentumreshuffling})
and (\ref{enum:recipe2}-\ref{enum:sudakov}), the shower will be characterized
by a tree of splittings with decreasing angles,
as depicted in figure~\ref{fig:typicalsh}.
At a given splitting vertex, the splitting angle will be typically smaller by
a factor $\alpha_S$ than its upstream angle. As the angles become small, they
will reach a point where the scale $t$ is of the order of
$\Lambda_{\tmop{QCD}}$, so that $\alpha_S \approx 1$, angles are no longer
ordered and the whole picture breaks down. At this stage the shower stops, and
some model of hadronization is needed in order to complete the description of
the formation of the final state. Notice also the role of the Sudakov form
factors of eq.~(\ref{eq:sudadef}). They suppress the configurations containing
lines with very large differences between upstream and downstream angles. In
fact, using eq.~(\ref{eq:alfas1loop}) we estimate
\begin{equation}
  \Delta_i (t', t'') \approx \exp \left[ - C \int_{t''}^{t'} \frac{d t}{t}
  \hspace{0.75em} \hspace{0.25em} \frac{1}{\log
  \frac{t}{\Lambda_{\tmop{QCD}^{}}^2} } \hspace{0.25em} \right] = \left(
  \frac{\log \frac{t''}{\Lambda_{\tmop{QCD}}^2}}{\log
  \frac{t'}{\Lambda_{\tmop{QCD}}^2}} \right)^C,
\end{equation}
which becomes very small if $t' \gg t''$. The behaviour of $\Delta$ as a
function of $t$ is shown in fig. \ref{fig:Sudaplot}.
As can be seen from eq.~(\ref{eq:sudadef}) and from figure \ref{fig:Sudaplot},
the Sudakov form factor suppresses the configurations that have no radiation
down to very small scales.

\subsection{Formal representation of a shower\label{sec:pictshower}}

In the following we will introduce some graphical notation for the
representation of a shower. We use the symbol
\begin{equation}
  \label{eq:showerdef}  \text{$\mathcal{S}_i (t, E)$} = \begin{array}{l}
 \raisebox{-0.7cm}[1cm][0.5cm]{\epsfig{file=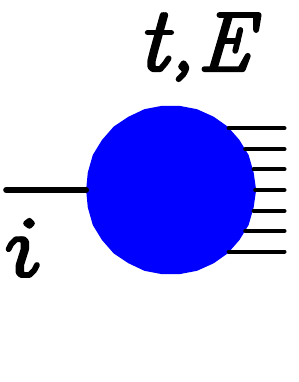,width=1.5cm}}
  \end{array},
\end{equation}
to represent the ensemble of all possible showers originating from parton $i$
at a scale $t$. The dependence of the shower upon the parton direction is not
explicitly shown, since we will not need it in the following. We can think of
$\mathcal{S}_i (t, E)$ as a function defined on the set $\mathcal{F} $of all
final states (by final state we mean here a set of partons with specific
momentum assignments), yielding the weight of the shower for that particular
final state. When writing
\begin{equation}
  \sum_{\mathcal{F}} \mathcal{S}_i (t, E),
\end{equation}
we mean sum over all final states $\mathcal{F}$, i.e. the total weight of the
shower attached to parton $i$. Of course, $\mathcal{F}$ is not a discrete set,
so, rather than a sum we should have a sum over the number and type of final
state particles and an integral over their momenta. Alternatively we may
imagine to divide the phase space into small cells, so that $\mathcal{F}$ can
be imagined as a discrete set, and the sum notation is appropriate.

\subsection{Shower equation}

We can easily convince ourselves that the rules given in items
(\ref{enum:recipe1}-\ref{enum:momentumreshuffling}) imply a recursive
equation, that is illustrated in the following graphical equation
\begin{equation}
\raisebox{-1.8cm}[1.8cm][1.8cm]{\epsfig{file=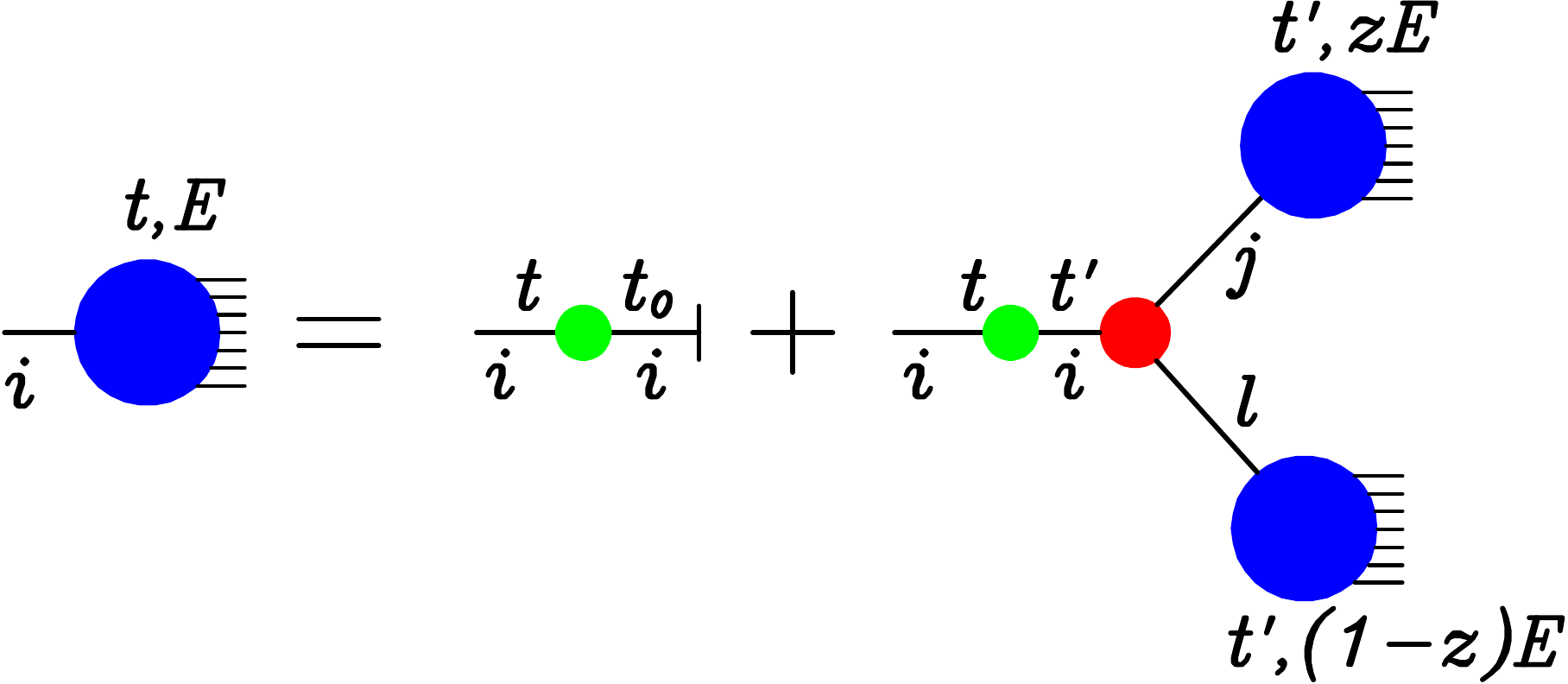,width=0.6\textwidth}}
     \label{fig:showereq}
\end{equation}
The meaning of the figure is quite intuitive: the ensemble of all possible
shower histories is obtained by adding the case in which no branching takes
place{\footnote{In this case the shower terminates with the given final state
parton. The hadronization model will take over when all showers are
terminated, building up the hadrons from the given set of coloured partons.
}}, to the case where one branching occurs, followed recursively by two
showers starting at smaller energies and scales. The small blobs along the
parton lines represent the Sudakov form factors, and the blob connecting the
$i, j, l$ partons is the splitting probability. Notice that the phase spaces
of the two independent showers, after the splitting, do not overlap in our
collinear approximation, because the angle at the vertex $t'$ is larger than
all subsequent angles.

The mathematical translation of eq.~(\ref{fig:showereq}) is given by the
equation
\begin{equation} 
   \text{$\mathcal{S}_i (t, E)$} = \Delta_i (t, t_0) \mathcal{S}_i (t_0, E)+
\nonumber
\end{equation}\begin{equation}
  \sum_{(j l)} \int_{t_0}^t \hspace{0.25em} \frac{dt'}{t'}
  \hspace{0.25em} \int_0^1 d z \int_0^{2 \pi} \frac{d \phi}{2 \pi}
  \hspace{0.75em} \frac{\alpha_S (t')}{2 \pi} P_{i, j l} (z) \hspace{0.25em}
  \hspace{0.25em} \Delta_i (t, t') \text{$\mathcal{S}_j (t', z E)$}
  \text{$\mathcal{S}_l (t', (1 - z) E)$}, \label{eq:showereq}
\end{equation}
where the two terms correspond to the terms in the figure: no branching, plus
one branching followed by two showers. $\mathcal{S}_i (t_0, E)$ represents the
final state consisting of the incoming particle $i$ alone, that has undergone
no branching (since no branching is possible below $t_0$). Notice that the
shower diagram for $\mathcal{S}_i (t_0, E)$ consist of a single line with the
Sudakov form factor $\Delta_i (t_0, t_0) = 1$, i.e. the shower assigns
probability 1 for particle $i$ to remain the same (i.e. to undergo no
branchings).

We can easily see that $\mathcal{S}$ satisfies the differential equation
\begin{eqnarray}
  t \frac{\partial \text{$\mathcal{S}_i (t, E)$}}{\partial t} & = & \sum_{(j
  l)} \int_0^1 d z \int_0^{2 \pi} \frac{d \phi}{2 \pi} \hspace{0.75em}
  \frac{\alpha_S (t)}{2 \pi} \hspace{0.25em} P_{i, j l} (z) \hspace{0.25em} 
  \text{$\mathcal{S}_j (t, z E)$} \text{$\mathcal{S}_l (t, (1 - z) E)$}
  \nonumber\\
  & + & \left[ - \sum_{(j l)} \int_0^1 d z \hspace{0.75em} \frac{\alpha_S
  (t)}{2 \pi} \hspace{0.25em} P_{i, j l} (z) \hspace{0.25em} \right]
  \mathcal{S}_i (t, E),  \label{eq:showdiff}
\end{eqnarray}
that arises because the derivative with respect to $t$ can act on the upper
limit of the integral in the second term of eq.~(\ref{eq:showereq}), giving
rise to the first term of eq.~(\ref{eq:showdiff}), or on the Sudakov form
factors in both terms of eq.~(\ref{eq:showereq}), giving rise to the square
bracket term in eq.~(\ref{eq:showdiff}). Eq.~(\ref{eq:showdiff}) is
particularly instructive. It has the following meaning: if we raise the scale
of the process by an infinitesimal amount, the shower has a larger probability
to split into two subshowers (the first term on the right hand side), and a
smaller probability to remain the same (the second term). By summing eq.
(\ref{eq:showdiff}) over all possible final state, and defining
\begin{equation}
  \mathcal{S}_i^{\tmop{inc}} (t, E) = \sum_{\tmop{final} \tmop{states}}
  \mathcal{S}_i (t, E),
\end{equation}
we see that $\mathcal{S}_i^{\tmop{inc}} (t, E)$ obeys the equation
\begin{eqnarray}
  t \frac{\partial \text{$S_i^{\tmop{inc}} (t, E)$}}{\partial t} & = &
  \sum_{(j l)} \int_0^1 \hspace{0.25em} dz \hspace{0.75em} \frac{\alpha_S
  (t)}{2 \pi} \hspace{0.25em} P_{i, j l} (z) \text{$S_j^{\tmop{inc}} (t, z
  E)$} \text{$S_l^{\tmop{inc}} (t, (1 - z) E)$} \nonumber\\
  & + & \left[ - \sum_{(j l)} \int_0^1 d z \hspace{0.75em} \frac{\alpha_S
  (t)}{2 \pi} \hspace{0.25em} P_{i, j l} (z) \hspace{0.25em} \right]
  S_i^{\tmop{inc}} (t, E) .  \label{eq:showdiffinc}
\end{eqnarray}
We immediately see that $S_i^{\tmop{inc}} (t, E) = 1$ satisfies the above
equation, and is also consistent with the obvious initial condition
$S_i^{\tmop{inc}} (t_0, E) = 1$. We thus
state the {\tmem{shower unitarity}} property
\begin{equation}
  \mathcal{S}_i^{\tmop{inc}} (t, E) = \sum_{\mathcal{\mathcal{F}}}
  \mathcal{S}_i (t, E) = 1 .
\end{equation}
This property is at the basis of the formulation of the shower Monte Carlo
algorithms. It has the following important consequence: \tmtextit{the total
cross section computed at the Born level is equal to the total multiparticle
cross section}. Of course, this statement holds in the approximation we are
working with. Since we are only considering collinear-enhanced corrections, we
should state more precisely that the net effect of collinear-enhanced
processes is one, when we sum over all processes. We also remind the reader
that this result also holds in QED. As known from textbooks QED, large soft
effects cancel in inclusive cross sections, leaving only small (i.e.
$\mathcal{O} (\alpha)$) corrections to the Born cross section. The same is
true also for collinear divergences, a fact that (as already remarked in
\ref{sec:simQED}) should be kept in mind when considering final stated with
electrons at the LHC.

It is also instructive to check unitarity by expanding the shower order by
order in $\alpha_S$. At order $\alpha_S$, for example, we may have at most a
single splitting, since each splitting cost a factor $\alpha_S$. When we sum
over all final states reached by parton $i$, we should thus consider only the
one and two parton final state. The weight of the one parton final state, at
order $\alpha_S$ is just the Taylor expansion of the Sudakov form factor at
order $\alpha_S$
\begin{equation}
  \label{eq:oneloalf} \Delta_i (Q, t_0) = 1 - \sum_{(j l)} \int_{t_0}^Q
  \frac{d t}{t} \int_0^1 d z \hspace{0.75em} \frac{\alpha_S}{2 \pi}
  \hspace{0.25em} P_{i, j l} (z) + \mathcal{O} (\alpha_S^2),
\end{equation}
while the total weight for a two parton final state is
\begin{eqnarray}
  &  & \int_{t_0}^Q \frac{d t}{t} \Delta_i (Q, t) \left[ \sum_{(j l)}
  \int_0^{2 \pi} \frac{d \phi}{2 \pi} \int_0^1 d z \hspace{0.75em}
  \frac{\alpha_S}{2 \pi} \hspace{0.25em} P_{i, j l} (z) \right] \Delta_j (t,
  t_0) \Delta_l (t, t_0) \nonumber\\
  &  & = \int_{t_0}^Q \frac{d t}{t} \sum_{(j l)} \int_0^1 d z \hspace{0.75em}
  \frac{\alpha_S}{2 \pi} \hspace{0.25em} P_{i, j l} (z) + \mathcal{O}
  (\alpha_S^2), 
\end{eqnarray}
that summed to eq (\ref{eq:oneloalf}) yields 1. At this point, one can see
that the form of the Sudakov form factor is dictated by the fact that
collinear singularities, according to the Kinoshita-Lee-Nauenberg theorem,
must cancel.

Shower unitarity makes it possible to write the branching process as a
sequence of independent branching processes (i.e. as a Markov chain). In fact,
after a branching, the total weight of the two newly initiated subshowers is
one, i.e. they do not influence that branching process we are considering.

\subsection{Shower algorithm for final state showers}

It is apparent now that the development of the shower can be computed
numerically using a simple probabilistic algorithm. We interpret
\begin{equation}
  \frac{\alpha_S (t')}{2 \pi} \hspace{0.25em} \frac{dt'}{t'} \hspace{0.25em}
  P_{i, j l} (z) \hspace{0.25em} dz \hspace{0.25em} \frac{d \phi}{2 \pi}
\end{equation}
as the elementary branching probability in the phase space element $d t', d z,
d \phi$. So
\begin{equation}
  \frac{\alpha_S (t')}{2 \pi} \hspace{0.25em} \frac{dt'}{t'} \hspace{0.25em}
  \int_0^1 d z P_{i, j l} (z)
\end{equation}
is the branching probability in the $d t'$ interval. Now we notice that,
dividing the $[t, t']$ interval into $N$ small subintervals of width $\delta
t$, calling $t_i$ the center of each subinterval, we have
\begin{equation}
  \Delta_i (t, t') = \prod_{i = 1}^N \left( 1 - \frac{\alpha_S (t_i)}{2 \pi}
  \hspace{0.25em} \frac{\delta t}{t_i} \hspace{0.25em} \int P_{i, j l} (z)
  \hspace{0.25em} dz \hspace{0.25em} \frac{d \phi}{2 \pi} \right),
\end{equation}
that is to say, the Sudakov form factor corresponds to the non-emission
probability in the given $[t, t']$ interval. The probability that, starting at
the scale $t$, the first branching is in the phase space element $d t', d z, d
\phi$, is then
\begin{equation}
  \label{eq:emissionProb} \Delta_i (t, t') \frac{\alpha_S (t')}{2 \pi}
  \hspace{0.25em} \frac{dt'}{t'} \hspace{0.25em} P_{i, j l} (z)
  \hspace{0.25em} dz \hspace{0.25em} \frac{d \phi}{2 \pi},
\end{equation}
i.e. is the product of the no-branching probability from the scale $t$ down to
$t'$ times the branching probability in the interval $d t', d z, d \phi$. This
is precisely equivalent to our shower recipe, if we remember that, because of
unitarity, the total weight associated to further branchings of partons $i$
and $j$ is 1.

One can easily set up an algorithm for the generation of the process:
\begin{enumeratealpha}
  \item Generate a hard process configuration with a probability proportional
  to its parton level cross section (for example, for the $e^+ e^- \rightarrow
  \tmop{hadrons}$ case the configuration consists of two back-to-back quarks,
  with energy $Q / 2$, distributed as $(1 + \cos^2 \theta) d \cos \theta d
  \phi$,). $Q$ is in this case the typical scale of the process.
  
  \item For each final state coloured parton, generate a shower in the
  following way:
  \begin{enumerateroman}
    \item Set $t = Q$
    
    \item \label{enum:genrand} Generate a random number $0 <
    r < 1$.
    
    \item Solve the equation $r = \Delta_i (t, t')$ for $t'$.
    
    \item If $t' < t_0$ then no further branching is generated, and the shower
    stops.
    
    \item \label{enum:mcbran1}If $t' \geqslant t_0$ then generate $j l$ and $z$
    with a distribution proportional to $P_{i, j l} (z)$, and a value for the
    azimuth $\phi$ with uniform probability in the interval $[ 0, 2 \pi]$.
    Assign energies $E_j = z E_i$ and $E_l = (1 - z) E_i$ to partons $j$ and
    $l$. The angle between their momenta is fixed by the value of $t'$. Given
    the angle and the azimuth $\phi$ (together with the fact that the sum of
    their momenta must equal to the momentum of $i$) the directions of $j$ and
    $l$ are fully reconstructed
    
    \item For each of the branched partons $j$ and $l$, set $t = t'$ and go
    back to step \ref{enum:genrand}.
  \end{enumerateroman}
\end{enumeratealpha}

\subsection{A very simple example}

The branching algorithm in a Shower Monte Carlo resembles closely the problem
of the generation of decay events from a radioactive source. We call $p d t$
the elementary radiation probability in the time interval $d t$. The
probability $\Delta (t')$ of having no radiation from time 0 up to time $t'$
is given by the product of no-radiation probability in each time subinterval
from 0 to $t'$
\begin{equation}
  \Delta (t) = (1 - p d t)^{\frac{t'}{d t}} = \exp [- p t']
\end{equation}
and the probability distribution for the first emission is
\begin{equation}
  \exp [- p t'] p d t' = - d \Delta (t') .
\end{equation}
Thus, the probability distribution for the first emission is uniform in
$\Delta (t')$; in order to generate the first emission at time $t'$, $0 < t' <
t$, we generate a random number $0 < r < 1$ and solve for $r = \Delta (t') /
\Delta (t)$.

\subsection{The inclusive cross section for single hadron production}

We will now compute the inclusive cross section for single hadron production,
and show that it obeys the Altarelli-Parisi equation for fragmentation
function. We begin by defining the fragmentation function
\begin{equation}
  \label{eq:showerdefsinc} D_i^m (t, x) = \frac{1}{\delta x} 
  \sum_{\mathcal{F} (m, x, \delta x)} \mathcal{S}_i (t, E) = \begin{array}{l}
   \raisebox{-0.8cm}[1.2cm][0.8cm]{\epsfig{file=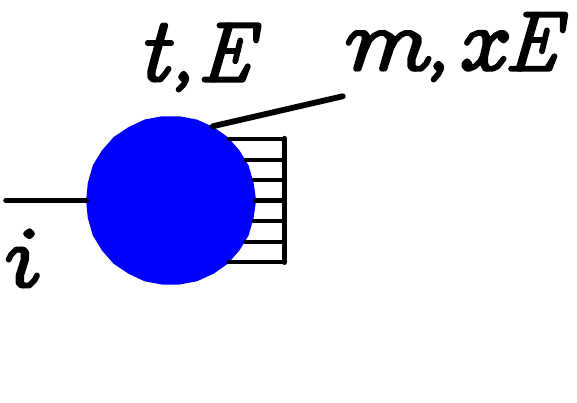,width=3cm}}
  \end{array},
\end{equation}
where $\mathcal{F} (m, x, \delta x)$ stands here for all final states having a
parton of type $m$ with energy between $x E$ and $(x + \delta x) E$. Notice
that $D_i^m (t, x)$ does not depend upon the absolute value of the energy,
since the shower recipe only involves energy fractions. It is easy to see that
the fragmentation function must obey the equation represented below
\begin{equation}
     \raisebox{-1.7cm}[1.7cm][1.7cm]{\epsfig{file=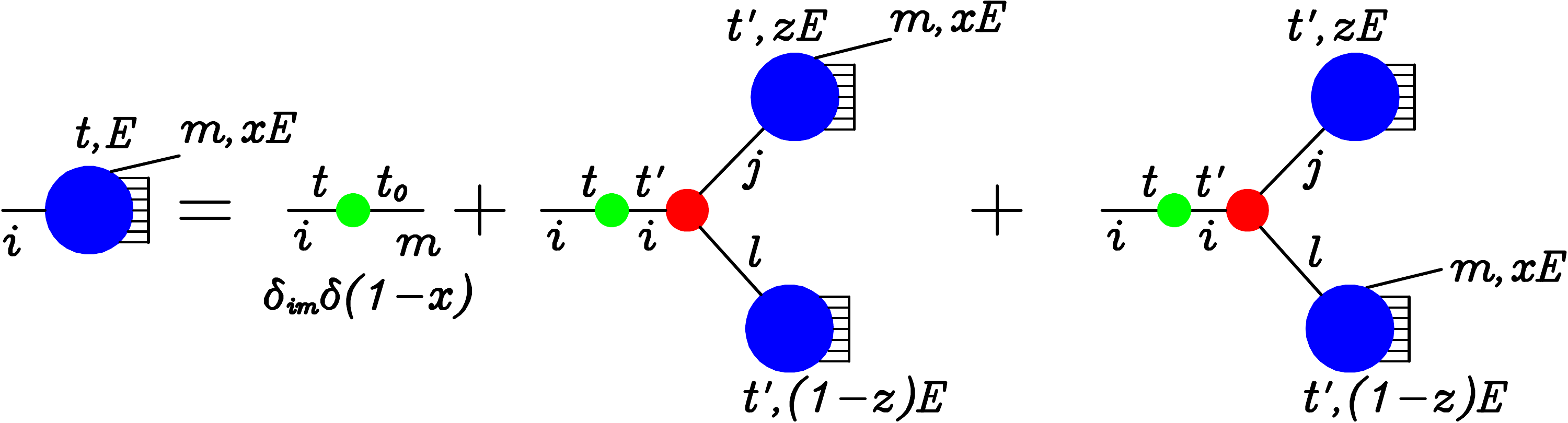,width=0.9\textwidth}}
     \label{fig:showersinc}
\end{equation}
The meaning of the equation is quite simple. If we want to keep a final state
particle of type $m$ and energy $x E$, the no-radiation term can contribute
only if $i = m$ and $x = 1$. In case a splitting takes place, particle $m$ can
be found in either of the two following showers. The graphical equation in
eq.~(\ref{fig:showersinc}) can be written more precisely as follows
\begin{eqnarray}
  D_i^m (t, x) & = & \Delta_i (t, t_0) \delta_{i m} \delta (1 - x) + \sum_{(j
  l)} \int_{t_0}^t \frac{dt'}{t'} \int_x^1 \frac{d z}{z} \hspace{0.75em}
  \frac{\alpha_S (t')}{2 \pi} \hspace{0.25em} \hspace{0.25em} P_{i, j l} (z)
  \hspace{0.25em} \hspace{0.25em} \Delta_i (t, t') D^m_j (t', x / z)
  \nonumber\\
  & + & \sum_{(j l)} \int_{t_0}^t \frac{dt'}{t'} \int_0^{1 - x} \frac{d z}{1
  - z} \hspace{0.75em} \frac{\alpha_S (t')}{2 \pi} \hspace{0.25em}
  \frac{dt'}{t'} \hspace{0.25em} P_{i, j l} (z) \hspace{0.25em}
  \hspace{0.25em} \Delta_i (t, t') D^m_l (t', x / (1 - z)) .  \label{eq:deq}
\end{eqnarray}
The presence of the $1 / z$ and $1 / (1 - z)$ on the middle member of eq.
(\ref{eq:deq}) is better understood if we imagine to multiply everything by
$\delta x$; we see then that $D (t, x)$ is multiplied by $\delta x$, and $D
(t', x / z)$, $D (t', x / (1 - z))$ are multiplied by $\delta x / z$ and
$\delta x / (1 - z)$ respectively, as the definition of $D$ suggests. As a
consequence of eq.~(\ref{eq:showdiff}), $D_i (t, x)$ must also satisfy the
equation
\begin{eqnarray}
  t \frac{\partial \mathcal{} D_i (t, x)}{\partial t} & = & \sum_{(j l)}
  \int_0^1 \hspace{0.75em} \frac{\alpha_S (t)}{2 \pi} \hspace{0.25em} P_{i, j
  l} (z) \hspace{0.25em} \frac{d z}{z} D_j (t, x / z) \nonumber\\
  & + & \sum_{(j l)} \int_0^1 \hspace{0.75em} \frac{\alpha_S (t)}{2 \pi}
  \hspace{0.25em} P_{i, j l} (z) \hspace{0.25em} \frac{dz}{z} \hspace{0.25em}
  D_l (t, x / (1 - z)) \nonumber\\
  & + & \left[ - \sum_{(j l)} \int_0^1 d z \hspace{0.75em} \frac{\alpha_S
  (t)}{2 \pi} \hspace{0.25em} P_{i, j l} (z) \hspace{0.25em} \right] D_i (t,
  x),  \label{eq:showdiffsinc}
\end{eqnarray}
Eq.~(\ref{eq:showdiffsinc}) is just another way of writing the
Altarelli-Parisi equations for fragmentation functions. Let us see in details
how this works. We replace $z \rightarrow 1 - z$ in the second term on the
right hand side of eq.~(\ref{eq:showdiffsinc}), and then use eqs.
(\ref{eq:APsinc}) to combine it with the first term. We get
\begin{eqnarray}
  t \frac{\partial D_i^m (t, x)}{\partial t} & = & \frac{\alpha_S (t)}{2 \pi}
  \hspace{0.25em} \sum_j \int_x^1 \frac{dz}{z} \hspace{0.25em} P_{i j} (z)
  \hspace{0.25em} \hspace{0.25em} D^m_j (t, x / z) \nonumber\\
  & + & \left[ - \sum_{(j l)} \int_0^1 d z \hspace{0.75em} \frac{\alpha_S
  (t)}{2 \pi} \hspace{0.25em} P_{i, j l} (z) \hspace{0.25em} \right] D_i (t,
  x) \nonumber\\
  & = & \frac{\alpha_S (t)}{2 \pi} \hspace{0.25em} \int_0^1 d z
  \hspace{0.25em} \left[ \frac{1}{z} \sum_j P_{i j} (z) \hspace{0.25em} D^m_j
  (t, x / z) \theta (z - x) - D_i^m (t, x) \sum_{(j l)} P_{i, j l} (z) \right]
  \nonumber\\
  & = & \frac{\alpha_S (t)}{2 \pi} \hspace{0.25em} \sum_j \int_x^1 \frac{d
  z}{z} \hspace{0.25em} \widehat{P}_{i j} (z) \hspace{0.25em} D^m_j (t', x /
  z) .  \label{eq:APfromshower}
\end{eqnarray}
In the last equality we have introduced the regularized Altarelli-Parisi
splitting functions $\hat{P}_{i j}$. They are defined as follows
\begin{eqnarray}
  \widehat{P_{}}_{q g} (z) & = & P_{q g} (z), \nonumber\\
  \widehat{P_{}}_{g q} (z) & = & P_{g q} (z), \nonumber\\
  \widehat{P_{}}_{q q} (z) & = & P_{q q} (z) - \delta (1 - z) \int_0^1 P_{q, q
  g} (z) d z, \nonumber\\
  \widehat{P_{}}_{g g} (z) & = & P_{g g} (z) - \delta (1 - z) \int_0^1 \left[
  P_{g, g g} (z) + P_{g, q \bar{q}} (z) \right] d z .  \label{eq:APreg}
\end{eqnarray}
It is easy to verify that the above definitions are equivalent to the usual
regularized Altarelli-Parisi splitting functions, defined in terms of the so
called ``+'' distributions
\begin{eqnarray}
  \widehat{P_{}}_{q q} (z) & = & C_F \left[ \frac{1 + z^2}{(1 - z)_+} +
  \frac{3}{2} \delta (1 - z) \right], \nonumber\\
  \widehat{P_{}}_{g g} (z) & = & 2 C_A \left[ \frac{z}{(1 - z)_+} + \frac{1 -
  z}{z} + z (1 - z) + \left( \frac{11}{12} - \frac{n_f T_f}{3 C_A} \right)
  \delta (1 - z) \right], 
\end{eqnarray}
by using the property
\begin{equation}
  \frac{1}{1 - z + \eta} - \log \frac{1}{\eta} \delta (1 - z) \Longrightarrow
  \frac{1}{(1 - z)_+} . \label{eq:pluspres}
\end{equation}

\subsection{Initial state radiation}

Until now, we have considered the problem of collinear splitting affecting
final state partons. The phenomenon of collinear splitting of initial state
partons is also relevant for hadronic collisions, and is commonly called
\tmtextit{initial state radiation} (ISR from now on). The reader familiar with
LEP physics will certainly remember the importance of QED ISR in $e^+ e^-$
collisions near the $Z$ peak. QCD \ ISR is fully analogous, from a formal
point of view, to QED ISR. There are, however, a few important differences:
\begin{itemizedot}
  \item The QCD coupling is much larger: thus QCD ISR is even more important.
  
  \item The QCD coupling grows for small momentum transfer. Thus we can never
  neglect ISR in QCD.
\end{itemizedot}
Because of these differences, while for QED initial state radiation at LEP it
was enough to work at one or two orders in the electromagnetic coupling, in
QCD one has to resort to an all order treatment. In other words, in QCD
initial state quarks and gluons \tmtextit{always} gives rise to an initial
state showers, in the same way as final state quarks and gluons
\tmtextit{always} manifest themselves as jets (i.e. as final state showers).

The treatment of initial state radiation in a shower Monte Carlo is very
similar to the case of final state radiation. In this case, the basic
factorization formula refers to the radiation from initial state particles
that give rise to some hard collision. In this case, after radiation, the
initial state acquires a spacelike virtuality, that is limited in magnitude by
the scale of the hard process. The factorization formula, however, has
essentially the same form
\begin{equation}
  d \sigma_j^{\tmop{ISR}} (p, \ldots) = \frac{\alpha_S}{2 \pi} \frac{d t}{t} d
  z P_{i j} (z) d \sigma_i (z p, \ldots),
\end{equation}
where now we consider a production process with a parton $j$ entering the
graph. The process is represented in the graph below
\begin{equation}
\raisebox{-2cm}[1.3cm][1.8cm]{\epsfig{file=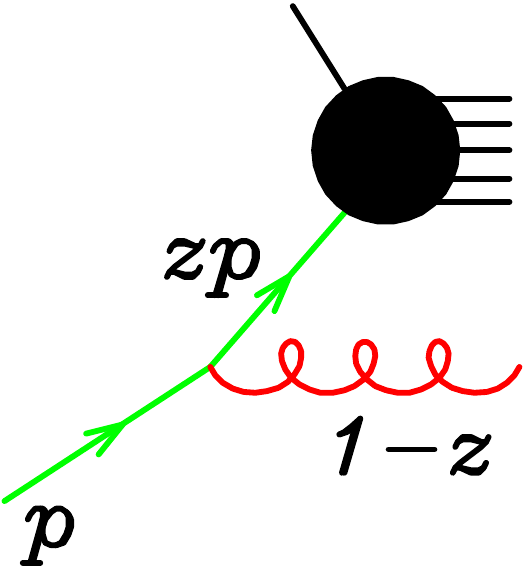,width=0.2\textwidth}}
     \label{fig:isr}
\end{equation}
In this case the initial parton is on shell, and the parton with reduced
momentum $z p$ acquires a negative virtuality. This is unlike the case of
final state radiation, where the virtuality is positive. Multiple initial
state radiation takes place with the virtuality ordered from small (absolute)
values (near the initial state parton) to large values (near the hard
scattering), limited by the hardness of the scattering process. In fact,
factorization holds as long as the virtuality of the parton entering the hard
scattering is negligible with respect to all the other scales entering the
hard scattering amplitude. On the other hand, the radiated partons (like the
gluon in (\ref{fig:isr})) can undergo further splitting with positive
virtualities. This yields the following picture for a shower arising from an
initial state parton in the collinear approximation
\begin{equation}
 \raisebox{-2cm}[1.3cm][2cm]{\epsfig{file=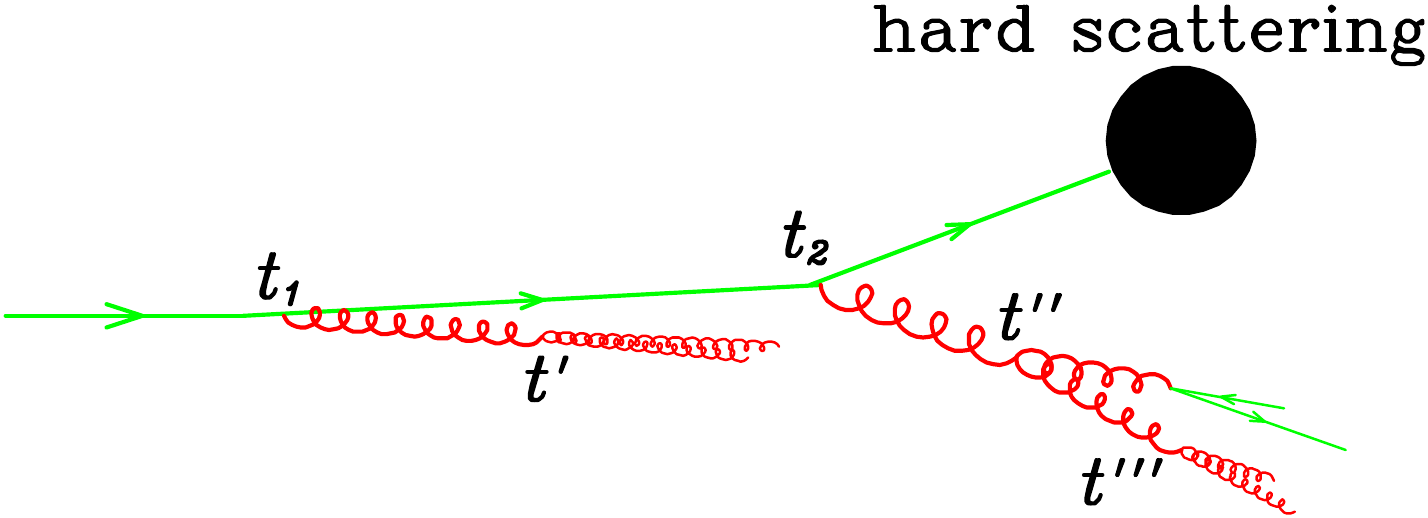,width=0.6\textwidth}}.
     \label{fig:isrshow}
\end{equation}
We have \ $t_1
< t_2 < Q$, and $t_1 > t'$, $t_2 > t'' > t'''$. The intermediate lines
between $t_1$ and $t_2$ and between $t_2$ and the hard scattering are
spacelike. All other intermediate lines are timelike.
The splitting functions and Sudakov form factors for initial state radiation
splittings are the same that enter in the final state radiation process
(differences arise only at the Next-to-Leading level). We now introduce a
notation for the initial state shower
\begin{equation}
  \mathcal{S}_i (m, x, t, E) = \begin{array}{l}
    \raisebox{-0.7cm}[1.3cm][0.7cm]{\epsfig{file=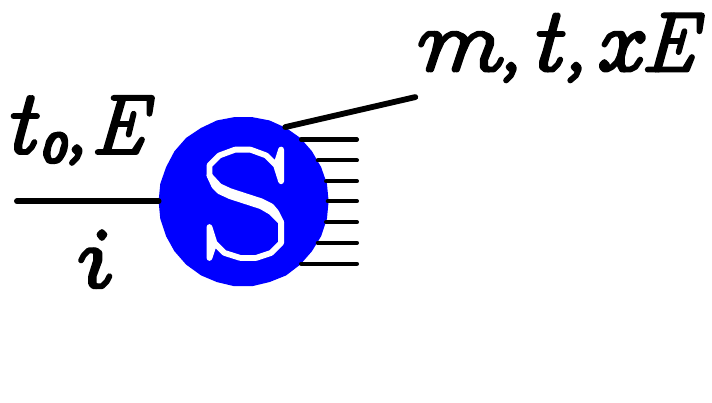,height=2cm}}
  \end{array} .
\end{equation}
The meaning of the notation is as follows: $\delta x _{} \mathcal{S}_i (m, x,
t, E)$ is a function on all possible states (yielding the weight of the shower
for such states) having a spacelike parton of type $m$ with energy between $x
E$ and $(x + \delta x) E$, and scale $t$.

The shower equation for the initial state (i.e., the spacelike shower) can be
represented with the following graphical equation
\begin{equation}
  \raisebox{-1.7cm}[1.7cm][1.7cm]{\epsfig{file=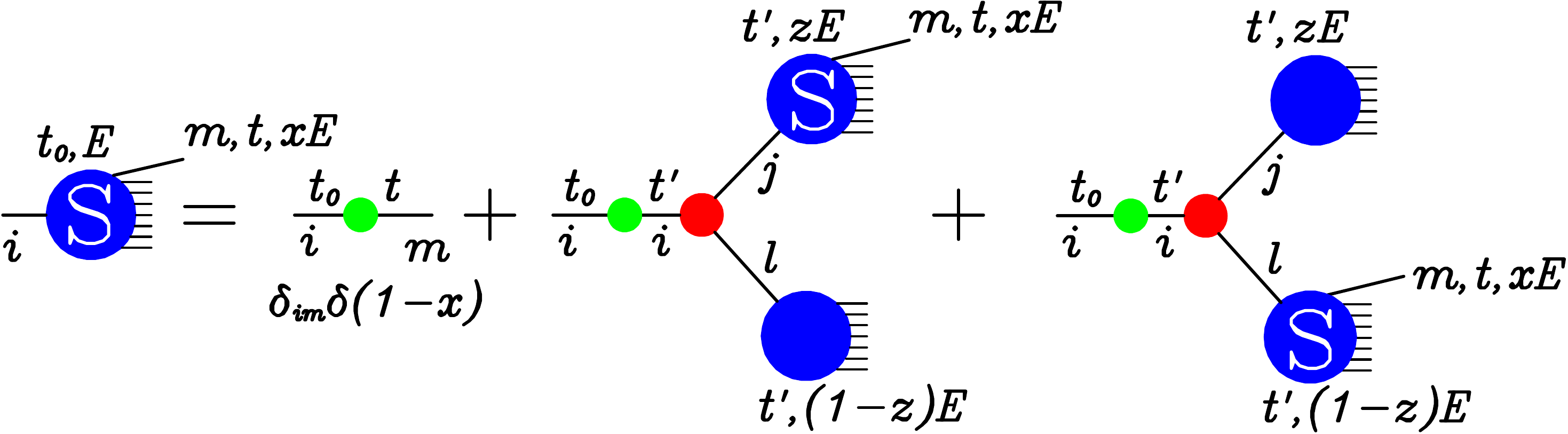,width=0.9\textwidth}}
  \label{fig:showereqisrfw}
\end{equation}
The blobs marked with S represent spacelike showers, while the solid blob
represents the timelike showers discussed in the previous subsections. Solving
this equation would correspond to the so called forward evolution solution of
the evolution equation. In modern Monte Carlo programs, it is preferred to
solve the evolution equation in the opposite direction, i.e. according to the
backward evolution method. The shower equation is then represented in an
equivalent way, but with a recursive procedure that starts at the high scale
instead of the low scale, as follows
\begin{equation}
     \raisebox{-1.4cm}[1.7cm][1.7cm]{\epsfig{file=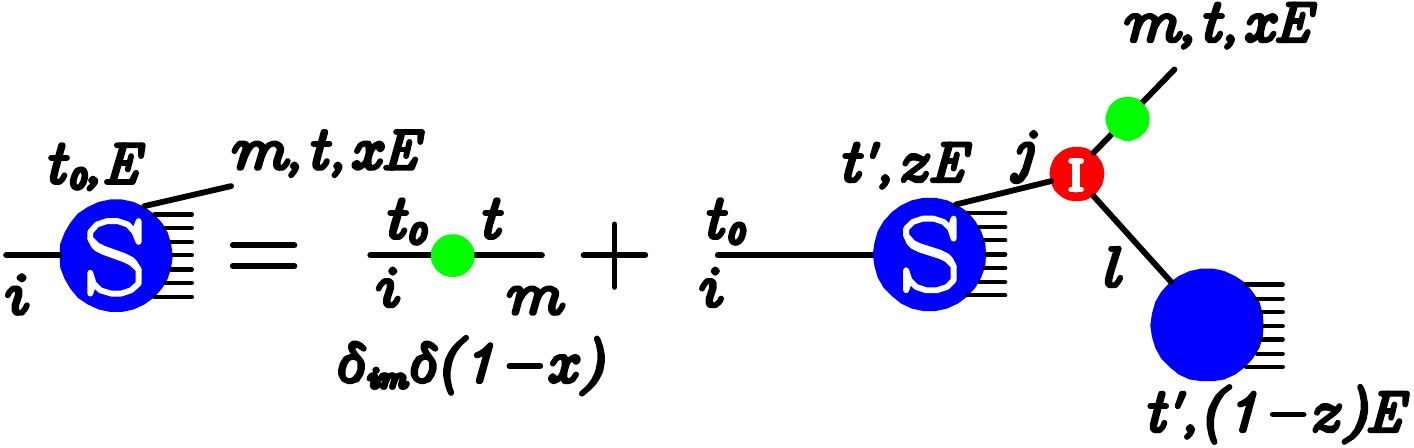,width=0.7\textwidth}}
     \label{fig:showereqisrbk}
\end{equation}
The blob marked with I at the splitting vertex is given by the inclusive
splitting kernel $P_{j m}$, instead of the exclusive one $P_{j, m l}$ (this is
because either branched parton can be spacelike). It is easy to convince
ourselves that the pictures in fig. \ref{fig:showereqisrfw} and
\ref{fig:showereqisrbk} represent the same object, with a different recursion
rule.

The probability for the first branching is obtained by summing over all final
states in the graphical equation of fig. \ref{fig:showereqisrbk}. This sums
yields 1 for the timelike blobs, as shown previously. Not so for the spacelike
blob, that yields
\begin{equation}
  \label{eq:shpdf}_{}  \sum_{\mathcal{F}} \mathcal{S}_i (m, x, t, E) =
  f^{(i)}_m (x, t),
\end{equation}
the (scale dependent) parton density function{\footnote{Since we are not yet
considering hadrons, our parton density is now the probability to find a
parton in a parton.}}, and the graphical equation of fig.
\ref{fig:showereqisrbk} yields
\begin{equation}
  f^{(i)}_m (x, t) = \delta_{m i} \delta (1 - x) \Delta_m (t, t_0) +
  \int_{t_0}^t \frac{d t'}{t'} \frac{d z}{z}  \sum_j f^{(i)}_j (z, t')
  \frac{\alpha_S (t')}{2 \pi}  \hat{P}_{j m} \left( \frac{x}{z} \right)
  \Delta_m (t, t'),
\end{equation}
and taking the derivative of both sides with respect to $t$ yields
\begin{eqnarray}
  t \frac{\partial f_m^{(i)} (x, t)}{\partial t} & = & \frac{\alpha_S (t)}{2
  \pi} \hspace{0.25em} \sum_j \int_x^1 \frac{d z}{z} \hspace{0.25em}
  \hat{P}_{j m} (x / z) \hspace{0.25em} f^{(i)}_j (z, t) \nonumber\\
  & + & \left[ - \sum_{(j l)} \int_0^1 d z \hspace{0.75em} \frac{\alpha_S
  (t)}{2 \pi} \hspace{0.25em} P_{i, j l} (z) \hspace{0.25em} \right] f_m^{(i)}
  (x, t), 
\end{eqnarray}
which is equivalent to the ordinary Altarelli-Parisi equation for the parton
densities. From fig. \ref{fig:showereqisrbk} and eq.~(\ref{eq:shpdf}) we find
the probability distribution for the first backward branching
\begin{equation}
  \label{eq:shfirstbr0} d P_{\tmop{first}} = \sum_j f^{(i)}_j (z, t')
  \frac{\alpha_S (t')}{2 \pi} P_{m j} (x / z) \Delta_m (t, t') \frac{d t}{t} 
  \frac{d z}{z}  \frac{d \phi}{2 \pi} . \text{}
\end{equation}
In order to generate the first branching, we must express eq.
(\ref{eq:shfirstbr0}) as a differential in $t'$. Using the Altarelli Parisi
equation, from eq.~(\ref{eq:shfirstbr0}) we obtain
\begin{eqnarray}
  \frac{d P_{\tmop{first}}}{d t'} & = & \frac{\partial f_m^{(i)} (t',
  x)}{\partial t} \Delta_m (t', t) + \left[ \sum_{(j l)} \int_0^1 d z
  \hspace{0.75em} \frac{\alpha_S (t)}{2 \pi} \hspace{0.25em} P_{i, j l} (z)
  \hspace{0.25em} \right] f_m^{(i)} (t', x) \Delta_m (t, t') \nonumber\\
  & = & \frac{\partial}{\partial t'} \left[ f_m^{(i)} (t', x) \Delta_m (t,
  t') \right] . 
\end{eqnarray}
Thus, the probability distribution for the first branching is uniform in
$f_m^{(i)} (t', x) \Delta_m (t, t')$. We just generate a random number $0 < r
< 1$, and then solve the equation
\begin{equation}
  \label{eq:backbr} r = \frac{f_m^{(i)} (t', x) \Delta_m (t, t')}{f_m^{(i)}
  (t, x)}
\end{equation}
for $t'$. Observe that the factor $f_m^{(i)} (t, x)$ in the denominator is
introduced to normalize the right hand side to 1 when $t' = t$. The Sudakov
form factor $\text{} \Delta_m (t', t)$ becomes very small when $t'$ become
small. Thus, the right hand side of eq.~(\ref{eq:backbr}) can become very
small, its smallest value being reached when $t' = t_0$. If $r$ is below the
smallest possible value, no branching takes place. Sometimes \ the equivalent
formula
\begin{equation}
  \exp \left[ - \sum_j \int_{t'}^t \frac{d t''}{t''}  \frac{\alpha_S (t'')}{2
  \pi} \int_x^1 \frac{d z}{z} P_{m j} (z) \frac{f^{(i)}_j (t'', x /
  z)}{f^{(i)}_m (t'', x)} \right] = \frac{f_m^{(i)} (t', x) \Delta_m (t,
  t')}{f_m^{(i)} (t, x)}
\end{equation}
is used.

We notice that, as in final state radiation, the Sudakov form factor
suppresses the $d t / t$ singularity for small values of $t$, thus yielding a
finite expression for the first emission probability.

\subsection{Shower algorithm for processes with incoming hadrons}

We can now formulate the full recipe for the generation of a process with
incoming hadrons.One can easily set up an algorithm for the generation of the
process:
\begin{enumeratealpha}
  \item Generate a hard process configuration with a probability proportional
  to its parton level cross section. This cross section includes now the
  parton density functions evaluated at the typical scale Q of the process
  
  \item \label{enum:fsrsh}For each final state coloured parton, generate a
  shower in the following way:
  \begin{enumerateroman}
    \item Set $t = Q$
    
    \item \label{enum:genrandf} Generate a random number $0 <
    r < 1$.
    
    \item Solve the equation $r = \Delta_i (t, t')$ for $t'$.
    
    \item If $t' < t_0$ then no further branching is generated, and the shower
    stops.
    
    \item \label{enum:mcbran2}If $t' \geqslant t_0$ then generate $j l$ and $z$
    with a distribution proportional to $P_{i, j l} (z)$, and a value for the
    azimuth $\phi$, with uniform probability in the interval $[ 0, 2 \pi]$.
    Assign energies $E_j = z E_i$ and $E_l = (1 - z) E_i$ to partons $j$ and
    $l$. The angle between their momenta is fixed by the value of $t'$. Given
    the angle and the azimuth $\phi$ (together with the fact that the sum of
    their momenta must equal to the momentum of $i$) the directions of $j$ and
    $l$ are fully reconstructed
    
    \item For each of the branched partons $j$ and $l$, set $t = t'$ and go
    back to step \ref{enum:genrandf}.
  \end{enumerateroman}
  \item \label{enum:isrstep}For each initial state coloured parton, generate a
  shower in the following way
  \begin{enumerateroman}
    \item Set $t = Q$
    
    \item \label{enum:genrandfc} Generate a random number $0
    < r < 1$.
    
    \item Solve the equation $r = \Delta_i (t, t')$ for $t'$.
    \[ r = \frac{f_i^{(h)} (t', x) \Delta_i (t, t')}{f_i^{(h)} (t, x)}, \]
    where $f^{(h)}$ is the parton density for the hadron where parton $i$ is
    found, and $x = E_i / E_h$ is the momentum fraction of the parton.
    
    \item If $t' < t_0$ then no further branching is generated, and the shower
    stops.
    
    \item \label{enum:mcbran3}If $t' \geqslant t_0$ then generate $j$ and $z$
    with a distribution proportional to $P_{i j} (z)$, and a value for the
    azimuth $\phi$, with uniform probability in the interval $[ 0, 2 \pi]$.
    Call $l$ the radiated parton, and assign energies $E_j = z E_i$ and $E_l =
    (1 - z) E_i$ to partons $j$ and $l$. The angle between their momenta is
    fixed by the value of $t'$. Given the angle and the azimuth $\phi$
    (together with the fact that the sum of their momenta must equal to the
    momentum of $i$) the directions of $j$ and $l$ are fully reconstructed
    
    \item For parton $j$, set $t = t'$ and go back to step \ref{enum:isrstep},
    \ref{enum:genrandfc}. For parton $l$, \ set $t = t'$ and go back to step
    \ref{enum:fsrsh}, \ref{enum:genrandf}.
  \end{enumerateroman}
\end{enumeratealpha}

\subsection{Soft divergences}

Besides having collinear divergences, QCD cross sections are also affected by
soft divergences, that are associated to gluons with small energy, even in the
case when the angles are not small. Soft and collinear divergences can take
place at the same time, giving rise to the so-called double-log singularities.
In the previous discussion we have only considered collinear singularities. We
have assumed that there is nothing special about the $z \rightarrow 1$ and $z
\rightarrow 0$ limits in the branching, that is to say, we have reasoned under
the false assumption that the splitting functions are all finite in these
limits. In particular, we have neglected the kinematic constraints that arise
in these regions. Let us assume, for example, that our $t$ variable is the
virtuality, and let us focus upon a single splitting at a scale $t$ and a
given value of $z$, that we assume to be the energy fraction. The two
splitting partons have energies $z E$ and $(1 - z) E$, so they form a system
with virtuality given by (neglecting their masses)
\begin{equation}
  2 z (1 - z) E^2 (1 - \cos \theta),
\end{equation}
where $\theta$ is the angle between the two partons. Thus, we must have
\begin{equation}
  z (1 - z) E^2 \geq t / 4,
\end{equation}
in order for the splitting to be possible. Thus, the $z$ integration is
(roughly) limited by
\begin{equation}
  \frac{t}{4 E^2} \leq z \leq 1 - \frac{t}{4 E^2} .
\end{equation}
If there are no soft singularities, this complication can be neglected,
because, under our assumptions, $t \ll E^2$ at any stage of the branching. In
fact, at the beginning of the shower $E \approx \sqrt{Q}$, and after each
branching $E$ is reduced by a factor of order 1, while $\sqrt[]{t}$ is reduced
by a factor of order $\alpha_S$. Thus the ratio $\sqrt[]{t} / E$ is of
subleading logarithmic magnitude with respect to 1. On the other hand, since
we do have soft singularities (i.e. the splitting functions are divergent for
$z \rightarrow 0$ and $z \rightarrow 1$) these region of subleading
logarithmic size can give contributions of order $1$. Furthermore, splittings
with small (or large) values of $z$ are enhanced, and one can no longer
conclude that the energy of the partons are reduced by a factor of order 1 for
each branching. In other words, in order to achieve logarithmic accuracy, soft
divergences should be accounted for in a proper way.

Since soft emission is associated with the production of low energy
particles, we expect them to have an important impact on the multiplicity of
hadrons in the final state, and a smaller impact on the energy flow in the
event. It is thus obvious that a correct treatment of soft singularities
(especially in the double logarithmic region) is important in order to have a
realistic description of the final state.

As discussed earlier, the choice of the hardness parameter $t$ affects the
treatment of soft divergences. Let us estimate the difference in the exponent
of the Sudakov form factor when we adopt the three different definitions of
the ordering parameter given in eqs. (\ref{eq:tvirt}), (\ref{eq:tpt}) and eq.
(\ref{eq:tangle}). If $t$ is to be interpreted as the virtuality of the
incoming line, then we must have $E^2 z (1 - z) \gtrsim t$, in order for eq.
(\ref{eq:tvirt}) to hold{\footnote{We are interested here into small values of
$\theta$, so it is fair to assume $\theta < 1$.}} for some value of $\theta$.
This yields a double logarithmic integral of the form
\begin{equation}
  \int \frac{d t}{t} \int_{t / E^2}^{1 - t / E^2} \frac{d z}{1 - z} \approx
  \frac{1}{2} \log^2 \frac{t}{E^2},
\end{equation}
the $1 / (1 - z)$ factor arising from the splitting functions. If instead $t$
is interpreted as \ the transverse momentum, then $E^2 z^2 (1 - z)^2 \gtrsim
t$, and we get
\begin{equation}
  \int \frac{d t}{t} \int_{\sqrt{t} / E}^{1 - \sqrt{t} / E} \frac{d z}{1 - z}
  \approx \frac{1}{4} \log^2 \frac{t}{E^2} .
\end{equation}
If $t$ is interpreted as the angle, we get yet another result
\begin{equation}
  \int \frac{d t}{t} \int_0^1 \frac{d z}{1 - z} \approx \log t \log
  \frac{E}{\Lambda} .
\end{equation}
In fact, if the ordering variable is proportional to the square of the angle,
the value of $z$ is not constrained by it, and we must impose a cutoff on $z$
in such a way that the energy of the final state particles cannot become
smaller than some typical hadronic scale $\Lambda$.

It turns out that, in order to treat correctly the double logarithmic region,
one should use as ordering parameter the angular variable $\theta$. This is a
profound result in perturbative QCD. It \ has also an intuitive explanation.
Suppose that we order the emission in virtuality. Soft emissions always yield
small virtuality. Thus, at the end of the shower, one has a large number of
soft emissions, essentially unrestricted in angle. But soft gluons emitted at
large angles from final state partons add up coherently.
\begin{figure}[tbh]
  \epsfig{file=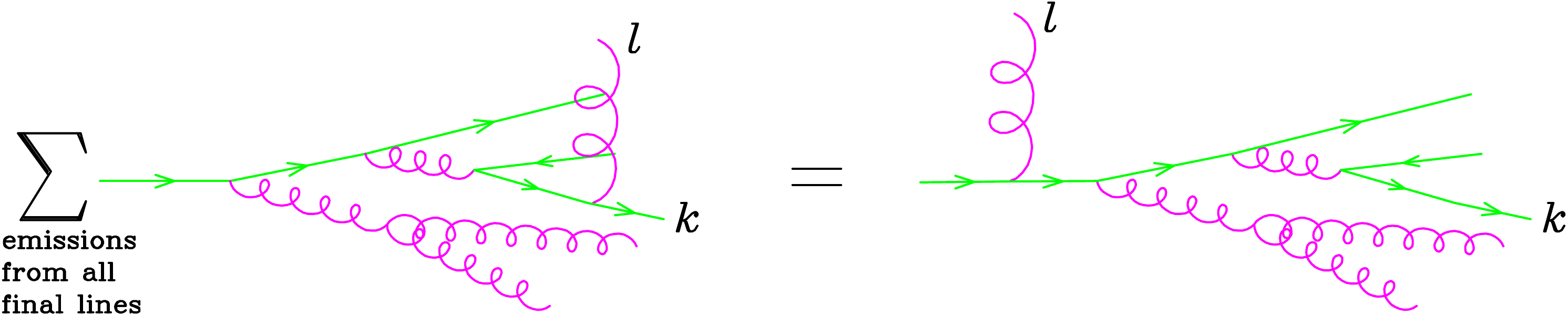,width=0.95\textwidth}
  \caption{\label{fig:softcoer}Soft emissions at large angle add coherently,
  i.e. they behave as if the emitter was the parton that originates the rest
  of the shower.}
\end{figure}
The soft gluons emitted from a bunch of partons with angular
separation that is smaller then the soft gluon emission angle sees all the
emitting partons as a single entity (see fig. \ref{fig:softcoer}). In other
words it is just as if the gluon was emitted from the parton that has
originated the rest of the cascade. Summarizing: if a parton is emitted at
large angle, and its energy is not small, then ordering in virtuality and
ordering in angle does not make any difference. If the parton energy is small,
the parton should be reordered by angle. Thus, ordering in angle from the
beginning gives the correct answer. Observe that angular ordering also emerges
naturally in case one has jets originating from the decay of a fast moving
neutral object, like, for example a relativistic $Z$. Angular ordering tells
us that radiation at angles larger than the angle of the two primary partons
in the decay of the $Z$ should be suppressed. But this must be the case, since
the radiation pattern from the $Z$ decay should be obtainable by considering
the $Z$ decaying in its own rest frame, and then boosting all decay products
with the $Z$ velocity. The effect of the boost is precisely to squeeze all
shower products towards the $Z$ direction, with the emission at large angle
from both primary partons being highly suppressed.

\subsection{Ordering variables: HERWIG and PYTHIA}

In HERWIG, the ordering variable is defined as $t = E^2 \theta^2 / 2$, where
$E$ is the energy of the incoming parton, and $\theta$ is the angle of the two
branched partons, carrying energies $z E$ and $(1 - z) E$. The Sudakov form
factor is defined as follows
\begin{equation}
  \Delta_i (t', t'') = \exp \left[ - \sum_{(j l)} \int_{t''}^{t'} \frac{dt}{t}
  \int_0^1 d z \theta (t z^2 (1 - z)^2 - t_0) \hspace{0.75em} \frac{\alpha_S
  \left( t z^2 (1 - z)^{2^{}} \right)}{2 \pi} \hspace{0.25em} \hspace{0.25em}
  P_{i, j l} (z) \hspace{0.25em} dz \right] .
\end{equation}
The argument of $\alpha_S$ is of the order of the transverse momentum. The
integral in $d z$ is always infrared divergent. An infrared cut-off is needed,
and is in fact provided by the $\theta$ function, that also avoids the region
where the argument of $\alpha_S$ becomes smaller than a given scale $t_0$, of
the order of $\Lambda_{\tmop{QCD}}$. If a parton of energy $E$ branches at a
scale $t$ into two partons of energies $z E$ and $(1 - z) E$, angular ordering
is achieved by choosing as the initial condition for subsequent branchings the
scales $t / z$ and $t / (1 - z)$.

The PYTHIA program has never adopted the angular ordering scheme. In PYTHIA,
virtualities are strictly ordered in the shower. This yields a more natural
kinematics, since virtuality is kinematically ordered in a branching process.
The lack of coherence, however, causes an unphysical increase in the number of
soft partons, so that the particle multiplicity in $e^+ e^-$ annihilation
processes does not have the correct growth with energy. The remedy in PYTHIA
is to veto branchings that violate angular ordering. This scheme (virtuality
ordering with angular ordering imposed by veto) yields the correct
multiplicity distributions. It can be understood as follows. Configuration
soft radiation at a large angle $\theta$ sum up coherently, their sum being
equivalent to a soft emission from the first parent parton that comes from a
branching at angles larger than $\theta$. Thus, many emissions become
equivalent to a single emission, which can be approximated to zero, as far as
the multiplicity is concerned. This is what PYTHIA does. It turns out that
PYTHIA, with the angular order constraint, reproduces well the energy
dependence of the multiplicity. On the other hand, the author is not aware of
any relevant output differences between PYTHIA and HERWIG due to the remaining
differences in the treatment of soft radiation.

Recently, new showering schemes have become available. In HERWIG$+ +$, new
showering variable have been introduce, that should be better from the point
of view of boost invariance properties of the shower. The new versions of
PYTHIA also offer an alternative showering scheme, ordered in transverse
momentum, that implements a variant of the so called {\tmem{dipole shower}}
approach, first implemented in the ARIADNE Monte Carlo.

\subsection{Flavour, colour and hadronization.}

The flavour flow in the collinear approximation is well defined. At the end of
a shower we find quarks and antiquarks with a given flavour. The flavour
content of the generated hadrons will depend to some extent upon the flavour
content of the partons at the end of the shower, in a way that depends
strictly upon the model of hadron formation.

The colour flow is not followed in the collinear approximation. In fact, the
factorization formula deals with colour averaged cross sections. On the other
hand, we know that final state hadrons are colour singlet. Whether or not we
need to take colour into consideration depends only upon the hadronization
model.

\subsubsection{Independent fragmentation}

The simplest hadronization model is the so called independent fragmentation
model. This model converts each final state quark $q$ of flavour $f$ into
hadrons. Each final state particle is treated independently from all the
others. One operates typically in the centre of mass of the parton system. One
picks up a random antiflavour $\overline{f'}$, to be associated with the
flavour $f$ to form a hadron with flavour $f \overline{f'}$. The momentum of
the hadron is taken to be a fraction $z$ of the momentum of the quark $q$,
with a probability dictated by a fragmentation function $F_{f'} (z)$, plus a
transverse momentum, of the order of a typical hadronic scale, typically
distributed according to a negative exponential. In order to conserve flavour
and momentum, a quark with flavour $f'$ is also generated, with momentum equal
to a fraction $(1 - z)$ of the initial quark momentum, and an appropriate
transverse momentum. The procedure is then continued with the left-over quark,
and it is stopped when the left over quark has momentum below a certain
threshold. Flavour is not conserved with this procedure, unless one deals in
some way with the left-over slow quarks. Also, the treatment of gluons is to
some extent arbitrary. One possible approach is to always force a gluon
splitting $g \rightarrow q \bar{q}$ at the end of the shower. In order to deal
with baryon production, quark flavours, also diquarks are introduced. One
assumes that a colour singlet baryon can be formed combining a quark and a
diquark.

Independent fragmentation ignores colour, and thus does not need any colour
information about the showered partons. On the other hand it has some clear
drawbacks, related to the arbitrarity in the choice of the hadronization
frame. Consider in fact the simple example of a virtual photon with a
relatively low invariant mass, decaying into a $q \bar{q}$ pair. We assume
that, because of the low mass, no parton is radiated by showering. It is clear
that the multiplicity of this event, in the independent fragmentation scheme,
depends upon the frame of reference in which we look at the event, the minimum
multiplicity being obtained in the photon CM frame. Of course, in this case we
may then decide to fragment the photon decay product in the photon rest frame,
i.e. in the frame of the colour singlet system formed by the $q \bar{q}$ pair.
But, in order to be consistent, every colour singlet system formed by final
state partons should be decayed in its own reference frame, and this
requirement is in conflict with the setup of independent fragmentation, where
a quark is decayed ignoring the kinematics of all other partons.

\subsubsection{\label{sec:largenc}Large $N_c$ colour approximation}

In order to deal more realistically with colour at the hadronization stage,
Shower Monte Carlo's adopt the so called large $N_c$ limit (also called planar
limit), $N_c$ being the number of colours (i.e. $N_c = 3$ in QCD). We should
thus think that the number of colour is large, and keep only the dominant
contribution in this sense.

The colour rules for the Feynman diagrams also become extremely simple in the
large $N_c$ limit. Colour and anticolour indices range from 1 to $N_c$. Each
oriented quark line is assigned a colour index and an antiquark line is
assigned an anticolour index (ranging from 1 to $N_c$). An oriented gluon is
assigned a pair of indices, corresponding to a colour and an anticolour. This
gives rise to $N_c^2$ gluons. We know that, in fact, there are $N_c^2 - 1$
gluons, since the combination $\sum_c c \bar{c}$ (with $c$ running over all
colours) is colour neutral (i.e. is a colour singlet). However, in the limit
when $N_c$ is considered to be large, one can replace $(N_c^2 - 1) \rightarrow
N_c^2$. Graphically, we may represent an oriented colour index with an arrow,
and an anticolour is represented by an arrow in the opposite direction. The
colour structure of a $q \rightarrow q g$, $g \rightarrow g g$ and $g
\rightarrow q \bar{q}$ splitting is shown in the following figure:
\begin{equation}
  \label{eq:planarrules}
    \raisebox{-1.2cm}[1.5cm][1.2cm]{\epsfig{file=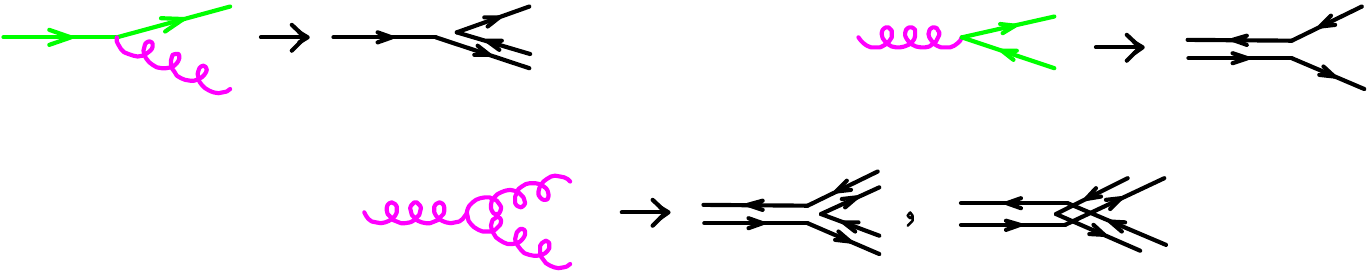,width=0.9\textwidth}}
\end{equation}
Notice that the two colour configurations associated to the gluon splitting
vertex turn into each other by exchanging the two final state gluons.

\begin{figure}[tbh]
  \begin{center}
    \epsfig{file=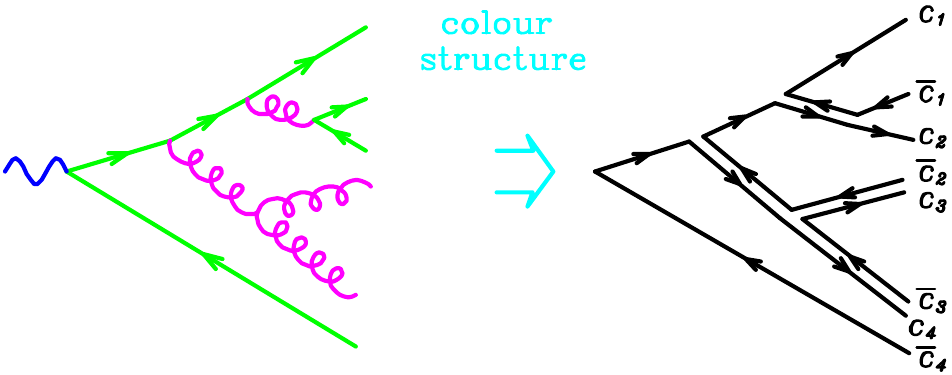,width=0.6\textwidth}
  \end{center}
  \caption{\label{fig:colourflow}Colour structure of the square of an
    amplitude in the large $N_c$ limit.}
\end{figure}

An illustration of the large $N_c$ limit of a contribution to the $e^+ e^-
\rightarrow \tmop{hadrons}$ cross section is given in fig.~\ref{fig:colourflow}.
The colour factor of the squared amplitude is obtained by summing
over the colour indices, i.e. there is a factor of $N_c$ for each colour
index. Notice that, when squaring the amplitude, interference terms are
suppressed by powers of $1 / N_c$. In fact, in order to have interference, two
colour indices must be the same, so that one looses a factor of $N_c$.

When assigning a planar colour configuration to a set of showered partons, one
begins by computing the Born level cross section in the large $N_c$ limit, for
each independent colour structure that is allowed, and chooses the initial
colour structure with a probability proportional to the corresponding
contribution. In our $e^+ e^- \rightarrow q \bar{q}$ example, there is only
one such colour structure, that assigns opposite colours to the quark and the
antiquark. Starting from the colour connections of the partons at the Born
amplitude, one reconstructs the colour connections of all partons in the
shower, according to the rules given in eq.~(\ref{eq:planarrules}). From the
figure, we can see that there is only one way to assign colour connections in
a $q \rightarrow q g$ or a $g \rightarrow q \bar{q}$ vertex. On the other
hand, there are two possible assignments in the $g \rightarrow g g$ splitting,
corresponding to the exchange of two final state gluons. In this case, one
chooses one of the two assignments with a $50\%$ probability. At the end of
the procedure, one obtains the colour connections of all partons in the
showered system.

Notice that, in the large $N_c$ limit, it is enough to know that the quark and
the antiquark are {\tmem{colour connected}}. One does not need to know which
specific colour is assigned to them. In fact, in the limit of large $N_c$, the
probability that two colour connected pairs of quarks have the same colour
index is suppressed by a $1 / N_c$ factor, and thus can be neglected.

\subsubsection{Cluster and string based fragmentation models}

The cluster and string fragmentation models are both based upon the
assignments of colour connections illustrated in section \ref{sec:largenc}.

In the cluster model, final state gluons are forced to split into
quark-antiquark pairs. Then one decays each colour connected quark-antiquark
pair independently. If the invariant mass of the colour connected pair is low
enough, one matches mass and flavour with a corresponding hadronic two-body
system (or with a resonance) with the same flavour. In angular ordered shower,
one can show that configurations with colour connected pairs with large
invariant mass are Sudakov suppressed (an effect known as
{\tmem{preconfinement}}).

In the string fragmentation model, colour connected partons are collected in a
system consisting of a quark, several intermediate gluons, and an antiquark.
For example, in figure \ref{fig:colourflow} there are two colour connected
system, one formed by the quark with colour $c_1$ and the antiquark with
colour $\bar{c}_1$, and the other one starting with the quark with colour
$c_2$, including the two final state gluons with colour $[ \bar{c}_2, c_3]$
and $[ \bar{c}_3, c_4]$ and ending with the antiquark $\bar{c}_4$. One then
imagines that a colour flux tube (i.e. a {\tmem{string}}) is stretched from
the quark to the antiquark of the colour connected system, going through each
intermediate gluon.

In the simplest case, the string is stretched between a quark and an
antiquark. The hadronic system is generated by pair creation by quantum
tunneling inside the string. In practice, at this stage the fragmentation
algorithm is similar to the independent fragmentation case. One goes to a
frame where the two string ends have opposite momenta, and, starting from each
string end, one has a fragmentation function to describe the probability to
generate a hadron carrying away a given fraction of the longitudinal momentum
of the string. To be more specific, let us assume that the string end has
flavour $f$. A hadron will be generated with flavour $f \overline{f'}$, and
the left over string will have a flavour $f'$ at his end. Unlike the case of
independent fragmentation, besides having a more reasonable description of the
role of colour in fragmentation, also flavour is treated consistently.

In the general case, with intermediate gluons in the colour connected system,
a similar procedure is adopted, with some care for the treatment of the kinks
in the string associated to the intermediate gluons.

It should be made clear that fragmentation models end up being one of the most
complex aspects of Shower Monte Carlo. The underlying theory (i.e. QCD) is
only used as a reasonable suggestions on certain features that the models
should have. The models have unavoidably a large number of parameters, that
are needed in order to represent faithfully the many final state features that
are observed in strong interactions.

\subsection{Dipole approach to Shower Monte Carlo}

\begin{figure}[tbh]
\begin{center}
  \epsfig{file=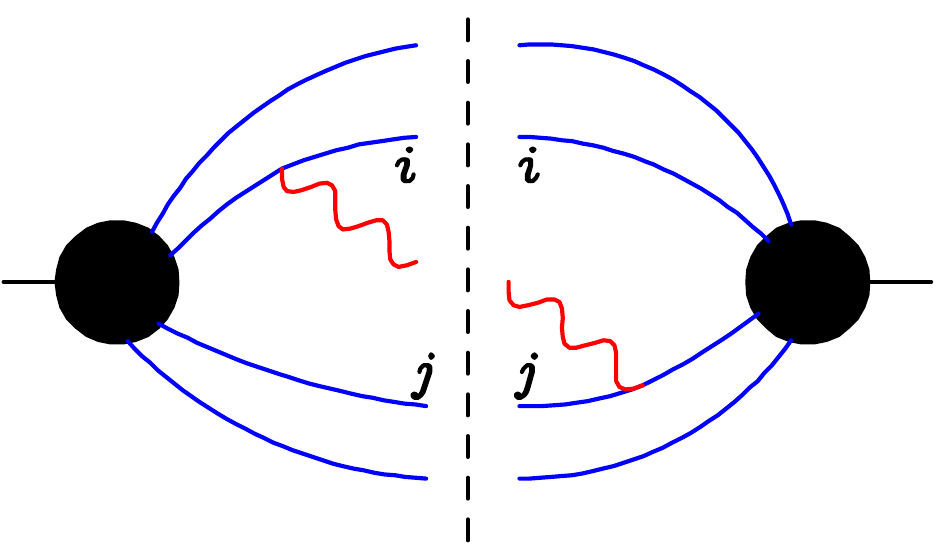,width=0.5\textwidth}
\end{center}
  \caption{\label{fig:qedsoft}The contribution of a single eikonal factor in
  QED. The above figure is a common way to represent the interference of two
  amplitudes: the amplitude on the left, times the complex conjugate of the
  amplitude on the right of the dashed line.}
\end{figure}

The historical development of shower algorithms has privileged the treatment
of collinear radiation. One first deals with collinear shower, and then fixes
the soft radiation. A different approach has also been pursued: one generates
first a soft shower, and then fixes the collinear region. In this approach one
begins with a formula for soft emission from the primary partons. Unlike
collinear singularities, soft singularities do not factorize in a simple way
in QCD. In order to illustrate this fact we begin by first considering QED,
where soft singularities do indeed factorize according to the formula
\begin{equation}
  \label{eq:softfact} |M_{n + 1 \gamma} |^2 \Rightarrow |M_n |^2 (4 \pi
  \alpha) \sum^n_{i, j = 1} Q_i Q_j \frac{p_i \cdot p_j}{(p_i \cdot k) (p_j
  \cdot k)} \hspace{0.75em},
\end{equation}
where $p_i$ are the momenta of the outgoing particles, and $Q_i$ their
electric charge in positron charge units, and $k$ is the momentum of the
emitted photon. Formula (\ref{eq:softfact}) holds as long as $k$ is much
smaller than all the amplitude momenta $p_i$. Thus, in QED, the emission of a
soft photon factorizes in terms of the original squared amplitude times the
sum of so called eikonal factors, associated to photon emission from a pair of
final state partons. This formula is also independent upon the spin of the
emitting particles; only their electric charge counts. When $i \neq j$ each
eikonal term comes from the interference of the photon emission amplitude from
partons $i$ and $j$, as represented graphically in fig.~\ref{fig:qedsoft}.
In QCD, soft emission still involves the same eikonal factors
that operate in QED. But the charges are replaced by colour matrices. So,
while in QED the contribution of a single eikonal factor (like the one
involving partons $i$ and $j$ in the figure) is always proportional to the
Born squared matrix element, in QCD it is proportional to a square of the Born
matrix elements where the colours of partons $i$ and $j$ have been scrambled.
These colour scrambled Born contributions are potentially different among each
other, so that simple, QED like factorization no longer holds. In order to
recover some manageable simplicity, one takes the large $N_c$ limits of QCD.
Planar soft emissions from a planar squared amplitude always amounts to add
one colour loop (i.e. to an extra factor of $N_c$). Thus, a planar
factorization formula holds in large $N_c$ QCD
\begin{equation}
  \label{eq:qcdplansoft} |M_{n + 1 g} (p_1, \ldots, p_n, k) |^2 \Rightarrow
  \left[ |M_n (p_1, \ldots, p_n) |^2 (4 \pi \alpha
  N_c) \sum_{\tmop{conn} .} \frac{p_i \cdot p_j}{(p_i \cdot k) (p_j \cdot k)}
  \hspace{0.75em} \right]^{\tmop{Symm}},
\end{equation}
where the sum extends over all colour connected final state partons, and
``Symm'' stands for symmetrization in the momenta of identical particles (the
planar squared amplitude not being symmetric). Thus, even in the planar limit,
soft factorization is not the same as in QED. It is however easily tractable,
since symmetrization is unnecessary (as long as one computes symmetric
observables).

In the dipole approach, one associates Sudakov form factors to dipoles, rather
than to partons, computes a no-radiation probability, and generates the
emission with a procedure similar to the one used in the single parton shower
approach. One generates a $t$ for each dipole, and then picks the hardest $t$
to decide which dipole is emitting. In the limit when the emitted gluon is
parallel to a final state parton, one adjusts the eikonal factors in such a
way that they become correct even if the energy of the emitted parton is not
small, in order to reproduce the Altarelli Parisi splitting probability. If
the emitting parton is a gluon, two dipoles can contribute to its emission,
and this has to be accounted for properly.

\section{Underlying event}

The hadronization model deals with final state partons, turning them into
hadrons. Also initial state partons require some treatment, in order to give a
realistic description of the physics of the hadronic remnants. First of all,
what we have introduced as the parton density to find a parton in a parton
(eq.~(\ref{eq:shpdf})) should be immediately interpreted as the probability to
find a parton in the incoming hadron. In the forward evolution scheme, this
would require to introduce an initial parton density at the scale $t_0$. In
the backward evolution scheme this is unnecessary: one compute the cross
section with the full pdf at the scale of the process, using standard pdf
parametrization. However, when the backward shower stops (i.e. a scale $t <
t_0$ is generated in the backward evolution formalism), we should provide some
model for the structure of the remaining part of the incoming hadron. This is
a subtle problem, that cannot be treated in a rigorous way in QCD. The crudest
approach one can think of, is to force initial state gluons at the end of the
shower to arise from a quark in backward evolution, then let the remaining
diquark in the incoming proton, carrying the left over momentum of the initial
hadron, hadronize with the remaining particles in the event. In other
approaches, if the backward shower stops with a gluon, the remaining quarks in
the incoming hadron are put into a colour octet state, and this system is
broken up with various rules, to yield objects that the hadronization
mechanism can handle. There is some evidence {\ref{chap:MBaUE}} that, in
order to represent the activity of the underlying event in a reasonable way,
the effect of multiparton interactions must also be included. In other words,
one must assume that the remnants of the incoming hadrons can undergo
relatively hard collisions. Even this phenomenon is implemented with
phenomenological models in Shower Monte Carlo programs. Among the ingredients
entering these models, one assumes that partons have a given transverse
distribution in a hadron. The cross section for secondary interactions is
assumed to be given by the partonic cross section with an appropriate cutoff
in transverse momentum. This cutoff has to be carefully tuned, since the
partonic cross section diverges as the cutoff goes to zero. The momentum of
the spectator partons has to be properly rescaled, to account for the momentum
taken away by the parton that initiates the spacelike shower. Flavour and
colour of the spectators has to be properly adjusted.

\section{Shower Monte Carlo resources}

Here I collect useful references to Shower Monte Carlo physics. First of all,
the pedagogic introductions in refs. {\cite{Webber:1986mc}},
{\cite{Ellis:1991qj}} and {\cite{Sjostrand:2006su}} offer an alternative
introduction to the one presented here.

In ref. {\cite{Dobbs:2004qw}} a primer on the main available Monte Carlo codes
and methods is given.

The PYTHIA manual {\cite{nSjostrand:2006za}} is a valuable source of
information on several aspects of Shower Monte Carlo physics. In the original
HERWIG paper {\cite{Marchesini:1992ch}}, more thorough discussion of the
problem of soft radiation can be found.

In the web page \url{http://www.hepforge.org/}, links to various Monte
Carlo programs, as well as to tools typically used in this framework (like jet
algorithms and the like) can be found.

\addtocounter{chapter}{1}
\newcommand{\beq}{\begin{equation}}
\newcommand{\eeq}{\end{equation}}
\newcommand{\qu}{u}
\newcommand{\qd}{d}
\newcommand{\qubar}{\bar u}
\newcommand{\qdbar}{\bar d}
\newcommand{\bfig}{\begin{center}\begin{picture}}
\newcommand{\efig}[1]{\end{picture}\\{\small #1}\end{center}}
\newcommand{\flin}[2]{\ArrowLine(#1)(#2)}
\newcommand{\wlin}[2]{\DashLine(#1)(#2){2.5}}
\newcommand{\zlin}[2]{\DashLine(#1)(#2){5}}
\newcommand{\glin}[3]{\Photon(#1)(#2){2}{#3}}
\newcommand{\lin}[2]{\Line(#1)(#2)}
\newcommand{\noi}{\noindent}
\newcommand{\sof}{\SetOffset}
\newcommand{\bmip}[2]{\begin{minipage}[t]{#1pt}\bfig(#1,#2)}
\newcommand{\emip}[1]{\efig{#1}\end{minipage}}
\newcommand{\putk}[2]{\Text(#1)[r]{$p_{#2}$}}
\newcommand{\putp}[2]{\Text(#1)[l]{$p_{#2}$}}
\newcommand{\bq}{\begin{equation}}
\newcommand{\eq}{\end{equation}}
\newcommand{\bqa}{\begin{eqnarray}}
\newcommand{\eqa}{\end{eqnarray}}
\newcommand{\nl}{\nonumber \\}
\newcommand{\eqn}[1]{eq. (\ref{#1})}
\newcommand{\eqs}[1]{eqs. (\ref{#1})}
\newcommand{\ibidem}{{\it ibidem\/},}
\newcommand{\vpb}{}
\newcommand{\p}[1]{{\scriptstyle{\,(#1)}}}
\newcommand{\gev}{\mbox{GeV}}
\newcommand{\tev}{\mbox{TeV}}
\newcommand{\mev}{\mbox{MeV}}
\newcommand{\ban}{\begin{eqnarray*}}
\newcommand{\ean}{\end{eqnarray*}}

\def    \eqnum          #1{(\ref{#1})}       
\def    \scite          #1{$^{\cite{#1}}$}     
  \newcommand{\ccaption}[2]{
    \begin{center}
    \parbox{0.85\textwidth}{
      \caption[#1]{\small{{#2}}}
      }
    \end{center}
    }
\newcommand{\BS}{\bigskip}
\def    \be             {\begin{equation}}
\def    \ee             {\end{equation}}
\def    \ba             {\begin{eqnarray}}
\def    \ea             {\end{eqnarray}}
\def    \nn             {\nonumber}
\def    \=              {\;=\;}
\def    \frac           #1#2{{#1 \over #2}}
\def    \ret            {\\[\eqskip]}
\def    \ie             {{\em i.e.\/} }
\def    \eg             {{\em e.g.\/} }
\def    \bentarrow      {\:\raisebox{1.1ex}{\rlap{$\vert$}}\!\rightarrow}
\def    \rd             {{\mathrm d}}    
\def    \Im             {{\mathrm{Im}}}  
\def    \bra#1          {\mbox{$\langle #1 |$}}
\def    \ket#1          {\mbox{$| #1 \rangle$}}

\def    \kev            {\mbox{$\mathrm{keV}$}}
\def    \mev            {\mbox{$\mathrm{MeV}$}}
\def    \gev            {\mbox{$\mathrm{GeV}$}}

\def    \nubar   {\bar{\nu}}
\def    \ubar   {\bar{u}}
\def    \dbar   {\bar{d}}
\def    \qbar   {\bar{q}}
\def    \bbar   {\bar{b}}
\def    \cbar   {\bar{c}}
\def    \cpbar   {\bar{c}'}
\def    \tbar   {\bar{t}}
\def    \Qbar   {\overline{Q}}

\def    \mq             {\mbox{$m_Q$}}  
\def    \mZ             {\mbox{$m_Z$} }
\def    \mZsq             {\mbox{$m_Z^2$} }
\def    \mW             {\mbox{$m_W$} }
\def    \mWsq             {\mbox{$m_W^2$} }
\def    \mH             {\mbox{$m_H$} }
\def    \mHsq             {\mbox{$m_H^2$} }
\def    \mt             {\mbox{$m_t$}}  
\def    \mtsq             {\mbox{$m_t^2$}}  
\def    \mb             {\mbox{$m_b$}}  
\def    \mbsq             {\mbox{$m_b^2$}}  
\def    \mqq            {\mbox{$m_{Q\bar Q}$}}
\def    \mqqsq          {\mbox{$m^2_{Q\bar Q}$}}
\def    \pt             {\mbox{$p_T$}}
\def    \et             {\mbox{$E_T$}}
\def    \xt             {\mbox{$x_T$}}
\def    \xtsq           {\mbox{$x_T^2$}}
\def    \ptsq           {\mbox{$p^2_T$}}
\def    \ptbsq           {\mbox{$p^2_{T,b}$}}
\def    \ptbbsq           {\mbox{$p^2_{T,\bar{b}}$}}
\def    \ptWsq           {\mbox{$p^2_{T,W}$}}
\def    \ptZsq           {\mbox{$p^2_{T,Z}$}}
\def    \mT             {\mbox{$m_T$}}
\def    \mTsq           {\mbox{$m^2_T$}}
\def    \etsq           {\mbox{$E^2_T$}}
        
\newcommand     \MSB            {\ifmmode {\overline{\rm MS}} \else 
                                 $\overline{\rm MS}$  \fi}
\def    \muf            {\mbox{$\mu_{\rm F}$}}
\def    \mug            {\mbox{$\mu_\gamma$}}
\def    \mufsq          {\mbox{$\mu^2_{\rm F}$}}
\def    \mur            {{\mbox{$\mu_{\rm R}$}}}
\def    \mursq          {\mbox{$\mu^2_{\rm R}$}}
\def    \mul            {{\mu_\Lambda}}
\def    \mulsq          {\mbox{$\mu^2_\Lambda$}}

\def    \bzero          {\mbox{$b_0$}}
\def    \as             {\ifmmode \alpha_s \else $\alpha_s$ \fi}
\def    \aem             {\mbox{$\alpha_{em}(\mZ)$}}
\def    \asb            {\mbox{$\alpha_s^{(b)}$}}
\def    \assq           {\mbox{$\alpha_s^2$}}
\def \oacube {\mbox{$ {\cal O}(\alpha_s^3)$}}
\def \oaemacube {\mbox{$ {\cal O}(\alpha\alpha_s^3)$}}
\def \oafour {\mbox{$ {\cal O} (\alpha_s^4)$}}
\def \oatwo {\mbox{$ {\cal O} (\alpha_s^2)$}}
\def \oaematwo {\mbox{$ {\cal O}(\alpha \alpha_s^2)$}}
\def \oaemas {\mbox{$ {\cal O}(\alpha \alpha_s)$}} 
\def \oas   {\mbox{$ {\cal O}(\alpha_s)$}}
\def\slash#1{{#1\!\!\!/}}
\def\rt1{\raisebox{-1ex}{\rlap{$\; \rho \to 1 \;\;$}}
\raisebox{.4ex}{$\;\; \;\;\simeq \;\;\;\;$}}
\def\ltap{\raisebox{-.5ex}{\rlap{$\,\sim\,$}} \raisebox{.5ex}{$\,<\,$}}
\def\gtap{\raisebox{-.5ex}{\rlap{$\,\sim\,$}} \raisebox{.5ex}{$\,>\,$}} 

\renewcommand\LambdaQCD{\Lambda_{\scriptscriptstyle \rm QCD}}

\def\naive{na\"{\i}ve}
\def\asp{{\alpha_s}\over{\pi}}
\def\half{\frac{1}{2}}
\def\herwig{{\small HERWIG}}
\def\herwigs{{\small HERWIG} \ }
\def\isajet{{\small ISAJET}}
\def\pythia{{\small PYTHIA}}
\def\grace{{\small GRACE}}
\def\amegic{{\small AMEGIC++}}
\def\vecbos{{\small VECBOS}}
\def\madgraph{{\small MADGRAPH}}
\def\comphep{{\small CompHEP}}
\def\ALPHA{{\small ALPHA}}
\def\ALPGEN{{\small ALPGEN}}
\def\ALPGENs{{\small ALPGEN} \ }
\def\alpgen{{\small ALPGEN}}
\def\alpgens{{\small ALPGEN} \ }
\def\ppbar{\mbox{$p \bar{p}$}}
\def\met{$\rlap{\kern.2em/}E_T$}

\newcommand{\ben}{\begin{enumerate}}
\newcommand{\een}{\end{enumerate}}
\newcommand{\bit}{\begin{itemize}}
\newcommand{\eit}{\end{itemize}}


%
%
%

\mchapter{Matrix Elements and Shower Matching}
{ Author: Mauro Moretti}
\vskip 0.3cm\noindent
{\it Revisors: Paolo Nason}
\vskip 1cm
\section{Introduction}
\label{sec:intro}

As discussed at length in the previous chapters, final
states with many hard jets will play an essential role
for LHC physics. 
These events will hide or strongly
modify all possible signals of new physics, which involve the chain
decay of heavy coloured particles, such as squarks, gluinos or the
heavier partners of the top, which appear in little-Higgs models.  Being
able to predict their features is therefore essential.
To this end it is crucial  to describe 
as accurately as
possible both the full matrix elements (ME) for the 
underlying hard
processes, as well as the subsequent development 
of the hard partons
into jets of hadrons.

It is therefore very important to design a strategy
to take advantage of the strength (and avoid the drawbacks)
of both fixed order
calculations and of Parton Shower-like evolution with
subsequent hadronization of the partonic event. 
  A given
$(n+1)$-jet event can be obtained in two ways: from the
collinear/soft-radiation evolution of an appropriate $(n+1)$-parton
final state, or from an $n$-parton configuration where hard,
large-angle emission during its evolution leads to the extra jet.  A
factorization prescription (in this context this is often called a
``matching scheme'' or ``merging scheme'') defines, on an
event-by-event basis, which of the two paths should be followed.  The
primary goal of a merging scheme is therefore to avoid double counting
(by preventing some events to appear twice, once for each path), as
well as dead regions (by ensuring that each configuration is generated
by at least one of the allowed paths).  Furthermore, a good merging
scheme will optimize the choice of the path, using the one
which
guarantees the best possible approximation to a given kinematics.

Here we shall briefly review two such merging approaches:
 the CKKW scheme
\cite{Catani:2001cc,Krauss:2002up}, and the MLM scheme \cite{matching}.
These two approaches are implemented in currently used
matrix element event generators, combined with parton 
showers tools, like SHERPA \cite{nGleisberg:2003xi,
Schalicke:2005nv}, ALPGEN \cite{alpgen}, 
MADGRAPH \cite{madgraph} and HELAC \cite{helac}.

Any merging algorithm is based on one or more 
{\em resolution parameters} which split the phase space
into two regions one of soft/collinear emissions to
be described by Parton Shower (PS) evolution and the other
one of hard and large angle emission to be described by
fixed order calculations. These resolution parameters
play the role of soft/collinear cut-off for fixed order
calculations and it is therefore crucial to assess the
(in)dependence of the algorithm on these parameters.
Notice that if both PS and ME descriptions would provide a
{\em perfect} description of QCD the final result
would be independent of the resolution parameters.

For the
CKKW scheme,
in the context of $e^+e^-\to jets$, it has been shown
\cite{Catani:2001cc} that the dependence on the resolution
parameter is shifted beyond the Next to Leading Log (NLL)
accuracy.

Such a proof in the context of $e p$ and $p p $ collisions
is missing and thus
for both CKKW (adapted to hadronic collision 
\cite{Krauss:2002up}) and for MLM scheme we don't have
any avaliable extimate of the dependence of the
final result on the resolution parameters.
{\em Ultimately, at present,
 such an extimate is possible only empirically}: one
has to study the effect of varying the resolution parameters
on the widest possible range.

A first series of studies to address both dependence on
the resolution parameters and the comparison of
 the two schemes
has been presented in \cite{Mrenna:2003if}.

The internal consistency of CKKW (as implemented in 
the SHERPA \cite{sherpa} event generator)
 inspired approach for 
hadronic collisions has been studied in
\cite{sherpawlhc,sherpawtev,sherpawwtev}
for Drell-Yan processes at the TEVATRON and at the LHC. 

The internal consistency for the MLM approach, as implemented
in the ALPGEN \cite{alpgen} event generator,
 has been
addressed in \cite{ttalpgen} for the process
$t \bar t$ plus jets. 

Monte Carlo event samples for associate productions
of jets and $W$ and $Z$ bosons and for jets productions at
the TEVATRON colliders have been compared with data
\cite{d0zjet_sh,d0dijet,messina} finding an overall
satisfactory agreement both for the shapes 
of the distributions and for {\em relative} jets 
multiplicities.

Finally an extensive set of comparisons among various
codes and matching prescriptions has been presented
in \cite{compare0,compare} where, in addition to 
a wide range of tests of internal consistency for the
variuos codes,
 a first attempt to assess some of the systematic
uncertainties associated to these approaches
($\alpha_S$ and PS scales) is presented.
In \cite{compare0,compare} results are presented
also for the event generator
ARIADNE \cite{Lonnblad:1992tz} and
for the L{\"o}nnblad matching prescription
\cite{Lonnblad:2001iq,Lavesson:2005xu} (a variant to 
CKKW adapted to the dipole emission approximation
which is the root of ARIADNE PS). 

\section{Matching}
\label{sec:matching}
Let's first try a sort of ``pedagogical'' introduction to the matching issue.
Our goal is to use the capability to compute fairly complex 
leading order (LO) matrix 
elements (ME) to describe hard QCD radiation and to complement
this description with showering, to include
soft and collinear corrections, and hadronization,
 allowing a realistic description of
the event.

The most simple approach is:

\bit

\item Use the ME to compute the {\em WEIGHT} of a given event.

\item Use the ME computation as a ``seed'' for the
Parton Shower
(PS) evolution: the PS needs as inputs the ME weight,
the event kinematic, the colour flow associated to the 
event. (As well as  the factorization and 
renormalization scales chosen for the ME calculation)

\eit

\begin{figure}[hbt]
\begin{center}
\epsfig{file=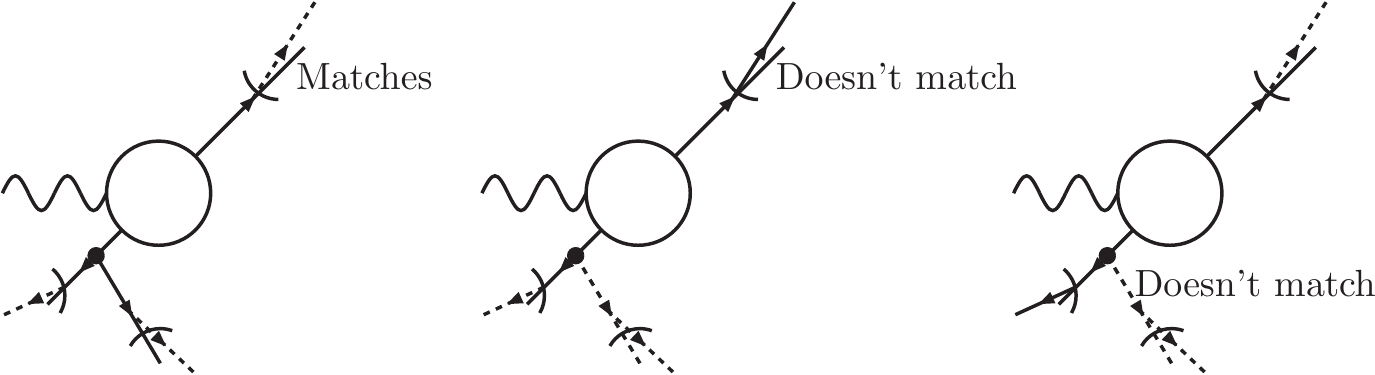,width=0.9\textwidth}
\end{center}
\ccaption{}{\label{dbcnt} 
Hadron production in $e^+ e^-$ collisions via
$\gamma^*$, $Z^*$ exchange. Example of double counting
in ME PS combination.
Wiggly line: $\gamma^*$, $Z^*$;
solid lines: ME (coloured ) partons; 
dashed lines: PS emissions. 
The same events obtained in three different ways. Left: hard emissions
from ME and soft/collinear ones from PS, Center: one soft emission
from ME and one hard emission from PS, Right: one collinear
 emission from ME and
one hard emission from PS.

 The second and the third one lack the
appropriate Sudakov suppression and lead to a divergent cross section.
The first one is the one we would like to retain.

Small arcs denote clusters used in MLM matching prescription.}
\end{figure}

This approach, however, leads
 to double counting: the same final state 
can arise in many different ways just swapping ME element generated partons
and shower generated partons as shown in fig.~\ref{dbcnt}. 

\begin{figure}[hbt]
\begin{center}
\epsfig{file=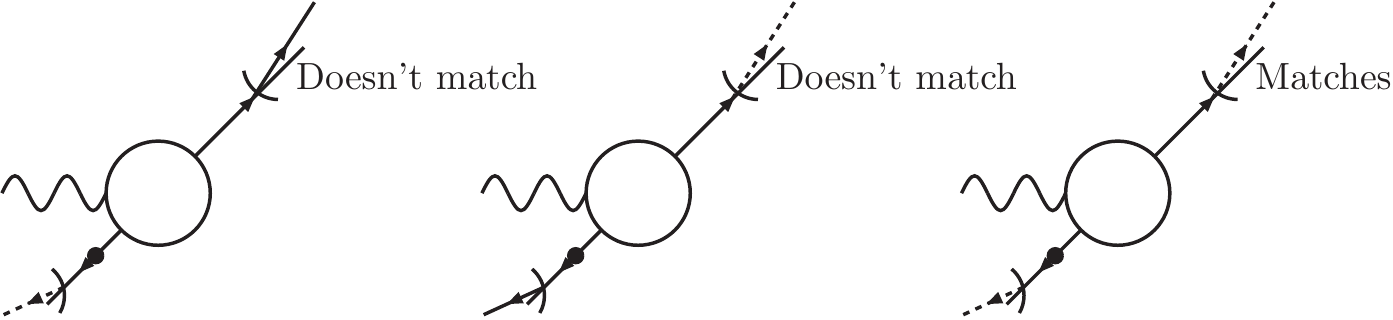,width=0.9\textwidth}
\end{center}
\ccaption{}{\label{dbcnt1}
Same symbols (and process) as in fig.~\ref{dbcnt}.
The same events obtained in three different ways.
Left: one collinear emission
from ME and soft/collinear emissions from PS, number of jets smaller than
number of ME partons; 
Center: one soft emission
from ME and soft/collinear emissions from PS, number of jets smaller than
number of ME partons; 
 Right: hard emissions from ME and
soft/collinear emissions from PS. The first and the
second one lack the
appropriate Sudakov suppression and lead to a divergent cross section.
The third one is the one we would like to retain.
}
\end{figure}

This effect is formally NLO (indeed any PS emission implies
an additional power of $\alpha_s$)
 and therefore
beyond the accuracy of our computation. 
However it opens the possibility
to particularly harmful events: soft and/or collinear 
ME partons 
toghether with hard shower emission to replace the missing
 hard jets,
as shown in figg.~\ref{dbcnt} and \ref{dbcnt1}.
The ME weight is {\em divergent} for soft/collinear emissions
and 
those events comes without the Sudakov suppression
 supplied by the 
showering algorithms and therefore leads 
to infrared and collinear sensitivity
(it's worth recalling that the {\em PS algorithm
doesn't modify the ME WEIGHT}, it simply dresses the
event with soft and collinear radiation).
Notice that, as thoroughly discussed in 
the previous chapters of these proceedings,
 soft/collinear emissions described by the PS
don't exibit the same unphysical behaviour: Sudakov form 
factors ensure that virtual effects are accounted 
for (in the NLL approximation) and thus enforce the
appropriate dumping of the singularities.

We are therefore forced to find a way to avoid double counting or at least
to push its impact below the accuracy of our prediction. The final goal
is to split the phase space in two regions: one, of soft and/or collinear
emissions, to be covered from the PS algorithm and the other one,
of hard and large angle emissions, to be described by the matrix element.
The separation among these two regions is achieved introducing one or more
``resolution parameters'' which discriminate among ``resolved'' jets
(to be described by the ME) and ``non resolved'' jets to be described
by the PS.
Notice that the solution has to fulfill three 
main requirements
\bit

\item It should avoid (minimize) double counting and ensure
full phase space coverage

\item It should ensure a smooth (as much as possible) transition
among the PS and ME description

\item It should ensure that the ME weight
is reweighted with the appropriate {\em Sudakov form factor},
where by appropriate we mean that it should reabsorb
the divergencies of the  ME weight. \footnote{
If we denote with ${\cal R}^{(n)}_{ME}$ 
the real  radiative correction to the
$n-$jets squared matrix element 
$X^{n}_{ME}$ 
with $l$ and $L$ the soft and collinear logarithms 
respectively and with $\xi$ an infrared/collinear
finite quantity we shall have
\[
{\cal R}^{(n)}_{ME}= 
X_{ME} \alpha_S (c_1 L l+ c_2 L + c_3 l)  
+ \alpha_S \xi 
\]
and the corresponding ``Sudakov form factor'' 
(to be used to reweight $X_{ME}$)
$\Delta$ has to be
\[
\Delta = \exp [ -\alpha_S (c_1 L l+ c_2 L + c_3 l)]
\]
Notice that with a wrong choice of $\Delta$ (different $c_j$)
one still obtains infrared/collinear finitness
({\em for the reweighted $X_{ME}$}), the 
result however will exhibit a strong dependence on
the chosen soft/collinear cut-off.
}

\eit

\section {Matching ME and PS: a practical perspective}

Let's now have a look at the practical implications
of the double countig problem.
One has in mind an event generator which combines the 
benefit of (fixed-order) ME calculation and showering
(+ hadronization).

Let's first attempt the more naive approach:

\bit

\item Use the ME to compute the {\em WEIGHT} of a given event.

\item Use the ME computation as input for the PS

\eit

One immediately faces the problem to determine
 the appropriate
parton level cuts required to build up the event sample. 
Notice that this is {\em mandatory} if one has final state
coloured partons (emitted by coulored partons): 
in the absence of cuts the ME {\em diverges}.

A first attempt
is to use, as parton level cuts, the same cuts used to define a jet in
the analysis.

Let's have a look at the consequences.
We analyze the answer of our event generator
(after ME computation and showering)
looking at {\em jets observables}.

To reconstruct {\em jets} out of final state
partons (namely
those found after the showering stage)
 we shall  use a simplified  cone algorithm
as provided by the
{\small GETJET} package \cite{getjet}, which represents a simplified
jet cone algorithm a la UA1.
Jets are defined requiring that
 jet $p_T$ has to be at least $20$ Gev,
the cone size is $R=0.4$ and the 
calorimeter coverage is $|\eta| < 2.5$.

Ultimately we shall study the signal 
 $p \bar p \to e^+ e^- + 2 \ jets$ 
{\em with at least two jets
with $p_T>40$ GeV and with $\Delta R > 0.7$} 
at the LHC COLLIDER.

We start by generating
$p \bar p \to e^+ + e^- + 2 \ partons$ ($parton\equiv g, u,d,c,s $)
with $p_T> 40$ Gev, $|\eta| <2.5$ and $\Delta R_{p_j p_k}<0.7$.
After ME computation the event is showered with PITHYA PS and the 
jets are reconstructed according to the chosen jet algorithm.

\begin{figure}
\begin{center}
\epsfig{file=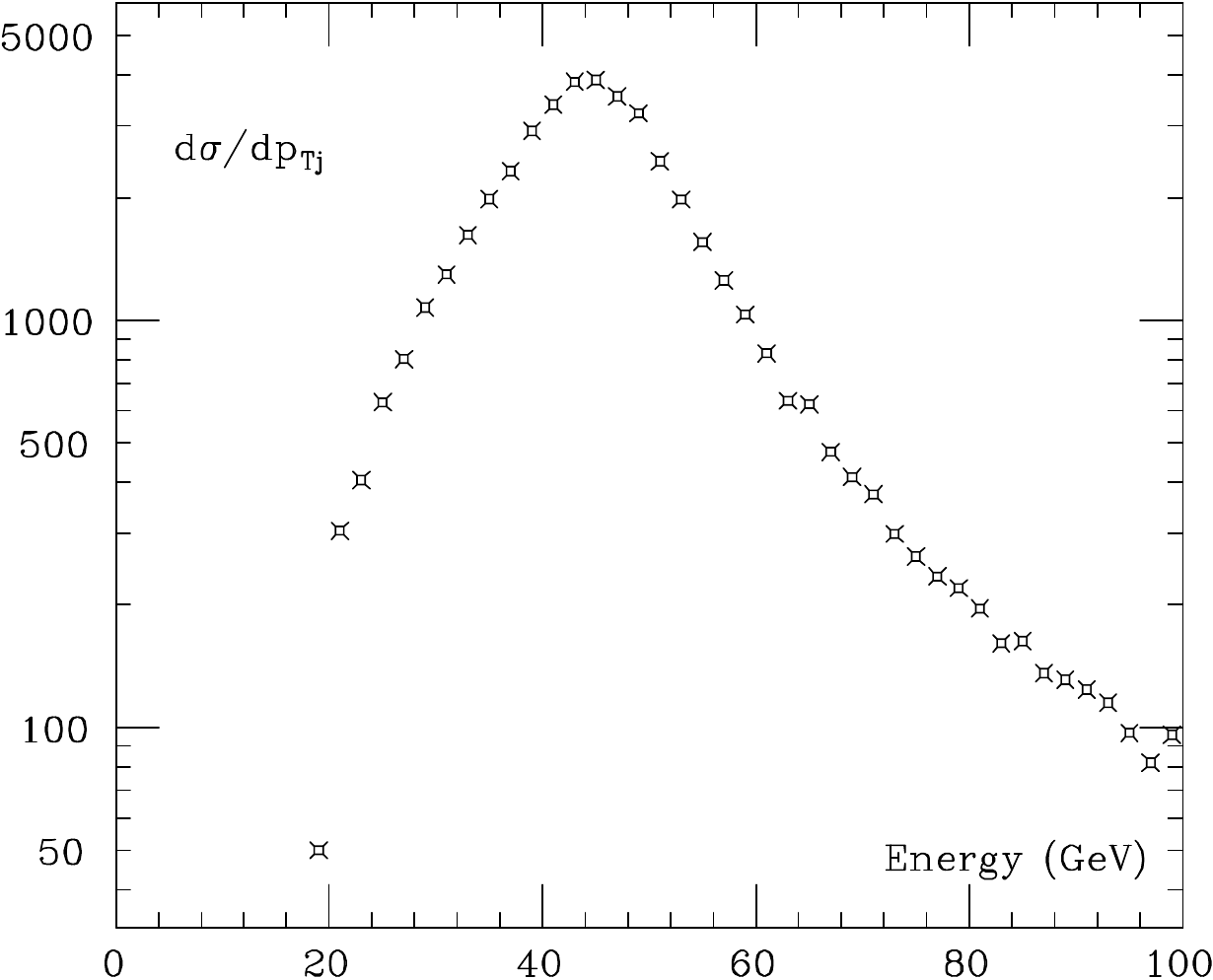,height=6.cm,width=0.48\textwidth} 
\epsfig{file=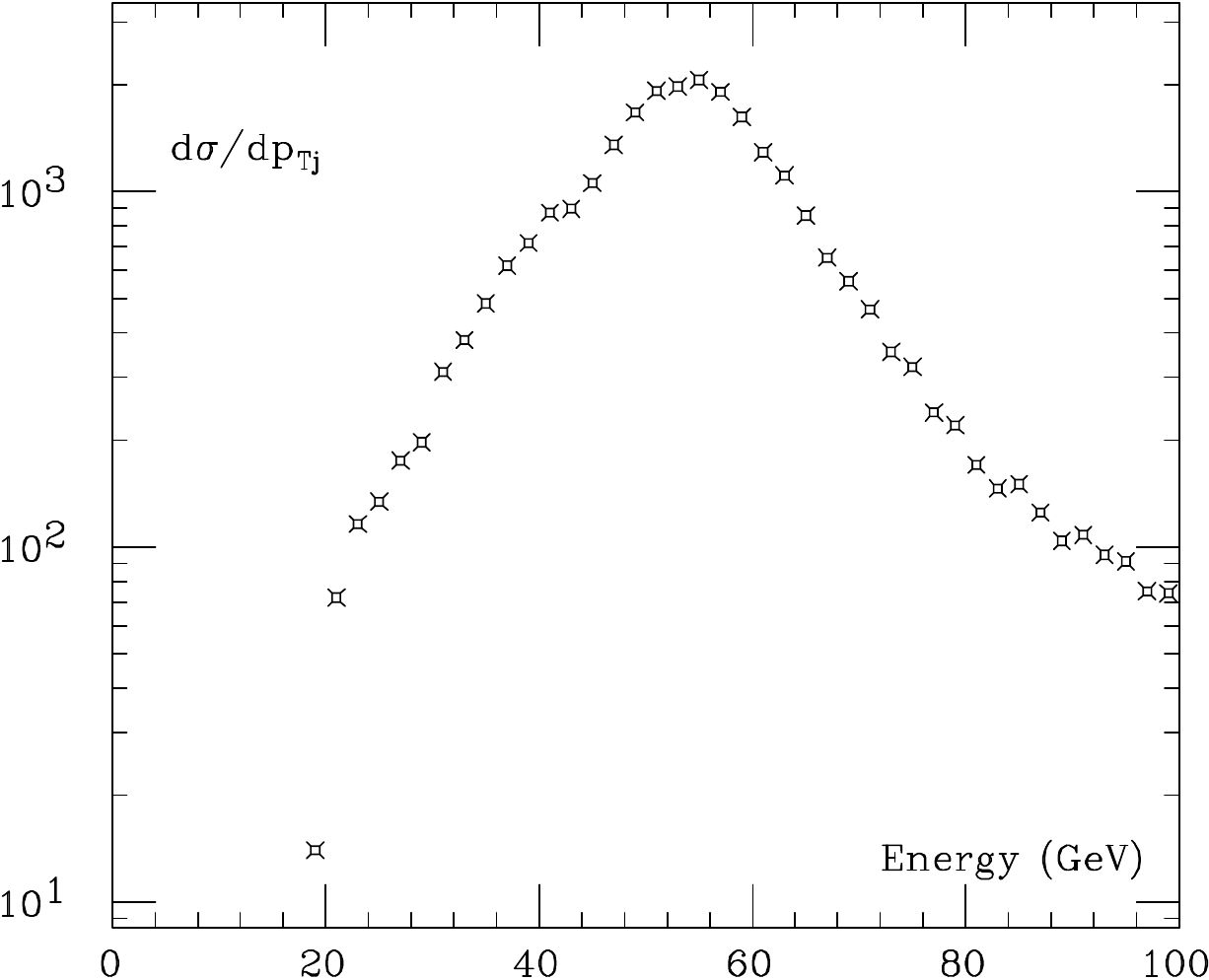,height=6.cm,width=0.48\textwidth} 
\\
\vskip 0.3cm
\epsfig{file=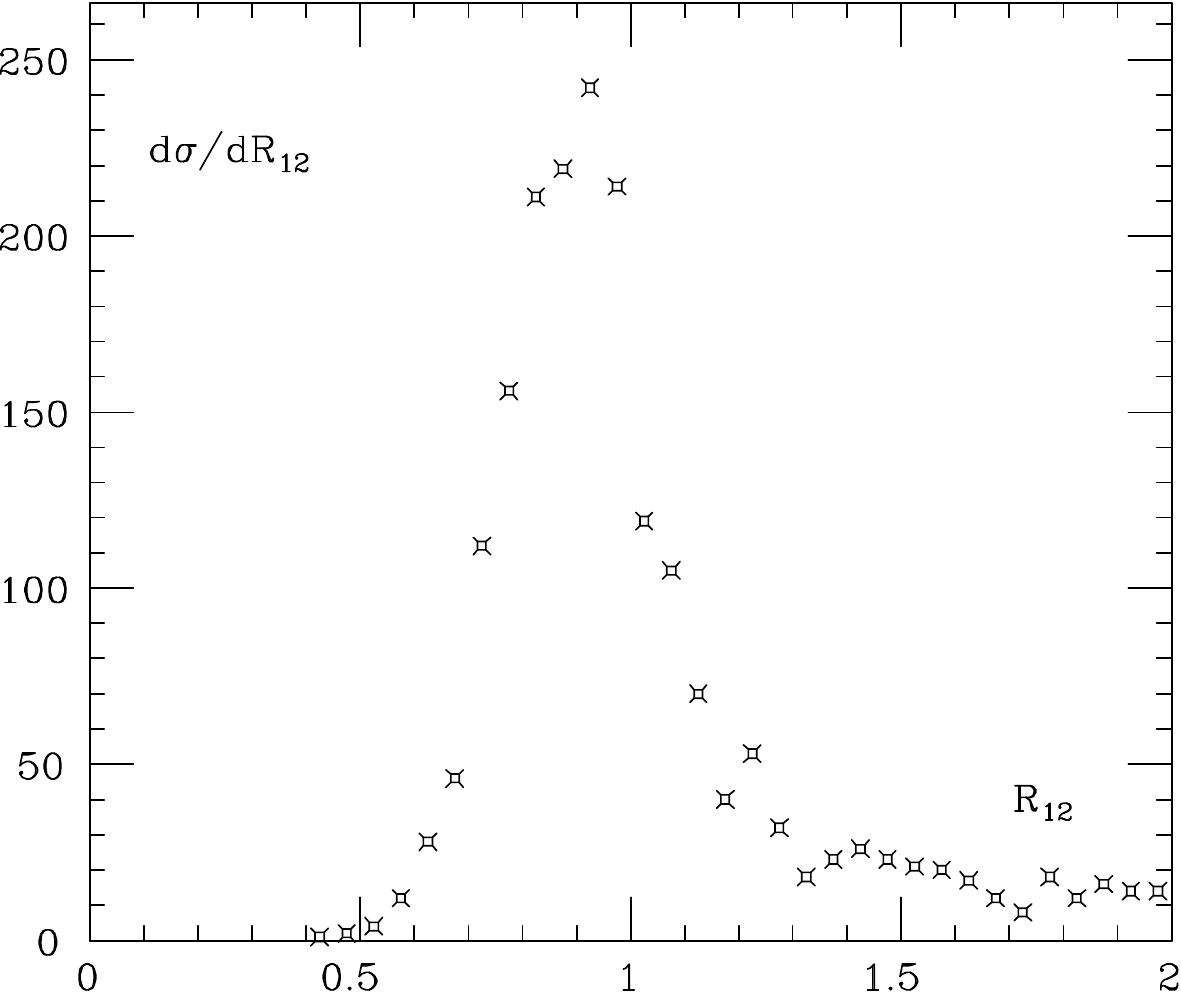,height=6.cm,width=0.465\textwidth}\hskip 0.2cm
\epsfig{file=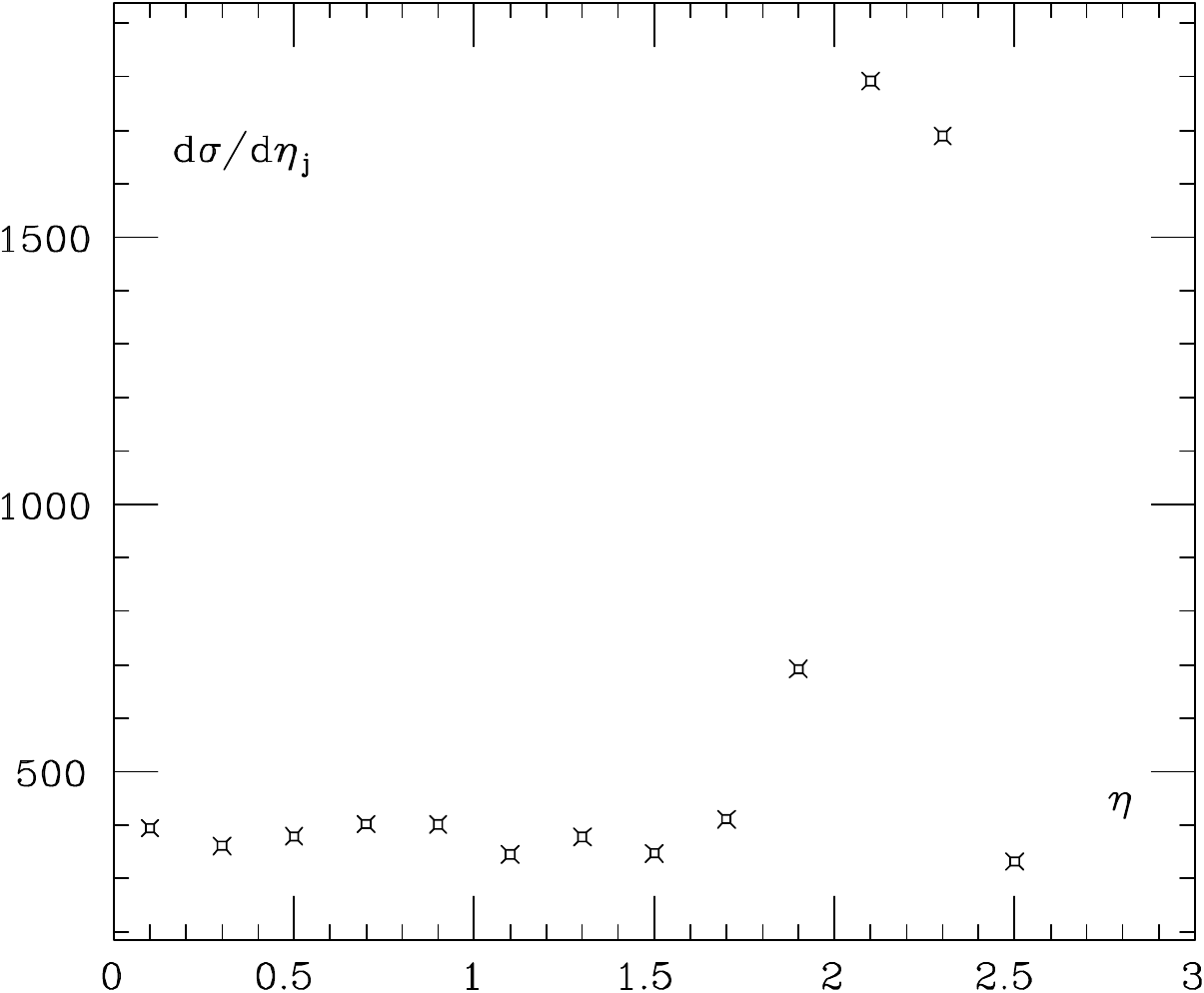,height=6.cm,width=0.475\textwidth} 
\end{center}
\caption{
\label{smear}
{\bf Upper-left panel:}  $p_T^{j_2}$ of the next to leading jet (jets ordered
in $p_T$) for {\em showered}
 events initiated by partonic event $Z^*/\gamma^* + 2 $ partons,
subject to the constraint $40 < p_{T2} < 50 $ GeV, $p_{T2}$ being 
the $p_T$ of the next to leading parton (partons ordered in $p_T$).
{\bf Upper-right panel:} Same as Upper-left panel but 
$50 < p_{T2} < 60 $ GeV
{\bf Lower-left panel:} distance $\Delta R_{12}$ among the two leading
jets for {\em showered}
 events initiated by partonic events $Z^*/\gamma^* + 2 $ partons,
subject to the constraint $0.7 <\Delta R_{12}^{partonic}< 1$.
{\bf Lower-right panel:} rapidity $| \eta^{j_1}| $ of the leading jet
for {\em showered}
 events initiated by partonic event $Z^*/\gamma^* + 2 $ partons,
subject to the constraint $2.5 > | \eta^1| > 2.0 $,  $\eta^1$ being the 
rapidity of the leading parton.
{\em All plots are for the LHC, and the normalizations are arbitrary}
}
\end{figure}

In fig.~\ref{smear} we display the $p_T$ of the 
second leading jet
(jets ordered according to $p_T$) for the events 
that, {\em at parton level (ME)}, have the second highest $p_T$ 
parton with a 
$p_T$ between 40 and 50 GeV
and with a $p_T$ between 50 and 60 GeV.
The effect of the shower is to smear the parton $p_T$:
some of the partons have their energy degraded by radiating energy,
other partons actually originate a harder jet collecting soft energy
(mostly originated by initial state radiation).
We are now facing a problem: by imposing generation cuts
 equal to the 
jet resolution
parameters we are loosing the contribution of ME partons with a $p_T$ just
below threshold which after showering would anyhow make up a jet
with a $p_T$ larger than that chosen in the analysis.
A similar ``edge effect'' occur for ``close'' ($R\simeq 0.7$),
see fig.~\ref{smear} or
large rapidity partons ($\eta \simeq 2.5$), see fig.~\ref{smear}.

\begin{figure}
\begin{center}
\hskip 0.4cm \epsfig{file=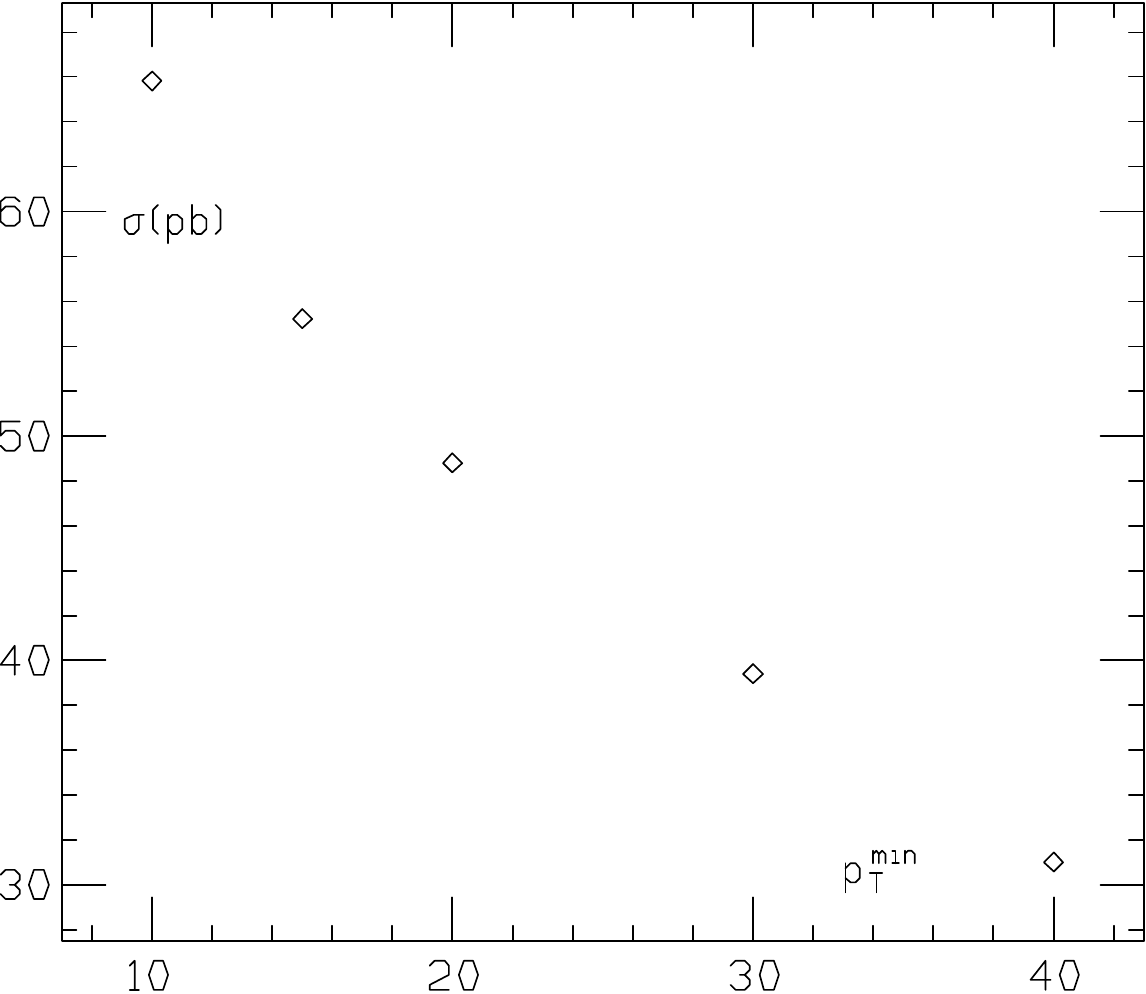,width=0.43\textwidth} 
\epsfig{file=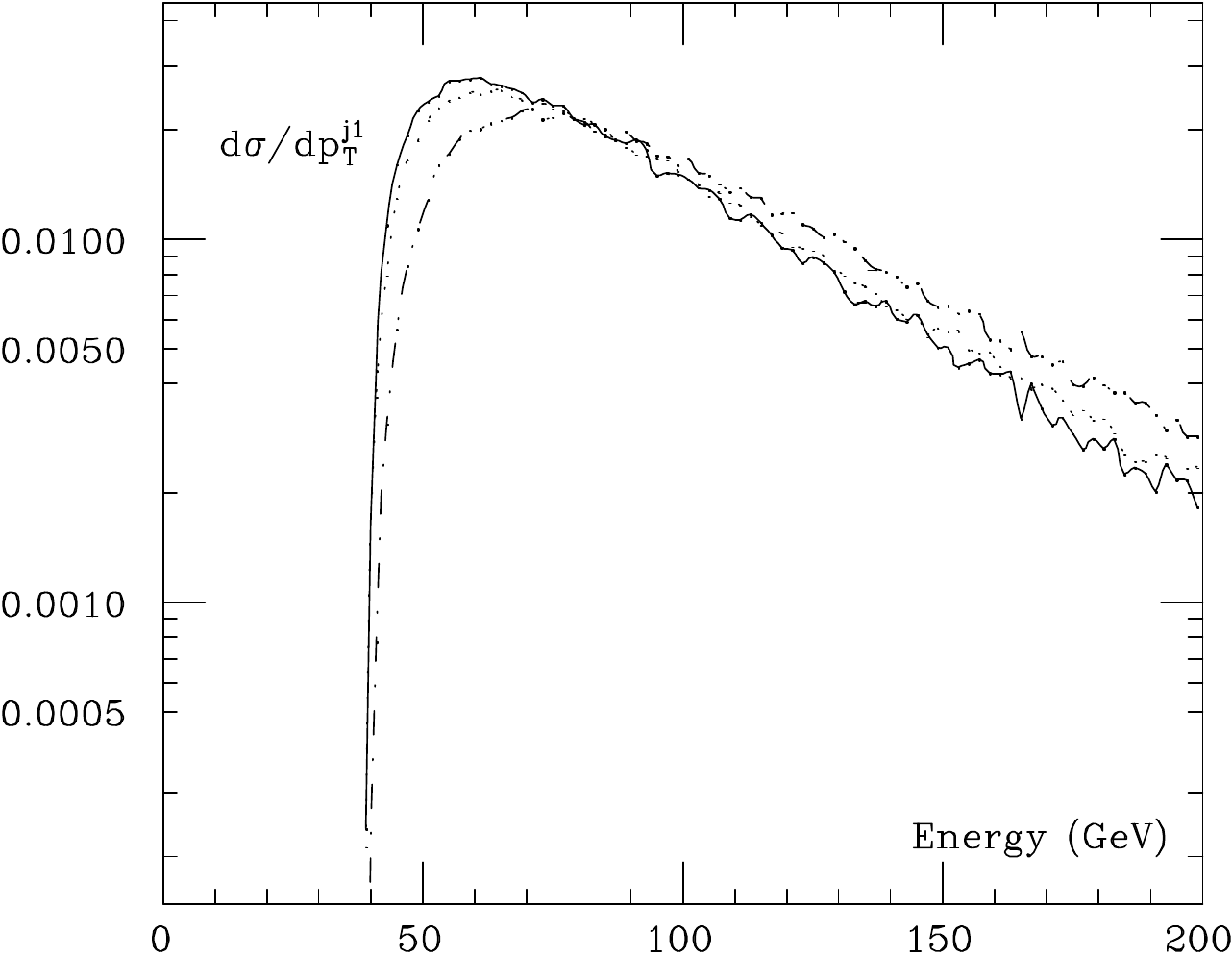,width=0.48\textwidth}
\\
\vskip 0.3cm
\epsfig{file=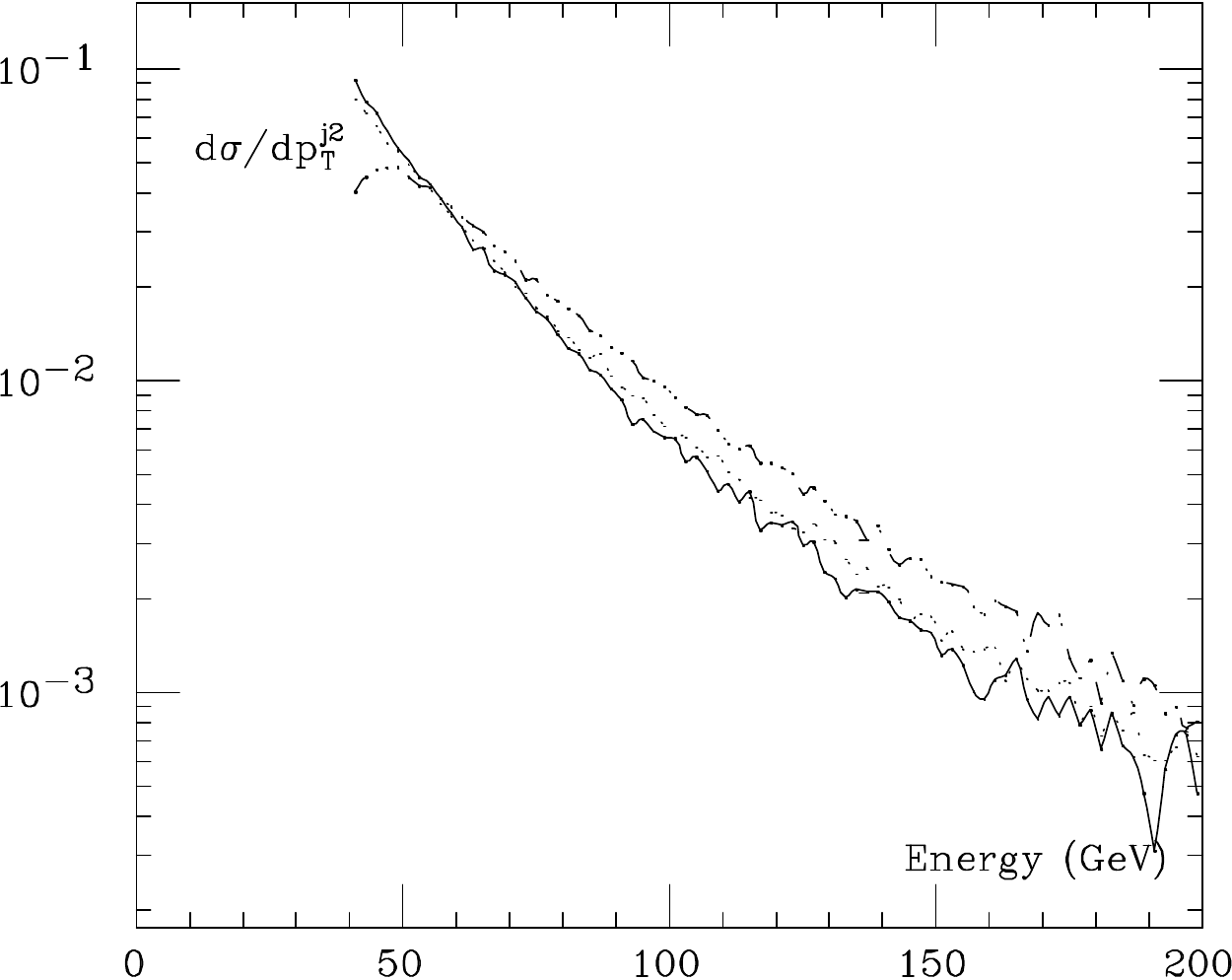,width=0.46\textwidth}
\epsfig{file=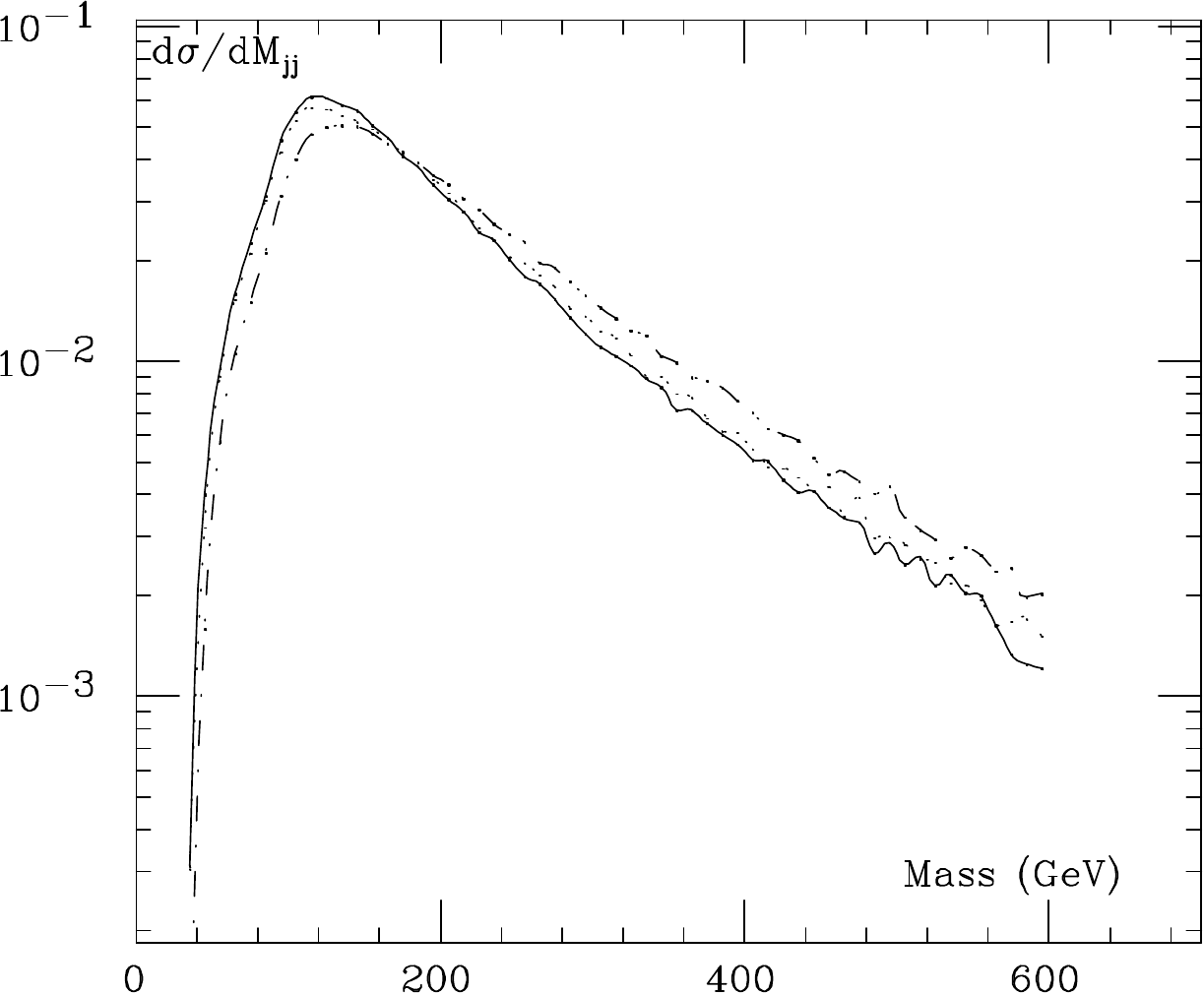,width=0.46\textwidth}\hskip 0.5cm
\end{center}
\caption{
\label{irclsns}
{\bf Upper-left panel:} 
Cross section in pb for $p p \to e^+ e^- + 2$ jets at the LHC as
a function of the {\em partonic $p_T$ at the generation level}.
Both jets, {\em after showering} are required to have $p_T > 40$ GeV,
$|\eta_j|<2.5$ and $\Delta R_{j_1 j_2} > 0.7 $.
{\bf Upper-right panel:}
transverse momentum $p_T^{j1}$ of the leading jet as a function
of parton level cuts. Continuos: $p_T^{part} > 10$ GeV; dots: 
$p_T^{part} > 20$ GeV; dot-dash: $p_T^{part} > 40$ GeV;
{\bf Lower-left panel:} 
transverse momentum $p_T^{j2}$ of the next to leading jet as a function
of parton level cuts.
{\bf Lower-right panel:}
invariant mass $m_{j_1 j_2}$ of the two leading jets as a function
of parton level cuts.
}
\end{figure}

An obvious solution to this problem is to soften the generation cuts.
In this way we loose efficency since many of the soft/collinear partons
don't originate resolved jets, however we recover the event which we
were missing in the previous analysis. 
We however face another
problem: our prediction is not stable against generation level
cuts. 
To see the effect we study the subsample of events that,
{\em after showering} have at least two jets with $p_T> 40 GeV$
and $\Delta R > 0.7$.
As it is seen in fig.~\ref{irclsns} the cross section increases
as parton level generation cuts are softened and also distributions are
affected. {\em Notice that resolution parameters for jets,
as well as the event selection criteria,
 are unchanged
and therefore the results, after showering should 
remain unchanged}.

The reason of this behaviour can be traced back to 
the problem of double counting associated with
soft/collinear ME emission. In the soft/collinear limit
the ME weight diverge, the PS can supply a hard 
and large angle
emission:\footnote {Notice that, if one or more ME parton
are ``soft'', there must be a corresponding number
of ``hard'' PS emissions in order to preserve the
number of ``observed'' hard jets}
 this is suppressed by a factor of $\alpha_S$
but enhanced by soft/collinear logarithms which (as opposite
to soft/collinear PS emission) {\em are not dumped} by
Sudakov suppression.

\section {Catani, Krauss, Kuhn and Webber algorithm}

A solution 
has been proposed in \cite{Catani:2001cc} in the context of
$e^+ e^-$ collisions. The dependence on the resolution parameter is
shifted beyond NLL. In \cite{Krauss:2002up} an extension of the procedure
to $ep$ and $pp$ environments has been proposed, without however a proof
that the dependence on the resolution parameter is below NLL.
This algorithm is implemented \cite {Schalicke:2005nv}
 in the {\tt SHERPA} MC \cite {nGleisberg:2003xi} and has been studied
in \cite {Mrenna:2003if} for {\tt HERWIG} and {\tt PITHYA} showers.

\subsection {PS and ME phase space boundaries}

The first ingredient of the algorithm is the {\em measure of
parton-parton separation}. To this pourpose the $k_\perp$ jet
algorithm \cite{Catani:1991hj,Catani:1992zp,Catani:1993hr}
is used: {\em the distance among   two final state partons} is defined
as
\beq
y_{ij} = \frac {2 \min \{E_i^2,E_j^2 \} (1-\cos\theta_{i,j})}{s}
\label {yij}
\eeq
$s$ being the center of mass squared energy, $E_{i,j}$ the parton
energies
and $\theta_{i,j}$ their relative angles.
{\em The ``distance'' between a parton and the incoming partons (the
beam) } is defined
as
\beq
y_i = \frac{p_{\perp i}^2} {s}
\label{yi}
\eeq
The separation among ME partons and PS partons is achieved
introducing a resolution parameter $Y_{sep}$ and
\bit

\item requiring that ME partons are {\em resolved}:
\[
y_{i,j}, \ y_i > Y_{sep}
\]

\item vetoing PS emissions at a scale harder than $Y_{sep}$

\eit

This ensures that a given phase space configuration is covered only
once

Notice that in the region described by the PS dead zones are still
present
and thus one has to choose $Y_{sep}$ in such a way to minimize
these effects in the regions relevant for the analysis of interest.

One could also use a different measure of the {\em parton-parton}
distance, it is however necessary that it preserve the properties
of $k_\perp$ algorithm if one wishes to retain NLL accuracy.

\subsection {Matching ME and PS weight }

The second key ingredient is {\em ME reweighting}. The ME weight is 
infrared and collinear divergent and thus will diverge as $Y_{sep}$
becomes small. On the other hand the PS is well behaved in this limit
due to soft and collinear emission resummation. The ME is thus
reweighted in order to ensure a smooth transition among ME and PS 
description:

\bit

\item for a given ME phase space point a {\em branching tree}
is reconstructed by clustering toghether the two
{\em closest partons}
 (according to  $y$ measure given in eqns.~(\ref{yij},\ref{yi})) 
and iterating the procedure until when the "leading order" process
is reached: for $p p \to W + \mathrm n-jets$ we proceed until $ q q' 
\to W$ is reached, for $p p \to t \bar t + \mathrm n-jets$ until 
$p p \to t \bar t$ is reached\footnote{some qualification is actually
required: if the scale of some QCD emission is larger than the typical
scale for the LO process the clustering is done in a different way. We
refer to \cite{Krauss:2002up} for a more thorough discussion.
}

\item for each branching reweight the squared ME by $\alpha_S(k_\perp)/
\alpha_S(Q_{ME})$

\begin{figure}[hbt]
\begin{center}
\epsfig{file=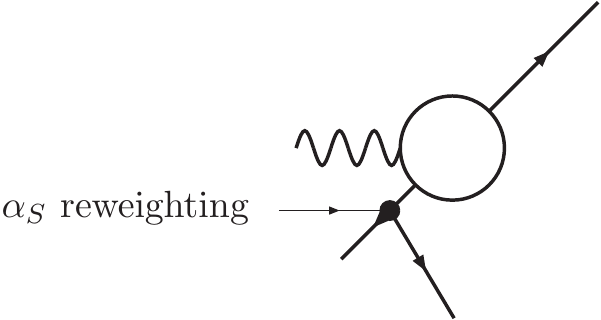,width=0.4\textwidth}
\end{center}
\ccaption{}{\label{asrw}
Same symbols (and process) as in fig.~\ref{dbcnt}.
ME final state partons (originating from small dots) are clustered
toghether to reduce the process to the leading order $2\to2$
process. Small dots represents would be branchings in a PS-like
 picture of the event. The $k_\perp$ separation among the
clustered partons is the appropriate PS scale for $\alpha_S$
evaluation at the given branching.
}
\end{figure}

\item to each internal and external line of the branching tree
associate the proper combination of {\em Sudakov}\footnote
{for a thorough discussion of Sudakov form factors meaning 
and definitions refer to the previous chapters of this 
proceedings.}
{\em form factor}:
defining
\ban
Y_{sep} & = & \frac{Q_1^2}{Q^2} \\
y_{j} & = & \frac{q_j^2}{Q^2} 
\ean
where $Q^2$ is the hard process scale,
to each {\em internal line}, connecting a branching at a scale
$q_j$ and a branching at a scale
$q_k$, associate a reweighting factor
\[
\frac{\Delta (Q_1,q_k)}{\Delta (Q_1,q_j)} 
\]
where
$\Delta(Q_1,q_j)$ is the appropriate Sudakov Form Factor.
To each external line, originating from a branching at a
scale $q_j$ associate a reweighting factor
\[
\Delta (Q_1,q_j)
\]

\item also the PS needs to be modified: 

\ben

\item 
{\em the scale }for the
PS evolution is given, for each parton, by the scale at which the
parton was produced (the hard process scale for initial state partons)

\item {
{\em resolved} PS emissions ($y> Y_{sep}$) are inhibited.
Notice that this is done simply {\em rejecting } those emissions
without affecting the event weight
}

\een

\eit

\subsection{Building the event sample}

\bit

\item Finally one has to build event samples
with up to $\infty$ ME partons 
(each normalized to the same luminosity, 
at least in principle)
and sum them up toghether.

\item
One obviously has to stop to some {\em finite} number
of ME partons.
The highest multiplicity sample needs to be treated
separately: for a given ME the smallest $k_\perp$
separation is computed and the PS is allowed to produce
branching up to this scale. In this way the higher parton
multiplicities are supplied by the shower emissions.

\eit

In \cite{Catani:2001cc} it is shown that, with the above
prescriptions, the NLL resummed exclusive $e^+ e^-
\to n $ jets is reproduced.

A few remarks are in order

\bit

\item the proof of NLL accuracy holds only for
$e^+ e^-$ collisions;

\item even in the $e^+ e^-$ framework, to achieve
 NLL accuracy,
 it is crucial that
the employed PS {\em correctly describes the soft sructure of
the ME, including interferences}: this is the case
for PS incorporating coherent branching like HERWIG
or based on dipole emission like ARIADNE but not
for virtuality ordered PS like PITHYA. Notice that
APACIC (SHERPA) provides both options: virtuality ordered
and angular ordered PS\footnote{actually the first emission
is not described by coherent branching 
} and thus it
provides the opportunity to study the numerical impact of
the two approaches.


\item ultimately the smoothness of the interpolation
must be judged inspecting the stability of the
relevant (for the analysis) ditribution over at least
a sizable range for the resolution parameter.

\item the sample with the highest multiplicity of
ME emissions is also the one with the larger systematics.
One should care to minimize its weight on the inclusive
 sample and anyway to check ``indipendence'' of
the predictions from the
maximum number of ME partons used to build up the sample.

\eit

{\em Let's finally add a few words of caution
\bit

\item NLL accuracy is already ensured by the PS standalone (if coherent
effects are included).

\item the ultimate goal of ME-PS merging is to correctly describe
{\em hard and large angle emissions} toghether with soft/collinear
resummation. This class of events is suppressed by at least
{\em two powers of log} and thus the proof of (\cite{Catani:2001cc})
doesn't ensure that they are dealt with correctly.

\item in particular if the Sudakov reweighting of 
{\em hard and large angle emissions} is not correct the hard tail of the
distributions will suffer of {\em LL dependence on the resolution parameter}
and thus of artificial enhancement/dumping.
\eit
}

\subsection{Implementation and comparison with TEVATRON data}

The CKKW algorithm for $pp$ collisions, according to the
proposal in (\cite{Krauss:2002up}), is implemented in
SHERPA \cite{sherpa} and has been studied in 
\cite{sherpawlhc,sherpawtev,sherpawwtev}. The overall consistency looks
good:

\bit

\item the overall rate is stable against sizable changes
of the resolution parameters.

\item the distributions doesn't show 
large discontinuities around
the resolution parameters.

\item stability is achieved with a moderately small number
of ME partons.

\item there is a nice agreement with {\tt MC@NLO} 
(\cite{mcnlo})

\eit

\begin{center}
\begin{figure}
\begin{center}
\epsfig{file=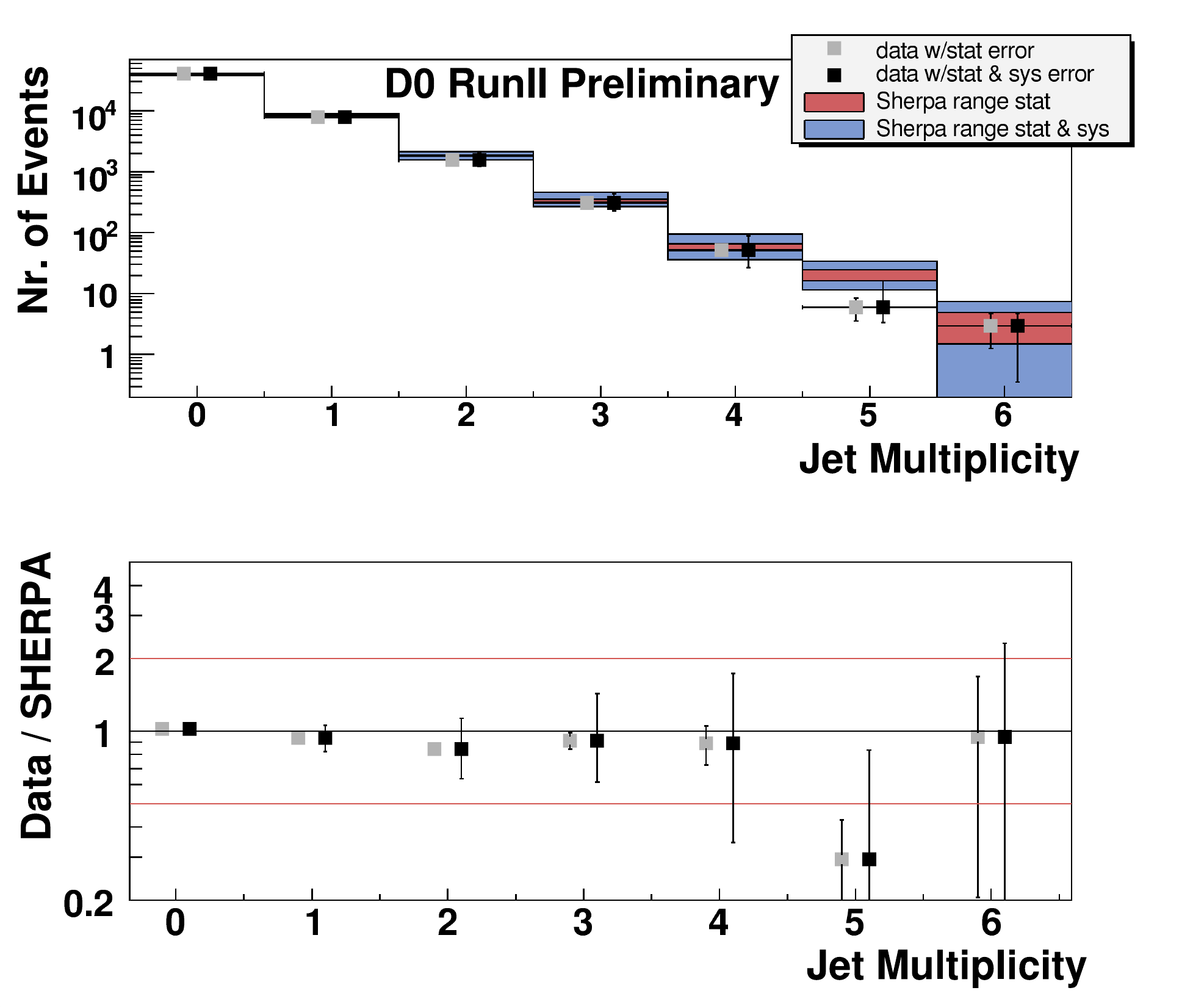,width=0.48\textwidth} 
\epsfig{file=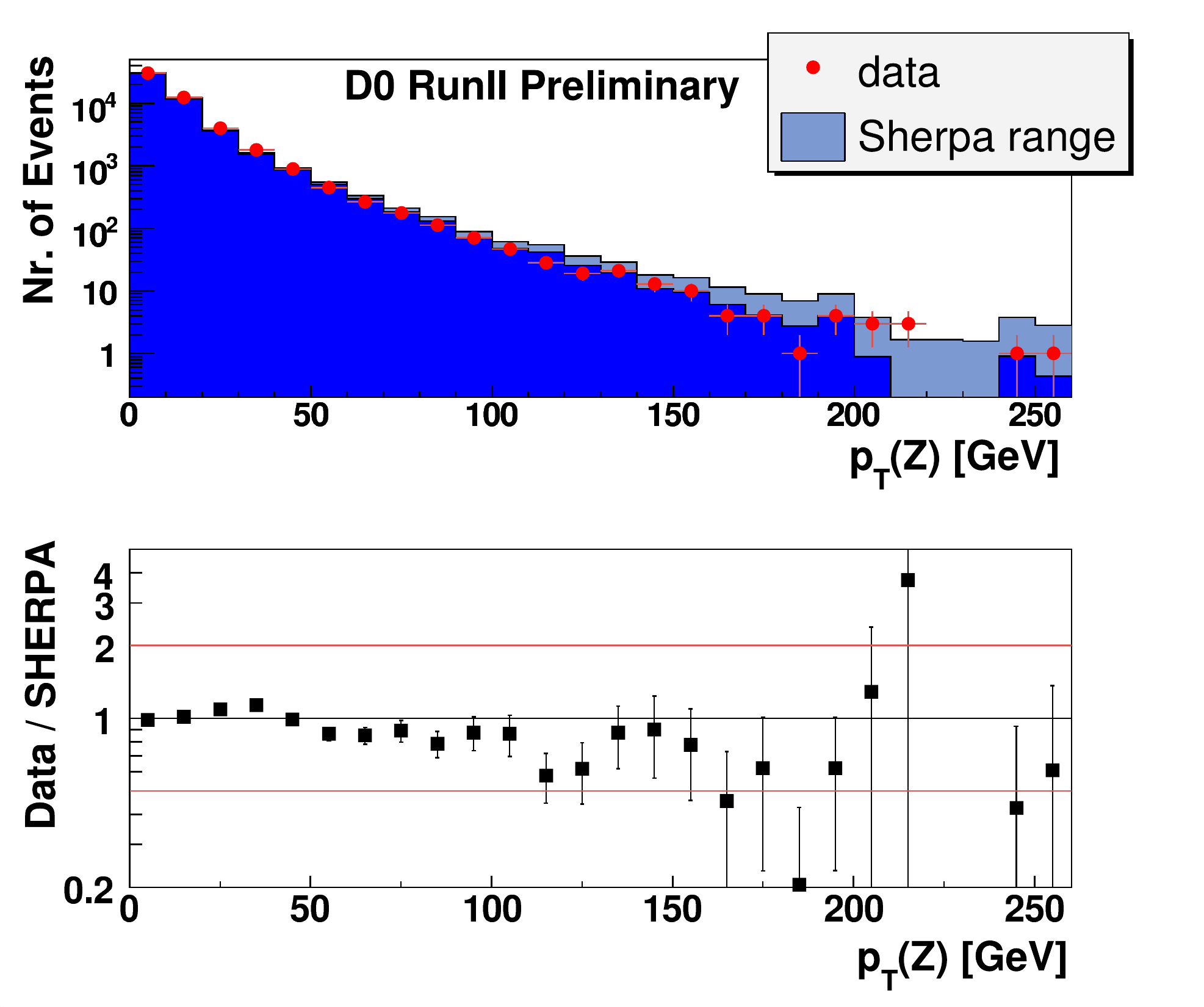,width=0.48\textwidth} 
\epsfig{file=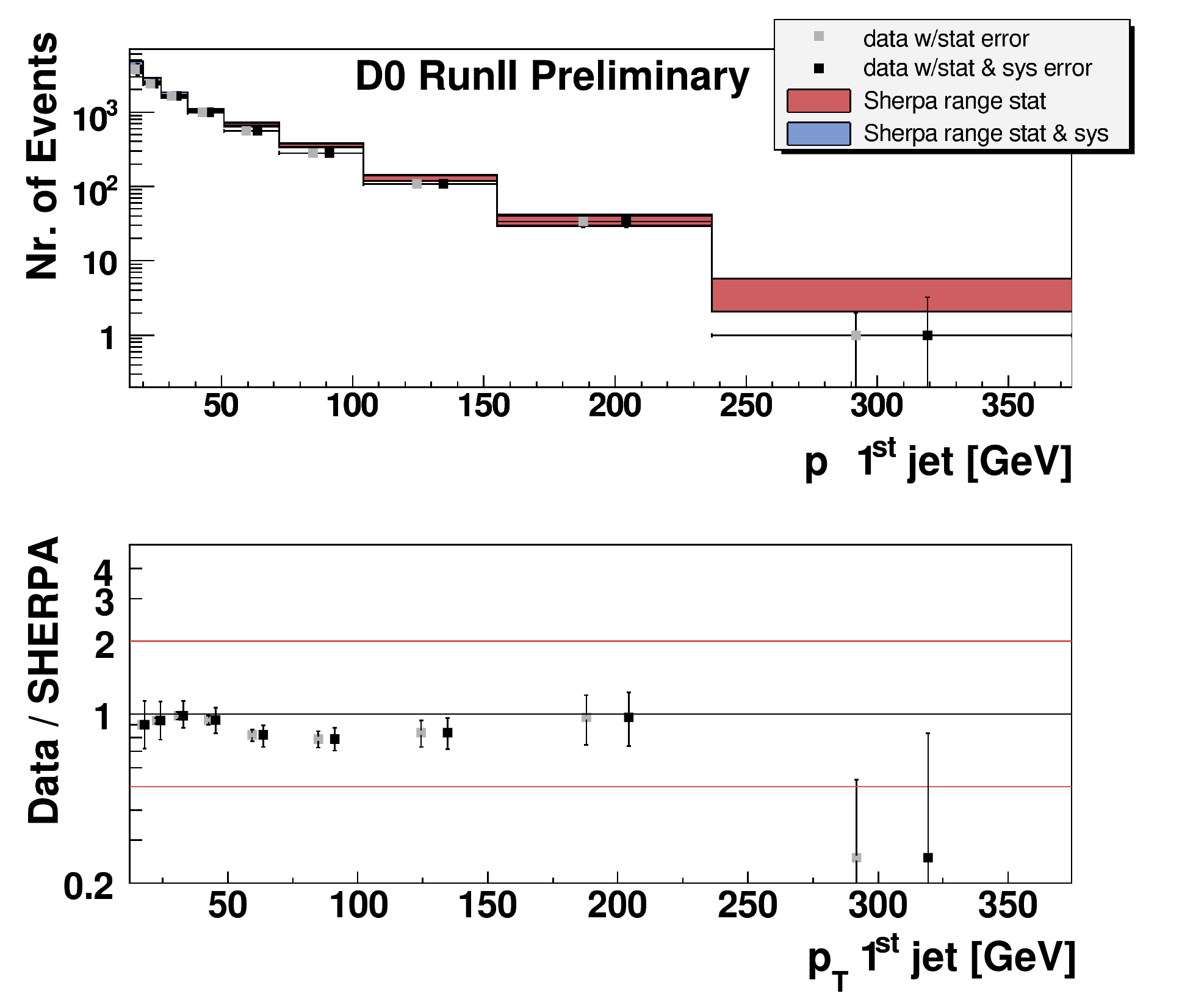,width=0.48\textwidth} 
\epsfig{file=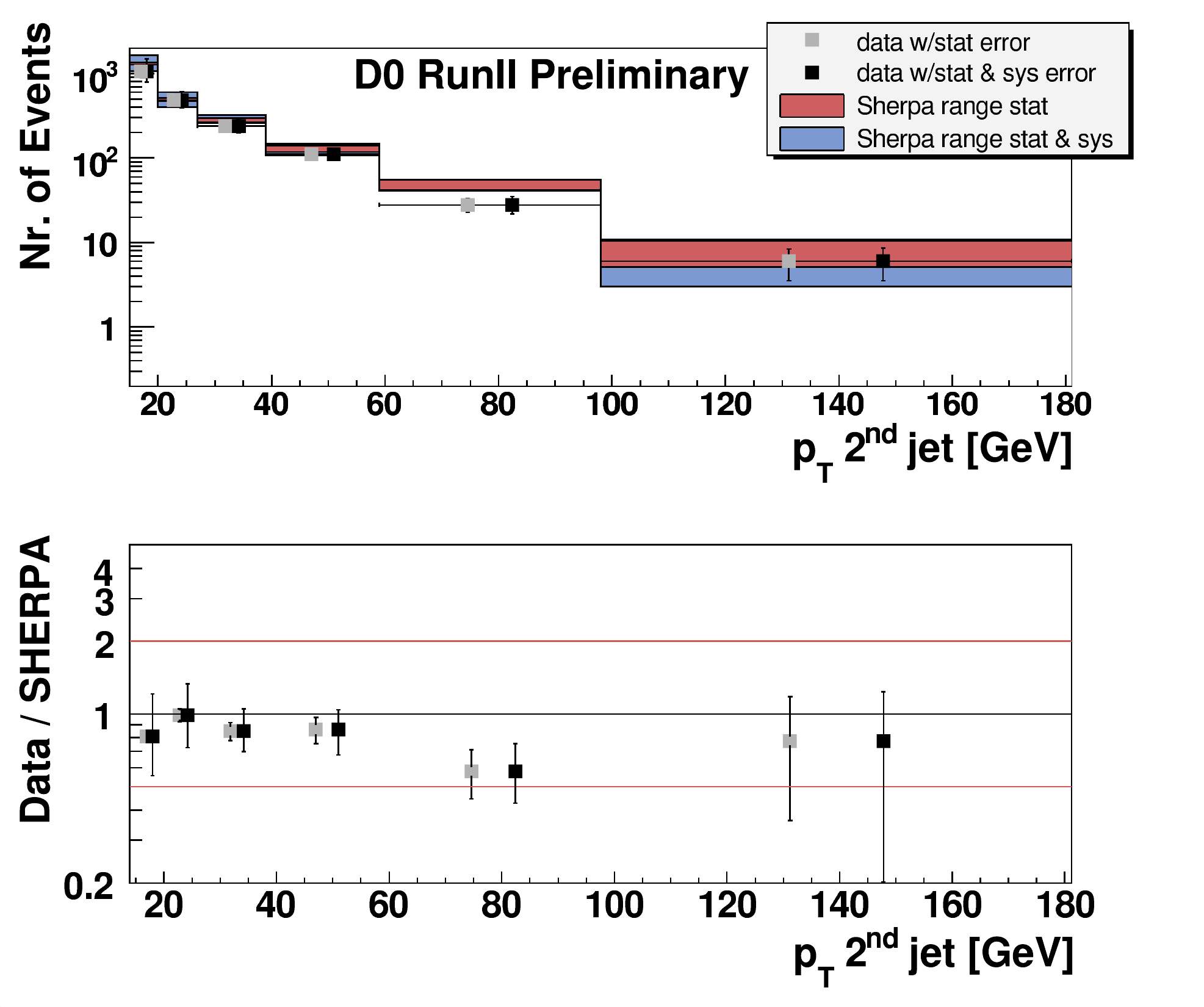,width=0.48\textwidth} 
\end{center}
\caption{
\label{sherpavszjtev}
{\bf Upper-left panel:}  Jet multiplicity in inclusive $Z$ production
{\bf Upper-right panel:} $p_T^{j_1}$ of the leading jet (jets ordered in
$p_T$ ) in $Z+ jets$ production
{\bf Lower-left panel:} $p_T^{j_1}$ of the leading jet (jets ordered in
$p_T$ ) in $Z+ jets$ production
{\bf Lower-right panel:}$p_T^{j_2}$ of the next to leading jet (jets ordered in
$p_T$ ) in $Z+ jets$ production
{\em All plots are for the Tevatron and the  normalization
of SHERPA prediction is fitted to the data. Both absolute values and
SHERPA to DATA ratio are shown.
Figures from \cite{d0zjet_sh}}
}
\end{figure}
\end{center}

There is ongoing experimental activity in testing SHERPA predictions
expecially for jet related quantities.
D0 collaboration has studied $Z+ jets$ production. 
A thorough account can be found in \cite{d0zjet_sh},
the overall agreement looks pretty good. 
In fig.~\ref{sherpavszjtev}
we show the comparison of SHERPA prediction and data for the
$p_T$ of the Z boson and of the two leading jets and for the jet multiplicity.

\begin{center}
\begin{figure}
\begin{center}
\epsfig{file=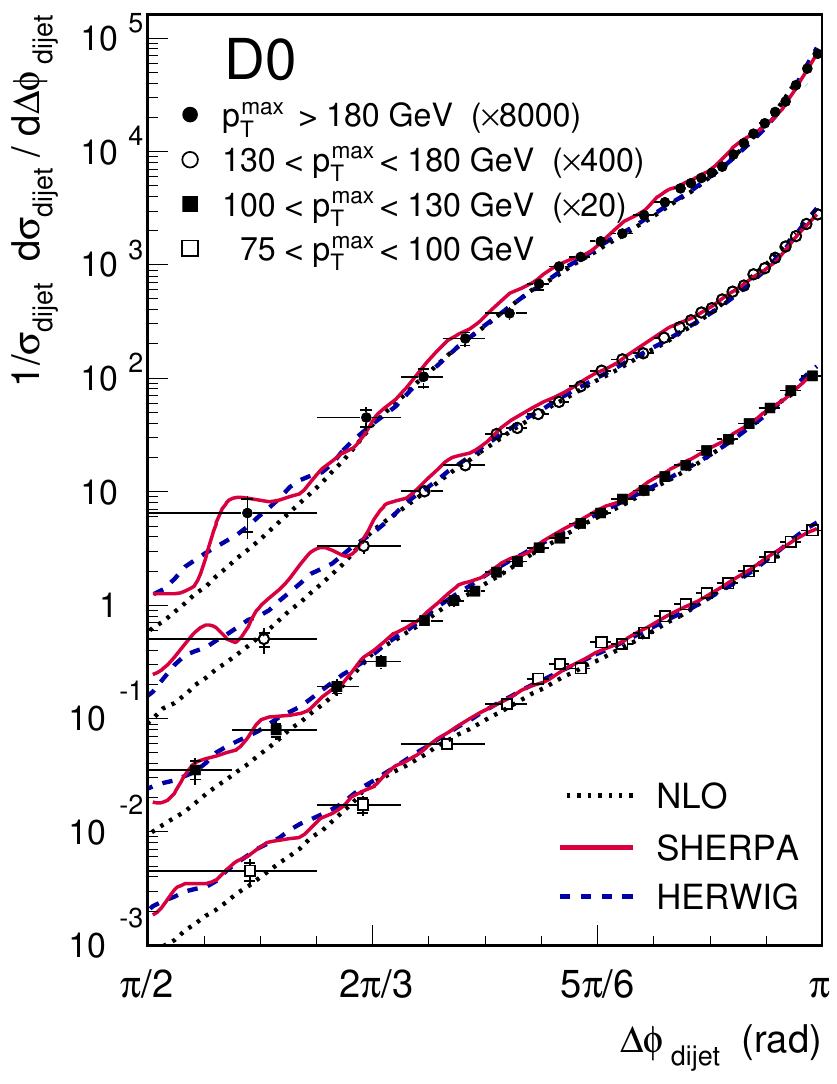,height=8.cm,width=10.0cm,angle=0} 
\end{center}
\caption{
\label{sherpadijettev}
Angular separation $\Delta\Phi$ in the transverse plane
among the two leading jets (jets ordered in $p_T$) in inclusive
two jets sample. D0 data versus SHERPA prediction are shown.
Figure from \cite {d0dijet}.
}
\end{figure}
\end{center}

D0 collaboration has also studied \cite {d0dijet}
dijet azimutal correlation in pure
jet sample and compared DATA to SHERPA predictions, again finding
good agreement as shown in fig.~\ref{sherpadijettev}.

\section {Michelangelo Mangano matching prescritpion}

An alternative prescription has been proposed by M.~Mangano in \cite{matching}.


The pourpose is to build up an inclusive event sample summing up
"exclusive" event samples with different {\em m-clusters} multiplicities.
Clusters are just  partons clustered toghether according to some
arbitrary jet finding algorithm and doesn't need to be identified with 
experimental 
jets, once the event sample is built the user can apply any kind of
analysys to the resulting events.

To produce an event sample with {\em m-clusters}

\bit

\item produce a sample of unweighted {\em partonic} events
with 
\[
P_T > P_{Tgen} \hskip 2truecm \Delta R_{j_1 j_2} > R_{gen}
\hskip 2truecm | \eta | < \eta_{gen}
\]
Notice that, in principle, $p_{Tgen}$, $R_{gen}$ and
$\eta_{gen}$ {\em are not parameters of the matching prescription}.
One should generate completely inclusive (no cuts at the
generation level) and, once the matching step is performed, unwanted 
kinematics configuration will be rejected. However this is not possible
since it will lead to null unweighting efficency and therefore one has
to find a satisfactory 
balance: generation cuts should
be soft enough to avoid edge effects and hard
enough to obtain a good unweigthing efficency.

\item for the given kinematic configuration a PS-like branching tree is
obtained
clustering (see fig.~\ref{asrw}) the final partons
according to $k_\perp$ \cite {Catani:1991hj,Catani:1992zp,Catani:1993hr}
 algorithm until when the LO process is obtained. Namely
for $p p \to t \bar t + m$-jets cluster until when
$p p \to t \bar t $ is reached, for 
$p p \to W + m$-jets cluster until when
$p p \to W $ is reached, ... Then at each
``branching'' assign the proper $\alpha_S(k_\perp)$ factor.
In this way the ME is reweighted to mimic more closely the
PS weight. This step is the same as in CKKW algorithm.

\item perform the shower and merge toghether the obtained
partons (ME + PS) to reconstruct cluster of partons according
to a jet finding algorithm, see figg.~\ref{dbcnt} and
\ref{dbcnt1} where small arcs denote clusters.
 In {\tt ALPGEN} Paige's {\tt GETJET}
algorithm is used. The minimum jet transverse momentum
$p_{Tmin}$, and separation $R_{min}$ toghether with
maximum rapidity $\eta_{max}$ are {\em the genuine
matching parameters}.
Notice that, to avoid edge effects due to the smearing of jet momenta
induced by the shower, matching parameters should be {\em harder}
than generation cuts
\[
p_{Tmin} > p_{Tgen}
\hskip 2truecm
R_{min} > R_{gen}
\hskip 2truecm
\eta_{max} < \eta_{gen}
\]
The larger the difference the smaller the edge effects and the
unweighting efficency. Actually for $\eta_{max}$ there is an additional
subtlety to be discussed later.

\item now reject the event if the number of clusters is not equal to the
number of ME generated partons. These events will be generated in 
other event samples with different parton multiplicities and this prescription
avoids double counting.
Notice that by performing the PS till the very end and
applying the rejection criteria to the final PS generated partons
we achieve, at least in the limit of no cuts at the generation level,
a net separation among PS and ME generated events, indeed there is
no chance that the same event can be generated by ME with different
multiplicities.

\item if the number of cluster is  equal to the
number of ME generated partons define the matching of a {\em parton}
and a {\em cluster} as follows. A parton matches a cluster
is the relative separation is smaller than $R_{min}$, namely
if the parton is inside the jet cone. If more than one parton
matches the same cluster (collinear ME partons)
 or if a parton doesn't match to any cluster (soft ME partons),
reject the event. With this prescription we avoid double counting and we
{\em reweight}
the ME with the appropriate {\em Sudakov form factor}
\footnote{Actually a residual infrared sensitivity is left:
a soft partons might accidentally fall inside the cone of
a cluster originated from a hard PS emission. This is suppressed
by the small avaliable phase space and in the studies performed insofar
we havn't found any appreciable effect even pushing the
generation cuts close the soft/colliner PS cut-off.
}. Indeed (with this prescription toghether with the requirement imposed
at the previous step)
a ME ``event'' will be accepted
according to the probability that the PS doesn't emit any ``hard''
(above the chosen resolution)
radiation \footnote{Actually the prescription overestimate
the Sudakov form factors: two ``soft'' partons can be clustered
even if they can't be traced back to a single splitting. If the 
resulting cluster is hard enough the event is vetoed. The lower
multiplicity sample will not return this PS history it will simply
return the contribution of the production and subsequent splitting
of the hard parton. This is again a phase space suppressed Log term.
}.
An important point has to be noticed here regarding
$\eta_{gen}$ and $\eta_{max}$. We have already noticed that
to avoid edge effects we should have $\eta_{gen} > \eta_{max}$.
There is an additional subtlety here. If one is not inclusive
in $ \eta_{max}$ we don't obtain the proper Sudakov form factor.
This is due to the fact that, not being inclusive in $\eta_{max}$ we
reweight the ME with the probability that the PS doesn't produce
any hard emission {\em inside the given rapidity range}. This
probability, with shrinking rapidity range, obviously approaches one
rather than the Sudakov form factor which we wish. Therefore
strictly speaking both $\eta_{gen}$ and $\eta_{max}$ should go to 
$\infty$. Taking smaller values increases the unweighting efficency and
again the actual choice
is a matter of balance among the 
increasing efficency and the
increasing systematic effects. Notice that whereas for $p_{Tgen}$
and $R_{gen}$ we are indeed {\em forced} to choose non zero values
to avoid null unweighting efficency, for $\eta_{gen}$ there is actually
a natural maximum allowed value once $p_{Tgen}$ is chosen and therefore,
at least in principle, it's possible to avoid completely this problem.

\item the cross section of the event sample is simply the input,
parton level, cross section times the ratio between
the number of accepted events and the total number of processed events.

\item we repeat the above steps for ME with 0
up to $\infty$ light quarks and jets and we sum up the various
event samples

\begin{figure}[hbt]
\begin{center}
\epsfig{file=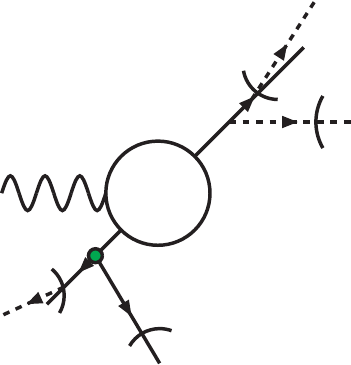,width=0.3\textwidth}
\end{center}
\ccaption{}{\label{inclus}
Same symbols (and process) as in fig.~\ref{dbcnt}.
Hard emissions from ME and one hard emission from PS, the
number of reconstructed clusters will be greater than the number
of ME partons. This event will be retained only in inclusive samples,
namely events initiated by ME with the highest 
particle multiplicity.
}
\end{figure}

\item actually, since it is impossibile to compute ME with an arbitrary
number of legs, we shall stop at a definite number $n_{max}$
of light quarks or gluons ($n_{max}= n_{light \ quarks}+ n_{gluons}$).
For the corresponding matrix element the matching procedure has to be
modified, to define an {\em inclusive event sample} (see fig.~\ref {inclus}),
 as follows

\ben

\item events with a number of reconstructed clusters {\em equal or larger}
than $n_{max}$ are accepted.

\item events are accepted only if $n_{max}$-ME partons match the
hardest m-clusters (ordinated by $p_T$).

\een

\eit

The major advantage of the above prescription is to be independent
from the PS algorithm which is employed and to require minimal
interaction with the  PS code itself: it is enough to have 
access to the final partonic configuration after the shower.

From a theoretical poin of view it has the disadvantage that a clean
classification of the Logarithmic
 structure accounted for or missed is very hard.
It's hard to work out a closed analytical form
for the ``Sudakov'' reweighting imposed by the algorithm
and ultimately it rests on the empirical evidence
provided by the smooth behaviour of the distributions
and their (in)dependence from the matching parameters. 
On the other hand it has the advantage that the 
``Sudakov'' reweighting is borrowed from the PS: assuming 
that indeed this is done exactly
(a strong and undemonstrated assumption), this would be the 
best possible recipe. In fact if, 
for the given kinematical configuration,
 the PS reproduces correctly
 the divergent structure of the ME the two descriptions will 
merge correctly, otherwise it 
will be anyway impossible to achieve simultaneously
a correct infrared/collinear damping and a smooth 
interpolation among PS and ME description.

\subsection{Implementation and comparisons with Tevatron
 data}

The algorithm described in the previous section
is implemented into the ME event generators
{\tt ALPGEN} \cite{alpgen}, {\tt HELAC}
\cite{helac} and {\tt MADGRAPH} \cite{madgraph}.

A fairly extensive exploration of the matching prescription,
for the case of $t \bar t + $jets production is reported in
\cite{ttalpgen}. 
The overall consistency looks good, the prediction is
stable against sizable variations of the matching parameters
and also the comparison with {\tt MC@NLO} description
is good, once the appropriate K-factor rescaling
is imposed.

The prescrition has also been tested against Tevatron data
mostly looking at jets productions.

\begin{center}
\begin{figure}
\begin{center}
\epsfig{file=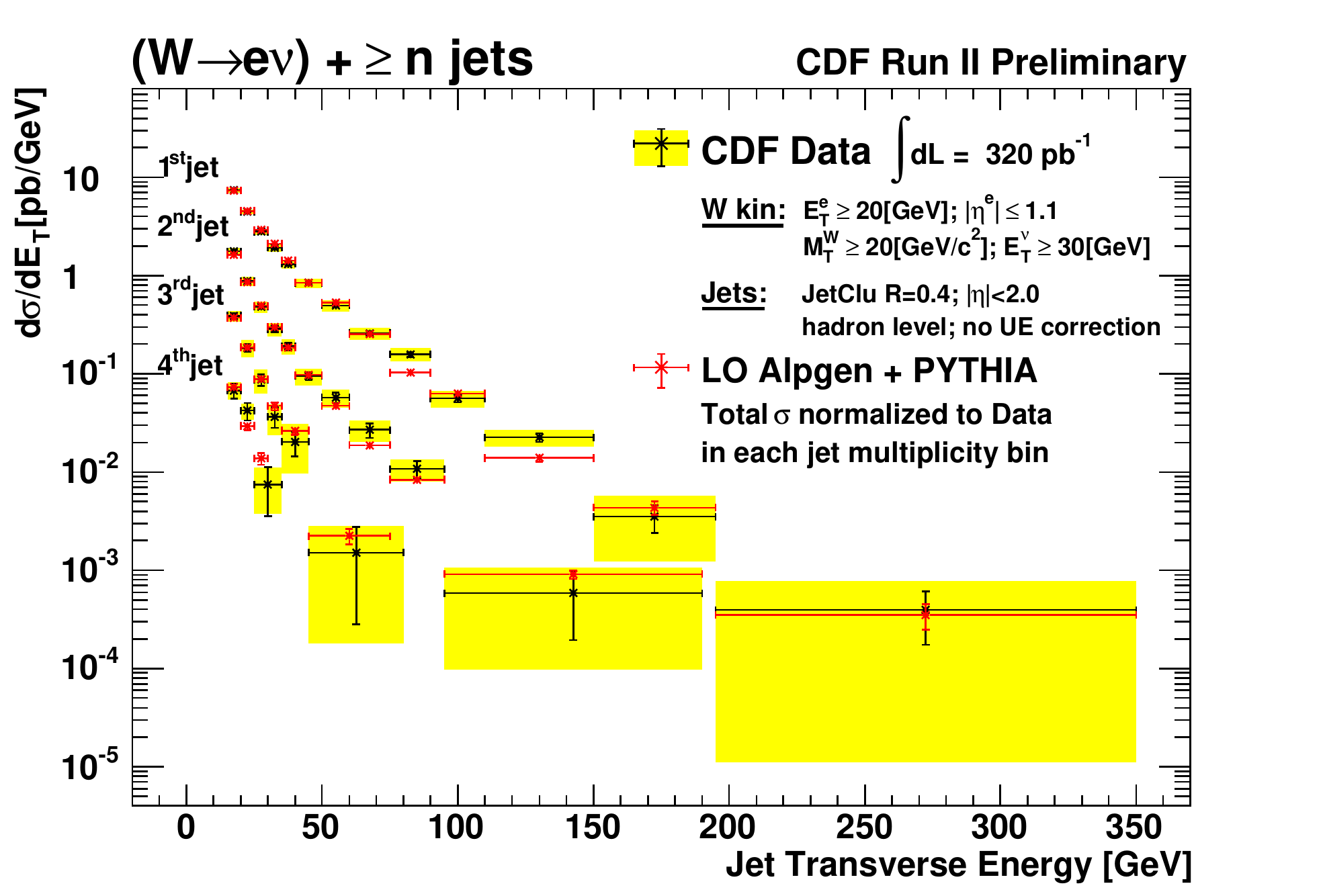,
height=7.cm,width=0.9\textwidth} 
\epsfig{file=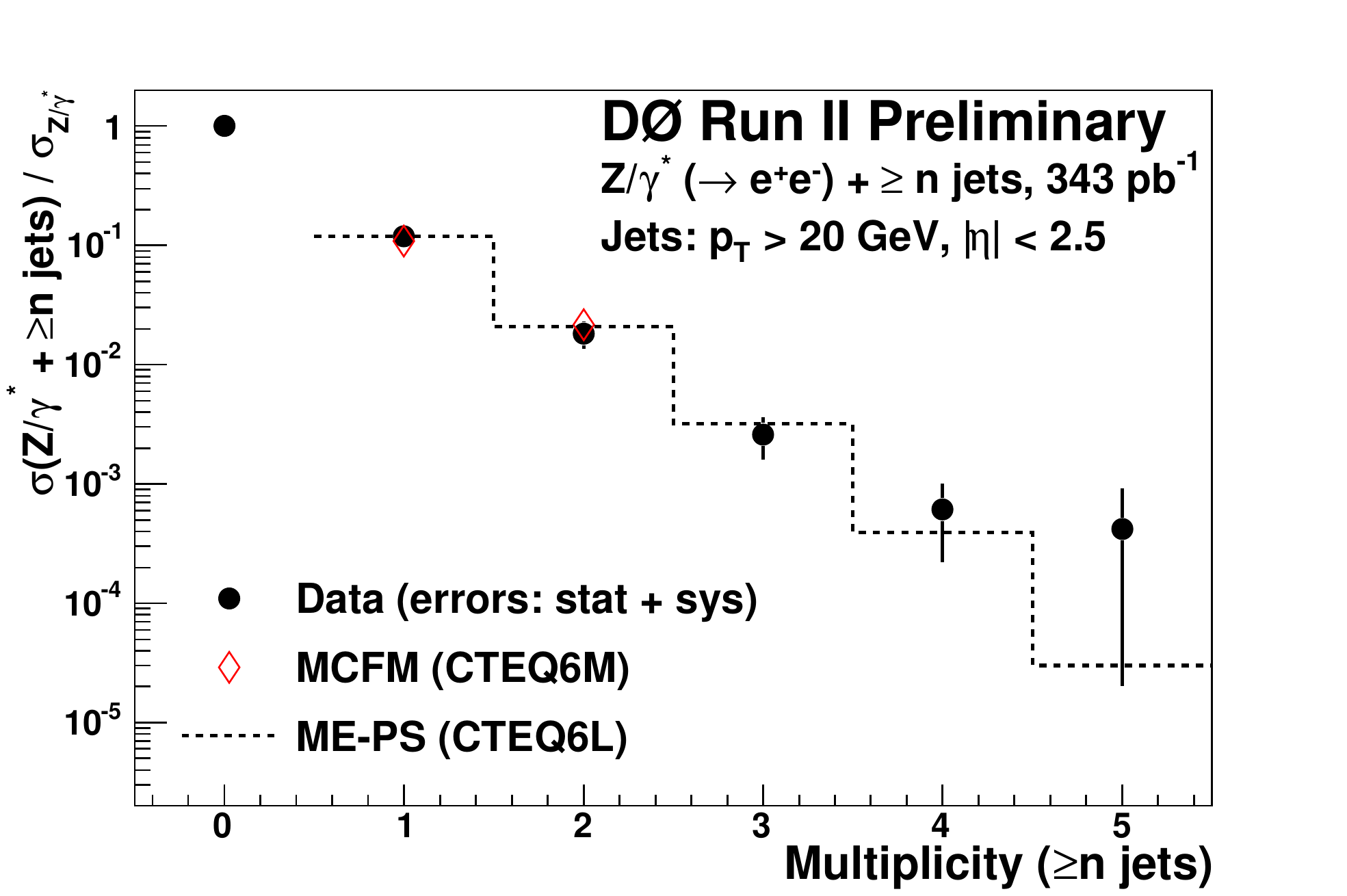,
height=7.cm,width=0.9\textwidth} 
\end{center}
\caption{
\label{wjcdf}
{\bf Left:} differential cross section 
$d\sigma(W\to e\nu+\geq n-{\rm jets})/d E_{T}^{jet}$ (Right) for the first, second, third and fourth inclusive 
jet sample.
Data are compared to 
{\tt Alpgen}+{\tt PYTHIA} predictions normalized to
the measured cross section in each jet multiplicity sample.
{\bf Right:} Measured cross section for $Z$+jets production
as a function of inclusive jet
  multiplicity compared to MADGRAPH + PITHYA. Absolute cross 
section normalized to data.
}
\end{figure}
\end{center}

CDF has looked at jets production \cite{messina}
in Drell-Yan processes
finding a satisfactory agreement between data and
 {\tt ALPGEN} +{\tt PITHYA} predictions, once MC predictions
are normalized to the data. Preliminary results are
shown in fig.~\ref{wjcdf} (left panel, from \cite{messina}).
Once the overall normalization is fitted to data
also jet multiplicities are well reproduced as shown
in fig.~\ref{wjcdf} (right panel from \cite{d0zjet}).

\begin{center}
\begin{figure}
\begin{center}
\epsfig{file=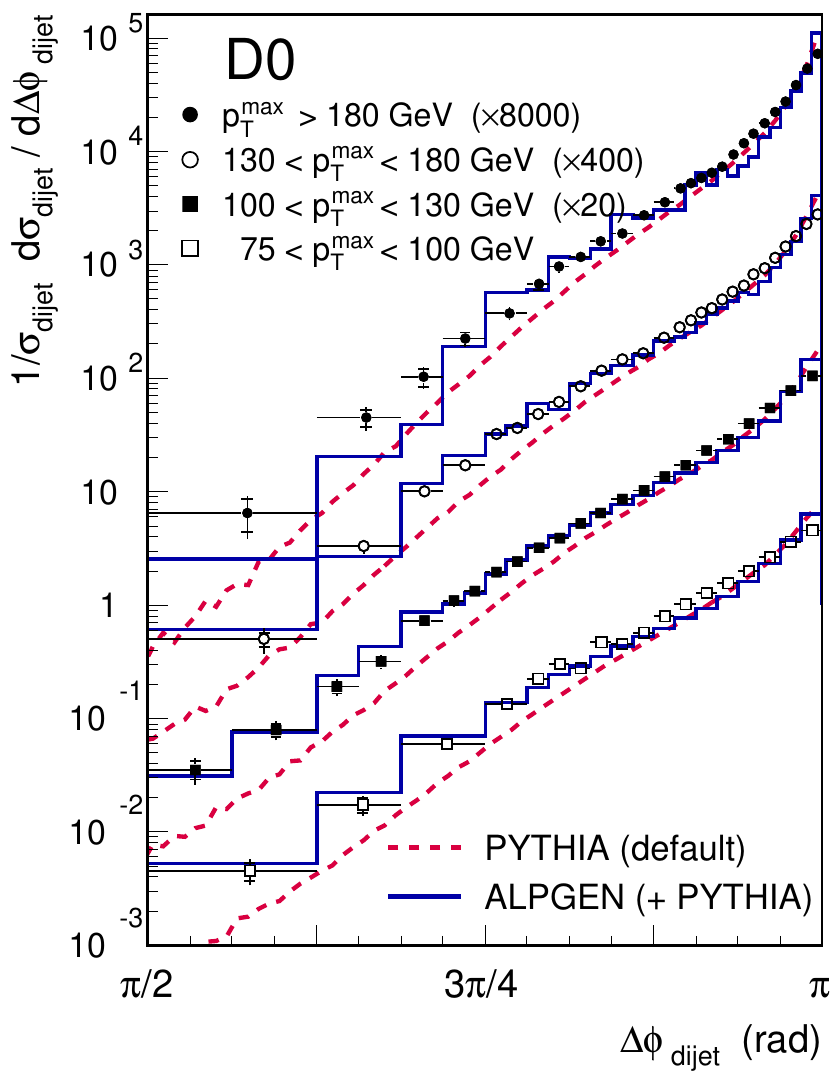,height=8.cm} 
\end{center}
\caption{
\label{alpgendijettev}
Angular separation $\Delta\Phi$ in the transverse plane
among the two leading jets (jets ordered in $p_T$) in inclusive
two jets sample. D0 data versus ALPGEN + PITHYA
 prediction are shown.
Figure from \cite {d0dijet}.
}
\end{figure}
\end{center}

D0 collaboration has studied \cite {d0dijet}
dijet azimutal correlation in pure
jet sample and compared DATA to ALPGEN+PITHYA predictions
 predictions, again finding
good agreement as shown in fig.~\ref{alpgendijettev}.

\section{Comparison among matching prescriptions}

We refer to \cite{Mrenna:2003if,compare} for a more
complete account of a detailed series of comparisons.

These comparisons have been performed for the Drell-Yan
process both at the Tevatron and at the LHC.

\begin{figure}
\begin{center}
\includegraphics[width=0.92\textwidth,clip]{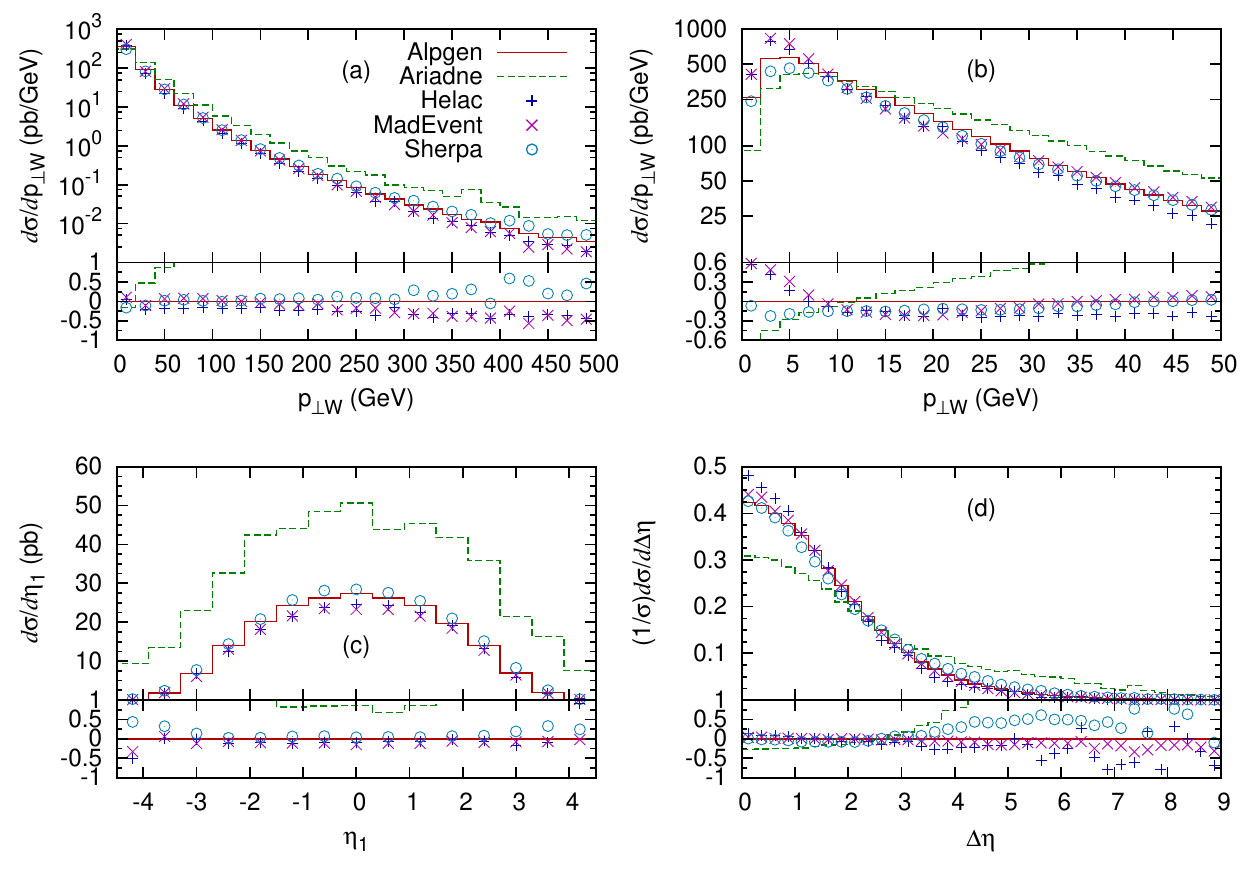}
\end{center}
\vskip -0.9cm \ccaption{}{\label{fig:ptw-lhc} (a) and (b) 
$p_\perp$
  spectrum of $W^+$ bosons at the LHC (pb/GeV).
 (c) $\eta$\/ spectrum
  of the leading jet, for $p_\perp^{\mathrm jet1}>100$~GeV;
 absolute normalization (pb).
  (d) Pseudo-rapidity separation between the $W^+$ and the leading
  jet, $\Delta\eta=|\eta_{W^+}-\eta_{{\rm jet}1}|$, for
  $p_\perp^{\mathrm jet1}>40$~GeV, normalized to unit area.
 In all cases the full
  line gives the ALPGEN results, the dashed line gives
 the
  ARIADNE results and the ``+'', ``x'' and ``o'' points give
  the HELAC, MADEVENT and SHERPA results respectively.
  In the lower frame relative deviation with respect to
ALPGEN predictions are shown.}
%
\vskip -0.3cm
\end{figure}

The overall agreement is relatively good as shown in
fig.~\ref{fig:ptw-lhc} (from \cite{compare})
and the differencies are compatible
with the effect of the factorization scale variation
for a LO calculation. ARIADNE exhibits larger
variation mostly due to the different approach to the shower
evolution.

\begin{figure}
\begin{center}
\includegraphics[width=0.92\textwidth,clip]{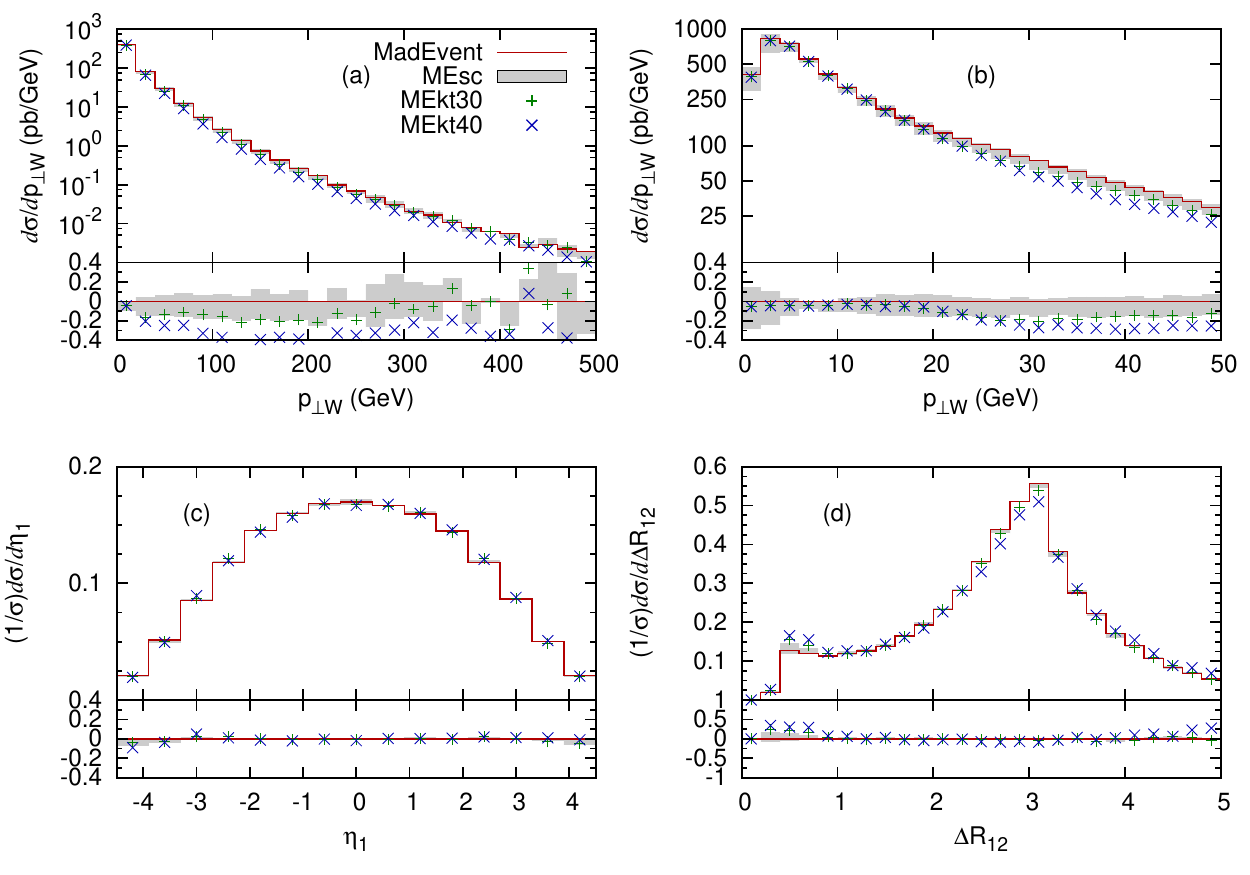}
\includegraphics[width=0.92\textwidth,clip]{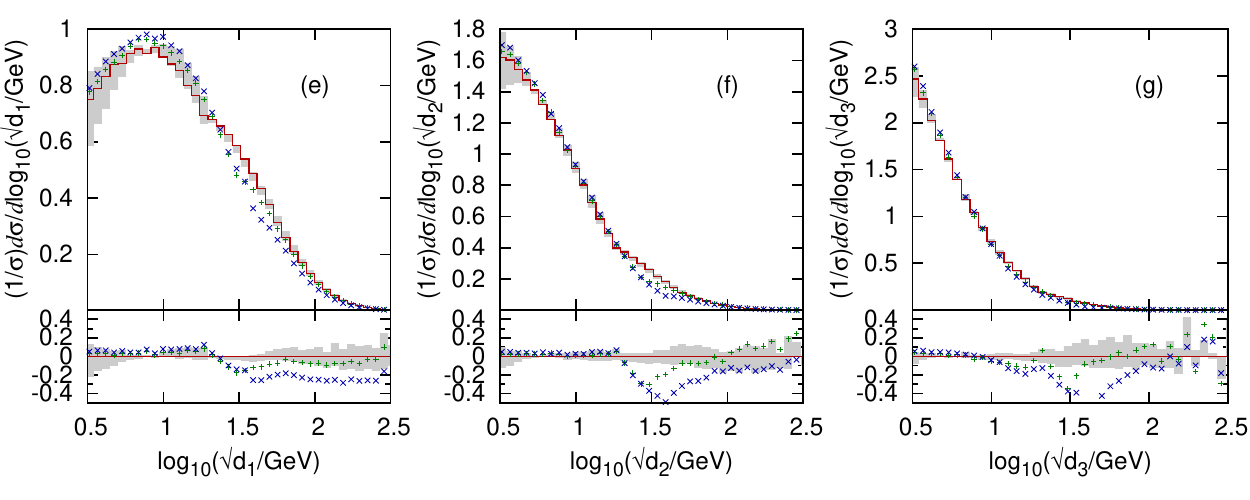}
\end{center}
\vskip -0.4cm
\ccaption{}{\label{fig:mad-ptw-lhc} MADEVENT
 systematics at the
  LHC. 
(a) and (b) show the $p_\perp$ spectrum of
  the $W$, (c) shows the pseudo-rapidity distribution of the leading jet,
  (d) shows the $\Delta R$\/ separation between the two leading jets,
  and (e)--(g) show the $d_i$ ($i=1,2,3$) spectra, where $d_i$ is the
  scale in a parton-level event where $i$\/ jets are clustered into $i-1$
  jets using the $k_\perp$-algorithm.
The full
  line is the default settings of MADEVENT, the shaded area
  is the range between MEscL and MEscH, while the points represent
  MEkt30 and MEkt40.
MEscL, MEscH, MEkt30 and MEkt40 are different
MADEVENT settings properly defined in \cite{compare}}
\end{figure}

In \cite{compare}
it is also provided an extimate of (at least some of)
the systematic uncertainties associated to the approach
varying $\alpha_S$ scale and, for SHERPA and ARIADNE,
also the PS scale. In fig.~\ref{fig:mad-ptw-lhc}, from
\cite{compare},
 we show
an example of this exploration for MADEVENT.

We address to \cite{Mrenna:2003if,compare} for a more
thorough discussion. Here we want just make a few remarks

\bit

\item As step zero, to gain confidence on an event sample,
one should first investigate the dependence on the 
resolution parameters looking at the impact of moving
away from the various codes default setting. We emphasize
once again that this is the {\em only way to extimate} this
 dependence since we lack an analitical extimate.

\item Some of the differencies among the various recipes
can be minimized adjusting the resolution parameters and/or
$\alpha_S$ scale. This doesn't make much sence in the 
absence of data. However once data are avaliable
all these parameters provide an handle to improve the
description of data.

\item Having performed step zero one should also move
to step one: investigating the impact of scale variation
on the prediction (expecially to assess the impact on
the {\em shapes} of the various observables).

\item As a final remark let's outline that if one is
interested in a fairly exclusive region of phase space
one should repeat the above steps for the region
of interest: {\em an overall stable and satisfactory picture for
$l^+ l^-$ production doesn't gurantee that the same
holds in the hard mass tail, say $m_{l^+ l^-} > 1$ TeV}.

\eit


\addtocounter{chapter}{1}

\mchapter{Jets at LHC}
{Authors:
Daniele Benedetti, 
Andrea Giammanco, 
Paolo Nason, 
Chiara Roda, 
Attilio Santocchia, 
Iacopo Vivarelli}\label{mcws:jets}
\vskip 1cm
This chapter deals with several aspects of jet physics at the LHC.
It is mostly based upon the study of ref.~\cite{leshouches2005}, and
thus many results that appear here are bound to become obsolete with time.
Nevertheless, we believe that this chapter condenses the main theoretical
and experimental problems that one encounters when studying jets
at hadron colliders.

In section \ref{sec:jetgeneral}, we formulate
the basic concepts of jets, as the manifestation of energetic coloured
particles in high energy reaction. The concept of infrared-safe jet
observables is discussed there.
In sec.~\ref{sec:jfa} the most popular jet finding algorithms
are introduced.

In section \ref{sec:optim} the study of \cite{leshouches2005}
on the optimization of the jet finding algorithm is reported.
Different algorithms are compared according to their ability to relate
jets to primary partons in the hard interaction. No detector effects
are considered in this section. Jets are reconstructed from the output
of a Shower Monte Carlo program. The goal of the optimization is to
find the optimal jet parameters (like, for example, the jet cone radius)
to be used. The quality criteria to use for the optimization are
defined as the goodness of the matching between jets and hard partons
emerging from the primary interaction,
as can be inferred from the Monte Carlo program. Although
this connection is only approximate, and, to some extent, Monte Carlo
dependent, it is certainly adequate to perform this task.

In section \ref{sec:calib} we discuss the problem of jet calibration.
The methods adopted for the definition of calorimeter jets are briefly
outlined, and the results of the calibration studies
of ref.~\cite{leshouches2005} are reported. The concepts of calibration
to the particle jet, and calibration to the parton level are illustrated
and discussed.

In sec.~\ref{sec:eflow}, the particle flow method for the reconstruction
of jets is discussed. The term particle flow (or energy flow) refers
to the use of other relevant information for jet reconstruction, other
than calorimetry, i.e. tracker and partiocle identification information.
These information can considerably improve the energy resolution,
in view of the fact that a large fraction of the energy of the jet is
carried by charged particles.

\section{Introduction}\label{sec:jetgeneral}
In high energy reactions,
quarks and gluons manifest themselves as jets of particles.
This fact has been discussed in many
places in these proceedings, and is due to the fact that collinear
and soft QCD radiation is a dominant process at high energy.
A quark or a gluon produced at a primary interaction will very often
radiate soft and collinear partons. Furthermore, only color neutral
hadrons can appear in the final state: quarks and gluons must undergo
strong non-perturbative interactions that lead to the formation of hadrons.
Thus, the concept of jet must be carefully defined in
order to simplify the interpretation of high energy events. It should
represent the footprint of a hard coloured parton. Ideally, a jet
should be in a one-to-one correspondence with a coloured
parton. In practice, this is possible only in an approximate sense.
A minimal requirement that we should make on the jet concept is
that it should at least be possible to use it to define and
compute cross sections.
\subsection{Infrared safe jet definitions}
Theoretical physicists have always advocated the use of jet
definitions that are calculable in perturbative QCD as a power
expansion in the strong coupling constant, with an accuracy that is
ultimatly limited by power suppressed corrections (i.e. by terms of
the order of a power of $\Lambda/Q$, where $\Lambda$ is a typical
hadronic scale and $Q$ is the scale involved in the jet
definition).
This requirement is met by jet definitions that allow for
the cancellation of infrared divergences in the cross section, the so
called IR-safe (for Infrared-Safe) jet definitions.
It turns out that, in order for the cancellation of infrared
divergences to take place, a QCD observable must have the
following properties:
\begin{itemize}
\item It should be collinear safe: this means that if the momenta
$p_1$, $p_2$ of two light final state particles form a small angle,
and we substitute the two final state particle with a pseudoparticle
with momentum $p_1+p_2$, the change in the observable becomes
tiny as the angle goes to zero.
\item It should be soft-safe: if the momentum of a light particle
becomes small, if we remove
that particle from the final state
the change in the observable should become tiny as the small
momentum goes to zero.
\end{itemize}
In the above definitions, the terms ``light'' refer to particles
with masses of the order of typical hadron masses. When we say
that the change in the observable should be ``tiny'',
we mean that it should be suppressed by a power of the mass of the particle
divided by the hard scale of the process\footnote{When heavy quarks like
charm and bottom are involved, depending upon the value of the
hard scale, they may or may not be considered light}. The corresponding
QCD calculation of the given observable is performed in terms of quarks
and gluons, rather than hadrons, where gluons and light quarks
are taken to be strictly massless (i.e. the light quark masses are neglected),
so that the terms ``light'' and ``tiny''  in the above definitions should be
replaced by ``massless'' and ``zero''.

Notice that if an observable is IR-safe,
it should not make much difference whether we define it in terms
of particles energies and directions, or in terms of 
energy deposition in calorimeter cells and the associated direction,
at least if we assume that we have ideal calorimetric energy measurement.
In fact, the particles entering a calorimetric cell are at relative
small angle, and so, if we merge them into a pseudoparticle, with
energy equal to the total energy deposited in the calorimeter,
the observable should not change much. Furthermore, particles with
very small energy, if removed, cause only a small change in the
energy deposited in the calorimeter cells.

In practice, an infrared safe definition of jets
yields results that are less affected by QCD effects,
the conditions listed above precisely requiring small sensitivity
to dominant QCD effects. In order to be able to compare a measured
cross section with a QCD calculation, infrared-safeness is a mandatory
requirement. We should stress, however, that there are measurements
where extracting a cross section is not so important, like, for example,
in the reconstruction of a mass peak or shoulder. One may
argue that in these cases, the sharpness of the peak should be pursued,
even at the price of giving up IR-safety.

\section{Jet finding algorithms}\label{sec:jfa}

The iterative cone algorithm had its origin in ref.~\cite{Huth:1990mi}, where an accord\footnote{The so called Snowmass accord
on jet definitions.} was reached for a jet algorithm that was satisfactory to both experimentalists
and theoreticians. A cone algorithm is characterized by a cone radius $R$ in the $\eta,\phi$ plane.
A stable cone is such that
\begin{equation}
\sum_{i\in {\rm cone}} E_T^{(i)}\Delta\eta^{(i)} = 0\,,\quad\quad
\sum_{i\in {\rm cone}} E_T^{(i)}\Delta\phi^{(i)} = 0\,,
\end{equation}
where $E_T^{(i)}$ is the transverse energy of the $i^{\rm th}$
particle or calorimetric tower, and
$\Delta\eta^{(i)},\Delta\phi^{(i)}$ are its distances in $\eta$ and
$\phi$ from the cone center.  Stable cones can be found by starting
with any cone, compute the ``center-of-weight'' of its transverse
energy distribution, and then iterating the procedure with a new cone
around its center of weight, untill the procedure stabilizes. The set
of all stable cones is obviously an infrared safe concept. However, it
would seem that, in order to find all stable cones, one should start
the stabilization procedure with cones centered in all possible
$\eta,\phi$ points, which seemed unfeasible at that time.
In the Snowmass accord, a compromise procedure is
adopted, where one takes all particles or towers with energy above a
certain threshold (i.e. seeds) as cone center from where one starts
the iteration procedure. Unfortunately, in this way IR-safety is
lost. Various attempts were maid in order to restore IR-safety, but
apparently, as long as we use seeds, all fixes are bound to fail at
some level, thus leading to an increasing complexity in the jet
definition. Very recently, a fast algorithm for the computation of
stable cones in a seedless approach has been found~\cite{Salam:2007xv},
the so called SISCone algorithm. It is concievable that LHC experiments
will move soon to this approach.

In the {\bf iterative cone algorithm} ({\bf ICA} from now on),
an $E_T$-ordered list of input
objects (particles or calorimeter towers) is created.  A cone of size
$R$ in $\eta,\phi$ space is cast around the input object having the
largest transverse energy above a specified seed threshold. The
objects inside the cone are used to calculate a proto-jet direction
and energy. The computed direction is used to seed a new
proto-jet. The procedure is repeated until stability is reached
(i.e. the energy of the proto-jet changes by less than 1\% between two
consecutive iterations and the direction of the proto-jet changes by
$\Delta R < 0.01$). When a stable proto-jet is found, all objects in
the proto-jet are removed from the list of input objects and the
stable proto-jet is added to the list of jets. The whole procedure is
repeated until the list contains no more objects with an $E_T$ above
the seed threshold. The cone size and the seed threshold are tunable
parameters of the algorithm.

An improvement over the ICA was introduced in CDF, in order not to
privilege too much the hardest seeds in the construction of the
jet. With this procedure, no particles were removed from the list. So,
at the end of the procedure there are overlapping jets. The following
merging-splitting procedure was adopted: if two jets share more than a
given fraction of the energy, they are merged into a single
jet. Otherwise, the energy is assigned to the closest (in $\eta,\phi$)
jet.

The {\bf midpoint-cone algorithm} ({\bf MCA} from now on)
 was designed to improve over the
iterative cone algorithm, by increasing the number of cone directions
from where stable coned are searched, thus moving closer to a seedless
approach.  It also uses an iterative procedure to find stable cones
(proto-jets) starting from the cones around objects with an $E_T$
above a seed threshold. No object is removed from the input list.
Then, a second iteration of the list of stable
jets is done. For every pair of proto-jets with distance less than the
cone diameter, a midpoint is calculated as the direction of the
combined momentum. All these midpoints are then used as additional
seeds to find more proto-jets. When all proto-jets are found, a
splitting and merging procedure is applied, starting with the highest
$E_T$ proto-jet. If a proto-jet does not share objects with other
proto-jets, it is defined as a jet and removed from the proto-jet
list. Otherwise, the transverse energy shared with the highest $E_T$
neighbouring proto-jet is compared to the total transverse energy of
this neighbour proto-jet. If the fraction is greater than a given
threshold $f$ (typically 50\%) the proto-jets are merged, otherwise
the shared objects are individually assigned to the closest
proto-jet. The procedure is iterated, always starting from the highest
$E_T$ proto-jet, until no proto-jets are left.  The parameters of the
algorithm include a seed threshold, a cone radius, the threshold $f$
mentioned above, and also a maximum number of proto-jets that are used
to calculate midpoints.

The {\bf inclusive $k_T$ jet algorithm} is a cluster-based jet
algorithm. The cluster procedure starts with a list of input objects,
stable particles or calorimeter cells. For each object $i$ and each
pair $(i, j)$ the following distances are calculated:
\begin{eqnarray}
d_i &=& (E_{T,i})^2 R^2 \nonumber \\
d_{i,j} &=& min(E_{T,i}^2,E_{T,j}^2) \Delta R_{i.j}^2 \ \ \ \mbox{with} \ \ \  \Delta R_{i.j}^2 = (\eta_i - \eta_j)^2 + (\phi_i - \phi_j)^2  \nonumber
\end{eqnarray}

where $R^2$ is a dimensionless arbitrary parameter.\footnote{
Sometimes, instead of transverse energy and pseudorapidity,
transverse momentum and rapidity are used. This makes a small
difference at the first iteration step, but can make a substantial
difference after a few steps, when the original input objetcts have
been replaced by massive clusters. In particular, the use of rapidity
makes the algorithm invariant under longitudinal
boosts.}
The algorithm
searches for the smallest $d_i$ or $d_{ij}$.  If a value of type
$d_{ij}$ is the smallest, the corresponding objects $i$ and $j$ are
removed from the list of input objects.  They are merged using one of
the recombination schemes and filled as one new object into the list
of input objects. If a distance of type $d_i$ is the smallest, then
the corresponding object $i$ is removed from the list of input objects
and filled into the list of final jets. The procedure is repeated
until all objects are included in jets.  The algorithm successively
merges objects which have a distance $R_{ij} < R$. It follows that
$R_{ij} > R$ for all final jets $i$ and $j$.

The cluster jet definition is IR-safe, and does not suffer from the
jet overlapping problem typical of the cone algorithms.

\begin{figure}[htb]
\begin{center}
\includegraphics[width=0.8\textwidth]{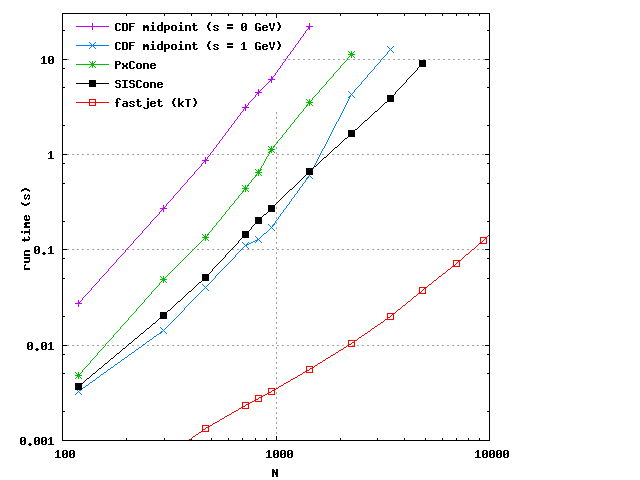}
\end{center}
\caption{ \label{fig:timings} Performance comparison for various jet algorithms.}
\end{figure}

The $k_T$~algorithm has found limited applications in hadron collider
physics, mostly due to algorithmic speed limitations, and partly
due to the fact that (unlike to cone algorithm) it is harder to
define a jet area, in order to subtract the effects of the underlying
event. This situation has recently changed. In
ref.~\cite{Cacciari:2005hq} a fast algorithm has been constructed.
A viable method for the subtraction of the underlying event has also
been suggested in \cite{Cacciari:2007fd}.
Thus, today it become feasible to use fully infrared safe
algorithms, which is in fact the current tendency. In
fig.~\ref{fig:timings} a comparison of the performance of different
algorithms is displayed.

In ref.~\cite{Salam:2007xv}, a more thorough discussion of how the ICA
and midpoint algorithms fail the IR-safety criteria is also given.

The code for SISCone and FastJet can be found in\newline
 {\tt
  http://projects.hepforge.org/siscone/} \newline
{\tt http://www.lpthe.jussieu.fr/\~{}salam/fastjet/}

\section{Optimization of the jet finding algorithms}\label{sec:optim}
This section summarizes the studies of Ref.~\cite{leshouches2005} on
the optimization of the jet finding algorithm.  This optimization is
defined in terms of quality criteria or quality markers, related to
the reconstruction efficiency of the complete kinematics of the
primary quark event topology. Physics effects like QCD
radiation, underlying event and pile up enlarge the error of the
reconstruction procedure. This study has been performed with simulated
particle information as input to the jet finding algorithms, and deals
with algorithmic and physics effects, independently of detector
specificities.

The scope of this study is to find the most efficient jet finding
setup in the presence of these effects, in order to maximise the
fraction of events for which all quarks are matched to reconstructed
jets, according to some predefined criteria. Hence, events suffering
from a large amount of hard gluon radiation will be rejected.



It has to be kept in mind that instrumental effects can, in principle,
alter significantly the conclusions of this study.  Work is currently
in progress in CMS for an analogous study with the full detector
simulation and reconstruction chain.

In the studies performed in the present work, only
the following jet reconstruction algorithms have been considered: the {\it iterative cone} algorithm (IC),
the inclusive $k_T$ algorithm ($k_T$) and the {\it MidPoint Cone} algorithm (MC)~\cite{Blazey:2000qt}.
For all jet finding algorithms, generated and stable final state
particles are used as input.
\subsubsection{Particle Jets}
We call ``particle jets'' those that can be reconstructed from
particles if one had a perfect detector (i.e. if one new the momentum
of all final state particles).  In simulated data, they are obtained
by applying the jet clustering algorithms to all stable particles
(charged and neutral) as obtained at the generator level after the
hadronization step, without considering any of the detector effects
(like calorimeter resolution or the sweeping from the magnetic
field\footnote{The minimum transverse momenta required to reach the
  calorimeter inner surface is about $350$~MeV for the ATLAS system
  and about $700$~MeV for the CMS system.}).  A particle jet includes
all particles. Thus, in simulated data, any particle emerging from the
hard scattering process or from the underlying event should be
included.  Some authors exclude the neutrinos from the list of input
particles, since they cannot give a signal in the detector, not even
in principle.
 In the present study, muons and neutrinos are excluded, and the
effects of the magnetic field are not taken into account.  All
particles are assumed to emerge from the primary vertex.

\subsection{The parton-jet connection}
Within the Shower Monte Carlo model of hard collisions, one has
access to the kinematics of partons
arising in the hard process, before the shower takes place.
One can therefore study the connection of the jet kinematics
to the parton kinematics, setup a method
to reconstruct the parton kinematics given the jet kinematics,
and associate an error with this procedure.

It is important to stress
that the parton-jet connection is not simply rooted in the physics of
hard processes. It may very well depend upon the
particular Shower Monte Carlo one is using. This is even more
apparent if one notices that in the dipole showering schemes
(like in ARIADNE, or in PYTHIA~6.4),
radiated partons arise from dipoles, i.e. from pairs of partons,
rather than from a single one. Furthermore, even in the framework of
traditional single-parton showers, the momentum reshuffling stage in the
shower (see chapter~\ref{ch:shower}) differs in different
implementation. This yields an explicitly different kinematic relation
between the four-momentum of a shower and the four-momentum
of the initial parton.

However, since the most important QCD processes are small angle or
soft emissions, at least as a first approximation the parton-jet
connection is universal.  Thus, parton-jet matching can be used to
device simple quality criteria to compare different jet finding
algorithms.

\subsection{Event generation}

For this study,
processes with two, four, six and eight primary quarks in the final state (dileptonic and single-leptonic top decays 
in $t \bar{t}$ events, single-leptonic and fully hadronic top decays in $t \bar{t} H$) have been considered. 

Proton collisions at 14 TeV have been generated at a luminosity of
2 $\times$ 10\textrm{$^{33}$} cm\textrm{$^{-2}$}s\textrm{$^{-1}$}. 
The $t\bar{t}$ events were generated using PYTHIA version 6.2~\cite{PYTHIA} and the
$t\bar{t}H$ events were generated with compHEP version 41.10~\cite{Pukhov:1999gg}, interfaced to
PYTHIA version 6.215 for showering and hadronization.
For the leptonic decays, only electrons and muons
are considered.

\subsection{Event selection and jet-quark matching}
\label{sec:EventSelection}
A realistic event selection (inspired by $t \bar{t}$ and $t \bar{t} H$
analyses) is applied. The reconstructed jets are required to have a
transverse energy larger than 20~GeV, and to be within the tracker
acceptance required for a proper $b$-tagging performance (in modern experiment 
the tracker generally reach $|\eta| \sim 2.5 $). Isolated signal leptons from the W-decay
are removed from the jet finding input. Only if the number of jets
passing these criteria is larger than or equal to the number of
primary partons the event is considered for the analysis.

An iterative procedure is used to match the reconstructed jets to the
generated quarks based on the $\Delta R$ distance in the
($\eta$,$\phi$) plane.  For each possible jet-quark couple the $\Delta
R$-value is calculated, and the smallest value is considered as a
correct jet-quark matching and is removed from the list for the next
iteration. When more jets have a minimal $\Delta R$-value with the
same quark, the couple with the lowest $\Delta R$-value is taken. This
procedure is iterated until all jets have their respective quark
match.

\subsection{Description of the quality markers }
\label{QMs}
In order to obtain an efficient reconstruction of the kinematics of
the primary partons, the selected jets should match both in energy and
direction the primary partons. Variables called quality markers are
defined to quantify the goodness of the event reconstruction from that
perspective. Although physics effects of pile-up, gluon radiation and
underlying event will degrade the overall event reconstruction
efficiency, it has to be reminded that in principle they can affect
differently the considered jet definitions.

\subsubsection{Event selection efficiency ``$\epsilon_{s}$''}
\label{subsec:seleff}

This efficiency is defined as the fraction of events that pass the
event selection, i.e. the events with a number of jets with $E_T > 20$
GeV and $|\eta| < 2.5$, greater or equal to the number of partons.
When the selection is applied on quark level (i.e. before the
shower), the efficiency is equal to 80\% for the two quarks final
state, 62\% for the four quarks final state, 61\% for the six quarks
final state and 52\% for the eight quarks final state.

\subsubsection{Angular distance between jet and parton ``Frac $\alpha_{jp}^{max}$''}
\label{subsec:alpha}
\begin{figure}[!ht] 
  \centering
\begin{tabular}{ll}
  \includegraphics[width=0.48\textwidth]{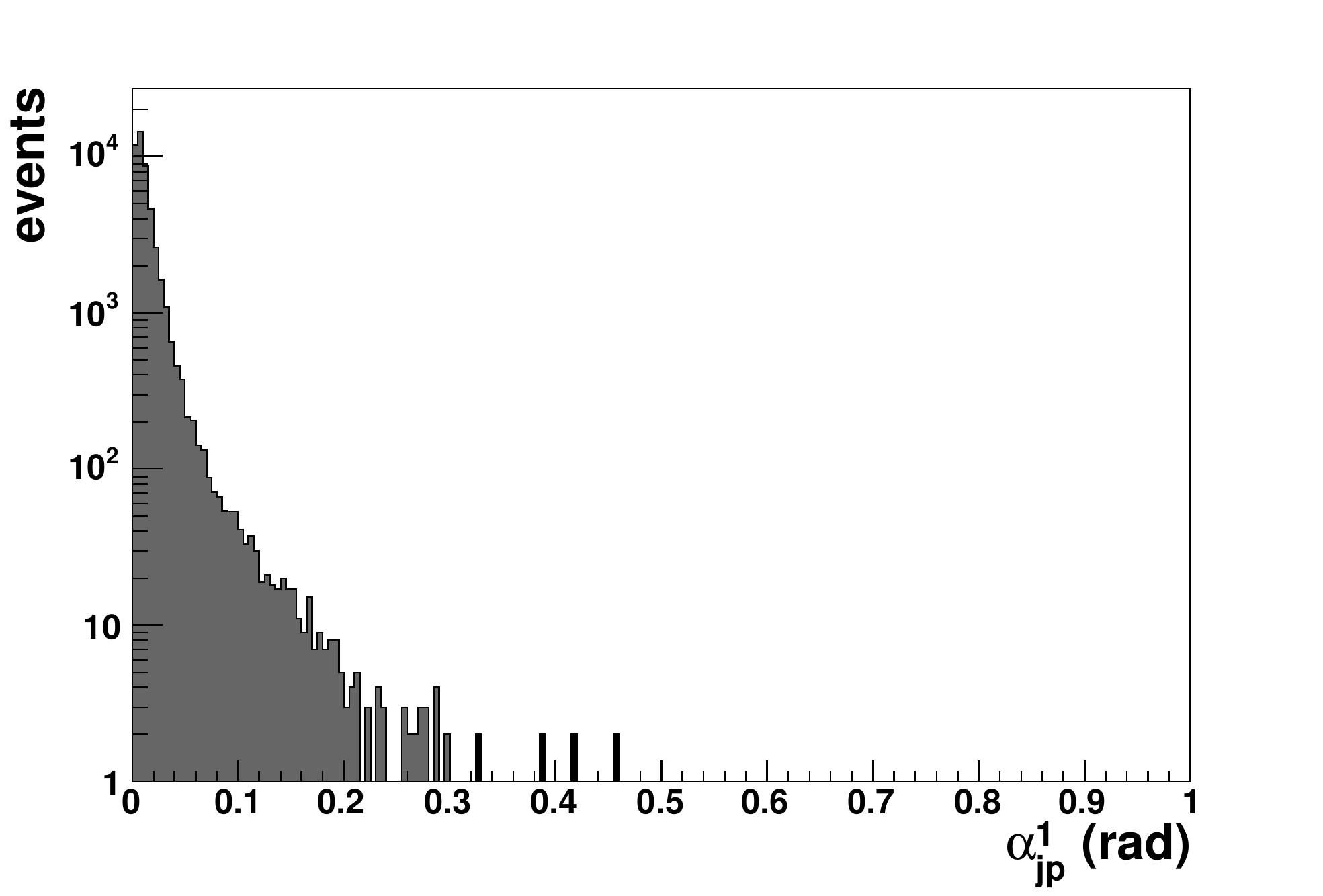} &
  \includegraphics[width=0.48\textwidth]{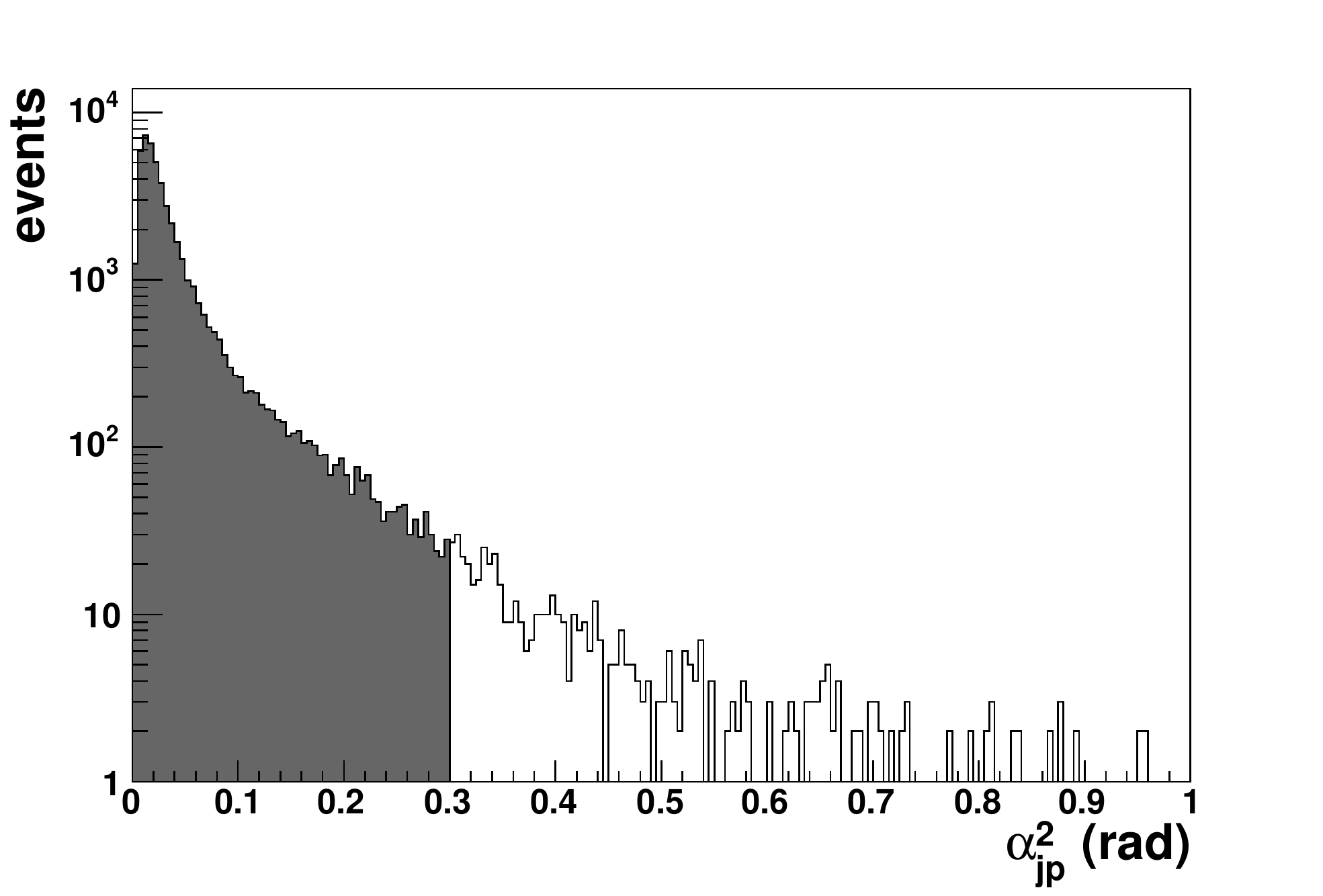} \\
  \includegraphics[width=0.48\textwidth]{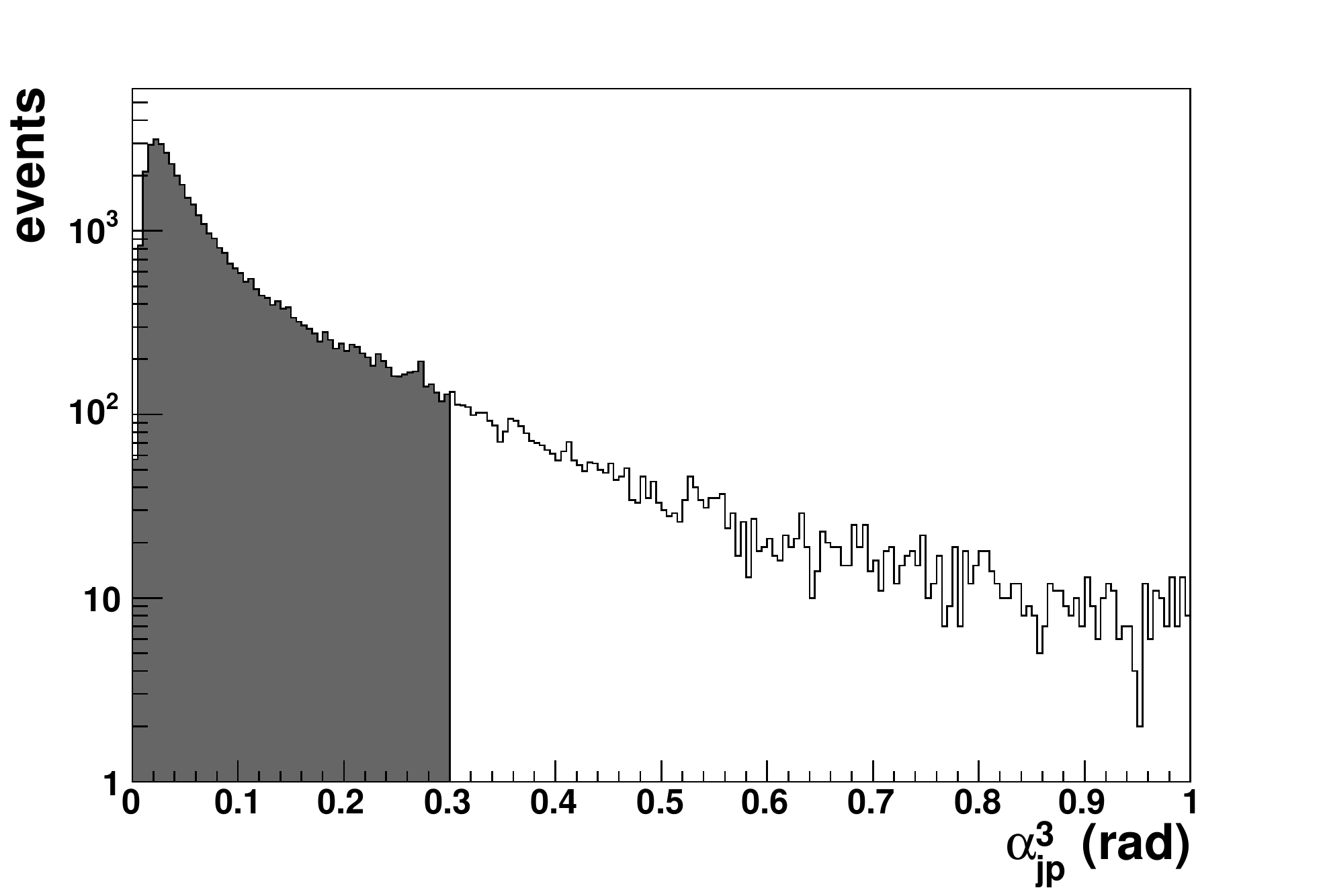} &
  \includegraphics[width=0.48\textwidth]{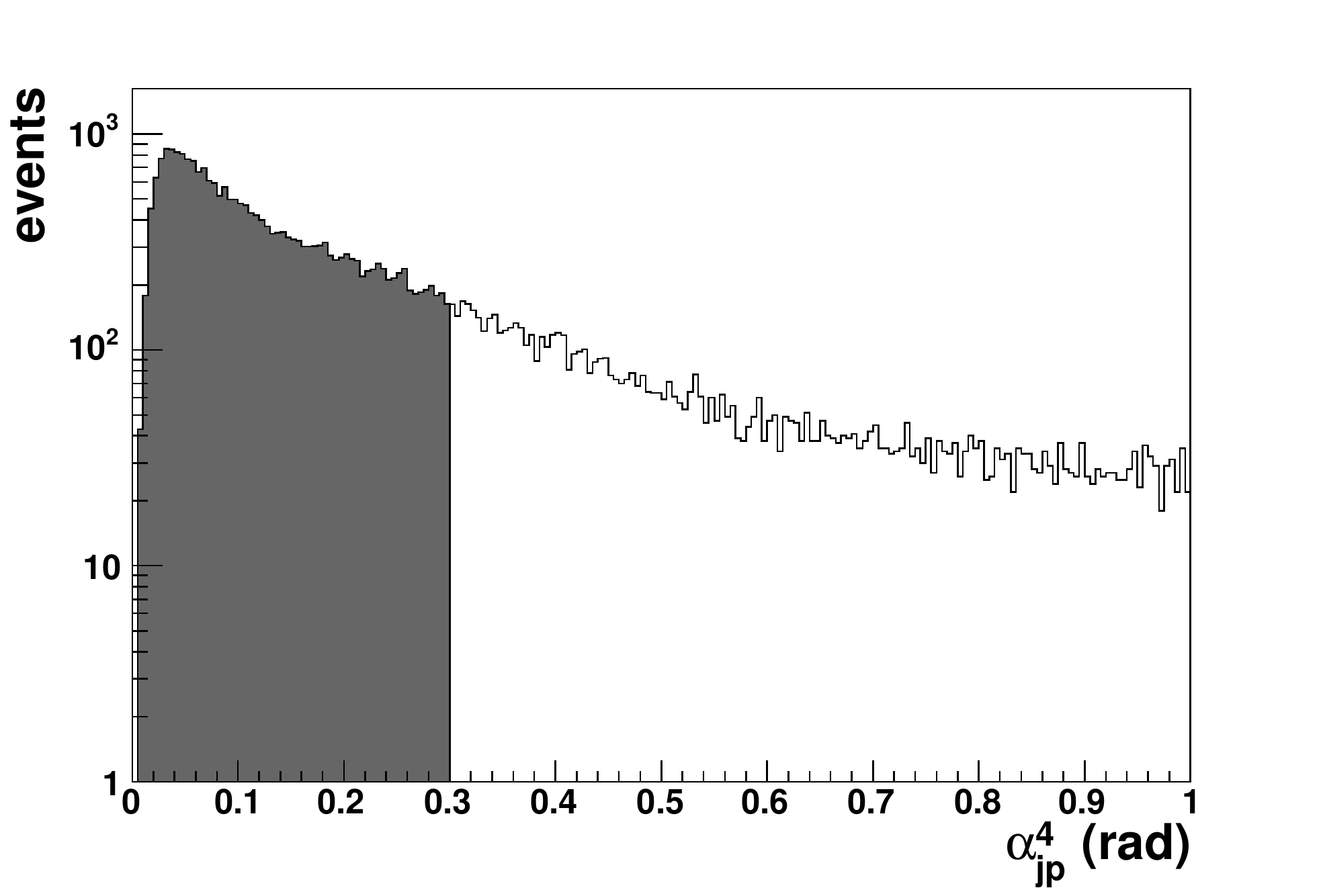} 
\end{tabular}
  \caption{Distributions of $\alpha_{jp}^{i}$  in increasing order
    for the IC algorithm with a cone radius of 0.4 in the case
    of a final state with four quarks. The 0.3 rad cut as
          discussed in the text is indicated.} 
  \label{maxalpha}
\end{figure}

A jet is considered to be well reconstructed, if the $\Delta R$ distance
between its direction and its best matched quark direction,
$\alpha_{jp}$, is sufficiently small. For each event, this results in a list of increasing $\alpha_{jp}^{i}$-values, $\{\alpha_{jp}^{1},...,\alpha_{jp}^{n}=\alpha_{jp}^{max}\}$, where $n$ is the amount of primary quarks in the considered event topology.
 Hence, $\alpha_{jp}^{max}$ is defined as the maximum $\alpha_{jp}^{i}$-value
of all $i$ jet-quark pairs in the event. The $\alpha_{jp}^{i}$ distributions for a four quarks final state are shown in Fig.~\ref{maxalpha}.
The last of these plots represents the  $\alpha_{jp}^{max}$
variable. 
To quantify the angular reconstruction performance of a particular jet definition, a quality marker is defined as the
fraction of events with a $\alpha_{jp}^{max}$ value lower than 0.3 and denoted as ``Frac~$\alpha_{jp}^{max}$''.

\subsubsection{Energy difference ``Frac $\beta_{jp}^{max}$''}
\label{efrac}
\begin{figure}[!h] 
  \centering
  \begin{tabular}{ll}
    \includegraphics[width=0.48\textwidth]{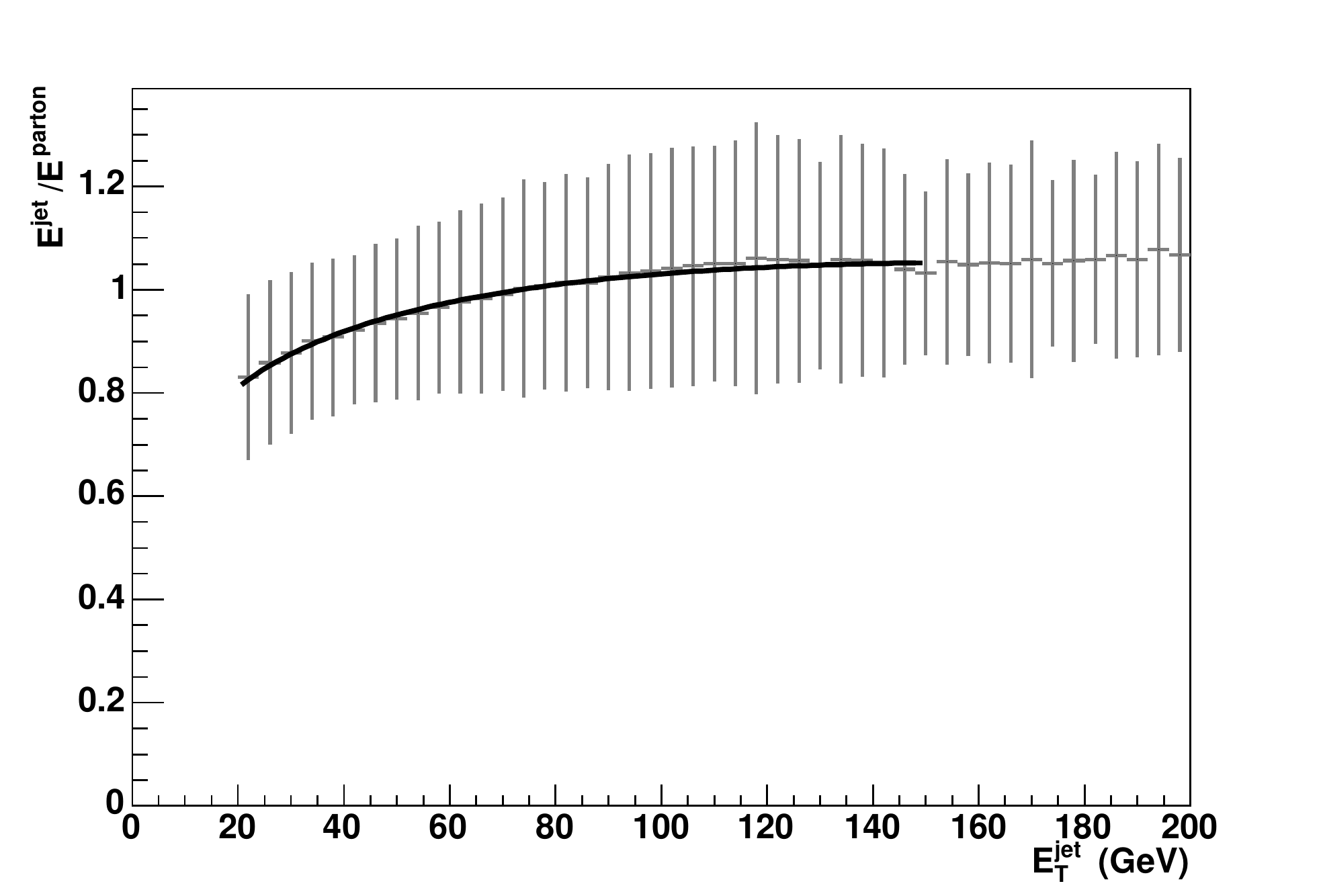} &
    \includegraphics[width=0.48\textwidth]{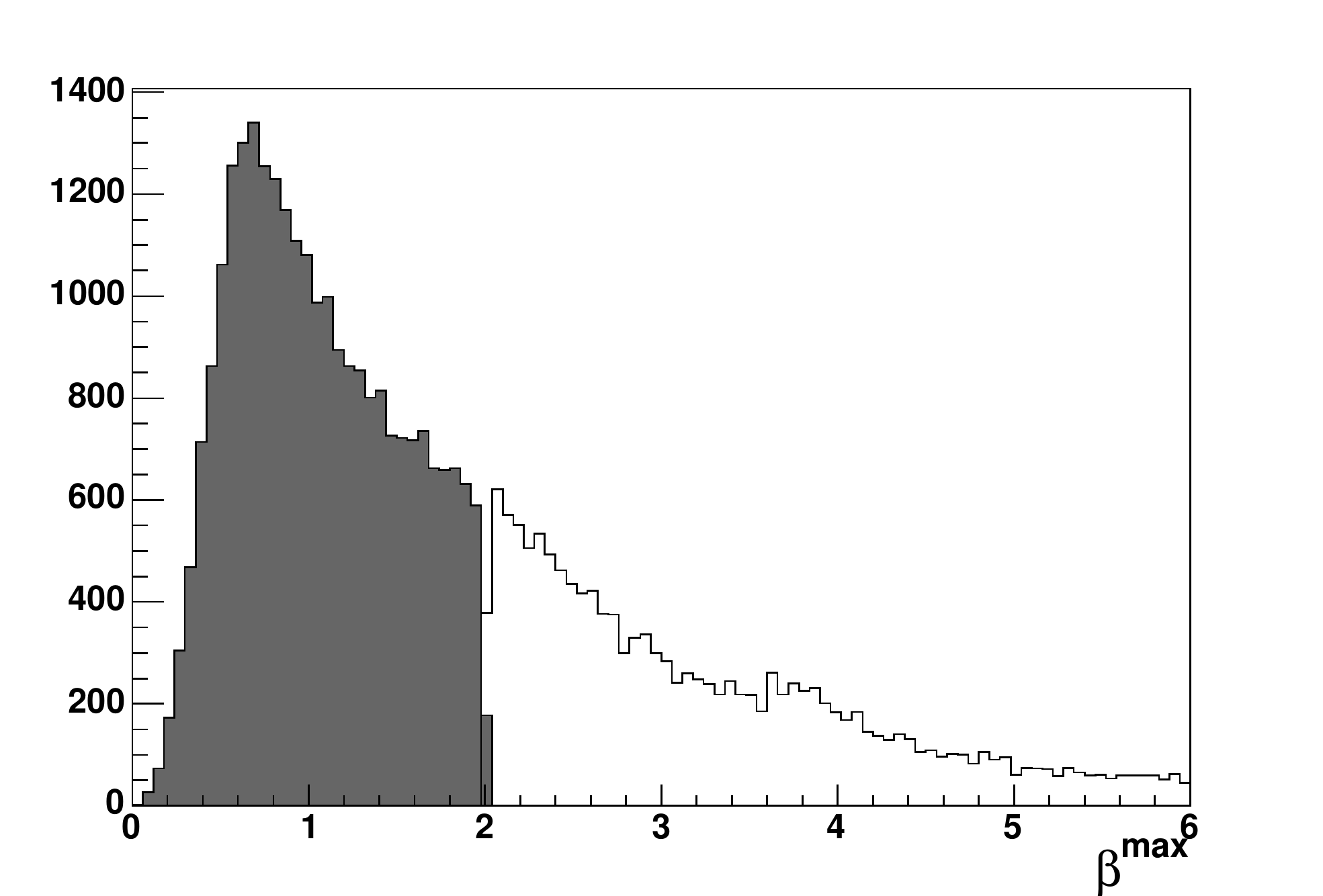} 
  \end{tabular}
  \caption{Left: example of a $\frac{E^{jet}}{E^{parton}}$
    vs. $E_{T}^{jet}$ curve for the IC algorithm with a cone
    radius of 0.4, applied on a final state with four primary quarks. The vertical bars illustrate the resolution. Right: distribution of $\beta_{jp}^{max}$  for the IC algorithm with a cone radius of 0.4, applied on a final
          state with four primary quarks.}
  \label{calibration}
\end{figure}

The reconstructed energy of the primary parton  is usually biased
(i.e. the reconstructed energy of the parton does not equal in the
average the energy of the jet) and
has a broad resolution.
Figure~\ref{calibration} (left) shows the average
fraction of the quark energy that is reconstructed for a specific  
algorithm as a function of the reconstructed transverse 
jet energy. Such a calibration curve can be interpreted as an
estimator for the expected reconstructed energy
\footnote{For this plot only well matched ($\alpha_{jp}<$0.3),  
non-overlapping jets were taken into account. For the iterative cone algorithm, a jet is considered
to be non-overlapping, if its $\Delta R$ distance to any other jet is
larger than twice the value of the cone radius parameter of the algorithm}
It is the aim of jet calibration studies to determine the average corrections
to be applied on the reconstructed jet energies. The remaining important component is the energy resolution:
after the reconstructed parton energy has been corrected for the bias, its difference from the jet
energy, in units of standard deviation, characterizes the quality of the reconstruction procedure
fot the given event.

The $\beta_{jp}^{i}$ values are defined for each primary quark $i$ as the distance from
the expected energy fraction (deduced from the fitted function in Fig.~\ref{calibration} left) in units of
standard deviations. For each selected event, the primary quark with the highest
$\beta_{jp}^{i}$ value, called $\beta_{jp}^{max}$ is considered to be
the one with the worst reconstruction performance from the energy point of view. An example for the $\beta_{jp}^{max}$ 
distribution is shown in Fig.~\ref{calibration} (on the right).  An energy related quality marker is defined as the
fraction of events with a $\beta_{jp}^{max}$ lower than 2 standard deviations, and denoted as ``Frac $\beta_{jp}^{max}$''.

\subsubsection{Combined variable ``Frac($\alpha_{jp}^{max}$+$\beta_{jp}^{max}$)''} 
\label{subsec:combined}
\begin{figure}[!ht] 
\begin{tabular}{ll}
    \includegraphics[width=0.48\textwidth]{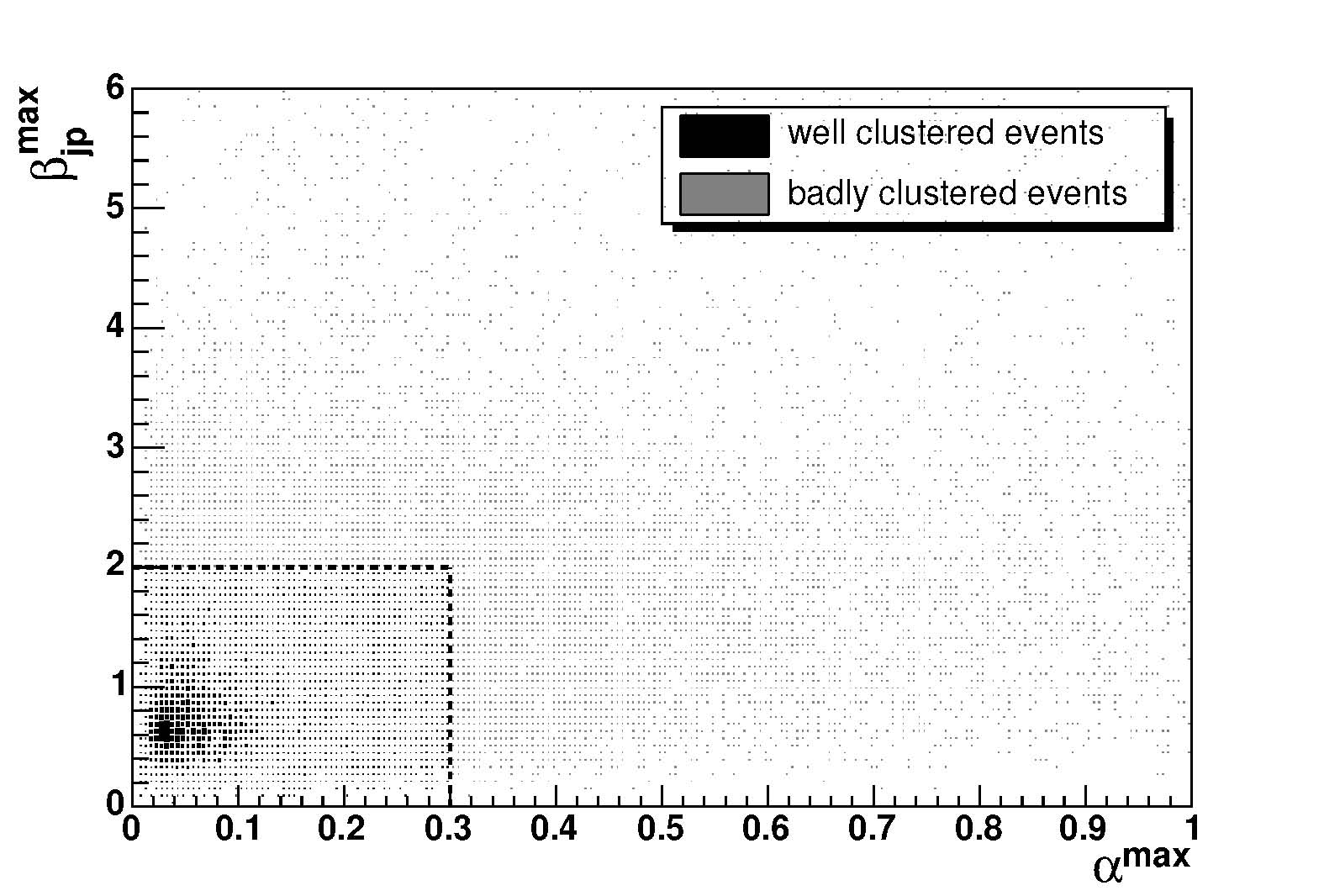} &
    \includegraphics[width=0.48\textwidth]{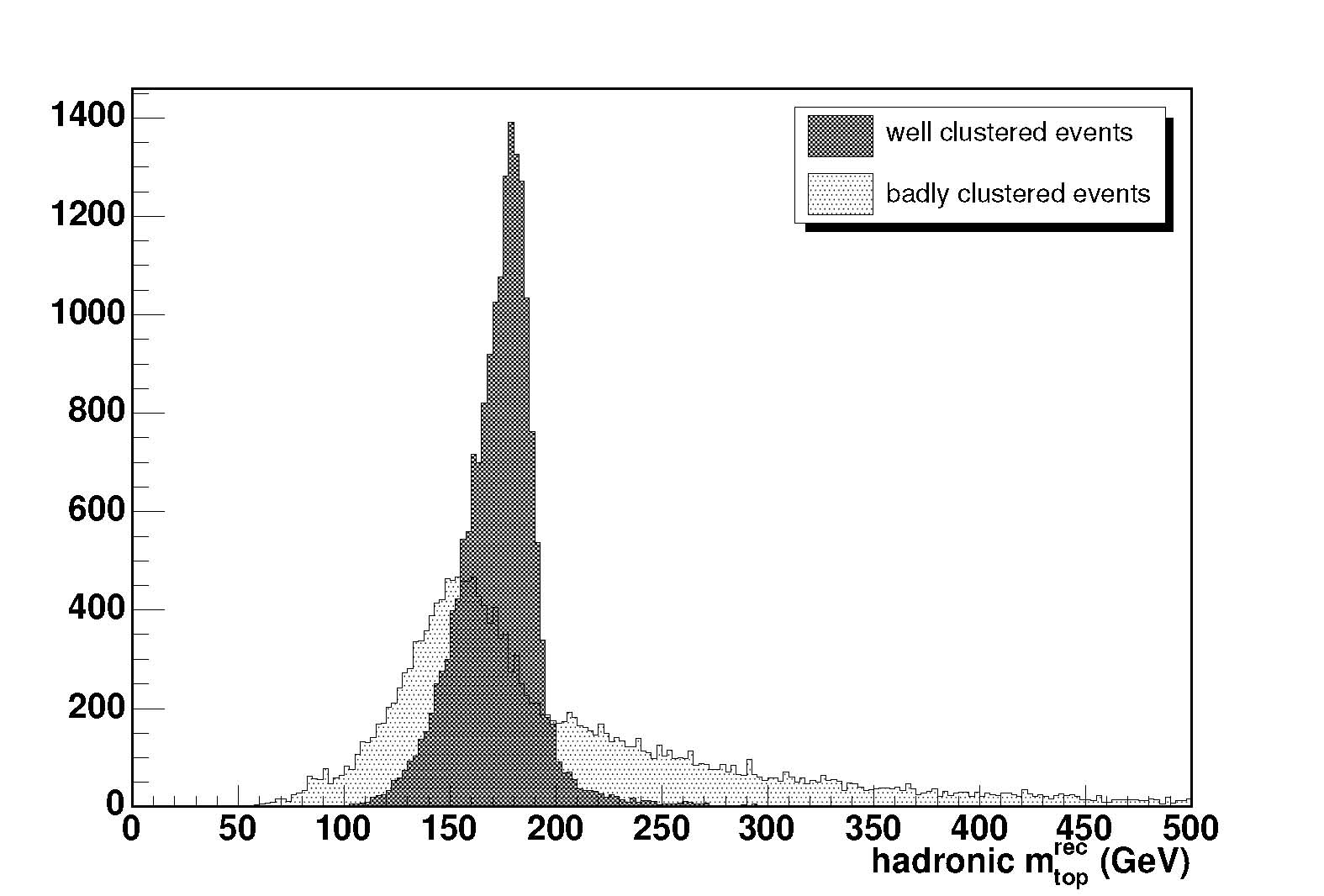} 
\end{tabular}
\caption{Left: box plot of $\beta_{jp}^{max}$
  vs. $\alpha_{jp}^{max}$ for the IC algorithm with a cone
  radius of 0.4, applied on a final state with four primary quarks. Right: distribution of the hadronic top quark mass, using jets clustered with the IC algorithm with a cone radius of 0.4,  applied on a final
  state with four primary quarks.} 
\label{topmass}
\end{figure}

This combined variable is defined as the fraction of events in which both the direction and
the energy of the $n$ primary quarks are well reconstructed following the   
definitions described above. The
correlation between  $\alpha_{jp}^{max}$ and  $\beta_{jp}^{max}$ is shown in 
Fig.~\ref{topmass} (left), where both quality criteria define a rectangular
area in which the kinematics of the primary quarks are sufficiently
well reconstructed from the analysis performance point of
view. As an illustration of the separation power of 
this combined variable,  the reconstructed spectrum of the hadronic
top quark mass in the semileptonic $t\bar{t}$ final state is shown in
Fig.~\ref{topmass} (right). 
The black histogram refers to the events in which the jets are reconstructed
with $\alpha_{jp}^{max} < 0.3 $ and $\beta_{jp}^{max} < 2$ (events inside the box of Fig.~\ref{topmass} left).
The grey histogram 
refers to the events in which the kinematics of the primary quarks are
badly reconstructed based on the combined variable (events outside the box of Fig.~\ref{topmass} left). 
 
\subsubsection{Overall quality marker "FracGood"}
The fraction of selected and well reconstructed
events, i.e. the selection efficiency $\epsilon_{s}$, multiplied by the combined variable
Frac($\alpha_{jp}^{max}$+$\beta_{jp}^{max}$) is defined as ``FracGood''.

\begin{figure}[!h] 
  \centering
\begin{tabular}{cc}
    \includegraphics[width=0.48\textwidth]{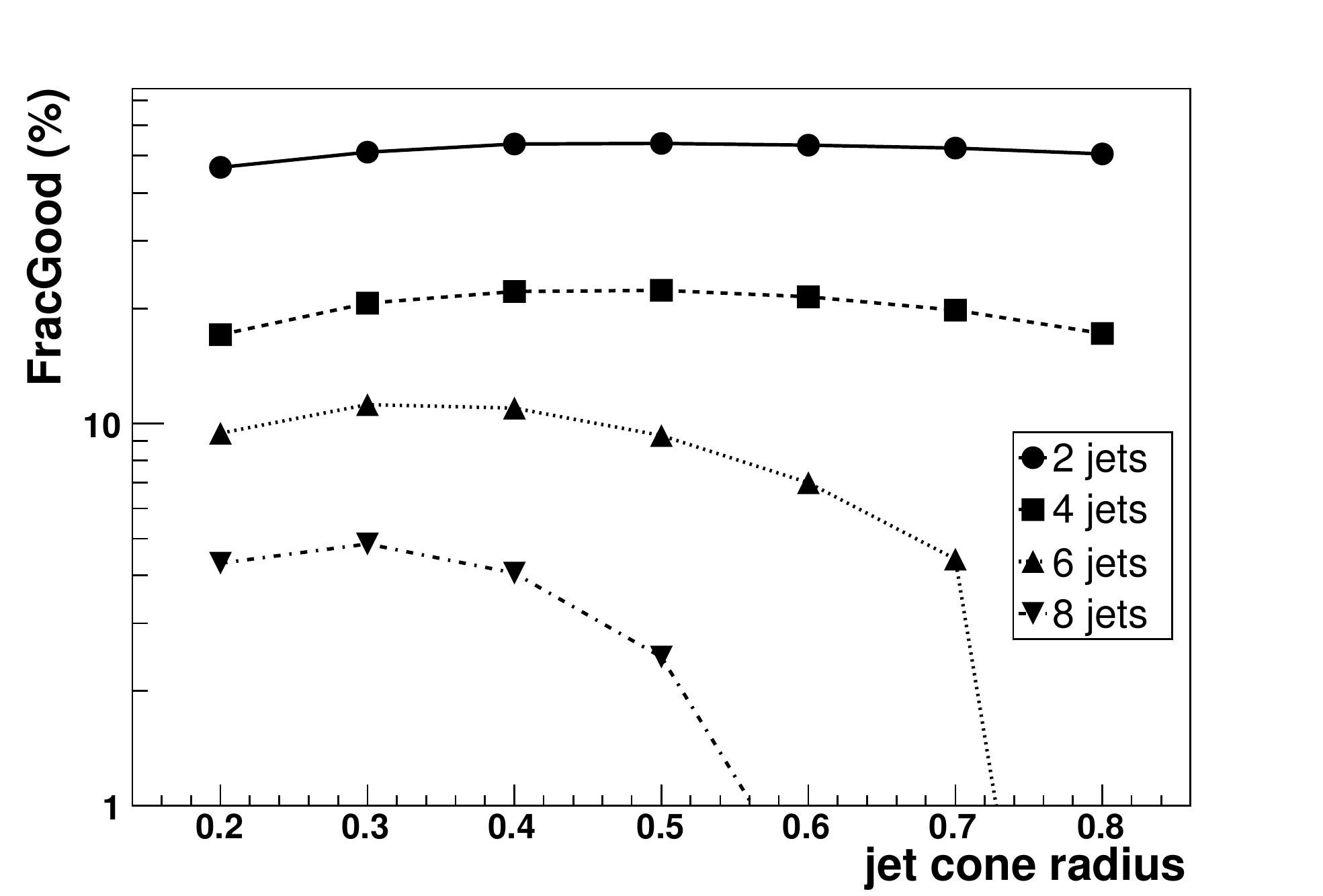}  &
    \includegraphics[width=0.48\textwidth]{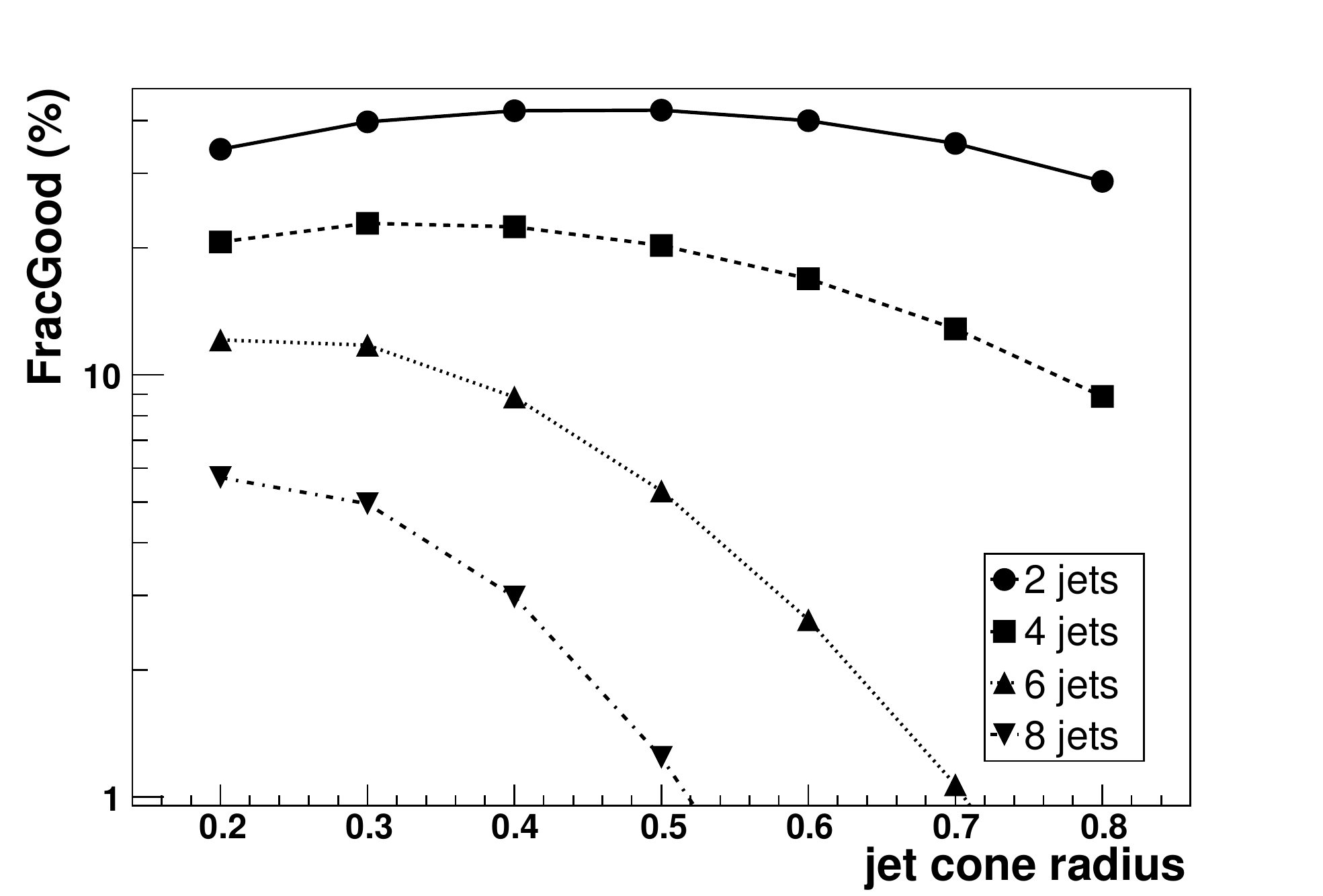} \\
    \multicolumn{2}{c}{\includegraphics[width=0.48\textwidth]{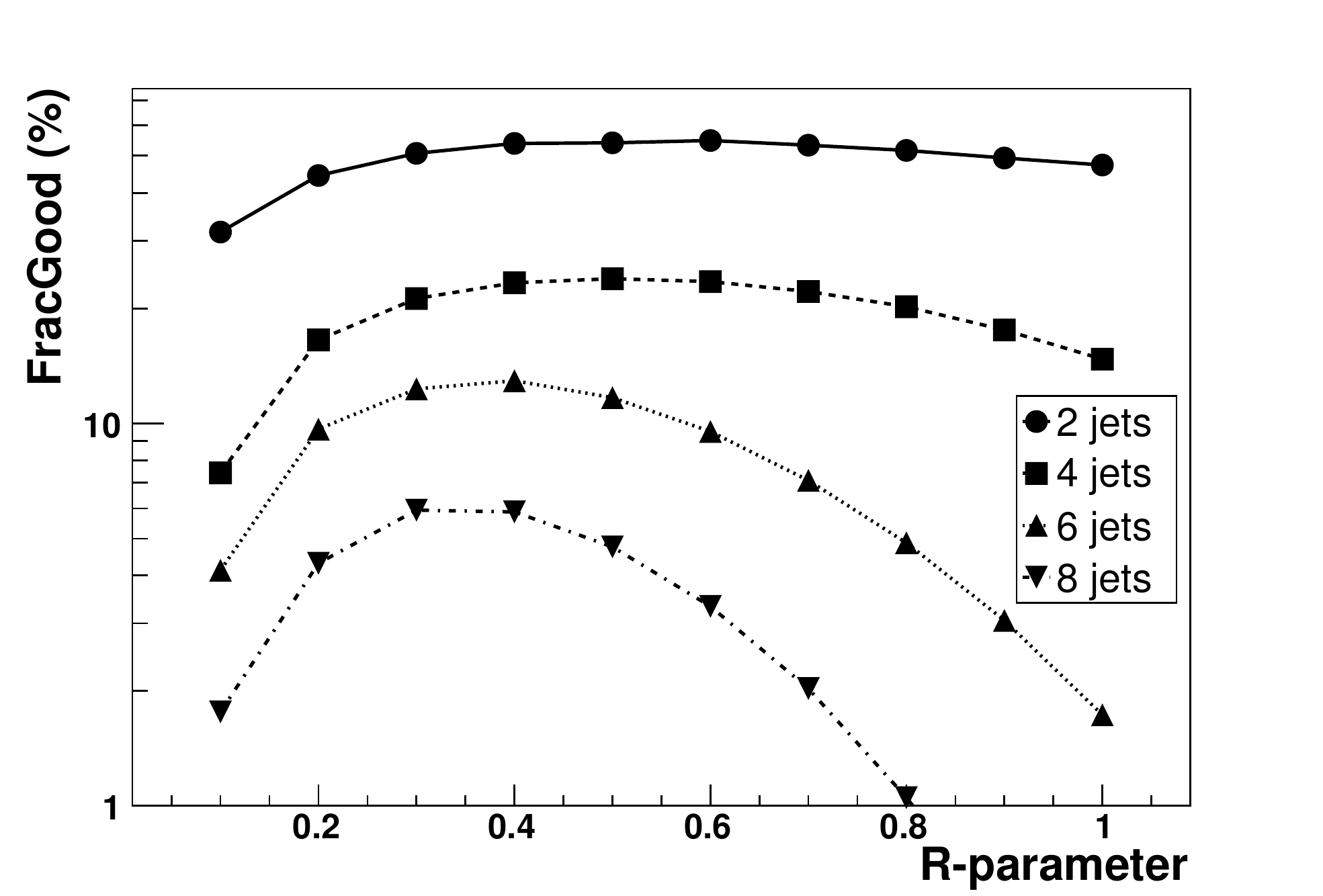}} 
\end{tabular}
  \caption{Top: Fraction of well clustered and selected events versus the cone radius (IC algorithm on the left and MC algorithm on the 
right). Bottom: Fraction of well clustered and selected events versus the R-parameter ({$k_{T}$} algorithm)}
  \label{Summary_FracGood}
 \end{figure}

This last quality marker is interpreted as an estimate for the
reconstruction efficiency of the kinematics of the primary quarks of the complete event, and therefore  used to  
compare different algorithms and setups. 
Fig.~\ref{Summary_FracGood} shows the ``FracGood'' variable as a function of the cone radius or the R-parameter for the 
three jet finding algorithms considered. It has to be remarked that a stronger
dependence as well as a larger optimal cone radius (or R-parameter) is however expected
when the jet input is changed from simulated to reconstructed particles and when the effects of the magnetic field
are taken into account. 
 
Although this variable gives a
powerful overall indication of a reasonable jet definition, it is sometimes useful to consider the partial information of the individual quality markers. Depending on the priorities of a specific physics analysis,
one would be interested in the average number of reconstructed jets, or
the energy resolution for non-overlapping jets, or the efficiency of the angular matching between primary quark and jet.  The
average number of jets gives 
an idea of the sensitivity to pile-up, underlying event, and the rate
of fake jets, while the energy resolution can be linked  
to the issue of jet calibration. 

\subsection{Results}
Table \ref{table1} summarizes the optimal parameter values for the three jet clustering algorithms, and for each of the 
considered event topologies.
For each optimal jet configuration, the respective estimate of the fraction of well reconstructed events is given.

\begin{table}[!h] 
\begin{center}
\begin{scriptsize}
\begin{tabular}{|c|c|c|c|c|c|c|}
\hline
&\multicolumn{2}{c|}{IC}&\multicolumn{2}{c|}{\textrm{$k_T$}} & \multicolumn{2}{c|}{MC}\\
&\multicolumn{2}{c|}{jet radius}&\multicolumn{2}{c|}{R-parameter} & \multicolumn{2}{c|}{jet radius} \\
\cline{2-7}
&$Value$&$FracGood$&$Value$&$FracGood$ &$Value$&$FracGood$ \\
\hline
2 quarks&0.5&53.9&0.6&54.9 &0.5&42.4 \\
4 quarks&0.5&22.3&0.5&23.8 &0.3&22.8 \\
6 quarks&0.3&11.2&0.4&12.9 &0.2&12.1 \\
8 quarks&0.3&4.85&0.3&5.93 &0.2&5.72 \\
\hline
\end{tabular}
\caption{Overview of the optimal parameter values with their respective estimate of the fraction of well reconstructed events.}
\label{table1}
\end{scriptsize}
\end{center}
\end{table}



\subsubsection{Robustness of the method against hard radiation \label{sec:radiation}}
\begin{figure}[!htb] 
  \centering
\begin{tabular}{ll}
    \includegraphics[width=0.48\textwidth]{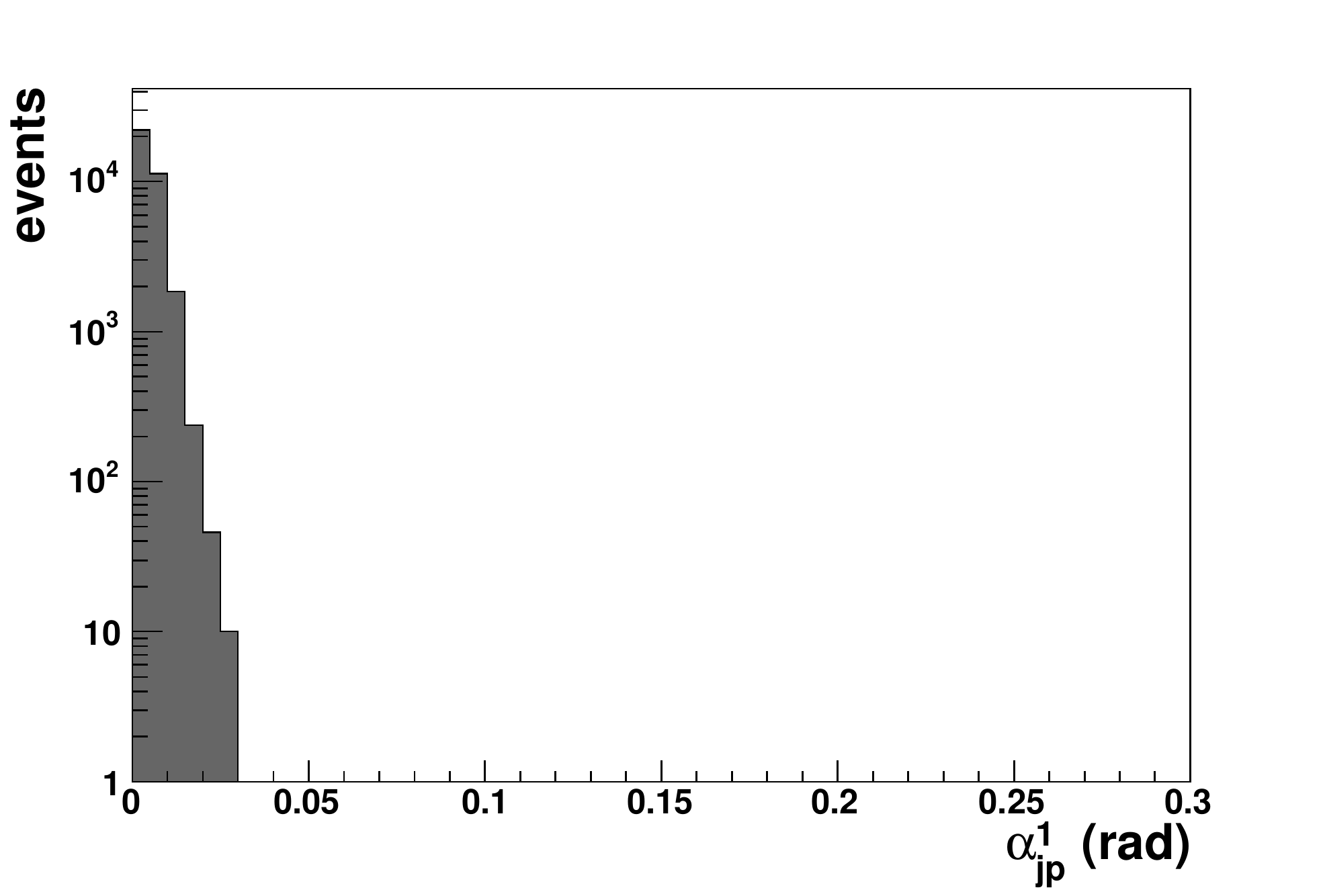}  &
    \includegraphics[width=0.48\textwidth]{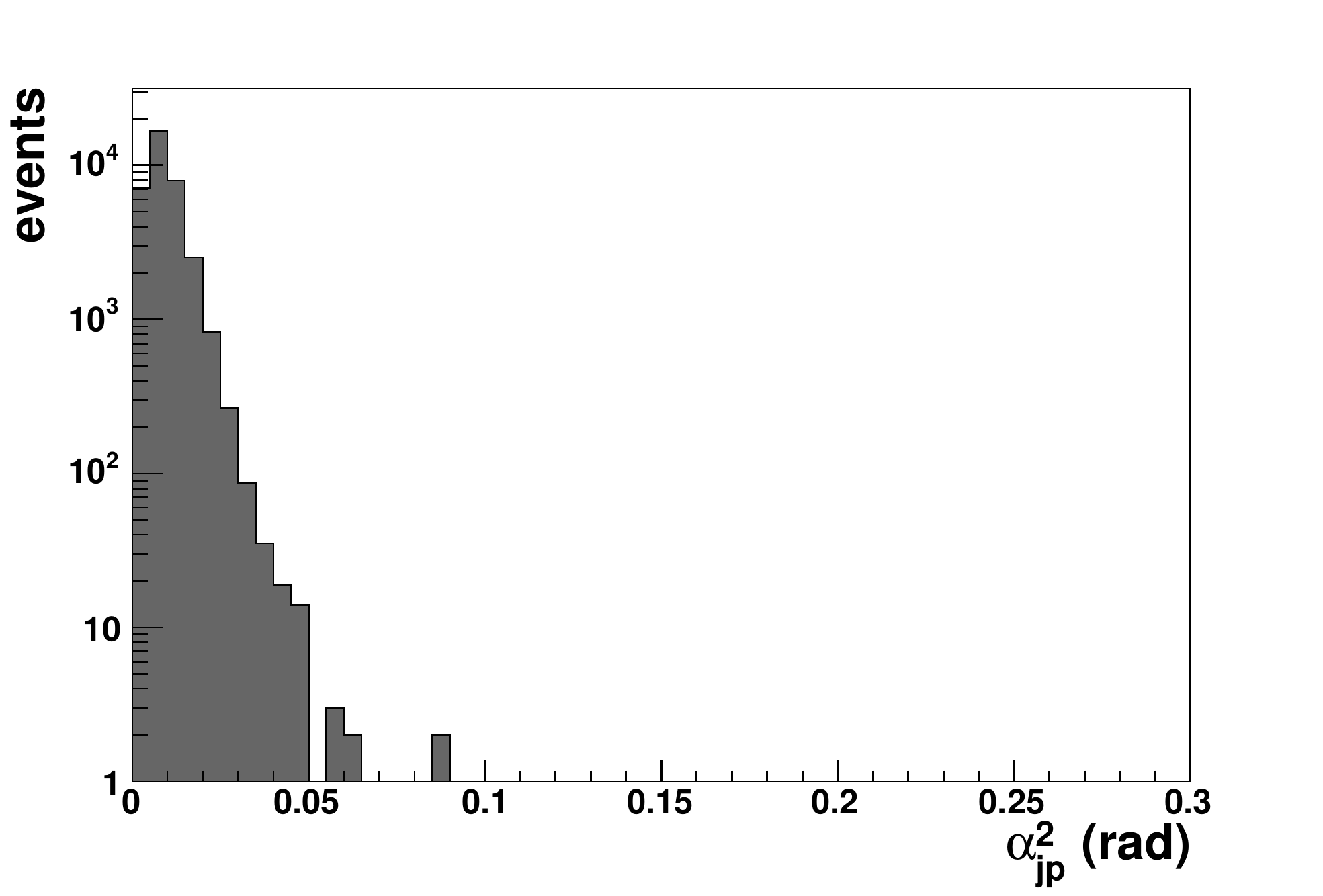} \\
    \includegraphics[width=0.48\textwidth]{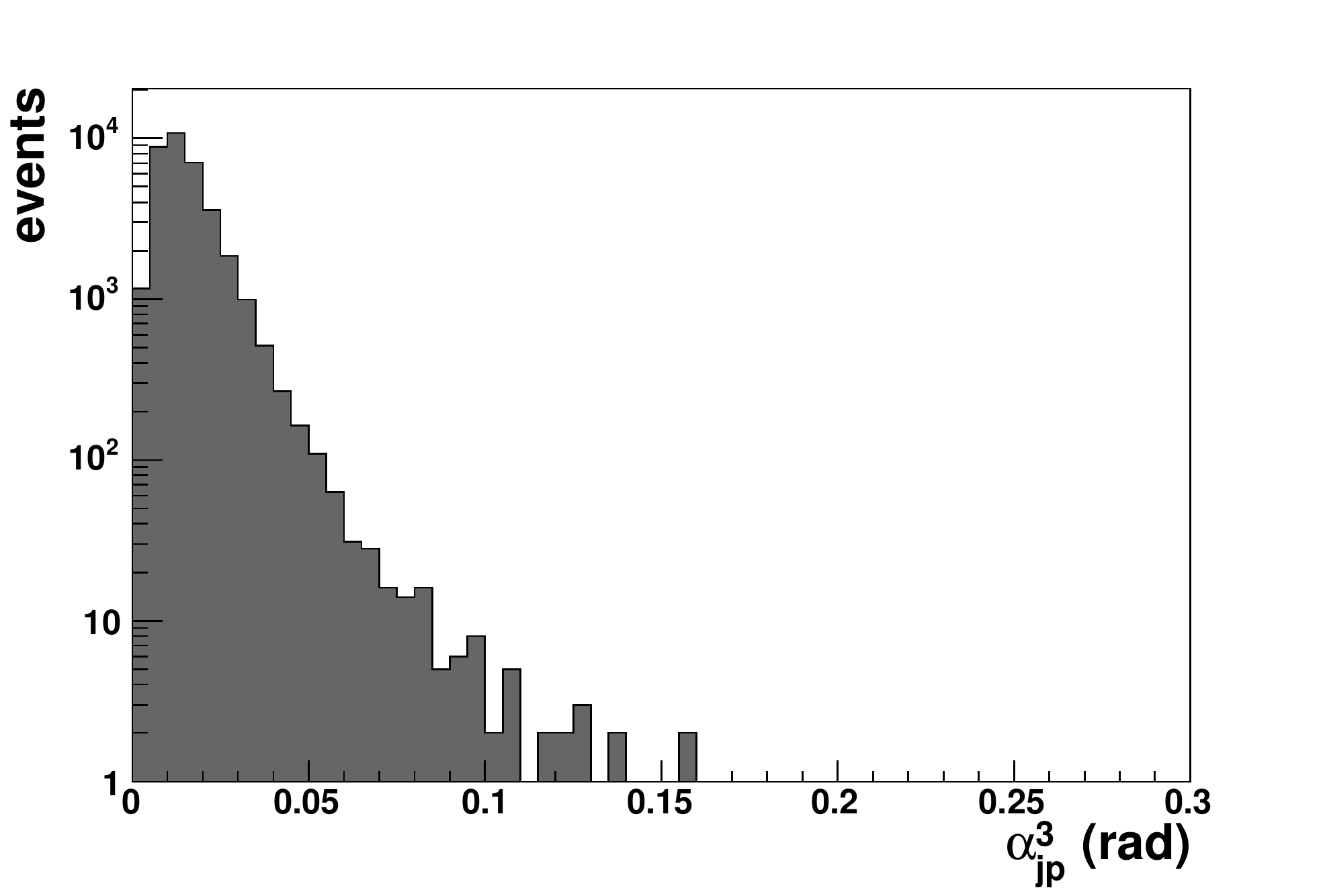} &
    \includegraphics[width=0.48\textwidth]{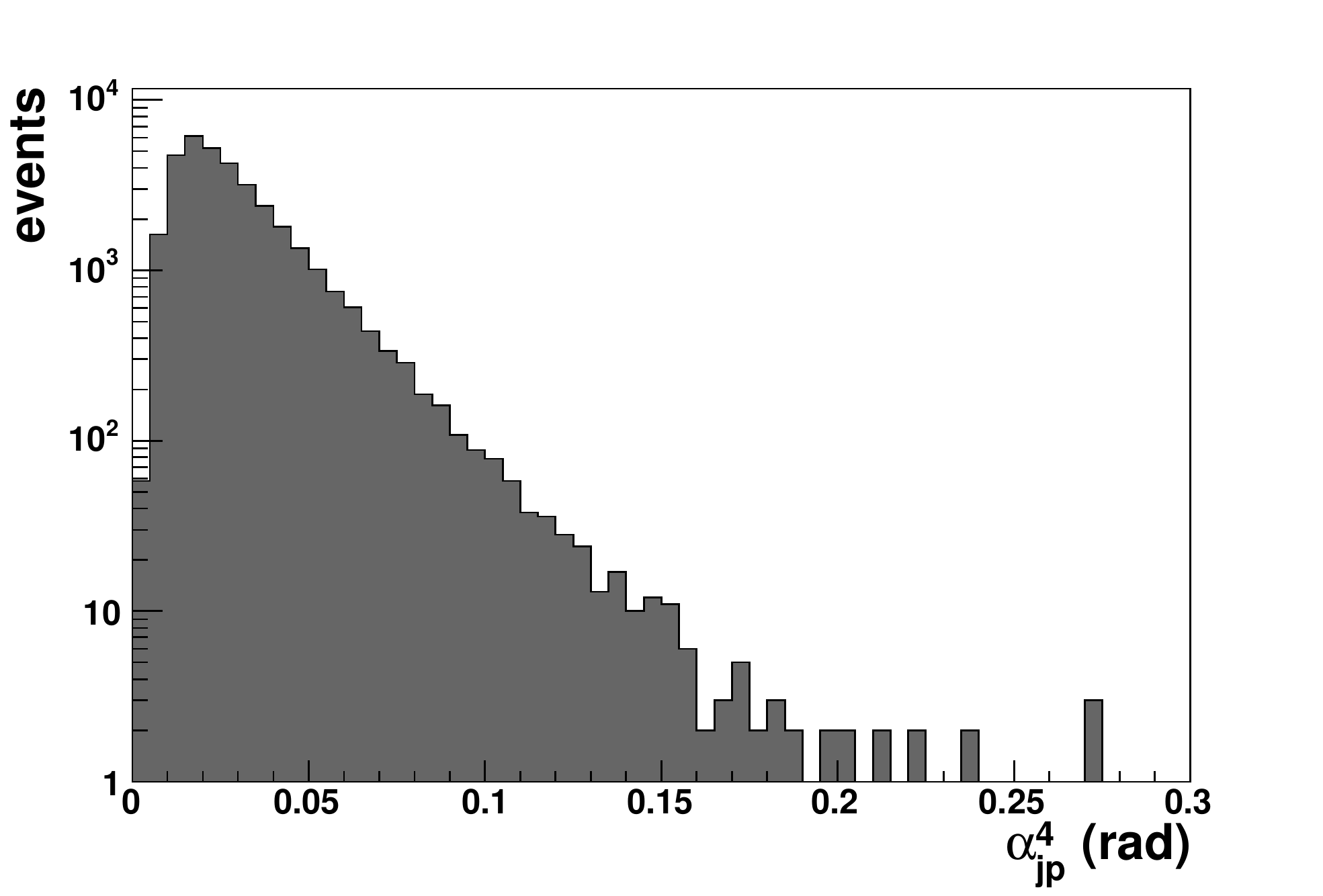} 
\end{tabular}
  \caption{Distributions of $\alpha_{jp}^{i}$  in increasing order of magnitude
          for the IC algorithm in the case of a
          final state with four primary quarks which do not radiate hard gluons.}
  \label{maxalphaNOFSR}
 \end{figure}

The sensitivity of the overall observations to
the radiation of gluons with a large transverse momentum relative to their mother quark, or from the initial state proton system, is investigated in the following. The distributions of
the $\alpha_{jp}^{i}$-values ordered by their magnitude within an event are shown in Fig.~\ref{maxalphaNOFSR} for a sample without initial and final state
radiation\footnotemark\footnotetext{PYTHIA parameters $MSTP$ $61$ and $71$ were switched off.}.

 This has to be compared directly to Fig.~\ref{maxalpha} which shows
the same plots including final state radiation.
Obviously, the long tails are not present in the case without
radiation which indicates that the $\Delta R$ cut of 0.3 for the worst jet is
not expected to have an effect in this case.
 The observation is
indeed, that the Frac($\alpha_{jp}^{max}$+$\beta_{jp}^{max}$) quality
marker has a flat distribution, but not the selection efficiency and
therefore the ``FracGood'' quality marker.

\begin{figure}[!htp] 
  \centering
  \begin{tabular}{ll}
    \includegraphics[width=0.5\textwidth]{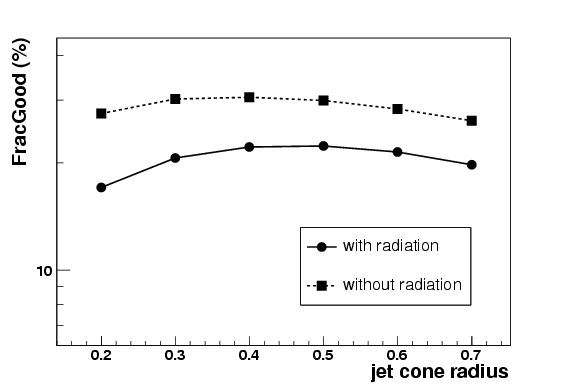} &
    \includegraphics[width=0.5\textwidth]{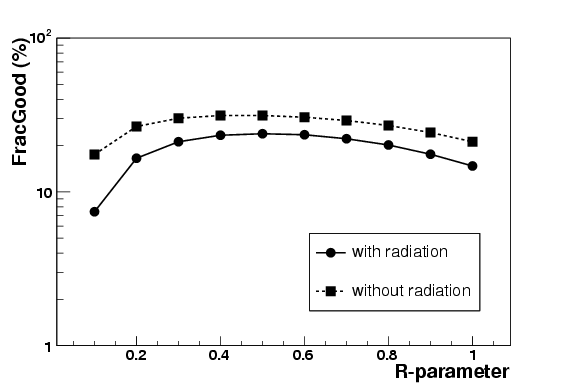} 
  \end{tabular}
    \caption{Left: influence of hard gluon radiation on the fraction
        of selected, well clustered events, as a function of the IC  
        cone radius in the case with four primary quarks in the final state. Right: influence of hard gluon radiation on the fraction
        of selected, well clustered events, as a function of the \textrm{$k_{T}$}  
        R-parameter in the case with four primary quarks in the final state.} 
    \label{radiationIC}
    \label{radiationKT}
\end{figure}

Fig.~\ref{radiationIC} (left) shows the
fraction of selected, well clustered semileptonic $t\bar{t}$ events with and without
initial and final state radiation for the {\it Iterative Cone}
algorithm. The addition of radiation results in an overall lower
efficiency, but the optimal cone radius and the shape of the curve
are robust. A similar observation was obtained for the inclusive
$k_T$ algorithm in Fig.~\ref{radiationKT} (right).

\begin{figure}[hbtp] 
    \includegraphics[width=0.5\textwidth]{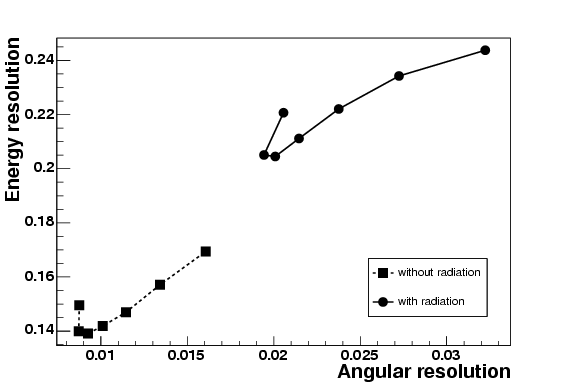} 
    \includegraphics[width=0.5\textwidth]{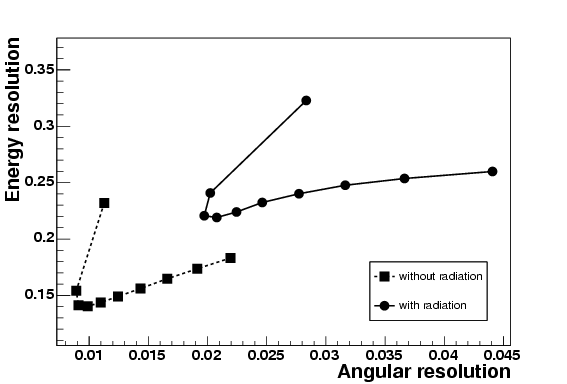} 
  \caption{Energy resolution versus angular resolution ($\Delta R$ distance between jet and quark) for the IC algorithm (left) and \textrm{$k_{T}$} algorithm (right)  in the case of four jets in the final state.
The curve are obtained by varying the parameter of the jet algorithm.
}
  \label{AngRes_vs_EnergyRes_NORAD}  
\end{figure}

In order to quantify the effect of radiation on the resolutions,
Fig.~\ref{AngRes_vs_EnergyRes_NORAD} shows the energy and angular resolution
are plotted together
for the {\it Iterative Cone} and the inclusive $k_T$ algorithm,
for the case with four partons in the final state. The curves are obtained
by varying the parameter of the jet algorithm.
The energy resolution is defined as the RMS divided by the mean value of the $E^{jet}/E^{quark}$ distribution, and the angular 
resolution is defined by the width of a gaussian fit to the symmetrized $\Delta R$ distribution.
As expected, the overall resolutions are better in the case without radiation, but the shape of the curves remains invariant.

\section{Jet Calibration} \label{sec:calib}

\subsection{Calorimeter Jets}\label{sec:calorimeterjets}
The calorimeter jets, or reconstructed jets\footnote{Although it has
  to be reminded that jets can be formed from other inputs, e.g., the
 Particle Flow objects (until very recent times, the slightly confusing
term ``Energy Flow'' was instead used in the literature).}
(see Sec.~\ref{sec:eflow}.), are obtained by applying the jet clustering
algorithm to the calorimeter signals.  Calorimeter signals are defined
by grouping the calorimeter cells to obtain a granularity best suited
to the scale of hadronic showers.

A considerable problem in the construction of calorimetric jets is noise.
In essence the output signal of a calorimeter cell, in the
absence of any energy deposition, has a continuous component superimposed
to electronic noise. The continuum component is subtracted from the
signal. A symmetric noise remains. Typical size of noise fluctuations
fake a signal of few hundred MeV.

The most common clusterization consists in assembling calorimeters
cells into towers in $(\eta,\phi)$ space. CMS builds towers of
dimension $(\Delta\eta\times\Delta\phi) = 0.087 \times 0.087$ (the
granularity of the hadronic section) in the central region, gradually
increasing in the end-cap and forward region, for a total of 4167
towers. The noise suppression algorithm consists in building the
towers using only those cells whose signals is higher than a predefine
energy threshold, whose value depends on the cell position in the
calorimeters, i.e. on the pseudorapidity and on the longitudinal
position (where longitudinal refers to the direction pointing to the
interaction region).  Various threshold schemes have been considered,
and the most used so far in the analyses uses 0.7~GeV and 0.85~GeV
thresholds for the Hadronic calorimeter barrel and outer section
respectively. In this scheme the noise contribution for a $\Delta R
=0.5 $ cone jet is equal to 1.4~GeV with a negligible loss of signal.

In ATLAS 6400 towers are built with a fixed dimension of
$(\Delta\eta\times\Delta\phi) = 0.1 \times 0.1$, corresponding to the
granularity of the central hadronic section. There is no noise
suppression applied by the tower builder algorithm.
 
A second and more evolved clusterization scheme has been developed to
obtain a good noise suppression while avoiding large biases in the
energy measurement.  This scheme consists of building
three-dimensional clusters associating neighboring cells which belong
to any calorimeter section\cite{NIMHEC}, with three minimum cell
thresholds: If a cell has energy higher than $T_{seed}$, it starts a
cluster, and all cells confining with it and having transverse energy
higher than $T_{neigh}$ are added to it. Finally, all contour cells
(i.e. cells confining with any of the cells included with the two
steps above) with transverse energy greater than $T_{cont}$ are added
to the cluster.  The defaults threshold values, applied to the
absolute cell energy, are $T_{seed}~=~4\sigma_{noise}$,
$T_{neigh}~=~2\sigma_{noise}$, $T_{cont}~=~0\sigma_{noise}$. The last
condition means that all contour cells are added
to the cluster.

The resulting clusters may contain one or more local maxima.
Eventually, the local maxima are interpreted as contributions from multiple
particles and a splitting procedure is applied to separate superimposed or
connected clusters.
A large reduction of noise is obtained if
three-dimensional clusters are used instead of the towers.



\subsection{Calibration}

The goal of jet calibration is to correct for various effects that degrade the 
measurement of the jet energy in the calorimeter.
These effects may be divided in two 
classes: detector driven effects (noise, non-compensation, cracks, 
dead-material, magnetic field effects, pile-up) and physics driven effects
(underlying event, showering effects, clustering). 
Many different strategies may be chosen to implement the jet calibration and
to check its performance and systematics. In the
next subsections the baseline strategies for the two experiments are 
discussed.
\begin{figure}[!h]
  \includegraphics[height=0.22\textheight]{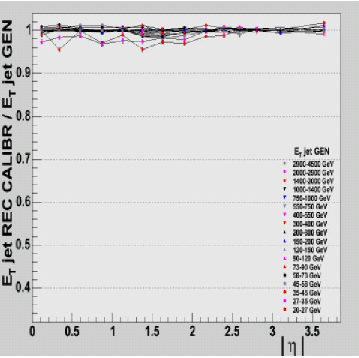}
  \hspace{3.5mm}
  \includegraphics[height=.22\textheight]{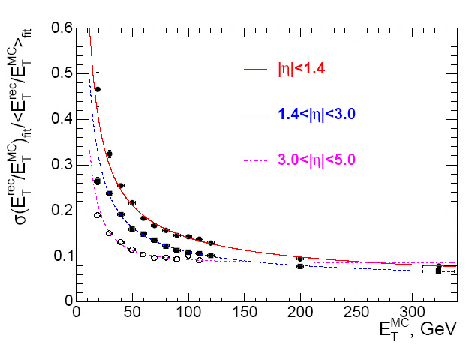}
\caption{ CMS Jet linearity after applying calibration (left) as a function 
of the 
particle jet pseudo-rapidity and in various particle jet energy ranges. Jet  
energy resolution resolution (right) as a function of particle jet energy 
in three ranges of pseudo-rapidity. Jets have been reconstructed with the 
IC algorithm with $\Delta R~=~0.5$~\cite{CMSPhysicsTDR}.}
\label{fig:CMSLinRes} 
\end{figure}

\subsection{Calibration to the Particle Jet}
\label{section:particlejetcalibration}
The degradation of the jet measurement performance 
caused by the detector effects may be corrected by applying weights that
calibrate the reconstructed jet to the particle jet. The idea to separate
detector and physics effect corrections is based on the fact that these
two classes of effects have different correlation to the jet kinematics.
\\
In order to obtain the calibration parameters, both ATLAS and CMS use QCD di-jet 
events generated with PYTHIA~\cite{PYTHIA} and simulated with the full detector descriptions.
Calorimeter and particle jets are matched on the base of 
their distance in the $(\eta,\phi)$ space.

In CMS the pseudo-rapidity range $|\eta|<4.8$ is divided into 16 regions. 
For each region the mean ratio of reconstructed jet transverse energy 
($E_T^{calo}$) to 
particle jet transverse energy ($E_T^{ptcl}$), 
$R_{jet}~=~E_T^{calo}/E_T^{ptcl}$, as a function of $E_T^{ptcl}$, is
approximated by a set of functions~\cite{MeasurementOfJetsCMS}.
Thus, let us stress that $E_T^{calo}$ is the jet $E_T$ obtained by
applying the jet finding algorithm to the calorimeter energy deposition,
which in turn is obtained by grouping valorimeter cells and
applying the noise reduction procedure
(as outlined in sec.~\ref{sec:calorimeterjets}) to
the output of the full simulation, with the magnetic field included.
With $E_T^{ptcl}$ (where $ptcl$ stands for ``particles'') we denote the
transverse energy obtained by applying the jet finding algorithm to the
particles generated by the Monte Carlo.
The values of $R_{jet}$ obtained are then used to correct the 
transverse jet energy. Since $R_{jet}$ is a function of $E_T^{ptcl}$, which 
is unknown in real data, an iterative procedure is used to obtain for each 
calorimeter jet energy 
the best estimate of the calibration parameter~\cite{CMSPhysicsTDR}. 
The linearity and the resolution obtained by applying this calibration
to a statistical independent sample of QCD di-jet events are 
shown in Figure~\ref{fig:CMSLinRes}.
The maximum deviation from linearity for the $E_T$ range [20 GeV - 4 TeV] 
is $\sim 5\%$. The energy resolution in the region $|\eta|<1.4$ is~:
\begin{equation}
\frac{\sigma(E_T)}{E_T} = \frac{1.25}{\sqrt{E_T(GeV)}} \oplus \frac{5.6}{E_T(GeV)} \oplus 0.03
\end{equation}.
\\
In ATLAS the calibrated jet energy is obtained by applying the weights ($w_i$) 
to the cell energies ($E_{cell}$) that compose the
jets:
\begin{equation}
E^{calib} = \sum_i w_i E_i
\end{equation}
The weights, which depend on the position and energy density 
of the cells, are extracted by minimizing a $\chi^2$ defined as~:
\begin{equation}
\chi^2 = \sum_j \left ( \frac{E_j^{calib}}{E_j^{ptcl}}  - 1 \right )^2 
\end{equation}
where the index $j$ runs on the ensemble of jets of all the events.
\begin{figure}[!t]
  \includegraphics[height=.24\textheight]{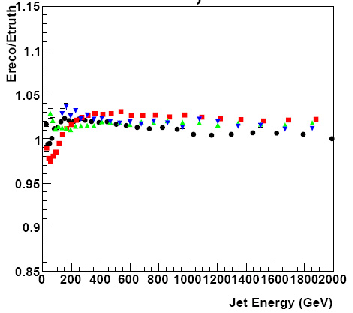}
  \hspace{3.5mm}
  \includegraphics[height=.24\textheight]{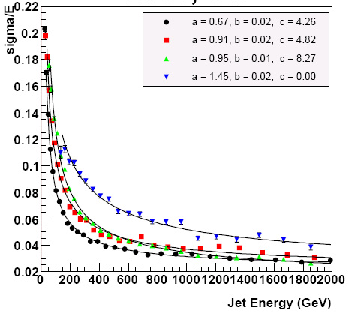}
\caption{ATLAS jet linearity (left) and resolution (right) 
after applying calibration as a function of the 
particle jet energy and in various pseudo-rapidity ranges ($|\eta|<0.7$~(black circles), $0.7<|\eta|<1.5$
(red squares), $1.5<|\eta|<2.5$ (green triangles), $2.5<|\eta|<3.2$ (blue
triangles)). Jets have been 
reconstructed with the $\Delta R~=~0.7$ cone algorithm.}
\label{fig:ATLASLinRes} 
\end{figure}
The dependence of the 
weight $w_i$ on the cell energy density is parameterized with a polynomial.
The basic idea behind this kind of calibration, which exploits the 
shower shapes, is that hadronic showers are diffuse while electromagnetic ones 
are dense. Therefore $w_i$ is typically larger than 1 for low cell 
energy densities and is around 1 for high cell energy densities. This is a
consequence of the fact that the ATLAS calorimeter (as the CMS one) is
non--compensating ({\em i.e.} it has different efficiency in the
measurement of the electromagnetic and hadronic part of the shower), and thus the
calorimeter response to hadrons is non--linear with the energy.
To understand the lower (and non--linear)
response of non--compensating calorimeters to hadrons, consider the following
three facts:
\begin{itemize}
  \item Part of the shower produced by hadrons in the calorimeter is
    electromagnetic. This is because of the decay of $\pi^0$ produced in the
    shower. 
  \item In non--compensating calorimeters, the efficiency of the measurement 
    of the electromagnetic and hadronic
    part of the shower are different ($e/h \neq 1$). This is mainly because part of the
    hadronic energy is lost in nuclear reactions to break the nuclei. 
  \item The electromagnetic fraction, {\em i.e.} the fraction of the
    shower energy carried by photons, depends on the energy of the impinging
    hadron.  This can be understood with the following, simplified
    model \cite{nwigmans}. Suppose a charged pion is impinging on the calorimeter: on the first
    hadronic interaction, mainly charged and neutral pions will be
    produced. On average, 1/3 of the energy will be carried by neutral
    pions. On the second stage, the fraction of the energy carried by $\pi^0$
    will be $f_{em} = 1/3 + 2/3 \cdot 1/3$. On the $n$--th stage, $f_{em} =
    1-(1- 1/3)^n$, where $n$, the maximum number of interactions, is energy
    dependent.
\end{itemize} 
This three facts together make the calorimeter response to hadrons non--linear.
Furthermore, since the fraction of produced neutral pions
undergoes large fluctuation, non-compensation also induces a worse
resolution in the jet energy measurement.
  
The linearity and resolution, as a function of 
the particle jet energy, obtained on a sample of QCD di-jet events 
for various pseudo-rapidity regions are shown on 
figure~\ref{fig:ATLASLinRes}.
The maximum deviation from linearity is within $2\%$ in the jet energy range 
[40~GeV - 2~TeV] and the resolution in the pseudo-rapidity region $|\eta|<0.7$
is equal to~:
\begin{equation}
\frac{\sigma(E)}{E} = \frac{0.67}{\sqrt {E(GeV)}} \oplus \frac{4.3}{E(GeV)} \oplus 0.02 
\end{equation}
The jet linearity, as estimated using a sample of events with different 
parton composition and topology, generated by HERWIG~\cite{HERWIG}, is also
well within $\pm 2\%$.

\subsection{Parton-level calibration}

Calibration to the parton jet can be implemented as a second step
in addition to particle jet calibration or 
as a single step which corrects for both 
detector and physics effect. ATLAS is presently considering the
first strategy, while CMS has implemented both~\cite{CMSpartonlevel}.

The definition of the parton jet energy is somehow artificial, since
partons cannot be defined as isolated objects (not even in the short
time scales of the hard interactions). Furthermore, as previously,
discussed, the association of a primary parton to a jet is unavoidably
dependent upon the Monte Carlo one is using.
It has been however widely used by previous experiments~\cite{CDF}.
It is fair to say that, with this method one can use the kinematics
of the reconstructed partons to look for mass peaks; however, the
method cannot yield an accurate mass measurement.

A first difference between particle and parton jet is caused 
by the smearing produced during final state radiation 
and fragmentation. Both phenomena generate
particles which may not be clustered into the particle jet. This results in
a fraction of the parton jet energy not 
attributed to the particle jet. In the case of 
cone clustering algorithms these losses are indicated as out-of-cone
losses. Second, some 
of the particles generated in the underlying event may fall in the jet 
region and be attributed to the particle jet although this contribution is not 
related to the parent parton jet. In this section some possible strategies 
to correct for these effects are discussed.

\begin{figure}[!t]
  \includegraphics[height=.24\textheight]{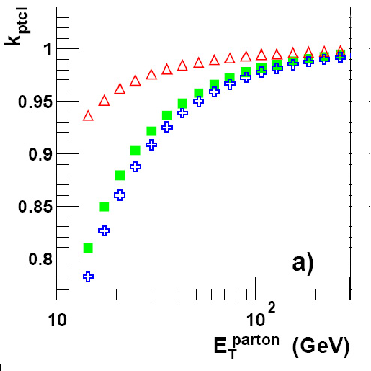}
  \hspace{3.5mm}
  \includegraphics[height=.24\textheight]{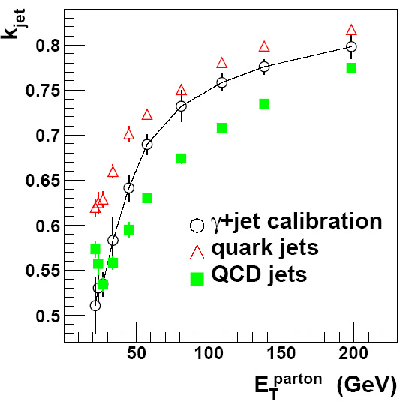}
\caption{Left: distributions of the mean value of $k_{ptcl}$ as a function of
transverse parton energy for QCD di-jets (green square), for quark jets (open
triangle) and for gluon jets (open crosses). Right: 
distributions of calibration coefficient 
obtained from $\gamma$+jets events (open circles) and their true value for 
generic QCD jet (full green squares) and quark jets (red triangles).
Jets are reconstructed with $\Delta R = 0.5 $ cone algorithm in the pseudo-rapidity
region $|\eta|<1.5$~\cite{CMSPhysicsTDR}.}
\label{fig:CMSGammaJet} 
\end{figure}

\begin{figure}[!t]
 \includegraphics[height=0.25\textheight]{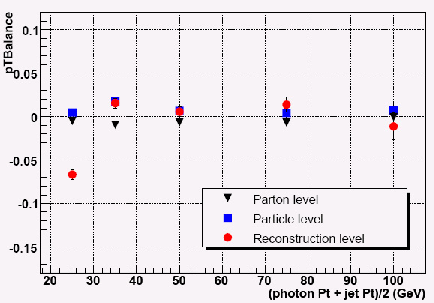}
\caption{
Distribution of  $pTBalance = (p_T^{jet} - p_T^\gamma)/p_T^\gamma$
as a function of $(p_T^{jet} + p_T^\gamma)/2$ obtained by ATLAS on a sample
of $\gamma$+jets events. The $pTBalance$ distribution is shown for 
calibrated calorimeter jets (full
red circles), particle jets (blue triangles) and partons 
(black squares)~\cite{ATLASGammaJet}. 
Jet have been reconstructed with $\Delta R = 0.7$ cone algorithm.}
\label{fig:ATLASGammaJet} 
\end{figure}

A first possibility, exploited by CMS, to obtain the parton jet energy scale is
to use simulated events and obtain a calibration constant 
$k_{ptcl} = E_T^{ptcl}/E_T^{parton}$
as a function of the transverse energy of the parton. 
In figure~\ref{fig:CMSGammaJet} (left) the values of $k_{ptcl}$
are 
shown for generic QCD jets and for gluon and quark generated jets separately. 
The scale uncertainty due to the different fragmentation of gluon and quark
generated jets is estimated by comparing the $k_{ptcl}$ values obtained
in the two cases. If $\Delta R=0.5$ cone jets are considered the calibration
coefficients differ by $5\%$ for $E_T = 40$ GeV~\cite{CMSGammaJet}.
\\
A second possibility to obtain calibration is to exploit kinematic constraints
from real data such as the W mass in $W \rightarrow jj$ decays or the $p_T$ balance in events where 
the jet is generated back-to-back with a well measured particle, either a
Z decaying to leptons or a $\gamma$. In this note 
studies using $\gamma$+jet events are discussed. 
\\
ATLAS and CMS 
plan to use these events in different ways. CMS exploits
the $p_T$ balance constraint to obtain the calibration from
calorimeter jet to parton jet while ATLAS plans to apply first
the calibration to particle jet and than use the $p_T$ balance
constraint as a further step to correct to the parton jet energy scale.
In the first phase of data taking the primary role of these events 
will be to help in understanding particle jet level calibration by 
comparing the data and Monte Carlo $p_T$ balance distributions.
\\
The selection of events in CMS requires a well isolated photon having a $\phi$
opening angle with the 
jet $\Delta\phi>172^o$~\cite{CMSPhysicsTDR,CMSGammaJet}. Events containing 
more than one jet with $E_T>20$~GeV are rejected. The main background is
given by QCD di-jet events where one jet is misidentified as a photon.
Background is suppressed well below the signal for $E_T^\gamma>150$~GeV. 
The ratio $k_{jet} = p_T^{calo}/p_T^\gamma$ is calculated as a function
of $p_T^\gamma$ and defines the calibration 
coefficients. The complication given by the presence of 
initial state radiation that spoils the $p_T$ balance constraint is 
partially overcome by defining, for each $p_T^\gamma$, the calibration 
coefficient to correspond to the most probable value of the  $p_T^{calo}/p_T^\gamma$
spectrum. The predicted values for the calibration coefficients and their 
true values ($k_{true} = p_T^{calo}/p_T^{parton}$) 
for quark jets and for jets from QCD background are shown in 
figure~\ref{fig:ATLASGammaJet}. At a transverse energy of 100~GeV a difference 
of about $10\%$ is observed between QCD jets and quark jets. It should be
noticed that this difference may be originated both by the different
fragmentation
spectrum of particles inside the jet and by the different out-of-cone losses.
The $p_T$ coverage of this channel after analysis cuts, 
indicates that, from a purely statistics evaluation,
with $10 fb^{-1}$ a $1\%$ statistical error is obtained up to a transverse
energy of 800~GeV in the central region. 
\\
The event selection of ATLAS also starts with the requirement of 
a well isolated photons with 
$E_T^\gamma>30$~GeV having an opening angle with respect to the highest 
$p_T$ jet in the event of $\Delta \phi > 168^o$~\cite{ATLASPhysicsTDR,ATLASGammaJet}. 
In order not to introduce a 
bias in the definition of the calibration coefficient due to the initial state
radiation, the binning is done in bins of $(p_T^\gamma+p_T^{jet})/2$. The 
calibration coefficient in each bin, as for CMS, is defined as the most probable
value of the $p_T$ balance spectrum. Distributions of the $p_T$ balance,
defined as $(p_T^{jet} - p_T^\gamma)/p_T^\gamma$, as a function of 
$(p_T^{jet} + p_T^\gamma)/2$ are shown in 
figure~\ref{fig:ATLASGammaJet}. The three curves correspond to 
the $p_T$ balance obtained using the
jet calibrated to the particle jet (as described in 
the previous section), the particle jet, 
and the parent parton. The balance obtained from particle jets and from
calibrated 
jets agree within $\pm 2\%$ indicating that the particle level calibration,
obtained on QCD di-jet events, may be applied also to different event 
topologies and different mixtures of partons. 
This result is somehow in 
disagreement with what is obtained by CMS (figure~\ref{fig:CMSGammaJet})
where a large difference
between quark and gluon jets is observed.
It should be noticed, however, that the 
different cone size and the different correction for energy inside the cone 
makes it difficult to better understand the significance of
this discrepancy.
We also notice that $\gamma+$jet at LHC is dominated by quark jets, while
the typical QCD jets are gluon jets.
 The particle level and parton 
level balance agree within $\pm 1\%$ indicating that 
underlying event contribution and the out-of-cone losses compensate 
each other to this level. Studies are ongoing to disentangle the two effects.
\\



\section{Energy Flow}
\label{sec:eflow}


Although the conceptual simplicity of calorimetric jets is a great
asset for very early calorimeter understanding and calibration, an
integration of the informations coming from the other detector
components can provide a substantial improvement in both the
measurement biases and the jet resolution.

In order to estimate the potential for improvement, one has to
consider that 65\% of the energy in an event is carried by charged
particles (including the decays of unstable neutral particles into
charged ones, the so called $V^0$'s, like $K^0_S\to\pi^+\pi^-$ and
$\Lambda^0\to p\pi$), 25\% by photons (including $\pi^0$ decays) and
only 10\% by long-lived neutral hadrons.  This means that ideally, if
all the photons were identified and corrected with specific
calibrations and all the charged particles were measured by the
tracking system, 90\% of the energy could be better known.  Additional
improvement comes from particle identifications: not only electrons
and muons would benefit from specific calibrations (since electrons
loose most of their energy in the electromagnetic calorimeter and the
muons deposit much less energy than hadrons in the calorimetric
systems) but also $V^0$ recognition (since the measured invariant mass
of the decay products can be replaced by the known mass of the
``mother'') and eventually the identification of the charged hadron as
pion, kaon or proton (since all the particles, in jet, in first
approximation are usually treated as pions, or even as massless
particles, but at momenta of the same order of the particle mass this
affects the energy measurement).

This ideal goal is made difficult by the unavoidable detector
inefficiencies (e.g., the least energetic charged particles never
reach the calorimeters due to the magnetic bending, so this part of
the jet energy is unrecoverable) and by the identification
ambiguities. Moreover, since the most important source of improvement
is the replacement of the calorimetric measurement with the tracking
information for charged hadrons, a critical factor is the ability of
1-to-1 association between tracks and calorimetric clusters, and this
is limited by the coarseness of the calorimeter.

\subsection{Energy Flow Algorithms in ATLAS}

Inside the ATLAS collaboration, two different approaches to the use of
the energy flow have been been studied.  The first one \cite{tovey}
(approach A in the following) builds EnergyFlow objects from
calorimeter towers and tracks and uses them as input objects for the
jet reconstruction algorithm, while the second \cite{froidevaux}
(approach B) applies energy flow techniques on reconstructed
jets. Both of them are at present somewhat limited by the {\em ad
  interim} solutions used inside ATLAS for the clustering. While at
present the standard clustering for jets is done only in the
$\eta$--$\phi$ space, the final clustering, which is under
development, will make use of the complete $\eta$--$\phi$-$r$
segmentation of the ATLAS calorimetry, thus allowing for 3D clusters,
more efficient in recognizing energy deposits belonging to a jet and
less sensitive to noise.

The aim of the approach A is to define consistently topologically
connected EnergyFlow objects. Each charged track seeds an EnergyFlow
object. The tracks are then associated to calorimeter clusters both in
the EM and in the HAD calorimeter extrapolating the track trajectory
(assumed to be helicoidal) and making a matching
in the $\eta$--$\phi$ space. The energy deposit expected for the
particle (given its identification and its momentum measured by the
tracker) is then subctracted from the calorimeter clusters. If the
remaining energy in the cluster is within 1.28 $\sigma_{noise}$ from
zero, the cluster is removed from the cluster list. The remaining
non--zero EM clusters seed EnergyFlow objects, the $\eta$--$\phi$
association is repeated and the expected energy deposits in the HAD
clusters is subctracted. The remaining HAD clusters seed EnergyFlow
objects.

Finally, EnergyFlow objects that are topologically connected (an EM cluster can be associated to more than one HAD cluster because of the bending of the magnetic field, for example) are grouped together in only one EnergyFlow object.   

\begin{figure}[tb]
\begin{center}
    \includegraphics[width=0.5\textwidth]{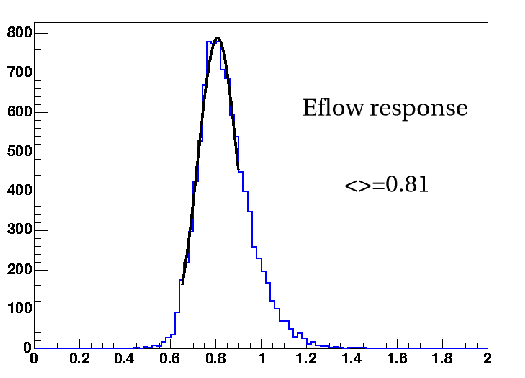} 
    \includegraphics[width=0.5\textwidth]{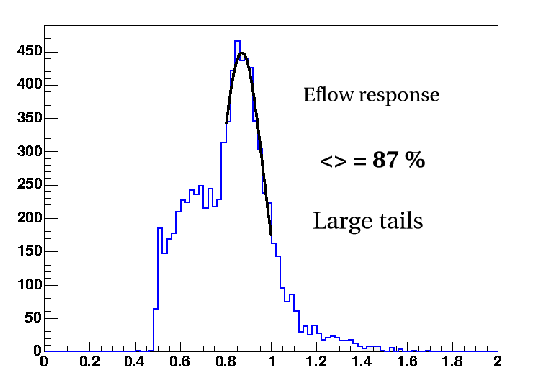} 
\caption {\small  {\em The ratio between the reconstructed and reference energy is considered for events with 3 particles in the final state ($\gamma, n, \pi^{\pm}$). The shape of the distribution is degraded as they get close (on the left: $\Delta R>0.1$, on the right: $\Delta R=0.05$).}}
\label{fig:dr}
\end{center}
\vspace{-1cm}
\end{figure}

Approach B considers as input for the Energy Flow algorithm the already reconstructed jets. The idea is to identify (within a jet) clusters generated from charged hadrons, photons, electrons and finally neutral hadrons. To do this, a first iteration is performed on EM clusters. The central cell of those  clusters that do not have a charged track pointing to them is chosen as a seed, and all the cells within $\Delta R = 0.0375$ are labelled as EMCL. Then an iteration over the tracks is performed, and all the cells within  $\Delta R = 0.0375$ from the track are labelled as CHRG. Finally, unassigned cells are labelled as NEUH. Ideally, EMCL should take into account photons, CHRG should account for charged pions, while NEUH should inlcude neutrons.

It has been already pointed out that the Energy Flow algorithms work at best with high granularity calorimeters and low multiplicity enviroment. If  the subtraction of the expected energy is performed on an isolated cluster, one can expect an improvement on the resolution. But as soon as the clusters are not well separated, the subtraction of the expected value does not lead to an improvement of the resolution. This can be seen for example in fig.~\ref{fig:dr}, where a ``jet'' composed by only three particles ($\gamma, n, \pi^{\pm}$)  is considered. If the particles are far away in the $\eta$--$\phi$ space (left plot), the distribution of the measured energy is well shaped, but as soon as the particles become close (right figure), the Energy Flow response loose its regularity.   
Therefore, a refined 3D clustering algorithm is mandatory to improve the performances of the Energy Flow algorithms in ATLAS.

\begin{figure}[tb]
\begin{center}
    \includegraphics[width=0.5\textwidth]{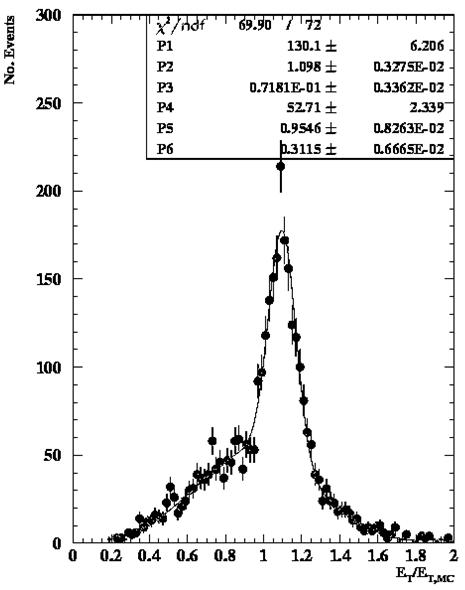} 
    \includegraphics[width=0.5\textwidth]{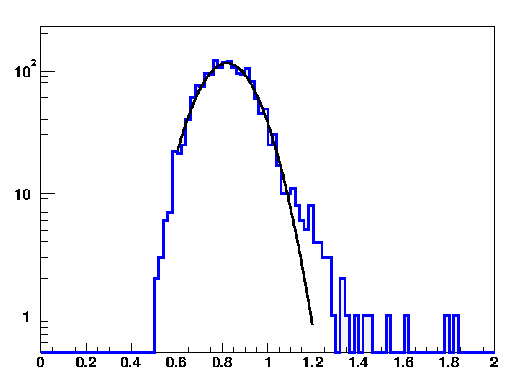} 
\caption {\small  {\em On the left: the ratio between the reconstructed and the reference energy for the approach A on 50 GeV jets. The $\sigma(E)/E$ on the core of the distribution is 7\%. On the right: The same for approach B for jets with energy between 20 and 60 GeV. The $\sigma(E)/E$ is 12--13\%. As a reference, the TDR resolution for jets at 50 GeV is 8--9 \%. }}
\label{fig:ATLresults}
\end{center}
\vspace{-1cm}
\end{figure}

Fig.~\ref{fig:ATLresults} shows the results of both the approaches discussed. Noise and pile--up are not included in the simulation. The left figure shows the current performances of approach A for 50 GeV jets. Two different contributions can be seen. The core of the distribution (whose $\sigma(E)/E$ is 7\%) shows the performances where the track subtraction has worked, while in the broad peak, it did not work. The right figures shows the performances of approach B on jets with energy between 20 and 60 GeV. While the distribution is much more regular, the peak is broader ($\sigma(E)/E \simeq 12-13\%$) with respect to the core of the left plot. For comparison, the resolution quoted in the TDR for 50 GeV jets (from the standard calotimeter measurement) is 8\%. The improvement of the clustering strategy could give an important improvement to the Energy Flow performances.

\subsection{Energy Flow Algorithms in CMS}

The improvement coming from the use of an Energy Flow technique is expected to be even more important for CMS than for ATLAS, due to their different detector designs: CMS has a more precise tracking system (thanks to the higher magnetic field and to the choice of using only pixel and microstrip silicon modules, while part of the ATLAS tracking system is constituted by the Transition Radiation Tracker (TRT), with coarser resolution), while the requirement of compactness makes its hadronic calorimeter less precise than the ATLAS counterpart.
For this reason, a big effort is currently under way in CMS for the development of an optimal Energy Flow algorithm (actually called ``Particle Flow'', since particle identification plays a big role in it), with a large dedicated development group.
This section presents only the first partial results towards this goal. Although these will be soon out of date and superseded by the complete algorithm, they show how much can be gained in CMS from the technique.

The simplest version \cite{noteKodolova} corrects the jet energy and direction
after its reconstruction by the jet-finding algorithm (that uses the calorimetric 
deposits only). 


The integration between Calorimeter and Tracking system measurements is performed by the 
EF algorithm through the following steps:
\begin{itemize}
\item Jets in the event are reconstructed by the calorimeter using an iterative cone 
algorithm. The jet object is defined by the collected energy and the direction.

\item In the event all tracks with $P_T > 0.9$ GeV and  $|\eta| < 2.4 $ are reconstructed and selected
 at the vertex in a cone $\Delta R$ around jet direction. The cone is the same  of the jet-finding algorithm.

\item For each track the impact point on the ECAL inner surface is extracted and extrapolated to 
the HCAL one.

\item The expected response of the calorimeter to each charged track is subtracted from the 
calorimetric cluster and track momentum is added.

\item Other low $P_T$ charged tracks, swept out of the jet cone definition by the magnetic field, 
are added to jet energy.
\end{itemize}


\begin{figure}[tb]
\begin{center}
  \centerline{
    \includegraphics[width=0.5\textwidth]{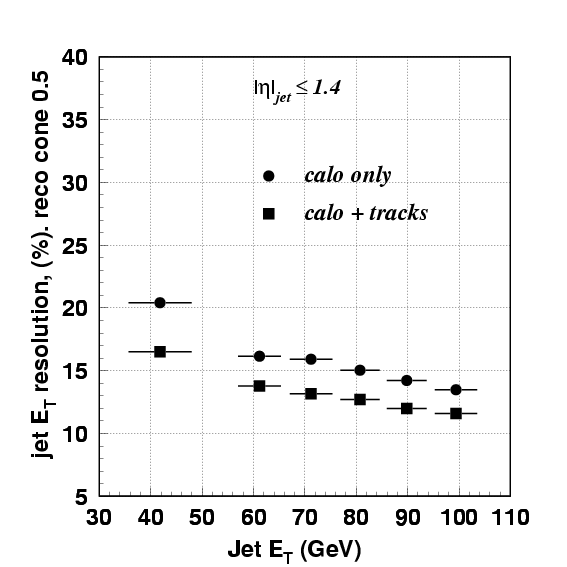} 
    \includegraphics[width=0.5\textwidth]{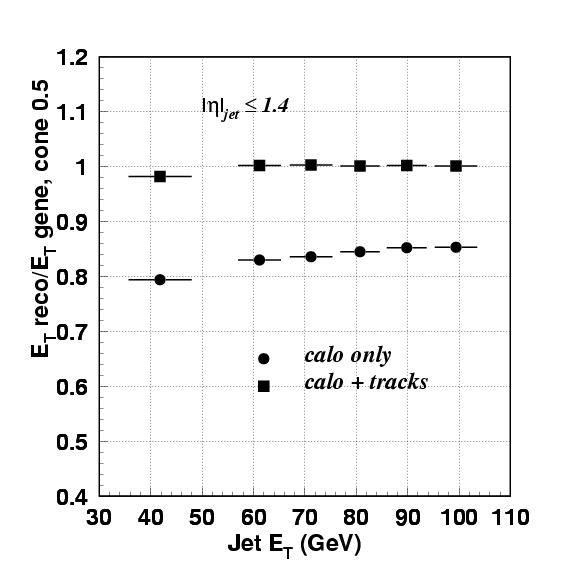} 
  }
\caption {\small  
  {\em Jet transverse energy resolution (left) and reconstructed jet transverse energy (right) as a function 
    of the generated jet transverse energy. Jets with $0<|\eta|<1.4$ (barrel) from a sample with low luminosity pile-up;
    reconstruction with calorimeter only (close circles), subtraction procedure of expected responses using library of 
    responses and out-of-cone tracks (close squares).
  }
}
\label{fig:CMSresultsBarrel}
\end{center}
\vspace{-1cm}
\end{figure}

The algorithm performance has been tested comparing Montecarlo\footnote{Montecarlo jets are reconstructed 
implementing the same jet-finding algorithm than for reconstructed jet with tracks information from the MC truth} 
and reconstructed jets, with and without EF applied. 
Di-jet events with $P_T$ between 80 and 120 GeV/c were generated with PYTHIA and fully simulated and reconstructed 
inside the CMS software framework \cite{OSCAR} \cite{ORCA}. Effects due to low luminosity 
($L = 2 \times 10^{33} cm^{-2}s^{-1}$) pile-up have been included. 
The resolution and the reconstructed jet energy fraction are shown for jets generated with
$|\eta| < 1.4 $ in fig.~\ref{fig:CMSresultsBarrel}. When the EF algorithm is  applied, the reconstructed 
jet energy fraction for 40 GeV generated jets increases form 0.80 to 0.99 and the same fraction  for 100 GeV jets 
increases from 0.85 to 1.00. The resolution improves by about 20-25\% as a result of adding the out-of-cone tracks.

\begin{figure}[tb]
\begin{center}
  \centerline{
    \includegraphics[width=0.5\textwidth]{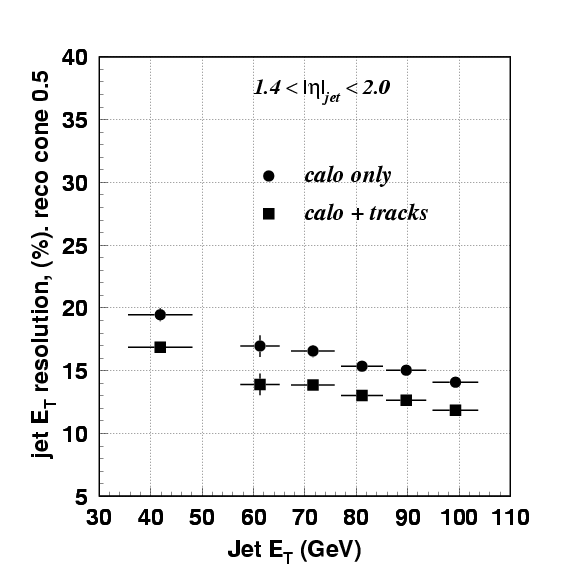} 
    \includegraphics[width=0.5\textwidth]{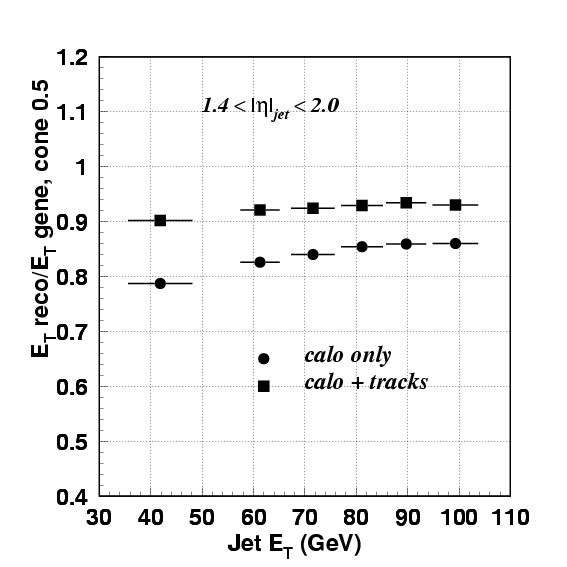} 
  }
\caption {\small  
  {\em Jet transverse energy resolution (left) and reconstructed jet transverse energy (right) as a function 
    of the generated jet transverse energy. Jets with $1.4<|\eta|<2.0$ (endcap) from a sample with low luminosity pile-up;
    reconstruction with calorimeter only (close circles), subtraction procedure of expected responses using library of 
    responses and out-of-cone tracks (close squares).
  }
}
\label{fig:CMSresultsEndcap}
\end{center}
\vspace{-1cm}
\end{figure}

In the endcap region (figs.~\ref{fig:CMSresultsEndcap}), jets with the same $\rm E_{\rm T}$ as in the 
barrel are more energetic and, in addition, the tracking efficiency is smaller in the endcap 
than in the barrel. Therefore, the tracker information is not relevant in the endcap above 80-90 GeV and 
is less rewarding for lower $\rm E_{\rm T}$ jets than in the barrel. Besides jets in the endcap are more affected
by pile-up than in the barrel.

\begin{figure}[tb]
\begin{center}
  \centerline{
    \includegraphics[width=0.5\textwidth]{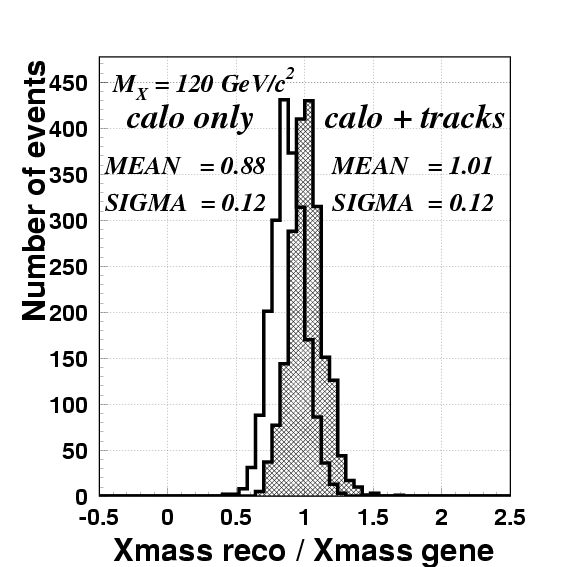} 
  }
\caption {\small  
  {\em Ratio of the reconstructed to the generated X mass with calorimeters only (empty histogram)
    and with calorimeter + tracks corrections (hatched histogram).
  }
}
\label{fig:CMSresonance}
\end{center}
\vspace{-1cm}
\end{figure}

The performance of the EF algorithm has been tested also on events with a 120 ${\rm GeV/c^2}$ X object 
decaying into light quarks with initial and final state radiation switched on.
The X mass is reconstructed from the two leading jets that are within $\rm R=0.5$ of the direction of the 
primary partons. The ratio of the X mass reconstructed to the X mass generated for calorimetry jets and 
calorimeter-plus-tracker jets is shown in Fig.~\ref{fig:CMSresonance}. 
The di-jet mass is restored with a systematic shift of about 1\% and the resolution is improved by 10\%. 
The ratio of the reconstructed to the generated X mass is 0.88 before corrections with tracks and 1.01 after 
corrections.

An improvement of the simple algorithm described above makes use of two cones with different 
size \cite{Santocchia}: a smaller one for the jet-finding step and a larger one for the out-of-cone 
charged tracks recovery step. The idea of two different cones is suggested by the fact that neutral tracks
release their energy basically along the jet direction , since they are not deflected by the magnetic field.
Therefore a small cone is sufficient to recover most of the neutral deposits in the 
calorimeter; the charged contribution to the jet energy is subsequently recovered by the 
tracker using a larger size cone. In this way, for the same amount of charged and neutral 
jet fragments recovered, the contamination by neutral deposit which do not belong to the jet 
(pile-up, underlying event, etc.) can be reduced.



\addtocounter{chapter}{1}
\renewcommand*{\eg}{\textit{e.g.},\ }
\renewcommand*{\ie}{\textit{i.e.},\ }
\newcommand*{\etal}{\textit{et al.\ }}
\renewcommand*{\gev}{\,\textrm{GeV}}
\newcommand*{\gevc}{\,\textrm{GeV/c}}
\newcommand*{\mevc}{\,\textrm{MeV/c}}
\renewcommand*{\tev}{\,\textrm{TeV}}
\renewcommand*{\pt}{$p_T$}
\newcommand*{\bigpt}{$P_T$}
\newcommand*{\UE}{``underlying event"}
\newcommand*{\BBR}{``beam-beam remnants"}
\newcommand*{\MB}{``min-bias"}
\newcommand*{\TR}{``transverse"}
\newcommand*{\TOW}{``toward"}
\newcommand*{\AWA}{``away"}
\newcommand*{\tmax}{``transMAX"}
\newcommand*{\tmin}{``transMIN"}
\newcommand*{\tdif}{``transDIF"}
\newcommand*{\LJ}{``leading jet"}
\newcommand*{\BB}{``back-to-back"}
\newcommand*{\Tone}{``transverse $1$"}
\newcommand*{\etone}{$E_T({\rm jet}\#1)$}
\newcommand*{\ptone}{$P_T({\rm jet}\#1)$}
\newcommand*{\Ttwo}{``transverse $2$"}
\newcommand*{\Jone}{jet\#$1$}
\newcommand*{\Jtwo}{jet\#$2$}
\newcommand*{\Jthree}{jet\#$3$}
\newcommand*{\Jfour}{jet\#$4$}
\newcommand*{\CJone}{chgjet$1$}
\newcommand*{\ptcj}{$P_T({\rm chgjet}1)$}
\newcommand*{\pthard}{$p_T({\rm hard})$}
\newcommand*{\ptmax}{$PT{\rm max}$}
\newcommand*{\ptmaxT}{$PT{\rm max}T$}
\newcommand*{\ptsum}{$PT_{sum}$}
\newcommand*{\etsum}{$ET{\rm sum}$}
\newcommand*{\pthcut}{$p_T\!>\!0.9\,{\rm GeV/c}$}
\newcommand*{\ptlcut}{$p_T\!>\!0.5\,{\rm GeV/c}$}
\newcommand*{\etcut}{$E_T\!>\!0.1\,{\rm GeV}$}
\newcommand*{\etacut}{$|\eta|\!<\!2$}
\newcommand*{\aveN}{$\langle\!N_{\rm chg}\!\rangle$}
\newcommand*{\avept}{$\langle\!p_T\!\rangle$}
\newcommand*{\etaphi}{$\eta$-$\phi$}
\newcommand*{\delphi}{$\Delta\phi$}
\newcommand*{\absdelphi}{$|\Delta\phi|$}
\newcommand*{\delphicut}{$|\Delta\phi|>150^\circ$}
\newcommand*{\ndenpt}{$dN_{chg}/{d\eta}{d\phi}dp_T$}
\newcommand*{\nden}{$dN_{chg}/{d\phi}{d\eta}$}
\newcommand*{\ptden}{$dPT_{sum}/{d\phi}{d\eta}$}
\newcommand*{\etden}{$dET_{sum}/{d\phi}{d\eta}$}
\newcommand*{\mpair}{$\mu^+\mu^-$}
%


\mchapter{Minimum Bias, Underlying Events and Multiple Interactions}
{ Authors: Filippo~Ambroglini, Paolo~Bartalini, Livio~Fan\`o, Lucia~Garbini,
Daniele~Treleani}\label{chap:MBandUE}\label{chap:MBaUE}

\noindent
{\it Revisors: Paolo~Nason}

\section{Introduction}

This chapter is sub-divided in four sections.
The next section gives the definition of \MB\ and \UE.
A brief review of the current status of the phenomenological studies and theoretical models is given in section~\ref{sec:models}.
The measurement plan at the LHC is described in section~\ref{sec:plan}, where the relevant observables sensitive to the examined processes are introduced by comparing different tunings of the most popular Monte Carlo models.

\section{Definition of the physics processes \label{sec:def}}

Events collected with a trigger that is not very restrictive are referred
to as minimum bias events (MB).  
The total proton-proton cross section is the sum of the elastic cross
section and the inelastic cross section.  
The inelastic cross section receives contributions from single and double
diffraction.  The remainder of the inelastic 
cross section is referred to as the ``hard core" component.  Minimum bias
events typically contain some single and 
double diffraction as well as most of the ``hard core"  component of the
inelastic cross section.  The ``hard core" component 
does not always correspond to a ``hard scattering".  Quite often the beam
and target hadrons ooze through each other 
and fall apart without any ``hard" collisions occurring in the event.  At
the Tevatron about $1\%$ of min-bias events contain a 
jet with $10\gev$ transverse energy.  At the LHC we expect the fraction of
MB events with a $10\gev$ jet to increase by more than a 
factor of $10$ from the Tevatron to about $12\%$.  We expect about $1\%$
of MB events at the LHC to contain a $20\gev$ jet.  
Understanding and modeling the jet structure of MB events is crucial at
the LHC because of the large amount of pile-up expected.

From an experimental point of view, in a hadron-hadron interaction with jets in the final state, the \UE\ is all the activity accompanying the 2 hard scattered outgoing jets.
It is impossible to separate these two components due to the lack of knowledge in modeling the underlying jet structure. Anyway one can use the topological structure of hadron-hadron collisions to define physics observables that are mostly sensitive to the underlying activity. The typical approach is to rely on particle and energy densities in \etaphi\ regions that are well separated with respect to the high $P_T$ objects (for example jets).
In shower Monte Carlo model, the \UE\ is a component of the process simulation that acts at the end of the showering and before the hadronization, in order to complete the process description taking into account soft components (hadronic remnants and multiple interaction). 

Huge progress in the phenomenological study of the underlying event in jet events has been achieved 
by the CDF experiment at the Tevatron~\cite{Affolder:2001xt,Acosta:2004wq}, using the multiplicity and transverse momentum spectra of charged tracks in different regions of the azimuth-pseudorapidity space, defined with respect to the direction of the leading jet. 
Regions that receive energy flow contributions mostly by the underlying event have been identified.
The CDF UE analysis showed that the density of particles in the UE in jet events is about  a factor of two larger than the density of particles in a typical Minimum Bias (MB) collision.  
This effect, referred to as "pedestal effect", is well reproduced only by varying impact parameters 
models with correlated parton-parton interactions. Simpler models seem to be ruled out. 
In general the most successful models predict an even more relevant difference between the MB and the UE activities at the LHC, with deep consequences on lepton and photon isolation, jet energy calibrations, etc.

\section{The QCD models and the Multiple Parton Interaction concept~\label{sec:models}}

In the years '80, the evidence for Multiple Parton Interaction (MPI) phenomena in the high-P$_{T}$ phenomenology of hadron colliders~\cite{afs_direct,ua2_direct,cdf_direct} suggested the extension of the same perturbative picture to the soft regime, giving rise to the first implementation of the MPI processes in a QCD Monte Carlo model~\cite{Sjostrand:1986ep}.

These models turned out to be particularly adequate to describe the MB
and the UE physics. In particular, the pedestal effect mentioned in
sec. 1.2 can be explained partly\footnote{A second important effect
that can contribute to the pedestal is the increase in initial state
radiation associated to the presence of a hard scattering} as an
increased probability of multiple partonic interaction in case a hard
collision has taken place (a hard scattering is more likely to be
present in a small impact parameter collision, which thus implies more
additional parton-parton interactions).

Examples of MPI models are implemented in the general purpose simulation programs 
PYTHIA~\cite{Sjostrand:2000wi}, HERWIG/JIMMY~\cite{Butterworth:1996zw} and SHERPA~\cite{Gleisberg:2003xi}. 
Other successful descriptions of UE and MB at hadron colliders are 
achieved by alternative approaches like PHOJET~\cite{Bopp:1998rc}, which was designed to describe rapidity gaps and diffractive
physics (relying on both perturbative QCD and Dual Parton Models). The purely phenomenological UE and MB description available in HERWIG~\cite{Corcella:2000bw} provides a very useful reference of a model not implementing multiple interactions.
The most recent PYTHIA version~\cite{Sjostrand:2004ef} adopts an optional alternative description of the colliding partons in terms of  correlated multi-parton distribution functions of flavours, colors and longitudinal momenta.

All these models have to be tested and tuned at the LHC, in particular for what concerns the energy dependent parameters.

\subsection{The SPS and Tevatron legacies}

The QCD models considered here are three different PYTHIA $6.4$ Tunes (with 2 different MPI models) and HERWIG (without MPI) as reference. The relevant parameters of the different PYTHIA Tunes are summarized in table~\ref{table_uetunes}.  

The main parameter of the PYTHIA tunes, $P_{T_{min}}$, is the minimum transverse momentum of the
parton-parton collisions; it effectively controls the average number of parton-parton interactions, hence the average particle multiplicity.
The studies reported in~\cite{lhc_workshop}, considering a homogeneous sample of average charged multiplicity measurements at six different center-of-mass energies ($\sqrt{s} =$ $50$, $200$, $546$, $630$, $900$ and $1800$~GeV) in the pseudo-rapidity region $|\eta|< 0.25$~\cite{average_ua5, average_cdf}, show that the power law expressed in the following Equation:

\begin{equation}
\displaystyle
P_{T_{min}}^{(\ \rm s \ )} ~ = ~ P_{T_{min}}^{(\ \rm s' \ )}
\left(\frac{s}{s'}\right)^{\epsilon}
\label{eq:ptcut_funct}
\end{equation}

\noindent holds for values of $\epsilon$ between $\simeq 0.08$ and $\simeq 0.10$ if post-HERA parton distribution functions are used. 

\begin{figure}[htbp]
\centering
\begin{minipage}[c]{.40\textwidth}
\centering
\includegraphics[scale=0.33, angle=90]{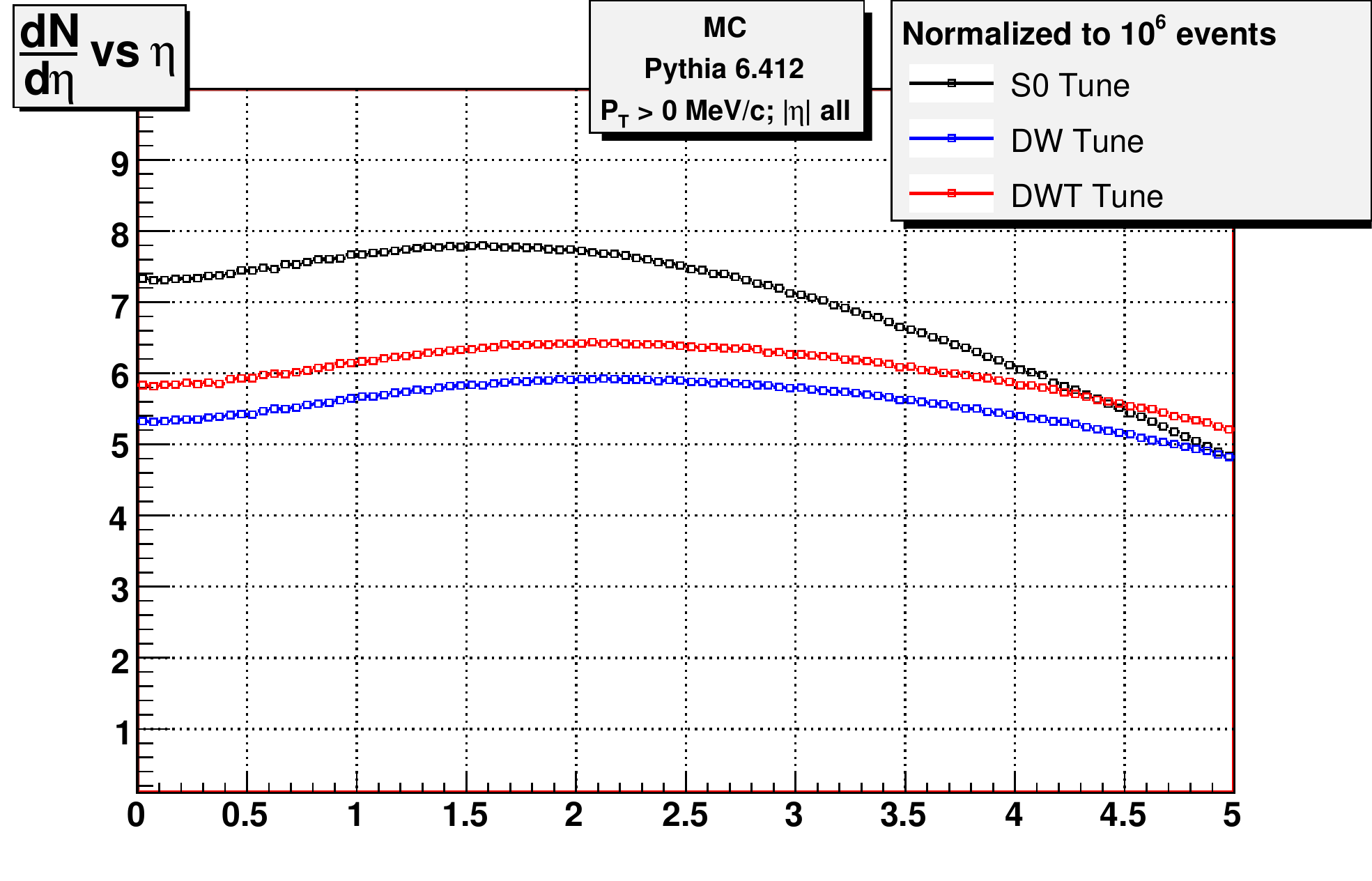}
\caption{Charged particle density distribution, $dN_{chg}/d\eta$,  for Minimum Bias events at LHC condition
with \textsc{Pythia}6.412 and Tune DW, DWT and S0.}
\label{fig:mbtunedndeta}
\end{minipage}
\hspace{9mm}
\begin{minipage}[c]{.40\textwidth}
\centering
\vspace{-5mm}
\includegraphics[scale=0.33, angle=90]{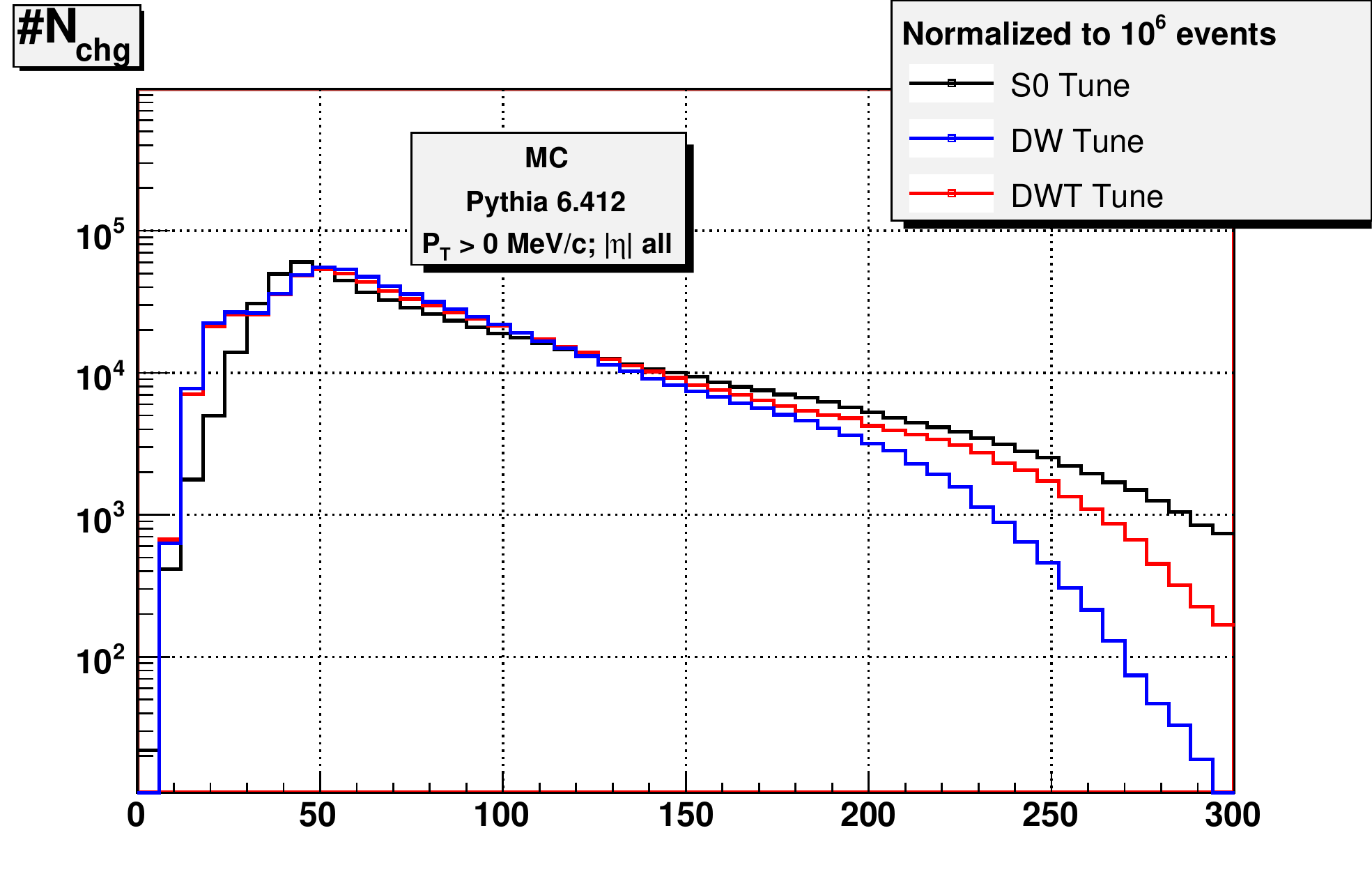}
\caption{Charged particle distribution for Minimum Bias events at LHC condition 
with \textsc{Pythia}6.412 and Tune DW, DWT and S0.}
\label{fig:mbtunencgh}
\end{minipage}
\end{figure}

All the considered PYTHIA tunes adopt varying impact parameter models with a continuous turn-off
of the cross section at $P_{T_{min}}$ and hadronic matter in the colliding hadrons described by two concentric Gaussian di\-stri\-bu\-tions\cite{Sjostrand:2006za}. These models were initially developed to reproduce the UA5 MB charged multiplicity~\cite{distr_ua5}. 
The variations of the impact parameter introduce correlations between the MPI, giving rise to a charged multiplicity shape which is basically the convolution of several Poissonians. This can be clearly seen in Fig.~\ref{fig:mbtunencgh}.

All the considered PYTHIA tunes describe the basic Tevatron UE phenomenology in a reasonable way.
One of the PYTHIA models is Tune DW~\cite{Acosta:2006bp}, a tune by R. Field which is similar to Tune A~\cite{Field:2005qt}, reproducing also the CDF Run 1 Z-boson transverse momentum distribution~\cite{Abe:1991pu}
Tune DWT~\cite{Acosta:2006bp} is identical to Tune DW at the Tevatron (\ie$1.96\tev$), but uses the same MPI energy dependence parameter as the ATLAS tune~\cite{Buttar:2004iy} ($\epsilon = 0.08$).
Tune S0~\cite{Skands:2007zg} also adopts the same energy dependence parameter as the ATLAS tune, however, In contrast to DW and DWT, it does adopt the new PYTHIA multiple interaction framework.

\begin{table}[!hbp]\footnotesize
\begin{center} 
\begin{tabular}{|l|c|c|c|c|c|}
\hline\hline  
Parameter (\textsc{Pythia} v.6412+) & A & ATLAS &DW & DWT & S0 \\
\hline
UE model MSTP(81)& 1& 1& 1& 1& 21\\
UE infrared regularisation scale PARP(82)& 2.0& 1.8 & 1.9& 1.9409& 1.85\\
UE scaling power with $\sqrt{s}$ PARP(90)& 0.25 & 0.16 & 0.25 & 0.16 & 0.16 \\
UE hadron transverse mass distribution MSTP(82) & 4 & 4 & 4 & 4 & 5\\
UE parameter 1 PARP(83) & 0.5 & 0.5& 0.5 & 0.5 & 1.6 \\
UE parameter 2 PARP(84) & 0.4 & 0.5 & 0.4 & 0.4 & n/a \\
UE total gg fraction PARP(86) & 0.95 & 0.66 & 1.0 & 1.0 & n/a \\
\hline
ISR infrared cutoff PARP(62) & 1.0 & 1.0 & 1.25 & 1.25&( = PARP(82) )\\
ISR renormalisation scale prefactor PARP(64) & 1.0 & 1.0 & 0.2 & 0.2& 1.0\\
ISR $Q^{2}_{max}$ factor PARP(67)& 4.0 & 1.0 &2.5 & 2.5 & n/a\\
ISR infrared regularisation scheme MSTP(70)& n/a & n/a & n/a & n/a & 2 \\
ISR FSR off ISR scheme MSTP(72) & n/a & n/a & n/a & n/a &  0 \\
\hline
FSR model MSTJ(41) & 2& 2& 2& 2 &($p_{t}-ordered$)\\
FSR $\Lambda_{QCD}$ PARJ(81)& 0.29 & 0.29 & 0.29& 0.29& 0.14\\
\hline
BR colour scheme MSTP(89) & n/a  & n/a  & n/a  & n/a  &1\\
BR composite $x$ enhancement factor PARP(79)& n/a & n/a & n/a & n/a & 2 \\
BR primordial $k_{T}$ width $<|k_{T}|>$ PARP(91)& 1.0 & 1.0 &2.1 & 2.1 & n/a \\
BR primordial $k_{T}$ UV cutoff PARP(93)& 5.0 & 5.0 & 15.0 & 15.0 &5.0\\
\hline
CR model MSTP(95) & n/a & n/a & n/a & n/a & 6 \\
CR strength $\xi_{R}$ PARP(78)& n/a& n/a & n/a & n/a & 0.2 \\
CR gg fraction (old model) PARP(85) & 0.9 & 0.33 & 1.0 & 1.0& n/a \\
\hline \hline  
\end{tabular}
\end{center}
\caption{Set of parameters defining the different versions of the PYTHIA 6.4 models adopted in this study. In all the configurations, the CTEQ5L parton distribution functions are considered.
The parameters are subdivided into five main categories: 
UE (underlying event), ISR (initial state radiation), FSR (final state radiation), 
BR (beam remnants), and CR (colour reconnections). 
The UE reference energy for all models is PARP(89)=1800GeV. 
GeV unit is adopted if applicable.\label{table_uetunes}}
\end{table}
%
%
\begin{figure}[htbp]
\centering
\begin{minipage}[c]{.40\textwidth}
\centering
\includegraphics[scale=0.33, angle=90]{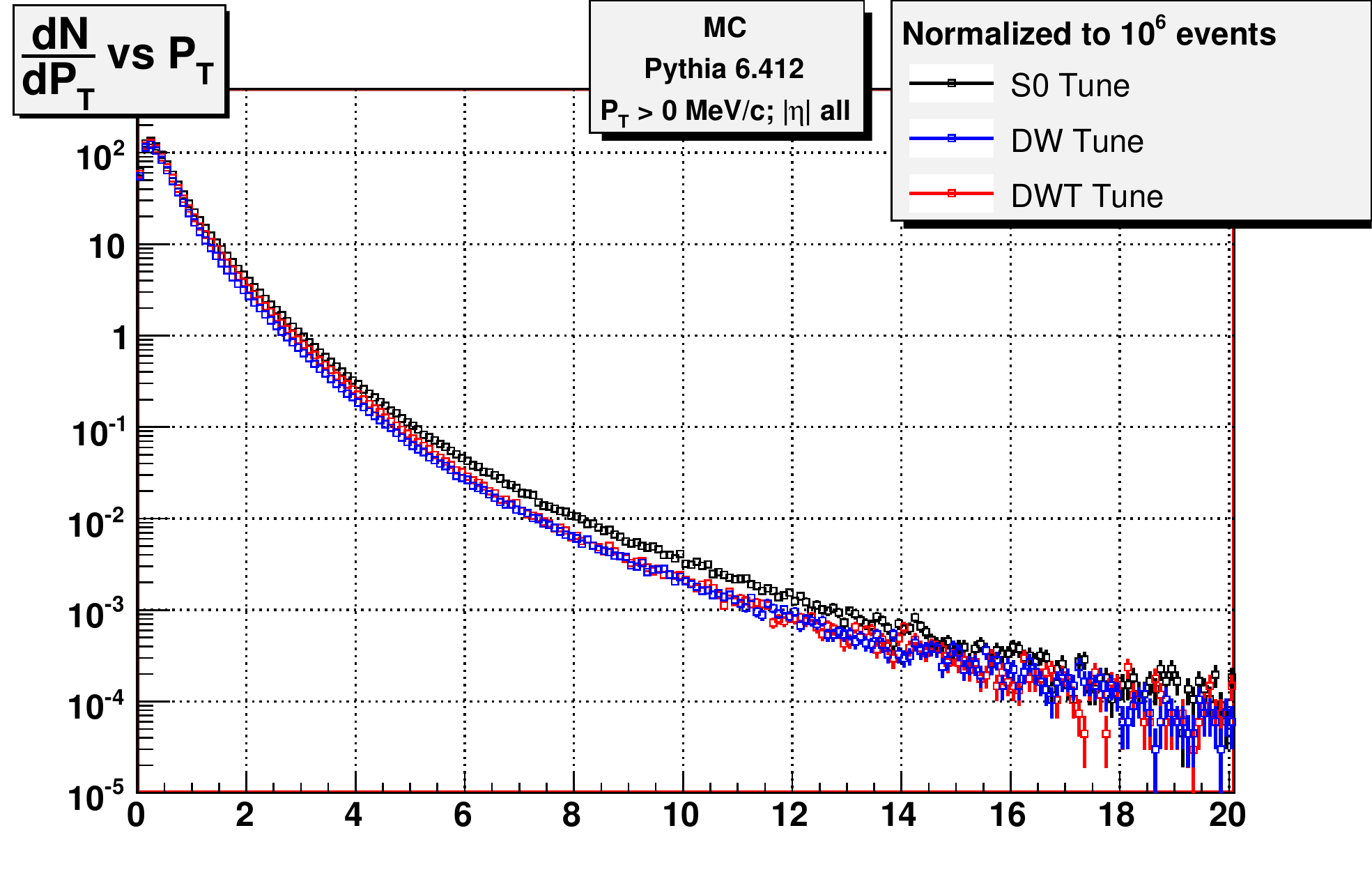}
\caption{Charged particle density distribution, $dN_{chg}/dP_{t}$, for Minimum Bias events at LHC condition 
with \textsc{Pythia}6.412 and Tune DW, DWT and S0.}
\label{fig:mbtunedndpt}
\end{minipage}
\hspace{9mm}
\begin{minipage}[c]{.40\textwidth}
\centering
\includegraphics[scale=0.33, angle=90]{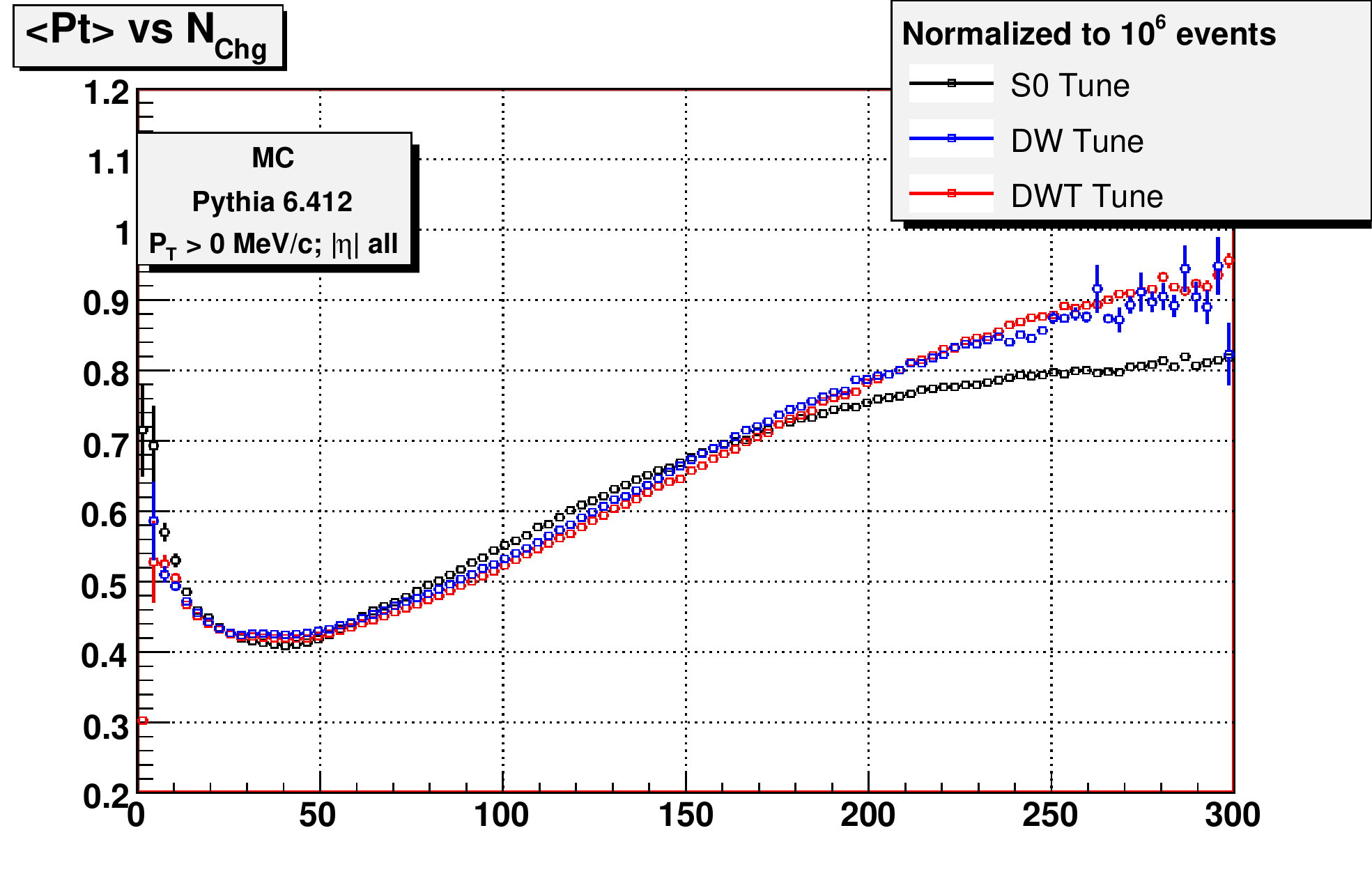}
\caption{Average $P_{t}$ as a function of chareg multeplicity, for Minimum Bias events at LHC condition
with \textsc{Pythia}6.412 and Tune DW, DWT and S0.}
\label{fig:mbtuneavgpt}
\end{minipage}
\end{figure}

\section{The Measurement plan at the LHC~\label{sec:plan}}

\subsection{The Basic Minimum Bias Observables}

One of the first results of LHC will be the measurement of the charged multiplicity and \pt\ spectrum in proton-proton collisions at $\sqrt{s} = 14$~TeV~\cite{ferenc}. 
The predictions of the considered PYTHIA tunes for these MB observables are reported in Fig.~\ref{fig:mbtunedndeta} and Fig.~\ref{fig:mbtunedndpt} respectively.

In Reference~\cite{average_cdf} the energy dependence of $dN_{ch}/d\eta$ at $\eta=0$ 
is fitted to older data using a linear and quadratic functions of $\ln(s)$. 
Using these fits to extrapolate at LHC energy would predict $dN_{ch}/d\eta =
6.11 \pm 0.29$ at $\eta=0$ (to be compared with the predictions of the models given by the intercept of the y axis and the curves of Fig.~\ref{fig:mbtunedndeta}).
\subsection{The Underlying Event as Observed in Charged Jet Events}

\begin{figure}[!Hhtb]
\begin{center}
\includegraphics[scale=0.7]{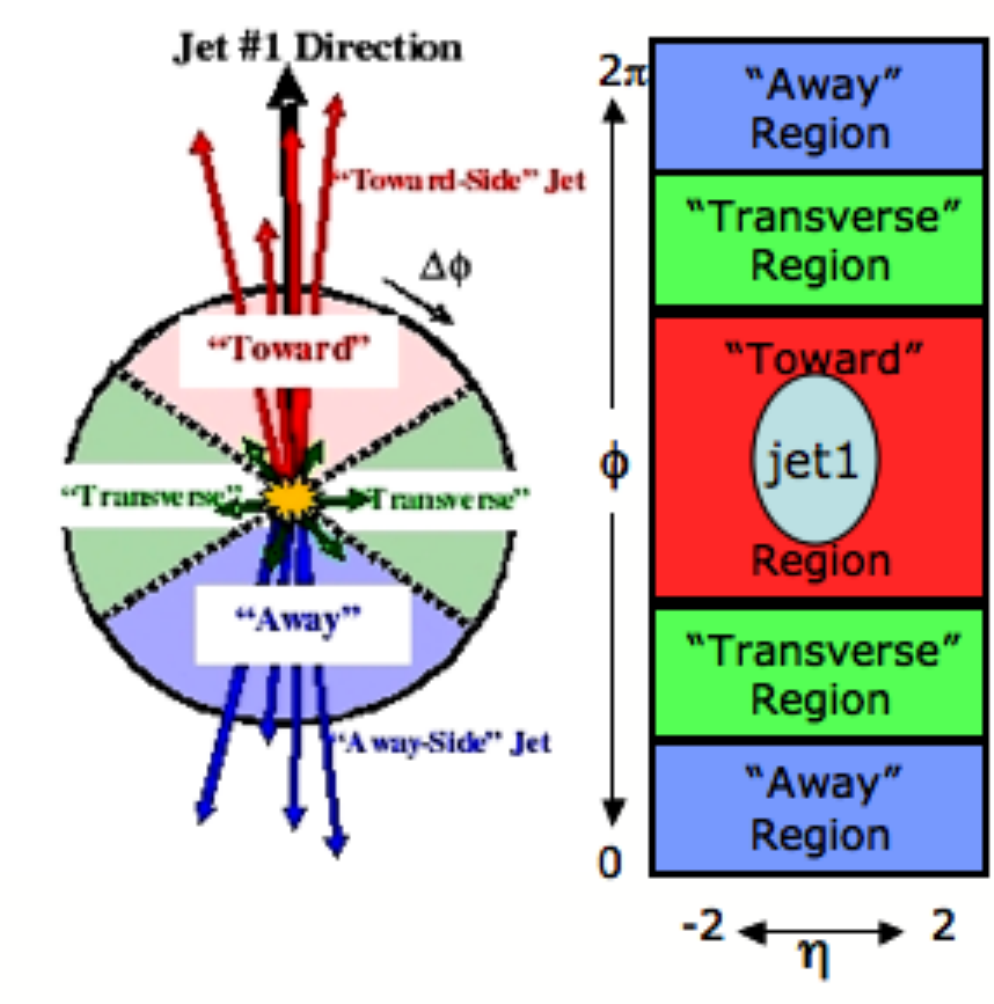}
\caption{\footnotesize 
Illustration of correlations in azimuthal angle $\phi$ relative to the direction of the 
leading charged particle jet (with cone size $R=0.5$) in the event, \CJone.  The angle 
$\Delta\phi=\phi-\phi_{\rm chgjet1}$ is the relative azimuthal angle between charged 
particles and the direction of \CJone.  The ``transverse" region is defined 
by  $60^\circ<|\Delta\phi|< 120^\circ$ and \etacut.  We examine charged particles in the 
range \etacut\ with \ptlcut\ or \pthcut.
\label{RDF_PTDR_fig1}}
\end{center}
\end{figure}

One can use the topological structure of hadron-hadron collisions to study the UE.  Furthermore, this 
can be done by looking only at the outgoing charged particles \cite{Affolder:2001xt}.  
Jets are constructed from the charged particles  using a simple clustering algorithm and then the direction of the leading charged particle jet is used  to isolate 
regions of the \etaphi\ space that are sensitive to the UE. 
As illustrated in Fig.~\ref{RDF_PTDR_fig1}, the direction of the 
leading charged particle jet, \CJone, is used to define correlations in the azimuthal angle, \delphi.  The 
angle $\Delta\phi=\phi-\phi_{\rm chgjet1}$ is the relative azimuthal angle between a charged particle and the direction of \CJone.  

The charged jet energy provide an indication of the energy scale of the event. 
Adopting the charged does allow to investigate the very low energy scale region (down to $P_{T} \rightarrow 0$ \gevc) which is not accessible to the calorimetric jets.
In other words, the charged jet does provide a better understanding of the systematic effects in the low $P_{T}$ limit, that can be interpreted in terms of very well understood quantities like the tracking efficiency and fake rate.
Another big advantage of the measurement relying on the charged tracks is its intrinsic insensitiveness to the pile up effect as only the charged particles coming from the primary vertex are retained in the computation of the UE observables.

The \TR\ region is almost perpendicular to the plane of the hard $2$-to-$2$ scattering and is therefore very 
sensitive to the UE.  We restrict ourselves to charged particles in the central 
region \etacut\ and consider two \pt\ thresholds, the nominal CMS cut \pthcut\ and a lower threshold with \ptlcut. 

Ultimately we would like to disentangle the hard initial and final state radiation (\ie multijet production) from the beam-beam remnants and MPI components.  This can be done by separating the 
various jet topologies.  First one considers events with at least one jet and uses the leading jet direction to define the \TR\ region 
(referred to as \LJ\ events).  Of course some of these \LJ\ events contain multijets that contribute to the activity in the \TR\ region.  
Next one considers \BB\ dijet events which are a subset of the \LJ\ events.  The \TR\ region for the \BB\ dijet events contains much 
less hard initial and final state radiation and by comparing the two classes of events one can learn about  gluon radiation as well as the beam-beam remnants and the MPI component.  In this note we will only discuss the \LJ\ events.

\begin{figure}[!Hhtb]
\begin{center}
\includegraphics[scale=0.3, angle=90]{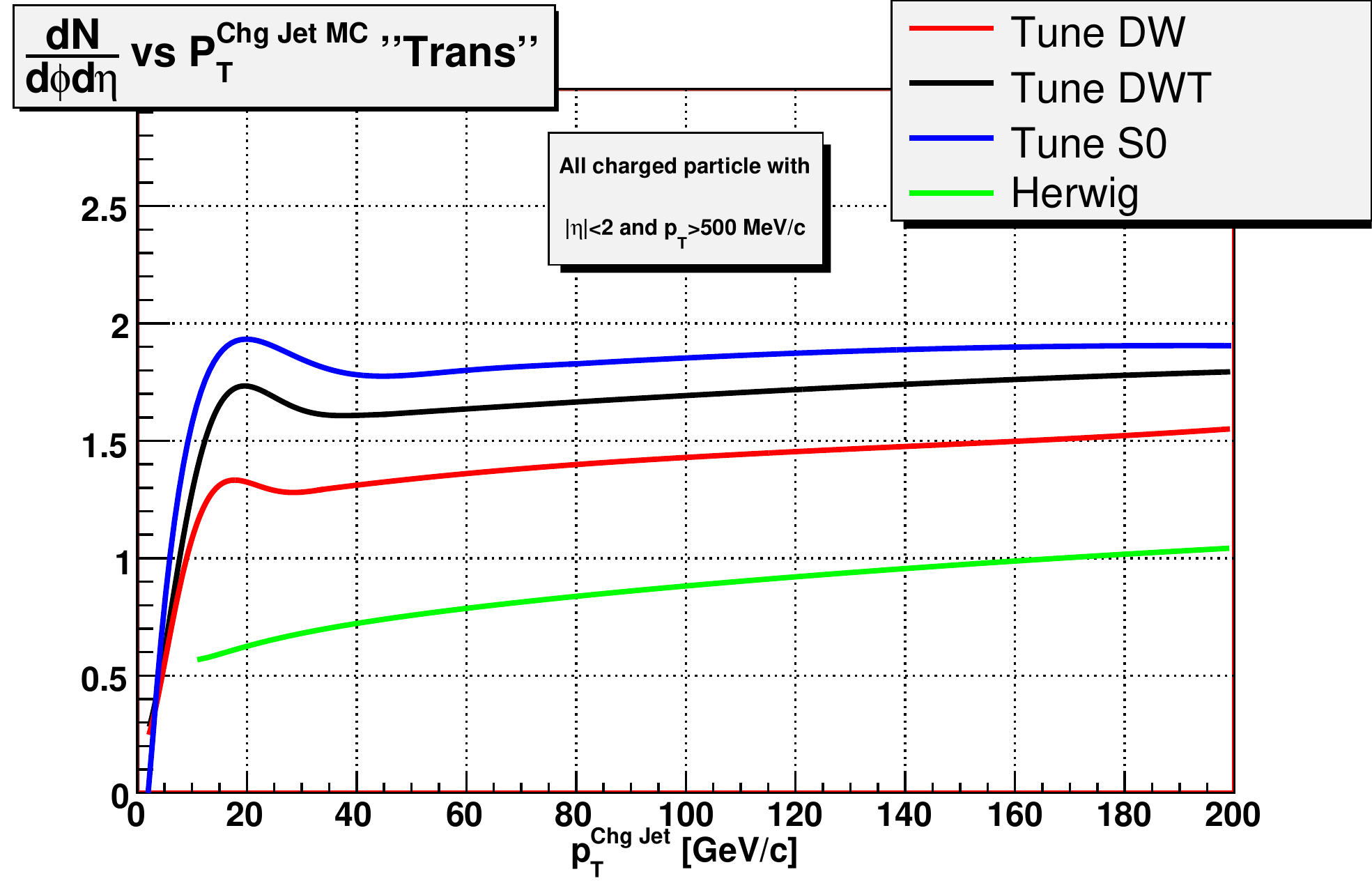}
\includegraphics[scale=0.3, angle=90]{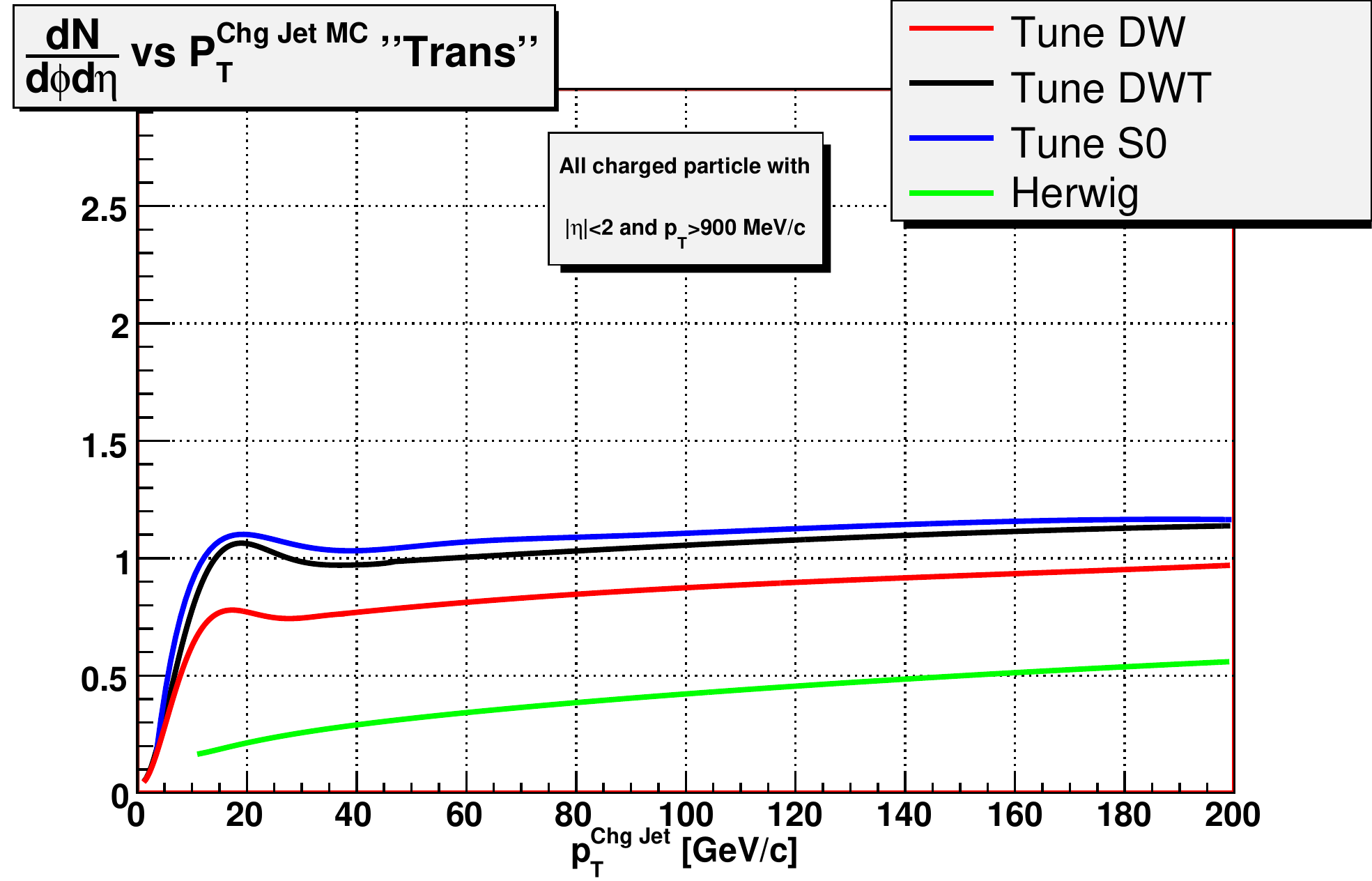}
\caption{\footnotesize 
QCD Monte-Carlo models predictions for charged particle jet production at $14\tev$.  
Observables in the \TR\ region. Average density of charged particles, \nden, with \etacut\ and \ptlcut\ (\textit{left}) or \pthcut\ (\textit{right}) versus the transverse momentum of the leading charged particle jet. 
The QCD models are HERWIG (without MPI) and three versions of PYTHIA $6.4$ 
(with MPI).
\label{RDF_PTDR_fig2}}
\end{center}
\end{figure}

\begin{figure}[!Hhtb]
\begin{center}
\includegraphics[scale=0.3, angle=90]{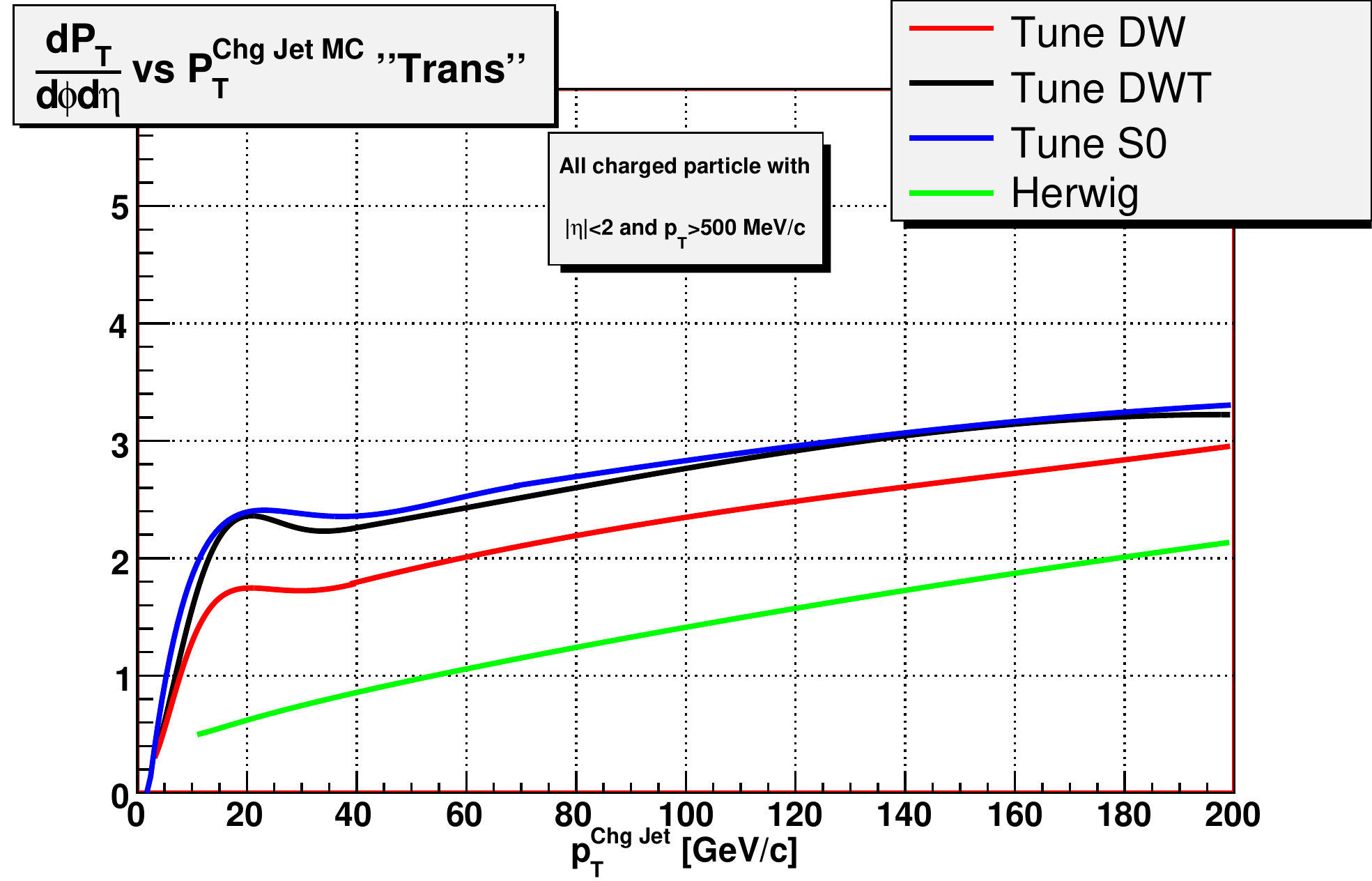}
\includegraphics[scale=0.3, angle=90]{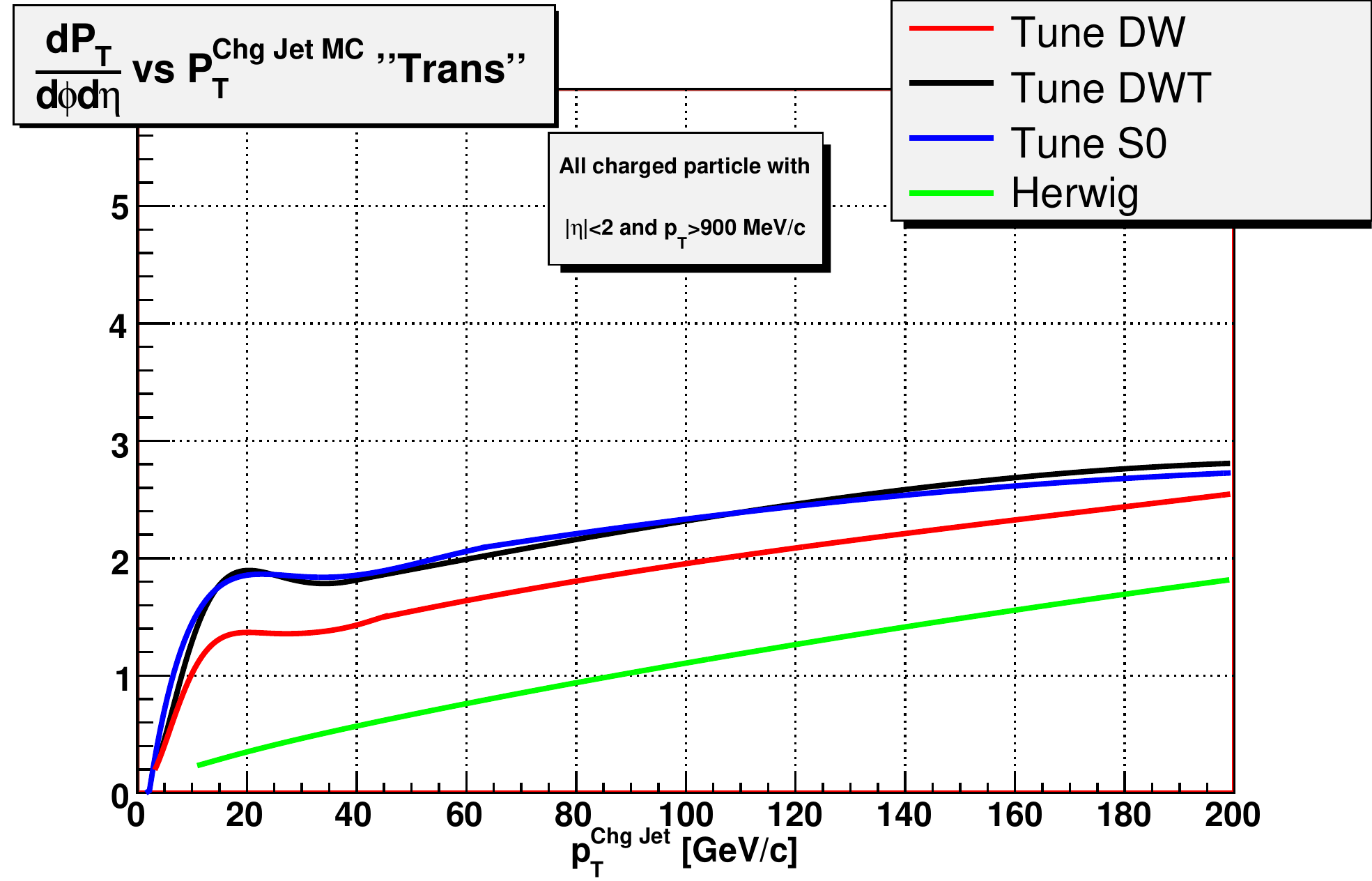}
\caption{\footnotesize 
QCD Monte-Carlo models predictions for charged particle jet production at $14\tev$.  
\TR\ region: average charged \ptsum\ density, \ptden, with \etacut\ and \ptlcut\ (\textit{left}) or \pthcut\ (\textit{right}) versus the transverse momentum of the leading charged particle jet.
The QCD models are HERWIG (without MPI) and three versions of PYTHIA $6.4$ 
(with MPI).
\label{RDF_PTDR_fig3}}
\end{center}
\end{figure}

The charged jet \pt\ range $0$ to $200\gevc$ shown in Figs.~\ref{RDF_PTDR_fig2} and~\ref{RDF_PTDR_fig3} is quite interesting.  
The three versions of PYTHIA (with MPI) behave much differently than HERWIG (without MPI). 
Due to the MPI the 
PYTHIA tunes rise rapidly and then reach an approximately flat 
``plateau" region at $P_T({\rm chgjet}1)\approx 20\gevc$.  
Then at $P_T({\rm chgjet}1)\approx 50\gevc$ they begin to rise again due to 
initial and final state radiation which increases as the $Q^2$ scale 
of the hard scattering increases. The rise is more evident for the high 
\pt\ threshold \pthcut. HERWIG has considerably fewer particles in the 
\TR\ region and predicts a steady rise over this region resulting from 
initial and final state radiation.  

Due to higher effective cut off in the $Q^2$ of the MPI, the Tune DW does achieve predictions which are around 25\% below with respect to the DWT and S0 for what concerns both the particle and energy densities. 
Even with a modest statistics, at the LHC we will be able to distinguish between these two different trends reflecting different choices of the energy dependent parameters in multiple interactions.

The S0 tune predicts a larger 
charged particle density in the \TR\ region than Tune DWT for 
\ptlcut.  However, the S0 and the DWT tunes have similar charged
particle densities in the \TR\ region for \pthcut. This is because the 
S0 tune has a slightly ``softer" charged particle \pt\ distribution than Tune DWT.  

S0 and DWT have very similar energy densities  in the \TR\ region, however there are interesting differences in shape: in particular S0 predicts a steeper rise with a flatter plateau at $P_T({\rm chgjet}1)\approx 20\gevc$ for both \ptlcut and \pthcut.  


Figures~\ref{RDF_PTDR_fig2} and Fig.~\ref{RDF_PTDR_fig3} show the QCD Monte-Carlo models predictions for the average density of charged particles, \nden, 
and the average charged \ptsum\ density, \ptden, respectively, in the \TR\ region for \etacut\ with \ptlcut\ and \pthcut\ versus the 
transverse momentum of the leading charged particle jet.  
The charged particle density is constructed by dividing the 
average number of charged particles per event by the area in \etaphi\ space (in this case $4\pi/3$).  The charged \ptsum\ density is 
the average scalar \pt\ sum of charged particles per event divided by the area in \etaphi\ space.  Working with densities allows 
one to compare regions of \etaphi\ space with different areas. 

\begin{figure}[!Hhtb]
\begin{center}
\includegraphics[scale=0.3, angle=90]{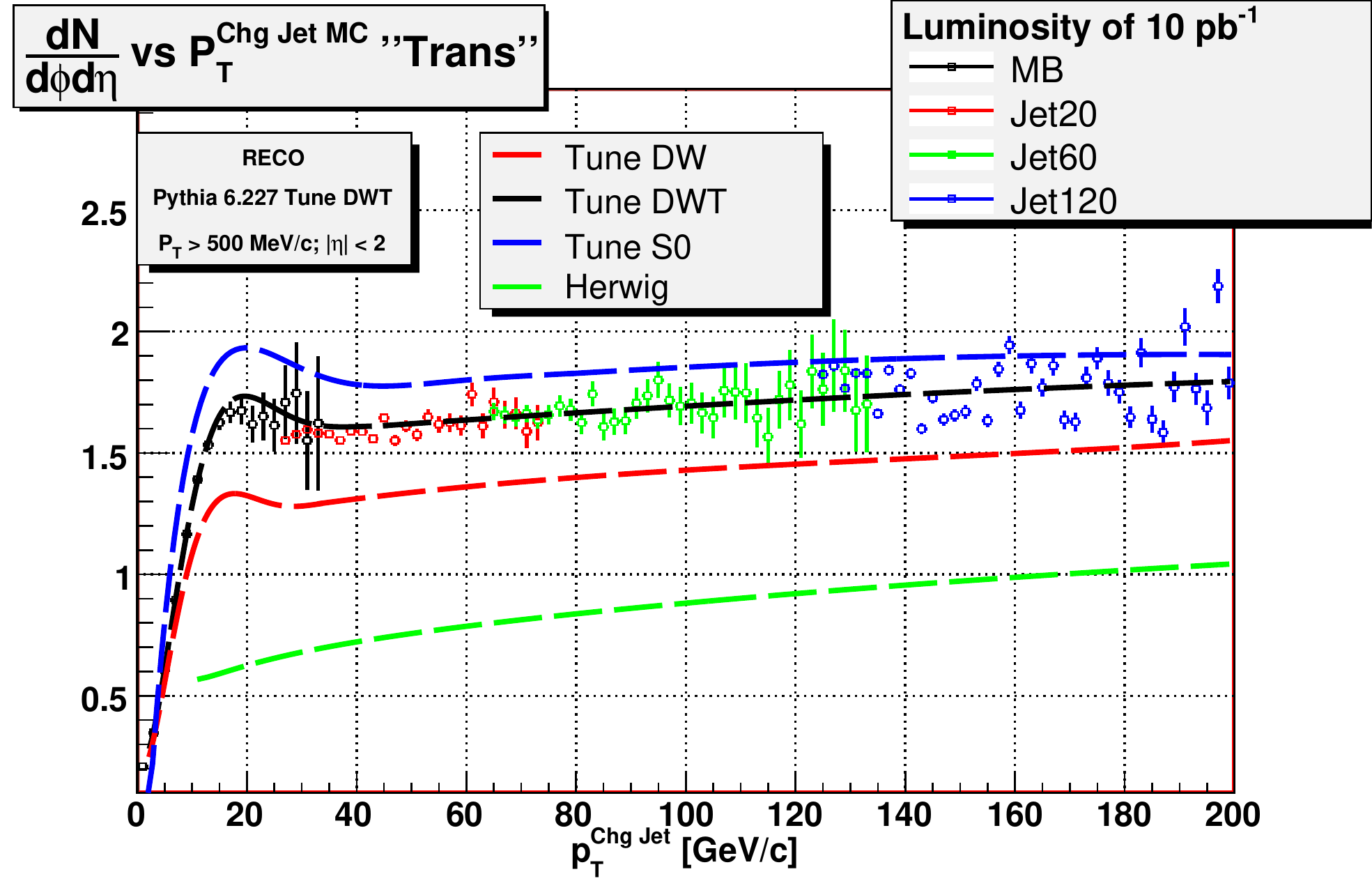}
\includegraphics[scale=0.3, angle=90]{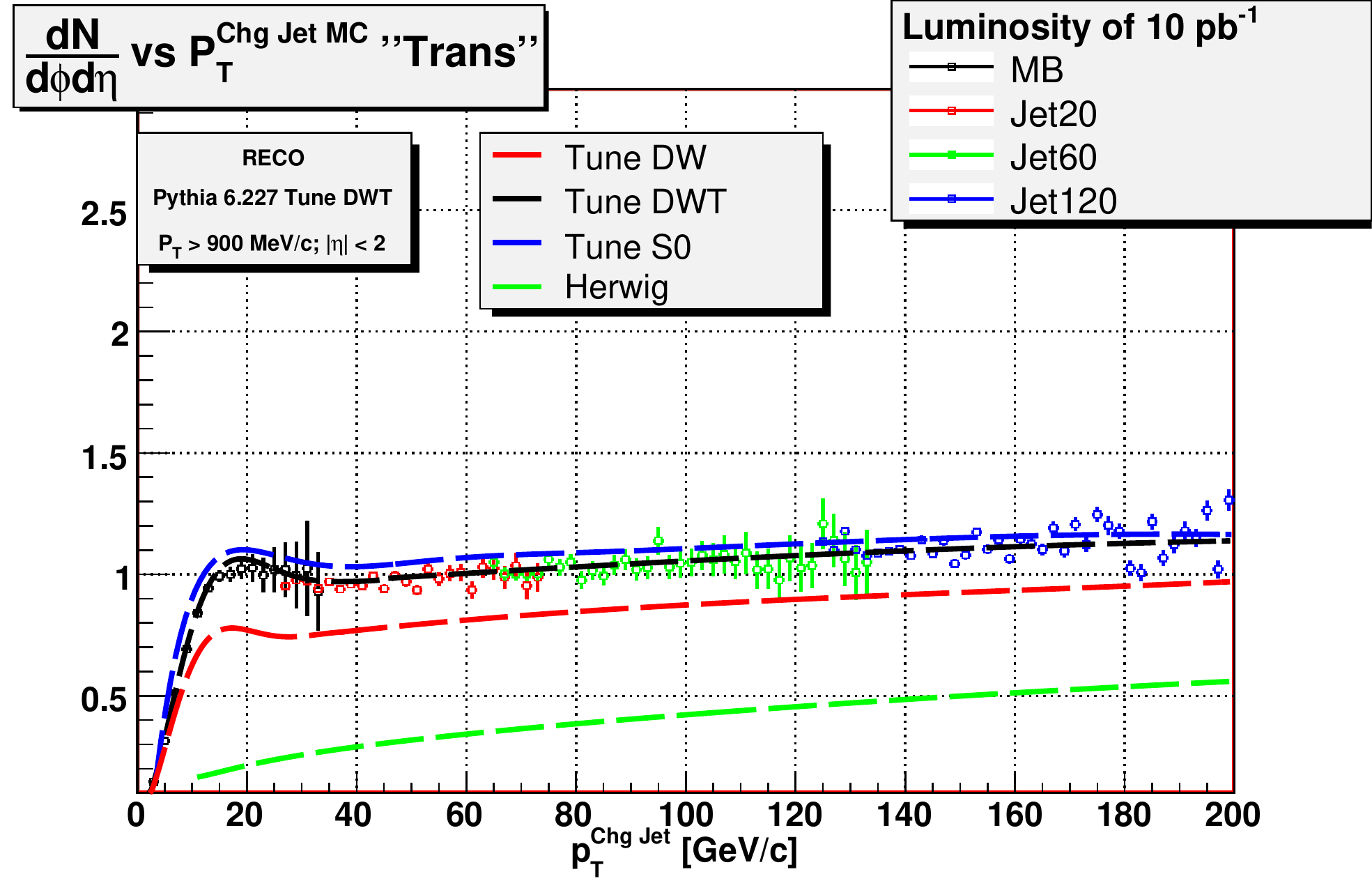}
\caption{\footnotesize 
QCD Monte-Carlo models predictions for charged particle jet production at $14\tev$.  
\TR\ region: average charged \ptsum\ density, \ptden, with \etacut\ and \ptlcut\ (\textit{left}) or \pthcut\ (\textit{right}) versus the transverse momentum of the leading charged particle jet.
The QCD models are HERWIG (without MPI) and three versions of PYTHIA $6.4$ 
(with MPI).
\label{RECO_fig1}}
\end{center}
\end{figure}

\begin{figure}[!Hhtb]
\begin{center}
\includegraphics[scale=0.3, angle=90]{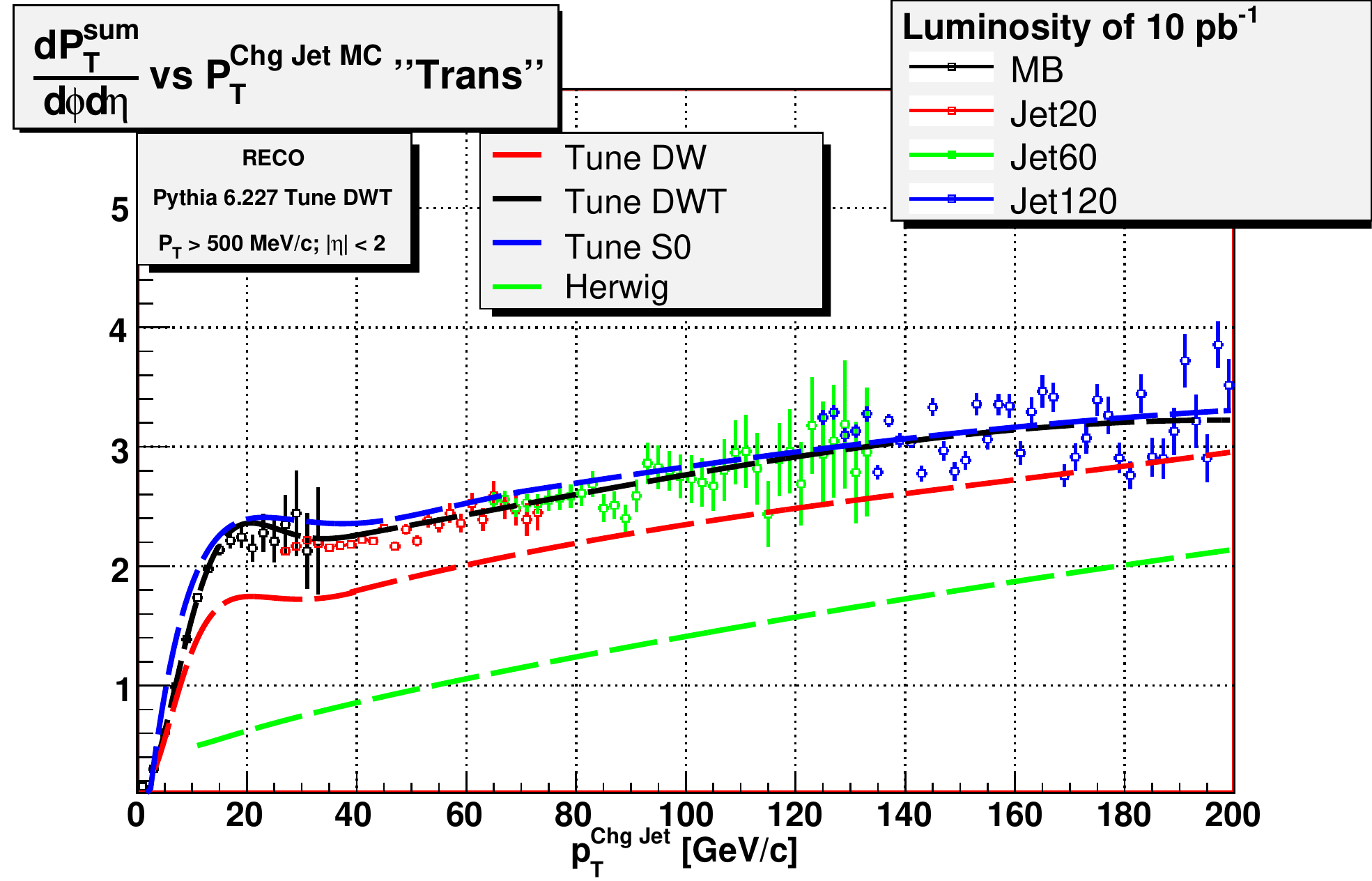}
\includegraphics[scale=0.3, angle=90]{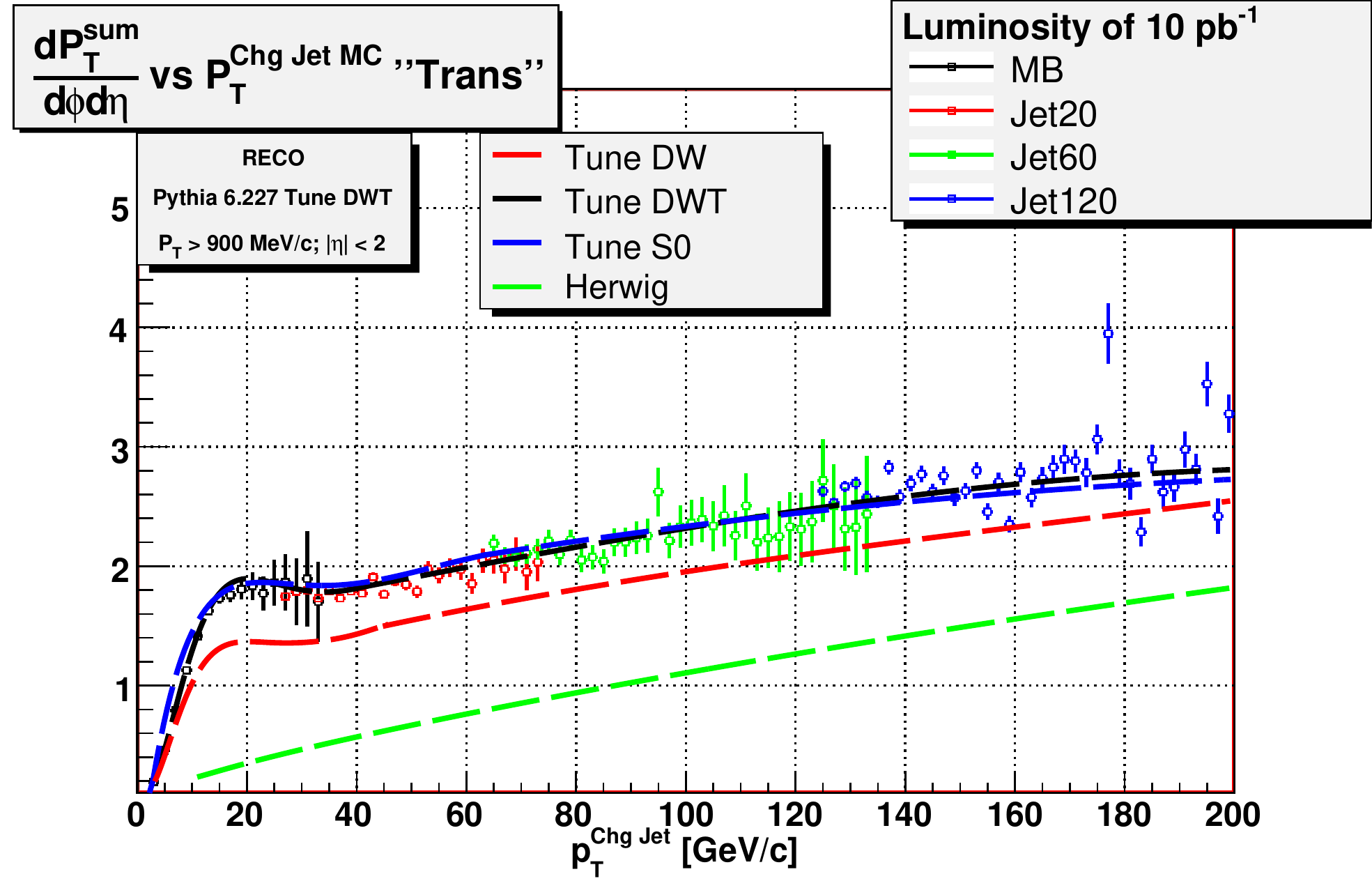}
\caption{\footnotesize 
QCD Monte-Carlo models predictions for charged particle jet production at $14\tev$.  
\TR\ region: average charged \ptsum\ density, \ptden, with \etacut\ and \ptlcut\ (\textit{left}) or \pthcut\ (\textit{right}) versus the transverse momentum of the leading charged particle jet.
The QCD models are HERWIG (without MPI) and three versions of PYTHIA $6.4$ 
(with MPI).
\label{RECO_fig2}}
\end{center}
\end{figure}

Figures~\ref{RECO_fig1} and Fig.~\ref{RECO_fig2} show the same quantities, \nden
\ and \ptsum\, for QCD Monte-Carlo models and superimposed the full simulation results for CMS experiment.
The reconstructed point are referred to 10 $pb^{-1}$ of low luminosity operation at LHC, without pile up.
The complete analysis is described elsewhere~\cite{ue_cms}.
Even with a reduced integrated luminosity, 10 $pb^{-1}$, it is possible to discriminate between different models taking the advantage to reconstruct tracks down to  \pt\ of 500 MeV/c.

\section{The Direct Observation of Multiple Partonic Interactions}

\begin{figure}[h]
\begin{center}
\includegraphics[scale=0.4]{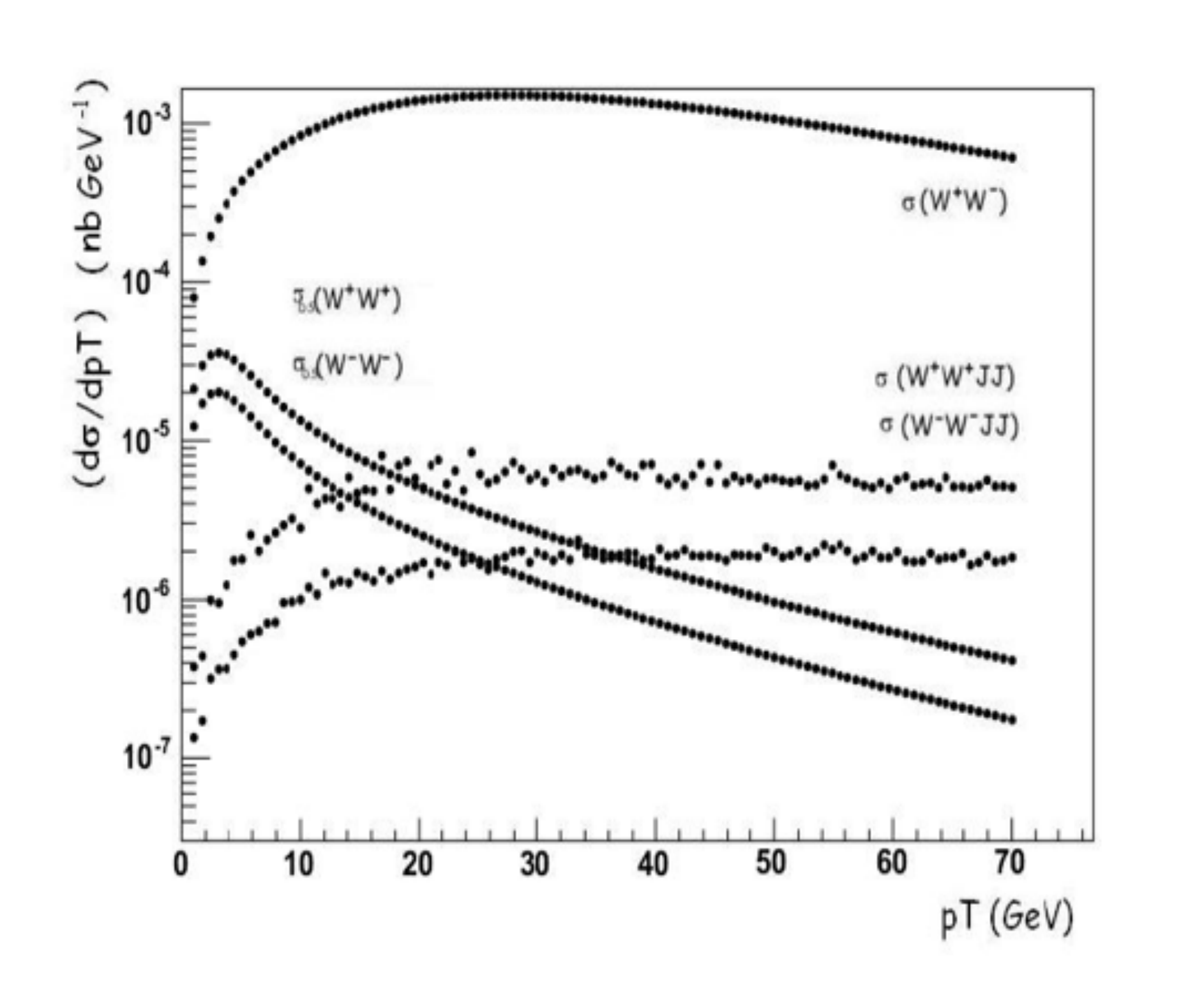}
\caption{differential cross section for same sign W production versus the minimum \pt\ of the boson pair. Contribution from double parton interactions are superimposed to those arising from single parton interaction processes.
$W^{+}W^{-}$ cross section is also drawn as reference.}
\label{hipt}
\end{center} 
\end{figure}

The final goal of the MPI study is to achieve a uniform and coherent description of MPI processes for both high- and the low-$P_{T}$ regimes. Recent theoretical progress in this field has been reported~\cite{dp1}. 
The cross section for a double high-$P_{T}$ scattering is parameterized as:\\
\begin{center}
$\sigma_{D}=\frac{m\sigma_{A}\sigma_{B}}{2\sigma_{eff}}$

\end{center}
where A and B are 2 different hard scatters, m=1,2 for indistinguishable or distinguishable scatterings respectively and $\sigma_{eff}$ contains the information about the spatial distribution of the partons~\cite{Paver:1982yp}~\cite{Ametller:1987ru}. In this formalism $m\sigma_{B}/2\sigma_{eff}$ is the probability that an hard scatter B occurs given a process A and this does strongly depend on the geometrical distribution of the partons inside the interacting hadrons.
The LHC experiments will perform this study along the lines of the CDF experiments~\cite{dp2}~\cite{dp3}), i.e. studying 3jet+$\gamma$ topologies. On top of that the extension to the study of same sign W production (Fig.~\ref{hipt}) is also foreseen.
Here we would like to propose an original study concentrating on the search for perturbative patterns in MB events looking for minijet pair production. 

\noindent Let's introduce the formalism for the study of MPI in mini-jet production. We re-write the inelastic cross section as the sum of one soft and one hard component.

\begin{equation}
\sigma_{inel}=\sigma_{soft}+\sigma_{hard}
\end{equation}

\noindent with $\sigma_{soft}$ the soft contribution to the
inelastic cross section $\sigma_{inel}$, the two contributions
$\sigma_{soft}$ and $\sigma_{hard}$ being defined through the
cutoff in the momentum exchanged between partons, $p_t^c$. Notice
that, differently from the case of the inclusive cross section ($\sigma_S$),
which is divergent for $p_t^c\to0$, both $\sigma_{hard}$ and all
exclusive contributions to $\sigma_{hard}$, with a given number of
parton collisions, are finite in the infrared limit.

\noindent
A simple relationship links the hard cross section to $\langle N\rangle$, i.e. the average number of partonic interactions:

\begin{equation}
\langle N\rangle\sigma_{hard}=\sigma_S
\end{equation}

\noindent
While the effective cross section $\sigma_{eff}$ turns out to be
linked to the dispersion $\langle N(N-1)\rangle$:

\begin{equation}
\quad\frac{1}{2} \langle N(N-1)\rangle\sigma_{hard}=\sigma_D
\end{equation}

\noindent
These relationships can be used to express $\sigma_{eff}$ in terms of the statistical quantities related to the multiplicity of partonic interactions: 

\begin{equation}
\langle N(N-1)\rangle=\langle N\rangle^2\frac{\sigma_{hard}}{\sigma_{eff}}
\end{equation}

\noindent
This last equation is particularly relevant from an experimental point of view.
Indeed, even with a reduced detector acceptance and detection efficiency, one can always measure the physical observable $\sigma_{hard}/\sigma_{eff}$ that accounts for the probability enhancement of having additional partonic interactions above the scale $p_t^c$.

We propose to perform this measurement counting the charged mini-jet pairs above a minimal scale $p_t^c$ in MB events. Mini-jets are reconstructed along the lines described in the previous sub-section.
First of all the charged jets are \pt-ordered.
A pairing criteria is introduced which is based on the maximum difference in azimuth between the charged jets. 
The pairing algorithm starts from the leading charged jet and associates the first secondary jet in the hierarchy that respects the criteria. The highest \pt\ of the pair is assumed to be the scale of the corresponding partonic interaction. The paired charged jets are removed from the list and the remnant charged jets are re-processed following the same steps.
One end-up with a list of paired charged jets. N is the number of charged pairs above the scale $p_t^c$.

\begin{figure}[h]
\begin{center}
\includegraphics[scale=0.33]{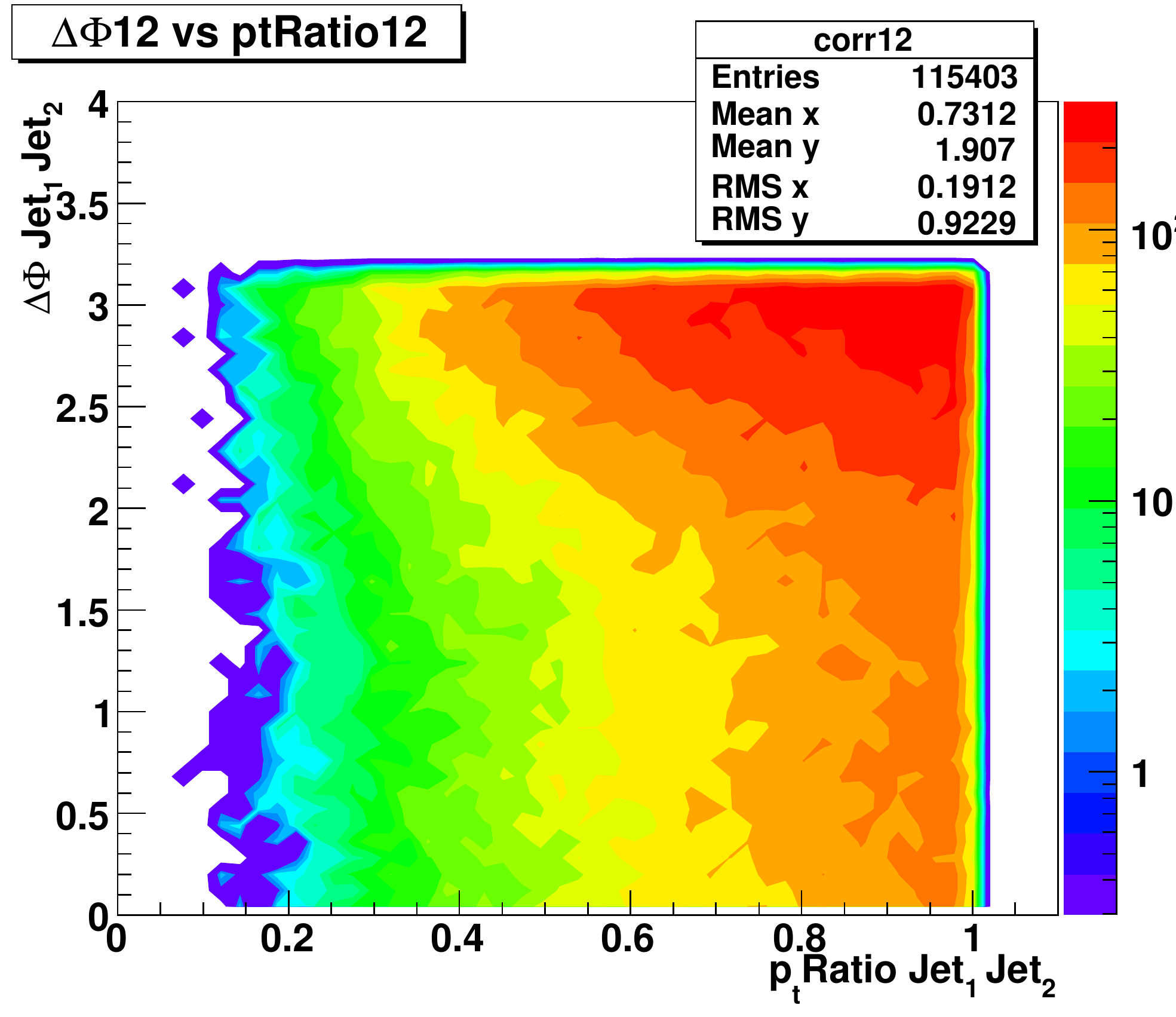}
\includegraphics[scale=0.33]{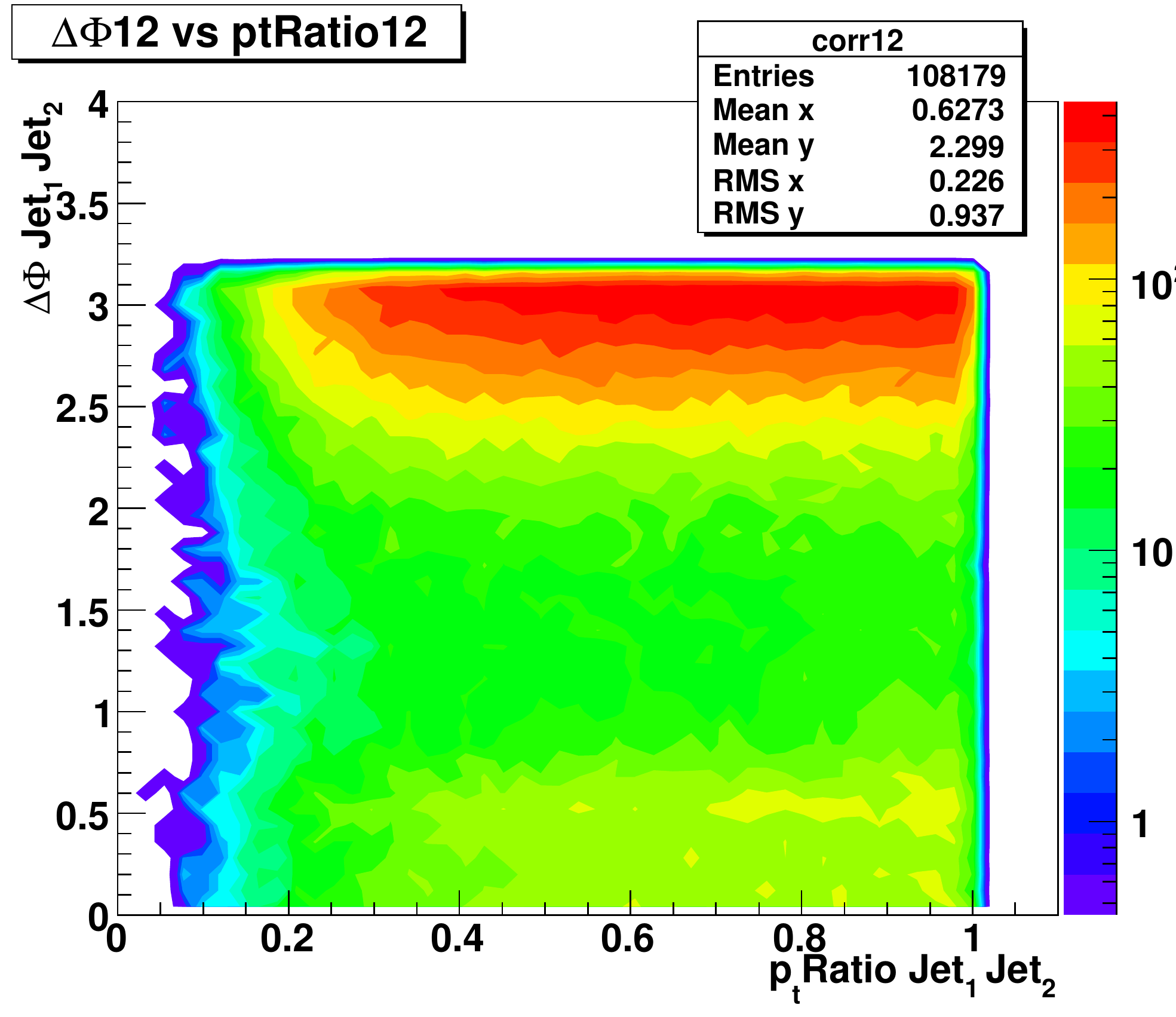}
\caption{Delta azimuth versus the \pt\ ratio between the first and the second charged jets in MB events at the LHC. 
Right plot is considered as a cross check for the pairing algorithm when Multiple Parton Interactions and radiation processes are switched off.
PYTHIA Tune S0 is considered.}
\label{fig:paired}
\end{center} 
\end{figure}

Fig.~\ref{fig:paired} shows the difference in azimuth versus the \pt\ ratio between the first and the second charged jet in the event. Right plot shows the case when both MPI and radiation are switched off to study the sensitivity of the pairing algorithm in a clean hard process. Two cuts have been set to define the pairs: $\Delta \phi>2.7$ and \pt\ $ratio >0.25$. 

\begin{figure}[h]
\begin{center}
\includegraphics[scale=0.33]{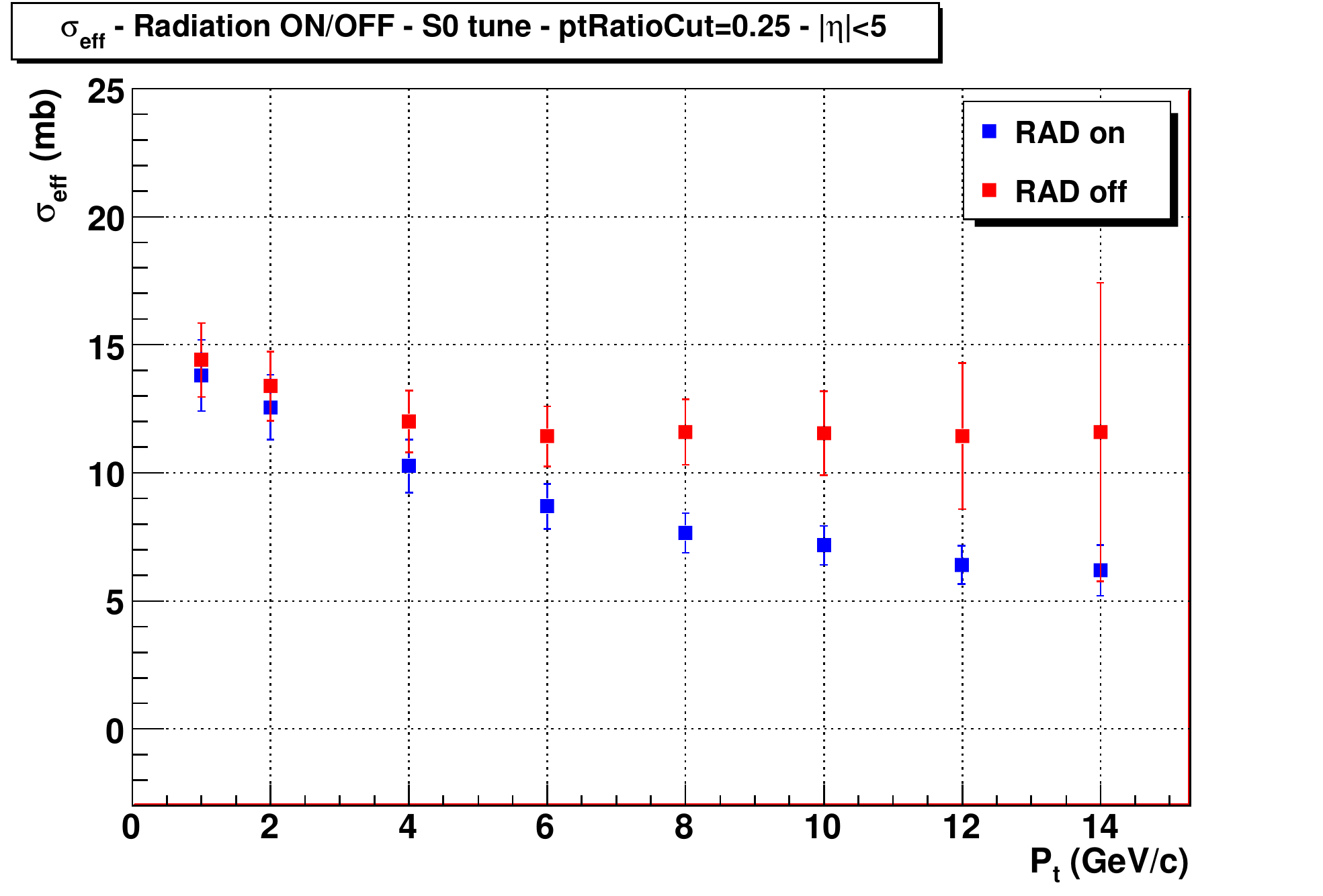}
\includegraphics[scale=0.33]{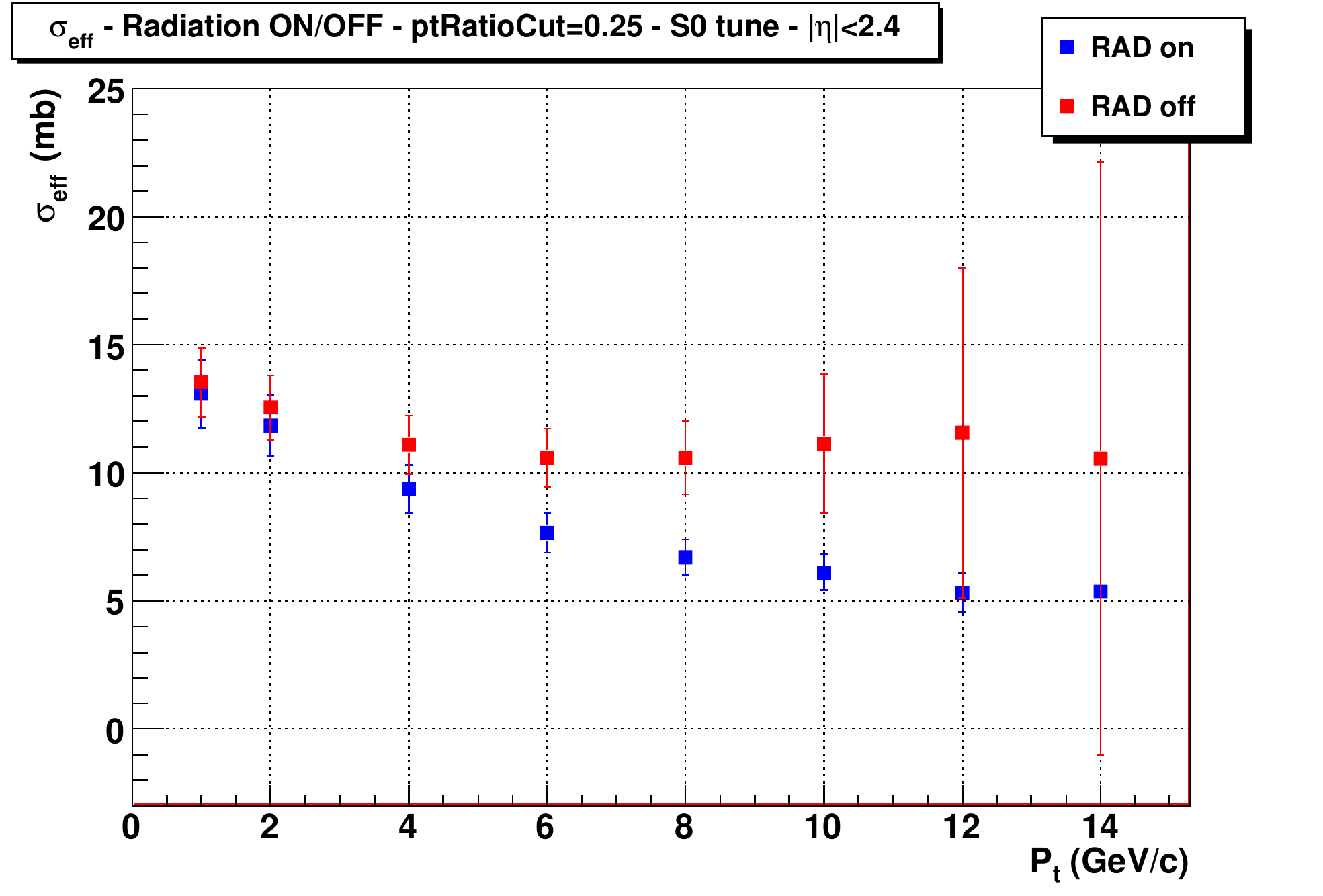}
\caption{Effective cross section in MB events at the LHC quoted for minijet processes in two different pseudorapidity ranges: $|\eta| \ < \ 5$ (left) and $|\eta| \ < \ 2.4$ (right) with and without radiation processes (blu and red). PYTHIA Tune S0 is considered.}
\label{fig:enh}
\end{center} 
\end{figure}

Fig~\ref{fig:enh} reports $\sigma_{eff}$ for two different pseudorapidity ranges $|\eta| < 5$ ($left$) and $|\eta| < 2.4$ ($right$). As expected $\sigma_{eff}$ does not depend on the detector acceptance. In the same figures is shown the sensitivity of the pairing algorithm to radiation coming from initial and final state (red points refer to the no-radiation case).

Notice that, while in the result of the simulation the effective cross
section does not depend on the acceptance of the detector,
one observes same dependence of $\sigma_{eff}$ on $p_t^{min}$ also after switching off the
radiation.
One should emphasize that this feature would not show up in the simplest
model of multiparton interactions, where the distribution in the number of
collisions, at fixed hadronic impact parameter, is a Poissonian. In this
case one would in fact obtain that the effective cross section is constant
not only as a function of the acceptance of the calorimeter, but also as a
function of the cutoff.
A cutoff dependent effective cross section might be produced by a
distribution in the number of collisions at fixed impact parameter
different from a Poissonian. It should be remarked that considering a
distribution, at fixed impact parameter, different from a Poissonian one
introduces correlations in the multiparton distributions additional to the
correlation in the transverse parton coordinates, taken into account by
the dependence of the average number of multiparton collisions on the
impact parameter. Observing a dependence of $\sigma_{eff}$ on $p_t^{min}$
one would hence provide evidence of further non trivial correlations
effects between partons in the hadron structure.
To trace back the origin of the dependence of $\sigma_{eff}$ on
$p_t^{min}$, observed in the simulation, one might notice that, in the
simplest uncorrelated Poissonian model, the impact parameter is chosen
accordingly with the value of the overlap of the matter distribution of
the two hadrons and independently on value of the cutoff $p_t^{min}$. In
Pythia, on the contrary, events are generated through a choice of the
impact parameter which is increasingly biased towards smaller values at
large $p_t$. The correlation induced in this way between the impact
parameter of the hadronic collisions and the scale of the interaction has
the result of decreasing the behavior of  $\sigma_{eff}$ at large
$p_t^{min}$.


\addtocounter{chapter}{1}
\newcommand{\newc}{\newcommand}

\mchapter{Early Standard Model physics at the LHC}
{ Marina Cobal, Giacomo Polesello, Roberto~Tenchini}
\vskip 1cm
\section{Introduction}
We will concentrate here on the first physics measurements that
the LHC experiments will be able to perform from the very beginning
of the data taking at 14~TeV.
\section{QCD measurements}
The hard scattering 
cross-section at the LHC is dominated by the production 
of QCD jets, which surpass by many orders of magnitude any other 
physics process. Therefore, at soon as the LHC switches on 
jet production will be observed at the LHC, even for extremely 
small integrated luminosity. For instance
the cross-section for jets with transverse momentum above 50~GeV 
is $\sim$25~$\mu\mathrm{b}$, i.e. for 
an integrated luminosity of 1~nb$^-1$, 25k such jets will be observed.\par
Jets will therefore be the main tool for understanding the 
detector performance, and already starting from luminosities
as low as a few $\mu\mathrm{b}^{-1}$, the LHC collaborations will 
use jets for e.g. equalizing the azimuthal response of the calorimeter,
and, through the exploitation of the jet balance in the transverse
plane, start chasing down the instrumental $E_T^{miss}$ sources.
Some very detailed considerations on the usage  the jet statistics 
collected with  the first nb$^{-1}$ of data is given in 
\cite{Green:2006fa}. \par
The further step, is the measurement of the jet cross-section,
and the comparison with the predictions of QCD calculation.
This will be the first benchmark of the ability of the LHC 
experiments for cross-section measurement, and possibly a first
window on discovery physics, as the high $p_T$ tails in the inclusive
jet cross-section  are sensitive to the presence of new physics,
and the invariant mass distribution of the two jets can show
the appearance of new physics under the form of resonances. 
The measurement is in principle simple, as high $p_T$ jets are 
easily identified the LHC detectors, and the statistics is enormous. 
In practice this is a difficult measurement, involving a large number
of uncertainties both from the theoretical and experimental point of
view. 
The complex issues related to the definition of the object
"jet", and to the determination of the correct energy calibration
for the optimisation of the jet energy response are the subject 
of another contribution. We will here limit ourselves to evaluate
the contribution of the most basic sources of uncertainty on the 
measurement of the jet cross-section.\par
\begin{figure}
\includegraphics[angle=-90,width=0.5\textwidth]{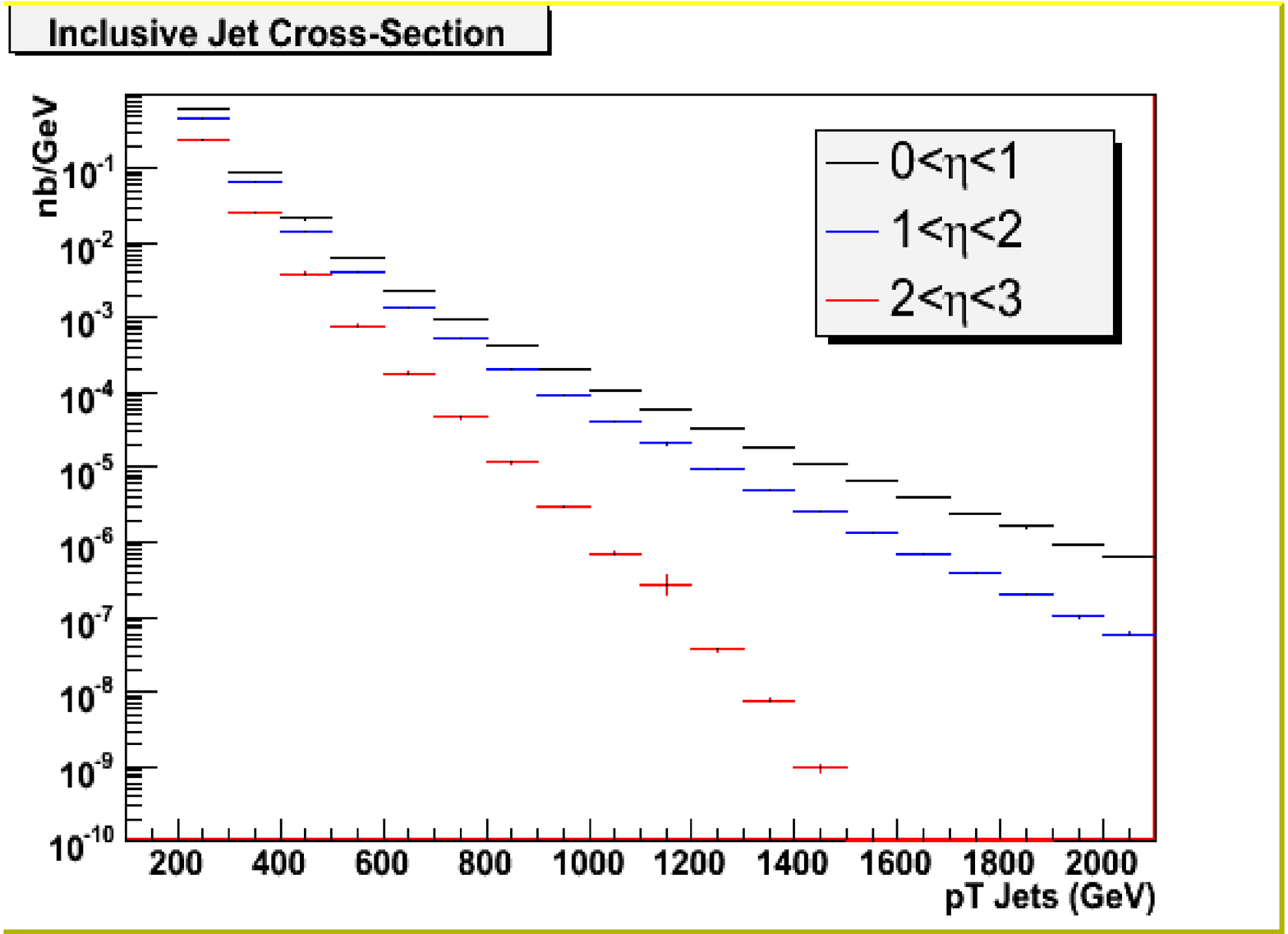}
\includegraphics[angle=-90,width=0.5\textwidth]{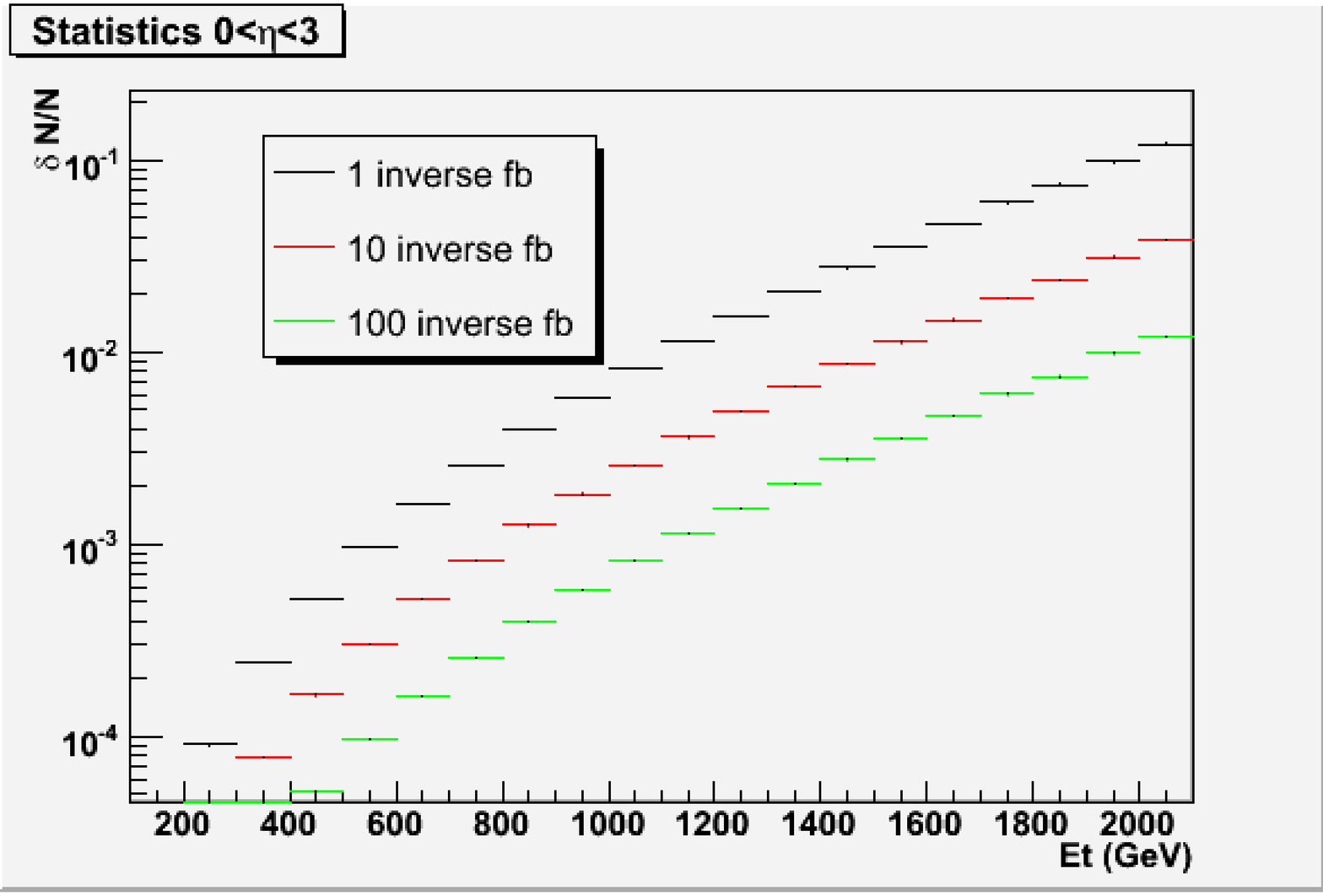}
\caption{\label{fig:clemens1} {Left: NLO jet cross section as a function of
$p_T$ at the LHC for three different rapidity ranges. Right:  
fractional statistical error at the LHC per hundred GeV bin in $p_T$ 
for three different scenarios of integrated luminosity. No trigger
selection assumed
}}
\end{figure}

The distribution of $d\sigma/dp_T$ calculated at NLO with the program
of \cite{Nagy:2001fj} for three different ranges of pseudorapidity
is shown in the left side of Fig~\ref{fig:clemens1},
from \cite{iacopo}.
One sees that with 1~$pb^{-1}$ jets with $p_T$ of
$\sim$800~GeV will be measured, and with 1~$fb^{-1}$ the kinematic 
range is extended up to $\sim 2$~TeV.\par
A more quantitative estimate of the achievable statistical error
is shown in the right side of Fig~\ref{fig:clemens1}, always from 
\cite{iacopo}, where naively for each bin in Jet $E_T$ the quantity
$\sqrt{N}/N$ is shown, where $N$ is the number of events per bin
for jets within $|\eta|<3$. A statistical error of $\sim1\%$ is expected
for a $p_T$ of 1~TeV and an integrated luminosity of 1~fb$^{-1}$.
The estimate is correct for high values of the jet $p_T$, however for
lower $p_T$ jets the statistical uncertainty will be determined 
by the online selection of the events. This issue is particularly 
relevant for the search of resonances in the jet-jet invariant 
mass, where an approximately uniform statistical error is 
desirable over a large range of invariant mass.
The argument goes as this: the technical and financial
limit on the number of events which 
can be selected and analyzed by each experiment is $\sim$100-200~Hz.
Now, even considering a low initial luminosity of 
$10^{32}~\mathrm{cm}^{-2}\mathrm{s}^{-1}$ a process with 
1~$\mu\mathrm{b}$ cross-section, such as the production of jets with 
100~GeV $p_T$ would saturate the trigger bandwidth. It is therefore 
necessary to adopt a flexible trigger strategy, evolving with 
luminosity, whereby jets are selected with multiple thresholds 
and the events selected with lower threshold are prescaled,
i.e. only a predetermined fraction of the events which would 
pass the trigger are actually written on mass storage. The
prescaling factor is defined as 1 over the fraction of accepted events.
\begin{figure}
\begin{center}
\includegraphics[width=0.6\textwidth]{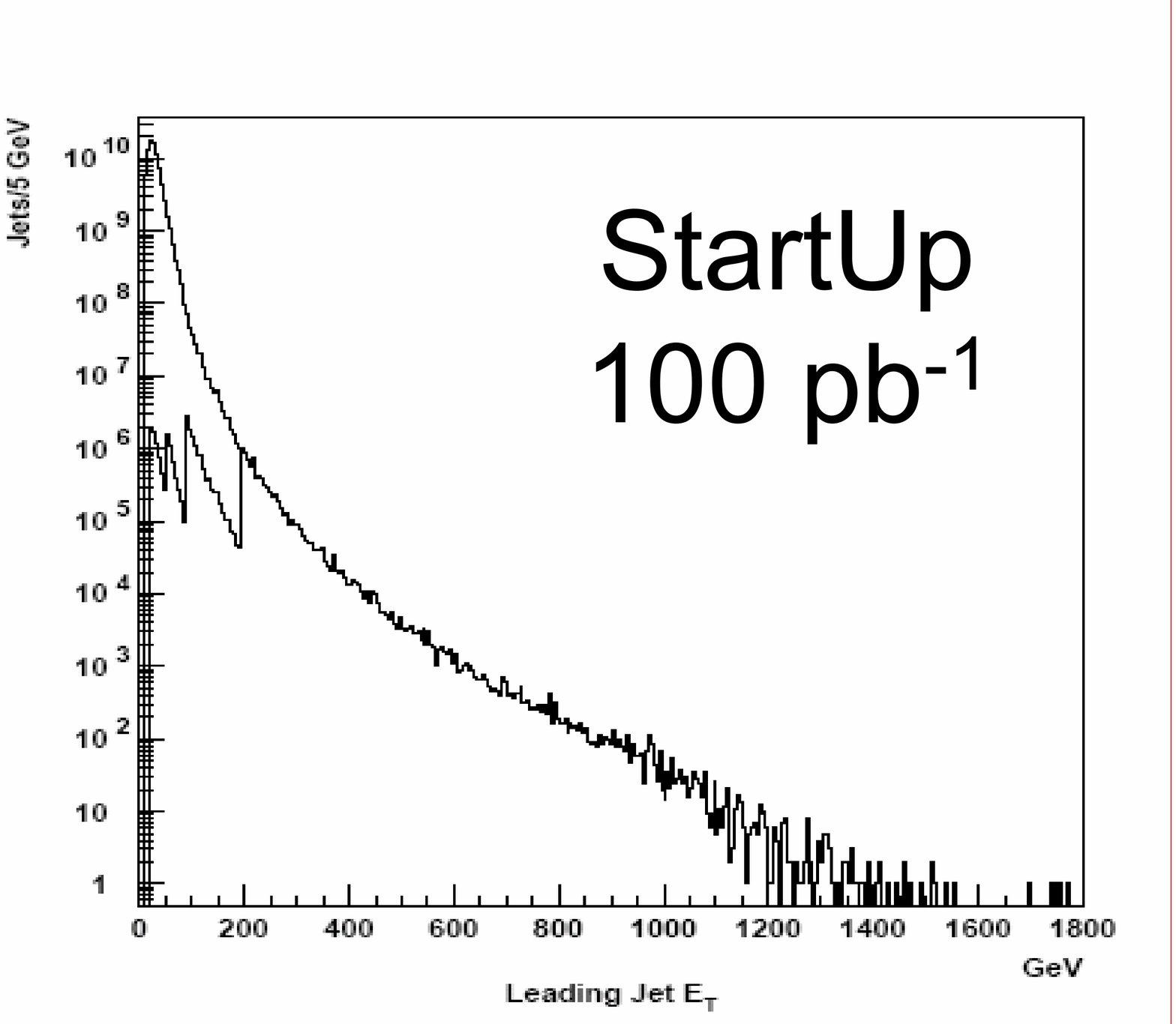}
\caption{\label{fig:tom}{Upper line: $p_T$ distribution of the
leading jet for jet events at the LHC. Lower line: $p_T$ distribution
of the leading jet accepted by the trigger for the multi-threshold
jet trigger scenario described in the text. The assumed instantaneous  
luminosity is $10^{31}~\mathrm{cm}^{-2}\mathrm{s}^{-1}$
}}
\end{center}
\end{figure}
\begin{figure}
\begin{center}
\includegraphics[width=0.6\textwidth]{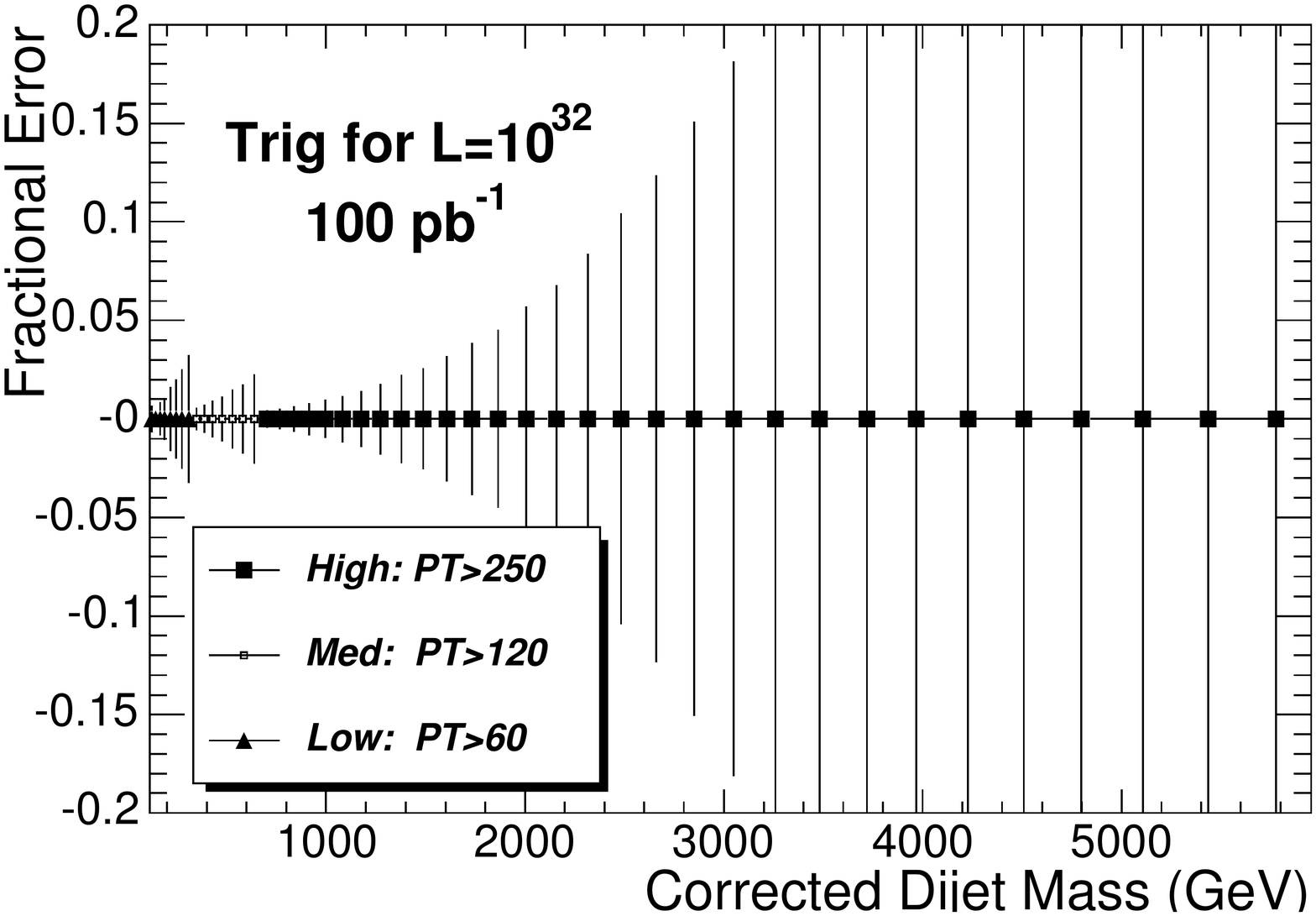}
\caption{\label{fig:cms1}{Fractional statistical error on the jet 
cross section in CMS for the multi-threshold jet trigger scenario 
described in the text. The assumed instantaneous
luminosity is $10^{32}~\mathrm{cm}^{-2}\mathrm{s}^{-1}$
}}
\end{center}
\end{figure}
For example, one may want to achieve at a luminosity of 
$10^{31}~\mathrm{cm}^{-2}\mathrm{s}^{-1}$ a jet trigger rate of
20~Hz, with an approximately flat rate for jets with 
$p_T<200$~GeV. A possible way of achieving this aim is 
selecting jets with a set of 6 thresholds: \{25, 50, 90, 170, 300, 400\}~GeV
with respective prescaling factors \{10k, 1k, 25, 1, 1, 1\}. The $p_T$
distribution for the accepted jets, is shown as the lower line in 
Fig.~\ref{fig:tom}, from \cite{chiara}. A similar effect is shown in 
a CMS study shown in Fig.~\ref{fig:cms1} from \cite{cmstdr2}.
The fractional statistical errors for the jet-jet invariant mass distribution 
are shown as a function of the jet-jet invariant mass for one 
month of data taking at $10^{32}~\mathrm{cm}^{-2}\mathrm{s}^{-1}$.
The distribution comes from the combination of three different jet trigger
thresholds: \{60, 120, 250\}~GeV with different levels of prescaling.\par
The precision of the comparison with the theoretical prediction 
will thus, up to a scle of a few TeV, 
be dominated by systematic effects, coming from two sources:
theoretical uncertainties in the prediction of the jet cross-section
and experimental uncertainties.\\
The jet cross-section, as explained in the introductory
chapter of this report  is calculated as
the convolution of of the Parton Distribution Functions (PDFs) 
with the partonic cross-section. From the jet studies at the Tevatron
the two main sources of theoretical uncertainty are: a) the 
uncertainty on renormalization($\mu_R$)/factorisation($\mu_F$),
arising from the perturbative calculation of the partonic cross-section at fixed order, and b) the uncertainty on the PDFs. The effect of the uncertainty
on $\mu_R$ and $\mu_F$ has been studied in \cite{iacopo}
by varying $\mu_R$ and $\mu_F$
independently between $0.5\times p_T^{max}$ and $2\times p_T^{max}$,
where $p_T^{max}$ is the transverse momentum of the leading jet.
The effect has little dependence on the jet $E_T$ and it induces an 
uncertainty of approximately 10\% at 1~TeV.\\
\begin{figure}
\begin{center}
\includegraphics[width=0.7\textwidth]{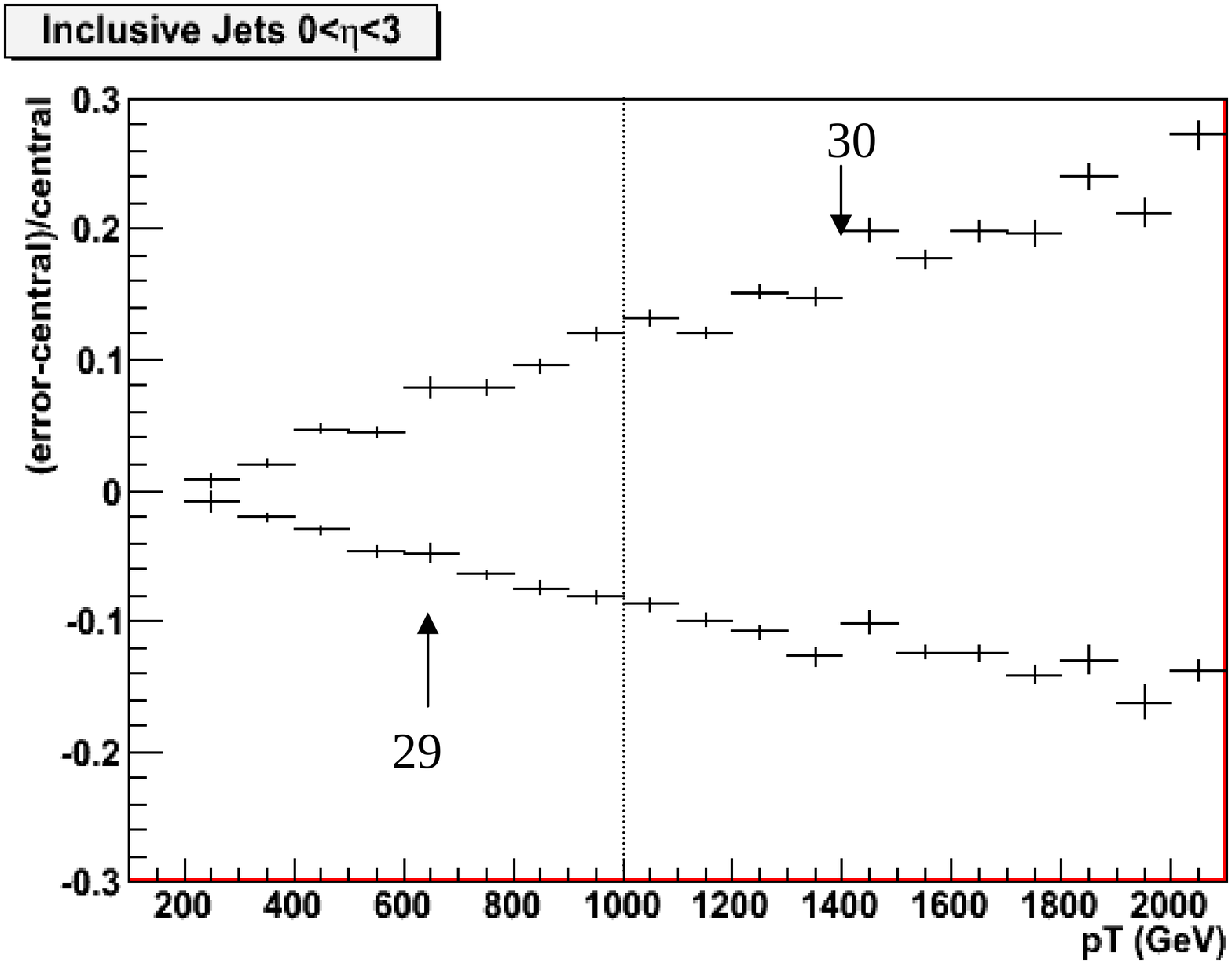}
\caption{\label{fig:pdf} {Fractional uncertainty  on the jet 
cross-section as a function of the jet $p_T$ due to the uncertainty 
on the PDF parametrisation. The PDFs used are CTEQ6M \cite{Pumplin:2002vw}, and
the error is evaluated using the LHAPDF scheme. 
}}
\end{center}
\end{figure}
\begin{figure}
\begin{center}
\includegraphics[width=0.5\textwidth,clip=true,viewport=14pt 14pt 286pt 268pt]{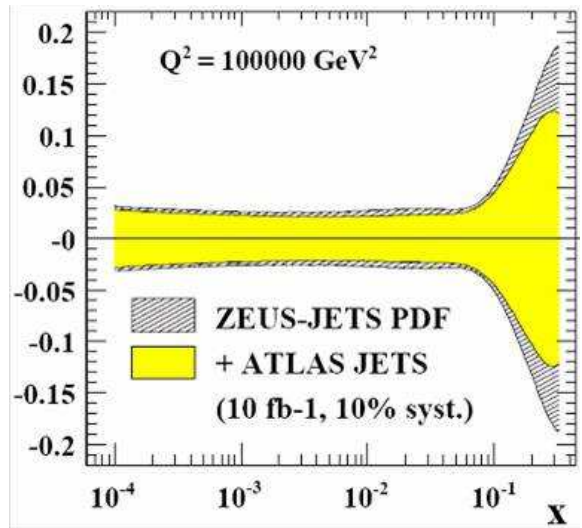}
\caption{\label{fig:claire}{
Uncertainty on the gluon PDF. The yellow band shows the effect of introducing
in the PDF fit the ATLAS jet data from inclusive jet cross-section for
$p_T>3$~GeV for $0<\eta<3$. The ATLAS psudo-data ssume an integrated 
luminosity of 10~fb$^{-1}$ and an uncorrelated sytematic uncertainty
of 10\% on the experimental cross-section.
}}
\end{center}
\end{figure}
The PDFs are not predicted by theory, but extracted from 
phenomenological fits to a mix of experimental results, 
dominated by experiments measuring the deep inelastic scattering
of leptons on hadrons. The distributions are then evolved through
the DGLAP equations to the $Q^2$ range relevant for the LHC.
This procedure has two main sources of uncertainty: the input
phenomenological function used for the fit of the experimental
data, which is different for the different groups performing
PDF fits, and  the propagation of the statistical and systematic errors
of the used data to the parameters of the PDF. The most recent
generations of PDFs provide a way of propagating the errors 
from the fit  to the cross-section calculation,
based on a common standard  called LHAPDF \cite{Whalley:2005nh}. 
In Fig~\ref{fig:pdf}
the extreme variations with respect to the central value 
for the structure function set CTEQ6M are shown as a function of the
$p_T$ of the jet. The resulting uncertainty is of order 15\%
for a jet $p_T$ of 1~TeV. 
This large uncertainty is due to the fact that the jet 
cross-section at high $p_T$ 
are determined by the value of the gluon structure function 
at high $x$ which is poorly determined by the available experimental 
data. A possibility is to use the ATLAS data to constrain 
the gluon structure function. Work in this direction using the 
rapidity distribution of W production is described in detail 
in the next section. In order to reach high values of $x$, though,
the jet data themselves must be used. In Fig.~\ref{fig:claire},
from \cite{clemens},
the uncertainty of the gluon structure function as a function of
$x$ is shown. The hatched band is the uncertainty from the ZEUS
PDFs, the grey (yellow) band would be the uncertainty if the ATLAS jet data
are incorporated into the ZEUS PDF fit. 
An uncorrelated experimental  systematic error of 10\%
on the ATLAS jet measurement is assumed in the fit. 
A significant improvement is observed, strongly 
dependent on the assumed experimental systematics. It might be argued that
using the LHC jet data to reduce the high-$x$ uncertainty would 
basically hide any signal of new physics into a redefinition
of the structure functions.
Indeed, there are two ways of selecting events where one of the two 
partons has a high $x$. The invariant mass of the two jets can 
be written as $m^2=x_1x_2s$ where $\sqrt{s}$=14~TeV at the LHC.
Therefore one can sample high $x$ either with central events ($x_1\sim x_2$)
at high invariant mass, or events with small 
invariant mass and large boost in one direction
($x_1\gg x_2$). New physics effects are 
expected to become visible for  high jet-jet invariant mass, 
therefore the constraints from events with high boost can be used to
reduce the PDF uncertainty in the high mass region without 
biasing the sensitivity to new physics.\par
\begin{figure}
\begin{center}
\includegraphics[angle=-90,width=0.6\textwidth]{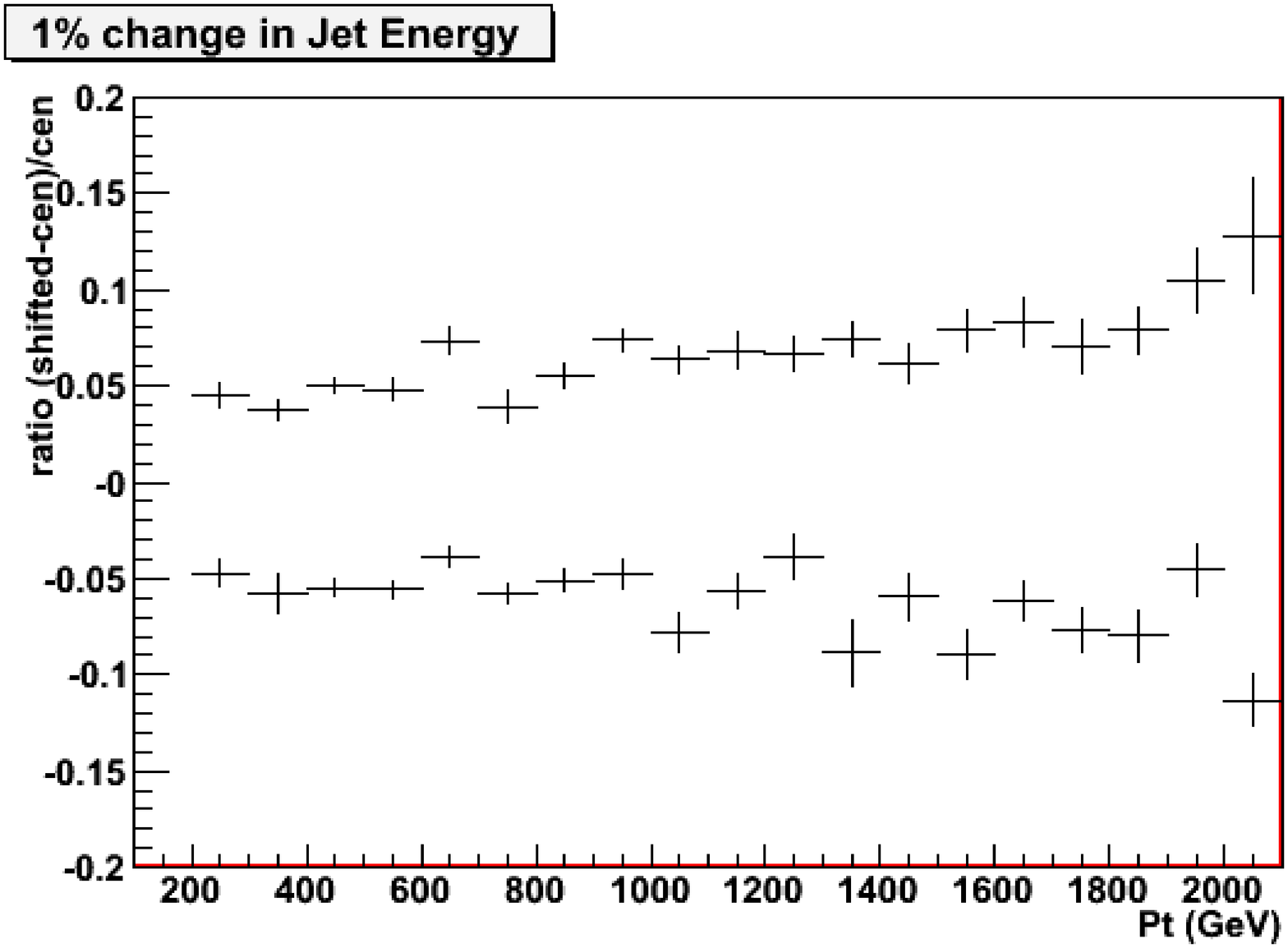}
\caption{\label{fig:atlasscale} { Relative variation of the jet cross-section
as a function of the jet $p_T$ for an assumed variation of 1\% on the
jet energy scale with respect to the nominal value.
}}
\end{center}
\end{figure}
\begin{figure}
\begin{center}
\includegraphics[width=0.8\textwidth]{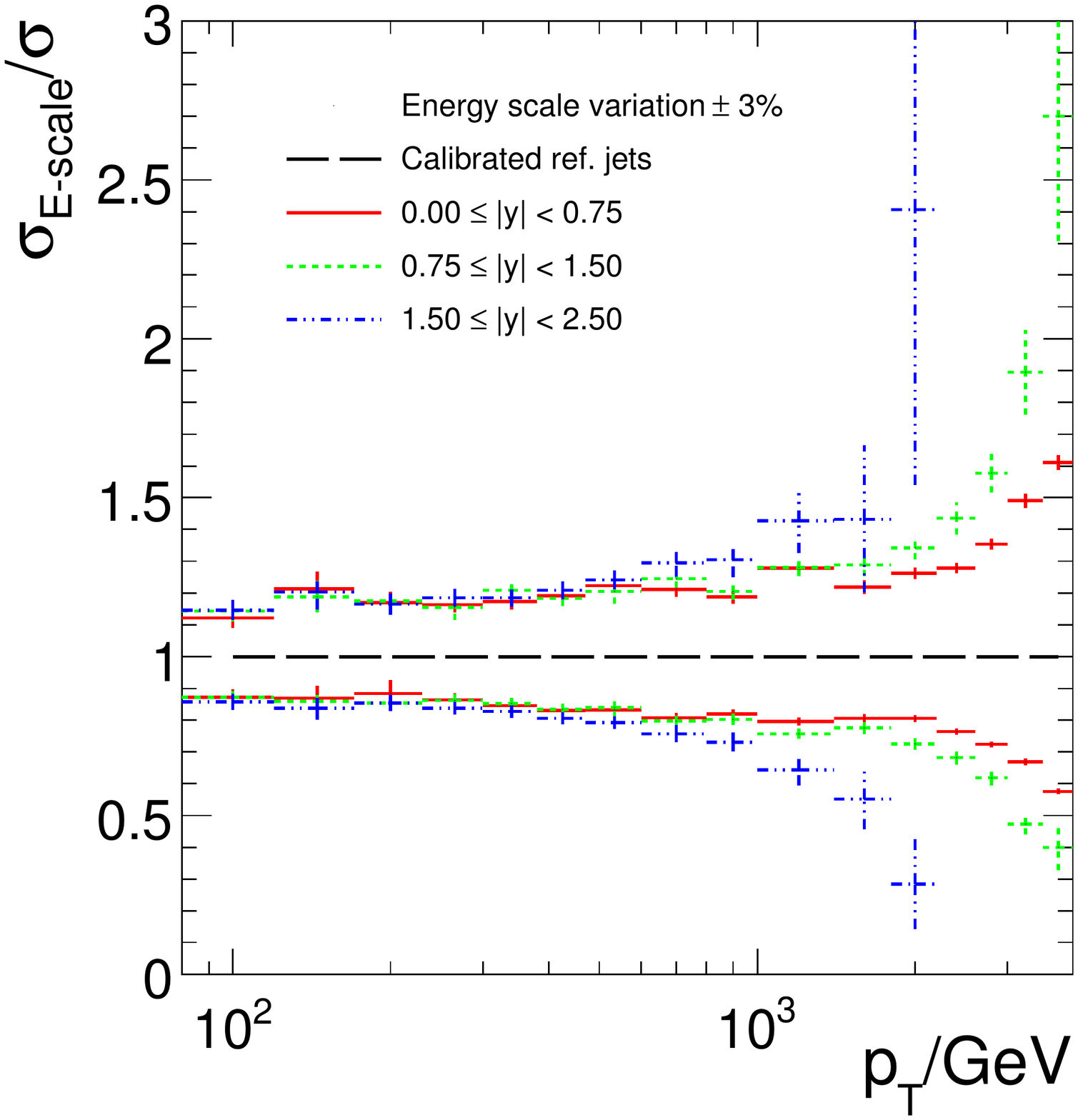}
\caption{\label{fig:cmsescale} {
Relative systematic uncertainties of the jet cross-section versus $p_T$ 
due to a change in jet energy scale of $\pm3\%$ for three bins in rapidity
in the CMS detector.
}}
\end{center}
\end{figure}

There are many possible sources of experimental errors
for the jet cross-section determination, for instance the uncertainty
on jet energy scale and jet resolution, uncertainty on the subtraction of 
the underlying event. For the Tevatron studies, the uncertainty
of the jet energy scale is the dominant factor. If the slope 
of the of the jet $p_T$ distributions goes approximately as $p_T^{-n}$,
for an uncertainty on the energy scale of, say,   1\% the uncertainty
on the cross-section is approximately $n\%$. For LHC jets $n$ 
is approximately 6 for a $p_T$ range between 200 and 1550~GeV, 
and the slope drastically increases for larger values of $p_T$.
This is shown in Fig.~\ref{fig:atlasscale} where 
for an 1\% change in jet energy the shift on the cross-section 
value is $\sim 6\%$, up to 1.5 TeV, rising to higher values
for higher $p_T$ when the $p_T$. 
An equivalent result is shown from a CMS study, Fig.~\ref{fig:cmsescale}
where a 3\% variation on the jet energy scale gives a $\sim$15-20\% uncertainty 
on the cross-section at low $p_T$, rising to $\sim$50\% for a 
$p_T$ of 4~TeV. It is therefore mandatory to control the
jet energy scale at the percent level up to a $p_T$ 
of a few TeV if we want the jet cross-section
measurement to be dominated by the theoretical uncertainties.
This is a hard requirement which will require a very intense work 
on the experimental data to be satisfied. 
\section{W, Z and Drell-Yan Physics}

The production cross sections of W and Z bosons at 14 TeV 
in hadronic collisions are large and their leptonic decays
are characterized by clear signatures. The cross section
for vector boson production at LHC, followed by leptonic
decay, is about 20 nb for the W and 2 nb for the Z.
Decays to electrons and muons will be detected in the very early
phase of the experiments, as the commissioning of electron and
muon triggers is expected to be relatively fast. Atlas and CMS
studies show that should be straightforward to obtain 
combined trigger and offline-selection efficiency around 50\%.
Because of the high rates these processes will play the r\^ole 
of standard candles for many other studies.
W and Z physics will start already with the first inverse picobarns
collected by the two experiments.
Decays to tau leptons require higher luminosities, because triggering
is based on more sophisticated criteria. Nevertheless, when the integrated
luminosity will reach one hundred picobarns or more, tau decays
will provide unvaluable information for the commissioning of the tau trigger.
Off-shell Drell-Yan leptonic decays have a considerable lower 
cross section, a few femtobarns are expected for dilepton invariant mass
larger than 1 TeV. 
During the first year of operation, the detection of WZ, ZZ and WW 
events will provide information
on important backgrounds to searches, while
W$\gamma$ and Z$\gamma$ will be important tools for physics commissioning.  
In the next paragraphs the main measurements that can be performed
with these processes, with total integrated luminosity lower than
1 $fb^{-1}$, are briefly described.

\subsection{W and Z decays to electrons and muons}
\label{WZemu}

The Tevatron experience has shown that W and Z decays to electrons
and muons can be selected with simple criteria and low background.
This is confirmed by the the Atlas and CMS detailed simulations
at 14 TeV. Since cross sections are high and good selection
efficiencies can be obtained, the crucial point in this case
is to design robust selections, with low dependence on experimental
systematic uncertainties. Therefore the main selection criteria
are aimed to select events in well defined geometrical acceptance region
and within the trigger acceptance.
The typical trigger thresholds for isolated electrons and muons will be set
at $p_T$ values around 20-30~GeV. For the electron channel
CMS is quoting~\cite{eth} efficiencies of 57\% and 26\% for
the Z and W bosons, respectively. In the muon channel~\cite{juan}
these become 52\% and 40\%, respectively. (Different fiducial regions
and trigger criteria are used in the electron and muon cases.)

Triggering of high $p_T$ isolated electrons and muons normally 
requires isolation criteria, i.e. a region around the lepton
(typically a cone) is defined and low activity (low total energy
in the calorimeters or low total $p_T$ of additional tracks)
is required in this {\it isolation region}.
This is a potential source of inefficiency and must be carefully controlled.
In order to study isolation effects the initial data at very low luminosity,
where the trigger criteria can be relaxed, can be used.
When the luminosity increases, di-lepton trigger streams, 
where isolation criteria are less strong or even abstent,
are a further tool for monitoring inefficiencies.

The trigger efficiency itself must be carefully studied. 
The methods for doing this are essentially three:

\begin{itemize}

\item Boostrap procedure. At very low luminosity very loose trigger criteria,
collecting events with {\it minimum bias} are set.
This allow to study inefficiencies in an unbiasied way with offline methods.
At higher luminosities the thresholds are raised to a value whose
efficiency is known from the first step. Higher thresholds values
are used, employing the same method, at even higher luminosities.

\item Orthogonal signatures. Detectors dedicated to triggering at LHC detectors
are often redundant and independent information, from two different subsystems,
can be used for a direct measurement of the efficiency. This is done
by counting double and single trigger rates.

\item Double physical objects. Events like, for instance, $Z \to \mu^+ \mu^-$ 
can be used to determine the muon trigger efficiency but triggering
on one muon and studying the unbiased behaviour of the other muon.
Particular care should be taken to take into account other 
physical sources of dimuons ($J/\psi \to \mu^+ \mu^-$ , etc.)

\end{itemize}

\noindent The study of the trigger efficiency will be one of the 
most important activity at the startup and during the lifetime
of the LHC experiments.

\subsection{W and Z cross-sections and the PDFs}

\begin{figure}
\centering
\includegraphics[scale=0.3]{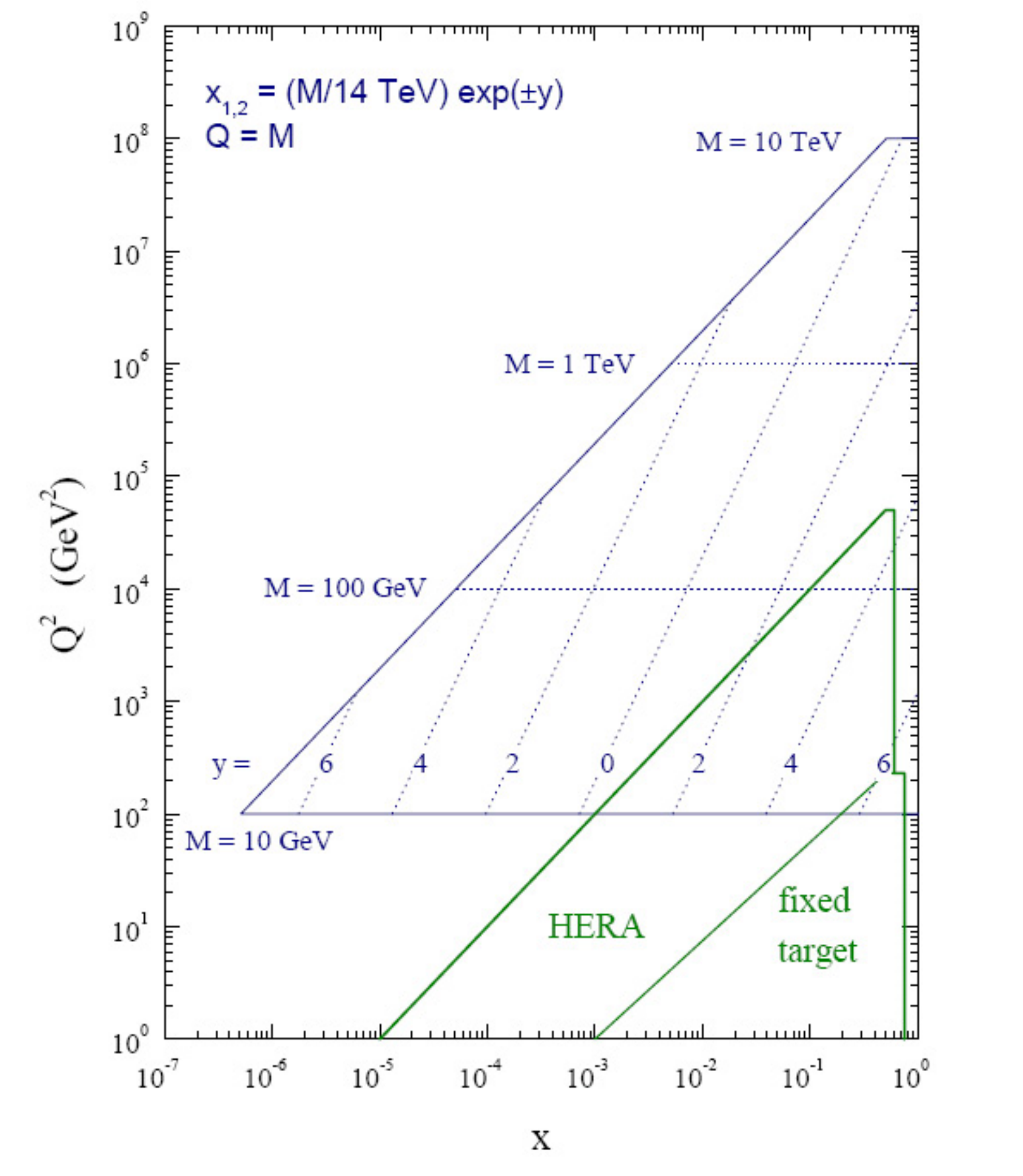}
\includegraphics[scale=0.3]{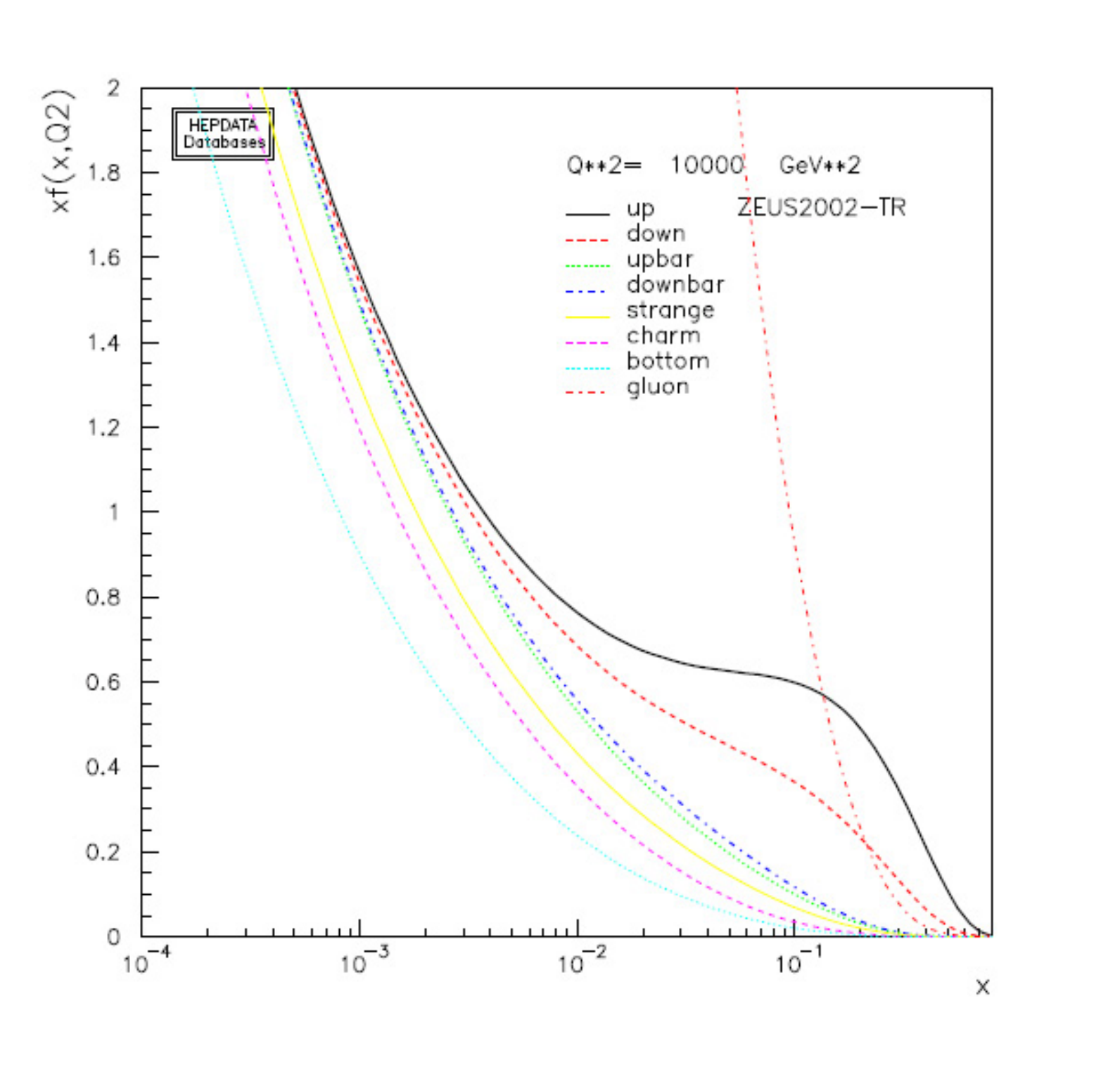}
\caption{Left: the region in the $Q^2$, $x$ plane exploread by LHC.
Right: example of PDF distribution at the electroweak scale
($Q^2=10,000$~GeV$^2$). }
\label{Stirling}
\end{figure}

It has been already mentioned in the introduction of this report,
that in hadronic collision the production cross section
can be described by the convolution of an {\it hard process}
with the parton density functions (PDFs).
In the $q \bar q \to W/Z$ process the momentum fraction of the
initial partons is given by $x_{1,2}=\frac{M_{W/Z}}{\sqrt(s)} \exp{\pm y}$
where the W (Z) mass is indicated by $M_{W/Z}$, the proton-proton
centre-of-mass is $\sqrt(s)$ and $y$ is the rapidity, already defined
in the introduction. 
The angular coverage of the apparatus is typically limited
in the pseudorapidity region $|\eta| < 2.5$ and since
in the relativistic limit the two variables ($y$ and $\eta$) are
equivalent the geometrical coverage translates in a rapidity
coverage in the plane described in Fig~\ref{Stirling}.

It is evident from this figure that the phase space region
explored by LHC is largely unknown and analysis of LHC data will be of
paramount importance in order to gain understanding.
The right-side figure shows the pdf for $Q^2 \approx M_{W/Z}$,
making evident that at the electroweak scale the gluons
are the dominant partons at LHC.

\begin{figure}
\centering
\includegraphics[scale=0.3]{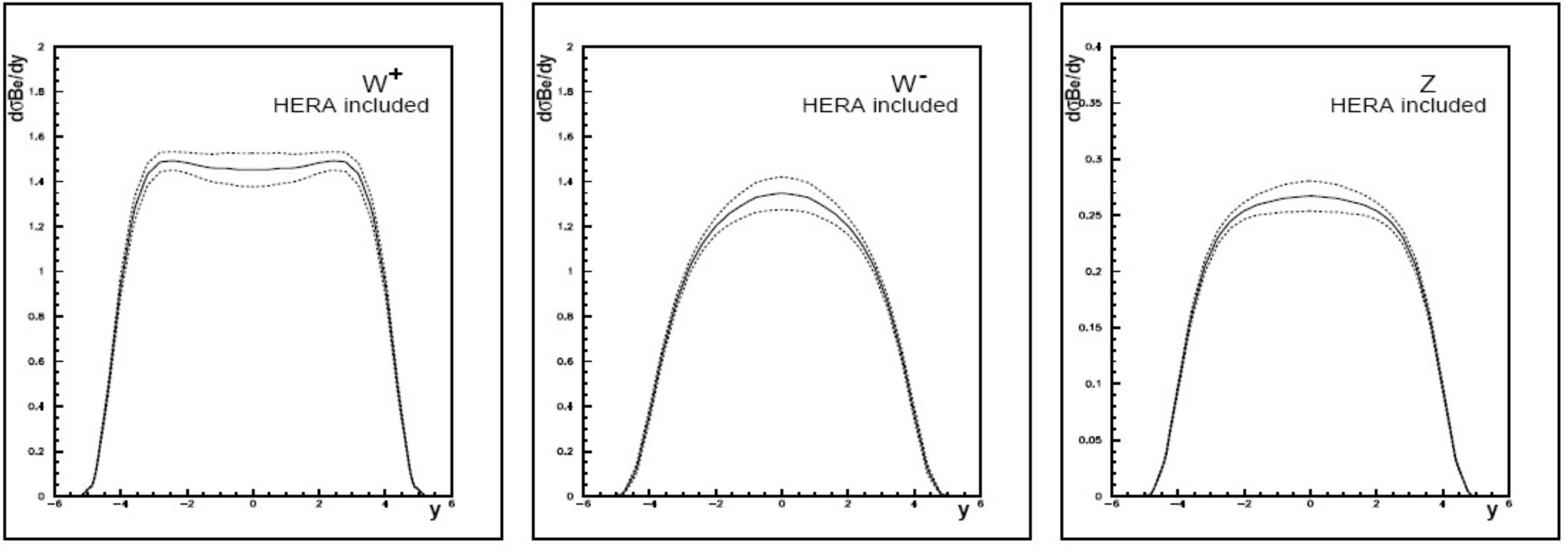}
\caption{The $W^+$, $W^-$ and $Z$ rapidity distributions
and their spread due to PDF uncertainties. The PDF fits
include recent HERA data.}
\label{Tricoli2}
\end{figure}

The present PDF fits~\cite{pdffits} are based on the data of
HERA and Tevatron. The resulting predictions for the $W^+, W^-$
and Z diffential cross sections are shown in Fig.~\ref{Tricoli2}.

The different $W^+$ and $W^-$ differential cross sections are due to the
structure of the weak charged current and to the presence
of the proton valence quarks. The $W^+$ cross section,
integrated in the apparatus acceptance,
is approximately 35\% higher than the $W^-$ one.
The uncertainty due to the limited knowledge of the PDFs 
on the total $W^+, W^-$ and Z cross section is around 5\% .

As the main source of uncertainty in the PDFs is due to the gluon
component, and this affects in the same way all vector bosons,
more robust predictions can be made when ratios are used:

\begin{eqnarray}
A_W=\frac{W^+ - W^-}{W^+ + W^-} \\
A_{ZW} = \frac{Z}{W^+ + W^-} \\
A_l = =\frac{l^+ - l^-}{l^+ + l^-}
\end{eqnarray}

\noindent where the last ratio concerns the leptons from $W^+ , W^-$ decay.
The ratios themselves can be used to constrain the quark PDFs
once LHC data will be there. A preliminary study~\cite{Tricoli}
indicates that already interesting improvements
can be obtained with 100 $pb^{-1}$.

\subsection{Measurement of the luminosity and parton luminosities}

The selections described in Section~\ref{WZemu} can be used to
measure the experimental W and Z cross sections through the
usual relation
 
\begin{equation}
\sigma = \frac{N-N_{bkg}}{\epsilon {\cal L}}
\end{equation}

\noindent where N is the number of selected events, $N_{bkg}$ is the
background contamination (expected to be small for these channels),
$\epsilon$ is the selection
efficiency and ${\cal L}$ is the integrated luminosity. 
The latter one can be calculated, with an uncertainty that is 
expected to be ${\cal O} (10\%)$ from the accelerator beam parameters:

\begin{equation}
{\cal L} = \frac{N^2 k f}{4 \pi \sigma_x \sigma_y} \times F
\end{equation}

\noindent where N is the number of protons in a bunch,
$k$ is the number of bunches,
$f$ is the beam revolution frequency (11 kHz at LHC),
F is a factor that accounts for the non-zero beam
crossing angle (about 0.9 at LHC) and 
$\sigma_x , \sigma_y$ are the horizontal and transverse
bunch widths at the interaction point.

A more precise determination of the luminosity is obtained
with dedicated forward detectors (roman pots) measuring
the rate of elastic scattering at very small transferred momentum.
From the optical theorem a relationship between
the rate of elastic scattering at zero transfer momentum
($\frac{d R_{el}}{dt}|_{t=0}$) and the total rate of $pp$
interactions ($R_{tot}$) can be obtained :

\begin{equation}
{\cal L} \frac{d R_{el}}{dt}|_{t=0} =
\frac{R^2_{tot}}{16 \pi^2} (1 + \rho^2)
\end{equation}

\noindent where $\rho$, amounting to about 0.1, is the ratio of 
the real to imaginary part of the elastic forward amplitude.
This method is potentially very precise, but requires a special
beam optics and low luminosity to avoid pile-up.

Alternatively the W and Z rates themselves can be compared to
the theoretical cross sections in order to extract the luminosity.
It is clear from previous Section that the main limitation would
be related to the knowledge of the PDFs, which are required
to compute the total W and Z expected cross sections.
Other theoretical uncertainties are related to the calculation
of the elementary $q \bar q \to W/Z$ process, in particular
to the knowledge of the NLO cross sections 
and of the EW corrections.

Another approach is to use the inclusive W and Z production 
to normalize other processes accessing the same parton phase space ($\Omega$).
Indeed the measurement of the W, Z rates correspond to the determination of the integral

\begin{equation}
\int_{\Omega} d x_1 d x_2 \sigma_{q \bar q \to W/Z} \times {\cal L} 
\times PDF(x_1,x_2, Q^2)
\end{equation}

\noindent which provides the so called "parton luminosity"~\cite{dittmar}.
By measuring cross section ratios the uncertainty on integrated luminosity
cancels out and the PDF uncertainty is greatly diminished.

\subsection{Determination of the W mass}

The precision electroweak measurements at LEP and SLC have shown that
the Standard Model is tested at one-loop level at the Z-pole.
On the other hand, the direct measurements of the W and top mass,
from LEP and Tevatron, provide an additional stringent test
and contribute to the global picture indicating a rather light
Higgs boson. Once the Higgs is discovered, it will be important 
to compare its mass with the predicted value: improving the
precision on the measurement of the W and top mass is therefore
important. The current W mass World Average has an uncertainty
of about 30 MeV~\cite{nlepewwg}.

The traditional method to determine the W mass at hadron colliders
is based on the measurement of the transverse mass 
\begin{equation}
m_T = \sqrt{2 p_T^l p_T^{\nu} (1- \cos\phi)}
\end{equation}
\noindent where $\phi$ is the angle, in the transverse direction,
between the lepton and the neutrino. The neutrino transverse
momentum, $p_T^{\nu}$, is assumed to be equal to the missing
transverse energy (missing $E_T$).  The transverse mass, compared to
other variables as the lepton transverse momentum ($p_T^l$),
has a reduced sensitivity
on the transverse motion of the W, but depends on the accuracy
of the $E_T$ measurement.

\begin{figure}
\centering
\includegraphics[scale=0.3]{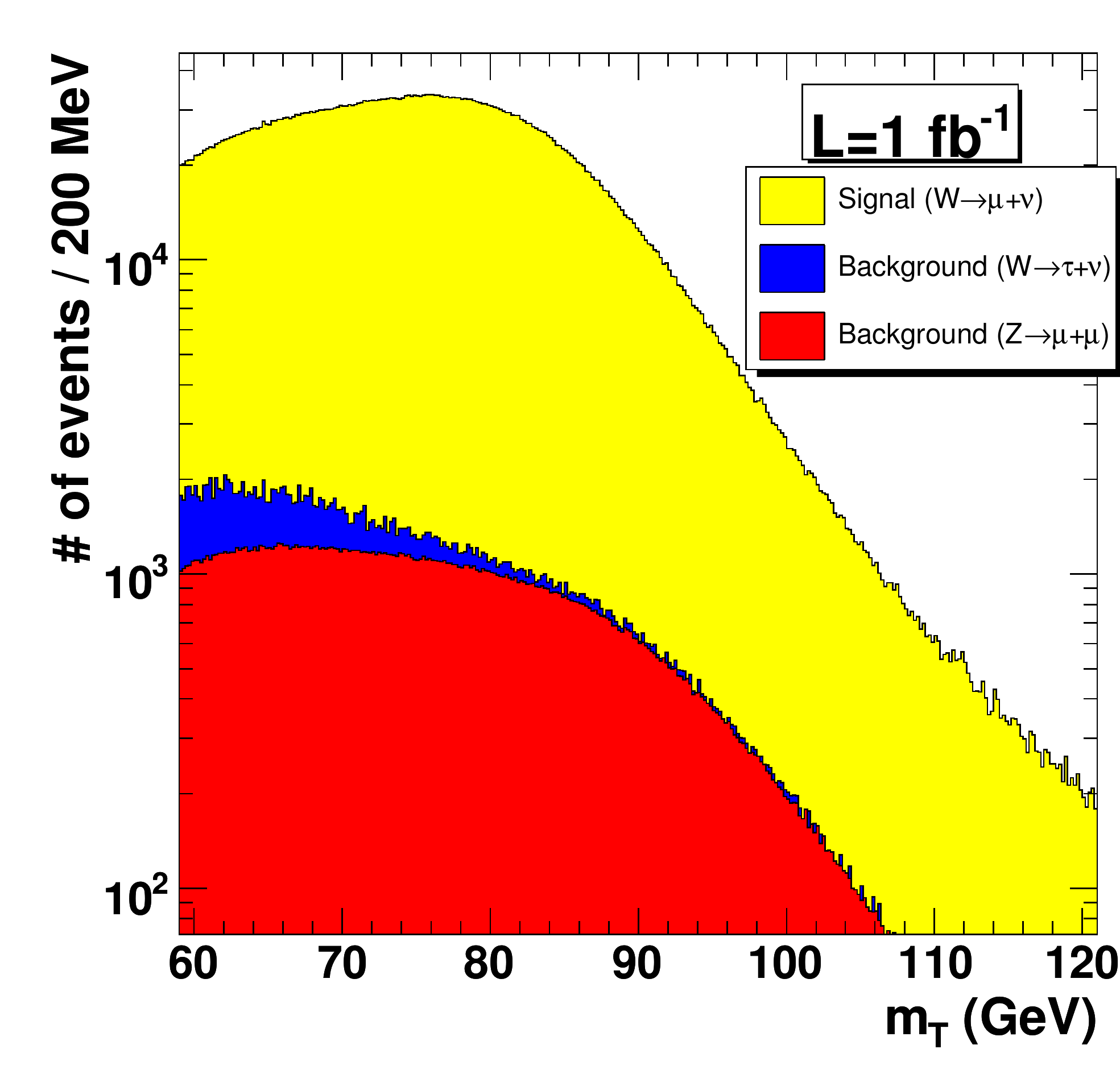}
\includegraphics[scale=0.43]{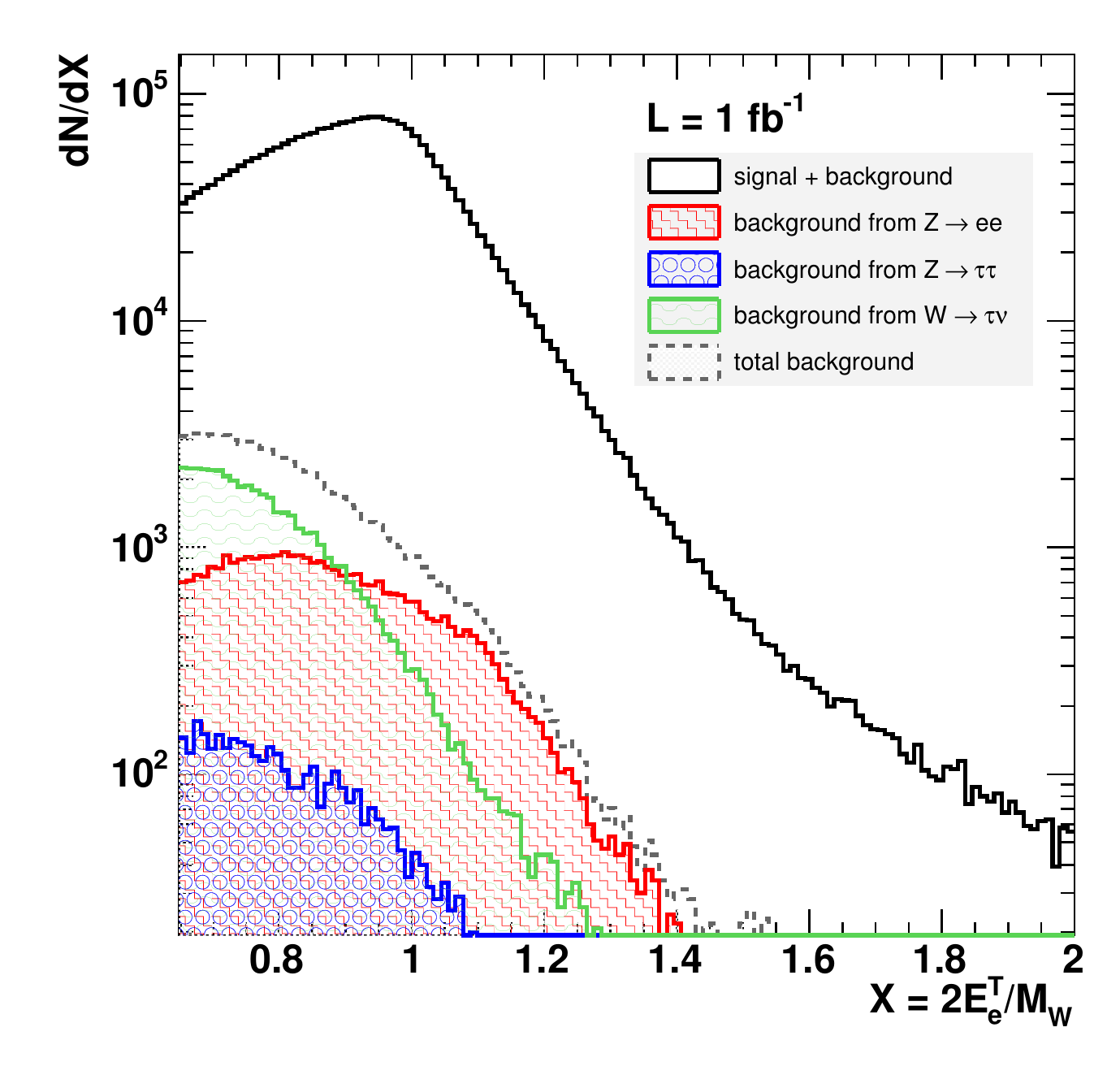}
\caption{Left: The electron scaled transverse energy
distribution for 1 $fb^{-1}$. The dominant W decay to electron
and the main backgrounds are shown.
Right: Transverse mass distribution for the same luminosity
in the muon channel.}
\label{fig_MW}
\end{figure}

With the huge W boson statistics expected at LHC, the statistical 
uncertainty on the W mass measurement will not be an issue;
using the transverse mass, for instance, an uncertainty of about 5 MeV
is expected with 1 $fb^{-1}$.
The measurement will be limited by systematic effects, in particular
by the knowledge of the lepton energy scale. 
Both experiments, Atlas and CMS, are developing techniques based 
on the large sample of Z leptonic decays to control the main
systematic effects, at the price of a larger statistical
uncertainty due to the lower Z production rate~\cite{AtMW}~\cite{CMSMW}.
For example the differential cross sections for a given $V=W,Z$ boson
observable $O^V$ can be used to set a relationship between W and Z bosons,
such as~\cite{Giele}

\begin{equation}
\frac{d \sigma^W}{d O^W} |_{predicted} =
\frac{M_Z}{M_W} R(X) \frac{d \sigma^Z}{d O^Z}|_{measured}
\end{equation}

\noindent where R(X) is given by theoretical calculations and the scaled
variable $O^Z = \frac{M_Z}{M_W} O^W$ is used.
Using this method, a Monte Carlo study~\cite{CMSMW} based on the electron channel
has shown that a statistical error of 40 MeV is foreseen at an integrated luminosity
of 1 $fb^{-1}$, with a similar instrumental uncertainty.
By combining two channels and two experiments one could potentially gain interesting information on the W mass already in the initial phase of LHC.

\subsection{Lepton pairs from Drell-Yan}

The production of lepton pairs from the process 
$q \bar q \to \gamma^*,Z \to \ell^+ \ell^-$,
usually called the Drell-Yan (DY) process ~\cite{DrellYan},
is dominated by the already-described on-shell Z production.
Above the Z pole, the DY cross section is steeply falling, as
can be seen from table~\ref{tab:DYAN}. The rapidity of the
lepton pair is related to the scaled momentum of the partons 
($x_{1,2}$) as $y = \ln\frac{x_1}{x_1}$ and the invariant mass
of the pair is $M_{ll}^2 = x_1 x_2 s$, where $s= 14$ TeV.
The production cross section can be written as
\begin{equation}
\frac{d^2 \sigma}{d M_{ll} d y} \approx 
\Sigma_{ij} (f_{i/p} (x_1) f_{j/p} (x_2) + (i \leftrightarrow j)) \hat\sigma
\end{equation}
\noindent where $f_{i/p} (x_k)$ is the probability to find 
a parton $i$ of momentum fraction $x_k$ in the proton and
$\hat\sigma$ indicates the $ij \to \ell^+ \ell^-$ subprocess.
At LHC the dominant $ij$ combinations are 
$u \bar u, \bar u u, d \bar d, \bar d d$, with the antiquarks
picked up from the sea.

\begin{table}
\begin{center}
\begin{tabular} {|c|c|c|c|}
\hline 
$M_{ll} \ge $ 160 GeV & $M_{ll} \ge $ 200 GeV  & $M_{ll} \ge $ 500 GeV  & $M_{ll} \ge $ 1 TeV \\
\hline 
5800 fb & 2500 fb & 100 fb & 6.6 fb \\
\hline 
\end{tabular}
\caption{Expected cross-sections for Drell Yan production at LHC. The cross
sections are computed with Pythia, using CTEQ5L for the PDFs, and are given for one lepton species. }
\label{tab:DYAN}
\end{center}
\end{table}

The measurement of Drell Yan production, below and above the Z-pole, 
will provide additional information on the PDFs. In the intial phase of LHC
the measurements, however, will be dominated by the low available statistics, 
especially for high-mass pairs. 
Trigger, reconstruction and selection efficiencies 
for high-mass pair are high~\cite{DYcms};  
the crucial experimental issues in the initial phase will be related
to the knowledge of the tracking alignment for the muon channel and to the knowledge
of the electromagnetic calorimeter calibrations for the electron channel.
High mass dilepton pairs provide a rich field 
of investigation for many new physics models, and searching for high-mass
dilepton peaks will be an important activity in the early phase of LHC.
A good control of alignments and calibrations will be important in order
to reduce the width of possible peaks,
opening the road to early discoveries. 

\subsection{Multiboson production}

In the initial phase of LHC, measurement of multiboson production
($WW$, $WZ$, $ZZ$, $W \gamma$ , $Z \gamma$) will constitute an
important step forward in understanding potential backgrounds to searches. 
At high luminosity these processes will allow improved measurements
of the Triple Gauge Couplings. The typical cross sections for
the five processes are given in Table~\ref{tab:multiboson}. Different
cross sections are expected at LHC for $W^+ Z$ ($W^+ \gamma$)
and $W^- Z$ ($W^- \gamma$), in the first case the yield
is typically 50\% higher. Triggering these events requires
the presence of an isolated lepton, and first measurements
will be made in fully leptonic channels with electrons and muons.
The proper W and Z leptonic branching ratios must be applied the 
cross sections given in Table~\ref{tab:multiboson}. 

\begin{table}
\begin{center}
\begin{tabular} {|c|c|c|c|c|}
\hline 
$WW$ & $WZ$ & $ZZ$ & $W \gamma$ & $Z \gamma$ \\
\hline 
120 pb & 50 pb & 16 pb & 350 fb & 35 fb \\
\hline 
\end{tabular}
\caption{Expected cross-sections for multiboson production at LHC. Typical NLO cross
sections, using CTEQ5L for the PDFs, are given. The $W^+$ and $W^-$
cross sections are summed over. The $W \gamma$, $Z \gamma$ cross sections
are given for $p_T^{\gamma} \ge 100$~GeV.}
\label{tab:multiboson}
\end{center}
\end{table}

\begin{figure}
\centering
\includegraphics[scale=0.3]{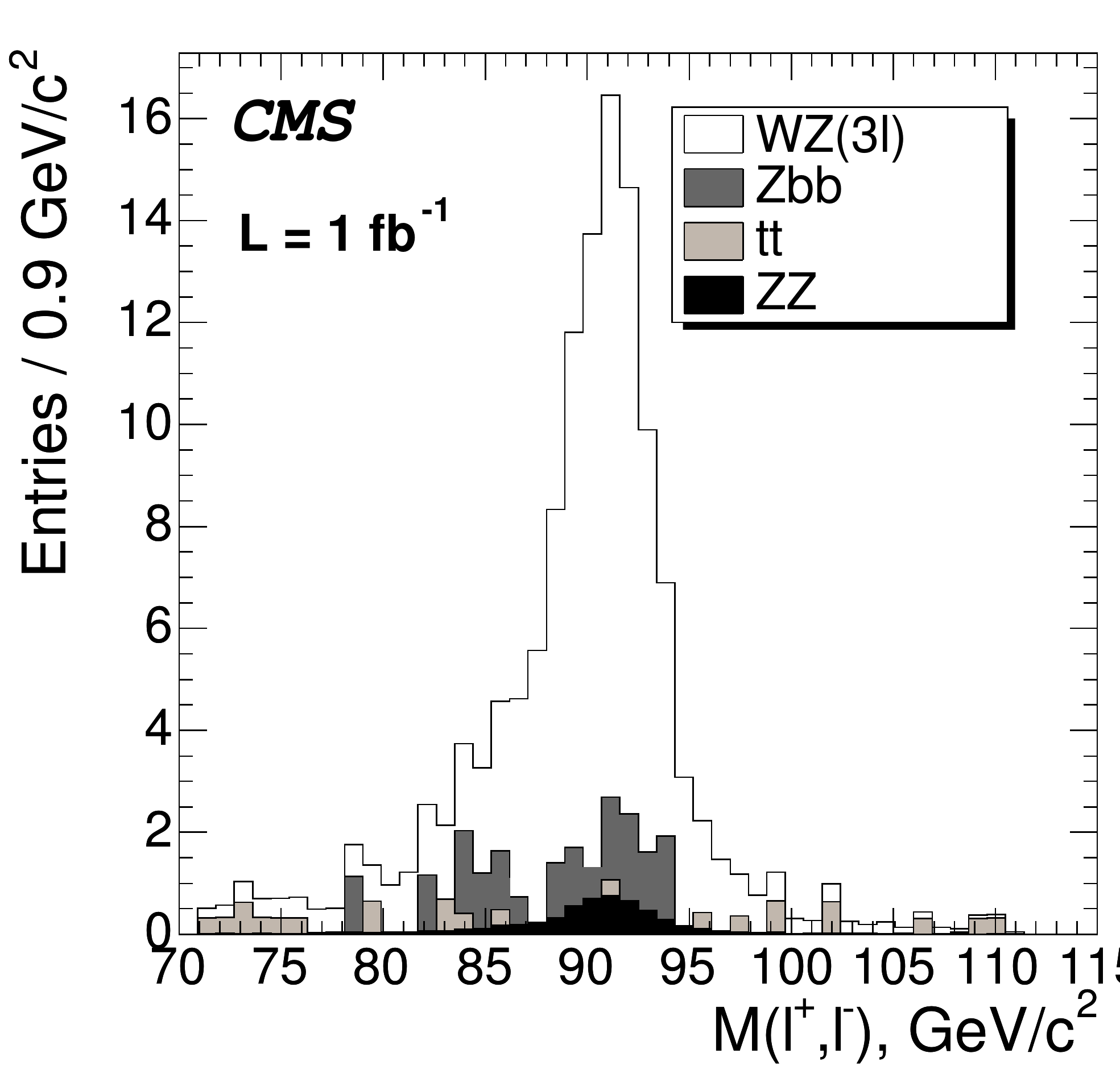}
\caption{Distribution of the $\ell^+ \ell^-$
invariant mass from the $W^{\pm} Z$ selection described
in Ref.~\cite{TGCcms}, for an integrated luminosity of 1 $fb^{-1}$.}
\label{fig:multib}
\end{figure}

A recent study has shown~\cite{TGCcms} that the $WZ$ process, 
in the three-lepton decay channel ($\ell = e, \mu$), can
already provide a distinctive signature at 1 $fb^{-1}$,
as shown in Fig.~\ref{fig:multib}. 
This measurement can provide a benchmark for multiboson production
already in the early phase of LHC. 

\section{Top quark physics}
\subsection{Introduction}
The top quark, discovered at Fermilab in 1995~\cite{topdiscovery}, completed the
three generation structure of the Standard Model (SM) and opened up the
new field of top quark physics. In hadron-hadron collisions the top quark is produced 
 predominantly through strong interactions, and as single $t$ or $\bar t$ in electroweak interactions. Top then decays rapidly without forming hadrons. The relevant
CKM coupling is already determined by the (three-generation) unitarity
of the CKM matrix.  Yet the top looks unique in between the other quarks because of its large
mass, about 35 times larger than the mass of the next heavy quark, and
close to the scale of electroweak (EW) symmetry breaking. This unique
property raises a number of interesting questions. Is the top quark
mass generated by the Higgs mechanism as the SM predicts and is its
mass related to the top-Higgs-Yukawa coupling? Or does it play an even
more fundamental role in the EW symmetry breaking mechanism? If there
are new particles lighter than the top quark, does the top quark decay
into them? Could non-SM physics first manifest itself in non-standard
couplings of the top quark which show up as anomalies in top quark
production and decays? 
\subsection{$t \bar t$ pair production and decay}
At LHC, top quarks will be mainly produced as unpolarised $t \bar t$ pairs via pair production mechanisms.
At the center-of-mass energies of 14 TeV the hard process $ gg \rightarrow t \bar t$ contributes to 90$\%$ of the total 
$t \bar t$ cross-section (the quark annihilation process accounts for the remaining 10$\%$) according to the large gluon
component in the proton parton distributions. 

 The cross-section for production at the LHC has been calculated up to 
NLO order including NLL soft gluon resummation, and results in about 833$\pm$100 pb, where the uncertainty 
reflects the theoretical error obtained from varying the renormalisation scale by a factor of two~\cite{scale}.
 This translates to 
83,000 top-quark pairs in a sample of 100 pb$^{-1}$ and of the order of 10$^7$ top quark pairs produced per 
year before any selection or detection criteria are applied.

We therefore expect to examine the top quark
properties with significant precision.  
In the SM the decay of top-quarks takes place almost exclusively 
through the t$\rightarrow$Wb decay mode. The experimental signature for $t \bar t$ pairs is therefore 
defined by how the two daughter W-bosons decay. A W-boson decays in about 1/3 of the 
cases into a lepton and a neutrino. All three lepton flavors are produced at equal rate. 
In the remaining 2/3 of the cases, the W decays into a quark-antiquark pair, and the abundance of 
a given pair is determined by the magnitude of the relevant CKM matrix elements. 
Specifically, the CKM mechanism suppresses the production of b-quarks as $|V_{cb}| \simeq 1.7 
\cdot  10^{-3}$. Thus the quarks from W-boson decay can be considered as a clean source of light quarks.

The following experimental signatures can be defined:
\begin{itemize}
\item {\it Fully leptonic:} it counts 1/9 of the $t \bar t $. Both W-bosons decay into a lepton-neutrino
pair, resulting in an event with two leptons, two neutrinos and two b-jets. This mode is identified 
 by requiring two high P$_T$ leptons and the presence of missing E$_T$, and it allows to obtain a clean 
 sample of top events. However, this sample has limited use in probing the top reconstruction capability 
of the experiment, due to the two neutrinos escaping.
\item {\it Fully hadronic:} represents 4/9 of all the $t \bar t $ decays. Both W's decay hadronically, which gives
 six jets in the event: two b-jets from the top decay and four light jets from the W boson decay. In this case, we 
 do not 
 have a high P$_T$ lepton to trigger, and the signal is not easily distinguishable from the abundant 
 SM QCD multi-jets production, which is expected to be order of magnitudes bigger than the signal. Another 
 challenging point of this signature is the presence of a high combinatorial background when reconstructing 
 the top mass.
\item {\it Semi-leptonic:} Again, 4/9 of the whole decays. The presence of a single high P$_T$ lepton 
 allows to suppress the SM W+jets QCD background. The P$_T$ of the neutrino can be reconstructed as it is the 
 only source of missing E$_T$ for signal events.
 A schematic view of the topology of these events is shown in 
fig.~\ref{fig:topology}
\end{itemize}
\subsection{Top studies at LHC}
As the LHC startup is approaching, both the ATLAS and CMS experiments have concentrated on studies 
  to be performed with a very low integrated luminosity, typically 
with only 10 or 100  pb$^{-1}$ of data. In this frame, two main analysis streams are the 
top mass and cross-section measurements. 
 Apart from the intrinsic value of these two measurements, it should not be forgotten that the 
 top pair production process will be valuable for the in-situ calibration of the LHC detectors during the
commissioning stage. 
 The early top samples selected will be a critical tool for many applications, for example they will be
 useful  to calibrate the jet energy scale (if one imposes the value of the reconstructed di-jet peak to 
be centered at the world average value of the W mass, a precision of about 1$\%$ can be reached~\cite{atlas_calib}). 
Top events can be used to estimate and calibrate the b-tagging efficiency. 
In addition, a top sample can be an excellent pool
to study the lepton trigger or to calibrate the missing E$_T$, using the W mass constraint in the event.
The relevant processes for any study which investigates the production and decay of $t \bar t$ events 
are the signal itself, but also the background from 
Drell-Yan (DY)+jets, dibosons, W/Z-boson+jets and QCD multi-jet production.
 In CMS (ATLAS) the Alpgen~\cite{nalpgen} (mc@nlo)~\cite{mcatnlo}) 
generator is used for the simulation of the $t \bar t$ signal. 
Both experiments use Alpgen for DY and W/Z+jets backgrounds. 
Di-boson and QCD background events are generated with Pythia~\cite{npythia}.  
In the simulations used for all the analyses covered in this part,  the limited understanding of the two detectors 
during the initial period of data taking has been taken into account. This, by using realistic scenarios of misalignment of the tracking systems and 
miscalibration of the calorimeters.\\ 
Before concentrating on the first top quark measurements which will be done, 
it is quite important to list and describe the main sources of 
systematic uncertainties present in all the analyses that will be described.\\
\subsubsection{Experimental systematic uncertainties }
The following uncertainties have been evaluated by the ATLAS experiment.\\
{\it Luminosity.}
At the LHC start-up in 2008 only a rough measurement of the machine parameters will be available.
The expected uncertainty on the luminosity during this phase will be of the order of 20-30$\%$. From
2009 onwards, a better determination of the beam profiles using special runs of the machine will lead
ultimately to a systematic uncertainty of the order of 5$\%$. Further, in 2009 the proposed ALFA detector
will come on-line to measure elastic scattering in the Coulomb-Nuclear interference region using special
runs and beam optics, determining the absolute luminosity with an expected uncertainty of the order of
3$\%$. The optical theorem, in conjunction with a precise external measurement of the total cross-section,
can achieve a similar 3$\%$ precision.\\
{\it Lepton identification efficiency.}
The lepton identification efficiency error is expected to be of the order of 1\% for electrons and muons for
the first 100 pb$^{-1}$ of integrated luminosity.\\
{\it Lepton trigger efficiency.}
The lepton trigger efficiency is measured from data using Z events. The uncertainty is expected to be of
the order of 1$\%$. \\
{\it Jet energy scale (JES).}
In the difficult hadron collision environment, the determination of the jet energy scale is rather challenging.
While several methods are proposed, such as using $\gamma$+jet events to propagate the electromagnetic
scale to the hadronic scale, the jet energy scale depends on a variety of detector and physics effects. This
includes non-linearities in the calorimeter response due, for example, to energy losses in ``dead'' material,
and additional energy due to the underlying event. Energy lost outside the jet cone can also affect the
measured jet energy. Effects due to the initial and final state radiation (ISR/FSR) modelling could also
affect the JES but they are evaluated separately. The ultimate goal in ATLAS is to arrive at
a 1\% uncertainty on jet energy scale though such performance is only reachable after several years of
study. To estimate the sensitivity of the analyses to the uncertainty on the jet energy scale in early data we
have repeated them while artificially rescaling the energies of the jets by $\pm$ 5$\%$. The resulting variation
in the analysis measurement (i.e. cross-section, mass etc.) gives a good measure of the
systematic uncertainty due to the jet energy scale.\\
{\it b-tagging uncertainties.}
The use of b-tagging in $t \bar t$ and single top events is essential in order to reduce the backgrounds, in
particular that from W+jets, and the combinatorial background when reconstructing the top. At the
beginning of data taking the b-tagging performance will need to be understood and $t \bar t$ events will be
used as a calibration tool for the determination of the b-tagging efficiency. To avoid having a large
dependence on the b-tagging efficiency in the early days of data taking we have studied methods to
extract the cross-section and the top mass without applying b-tagging. The uncertainty on the b-jet
efficiency is currently estimated to be of the order of $\pm$5$\%$.\\
{\it ISR and FSR systematics.}
More initial and final state QCD radiation increases the number of jets and affects the
transverse momentum of particles in events. Selection cuts for top events include these quantities, therefore
ISR and FSR will have some effect on the selection efficiency. In order to evaluate the effect of
the ISR and FSR systematics, several studies have been performed using the AcerMC~\cite{acermc}  generator
interfaced with the PYTHIA~\cite{npythia} parton showering.
Samples of $t \bar t$ and single top events with separate variations of the PYTHIA ISR and FSR parameters
have been generated. The study was limited to parameters which have been shown to have the biggest
 impact on event properties at the generator level. The choices of the parameters depend on the analysis\\
{\it Parton density uncertainties.}
The systematic error due to the parton density functions (PDF) uncertainties is evaluated on $t \bar t$ signal
samples. Both the PDF error sets CTEQ6M and MRST2002 at NLO are used. Both sets
have positive and negative error PDFs. In order to evaluate the systematic effect on an observable,
 the  approach  proposed in reference~\cite{pdf} has been adopted.\\


\subsubsection{Top mass determination}
\begin{figure}[!ht]
  \begin{center}
    \includegraphics[width=6.5cm]{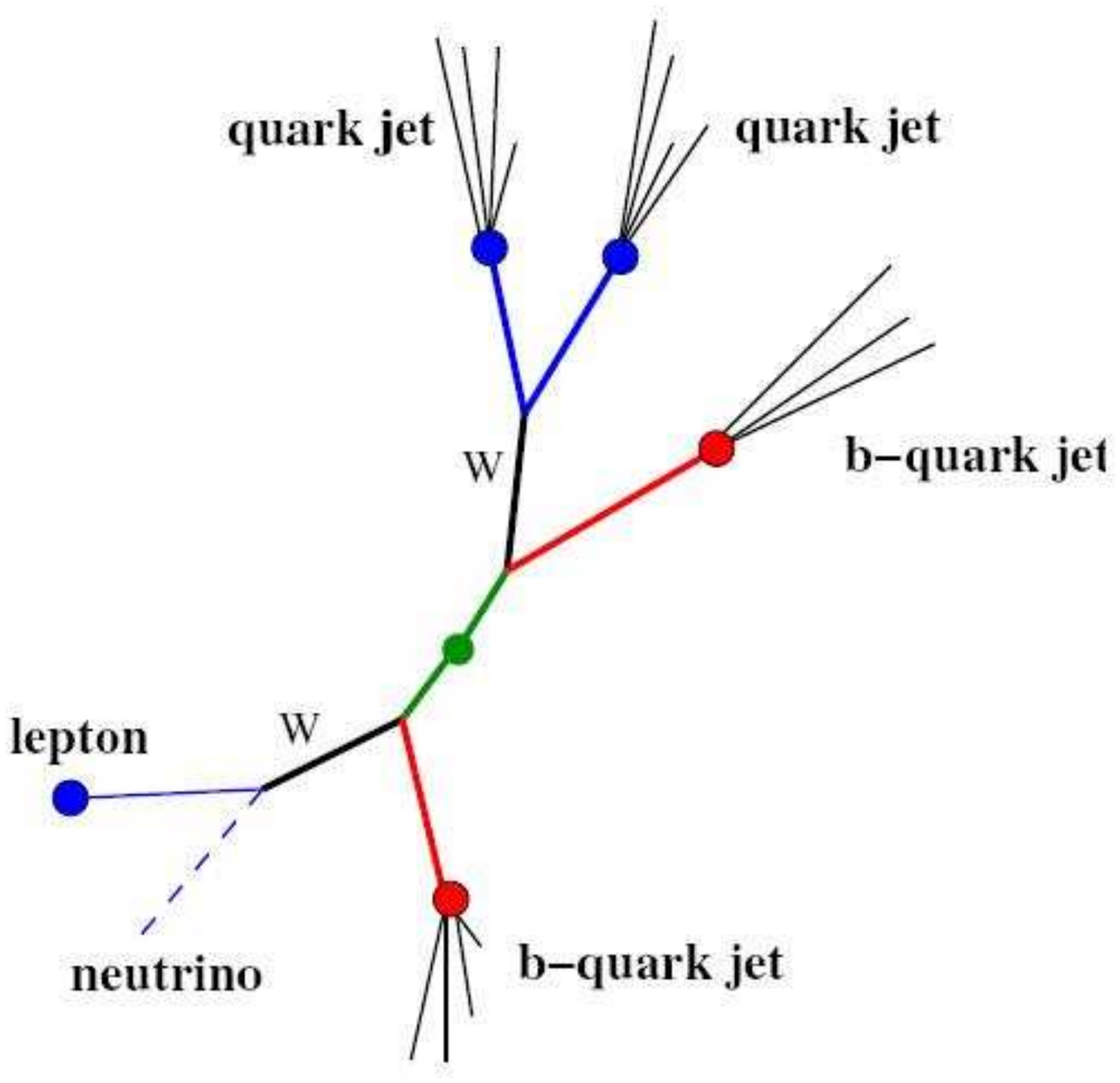}
    \includegraphics[width=6.5cm]{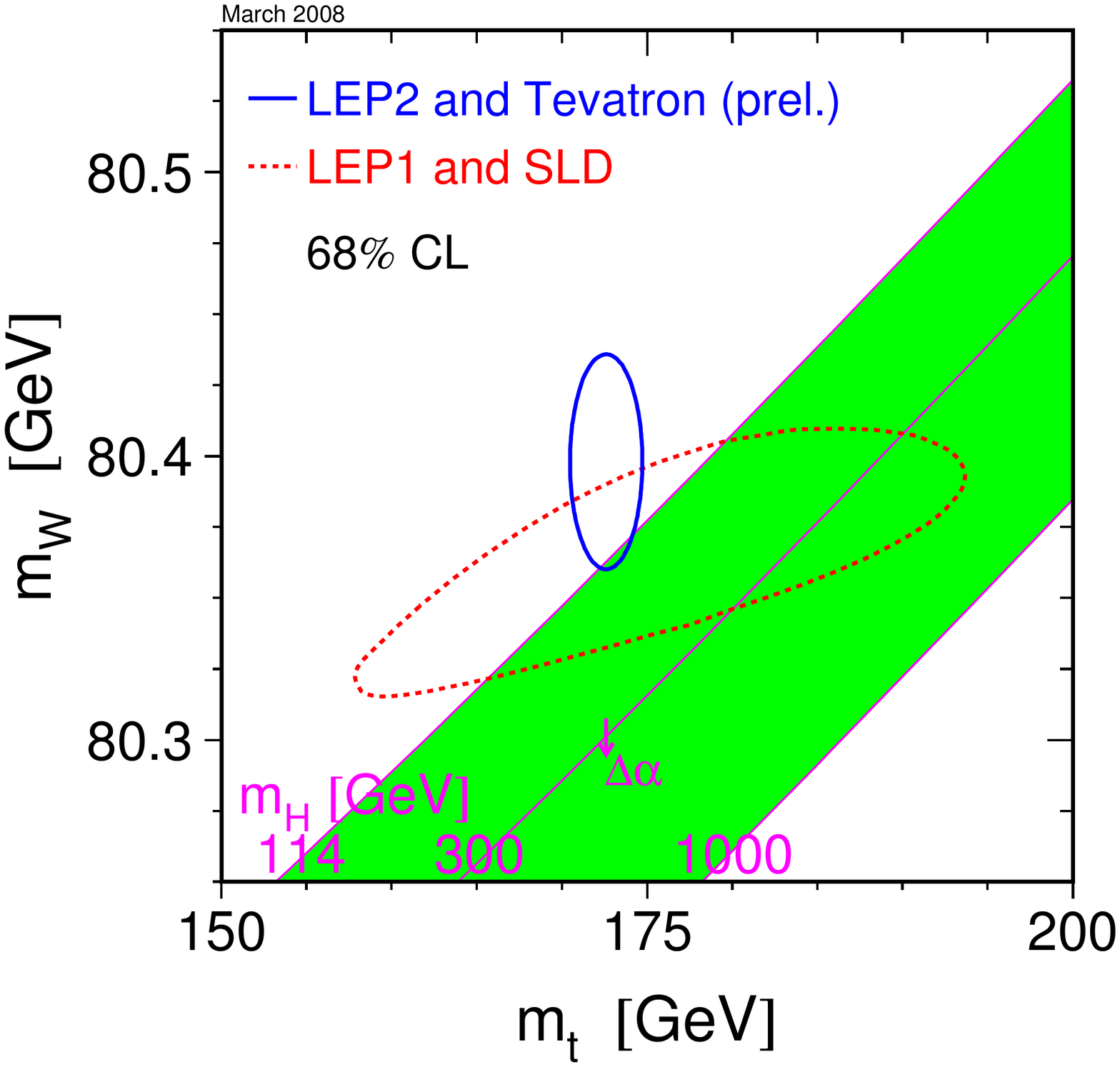}
  \end{center}
  \caption{Left plot: Schematic event topology of a single lepton top event. Right plot: 
Allowed region in the (M$_W$,M$_{top}$) plane.}
  \label{fig:topology}
\end{figure}

EW precision observables in the SM and in the Minimal Supersymetric
Standard Model (MSSM) depend on the value of the top mass (M$_{top}$); therefore, a high accuracy in
the measurement of M$_{top}$ is needed for consistency tests of the SM, constraints on
the Higgs mass (M$_H$) within the SM and a high sensitivity to physics beyond the SM. The
most important M$_{top}$-dependent contribution to the EW observables arises via the
one-loop radiative correction term $\Delta r$ [1], related to the W mass through the following relation :
$M_W$ = $[(\pi \alpha)/(\sqrt{2} G_F sin^2\theta_W)] \cdot (1 +\Delta r)$. M$_{top}$ appears in $\Delta r$ 
via terms proportional to M$^2_{top}$/M$^2_Z$, while the
Higgs mass gives rise to terms proportional to log(M$_H$/M$_Z$): the dependence on M$_H$ 
 is much weaker than the dependence on M$_{top}$.
The relation thus obtained is used as an indirect estimate of M$_{Higgs}$, relying on W
boson and top quark masses measurements as accurate as possible. The current value for  
M$_{top}$ = 172.6 $\pm$ 1.4. The
allowed region in the (M$_W$, M$_{top}$) plane is
 displayed in the right plot of fig.~\ref{fig:topology}, for different 
M$_{Higgs}$, in the SM and in the MSSM.

In order to ensure a similar contribution to the indirect measurement of the Higgs mass, the
precision on M$_W$ and M$_{top}$ must fulfill the following relation: 
$\Delta M_{top}$ $\simeq$ 0.7$\cdot$10$^{-2}$ $\Delta M_W$. 
At LHC, we expect to reach an accuracy of 15 MeV on M$_W$ and 1~GeV on M$_{top}$ . With these precision
measurements, the relative precision on a Higgs boson mass of 115~GeV would be of the order
of 18$\%$.

The lepton plus jets channel will provide a large and clean sample of top events and is probably
the most promising channel for an accurate measurement of M$_{top}$. The main background
 is due to W+jets and Z+jets from QCD, single top and $t \bar t$ events with a different decay mode, di-boson events.
The QCD production of $pp \rightarrow$ $b \bar b$ is characterized by
a cross-section of about 100 $\mu$b, and can therefore be an important
background for the signal. Requiring the presence of a high P$_T$
lepton and missing energy can reduce its contribution, but since the
cross-section difference with the signal is so important, there might
be QCD events with a fake lepton and/or bad missing energy
reconstruction that may pass these requirements as well.
The rate for extra (medium) electrons  per jet is roughly 
1.0 $\cdot 10^{-3}$ and is divided between semi-leptonic B(D) decays and 
true fakes, i.e hadronic objects identified as electrons. 
The origin of extra isolated muons is dominated by semi-leptonic 
B decays, i.e. by the presence of hard b-quarks. The isolated muon rate 
per b-parton reaches a few times $10^{-3}$ for b-parton momenta around 
40~GeV, while the fake rate is only a few times $10^{-5}$.
By studying their origin and dependence on jet/parton kinematics like 
the P$_T$, $\eta$, jet multiplicity and quark content of the jet, 
 an estimate of the fraction of multi-jet events that will 
pass the lepton requirement in the event selection can be obtained. 
The validity of this approach has been checked using a large sample of di-jet events 
at various transverse momenta.
One of the strategy developed by ATLAS to measure M$_{top}$, is to select events 
by requiring one isolated lepton (e or $\mu$) with P$_T$ $>$ 20~GeV,
missing E$_T$ $>$ 20~GeV, and at least four jets with P$_T$ $>$ 40~GeV, of which two of them are
required to be tagged as b-jets.
After these cuts, a S/B$\simeq$5 is obtained~\cite{atlastopmass}.

M$_{top}$ is then estimated from the reconstruction of the invariant mass of a three-jet
system: the two light jets from the W and one of the two b-jets. The determination of this combination
of three jets proceeds in two steps : the choice of the two light jets, and the choice of the
b-jet associated to the reconstructed hadronic W.
Events kept after the selection described above have at least two light jets above a given threshold
on their transverse momentum. In a first step, the hadronic W candidates are selected in a mass
window of $\pm$5$\sigma_{mjj}$ around the peak value of the distribution of the invariant mass of the light jet
pairs, made with events with only two light jets ($\sigma_{mjj}$ is the width of this distribution).
In order to reduce the incidence of a light-jet energy mis-measurement (due to the energy lost out
of cone) on the precision of the top mass measurement, an in-situ calibration of these jets is performed,
through a $\chi^2$ minimization procedure. This minimization is applied event by
event, for each light-jet pair combination. The $\chi^2$  is the
sum of three terms: the first (and leading) one corresponds to the constrain of the jet pair invariant
mass m$_{jj}$ to the PDG W mass; the others correspond to the jet energy correction factors,
$\alpha_i$ (i =1,2), to be determined by this minimization which includes
 the resolution on the light jet energy ($\sigma_i$ (i =1;2)).
The $\chi^2$ is minimized, event by event, for each light jet pair; the light jet pair j$_1$
, j$_2$ corresponding to the minimal $\chi^2$ is kept as the hadronic W candidate. 
This minimization procedure also leads
to the corresponding energy correction factors $\alpha_1$, $\alpha_2$. The hadronic W is then reconstructed with
the light jets chosen by this $\chi^2$ minimization.

\begin{figure}[!tb]
  \begin{center}
    \includegraphics[width=6.5cm]{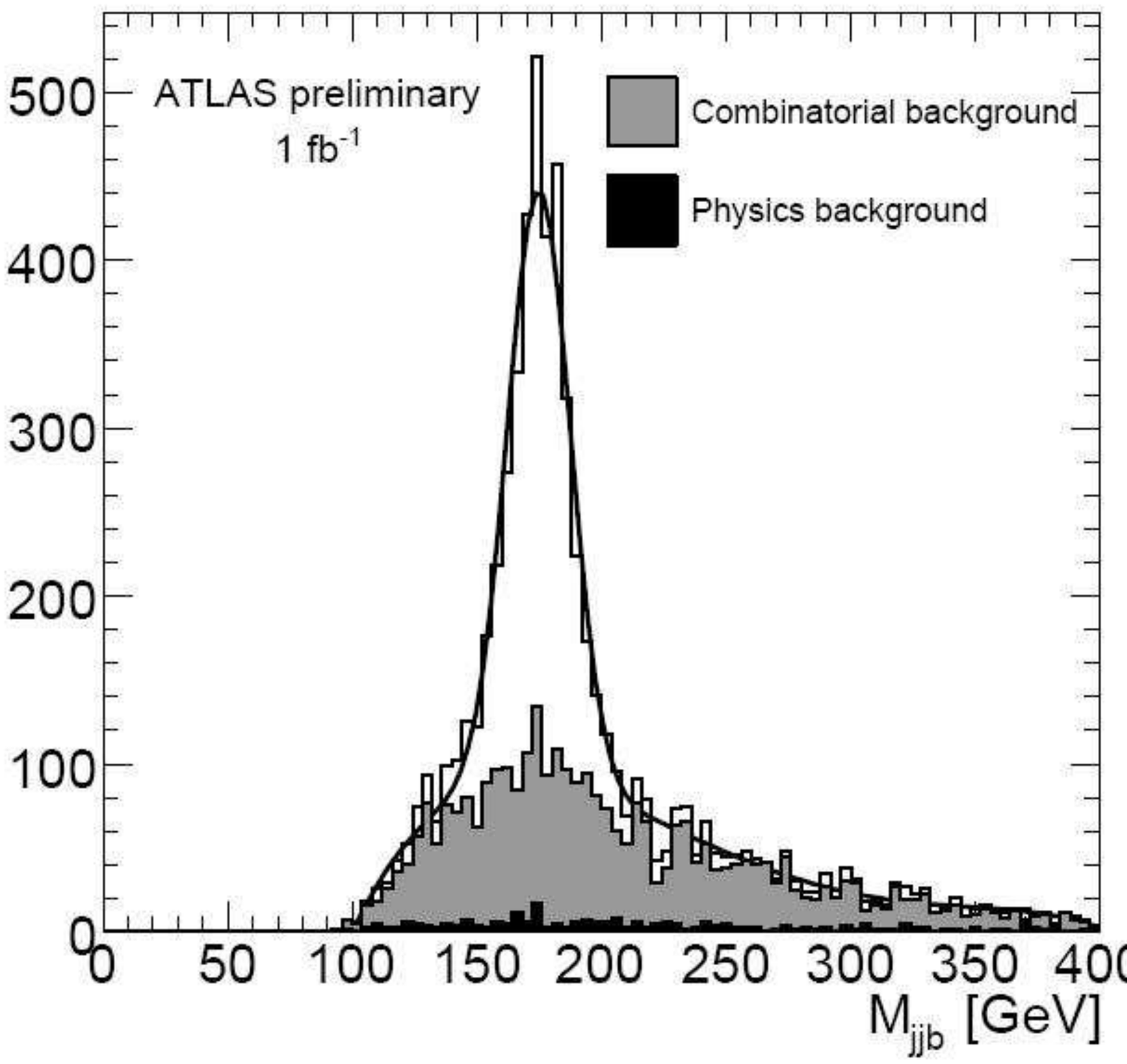}
    \includegraphics[width=6.5cm]{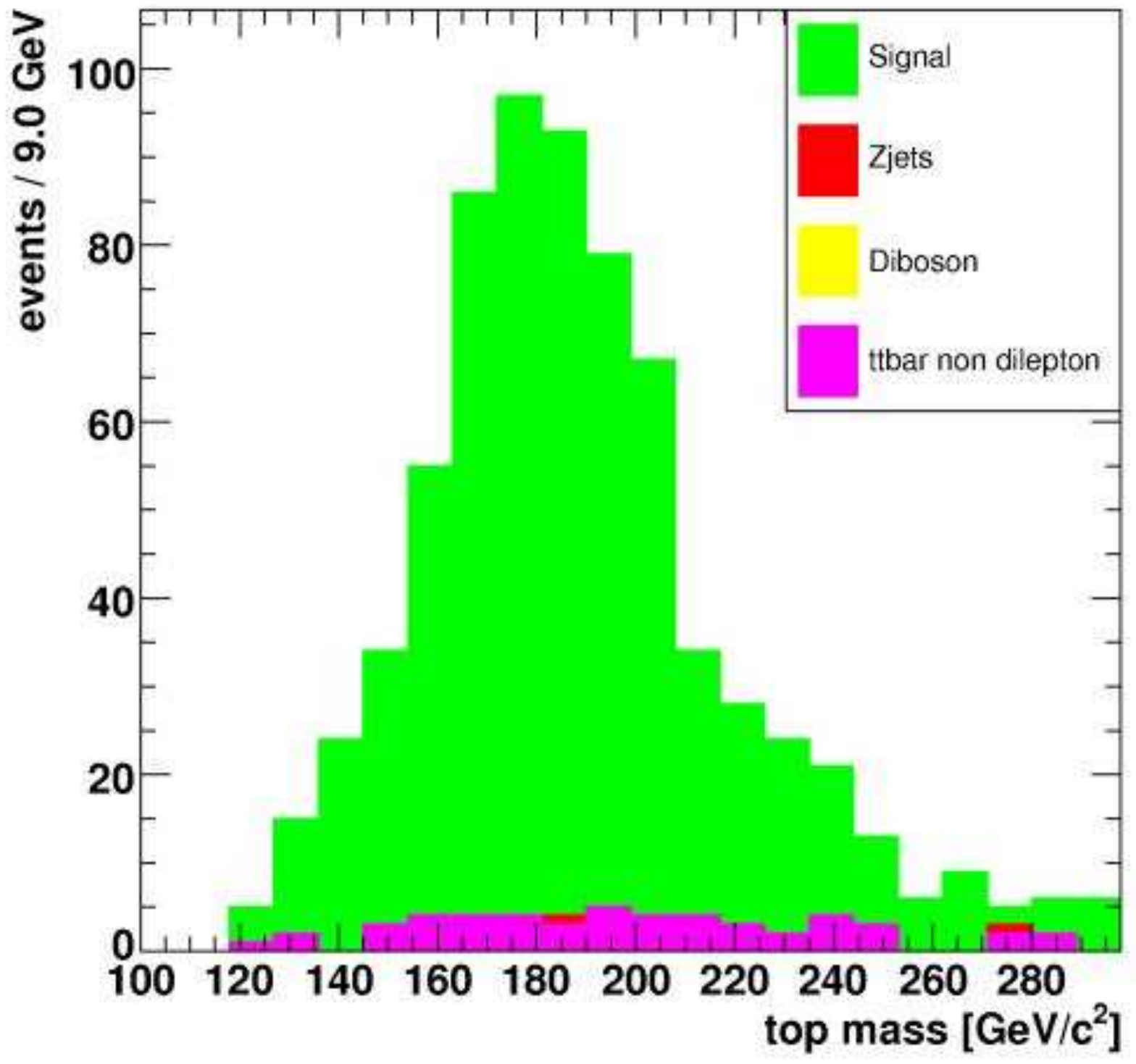}
    \caption{Left plot: Reconstructed top mass in the single lepton channel (ATLAS).
    Right plot: Reconstructed top mass in the di-lepton channel (CMS).}
    \label{fig:Topmass}
  \end{center}   
\end{figure}

Several methods have been investigated to choose the b-jet among the two candidates, and
the one giving the highest purity has been kept: the b-jet associated to the hadronic W is the one
leading to the highest P$_T$ for the top.
The reconstructed three jets invariant mass is shown in the left plot of fig.~\ref{fig:Topmass}, fit to
the sum of a Gaussian and a polynomial. For 1 fb$^{-1}$, the Gaussian fit has its mean at 175 $\pm$ 0.2~GeV
and a width of 11.6 $\pm$ 0.2~GeV.

An alternative method for the top mass measurement in the lepton plus jets channel consists
in reconstructing the entire $t \bar t$ final state, in order to reduce the systematic error due to FSR. The
hadronic part is reconstructed in a similar way to the previous section. The leptonic side can not be
directly reconstructed due to the presence of the undetected neutrino, but can be estimated with these three
steps: 1) assuming that P$_T$($\nu$ )=missing E$_T$ 2) evaluating P$_z$($\nu$ )
by constraining the invariant mass of the lepton-neutrino system to the
PDG W mass value: this kinematic equation leads to two P$_z$($\nu$ )solutions 3) associating the remaining 
b-jet to the reconstructed Ws.
The systematics errors for the top mass reconstruction are listed in Table~\ref{tab:sys}, assuming 1 fb$^{-1}$ of 
 integrated luminosity.
\begin{table}[!h]
 \begin{center}
 \begin{tabular}{ | c | c | c |}
    \hline     
    \hline    
    Source of uncertainty       &  Hadronic top              \\
                                &  $\delta$M$_{top}$ (GeV/c$^2$)  \\ 
     \hline                                                               
 Light jet energy scale (1$\%$) &  0.2                              \\
 b-jet energy scale  (1$\%$)    &  0.7                                    \\
 b-quark fragmentation          &  $<$0.1                                  \\
     ISR/FSR                    &  $\simeq$ 0.3                             \\
     \hline
\hline
 \end{tabular}
 \end{center}
 \vspace{-1em}
 \caption{Systematic errors on the top mass measurements in the lepton+jets channel, for 1 fb$^{-1}$ of integrated 
luminosity.}
 \label{tab:sys} 
 \end{table}

Requiring missing E$_T$, two high P$_T$ leptons and 2 b-tagged jets, and applying a Z-mass veto, 
one can also reconstruct the top mass in the di-lepton channel, as done by CMS~\cite{cmstopmass}. In this case 
the event is under-constrained, so that M$_{top}$ and the longitudinal direction of the neutrinos have 
 to be assumed. The system has to be solved analytically, by generating many Monte Carlo samples with 
 different top masses, stepping in top mass values between 100 and 300~GeV, and weighting the event solutions 
according to the missing E$_T$ measured and the expected neutrino distributions. The algorithm ends with the choice 
 of the most likely M$_{top}$. The mass spectrum for the most likely solution is shown in the right plot of 
 fig.~\ref{fig:Topmass}. The overall uncertainty is of about 4.5 (1.2)~GeV for less than 1 (10)fb$^{-1}$ of data, 
mainly due to the effect 
 of ISR and FSR and of the JES.

In conclusion, with the current simulations, M$_{top}$ is expected to be measured with high accuracy,
already using 1 fb$^{-1}$ of data.
For data commisioning without the use of b-jets tagging, a mass accuracy of 3.5~GeV is expected, assuming an initial
JES uncertanty of 5$\%$. Good consistency can be validated between channels already at low integrated luminosity.
At higher luminosities independent mass measurements can be made which are less
sensitive to jet modelling.


\subsubsection{Top cross section determination}
The determination of the top pair production cross-section 
is linked to the intrinsic properties of the top quark and its electroweak interactions. 
Cross-section measurements are also an important test of perturbative QCD 
at high P$_T$, as non-SM top quark production (for example 
resonant top-quark production) can lead to a significant increase of the cross-section.
New physics may also modify the cross-section differently in various decay channels, as
for example predicted by Supersymmetric models~\cite{gunion} with charged Higgs particles, or super partners 
 to the top-quark. 
Presently, the measurements performed at Tevatron are in good agreement
with the theoretical predictions.  
With the collected luminosity of  1~fb$^{-1}$ the errors
have been sizeably reduced and in some of the decay channels 
reached the $15\%$.
From a combination of all results an experimental error of the order of the 
theoretical error is expected. 
 
\begin{figure}[!htb]
  \begin{center}
    \includegraphics[width=6cm]{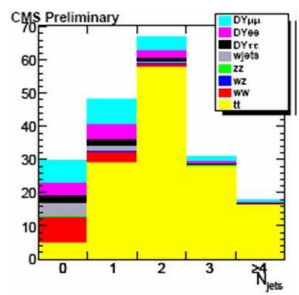}
    \caption{Jet multiplicity distribution for the di-lepton analysis developed by CMS with 10 pb$^{-1}$, 
for the three channels combined. The $ee$ and $\mu\mu$ channels introduce a significant 
 Drell-Yan background component.}
    \label{fig:topcms10pb}
  \end{center}   
\end{figure}

CMS proved that by requiring two opposite-charge leptons with P$_T$ $>$ 20~GeV, missing E$_T$ $>$ 20 (30)~GeV if 
looking at e$\mu$ (ee,$\mu\mu$), and imposing 
a  cut on the dilepton mass, to remove the $Z$'s background, a top peak can be identified  
already with only 10 pb$^{-1}$ of data~\cite{cmscrossone}. 
The statistical error is of the order of 9$\%$ and the systematics are expected
 to be similar. Fig.~\ref{fig:topcms10pb} shows the jet multiplicity distribution for 
the $t \bar t$ signal events and 
all the relevant backgrounds.\\
ATLAS developed a robust analysis for the first 100$^{-1}$ of
 data in the semi-leptonic $t \bar t$ decay channel~\cite{atlasxsec}. The strategy is based upon the
 attempt to identify top events without utilizing the
full b-tagging capabilities. This is brought about by the fact
that efficient b-tagging is non trivial and implies to have reached
a precise level of alignment of the tracking detector, a situation which
will probably require several months of data taking.
Such an analysis solely relies on the measurement of jets, leptons and missing E$_T$, and requires a
functioning lepton triggering system.

Events are selected requiring one lepton (electron or muon) with P$_T$ $>$ 20~GeV, missing E$_T$ $>$ 20~GeV, at 
least four jets with P$_T$ $>$ 20~GeV of  which at  least three jets with P$_T$ $>$ 40~GeV.
A top quark decay candidate is defined as the three-jet combination of all jets, which has 
the highest transverse vector sum momentum.
One can exploit additional information: every 3-jet combination that originates from a top decay
also contains a 2-jet combination that originates from a W
decay. An unbiased W mass distribution is preferred in the
analysis, for which we choose not to pick/define one particular
W di-jet pair out of the three permutations, but rather require that at least 
one of the three di-jet invariant masses is within 10~GeV of the reconstructed mass 
of the W. This selection will be referred to as the W mass constraint selection.

A number of background processes have been considered. The most dominant
expected background is the W+jets, but also single top production and
others are sizeable. After the W-mass constraint a S/B of about 4 is reached.

\begin{figure}[!htb]
  \begin{center}
    \includegraphics[width=6.75cm]{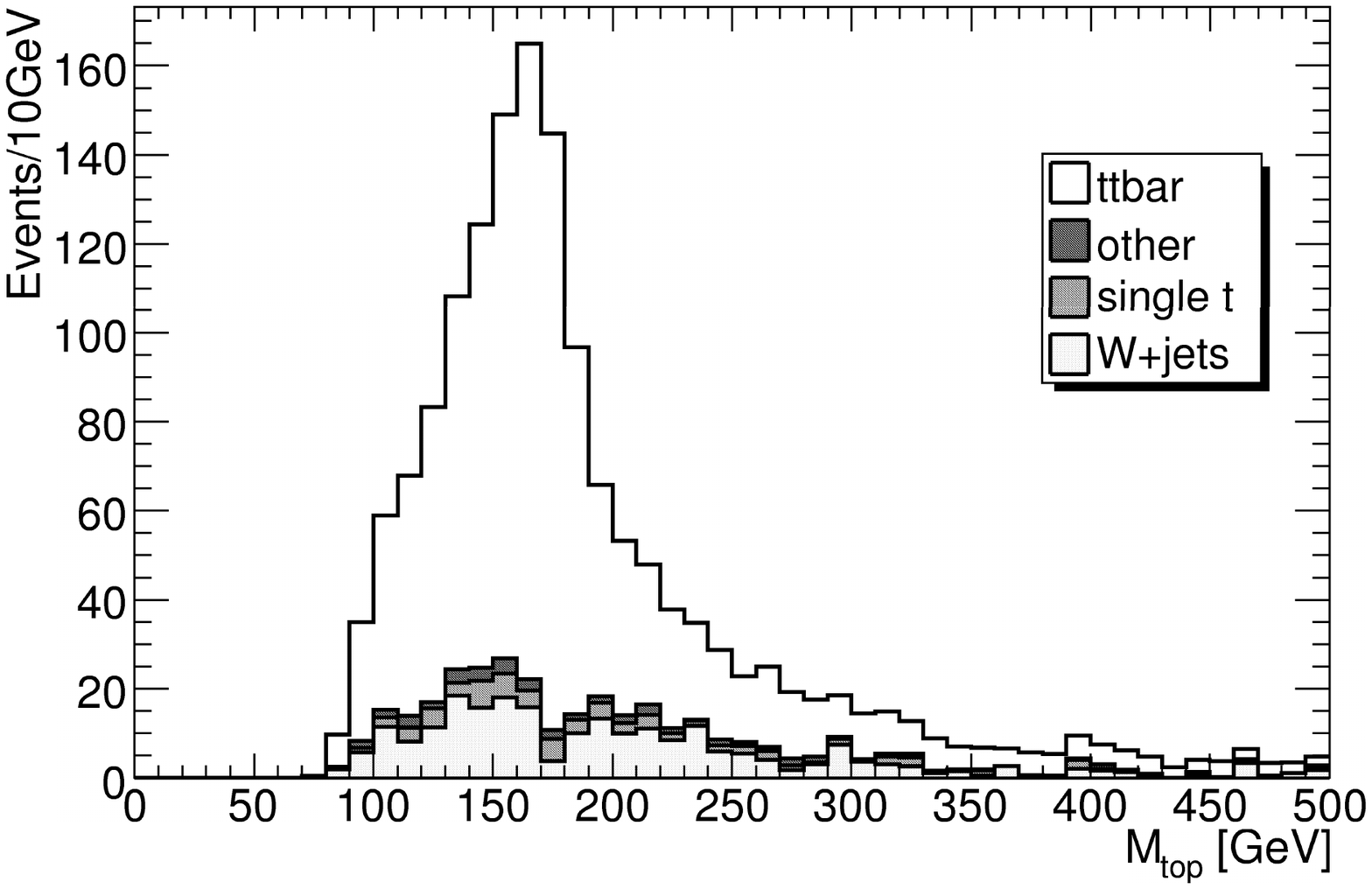}
    \includegraphics[width=6.75cm]{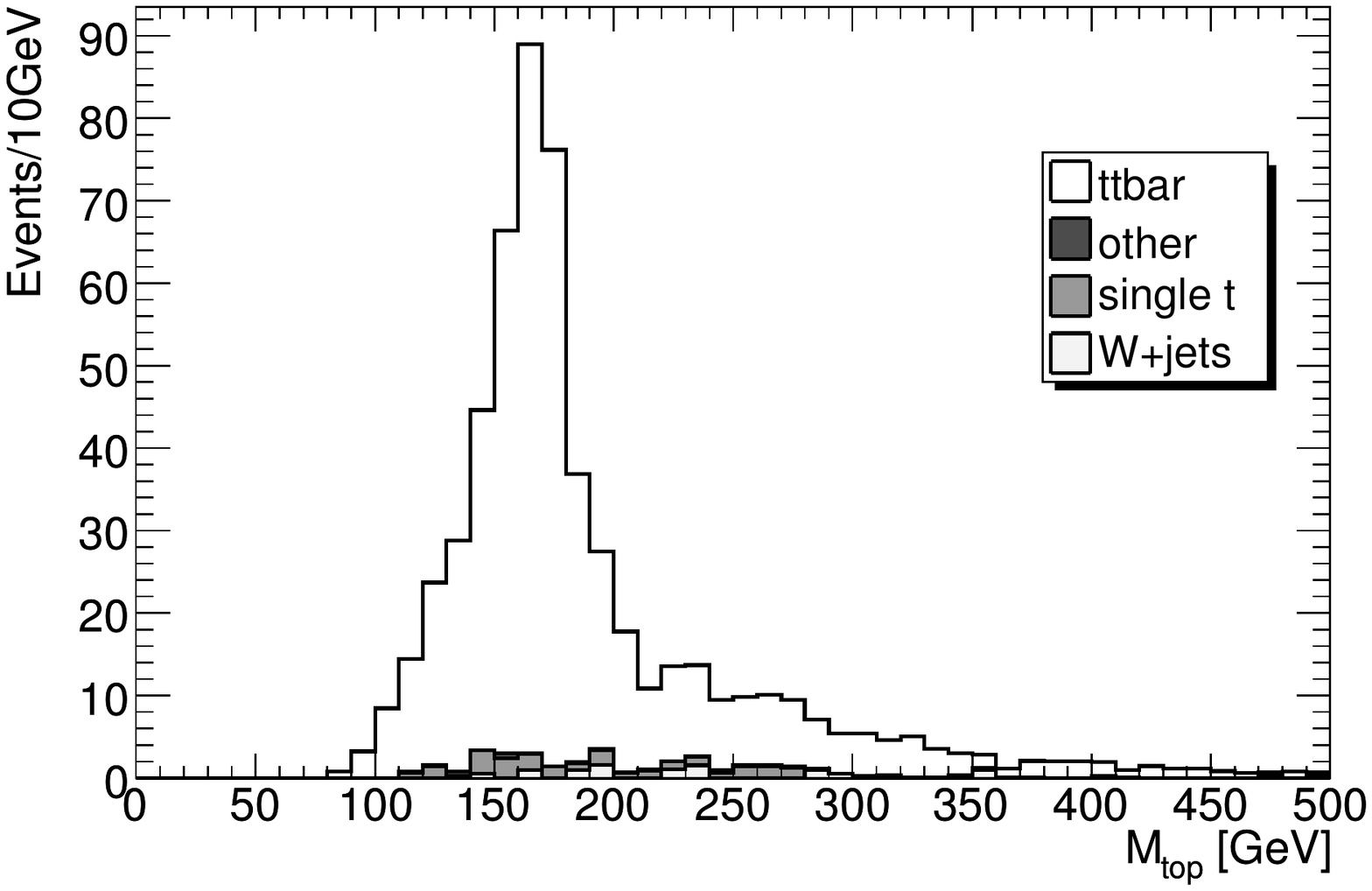}
    \caption{Left plot: ATLAS reconstructed three-jet mass for $t \bar t$ , single top and W + jet events for the default
electron selection + the W-boson mass constraint. Right plot: same distribution but requiring
        one or two jets tagged as coming from a $b$-quark.}
    \label{fig:topmassxsec}
  \end{center}   
\end{figure}

The distribution of the invariant mass of the three-jet combination
that forms our hadronic top-quark candidate with the default selection
and with the backgrounds added together, is shown in the left plot of
fig.~\ref{fig:topmassxsec}. The events where the correct top-quark
pair was chosen are clearly visible as the mass peak
on top of a smooth background distribution. 
The $t \bar t$ cross-section can be obtained by performing a counting experiment:
\begin{center}
$ \sigma(p\bar{p} \rightarrow t\bar{t}) =  (N_{\rm obs} - 
N_{\rm bkg}) / A_{\rm{tot}} {\cal L}$. \\
\end{center}
where $N_{\rm bkg}$, the number of background events estimated from Monte Carlo simulations and/or
data samples, is subtracted from $N_{\rm obs}$, the number of observed events meeting the selection criteria 
of a top-event signature. This difference is
divided by the integrated luminosity ${\cal L}$ and the total acceptance
$A_{\rm tot}$.
$A_{\rm tot}$ includes the geometric acceptance as well as trigger
efficiency and event selection efficiency and is slightly dependent 
on M$_{top}$. The advantage of using event counts in the commissioning
phase is that early on, the Monte Carlo simulations will presumably not predict the shapes
of distributions very well.
With the first 100$^{-1}$ of data, we expect to reach the following accuracies (for the default selection + 
 the W-boson mass constraint, and for the combined electron and muon channels):
\begin{eqnarray}
 &\Delta \sigma = & 3 (\mbox{stat}) \pm 16 (\mbox{syst}) \pm 3. (\mbox{pdf}) \pm 5 (\mbox{lumi}) \\ 
\end{eqnarray}
The main sources of systematics are the ISR and FSR as well as the JES.  
Once there will be a reliable algorithm for the identification of the jets coming from a b-quark, 
the b-tagging will greatly help in improving the S/B. 
Requiring one or two 
 b-tagged jets improves the purity of the sample by more than a factor of 4, 
 while the signal efficiency is only reduced by a  factor of 2. 
In fig.~\ref{fig:topmassxsec}, the reconstructed 3-jets mass 
is shown when one or two b-tagged jets are required for the default selection (left plot) and for 
the default selection plus the W-boson contraint (right plot).
To reconstruct the top mass, we find the three jets combination with the highest possible P$_T$, 
obtained by requiring that 
one and only one of the three jets is a b-jet.
When the W-boson mass constrain is applied, it's applied on the two jets among those three which are not b-tagged.
Thus if the maximum triple found above is such that the two non-b-jets don't combine to give a W, 
 that event is rejected. 
 
 The statistical error on the cross-section which is obtained by requiring one or two b-tagged jets is 4.5$\%$.
 The systematic error due to the jet energy scale is in this case of 4.9$\%$ about, while a 
 wrong normalization of the W+jets background by a factor of 20$\%$, 50$\%$ or even two, brings a systematic error on 
 the cross-section of 3.4$\%$, 4.7$\%$ and 6.9$\%$ respectively. 
For the b-tagging efficiency, the various on-going studies seem to indicate that a 5$\%$ relative error on 
the b-tagging efficiency is what one should have with 100 pb$^{-1}$, 
for the usual efficiencies around 50-60$\%$.

\subsection{Electroweak single top production and decay} 
\begin{figure}[!htb]
  \begin{center}
    \includegraphics[width=12cm]{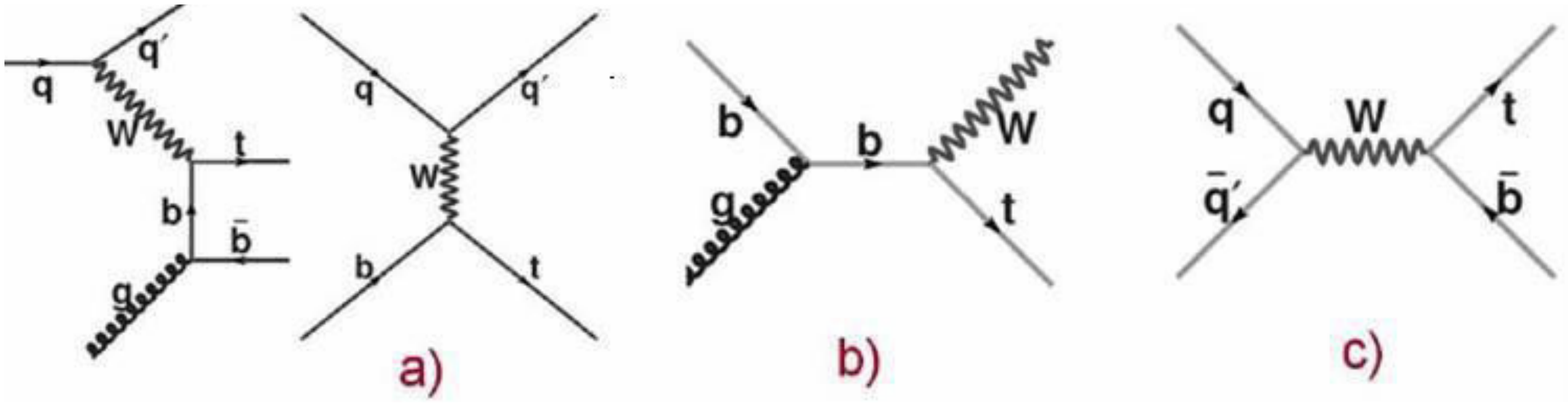}
    \caption{Main graphs corresponding to the three production mechanisms of single-top
events: (a) t-channel (b) Wt associated production (c) s-channel.}
    \label{fig:singletopgraph}
  \end{center}   
\end{figure}

In the SM three production modes are available for single top events, distinguished
by the virtuality of the W boson coupled to the top (see fig.~\ref{fig:singletopgraph})

Just recently, the D0 experiment gave evidence of single top 
events~\cite{d0singletop}, but 
 LHC will provide much higher statistics for all the three
channels (the production of single top quarks will account for a third of the top pair production), allowing 
the observation also of the Wt production mode, and a more precise study of
the single top phenomenology.
The study of single top production provides a unique possibility to investigate some aspects of
top quark physics that cannot be studied in $t \bar t$ production. In particular,
 the only way to measure directly V$_{tb}$ (CKM matrix element), to investigate the tWb vertex structure and 
 the FCNC coupling directly in the production processes, and to search for possible manifestation of New Physics beyond SM
 such as anomalous couplings and s-channel resonances.
Moreover, the single top quark production presents an irreducible background to several searches
for SM and New Physics signals (for example Higgs boson searches in the associated
production channel) and may provide additional measurements of M$_{top}$ and of the
top quark spin, together with the top pair channel.
The EW single-top-quark production rate at the LHC is also calculated in the SM to
the NLO level of accuracy for all three production mechanisms. The computed NLO cross-sections
for the t-channel, the tW and the s-channel are respectively 240, 60 and 10 pb.
The three single-top processes result in quite distinct final states, leading to the definition of
specific analyses in each case, making use of differences in jet multiplicity, number of b-tagged
jets required, as well as angular distributions between lepton and/or jets present in the final states.
Besides, important differences subsist in the level of backgrounds that are faced in the various
analyses, leading to the development of tools dedicated to the rejection of specific backgrounds.

Similarly to the situation at the Tevatron,
the selection of single top events will suffer from the presence of both W+jets and $t \bar t $ background, which
are produced at much higher rates. Thus, careful approaches devoted to the understanding of these
backgrounds in terms of shape and normalization performed directly from data will have to be defined.
Besides, single top analyses will be very early dominated by the systematic
uncertainties, and will require a good control of b-tagging tools and a reliable determination of the jet
energy scale.

\subsubsection{t-channel}

CMS has performed a study for 10 fb$^{-1}$ of integrated luminosity, with the pile-up expected for a 
 luminosity of 10$^{33}$. 
They assume to extract only the cross-section, with a simple counting experiment 
and without the use of any multivariate methods~\cite{cmssingletop}.
The generators which have been used for the signal are: SingleTop~\cite{SingleTop} and TopRex~\cite{Toprex}.
ATLAS explored the case of 1fb$^{-1}$ of integrated luminosity, with no pileup~\cite{atlassingletop}.
A cut-and-count analysis consitutes a baseline; more complex multivariate methods have been developed 
in addition to get a better background rejection.
The AcerMC Monte Carlo has been employed to generate the signal events. 
ATLAS made use of the fact that there are similar features in the three channels: a common pre-selection
 is therefore possible to reduce backgrounds. This pre-selection
 requires exactly one isolated high P$_T$ lepton, from 2 to 4 jets, one of which is tagged ad a b-jet, missing 
E$_T$ $>$ 20~GeV. The efficiency for a single-top signal is 9-10$\%$ (10-12 $\%$) for electron (for muons).
With these cuts, the rejection of W+jets is of order O(10$^4$), while for $t \bar t$ is O(20). 
As shown in fig.~\ref{fig:singletopgraph} (left graph) for the t-channel, the final partons (b-quark from top-quark decay, 
the charged lepton and
light quark) have relatively large transverse momenta. However, an additional b-quark is produced
with small transverse momentum. This will make very difficult to identify the low P$_T$ jet originating
from this quark and tag it as b-jet. Another specific feature of the t-channel single top events is the
production of a light jet in the forward/backward direction. 
A cut on b-tagged jet p$_T$ $>$ 50~GeV reduces the W+jets significantly, while a cut on the
hardest light jet $|\eta|$ $> 2$.5 can reject $t \bar t$ events.
With this simple cut and count analysis a S/B value of 0.37 is reached. The statistical error on the cross-section measurement 
is around 5$\%$, while the systematics (b-tagging, JES scale, ISR/FSR) reach 44.7$\%$. 
The left plot in fig.~\ref{singletopana} shows the number of jets for single top candidates in the t-channel 
and for the relevant backgrounds.
By applying a more sophisticated multivariate analysis (Boosted Decision Tree), this last one can be reduced by a factor of 2 
 about.

\subsubsection{Wt channel}

From the theoretical point of view the definition of the Wt signal is not trivial, since at NLO
it mixes with $t \bar t$. The final state is very similar to $t \bar t$ production, except for the presence of 
one less b-jet: jet counting is therefore critical.
Since it is not possible to achieve a good S/B, a correct background normalization from data will be important,
 to avoid large systematic uncertainties.
CMS selects the events by requiring exactly one lepton (e or $\mu$), one b-jet and two light quark jets, and missing E$_T$.
The correct (Wb) pairing is obtained from a Fisher discriminant using variables like the P$_T$(b+W), $\Delta$R(W,b) and 
the product of the b-quark and W charges.

\subsubsection{s-channel}
\begin{figure}[!tb]
  \begin{center}
    \includegraphics[width=6.5cm]{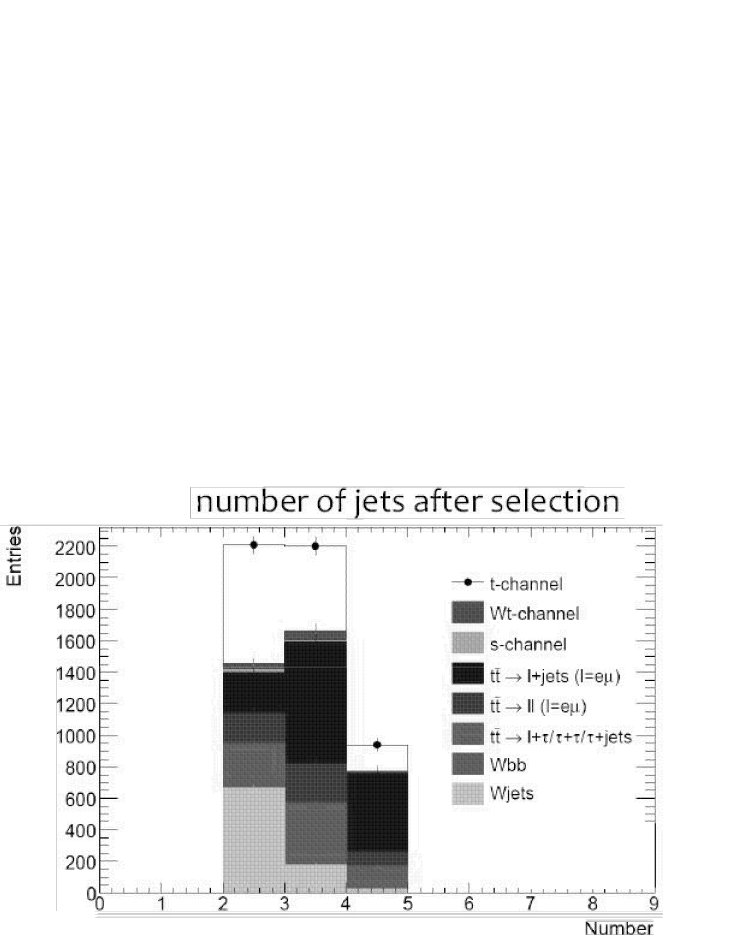}
    \includegraphics[width=6.5cm]{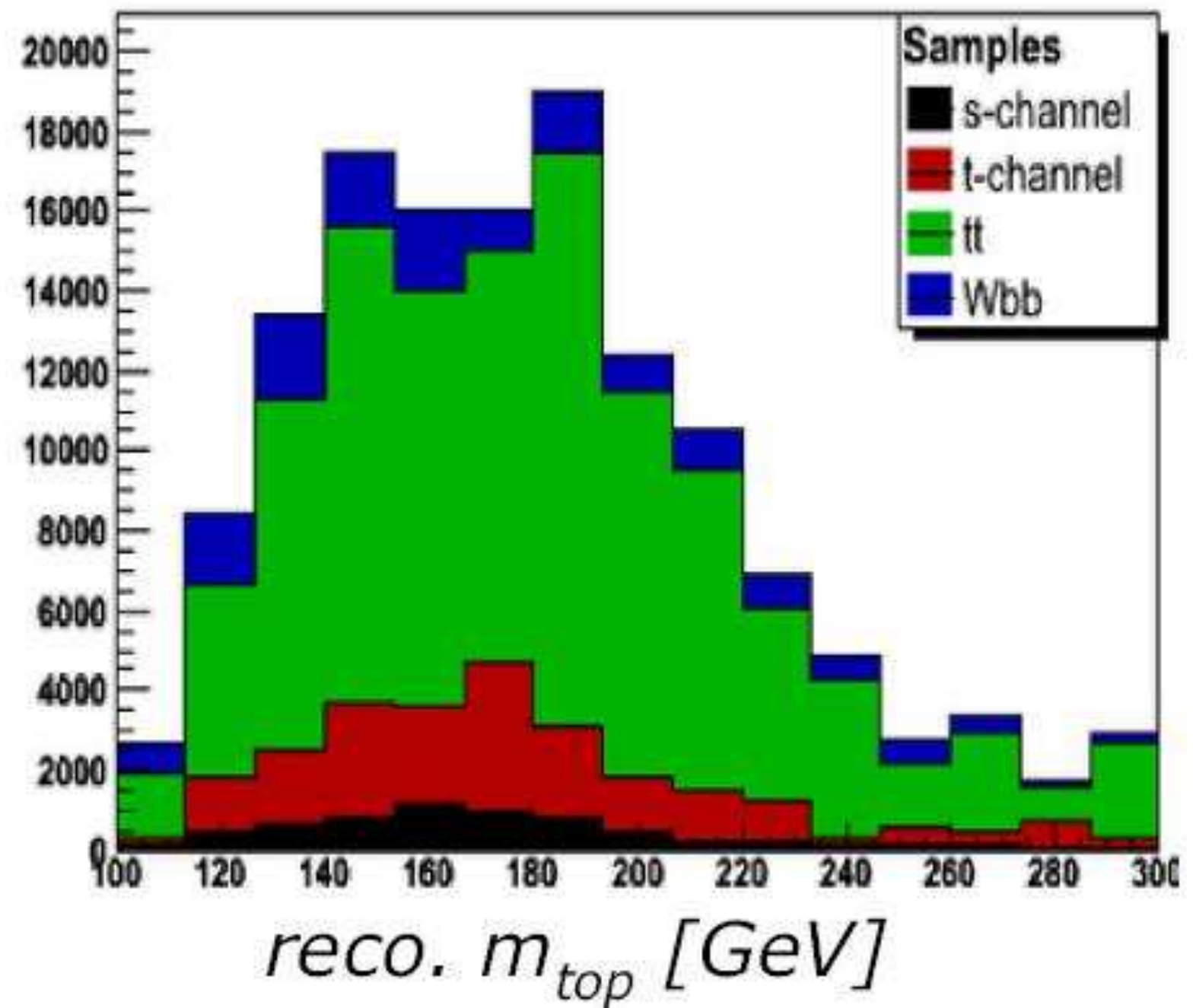}
    \caption{Left plot: Number of jets for single top candidates in the t-channel and for the relevant backgrounds (ATLAS).
    Right plot: Reconstructed mass for single top candidates in the Wt channel and for the relevant backgrounds (CMS).}
    \label{singletopana}
  \end{center}   
\end{figure}

The identification of s-channel events will be much harder at LHC than at Tevatron, as the relative cross-section is much 
 smaller. The CMS selection requires one isolated lepton (e or $\mu$), exactly two jets, both b-tagged, missing E$_T$, 
 and cuts on the transverse mass of the reconstructed W, on M$_{top}$, on P$_T$(top), on $\Sigma_T$ and on H$_T$.
The uncertainty which can be reached on a cross-section measurement with 10 fb$^{-1}$ of data, is of 18$\%$ (statistical) and 
 31$\%$ (systematics), not including the error coming from the luminosity measurement. 
The right plot in figure~\ref{singletopana} shows the reconstructed mass for single top candidates in the Wt channel 
and for the relevant backgrounds (CMS)

In a context of low S/B, the use of sophisticated tools like
likelihoods and Boosted Decision Trees, appears very useful if one wants to reach evidence of the signal
with the early data or to determine precisely their cross-section. These techniques, which are now of
common use at the Tevatron, will require the use of reliable event samples for modeling signal and
backgrounds, that will presumably be produced from the data. The analyses should also be optimized 
with respect to the total level of systematic uncertainty, which will be the main limiting factor for 30 fb$^{-1}$ 
measurements.
Finally, a precise determination of single top cross-sections can be achieved for a few fb$^{-1}$ in the
t-channel and the Wt-channel , while for the s-channel, higher statistics will be required. Their interpretation
in terms of new physics should thus come at a later stage, once the systematic effects are under
control.

\subsection{Top properties}
The sensitivity which can be reached at the LHC in the measurement of many top properties, like  
the top charge, the spin and spin correlations, the rare top decays
associated to flavour changing neutral currents (FCNC: t$\rightarrow$ qX, with X = $\gamma$,Z,g) and the $t \bar t$ resonances,
 has been studied. 
ATLAS measured the precision of these measurements which can be obtained with 1 fb$^{-1}$ of integrated luminosity
~cite{atlasproperties}. 
For the tests of physics beyond the SM
associated with the production of top quarks, the 95$\%$ CL limit (in the absence of a signal) was also derived.
Several sources of systematic errors were considered using an approach common to all studies, as mentioned at the beginning.
 Few examples follows: the sensitivity of the ATLAS experiment to the top quark charge measurement is such that 
already with 1 fb$^{-1}$ (using the semi-leptonic b-decay) it is possible to distinguish with
a 5$\sigma$ significance, between the SM scenario (q=2/3) and the alternative (q=4/3). 
A complete study of the precision reachable on the W polarisation fractions F$_0$, F$_L$ and F$_R$ (respectively 5$\%$, 12$\%$ and 0.03)
  and the $t \bar t$ spin correlation parameters A and A$_D$ (50$\%$ and 34$\%$) has been performed in the 
semileptonic $t \bar t$ channel. Reconstructed
and corrected angular distributions are used to extract polarisation measurements.
Expected limits on the top quark rare decays through FCNC processes (t $\rightarrow$ qZ, q$\gamma$,qg) 
were set at 95$\%$ CL in the absence of signal .
The discovery potential of the ATLAS experiment for the $t \bar t$ resonances decaying in the semileptonic
channel, have been studied as a function of the resonance mass. Using this information, Kaluza-Klein gluon
resonances with masses up to 1.5 TeV can be excluded with 1 fb$^{-1}$ of data.


\addtocounter{chapter}{1}

%
%
\hyphenation{di-gi-ti-zer}
\hyphenation{ex-tra-po-la-ted}
\hyphenation{si-mu-la-tion}

\def\perx{$\times$}
\def\perdot{$\cdot$}
\def\magg{$>$}
\def\min{$<$}
\def\cca{$\sim$}
\def\pT{$p_T$}
\def\ET{$E_T$}
\def\SpT{$\sum{p_T}$}
\def\SET{$\sum{E_T}$}
\def\keV{$keV$}
\def\MeV{$MeV$}
\def\GeV{$GeV$}
\def\GeVc{$GeV/c$}
\def\GeVcc{$GeV/c^2$}
\def\kHz{$kHz$}
\def\MHz{$MHz$}
\def\GHz{$GHz$}
\def\HO{$H^0$} 
\def\mH{$M_H$}
\def\mZ{$M_Z$}
\def\bb{$b\bar{b}$}
\def\ttbar{$t\bar{t}$}
\def\Zbb{$Zb\bar{b}$}
\def\ZZ{$ZZ$}
\def\ZZs{$ZZ^{(\ast)}$}
\def\meme{$2\mu2e$}
\def\mm{$\mu^{+}\mu^{-}$}
\def\ee{$e^{+}e^{-}$}
\def\mmee{$\mu^{+}\mu^{-}e^{+}e^{-}$}
\def\ra{$\rightarrow$}
\newcommand{\decay}[2]{#1\ra#2}
\newcommand{\Decay}[3]{#1\ra#2\ra#3}
\newcommand{\Hdec}[1]{$H$\ra#1}
\newcommand{\HDec}[2]{$H$\ra#1\ra#2}
\newcommand{\HOdec}[1]{\HO\ra#1}
\newcommand{\HODec}[2]{\HO\ra#1\ra#2}
\def\HZZ{\Hdec{\ZZ}}
\def\HWW{\Hdec{$WW$}}
\def\HZZllll{\HDec{\ZZ}{$4\ell$}}
\def\HZZmeme{\HDec{\ZZ}{\meme}}
\def\HZZmmee{\HDec{\ZZ}{\mmee}}
\def\Hllll{\Hdec{4\ell}}
\def\Hmeme{\Hdec{\meme}}
\def\Hmmee{\Hdec{\mmee}}
\def\ZZmeme{\decay{\ZZ}{\meme}}
\def\ZZmmee{\decay{\ZZ}{\mmee}}
\def\Zbbmeme{\decay{\Zbb}{\meme}}
\def\Zbbmmee{\decay{\Zbb}{\mmee}}
\def\tt2m2e{\decay{\tt}{\meme}}
\def\ttmmee{\decay{\tt}{\mmee}}
\def\LLumi{$\mathcal{l}=2\cdot 10^{33}\;cm^{-2}s^{-1}$}
\def\HLumi{$\mathcal{l}=10^{34}\;cm^{-2}s^{-1}$}
\def\SULumi{$\mathcal{l}=2\cdot 10^{32}\;cm^{-2}s^{-1}$}
\def\EWgroup{$SU(2)_I \otimes U(1)_Y$}
\newcommand\sups[1]{\textsuperscript{\rm #1}}

\mchapter{The Standard Model Higgs Boson}
{ Authors: Sara Bolognesi, Chiara Mariotti and Daniele Trocino}
\vskip 0.3cm\noindent
{\it Revisors: Barbara Mele, Paolo Nason}
\vskip 1cm
\section{Higgs Boson Mass}
The Higgs boson mass is the only yet unknown free parameter of the SM. The Higgs in fact has never been observed 
experimentally and its mass cannot be predicted by the SM. It depends on the parameters $v$ and $\lambda$, but 
while the former can be estimated by its relation with the constant $G_F$ of Fermi's theory, the latter is 
characteristic of the field $\phi$ and cannot be determined other than measuring the Higgs mass itself.
Both theoretical and experimental constraints exist, including those from direct search at colliders, 
in particular LEP.

\subsection{Theoretical constraints}
\begin{figure}[htbp]
 \centering
 \includegraphics[width=0.7\linewidth]{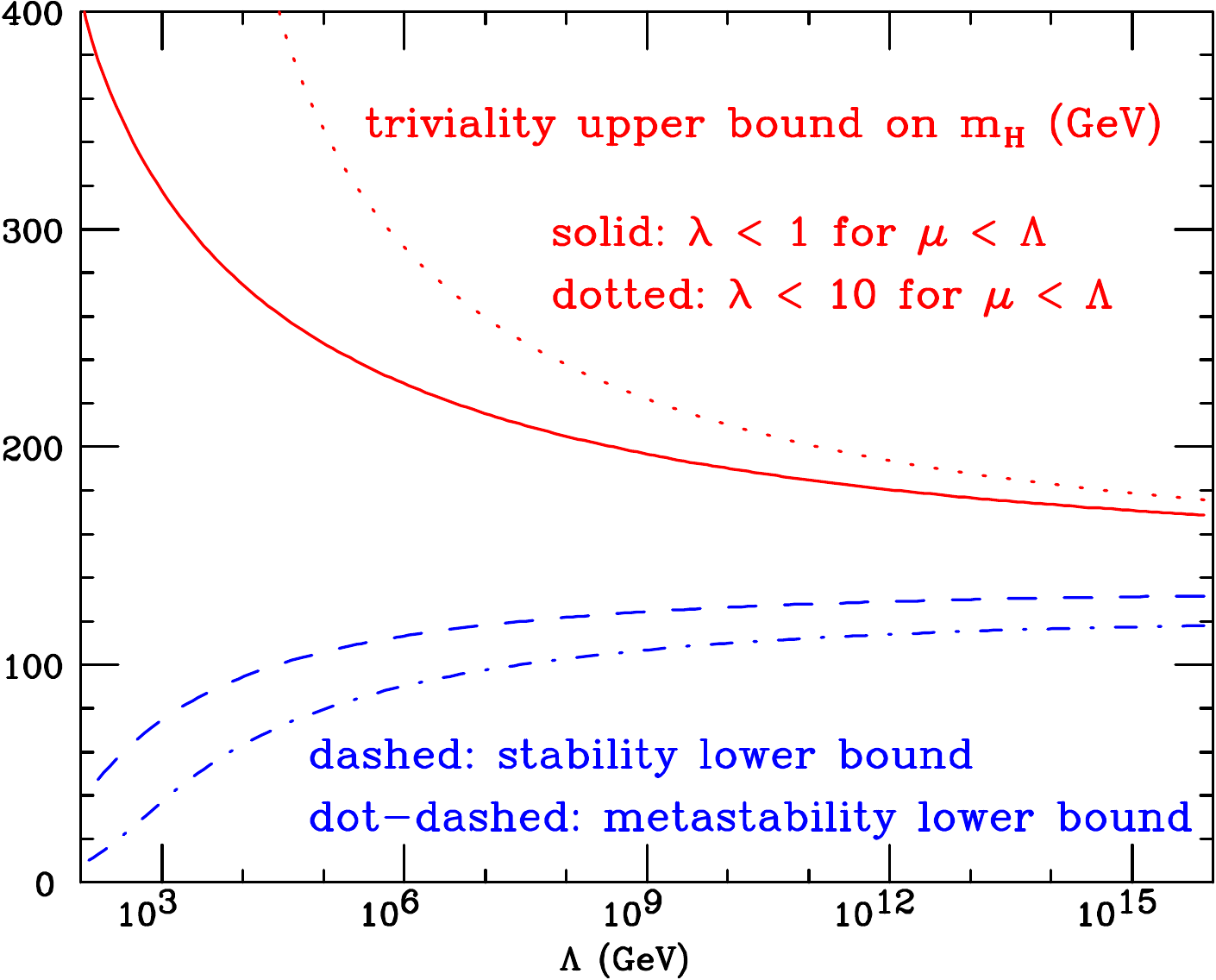}
 \caption{\textit{Red line}: \textit{triviality} bound (for different upper limits to $\lambda$); \textit{blue line}: 
\textit{vacuum stability} (or \textit{metastability}) bound on the Higgs boson mass as a function of the new physics 
(or cut-off) scale $\Lambda$~\cite{bib:TheorConstraints}.}
 \label{fig:TrivialVacuumStabil}
\end{figure}

Theoretical constraints to the Higgs mass value~\cite{bib:TheorConstraints} can be found by imposing the energy 
scale $\Lambda$ up to which the SM is valid, before the perturbation theory breaks down and non-SM phenomena emerge.
The upper limit is obtained requiring that the running quartic coupling of Higgs potential $\lambda$ remains 
finite up to the scale $\Lambda$ (\textit{triviality}). A lower limit is found instead by requiring that $\lambda$ 
remains positive after the inclusion of radiative corrections, at least up to $\Lambda$: this implies that the 
Higgs potential is bounded from below, i.e. the minimum of such potential is an absolute 
minimum (\textit{vacuum stability}). A looser constraint is found by requiring such minimum to be local, 
instead of absolute (\textit{metastability}).
These theoretical bounds on the Higgs mass as a function of $\Lambda$ are shown in Fig.~\ref{fig:TrivialVacuumStabil}.

If the validity of the SM is assumed up to the Plank scale $(\Lambda \sim 10^{19}\;GeV)$, the allowed Higgs mass range 
is between 130 and 190 \GeVcc, while for $\Lambda$\cca$1\;TeV$ the Higgs mass can be up to 700 \GeVcc. On the basis of 
these results, the LHC has been designed for searches of the Higgs boson up to masses of \cca$1\;TeV$. 
If the Higgs particle is not found in this mass range, then a more sophisticated explanation for the EWSB mechanism 
will be needed.

\subsection{Experimental constraints}
\begin{figure}
 \centering
 \includegraphics[width=0.8\linewidth]{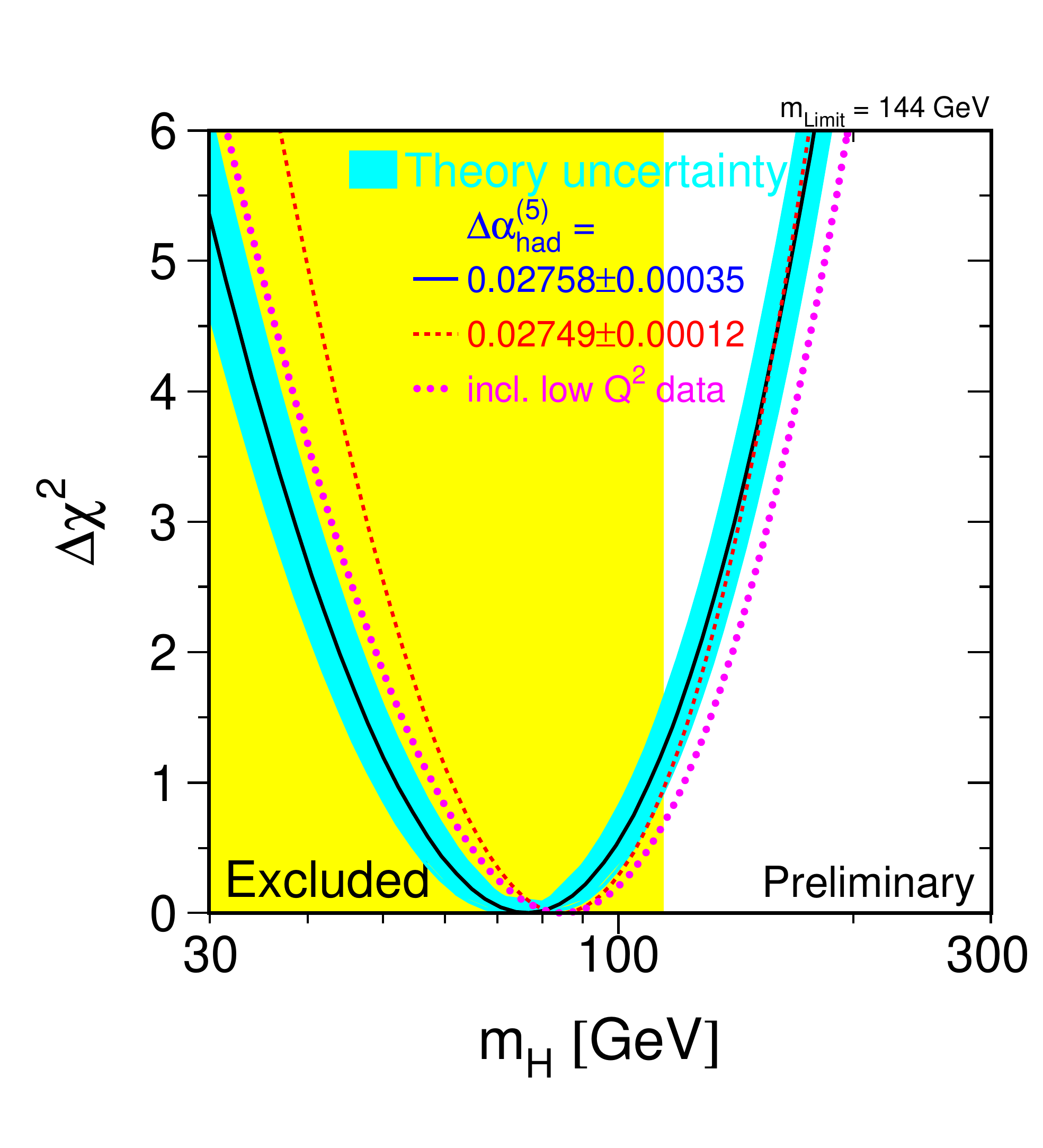}
 \caption{$\Delta \chi^2$ of the fit to the electroweak precision measurements of LEP, SLC and Tevatron as a 
function of the Higgs mass (July 2007). The solid line represents the result of the fit and the blue shaded 
band is the theoretical error from unknown higher-order corrections. 
The yellow area represents the region excluded by direct search.}
 \label{fig:GlobalEWFit}
\end{figure}

Bounds on the Higgs mass are also provided by measurement at LEP, SLC and Tevatron~\cite{bib:ExperConstraints_1} 
(updated at July 2007). A lower bound at 114.4 \GeVcc{} (at $95\%$ C.L.) has been established by direct 
searches at LEP~\cite{bib:ExperConstraints_2}. Moreover, since the Higgs boson contributes to radiative corrections, 
many electroweak observables are logarithmically sensitive to \mH{} and can thus be used to constraint its mass. 
All the precision electroweak measurements performed by the four LEP experiments, SLD, CDF and D$\emptyset$ 
have been combined together and fitted, assuming the SM as the correct theory and using the Higgs mass as free 
parameter. The result of this procedure is summarized in Fig.~\ref{fig:GlobalEWFit}, 
where $\Delta\chi^2=\chi^2-\chi^2_{min}$ is plotted as a function of \mH.
The solid curve is the result of the fit, while the shaded band represents the theoretical uncertainty 
due to unknown higher order corrections.

As of Summer 2007, the indirectly measured value of the Higgs boson mass, corresponding to the minimum of the curve, 
is $M_H=76^{+33}_{-24}\;GeV$ (at $68\%$ C.L. for the black line in Fig.~\ref{fig:GlobalEWFit}, thus not taking 
the theoretical uncertainty into account). An upper limit of $144\;GeV$ can also be set 
(one-sided $95\%$ C.L.) including the theoretical uncertainty; this limit increases to 182 \GeVcc{} 
when including the direct search limit of 114.4 \GeVcc.

Such results are obviously model-dependent, as the loop corrections take into account only contributions 
from known physics. This result is thus well-grounded only within the SM theory and has always to be 
confirmed by the direct observation of the Higgs boson.

\section{Standard Model Higgs Boson search at LHC}
\label{sec:HiggsProductionDecay}
The experiments at the LHC will search for the Higgs boson within a mass range going from 100 \GeVcc{} 
to about 1 $TeV$. In this section, the main Higgs boson production and decay processes are described, 
in order to determine the most promising channels for the Higgs discovery at LHC.

While the Higgs boson mass is not predicted by the theory, the Higgs boson couplings to the fermions and bosons are 
predicted to be proportional to the corresponding particle masses (for fermions) or squared masses (for bosons). 
For this reason, the Higgs boson production and decay processes are dominated by channels involving the coupling 
of Higgs boson to heavy particles, mainly to $W^\pm$ and $Z$ bosons and to the third generation of fermions. 
For what concerns the remaining gauge bosons, the Higgs boson does not couple to photons and gluons at tree 
level, but only by one-loop graphs where the main contribution is given by $t$ loops for 
the $gg$\ra$H$ channel and by $W^+W^-$ and $t$ loops for the $\gamma\gamma$\ra$H$ channel.

\section{Higgs boson production}

\begin{figure}[htb]
  \begin{center}
    \includegraphics[width=0.35\linewidth]{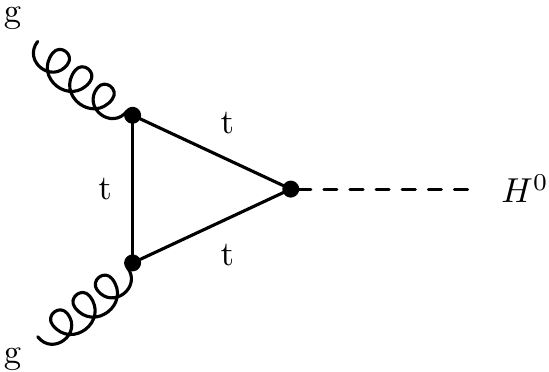}
    \hspace{0.5cm}
    \includegraphics[width=0.35\linewidth]{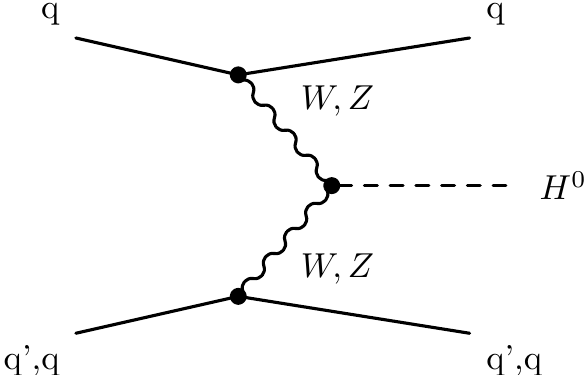}\\
    \makebox[\linewidth]{\hspace{\stretch{1.}} (a)\hspace{\stretch{1.5}} (b)\hspace{\stretch{1.2}}} 
    \includegraphics[width=0.35\linewidth]{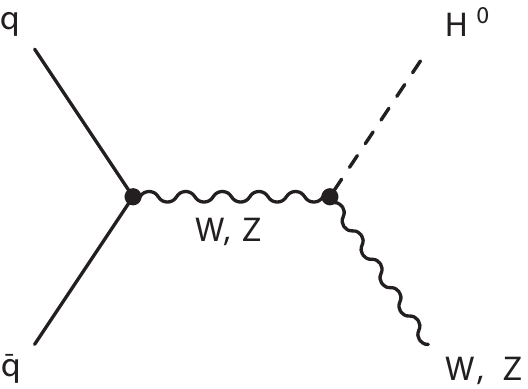}
    \hspace{0.5cm}
    \includegraphics[width=0.35\linewidth]{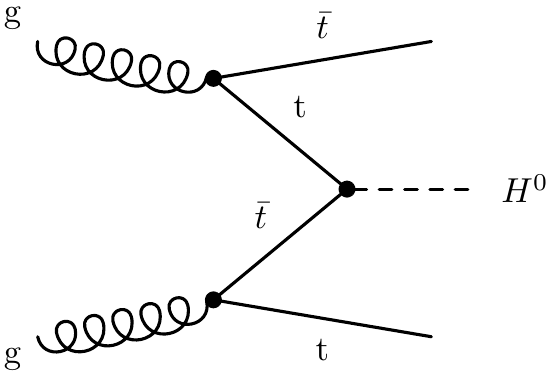}
    \makebox[\linewidth]{\hspace{\stretch{1.}} (c)\hspace{\stretch{1.5}} (d)\hspace{\stretch{1.2}}} 
    \caption{Higgs boson production mechanisms at tree level in proton-proton collisions: (a) gluon-gluon fusion; 
(b) $VV$ fusion; (c) $W$ and $Z$ associated production (or \textit{Higgsstrahlung}); 
(d) $t\bar{t}$ associated production.}
    \label{fig:HiggsProductionChannels}
  \end{center}
\end{figure}

\begin{figure}[hbt]
  \begin{center}
    \includegraphics[width=0.6\linewidth,angle=270]{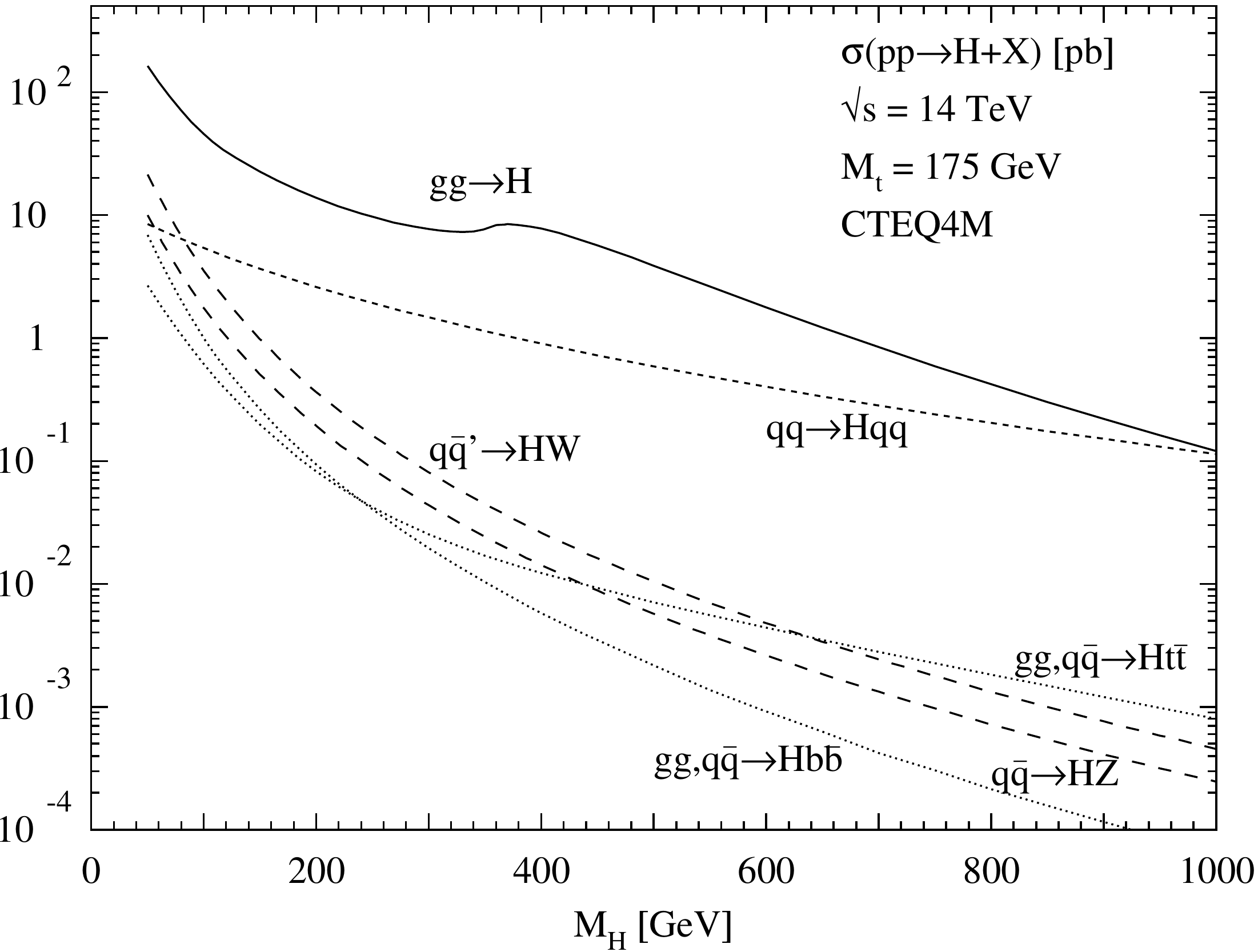}
    \caption{Higgs boson production cross sections at $\sqrt{s}=14\;TeV$ as a function of the Higgs boson mass.
             The cross sections are calculated using HIGLU and other programs~\cite{bib:higlu}; 
             they contain higher order corrections and the CTEQ6m~\cite{bib:cteq} p.d.f. has been adopted.}
    \label{fig:HiggsProductionSigmas}
  \end{center}
\end{figure}

The main processes contributing to the Higgs boson production at a hadron collider are represented by the Feynman 
diagrams in Fig.~\ref{fig:HiggsProductionChannels}. The corresponding cross sections for a center of mass 
energy $\sqrt{s}=14\;TeV$, corresponding to the design value at the LHC, are shown 
in Fig.~\ref{fig:HiggsProductionSigmas}.

\subsection{Gluon-gluon fusion}
The $gg$ fusion is the dominating mechanism for the Higgs boson production at the LHC over 
the whole Higgs boson mass spectrum. The process is shown in 
Fig.~\ref{fig:HiggsProductionChannels}(a), with a $t$ quark-loop as the main contribution. 

The cross section for the basic gluon to Higgs boson process is~\cite{bib:gluon-fusion}

\begin{equation}
\sigma(gg\rightarrow H)=\frac{G_{\mu}\alpha_{S}^{2}(\mu_{R}^{2})}{288\sqrt{2}\pi}
\left|\frac{3}{4}\sum_{q}A^{H}_{1/2}(\tau_{Q})\right|^{2}\, ,
\label{gluon fusion}			
\end{equation}
where $A^{H}_{1/2}(\tau_{Q})$ with $\tau_{Q}=M_H^{2}/4 m_q^{2}$ is a form factor~\cite{bib:higgs-djuadi}.

The lowest order cross section has large corrections from higher order QCD diagrams. 
The increase in cross section from higher order diagrams is conventionally defined as the $K$-factor

\begin{equation}
K=\frac{\sigma_{NLO}}{\sigma_{LO}}
\end{equation} 
where LO (NLO) refer to leading (next-to-leading) order results. The $K$-factor for gluon fusion is evaluated 
in Ref.~\cite{bib:spira-NLO} with a next-to-leading order calculation and it results \cca2. 

The value of the cross section including the $K$-factor has two main uncertainties. The first is from the gluon 
structure function which still has large uncertainty in the low $x$ region. The cross section using a large set 
of todays best available structure functions was calculated in Ref.~\cite{bib:K-factor-gluon-fusion} and the 
results differ by around $20\%$ which can be taken as the theoretical uncertainty from the gluon structure function. 
At the time of data taking for LHC it can be expected to have much better structure functions available with 
data from HERA and the Tevatron.

Next-to-next-to leading order calculations are also available and show a further increase of about $10\%$ 
to $30\%$. Other sources of uncertainty are the higher order corrections ($10\div20\%$ estimated). 

The production of the Higgs boson through gluon fusion is sensitive to a fourth generation of quarks. 
Because the Higgs boson couples proportionally to the fermion mass, including a fourth 
generation of very heavy quarks will more than double the cross section. 

\subsection{Vector boson fusion}
The $VV$ fusion (Fig.~\ref{fig:HiggsProductionChannels}(b)) is the second contribution to the Higgs 
boson production cross 
section. It is about one order of magnitude lower than $gg$ fusion for a large range of \mH{} values and the two 
processes become comparable only for very high Higgs boson masses ($\mathcal{O}(1\;TeV)$). However, this channel is 
very interesting because of its clear experimental signature: the presence of two spectator jets with high 
invariant mass in the forward region provides a powerful tool to tag the signal events and discriminate the 
backgrounds, thus improving the signal to background ratio, despite the low cross section. Moreover, both 
leading order and next-to-leading order cross sections for this process are known with small uncertainties 
and the higher order QCD corrections are quite small.


\subsection{Associated production}

In the \textit{Higgsstrahlung} process (Fig.~\ref{fig:HiggsProductionChannels}(c)), the Higgs boson is produced 
in association with a $W^\pm$ or $Z$ boson, which can be used to tag the event. The cross section for this process 
is several orders of magnitude lower than $gg$ and $VV$ fusion ones. The QCD corrections are quite large and the 
next-to-leading order cross section results to be increased by a factor of $1.2\div1.4$ with respect to the leading 
order one.

The last process, illustrated in Fig.~\ref{fig:HiggsProductionChannels}(d), is the associated production of a Higgs 
boson with a $t\bar{t}$ pair. Also the cross section for this process is orders of magnitude lower than those of 
$gg$ and $VV$ fusion, but the presence of the $t\bar{t}$ pair in the final state can provide a good experimental 
signature. The higher order corrections increase the cross section of a factor of about 1.2.

\section{Higgs boson decay}
\begin{figure}[hbt]
  \begin{center}
    \includegraphics[width=0.5\linewidth,angle=270]{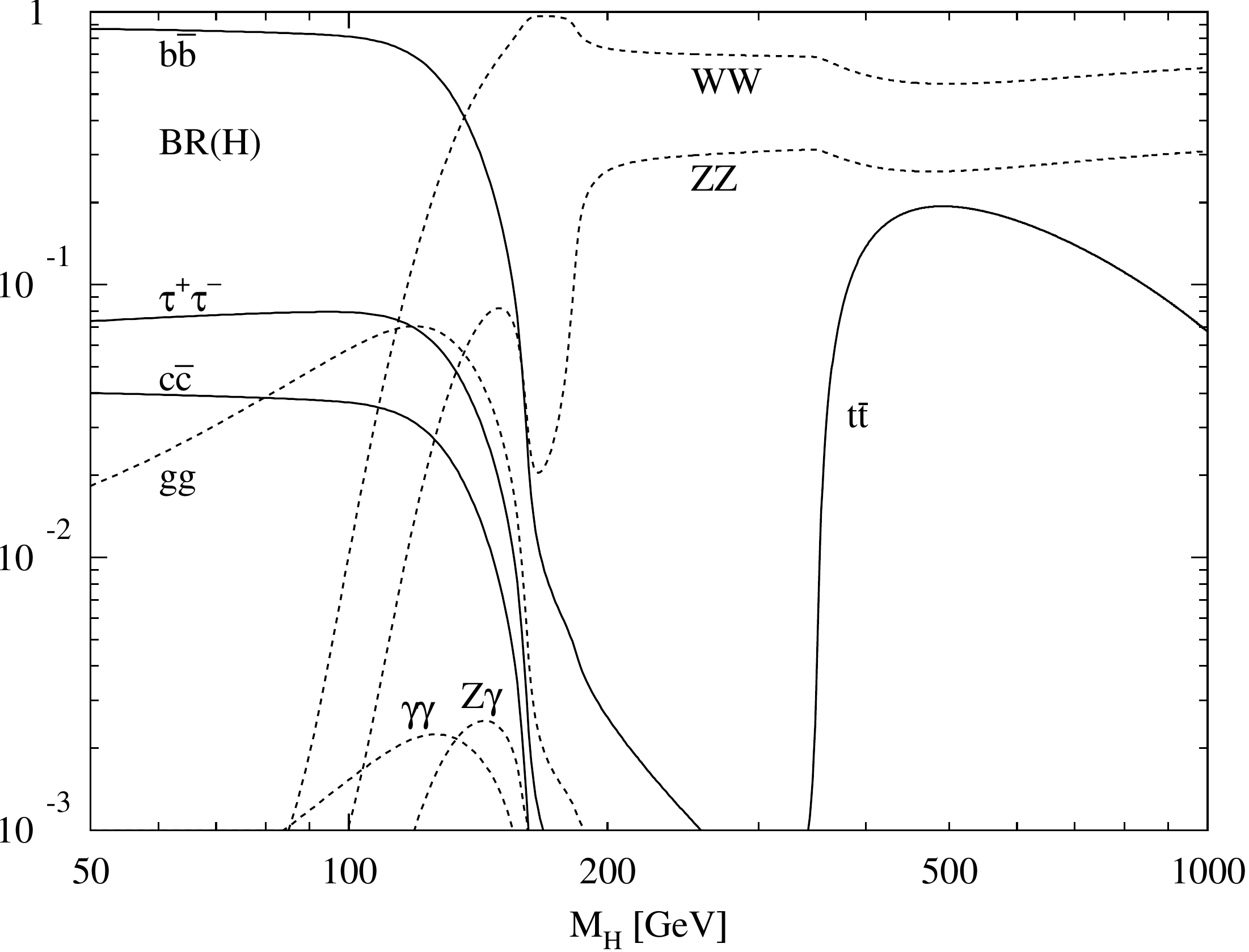}
    \caption{Branching ratios for different Higgs boson decay channels as a function of the Higgs boson mass. 
     They are calculated with the program \texttt{HDECAY}~\cite{bib:hdecay} which includes the dominant higher
     order corrections to the decay width.}
    \label{fig:HiggsDecayBranchingRatios}
  \end{center}
\end{figure}

The branching ratios of the different Higgs boson decay channels are shown in Fig.~\ref{fig:HiggsDecayBranchingRatios} as 
a function of the Higgs boson mass. Fermionic decay modes dominate the branching ratio in the low mass region 
(up to \cca150 \GeVcc). In particular, the channel $H$\ra$b\bar{b}$ has the highest branching ratio 
since the $b$ quark is the heaviest fermion available. When the decay channels into vector boson pairs 
open up, they quickly dominate. A peak in the $H$\ra$W^+W^-$ decay is visible around 160 \GeVcc, when the 
production of two on-shell $W$'s becomes possible and the production of a real $ZZ$ pair is still not allowed. 
At high masses (\cca350 \GeVcc), also $t\bar{t}$ pairs can be produced.

As shown in Fig.~\ref{fig:HiggsDecayBranchingRatios}, the branching ratios change dramatically across the possible 
range of the Higgs boson mass requiring different strategies for the different Higgs boson mass range.
The most promising decay channels for the Higgs boson discovery do not only depend on the corresponding branching ratios, 
but also on the capability of experimentally detecting the signal rejecting the backgrounds. 
Fully hadronic events are the most copious final states from Higgs boson decays. These decays can not be easily 
resolved when merged in QCD background, therefore topologies with leptons or photons are preferred, 
even if they have smaller branching ratio. 

Such channels are illustrated in the following, depending on the Higgs boson mass range.

\subsection{Low mass region}
Though the branching ratio in this region is dominated by the Higgs boson decay into $b\bar{b}$, the background 
constituted by the di-jet production (more tan six order of magnitude higher than the signal) 
makes quite difficult to use this channel for a Higgs boson discovery. 
Some results from this channel can be obtained when the Higgs boson is produced in association with a $t\bar{t}$ 
or via \textit{Higgsstrahlung}, since in this case the event has a clearer signature, despite its low cross section.

The most promising  way of identifying a Higgs boson in the low mass region is to select the decay channel 
$H$\ra$\gamma\gamma$. In spite of its lower branching ratio (around $10^{-3}$), 
the two high energy photons constitute a very clear signature, which only suffers from 
the $q\bar{q}$\ra$\gamma\gamma$ and $Z$\ra$e^+e^-$ backgrounds or jets faking photons.
The expected signal to background ratio is $10^{-2}$, which make this channel much more 
attractive than the $b \bar b$ channel.

\subsection{Intermediate mass region}
For a Higgs boson mass value between  $130 GeV/c^2 \leq M_H \leq 2M_Z$, the Higgs boson decays into $WW^{(\ast)}$ 
and $ZZ^\ast$ open up and 
their branching ratios quickly increase.  Thus the best channels in this mass region are 
$H$\ra$WW^{(\ast)}$\ra$2\ell2\nu$ and $H$\ra$ZZ^{\ast}$\ra$4\ell$ with only one vector boson on-shell.

The branching ratio of $H$\ra$WW^{(\ast)}$ is higher, because of the higher coupling of the Higgs boson to 
charged current with respect to neutral current. Moreover, this decay mode becomes particularly important 
in the mass region between $2M_W$ and $2M_Z$, where the Higgs boson can decay into two real $W$'s (and not yet 
into two real $Z$'s): its branching ratio is \cca1. Anyway, in such channel  because of the 
presence of the two $\nu$'s in the final state, the Higgs boson mass cannot be reconstructed. 
Such measurement can be performed instead when one $W$ decays leptonically and the other one decays in two 
quarks. But, in this case, the final state suffers from the high hadronic background.

The decay $H$\ra$ZZ^{\ast}$\ra$4\ell$, despite its lower branching ratio, offers a very clear experimental 
signature and high signal to background ratio. Furthermore, it allows to reconstruct the Higgs boson mass with 
high precision. Therefore, this channel seems to be the best candidate for a Higgs boson discovery in this mass range.

\subsection{High mass region}
This region corresponds to Higgs boson mass values above the 2\mZ{} threshold, where the Higgs boson can decay 
into a real $ZZ$ pair. Though the $H$\ra$ZZ$ width is still lower than $H$\ra$WW$ one, a decay into four 
charged leptons (muons or electrons) is surely the ``golden channel'' for a high mass Higgs boson discovery.

The upper mass limit for detecting the Higgs boson in this decay channel is given by the reduced production rate 
and the increased width of the Higgs boson. As an example, less than 200 Higgs particles with $M_H = 700 GeV/c^2$ will 
decay in the $ H\rightarrow Z Z\rightarrow 4\ell$ channel in a year at high luminosity and the large width 
will increase the difficulty to observe the mass peak. 

In order to increase the sensitivity to a heavy Higgs boson production, decay channels with one boson 
decaying into jets or neutrinos can be also considered.
The decay channel $ H\rightarrow W W\rightarrow\ell\nu_{\ell}\textrm{jj}$, where j denotes a jet from a quark in the 
W decay, has a branching ratio just below $30\%$, yelding a rate some 50 times higher than the four lepton channel 
from $ H\rightarrow Z Z$ decays. 
The decay channel $ H\rightarrow Z Z \rightarrow \ell \bar{\ell} \nu_{\ell^{'}}\bar{\nu}_{\ell^{'}}$ which has a six times 
larger branching ratio than the four lepton channel could also be interesting.

\subsection{Higgs boson total decay width}
\begin{figure}[hbtp]
  \begin{center}
    \includegraphics[width=0.6\linewidth]{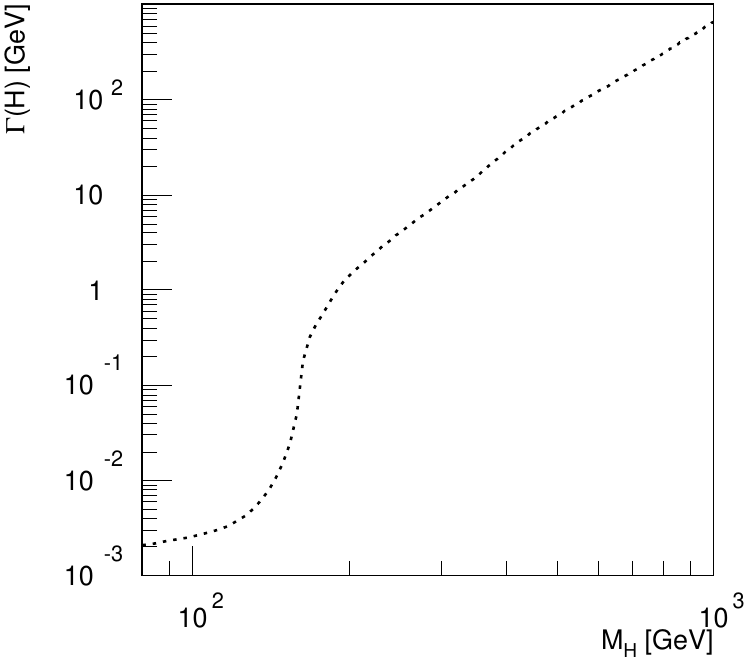}
    \caption{Higgs boson total decay width as a function of the Higgs boson mass.}
    \label{fig:HiggsWidth}
  \end{center}
\end{figure}

The total width of the Higgs boson resonance is shown in Fig.~\ref{fig:HiggsWidth} as a function of \mH. 
Below the $2M_W$ threshold, the Higgs boson width is 
of the order of the $MeV$, then it rapidly increases, but remains lower than 1 \GeVcc{} up to $M_H\simeq200\;GeV$: 
the low mass range is therefore the most challenging region, because the Higgs boson width is dominated by the 
experimental resolution.

In the high mass region (\mH\magg2\mZ), the total Higgs boson width is dominated by the $W^+W^-$ and $ZZ$ partial 
widths, which can be written as follows:
\begin{eqnarray}
 \Gamma(H\rightarrow W^+W^-) & = & \frac{g^2}{64\pi} \frac{M_H^3}{M_W^2} \sqrt{1-x_W} \left( 1-x_W+ \frac{3}{4} x_W^2 \right) \label{eq:WWPartialWidth} \\
 \Gamma(H\rightarrow ZZ) & = & \frac{g^2}{128\pi} \frac{M_H^3}{M_W^2} \sqrt{1-x_Z} \left( 1-x_Z+ \frac{3}{4} x_Z^2 \right) \label{eq:ZZPartialWidth}
\end{eqnarray}
where $$x_W=\frac{4M_W^2}{M_H^2}\, , \;\;\;\;\;\;\;\; x_W=\frac{4M_Z^2}{M_H^2}\, .$$
As the Higgs boson mass grows, $x_W,\;x_Z\rightarrow 0$ and the leading term in Eqs.~\ref{eq:WWPartialWidth} 
and~\ref{eq:ZZPartialWidth} grows proportional to $M_H^3$. Summing over the $W^+W^-$ and $ZZ$ channels, 
the Higgs boson width in the high mass region can be written as
\begin{equation}
 \Gamma(H\rightarrow VV) = \frac{3}{32\pi} \frac{M_H^3}{v^2}\,.
 \label{eq:VV_Width}
\end{equation}
From Eq.~\ref{eq:VV_Width}, it results that $\Gamma_H \simeq M_H$ for $M_H \simeq 1\;TeV$. When \mH{} 
becomes larger than a $TeV$, therefore, it becomes experimentally very problematic to separate the Higgs boson 
resonance from the $VV$ continuum. Actually, being the resonance width larger than its own mass, the Higgs boson 
cannot be properly considered as a particle any more. In addition, if the Higgs boson mass is above 1 $TeV$, 
the SM predictions violate unitarity. All these considerations suggest the $TeV$ as a limit to the Higgs
boson mass: at the $TeV$ scale at least, the Higgs boson must be observed, or new physics must emerge.

\section{The Higgs search from  the first fb$^{-1}$ to 100 fb$^{-1}$} 

In this session the discovery strategies for the Standard Model Higgs boson are
presented with  the  focus on the results with 1, 10 and
30 $fb^{-1}$, which should correspond respectively to about one year of
data taking at the start-up luminosity and the first year and three
years at low luminosity ($2\times 10^{33} cm^{-2} s^{-1}$). 
Finally the significances for a Higgs discovery after 100~fb$^{-1}$ are
summarized.

\section{$H\rightarrow \gamma\gamma$}
\begin{figure}
\begin{center}
    \includegraphics[width=0.6\linewidth]{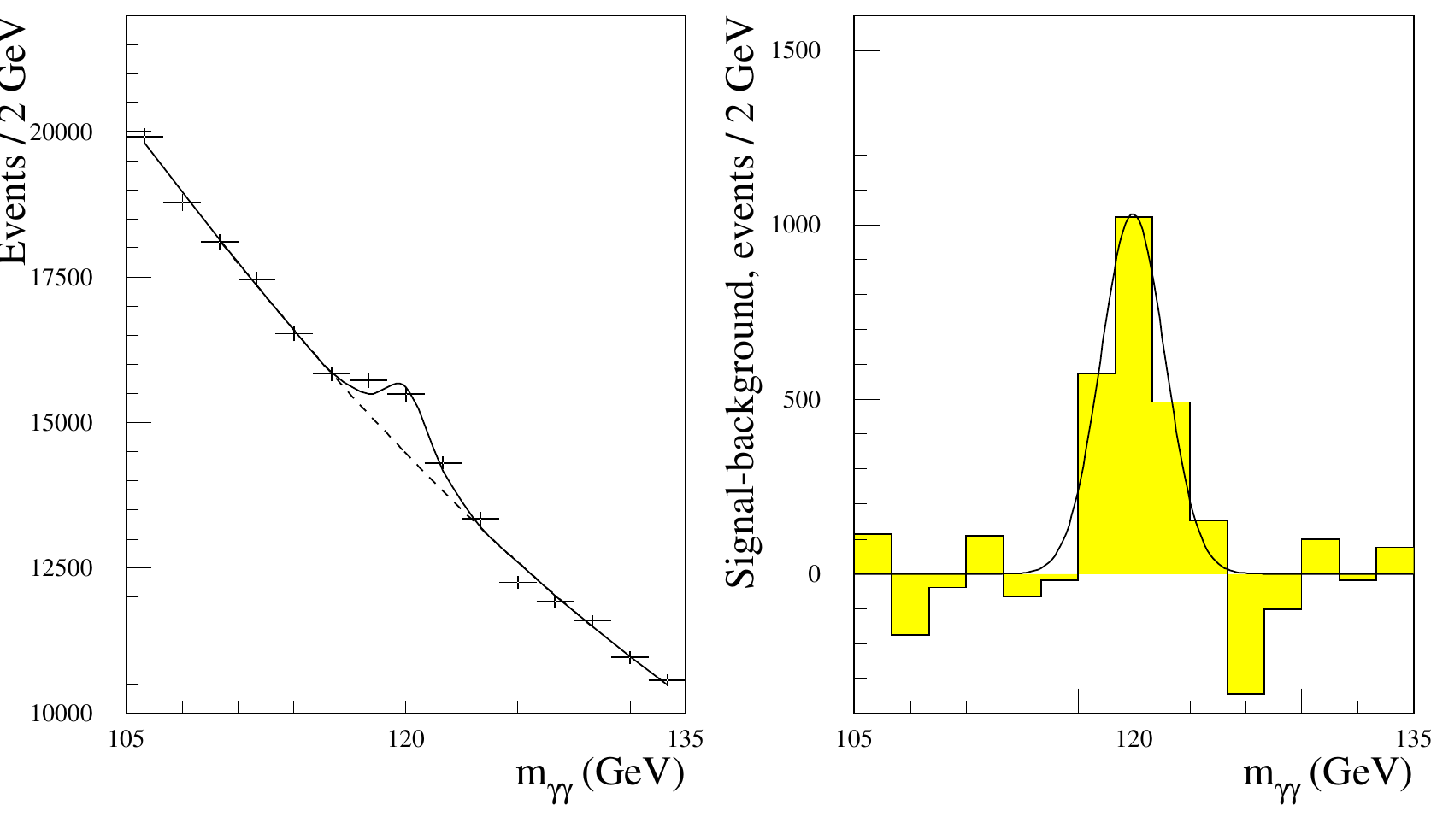}
\caption{The $\gamma\gamma$ mass spectrum from ATLAS simulation 
with an integrated luminosity of 100 fb$^{-1}$. On the right 
after the background subtraction.}
\label{gammagamma}
\end{center}
\end{figure}

If $M_H=100-140$ GeV/c$^2$ the decay with the highest probability
to observe it in $\gamma\gamma$. 
Even if the BR is very low, NLO BR ($\simeq$ 0.002), we expect a narrow 
peak in 2 photon invariant mass (see Fig.~\ref{gammagamma}).
The amount of background is very high: Drell-Yan $e^+e^-$,
$pp\rightarrow \gamma\gamma$ (irreducible), $pp\rightarrow jets
+\gamma$ and $pp\rightarrow jets$ where one or more jets are
misidentified as $\gamma$ (reducible).
In particular this last kind of background has a big dependence on the
detector performance and it involves not well known QCD physics.
Therefore there is a great deal of uncertainty in the benchmark
estimate of significance and of needed luminosity (shown in
Figures~\ref{all_channels_results} and \ref{atlas_low_lumi}.
However this will not be a systematic error on real data since the
background will be precisely measured from the data themselves,
exploiting the big $M(\gamma\gamma)$ sidebands signal free ($\simeq$ 1\%).

For this channel the resolution on the electromagnetic calorimeters is critical,
as it is the amount of material in front of them. 
In ATLAS and CMS there is about 1.5 $X_0$ of material in front of the electromagnetic
calorimeter, that makes 50\% of the electrons and photons loose more than
50\% of their energy. 

At LHC the longitudinal spread of the interaction vertexes is of 53 mm resulting in almost 
2 $Gev/c^2$ smearing on $M_H$. The charged tracks in the event will allow the determination 
of the primary vertex with a 5 mm precision in most of the signal events.

\section{$H\rightarrow b \bar b$}
Experiments are putting a lot of effort in the search for a Higgs boson decaying into
b quarks, in order to have an alternative channel with respect to the photonic one
for low Higgs masses. The background is the production of  $b \bar b$ and $ t \bar t$ pairs
than it is impossible to think to be able to isolate a Higgs produced via gluon fusion.
The Higgs boson production via $ t \bar t$ fusion presents a possible solution.
Three possible final state are taken into consideration: the fully leptonic: 
$H \rightarrow \bar{b} b, t \rightarrow \ell \nu b, ~\bar t \rightarrow \ell \nu \bar{b}$, the semi-leptonic:
$H \rightarrow \bar{b} b, t \rightarrow  \bar{q} q b, \bar t \rightarrow \ell \nu \bar{b}$ and the fully hadronic:
$H \rightarrow \bar{b} b, t \rightarrow  \bar{q} q b, \bar t \rightarrow \bar{q} q \bar{b}$
This signal will be recognized due to the presence of a high $p_T$ lepton from one of the
two W and missing energy and 4 b-tagged jets (of which two from the Higgs).
The background will be high to due the presence of many jets in the event. 
The major backgrounds are the production of $ttbb$, $Zbb$, $tt+Njets$ and multi-jets QCD events.
The main sources of uncertainty are the MC predictions, the jet energy scale 
the b-tagging efficiency.
 
A novel study for the search for a Higgs into $b\bar b$ 
\cite{hbbgamma} considers the production via Vector Boson Fusion
in association with a photon. The final state is then characterized by two forward/backward
jets, two b-jets and a central high $p_T$ photon. 
The additional request of a photon in the detector
increases drastically the signal over background ratio. Studies after detector 
effects are under way. 

\section{$H\rightarrow ZZ(^*) \rightarrow 4l$}
These channels are very promising for the Higgs detection in the mass
range 130~$GeV$-500~$GeV$, with the exception of a small interval near 
160~$GeV$ where the $H\rightarrow ZZ(^*)$ branching ratio (BR) has a
big drop due to the opening of the $WW$ on-shell production. 

The main backgrounds are:  $t\bar{t}$ ($\sigma \simeq$ 840 $pb$),
$Zb\bar{b}$ ($\sigma \simeq$ 280 $pb$) usually it is generated with CompHEP
($gg$ and $q \bar q$ initial state),
 and the irreducible $ZZ(^*)/\gamma^*$ ($\sigma\simeq$ 30 $pb$) generated as well with CompHEP 
(both $t$ ans $s$ channel)
to be compared with the $H\rightarrow ZZ(^*)$ cross section of about 10-50 $pb$.
$Zcc$ has been found to be negligible.  

The trigger and the offline cuts applied in the analysis rely on the presence of
isolated charged leptons coming from the primary vertex and with high transverse momentum.
The $Z$ mass peak is also a powerful feature: more than 50\% (80\%) of
the events have at least one on-shell $Z$ for $M(H)>115~(150)~GeV$.
Requiring lepton isolation and cutting differently on the sorted $p_T$ of the leptons
the reducible background become negligible.
The irreducible background can be partly cut away with cuts on angular variables 
(that are Higgs Mass dependence).

The studied final states are $2e2\mu$, $4\mu$ and $4e$. The first
has the biggest BR while the second is the cleanest one. The main
concern of the last channel is the presence, for low Higgs masses, of
very soft electrons, well below the range for
which the reconstruction will be best controlled via single $Z$
and $W$ measurements. 
The main systematics error sources are: the choice of the PDF and the QCD scale, 
the NLO versus the LO dynamics, the isolation cut and its efficiency, the electron reconstruction
efficiency, the energy and momentum scale and the charge identification.
It is important to normalize with the data itself by using two other control samples:
the Drell-Yan and the side bands of the Higgs spectra that suffer from low statistics.  

In Figures~\ref{all_channels_results} and \ref{atlas_low_lumi} the luminosity needed for a $5\sigma$
discovery and the significance achievable with 30 $fb^{-1}$, 
combining the three possible final states, is plotted as a function of the Higgs mass. 

\section{$H\rightarrow WW(^*) \rightarrow l\nu l\nu$}
\label{H_WW_lnln}
The leptonic decays of both the $W$ in the $ee$, $e\mu$ and
$\mu\mu$ combinations have been studied.
The signal has a cross section of 0.5-2.3 $pb$ with a peak at
$M(H)=160~GeV$ but does not present an invariant mass peak due to presence
of the two neutrinos. 
This channel present a very clean signature: 2 isolated high $p_T$ leptons
pointing to the primary vertex and high missing energy and no hadron activity.
The main backgrounds are single and double top
production ($\sigma \simeq$ 90 $pb$) and double boson production ($\sigma \simeq$ 15
$pb$), considering only the fully leptonic decays.
The Drell-Yan background after the full selection should be less
than 2\% of the total background (there is no high missing energy).
Figures~\ref{all_channels_results} and \ref{atlas_low_lumi} shows the luminosity needed for a
$5\sigma$ discovery and the significance obtained with 30 $fb^{-1}$ as
a function of the Higgs mass.

\begin{figure}
\begin{center}
    \includegraphics[width=0.6\linewidth]{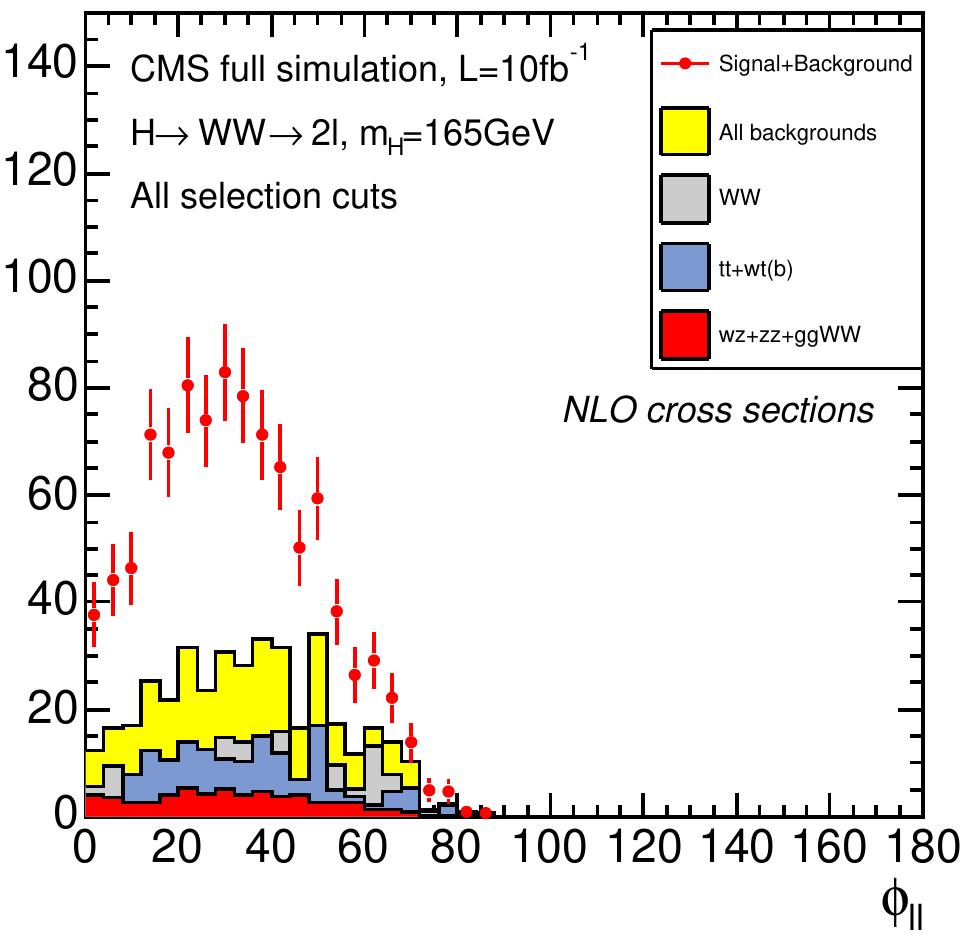} 
\caption{Distribution of the angle between the charged leptons in the
  transverse plane after all selection cuts (excluding the one on
  $\phi_{ll}$), in $H\rightarrow WW(^*) \rightarrow l\nu l\nu$ signal
  (with $M(H)=165~GeV$) and in various backgrounds,
  with 10 $fb^{-1}$ of integrated luminosity.}
\label{WWlnln_phi_ll}
\end{center}
\end{figure}

The final state selection relies mainly on the request of high missing
energy ($>50~GeV$) and on a central jet veto.
The main kinematic peculiarity of this channel is the closeness of the
two charged leptons due to the fact the Higgs boson is a scalar and that the
V-A structure of the weak interaction. For this channel the 
correct simulation of the spin correlation matters (see Figure~\ref{WWlnln_phi_ll}) 
The absence of the Higgs peak 
requires an high signal over background ratio and a good
control of the background shape.
Therefore a procedure to normalize the background from the data is
necessary: a different signal free region for each background has been
defined varying the analysis cuts. The uncertainties for the various
backgrounds are between 15\% and 20\%, with the
exception of single top and $gg\rightarrow WW$ processes for which it's
not possible to find a good normalization region so that the
systematics ($\simeq$ 30\%) are dominated by MC theorethical errors.

\section{Vector Boson Fusion Production} 
During the last years a lot of effort has been put on the Vector Boson
Fusion production channels. These channels present a very clear signature
given by the two forward and backward jets. The presence of these two jets
together with the decay products of the Higgs allows a good rejection of 
dominant background coming from V+n jets, VV+n jets and $t \bar t$ production.
The forward and backward jets tend to be well separated in pseudo-rapidity and 
to have a very high invariant mass. 

More generally the Vector Boson Fusion cross section (with or without a production
of a Higgs particle) is an extremely interesting
process to study  because the cross section
$\sigma(pp\rightarrow VVjj)$ and the polarizations of the $VV$
pair depend sensitively on the presence or absence of a light Higgs in the
physical spectrum.
If a massive Higgs boson exists, a resonance will be observed in the $VV$ invariant mass
spectrum in correspondence of the Higgs mass.
In absence of the Higgs particle, the Standard Model (SM) predicts that the
scattering amplitude of longitudinally polarized vector boson grows linearly with $s$
and violates unitarity at about 1--1.5 TeV. 
As a consequence, the measurement of the cross section at large $M(VV)$ could
provide information on the existence of the Higgs boson independently of its direct
observation. 
In particular, absence of strong interactions in high energy boson-boson scattering
could provide a strong incentive to probe harder 
for a light Higgs, which will require several years of data taking for a reliable
discovery.
But even if a Higgs particle is discovered, in this or other channels, it will be necessary
to verify that indeed longitudinally polarized vector bosons are weakly coupled
by studying boson-boson scattering in full detail.

The following vector boson fusion processes can be studied at LHC:
\begin{eqnarray}
qq \rightarrow qqVV \rightarrow qqVZ \rightarrow qqqq\mu\mu / ee \nonumber\\
qq \rightarrow qqVV \rightarrow qqVW \rightarrow qqqq\mu\nu / e\nu.\nonumber
\end{eqnarray}
They offer a clear experimental signature, because of the presence of high $p_T$ leptons from the
$W$ or $Z$ decay, together with the highest branching ratio among the final states which are possible
to reconstruct in an hadronic environment.
In fact boson-boson scattering with a totally hadronic final state cannot be isolated from the non 
resonant QCD backgrounds whose cross section is much higher. 

Final states where both bosons decay leptonically have been also analyzed:
\begin{eqnarray}
qq \rightarrow qqVV \rightarrow qqZZ \rightarrow qq\mu\mu\mu\mu / qqeeee \nonumber\\
qq \rightarrow qqVV \rightarrow qqZW \rightarrow qq\mu\mu\mu\nu \nonumber\\
qq \rightarrow qqVV \rightarrow qqW^{\pm} W^{\pm} \rightarrow qq\mu^{\pm}\nu\mu^{\pm}\nu \nonumber
\end{eqnarray}
They have a small rate but a very clear signature. Moreover in these channels both of the
outgoing bosons can be unequivocally recognized as $W$ or $Z$ and this can
be useful for the study of the cross section behavior at high boson-boson invariant mass. 
Finally in the $qqVV \rightarrow qqW^{\pm}W^{\pm} \rightarrow qq\mu^{\pm}\nu\mu^{\pm}\nu$
process the exact $VV$ invariant mass cannot be reconstructed so
an appropriate kinematic treatment is necessary. Nevertheless the
study of this final state in the high invariant mass region is very
promising because in the $W^{\pm}W^{\pm} \rightarrow W^{\pm}W^{\pm}$ process the
enhancement of the cross section due to the unitarity violation in the
no-Higgs case is large.

All the listed channels have similar kinematic behavior.
The particles in the final state coming from the decay of a $W$ (or a $Z$)
are expected to have quite high transverse momentum ($p_T$) and to be
mostly produced centrally in the detector, 
i.e. at low absolute value of pseudo-rapidity~($\eta$).
On the contrary, the two quarks that have radiated the vector bosons, 
the so called ``spectator quarks'',
tend to go in the forward/backward regions at high $\vert\eta\vert$ and they have 
very large energy and $p_T$.
Thanks to their peculiar kinematic pattern, the presence of these two spectator quarks 
is essential to tag the $VV$ fusion events as a six fermions final state,
therefore they also are called ``tag quarks''.

The most problematic background for the vector boson fusion signal in
the semi-leptonic final state is the production of a single $W$ (or $Z$) in
association with n jets (n=2,3,4,5) which has a huge cross section (of
the order of nanobarns). The background most difficult to
reject in the totally leptonic channel is instead the production of a
couple of bosons in association with n jets (n=0,1,2,3) with a cross
section of some picobarns. Another
potentially dangerous background with a big cross section is the QCD production of top
pairs ($\simeq 200 $ pb). Lastly, the irreducible background coming from
$\bar{t}t$ Electroweak production, Triple and Quartic Gauge Coupling (TCG,QGC)
and non resonant 6 fermion final state, must be considered, which
has a cross section of the same order as the signal and a very similar
kinematic behavior.

For a complete overview of one of the analysis see e.g. 
~\cite{nostra_nota}. In the following section some of the final state
already accessible at low luminosity will be addressed as an example.
In general, the discovery of the Higgs mass peak up to 500 GeV should
require something more than 100 fb-1, while in the case of absence of
the Higgs, due to the quite poor signal over background ratio, 
it is still difficult to say how long it will take
to be able to relay only on the high $M(VV)$ region to understand the mechanism 
which breaks the symmetry, and high luminosity will surely be needed.

\subsubsection{$qqH$ with $H\rightarrow WW$}
The analysis of the fully leptonic decay channel ($qql\nu l\nu$ final
state) is similar to that described in the previous section
(Sec.~\ref{H_WW_lnln}). This process has a lower cross section (50-250
$fb$) but the presence of the two additional quarks from the VBF, 
with high energy and pseudorapidity, can be exploited to
disentangle the signal from the background.

The semileptonic decay channel ($qqqql\nu$ final state) has the
advantage of a higher BR 
and it allows to reconstruct the Higgs mass peak.
On the other hand it suffers from very high background: double
top ($\sigma\simeq$ 840 $pb$), single top ($\sigma\simeq$ 100 $pb$), double boson
plus jets ($\sigma\simeq$ 100 $pb$) and single boson plus jets
($\sigma$ bigger than
1~$mb$), to be compared with the $qqH\rightarrow qqWW$ cross section 
of about 0.6-2.7 $pb$. Thus strong cuts are necessary and
this implies a good knowledge of the physics involved. However the
cross sections of the multiple jets processes at the LHC scale are not yet very well
known and they will be measured precisely only from the LHC data themselves.
Moreover many systematics about the jets detection and reconstruction
are still quite uncertain, they can be understood and measured 
only from the data.

The preliminary estimation of the significance with 30 $fb^{-1}$ is
shown in Figures~\ref{all_channels_results} and \ref{atlas_low_lumi}(left). 

\subsubsection{$qqH$ with $H\rightarrow \tau\tau$}
This channel has been analyzed with one $\tau$ decaying into leptons
and the other $\tau$ into hadrons ($\sigma \simeq$ 50-160 $pb$).
The irreducible backgrounds are the QCD and EW production of two
$\tau$ leptons from $Z/\gamma^*$ with associated jets (QCD
2$\tau$+2/3 jets $\sigma \simeq$ 1.6 $pb$, EW 2$\tau$+2 jets $\sigma \simeq$ 230
fb). The reducible backgrounds considered are the $W$+ multi-jet
production ($W$+3/4 jets $\sigma \simeq$ 14.5 pb with $W\rightarrow \mu\nu$) 
and $t\bar{t}$ events ($\sigma \simeq$ 86 pb with $W\rightarrow l\nu$), in which one of the jets can be
misidentified as a $\tau$-jet.

\begin{figure}
\begin{center}
    \includegraphics[width=0.6\linewidth]{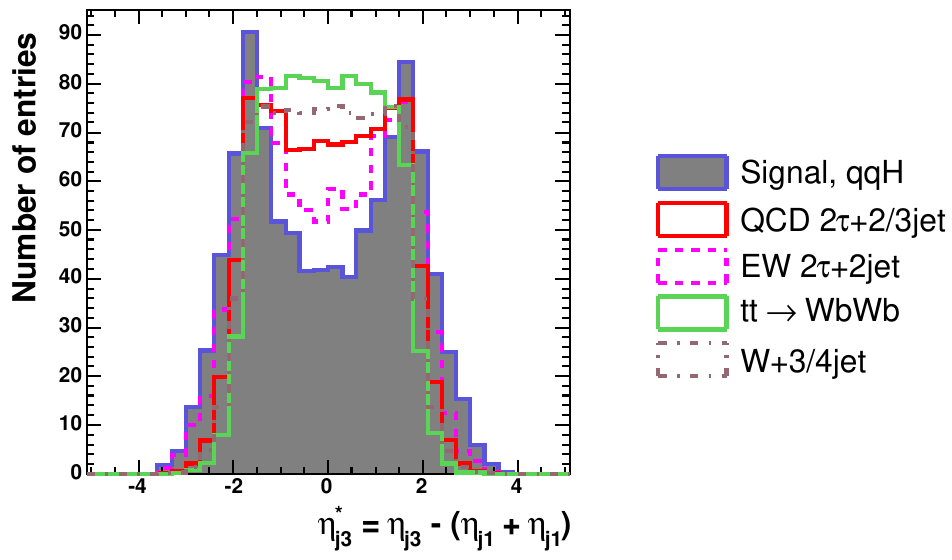} 
\caption{The pseudorapidity ($\eta$) distribution of the 3rd jet with
  respect to the two forward jets, $\eta^*_{j3} = \eta_{j3} -
  (\eta_{min} + \eta_{max}) $, for the VBF $H\rightarrow \tau\tau$
  signal and the various backgrounds in CMS. The total number of entries is
  normalized to 1000 events.}
\label{Htt_gap}
\end{center}
\end{figure}

This analysis has to reconstruct a very complex final state.
The hadronically decaying $\tau$ is reconstructed from a little ($\Delta
R=0.4$) isolated jet. A very low impurity (2.7\%) is obtained thanks
to the selection cuts, costing a low reconstruction efficiency
(30\%). The energy resolution on the reconstructed $\tau$ is 11.3\%.
The leptonically decaying $\tau$ is recognized from the electron or
muon with highest transverse momentum, requiring $p_T>$ 15 $GeV$.
The $\tau$ energies are calculated using collinear approximation of
visible part of $\tau$ decay products and neutrinos.
A raw (not calibrated) missing transverse energy (MET) greater than 40 $GeV$
is required. The MET resolution after all corrections is 20\%, this is
the largest contribution to the Higgs mass resolution.
Finally the presence of the two quarks emitting the
bosons in the VBF process can be exploited: they have very high
energy and high rapidity gap (as shown in Figure~\ref{Htt_gap}) 
because there is not color exchange between them, being produced trough an EW process.
After having removed the $\tau$ jet and the two VBF jets, a central
jet veto is applied using a Monte Carlo jet energy calibration.

\begin{figure}
\begin{center}
\includegraphics[width = 6.9 cm]{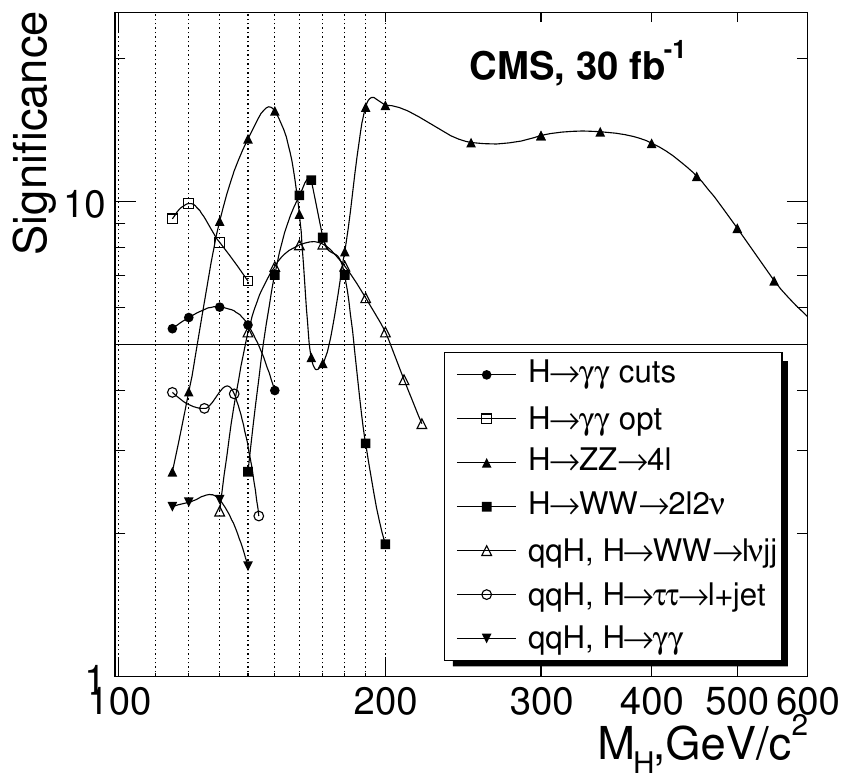} 
\includegraphics[width = 6.9 cm]{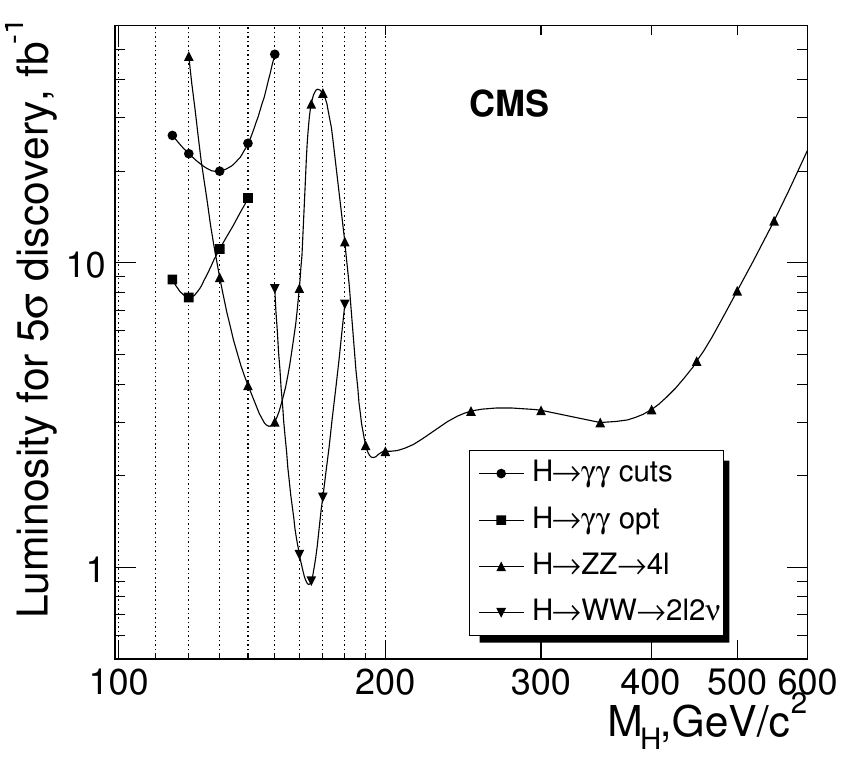} 
\caption{Significance achievable with 30 $fb^{-1}$ (left) and
  luminosity needed for a $5\sigma$ discovery (right) in the various
  channels as a function of the Higgs mass with the CMS detector \cite{cms_ptdr}.}
\label{all_channels_results}
\end{center}
\end{figure}

\begin{figure}
\begin{center}
\includegraphics[width = 6.9 cm]{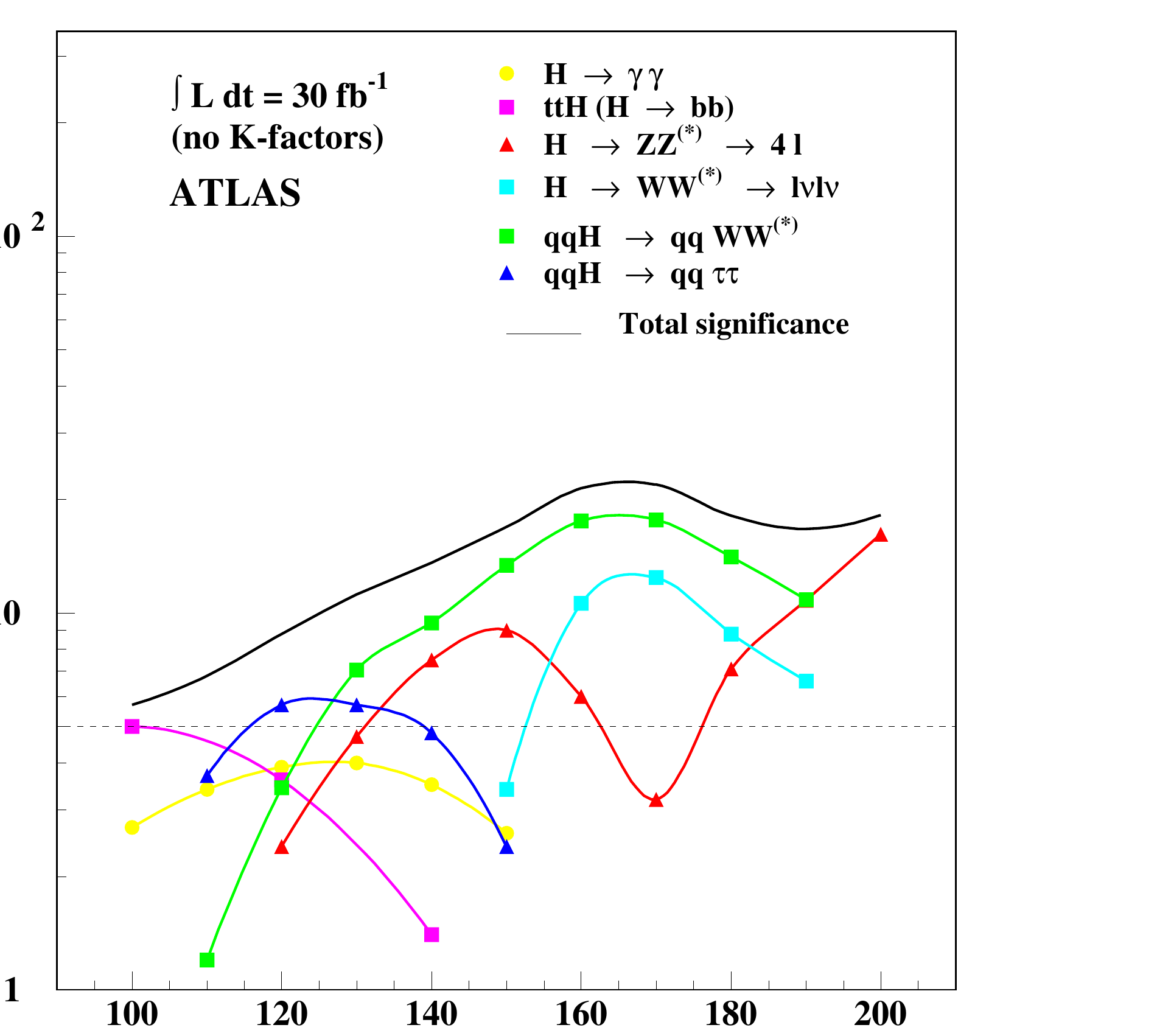} 
\includegraphics[width = 6.9 cm]{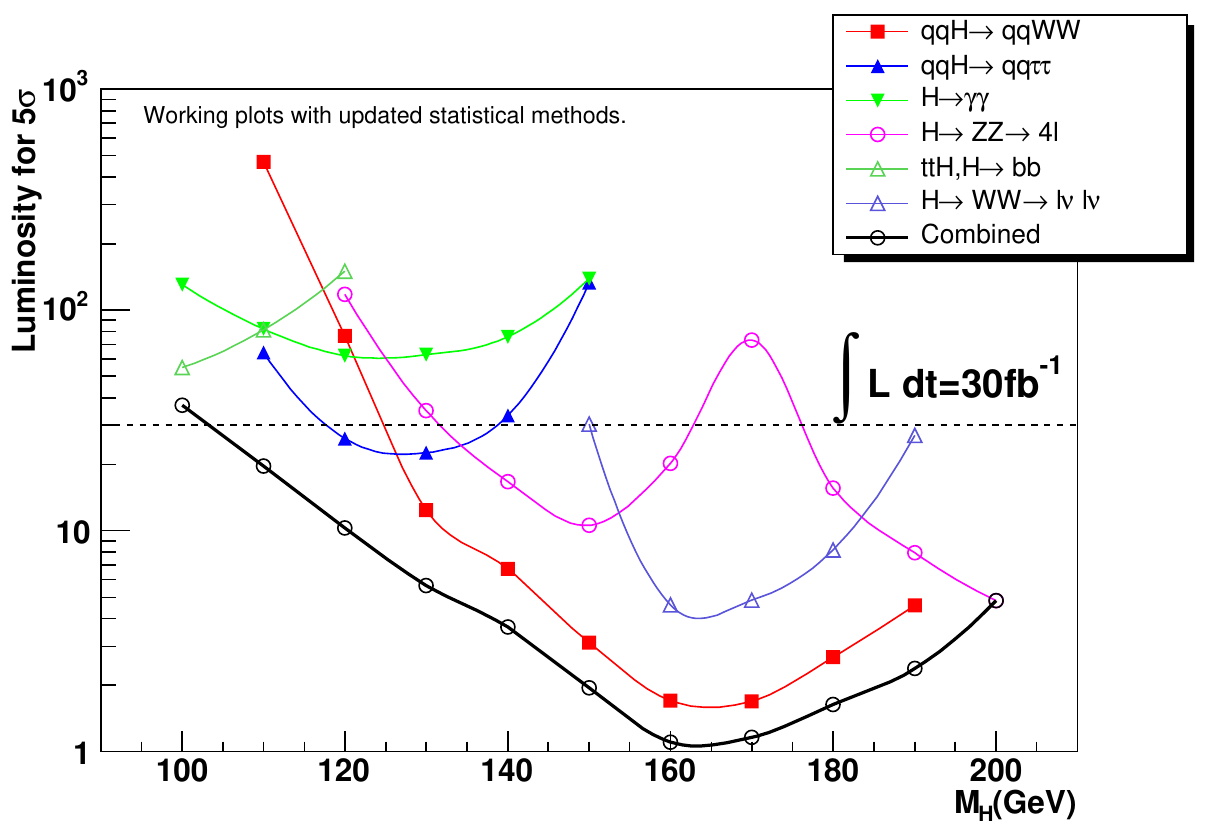} 
\caption{Significance achievable with 30 $fb^{-1}$ (left) and
  luminosity needed for a $5\sigma$ discovery (right) in the various
  channels as a function of the Higgs mass with the ATLAS detector.}
\label{atlas_low_lumi}
\end{center}
\end{figure}

The significance exceeds 3$\sigma$ with 30 $fb^{-1}$, as reported in
Figures~\ref{all_channels_results} and \ref{atlas_low_lumi}(left). The number of events is measured
directly from the data fitting the $M(\tau\tau)$ distribution.
The uncertainty on the number of background events (7.8\% with 30
$fb^{-1}$) is computed from its spread in 10.000 toy Monte Carlo data
distributions generated following the fit results.

\subsection{The Higgs signal Significance at LHC}
Figure~\ref{all_channels_results} and figure~\ref{atlas_low_lumi} (see ref. \cite{cms_ptdr}
and \cite{atlas_ll}) are a good
summary of the CMS and ATLAS potential for the Higgs discovery with low luminosity.

\begin{figure}
\begin{center}
\includegraphics[width = 6.9 cm]{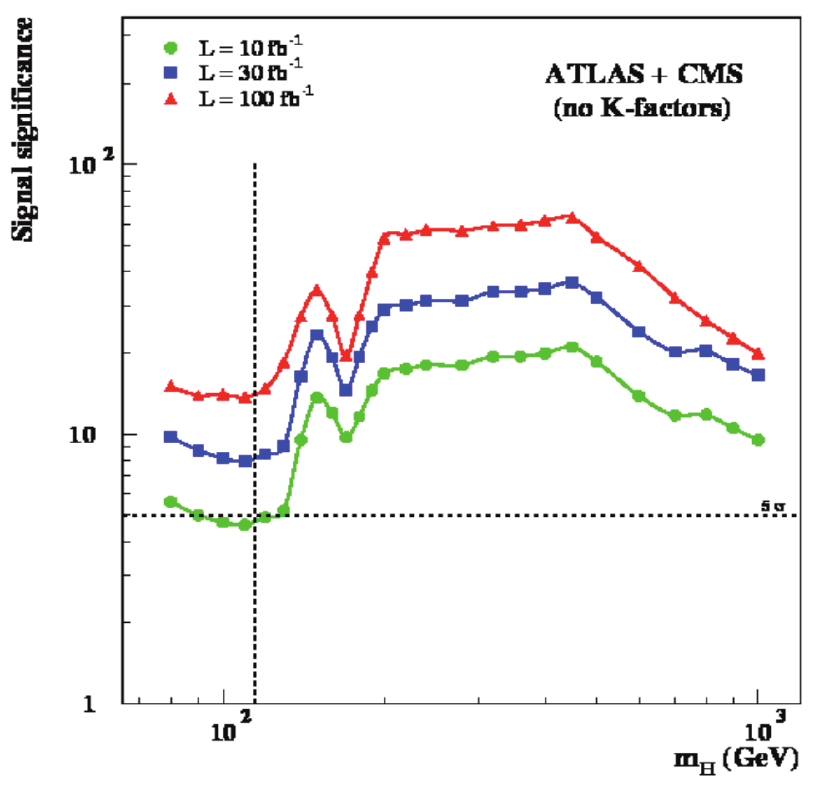} 
\caption{Significance achievable with 10, 30 and 100 $fb^{-1}$ for the two
 experiment combined together.}
\label{LHC_results1}
\end{center}
\end{figure}

\begin{figure}
\begin{center}
\includegraphics[width = 8 cm]{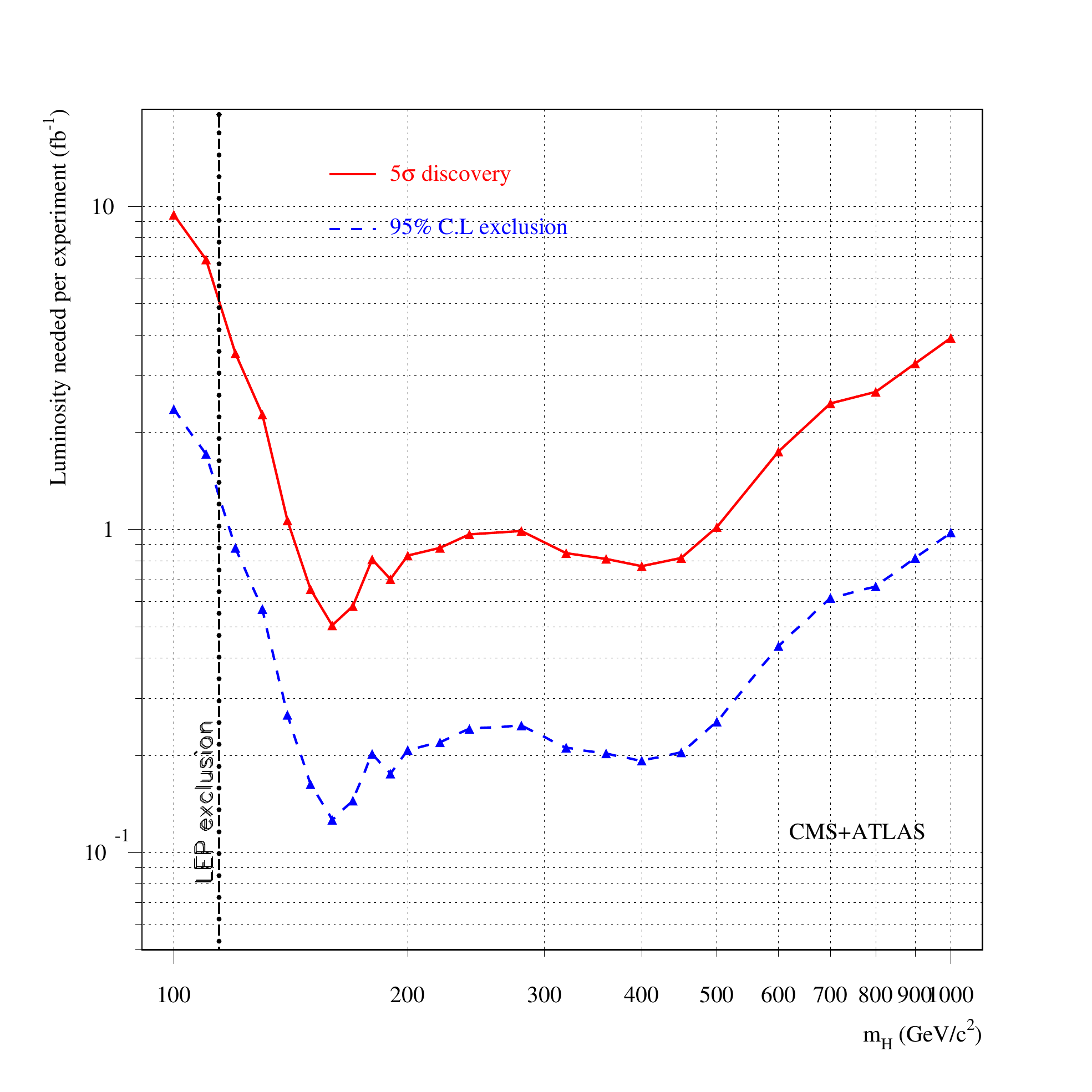} 
\caption{The needed luminosity for an exclusion at 95\% CL or for a discoryat 5$\sigma$}
\label{LHC_results2}
\end{center}
\end{figure}

The various channels will be combined to get a coherent picture. Depending on the Higgs mass
and on the cross section~$\times BR$ only 10 fb$^{-1}$ or few years of data taking
will be necessary to get a undiscussed significance. 
Figure~\ref{LHC_results1} shows instead the 
ATLAS and CMS potential for the Higgs discovery including as well one year at high luminosity.
In figure~\ref{LHC_results2} the needed luminosity for a 95\% exclusion and for a 5$\sigma$ 
discovery are shown as a function of the Higgs boson mass.
It should be noticed that a careful preliminary work must be
done in order get those results: the first data will be used to study
the detector systematics (in particular the control of the jets
response and of the MET resolution will be difficult at the beginning)
and to measure the cross sections of multi-jets background processes
(mainly $t\bar{t}$ and single and double boson production in association with jets).

\addtocounter{chapter}{1}




%
\mchapter{WW Scattering}
{ Authors: Alessandro Ballestrero and Ezio Maina}
\vskip 1cm
The Standard Model (SM) has passed with flying colours about twenty years
of comparisons with precision electroweak data\cite{lepewwg}.
However at present we don't
have yet any direct experimental evidence for the mechanism which breaks the 
$SU(2)_L \times U(1)_Y$ down to $U(1)_{EM}$. In the SM this is
accomplished by a single scalar doublet which also provides masses to all
fermions.
Despite its simplicity, elegance and spectacular succes the SM leaves a number
of unanswered questions \cite{reviews}:
\begin{itemize}
\item Why Electroweak Symmetry Breaking (EWSB) occurs and why at the weak scale
      $v\approx 250$ GeV.
\item Why fermions have the experimentally measured masses. Why three families.
\end{itemize}
and a number of shortcomings:
\begin{itemize}
\item It involves fundamental scalars, while none such particle has been
observed (In a supersymmetric world however fundamental scalars would be quite
natural).
\item If the theory has to be valid up to the GUT
or Planck scale, the parameters of the theory have to be fine-tuned in order to
keep the electroweak scale low instead that at the large mass scale.
\item Scalar theories, if they have to make sense, that is if their running
couplings must remain finite, at arbitrary large energies, are trivial: the
quartic coupling at low energies must be zero.
\item It generates a cosmological constant about 50 orders of magnitude larger
than the experimental upper bound.
\end{itemize}

Several theories have been proposed, which solve at least some of the above
problems. Schematically one can group them in four categories \cite{sd}:
\begin {description}
\item{-} {\it Supersymmetry.} The Higgs sector consists of two Higgs doublets 
which result in 5 Higgs particles: 2 neutral, two charged and one pseudoscalar. 
The lightest neutral Higgs is predicted to be not heavier than about 160 GeV.
\item{-} {\it Little Higgs.} In these models there is an expanded gauge 
symmetry at the TeV scale that contains the standard $SU(2)xU(1)$.
An approximate global symmetry prevents the Higgs from obtaining quadratically
divergent mass at one loop. The Higgs boson is a pseudo-Golstone boson
resultimg from spontaneous breaking of the approximate symmetry and it
is therefore light. These models contain new heavy gauge bosons whose mass
is of the order of the TeV. 
\item{-} {\it Dynamical symmetry breaking.} EWSB
arises in these theories from chiral symmetry breaking of a new strong 
interacting gauge sector. The role of the Higgs is played by a condensate
of new heavy quarks (techniquarks). The oldest version of these theories, 
Technicolor,  dates back to 1976 but it was incompatible with electroweak data. 
Then Extended Techincolor was  intended to explain also the problem of quarks 
and  leptons flavour but this induced problems in preventing flavour changing 
neutral corrents. Successive versions went under the name of Walking  
Technicolor (with different  scales of chiral symmetry breaking). 
For a modern realization of these kind of theories see ref. \cite{sanni} 
\item{-} {\it Higgless models.} Models with extra dimensions can generate 
EWSB from boundary conditions on the brane of the 
extra dimensions. The Higgless models all contain a tower of Kaluza Klein 
particles with the quantum numbers of the SM gauge bosons. These particles
take part in the elastic scattering amplitudes and restore
unitarity as the Higgs does in the SM.
 
\end {description}

The last two groups of theories, as any theory in which there is no 
elementary Higgs particle or this is much heavier than  1 TeV, give rise to 
strong scattering among heavy bosons at high energy, which results in 
predictions markedly different from the SM case and in possible formation 
of resonances. There are also interesting theories with
pseudo Goldstone Higgs in which, even though a low mass Higgs is present,
strong scattering between bosons at high energies is predicted \cite{conti}.
    
\begin{figure}[bht]
\begin{center}
\mbox{\epsfig{file=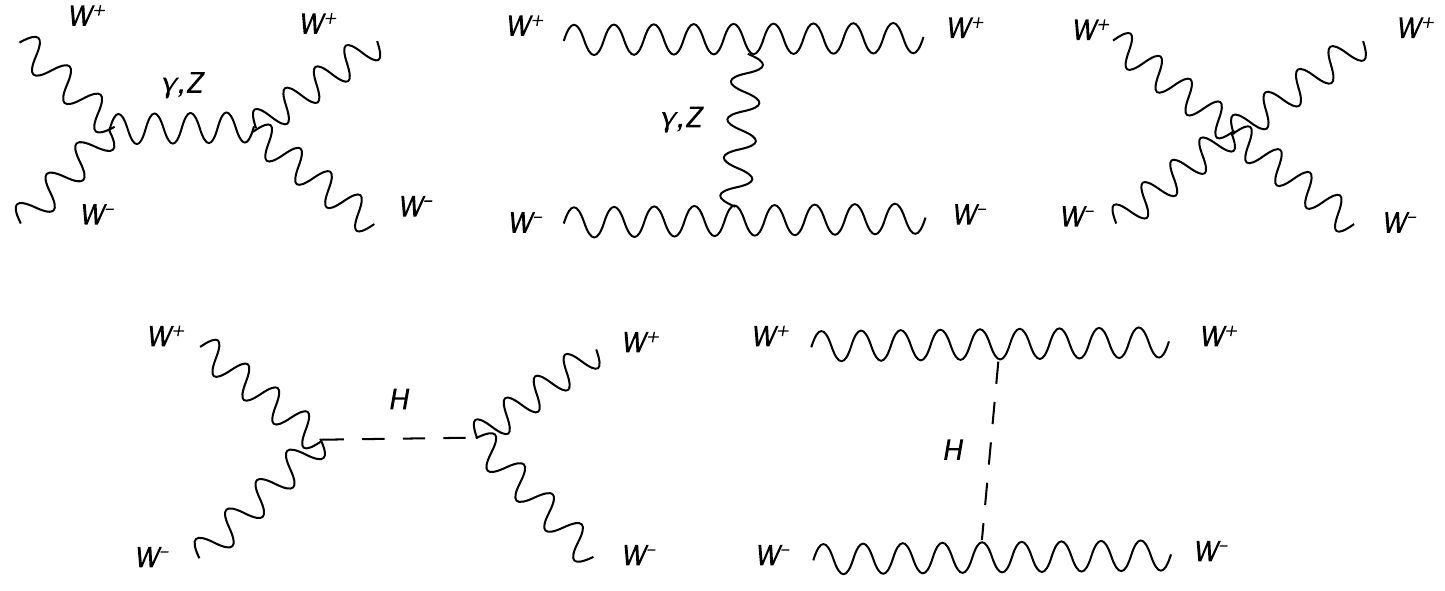,width=12.cm}}
\caption{Vector boson scattering process.}
\label{VV-diag}
\end{center}
\end{figure}

The centrality of WW scattering to the exploration of EWSB stems from 
the issue of cancellation of high energy divergences.
Any scattering amplitude in a consistent quantum mechanical theory must respect
the unitarity of the $S$
matrix, which is equivalent to the conservation of total probability. 
This implies that no amplitude can indefinitely grow with energy.
The reaction which best exemplifies the relationship between unitarity and EWSB
is the scattering among longitudinally polarized vector bosons.
The Feynman diagrams for $W^+W^-\rightarrow W^+W^-$ are shown in
Fig.(\ref{VV-diag}).
The polarization vectors of a transversely/longitudinally $(T/L)$ polarized
$W$ boson traveling along the  $\hat{z}$ axis are:
\begin{equation}
\epsilon_T = \left( 0;\pm \frac{1}{\sqrt{2}},\frac{-i}{\sqrt{2}}\right) \qquad
\epsilon_L = \frac{1}{M_W}\left( \vert \vec{k} \vert ; 0, 0, E_W \right)\qquad
\vec{k/}/\hat{z}
\end{equation}
\noindent
so that for $E_W \gg M_W$ $\epsilon^\mu_L = k^\mu/M_W + O(M^2_W/E^2_W)$.
Therefore

\begin{equation}
\epsilon^{W^+}_L \cdot \epsilon^{W^-}_L \approx 
\frac{k_{W^+}\cdot k_{W^-}}{m^2_W} = \frac{s}{2 m^2_W}
\end{equation}
and the leading behaviour of each diagram $D_i$ in the top row of
Fig.(\ref{VV-diag}) is:
\begin{equation}
D_i \propto \frac{k_{W^+}\cdot k_{W^-}}{m^2_W} 
                \frac{k_{W^+}\cdot k_{W^-}}{m^2_W} 
= \frac{s^2}{4 m^4_W}
\end{equation}
However the terms proportional to $s^2$ cancel when we sum the five diagrams
in the top row,
leaving an amplitude proportional to $s$. This unacceptable 
behaviour is canceled by the sum of the two Higgs exchange diagrams leaving an
amplitude which tends to a constant in the high energy limit.

The linear rise with $s$ of the 
$WW$ scattering amplitude in the absence of the Higgs,
which is predicted by the Low Energy Theorem
(LET) \cite{LowEnergyTheorem},is completely analogous
to the threshold behavior of the pion-pion amplitude. In both cases it is a
consequence of their nature of Goldstone bosons.

It should be noticed that the issue of bad high energy behaviour of electroweak
amplitudes for longitudinally polarized vector bosons and its link with the
Higgs boson is completely general.
For instance the amplitude for $e^+e^- \rightarrow  W^+_LW^-_L$ with
{\it massive} electons grows as $\sqrt{s}$ in the absence of the Higgs and only
when the Higgs exchange graph is included the amplitude displays an acceptable
high energy limit.

The analysis of the interactions among longitudinally polarized bosons
is simplified by the Equivalence Theorem which states 
that for any longitudinally polarized boson $V^i_L$ and corresponding Goldstone
boson $\omega^i$:
\begin{equation}
A(V^1_L \dots V^N_L \rightarrow V^1_L \dots V^{N^\prime}_L ) = 
(i)^N(-i)^{N^\prime}
A(\omega^1 \dots \omega^N \rightarrow \omega^1 \dots \omega^{N^\prime} ) 
+O\left(\frac{M^2_V}{s}\right)
\end{equation}

The Equivalence Theorem, in addition to make the calculation of high energy reactions among 
longitudinal vector boson easier, is physically interesting because it allows
to link boson boson scattering with pion pion scattering at low energies.
This is because the Lagrangian of the Goldstone bosons,
is identical to the linear
$\sigma$-model which has been used to describe  pion-pion
interactions (Veltman \cite{reviews}). 
In the limit $m_H\rightarrow \infty$ the linear sigma model
leads to the non--linear sigma model
\cite{NonLinearSigma} in which the effects of the heavy Higgs appear
via an infinite tower of non-renormalizable terms of progressively higher
dimension. These terms are multiplied by appropriate inverse powers of the heavy
Higgs mass in order to keep the overall mass dimension to four as required.   
More generally one use this approach to parametrize any new physics effect, with
the scale of the onset of new physics $\Lambda$ in place of the Higgs mass. 
This leads to the so called Electroweak Chiral Lagrangian (EChL).
In this effective
field theory corrections to observables generated by new physics can be computed
systematically truncating the series in $E/\Lambda$ at some fixed order, where $E$
is the relevant energy of the experiment.
 This procedure is equivalent to taking into account only operators
up to a fixed dimension and is valid for $E<<\Lambda$.

\begin{figure}[th]
\begin{center}
\mbox{
{\epsfig{file=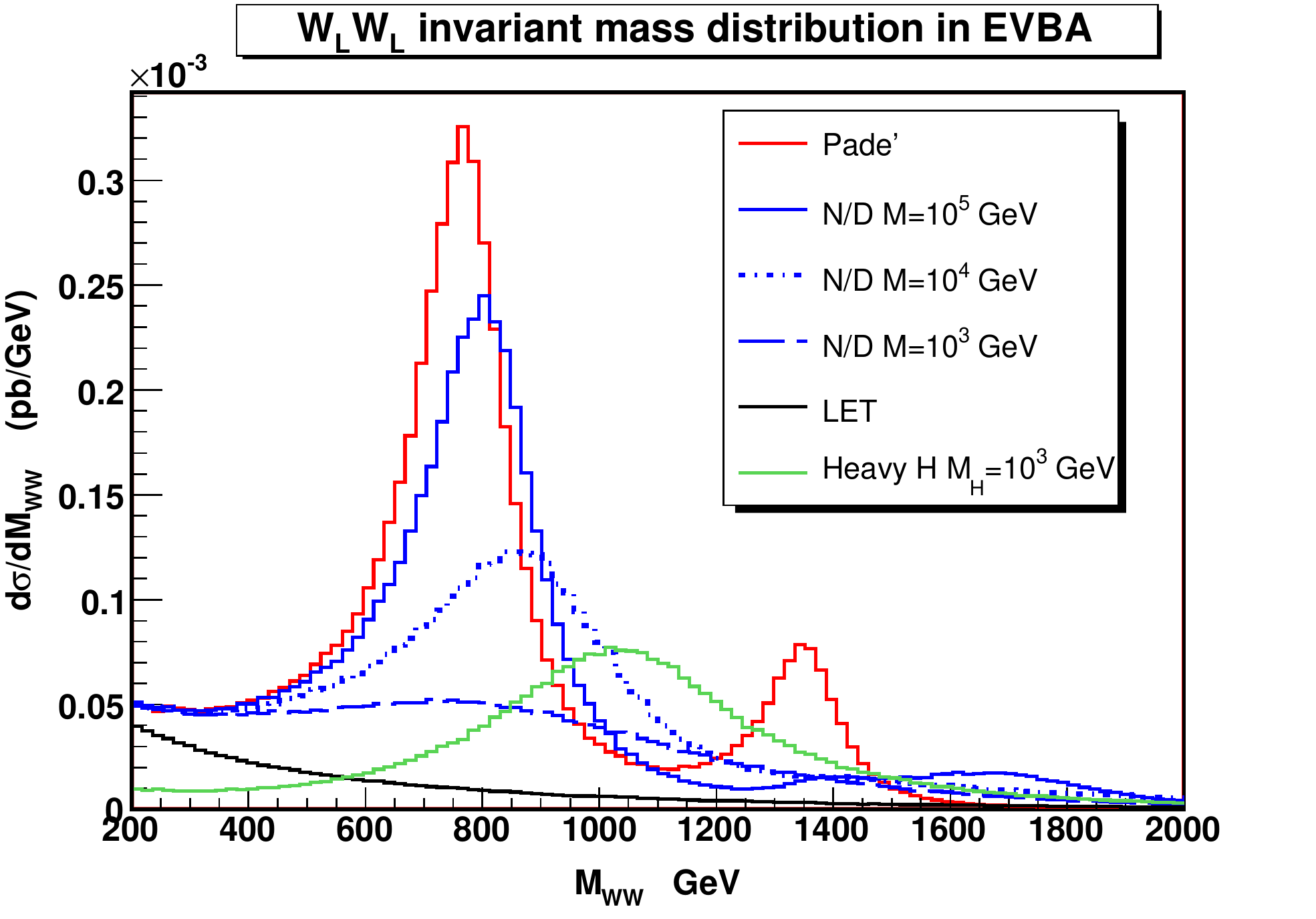,width=10cm}}
}
\caption{Invariant mass distribution of the 
two final state W's in $W_L W_L\rightarrow W_LW_L$
with different unitarization procedures.
For comparison the cross section for a Higgs boson of 1 TeV and the
LET prediction are also shown. Notice that in this approximation the SM cross
section with a light Higgs would be zero.} 
\label{WWMassDistribution}
\end{center}
\end{figure}

In $WW$ scattering, in the absence of a relatively light Higgs boson,
tree-level unitarity is violated at about 1 TeV (Typically other processes 
clash with unitarity at a much larger energies),
therefore either the Higgs must exist or some other mechanism must intervene
at about the TeV scale and play the same role in taming the divergent
behaviour of high energy ampitudes. Hence these processes are
the ideal testing ground for the mechanism of EWSB.

However at the LHC, or any other collider, no beam of on shell EW bosons will be
available. Boson boson interactions will be initiated by
the emission of spacelike virtual bosons from the incoming quarks. 
These bosons will eventually scatter among themselves and finally decay.
These kind of events is characterized by the presence of two energetic jets in
the forward and backward direction and by high $p_T$ jets or leptons in the
central part of the detector from the decay of the final state bosons. 
It is by studying these Vector Boson Fusion (VBF) events that we hope to obtain
clues about the behaviour of boson-boson scattering.

The analogy with pion-pion scattering suggests to apply the same
unitarization techniques
which have proved useful in low-energy QCD also in high-energy boson scattering.
There are various methods to construct amplitudes which satisfy the unitarity
constraints on the basis of the first few terms in the perturbative expansion.
Unfortunately the different methods result in different predictions for 
boson-boson scattering. This is illustrated in Fig.(\ref{WWMassDistribution})
which is based on the paper of Butterworth{\it et al.} \cite{unitarization}
to which we refer for the details.
In Fig.(\ref{WWMassDistribution}) we present the predictions at the LHC
for the Pade' and
N/D method (for three values of the mass parameters M) for a specific set of
values of the coefficient of the dimension--4 operators  in the EChL.
For comparison we also show  the cross
section for a Heavy Higgs of 1 TeV and the result from the LET
amplitude which corresponds to the infinite Higgs mass case.
Notice that the growth proportional to $s$ of the LET amplitude is completely
swamped by the decresase of the PDF distribution functions at large $x$. 
The results of Fig.(\ref{WWMassDistribution}) have been obtained in the 
Equivalent Vector Boson Approximation (EVBA) 
\cite{EVBAref} and only include longitudinal $W$'s.
The EVBA provides a particularly simple and
appealing framework in which the cross section for the
full process is approximated by the convolution
of the cross section for the scattering of on shell vector bosons times
appropriate
distribution functions which can be interpreted as the probability of the
initial state quarks to emit the EW bosons which then interact.
This approach relies on the neglect of all diagrams which do not include
boson boson scattering subdiagrams and on a suitable on-shell projection
for the scattering set of diagrams.
It has been shown \cite{ABBM06}
that in general EVBA results and their relationship to
exact results depend quite sensitevely on the set of cuts
which need to be applied in order to obtain a finite result when photon exchange
diagrams are included.
Therefore, it is extremely difficult to extract from the EVBA
more than a very rough estimate of the actual behaviour of the Standard Model
predictions for boson boson scattering.

\begin{figure}[th]
\begin{center}
\mbox{
{\epsfig{file=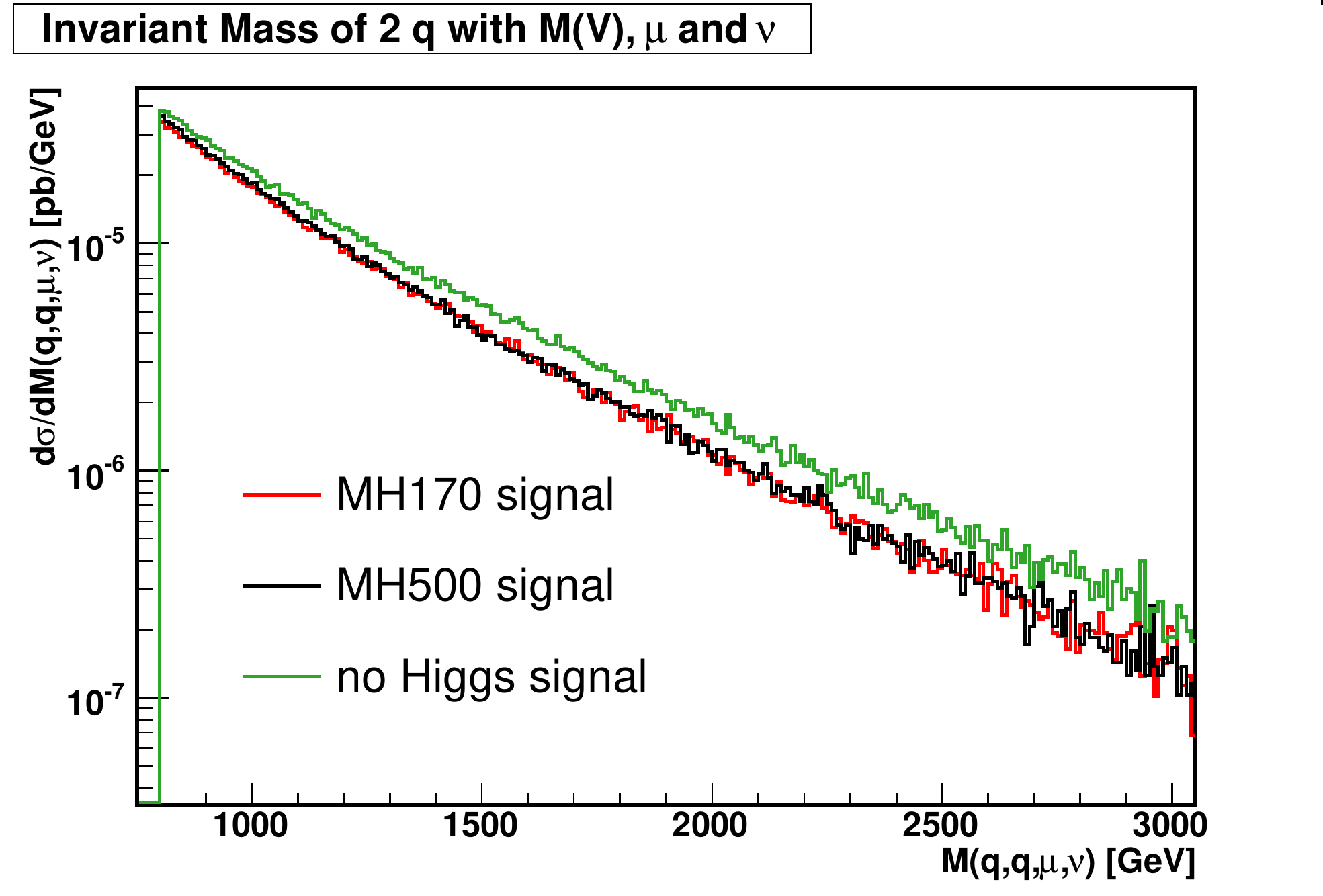,width=10cm}}
}
\caption{The invariant mass of the two vector bosons
In green (full line) 
for the no-Higgs case, in black (dashed) for M(H)=500 GeV and in red
for M(H)=170 GeV.
} 
\label{WWMassHighEnergy}
\end{center}
\end{figure}

Whether or not the LHC will be able to determine the details of  EWSB depends on
the mechanism Nature has chosen to accomplish the task. To put things in
perspective the peak at about 800 GeV in the Pade' unitarization scheme in 
Fig.(\ref{WWMassDistribution}) corresponds to a cross section of about 150 $fb$.
At high luminosity, $L=100 fb^{-1}/year$, and with BR($WW\rightarrow l\nu
jj$)=8/27 this corresponds to about 4400 semileptonic events per LHC year, which
will be very difficult to miss.
If on the other hand we assume the unfavourable scenario of a very heavy Higgs,
the signal to search for is an increase in the cross section for $qqVV$ events
at large $VV$ invariant mass without resonant structures.
The corresponding mass distribution is compared in Fig.(\ref{WWMassHighEnergy})
with the distribution obtained with a light Higgs.
Fig.(\ref{WWMassHighEnergy}) has been produced using a full six fermion event
generator \cite{Accomando:2005hz} which includes all possible polarizations of
the vector bosons, after standard acceptance cuts. With the help of a Neural Net
analysis it has been shown that the event rate in the no Higgs case can be about
twice the rate for a light Higgs, with about 200 events per high luminosity year
in the no Higgs case.

It is clear that the investigation of the mechanism of EWSB will be among the
priorities of the
LHC physics agenda, in parallel with the quest for the Higgs with which it is
intimately related, as we have seen. Soon data will start guiding our
speculations and efforts and some answers will hopefully start to emerge.


\addtocounter{chapter}{1}
%
%
%
%
%
%
%
%
\def\eq#1{{eq.~(\ref{#1})}}
\def\eqs#1#2{{eqs.~(\ref{#1})--(\ref{#2})}}
\def\fig#1{{fig.~(\ref{#1})}}
\def\sec#1{{sec.~(\ref{#1})}}
\def\fig#1{{fig.~(\ref{#1})}}
\def\tab#1{{tab.~(\ref{#1})}}
\def\tabs#1#2{{tabs.~(\ref{#1})--(\ref{#2})}}
\def\vev#1{\left\langle #1\right\rangle}
\def\abs#1{\left| #1\right|}
\def\mod#1{\abs{#1}}
\def\wh{{\cal W}}
\def\bh{{\cal B}}
\def\wtu{\wh^{\mu \nu}}
\def\wtd{\wh_{\mu \nu}}
\def\btu{\bh^{\mu \nu}}
\def\btd{\bh_{\mu \nu}}
\def\Im{\mbox{Im}\,}
\def\Re{\mbox{Re}\,}
\def\Tr{\mbox{Tr}\,}
\def\Str{\mbox{Str}\,}
\def\Det{\mbox{Det}\,}
\def\etal{{\it et al.}}
\def\ie{{\it i.e. }}
\def\di{\mbox{d}}
\def\ltap{\ \raisebox{-.4ex}{\rlap{$\sim$}} \raisebox{.4ex}{$<$}\ }
\def\gtap{\ \raisebox{-.4ex}{\rlap{$\sim$}} \raisebox{.4ex}{$>$}\ }
\def\al{\alpha^{\prime}}
\def\a{& \hspace{-7pt}}
\def\dslash{\raisebox{1pt}{$\slash$} \hspace{-7pt} \partial}
\def\hbar{\hspace{0pt}\raisebox{1pt}{$-$} \hspace{-7pt} h}
\def\Aslash{\hspace{3pt}\raisebox{1pt}{$\slash$} \hspace{-9pt} A}
\def\Dslash{\hspace{3pt}\raisebox{1pt}{$\slash$} \hspace{-9pt} D}
\def\c{\hspace{-5pt}}
\def\Z{{\bf Z}}
\def\5{\overline 5}
\def\S{{\cal{S}}}
\def\oo{$\hat{{\mbox{o}}}$\hspace{-0pt}}
\newenvironment{2figures}[1]{\begin{figure}[#1] 
  \begin{center}
    \begin{tabular}{p{.99\textwidth}p{.01\textwidth}} }
 {  \end{tabular}
  \end{center} 
 \end{figure}
}
\renewcommand{\be}{\begin{equation}}
\renewcommand{\ee}{\end{equation}}
\newcommand{\bea}{\begin{eqnarray}}
\newcommand{\eea}{\end{eqnarray}}
\renewcommand{\nn}{\nonumber}
\newcommand{\spav}[1]{\parbox{1mmhe}{\vspace*{#1}}}
\newcommand{\spao}[1]{\mbox{\hspace{#1}}}
\newcommand\tw[2]{
 \Bigg[\hspace{-1pt}\raisebox{1pt}
 {$\begin{array}{c}
 \displaystyle{#1} \\ \displaystyle{#2}
 \end{array}$}
 \hspace{-1pt}\Bigg]}
\newcommand\pertw[4]{
 \Bigg[\hspace{-1pt}\raisebox{1pt}
 {$\begin{array}{c}
 \displaystyle{#1} \\ \displaystyle{#2}
 \end{array}$}
 \hspace{0pt}\Bigg|\hspace{0pt}\raisebox{1pt}
 {$\begin{array}{c}
 \displaystyle{#3} \\ \displaystyle{#4}
 \end{array}$}
 \hspace{-1pt}\Bigg]}
%
%
\mchapter{Gauge boson scattering at the LHC
without a light Higgs boson}
{ Authors: Marco~Fabbrichesi, Alberto~Tonero and Luca~Vecchi}
\vskip 1cm
%
\section{Motivations} 
\label{sec:mot}
A common prediction of weakly coupled models like the standard model (SM)  and minimal SUSY  as well as strongly coupled composite models of the Higgs boson is that  the breaking of the electro-weak (EW) symmetry is  due to a light---that is, with a mass less than 300 GeV---Higgs boson.

What happens if  the LHC will not discover any light Higgs boson? Most likely, 
 this would mean that the EW symmetry must be broken by  a new and strongly interacting sector.

In this scenario---in which there is no SUSY and no light (fundamental or composite) Higgs boson to be seen---it becomes particularly relevant to analyze the physics of gauge boson scattering---$WW$, $WZ$ and $ZZ$---because it is here that  the strongly interacting sector should manifest itself most directly. For a short review, see these proceedings~\cite{BM}. 

Gauge boson scattering  in this regime looks  similar in many ways to $\pi\pi$ scattering in QCD and similar techniques can be used. 
The natural language  is that of the effective electro-weak lagrangian introduced in \cite{eff-lag}. This lagrangian contains all dimension four operators for the propagation and interaction of the Goldstone bosons of the breaking of the global $SU(2)\times U(1)$ symmetry. If we knew the coefficients of these operators we could predict the physics of gauge boson scattering at the LHC. Unfortunately the  crucial coefficients  do not enter directly in currently measured observables. We do not know their values   and constraints on them can only be inferred by their effect in small loop corrections to the EW observables. Accordingly they are rather weak. In addition,  even though the LHC will explore these terms directly, its sensitivity is not as good as we would like it to be and an important range of  values will remain unexplored.

This lack of predictive power can  be ameliorated if we assume  some model of the strong dynamics responsible of the electro-weak symmetry breaking. In this case, additional relations among the coefficients can be found and used to relate them to known constraints.
Our strategy is therefore to use our prejudices---that is,  model-dependent relationships among the coefficients of the effective lagrangian---plus general constraints coming from causality and analyticity of the amplitudes to see what values the relevant coefficients of the effective electro-weak lagrangian  can assume without violating any of the current bounds. 

We are aware that in many models the relations among the coefficients we utilize  can be made weaker and therefore our bounds will not apply. Nevertheless we find it useful to be as conservative as possible and explore---given what we know from electro-weak precision measurements and taking the models at their face values---what can be said about gauge boson scattering if  electro-weak symmetry is  broken by a strongly interacting sector.
Within this framework, we find that the crucial coefficients  are bound to be  smaller than the expected sensitivity of the LHC and therefore  they will be probably not be detected directly.

This is  not the end of the story though. The cutoff scale of the effective theory is given by the energy at which unitarity is lost. This is around 1.3 TeV in the case of the electro-weak theory as described by the effective lagrangian at the tree level. Unitarity is recovered after introducing additional states which are the Higgs boson in the case of the standard model while they are resonances made of bound states of the strongly interacting sector in our case.
On a more practical level,  there exist unitarization procedures that move the scale at which unitarity is lost  to higher values and we will consider one of them. It is characteristic  of these procedures to automatically include the necessary resonances in the spectrum.  The presence of  resonances is particularly interesting if the coefficients of the effective lagrangian cannot be measured. They may well be the only signatures of the strongly interacting sector accessible at the LHC. We discuss in same detail the most likely masses  and widths of these resonances and their experimental signatures.

\vskip1.5em
\section{Gauge boson scattering} 
\label{sec:con}
Consider the case in which the LHC will not find any  new particle propagating under an energy scale $\Lambda$ around 2 TeV. By new we mean those particles, including  the scalar Higgs boson, not directly observed yet.
In this case---since $\Lambda\gg m_W$---the physics of gauge boson scattering is well described by the standard model (SM) with the addition of the effective  lagrangian containing all the possible electro-weak (EW) operators for the
Goldstone bosons (GB)---$\pi^a$, with $a=1,2,3$---associated to 
the $SU(2)_L\times U(1)_Y\rightarrow U(1)_{em}$ symmetry breaking.
The GB are written as an $SU(2)$ matrix
\be
U=\exp\left(i\pi^a\sigma^a/v\right)\, ,
\ee
where $\sigma^a$ are the Pauli matrices and  $v=246\;\hbox{GeV}$ is the electro-weak vacuum.
The GB couple to the EW gauge and fermion
fields in an $SU(2)_L \times U(1)_Y$ invariant way.
As usual, under a local $SU(2)_L\times U(1)_Y$ transformation $U\rightarrow L U R^\dagger$, with $L$ and $R$ an $SU(2)_L$ and $U(1)_Y$ transformation respectively. The EW precision tests require an approximate $SU(2)_C$ custodial symmetry to be preserved and therefore we assume $R \subset SU(2)_R$.

The most general lagrangian respecting the above symmetries, together with $C$ and $P$ invariance, and up to dimension 4 operators is  given in the references in~\cite{eff-lag} of which we mostly follow the notation: 
\begin{eqnarray} 
 {\cal L} &=& 
 \frac{v^2}{4} \Tr [(D_{\mu}U)^\dagger(D^{\mu}U)] + \frac{1}{4}a_0g^2v^2[Tr(TV_\mu)]^2 
                + \frac{1}{2}a_{1}g g' B_{\mu\nu}Tr(T W^{\mu\nu}) \nn \\
                & + &\frac{1}{2}i a_{2}g' B_{\mu\nu} Tr(T \left[ V^{\mu},V^{\nu}\right]) + ia_{3}gTr(W_{\mu\nu}[V^{\mu},V^{\nu}]) \nn \\ 
               &+& a_{4} [ Tr(V_{\mu}V_{\nu})] ^2 + a_{5} [ Tr(V_{\mu}V^{\mu})] ^2 + a_6Tr(V_\mu V_\nu)Tr(TV^\mu)Tr(TV^\nu) \nn\\ 
               &+& a_7Tr(V_\mu V^\mu)Tr(TV_\nu)Tr(TV^\nu) + \frac{1}{4}a_{8}g^2 [Tr(T W_{\mu\nu})]^2 \nn \\
              &+& \frac{1}{2}ia_9Tr(TW_{\mu\nu})Tr(T[V^\mu,V^\nu]) + \frac{1}{2}a_{10}[Tr(T V_\mu)Tr(TV_\nu)]^2 \nn \\
              &+& a_{11}g\epsilon^{\mu\nu\rho\lambda}Tr(TV_\mu)Tr(V_\nu W_{\rho\lambda})
               \label{lag} \, .   
 \end{eqnarray} 
In (\ref{lag}), $V_\mu = (D_{\mu}U)U^\dagger$, $T = U\sigma^{3}U^\dagger$ and 
\be
D_\mu U = \partial_\mu U + i\frac{\sigma^k}{2}W^k_\mu U - ig'U\frac{\sigma^3}{2}B_\mu\, ,
\ee
with $W_{\mu\nu}=\sigma^kW^k_{\mu\nu}/2 = \partial_\mu W_\nu - \partial_\nu W_\mu + ig[W_\mu,W_\nu]$ is expressed in matrix notation.

This lagrangian, as any other effective theory, contains arbitrary coefficients, in this case called $a_i$, which have to be fixed by experiments or by matching the theory with a UV completion. 
The coefficients $a_2,a_3,a_9,a_{11}$ and $a_4,a_5,a_6,a_7,a_{10}$ contribute at tree level to the gauge boson scattering and represent anomalous triple and quartic gauge couplings respectively. They are not directly bounded by experiments.
On the other hand, the coefficients $a_0$, $a_1$ and $a_8$ in (\ref{lag}) are related to the electro-weak precision measurements parameters $S$, $T$ and $U$~\cite{peskin} and therefore
directly constrained by LEP precision measurements.\footnote{The authors of~\cite{barbieri} defined the complete set of EW  parameters which includes---in addition to  $S$, $T$ and $U$---$W$ and $Y$. These latter come from $O(p^6)$ terms and can be neglected in the present discussion.}

\subsubsection{Precision tests, custodial symmetry and the effective lagrangian}

The EW precision measurements test  processes in which  oblique corrections  play a dominant role with respect to the vertex corrections. This is why we can safely neglect the fermion sector (in our approximate treatment) and why the parameters 
$S$, $T$, $U$, $W$ and $Y$ represent such a  stringent phenomenological set of constraints for any new sector to be a candidate for EW symmetry breaking (EWSB). The good agreement between experiments and a single fundamental Higgs boson is encoded in the very small size of the above EW precision tests parameters. The idea of a fundamental Higgs boson is perhaps the most appealing  because of its extreme economy but it is not the only possibility  and what we  do here is to consider some strongly interacting new physics  whose role is providing masses for the gauge bosons in place of the Higgs boson. 

To express the precision tests constraints in terms of bounds for the coefficients of the 
low-energy lagrangian in \eq{lag} we have to take into account that the parameters $S$, $T$ and $U$ are defined as deviations from the SM predictions evaluated at a reference value for the Higgs and top quark masses. Since we are interested in substituting the SM Higgs sector, we keep separated the contribution to $S$ of the Higgs boson and write
\be
S_{H} + S = S_{EWSB} \, ,
\ee
and analog equations for $T$ and $U$. 
The contributions coming from the SM particles, including the GB, are not relevant because they appear on both sides of the equation. $S_H$ is given  by diagrams containing at least one SM Higgs boson propagator while $S_{EWSB}$ represents the contribution of the new symmetry breaking sector, except for contributions with GB loops only.
We thus find that, in the  chiral lagrangian (\ref{lag}) notation,
\bea
S_{EWSB} &=& -16\pi a_1\nn\\
\alpha_{em}T_{EWSB} &=& 2g^2a_0\nn\\
U_{EWSB} &=& -16\pi a_8 \label{a-STU}
\eea
The coefficients $a_0$, $a_1$ and $a_8$  typically have a scale dependence (and the same is true for $S_{H}$, $T_{H}$ and $U_{H}$) because they renormalize the UV divergences of the GB loops which yields a renormalization scale independent $S$, $T$ and $U$.  
One expects by dimensional analysis that $U\sim (m_Z^2/\Lambda^2)T \ll T$ and therefore $U$ is typically ignored. The relationships (\ref{a-STU}) have been used in \cite{bagger} to study the possible values of the effective lagrangian coefficients in the presence of SM Higgs boson with a mass larger than the EW precision measurements limits.

Using the results of the analysis presented in~\cite{barbieri}, taking as reference values  $m_H = 115$ GeV, $m_t = 178$ GeV and summing the 1-loop Higgs contributions, we obtain:
\bea
S_{EWSB} &=& -0.05 \pm 0.15 \nn \\ 
\alpha_{em} T_{EWSB} &=& (0.3 \pm 0.9)\times10^{-3} \label{S}
\eea
at the scale $\mu=m_Z$. We shall use these results  to set constraints to the coefficients of the effective lagrangian (\ref{lag}).

The smallness of the parameter $T$ can be understood as a consequence of an approximate symmetry of the underlying theory under which the matrix $U$ carries the adjoint representation. In fact, if we require a global $SU(2)_L\times SU(2)_R\rightarrow SU(2)_C$ pattern the $T=U\sigma^3U^\dag$ operator would not be present in the non-gauged chiral lagrangian. The gauge interactions break explicitly this symmetry through $SU(2)_R\supset U(1)_Y$ (and consequently by $SU(2)_C\supset U(1)_{em}$) thus producing a non-vanishing $T$ parameter as a small loop effect proportional to $g'^2$. Moreover, any new EWSB sector must eventually be coupled with some new physics responsible for the fermions masses generation and thus requiring a breaking of the $SU(2)_C$.
Due to this approximate symmetry we expect the couplings $a_{0,2,6,7,8,9,10,11}$ to be subdominant with respect to the custodial preserving ones. 

Most of the strongly coupled theories have large and positive $S_{EWSB}$ and the assumption that this sector respects an exact custodial symmetry is in general in contrast with  smaller values of the $S$ parameter. In fact, a small deviation from the point $T_{EWSB} = 0$ can lead to a negative correction of the same order in the $S$ parameter. Using the effective lagrangian formalism and going to the unitary gauge we find
\bea
S_{EWSB} &=& \frac{4}{\alpha_{em}}\left( s_W^2 \Delta_{Z} - c_W^2 \Delta_{A} \right)\nn \\
U_{EWSB} &=& -\frac{8 s_W^2}{\alpha_{em}}\left( \Delta_{Z} + \Delta_{A} \right)
\eea
where the $\Delta_{A,Z}$ are the shifts in the photon and $Z^0$ kinetic terms due to new physics---once the shifts in the $W$ propagators have been rescaled to write its kinetic term in the canonical way~\cite{holdom}. If a new theory has $\Delta = \Delta^0 + \hat{\Delta}$ with $\Delta^0$ a custodial symmetric term and $\hat{\Delta}$ small custodial-symmetry breaking term satisfying $s_W^2 \hat{\Delta}_{Z} - c_W^2 \hat{\Delta}_{A} = -\varepsilon\alpha_{em}$ then $S_{EWSB} = S^0 - 4\varepsilon$ and $U_{EWSB} =  O (\varepsilon)$.
This result agrees with the experiments: a large and positive $S$ can only be consistent with data if $T$ is greater than zero. 

Bearing the above arguments in mind, we can, in first approximation, consider the custodial symmetry to be exact and therefore discuss only  those terms in the lagrangian (\ref{lag}) that are invariant under this symmetry. Gauge boson scattering is then dominated by only two coefficients: $a_4$ and $a_5$. 

\subsubsection{Scattering amplitude}

Being interested in the EW symmetry breaking sector, we will mostly deal with longitudinally polarized vector bosons scattering because it is in these processes that the new physics plays a dominant role.
We can therefore make use of the equivalence theorem  (ET) wherein
the longitudinal $W$ bosons are replaced by the Goldstone bosons~\cite{ET}. This approximation is valid up to orders $m_W^2/s$, where $s$ is  the center of mass (CM) energy, and therefore---by also including the assumptions underlaying the effective lagrangian approach---we require our scattering amplitudes to exist in a range of energies such as $m_W^2\ll s\ll \Lambda^2$.

Assuming exact $SU(2)_C$, the  elastic scattering of gauge bosons  is described by a single amplitude $A(s,t,u)$. Up to $O(p^4)$, and by means of the lagrangian (\ref{lag}) we obtain~\cite{GL} 
\bea
A(s,t,u) & =& \frac{s}{v^{2}} \label{amp} \\
& +&\frac{4}{v^{4}}\left[ 2a_{5}(\mu )s^{2}+a_{4}(\mu )(t^{2}+u^{2})+\frac{{1}}{(4\pi )^{2}}\frac{{10s^{2}+13(t^{2}+u^{2})}}{72}\right] \nn \\
 & -&\frac{1}{96\pi ^{2}v^{4}} \left[ t(s+2t)\log (\frac{-t}{\mu ^{2}})+u(s+2u)\log (\frac{-u}{\mu ^{2}})+3s^{2}\log (\frac{-s}{\mu ^{2}})\right]\nn
 \eea
where $s,t,u$ are the usual Mandelstam variables satisfying $s+t+u=0$ which in the CM frame and for any $1+2\rightarrow1'+2'$ process  can be expressed as a function of  $s$ and  the scattering angle $\theta$ as $t=-s(1-\cos\theta)/2$ and $u=-s(1+\cos\theta)/2$.

The couplings $a_{4,5}(\mu)$ appearing in~(\ref{amp}) are the effective coefficients renormalized using the minimal subtraction scheme and they differ by an additive finite constant from those introduced in~\cite{GL}. In the latter non-standard renormalization, the numarator of the one loop term in the first bracket of~(\ref{amp}) is shifted from $10 s^2+13 (t^2+u^2)$ to $4 s^2+7 (t^2+u^2)$.

The GB carry an isospin $SU(2)_C$ charge $I=1$ and we can express any process in terms of isospin amplitudes $A_I(s,t,u)$ for $I=0,1,2$:
\begin{eqnarray}
A_{0}(s,t,u) & = & 3A(s,t,u)+A(t,s,u)+A(u,t,s)\nn \\
A_{1}(s,t,u) & = & A(t,s,u)-A(u,t,s)\nn \\
A_{2}(s,t,u) & = & A(t,s,u)+A(u,t,s) \label{isoamp} \, .
\end{eqnarray}

From the above results, we obtain the amplitudes for the scattering of the physical longitudinally polarized gauge bosons as follows:
\bea
A(W^{+}W^{-}\to W^{+}W^{-}) & = & \frac{1}{3}A_{0}+\frac{1}{2}A_{1}+\frac{1}{6}A_{2}\nn \\
A(W^{+}W^{-}\to ZZ) & = & \frac{1}{3}A_{0}-\frac{1}{3}A_{2} \nn
 \\
A(ZZ\to ZZ) & = & \frac{1}{3}A_{0}+\frac{2}{3}A_{2}\nn \\
A(WZ\to WZ) & = & \frac{1}{2}A_{1}+\frac{1}{2}A_{2}\nn \\
A(W^{\pm }W^{\pm }\to W^{\pm }W^{\pm }) & = & A_{2}\, .\label{Wamp}
\eea

It is useful to re-express the scattering amplitudes in terms of partial waves
 of definite angular momentum $J$
and isospin $I$ associated to the custodial $SU(2)_C$ group. 
These partial waves are denoted $t_{IJ}$ and are defined,
in terms of the  amplitude $A_I$ of (\ref{isoamp}), as
\be
  t_{IJ}=\frac{1}{64\,\pi}\int_{-1}^1\,d(\cos\theta)
\,P_J(\cos\theta)\,A_I(s,t,u)\;.
\ee
Explicitly we find:
\begin{eqnarray}
 t^{(2)}_{00}&=&\frac{s}{16\,\pi v^2}, \nn \\
 t^{(4)}_{00}&=&\frac{s^2}{64\,\pi v^4}
\left[\frac{16(11a_5+7a_4)}{3}
+\frac{101/9-50 \log(s/\mu^2)/9+4\, i\,\pi}{16\,\pi^2}
\right],\nonumber\\
 t^{(2)}_{11}&=&\frac{s}{96\,\pi v^2},  \nn \\
 t^{(4)}_{11}&=&\frac{s^2}{96\,\pi v^4}\left[4(a_4-2a_5)
+\frac{1}{16\,\pi^2}\left(\frac{1}{9}+\frac{i\,\pi}{6}\right)
\right],\nonumber\\
 t^{(2)}_{20}&=&\frac{-s}{32\,\pi v^2}, \nn \\
 t^{(4)}_{20}&=&\frac{s^2}{64\,\pi v^4}\left[\frac{32(a_5+2a_4)}{3} 
+\frac{273/54-20\log(s/\mu^2)/9
+i\,\pi}{16\,\pi^2}
\right] \, ,
\label{pertamplis}
\end{eqnarray}
where the superscript refers to the corresponding power of momenta.

The contributions from $J\geq2$ starts at order $p^4$ and turn out to be irrelevant for our purpose.
The $I=1$ channel is related to an odd spin field due to the Pauli exclusion principle. 
The $(I=2, J=0)$ channel has a dominant minus sign which, from a semiclassical perspective, indicates that this channel is repulsive and we do not expect any resonance with these quantum numbers. 

The effective lagrangian (\ref{lag}) and gauge boson scattering were extensively discussed  in~\cite{ww1}.

\subsubsection{Unitarity violation}

The amplitudes (\ref{amp}) (or, equivalently (\ref{pertamplis})) show that, for $s\gg m_W^2$, the elastic scattering of two longitudinal polarized gauge bosons is observed with a probability that increases with the CM energy $s$. We  expect that for sufficiently large energies the quantum mechanical interpretation of the $S$-matrix will be lost. This fact can be restated more formally in terms of the partial waves defined in \eq{pertamplis}. The unitarity condition for physical values of the CM energy $s<\Lambda^2$ can be written as
\be
\hbox{Im}\, t_{IJ}(s) =\mid t_{IJ}(s)\mid^2\, ,
\label{unitarity}
\ee
which, up to order $p^4$ terms, reads $\hbox{Im}\, t_{IJ}^{(4)}(s) = \mid t_{IJ}^{(2)}(s)\mid^2$. This relation leads to an upper bound on the cut-off scale $\Lambda$ above which the theory is no more unitary.  A necessary condition to satisfy is therefore that $Re(t_{IJ})<1/2$, which at tree level yields $\Lambda<$ 1.3 TeV. This constraint holds irrespective of the value of the $a_i$ and is even lower when loops are included.
We explicitly show the unitarity bound thus obtained as a dashed line in the plots presented below in Figures (\ref{fig3}) and (\ref{fig4}) at the end of the paper.

\subsection{Limits and constraints}

If we knew all the coefficients of the lagrangian (\ref{lag}), and $a_4$ and $a_5$ in particular, we could fully predict gauge boson scattering at the LHC.  We therefore turn now to  what is known about them in order to review all current constraints on their possible values and compare them with   the limits on their values which are going to be explored given the expected LHC sensitivity. As we shall see,  these two crucial coefficients are poorly  known quantities which furthermore will not be fully explored at the LHC.

\subsubsection{LHC sensitivity}

First of all, let us consider the capability of the LHC of exploring the coefficients $a_4$ and $a_5$ of the effective lagrangian (\ref{lag}). This has been discussed most recently  in~\cite{eboli} by comparing cross sections with and without the operator controlled by the corresponding coefficient. They consider scattering of $W^+W^-$,  $W^\pm Z$ and $ZZ$ ($W^\pm W^\pm$ gives somewhat weaker bounds) and report limits (at 99\% CL) that we take here to be
\bea
-7.7 \times 10^{-3} \leq a_4 \leq 15 \times 10^{-3} \nn\\
-12 \times 10^{-3} \leq a_5 \leq 10 \times 10^{-3} \, . \label{limLHC}
\eea
The above limits are obtained considering as non-vanishing only one coefficient at the time.  It is also possible to include both coefficients together and obtain a combined (and slightly smaller) limit. We want to be conservative and therefore use (\ref{limLHC}).  Comparable limits were previously found in the papers of ref.~\cite{pre-eboli1}.
 
To put these results in perspective, limits roughly one order of magnitude better can be achieved by a linear collider~\cite{LC}.

\subsubsection{EW precision measurements: indirect bounds}

Bounds on the coefficients $a_4$ and $a_5$ can be obtained by including their effect (at the one-loop level) into low-energy and $Z$ physics precision measurements. They are refereed as indirect bounds since they only come in at the loop level.

As expected, these bounds turn out to be rather weak~\cite{eboli} :
\bea
-320 \times 10^{-3} \leq a_4 \leq 85 \times 10^{-3} \nn \\
-810 \times 10^{-3} \leq a_5 \leq 210 \times 10^{-3} \,  \label{indirect}
\eea 
at 99\% C.L. and for $\Lambda=2$ TeV. Comparable bounds were previously found in the papers in ref.~\cite{pre-eboli2}. As before, slightly stronger bounds can be found by a combined analysis.

Notice that  the $SU(2)_C$ preserving triple gauge coupling $a_3$ has not been considered in the  computations leading to the previous limits. Once its contribution is taken into account, the LHC sensitivity and the indirect bounds presented here are  slightly modified although the ranges shown are not  changed drastically. 

\begin{figure}[h]
\begin{center}
\includegraphics[width=5in]{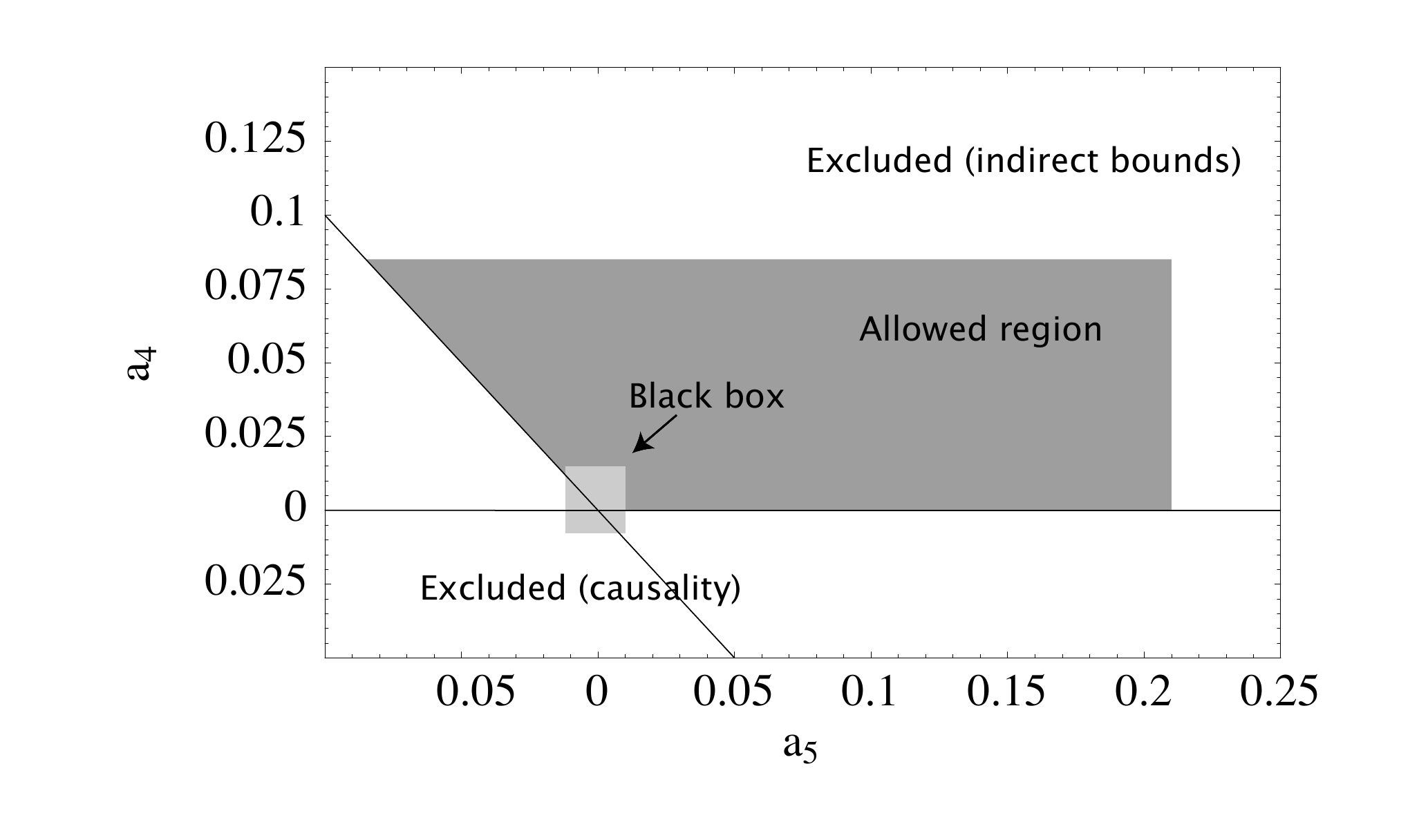}
\caption{\small  The region of allowed values in the $a_4$-$a_5$ plane (in gray) as provided by combining indirect bounds and causality constraints. Also depicted, the region below which LHC will not able to resolve the coefficients (Black box). \label{fig1}}
\end{center}
\end{figure}

\subsubsection{Unitarity, analyticity and causality}

The requirement of  unitary of the theory, as we have seen, forces the cut off of the lagrangian (\ref{lag}) to be $\Lambda\leq1.3$ TeV but does not impose any constraint on the coefficients $a_{i}$. Other fundamental assumptions like causality and analyticity of the $S$-matrix do give rise to interesting constraints.

In particular, the causal and analytic structure of the amplitudes leads to bounds on the possible values the two coefficients $a_4$ and $a_5$ can assume. This is well known in the context of chiral lagrangians for the strong interactions~\cite{pham} and can be extended with some caution to the weak interactions.
It can be shown in fact that the second derivative with respect to the center of mass energy of the forward elastic scattering amplitude of two GB is bounded from below by a positive integral of the total cross section for the transition $2 \pi \rightarrow everything$. The coefficients $a_4$ and $a_5$ enter this amplitude and one can use the mentioned result to bound them. 

The most stringent bounds come from the requirement that the underlying theory respects causality~\cite{vecchi}. 
The causality bound can be understood by noticing that, given a classical solution of the equations of motion, one can study the classical oscillations around this background interpreting the motion of the quanta as a scattering process on a macroscopic object~\cite{Rattazzi}. If the background has a constant gradient, the presence of superluminal propagations sum up and can in principle become manifest in the low-energy regime. 
Following the argument in~\cite{Rattazzi}, we obtain the free equations of motion by considering oscillations around one of the possible backgrounds $\pi_{0}=\sigma C_\mu x^\mu$, where $\sigma$ is a general direction is isospin space. They can be written as
\be 
p^2\left( 1+O(a)\right) + \frac{a}{v^4}\left( C\cdot p\right)^2 = 0\, ,  
\ee
with $a=a_{4}$ or $a=a_4+a_5$. In this derivation we made use of the assumption $C^2\ll \Lambda^4$ which is necessary to ensure a perturbative expansion in the framework of the effective theory. 
The above relations imply a subluminal group velocity only in the case $a\geq0$. 
These classical results can be implemented in a quantum framework provided we take into account that all of the coefficients $a_i$ are formally evaluated at a scale $\mu< \Lambda$ through a matching procedure between the UV theory and the lagrangian (\ref{lag}). 

In conclusion, the causality constraints can  be taken to be
\bea
a_4(\mu) &\geq& 0\nn \\
a_4(\mu) + a_5(\mu) &\geq& 0 \, . 
\label{unitbound}
\eea

Notice that the constraints in \eq{unitbound} remove a quite sizable region (most of the negative values, in fact) of values of the parameters $a_4$ and $a_5$ allowed by the indirect bounds alone.  Fig.~\ref{fig1} summarizes the allowed values in the $a_4$-$a_5$ plane and compare it with LHC sensitivity.

\subsection{EW precision measurements: direct (model dependent) bounds}
\label{sec:db}

Given the results in Fig.~\ref{fig1}, we can ask ourselves how likely are the different values for the two coefficients $a_4$ and $a_5$ among those within the allowed region. Without further assumptions, they are all equally possible and no definite prediction is possible about what we are going to  see at the LHC. 

In order to gain further information, we would like to find  relationships between these two coefficients and between them and those of which the experimental bounds are known. In order to accomplish this, we have to introduce some more specific assumptions about the ultraviolet (UV) physics beyond the cut off of the effective lagrangian.  We do it in the spirit of using as much as we know in order to guess what is most likely to be found. 

As a first step,  simple relations  for $a_4$ and $a_5$ are found by means of assuming that their values are dominated by the integration of particles with masses larger than the cut off. It is what is successfully done in QCD, following vector meson dominance, and estimating the coefficient of the chiral lagrangian by integrating out the $\rho$ meson.

The spin 1, $I=1$ particles can be introduced as gauge vectors of a hidden local symmetry and in this case $a_{4}=-a_{5}>0$. The integration of a scalar $I=0$ particle  gives $a_{5}>0$ and $a_{4}=0$.  Scalar $I=2$ particles give $a_{4}=-3a_{5}>0$. Massive spin 2 particles yield, for the isoscalar  $a_{4}=-3a_{5}>0$, while for the $I=2$ $a_{5}>0$ and $a_{4}=0$. 

This kind of matching is what we would expect from a weakly coupled model or even from a strongly coupled theory in a large-$N$ approximation. This exercise provides us with  some insight into the possible and most likely values for the coefficients. In particular we can see  the characteristic relations between $a_4$ and $a_5$ depending on the different quantum numbers of the resonance being integrated.

A further step consists in assuming a specific UV completion beyond the cut off of the effective lagrangian in \eq{lag}. The two most likely scenarios which can be studied with the effective lagrangian approach are a confining theory (essentially the gauge sector of a strongly interacting model of a rescaled QCD) and the strongly coupled regime of a  model like the SM Higgs sector in which the Higgs boson is heavier than the cut off. For each of these two scenarios it is possible to derive more restrictive relationships among the coefficients of the EW lagrangian and in particular we can relate parameters like $a_0$ and $a_1$ to $a_4$ and $a_5$. These new relationships make possible to use EW precision measurements to constrain the possible values of the coefficients $a_4$ and $a_5$.

\subsubsection{Large-$N$ scenario}

This scenario is based on a new $SU(N)$ gauge theory coupled to new fermions charged under the fundamental representation. By analogy with QCD these particles  are invariant under a flavor chiral symmetry containing the gauged $SU(2)_L\times U(1)_Y$ as a subgroup. 
Let us consider the case in which no other GB except the 3 unphysical ones are present and therefore the chiral group has to be $SU(2)_L \times SU(2)_R$, with $U(1)_Y\subset SU(2)_R$. The new strong dynamics  leads directly to  EWSB through the breaking of the axial current under the confining scale around $4 \pi v$ and to the appearance of an unbroken $SU(2)_{L+R}=SU(2)_C$ custodial symmetry. Following these assumptions, there are no bounds on the new sector from the parameter $T$  and the relevant constraints come from the $S$ parameter only.\footnote{We are not concerned here with the fermion masses and therefore we can bypass most of the problems plaguing technicolor models.}

At energies under the confining scale, the strong dynamics can be described in terms of the hadronic states. Their behavior can be simplified by making use of the large-$N$ approximation. The main result is that the resonances appearing as low-energy degrees of freedom have negligible self-interactions with respect to the couplings to the GB. This limit turns out to be a good approximation of  low-energy QCD even if $N$ is not large.

The large-$N$ approximation allows us to readily estimate the coefficients of the effective lagrangian. At the leading order, we find that $a_4$ and $a_5$ are finite and (by transforming the result of~\cite{N} for QCD)
\be
a_4 = - 2 a_5 = - \frac{1}{2} a_1 \, ,
\ee
which provide us with the link between gauge boson scattering and EW precision measurements---the coefficient $a_1$ being directly related to the parameter $S$ as indicated in \eq{a-STU}.

In a more refined approach, the  non-perturbative effects have been integrated out giving rise to a constituent fermion mass and a gauge condensate. The chiral lagrangian is a consequence of the integration of these massive states. The result becomes~\cite{bijnens}:
\bea
a_{4} &=& \frac{N}{12(4\pi)^2} \nn \\
a_{5} &=& -\left(\frac{1}{2} + \frac{6}{5}\langle G^2\rangle \right)a_{4}\, , \label{QCD1}
\eea
where $\langle G^2\rangle$ is an average over gauge field  fluctuations. The latter is a positive and order 1 free parameter that encodes the dominant soft gauge condensate contribution  which there is no reason to consider as a negligible quantity. Without these corrections the result is equivalent to those obtained considering the effect of a heavier fourth family. Causality requires $\frac{6}{5}\langle G^2\rangle \leq \frac{1}{2}$ and therefore we will consider values of $\langle G^2\rangle$ ranging between $0<\langle G^2\rangle<0.5$.

The coefficients $a_i$ are scale independent at the leading order in the $1/N$ expansion.

The $S$ parameter gives stringent constraints on $N$:
\be
S_{EWSB} = \frac{N}{6\pi}\left(1+\frac{6}{5}\langle G^2\rangle\right)
\ee
which is slightly increased by the strong dynamics with respect to the perturbative estimate, in good agreement with the non-perturbative analysis given in~\cite{peskin}. From the bounds on $S_{EWSB}$, we have $N$ $<$ 4 ($2\sigma$) and N $<$ 7 ($3\sigma$) respectively. 

The relevant  bounds on $a_4$ is then obtained via $a_1$ and yields
\be
0<a_4 < \frac{S_{EWSB}}{32\pi}\, . \label{QCD2}
\ee
We are going to use the bounds given in \eq{QCD1} and \eq{QCD2}. 

Taking  $a_1$ at the central value of $S_{EWSB}$ gives $a_4<0$, which is  outside the causality bounds. This is just a reformulation in the language of effective lagrangians of the known disagreement with EW precision measurements of most models of strongly interacting EW symmetry breaking.

We expect vector and scalar resonances to be the lightest states. The high spin or high $SU(2)_C$ representations considered earlier are typically bound states of more than two fermions and therefore more energetic. Their large masses make their contribution to the $a_i$ coefficients subdominant. 

The relations (\ref{unitbound}) and (\ref{QCD1}) satisfied by the model imply that $-a_4<a_5<-a_4/2$, an indication that  scalar resonances give contributions comparable with the vectorial ones in the large-$N$ limit. If vectors had been the only relevant states, the relation would have been $a_4=-a_5$. 

It is useful to pause and  compare this result with that in low-energy QCD. 

Whereas in the EW case we find that the large-$N$ result indicates the importance of having low-mass scalar states, the chiral lagrangian of low-energy QCD has the corresponding parameters  $L_1$ and $L_2$ saturated by the vector states alone. This vector meson dominance is supported by the experimental data and  in agreement with the large-$N$ analysis, which in  the case of the group $SU(3)$ is different from that of the EW group $SU(2)\times U(1)$.

Even though the scalars have little  impact on the effective lagrangian parameters of low-energy QCD, they turn out to be relevant  to fit the data at energies larger than the $\rho$ mass where the very wide $\sigma$ resonance appearing in the amplitudes is necessary~\cite{sannino}.
One may ask if something similar applies to the EWSB sector, it being described by a similar low-energy action. This can be  seen by looking at the contribution of a single vector to the tree-level fundamental amplitude:
\be
A(s,t,u) = \frac{s}{v^2} - \frac{3M_V^2 s}{\hat{g}^2 v^4} + \frac{M_V^4}{\hat{g}^2 v^4}\left(\frac{u - s}{t - M_V^2} + \frac{t - s}{u - M_V^2} \right) 
\ee
with $\hat{g}$ (not to be interpreted as a gauge coupling) and $M_V^2$ representing the only two parameters entering up to order $p^4$.
The limit $s \ll M_V^2$ corresponds to integrate the vector out and gives the low energy theorem with the previously mentioned $a_{4} = -a_{5} = 1/(4\hat{g}^2)$, while the opposite limit $s\gg M_V^2$ is not well defined. 
The condition $M_V^2=\hat{g}^2v^2/3$ erases the linear term but cannot modify the divergent behavior of the forward and backward scattering channels. In fact we still find the asymptotic form $t_{00}(s) \simeq \hat{g}^2/(36\pi)\log(s/M_V^2)$ which has to be roughly less than one half to preserve unitarity.
This shows why models with only vector resonances cannot move the UV cut off too far from the vector masses, as opposed  to what happens in the case of scalar particles.

The larger dark triangle in Fig.~\ref{fig2} shows  the allowed values for the coefficients $a_4$ and $a_5$ as given by  \eq{QCD1} and \eq{QCD2}. The gray background is drawn according to the causality constrain which is assumed  scale independent  to be consistent with the leading large-$N$ result.

\subsubsection{Heavy-Higgs scenario}

This scenario is a bit more contrived than the previous one and a few preliminary words are in order.

A scalar Higgs-like particle violates unitarity for masses of the order of 1200 GeV~\cite{willenbrock}. Moreover, the mass of the Higgs is proportional to its self coupling and from a naive estimate we expect the perturbation theory to break down at $\lambda \sim 4 \pi$, that is $m_H\sim1300$ GeV. 
What actually happens in the case of a non-perturbative coupling is not known. Problems connected with triviality are not rigorous in non-perturbative theories and therefore the hypothesis of a heavy Higgs cannot be ruled out by this  argument.  

As long as we intend such a heavy Higgs boson only as a modeling of the UV completion of the EW effective lagrangian, we can   study this scenario by assuming a Higgs mass between 2 and 2.5 TeV. Even though we cannot expect the perturbative calculations to be  reliable at these scales, they may still  provide some insight  into   the strongly interacting behavior.

The effective lagrangian parameters in the case of a heavy Higgs can be computed by retaining only the leading logarithmic terms to yield: 
\be
a_4 = - a_1 \quad \mbox{and} \quad a_4 = 2 a_5 \, ,
\ee 
which contains the link between gauge boson scattering and the coefficient $a_1$ we need.
A more complete computation~\cite{herrero} gives
\bea
a_4(m_Z)&=&-\frac{1}{12}\frac{1}{(4\pi)^2}\left( \frac{17}{6}-\log\frac{m_H^2}{m_Z^2}\right)\nn \\ 
a_5(m_Z)&=&\frac{v^2}{8m_H^2}-\frac{1}{24}\frac{1}{(4\pi)^2}\left( \frac{79}{3}-\frac{27\pi}{2\sqrt{3}}-\log\frac{m_H^2}{m_Z^2}\right)
\eea
and
\be
S_{EWSB}=\frac{1}{12\pi}\left( \log\frac{m_H^2}{m_Z^2}-\frac{5}{6}\right) \, .
\ee

The causality constrain (\ref{unitbound}) applied to the above equations implies a bound on the possible values of the cutoff $\Lambda$ compared to $m_H$. An effective lagrangian cutoff consistent with LHC physics yields a Higgs mass at least of the order of 2 TeV. 

Putting these equations together,  we obtain:
\bea
a_4 = \frac{1}{16\pi}\left( S_{EWSB}-\frac{1}{6\pi}\right) \nn \\
 a_4 = 2a_5 - \frac{v^2}{4m_H^2} + \frac{1}{12}\frac{1}{(4\pi)^2}\left(\frac{141}{6}-\frac{27\pi}{2\sqrt{3}}\right) 
\eea
As before in the large-$N$ scenario, the central value of $S_{EWSB}$ yields a value of $a_4$ outside the causality bounds.

At this point we can collect these results with those of the previous section and conclude that in both scenarios under study, the limits on the coefficients $a_4$ and $a_5$ are well below LHC sensitivity (compare Fig.~\ref{fig1} and Fig.~\ref{fig2}). If this is the case,
the  LHC will probably not be able to resolve the value of these coefficients because they are too small to be seen. It goes without saying that this can only be a provisional conclusion in as much as in many models the relations among the coefficients we utilize  can be made weaker by a variety of modifications which make the models more sophisticated.  Accordingly, our bounds will not apply and the LHC may indeed measure $a_4$ or $a_5$ and we will then know that the UV physics is not described by the simple models we have considered.

\subsubsection{A comment about Higgsless models}

Higgsless models~\cite{higgsless} have been proposed to solve the hierarchy problem. They describe a gauge theory in a 5D space-time that produces the usual tower of massive vectors on the 4 dimensional brane (our world). The lightest Kaluza-Klein modes are  interpreted as the $W^{\pm}$ and $Z^0$ while those starting at a mass scale $\Lambda$, represent a new weakly coupled sector. 

The scale of unitarity violation is automatically raised to energies larger than 1.3 TeV because the term in the amplitude  linearly increasing with the CM energy $s$ is not present in these models.
Every 5D model, whatever the curvature, has this property and  fine tuning is neither required nor possible. 
For this reason,  a saturation of the unitarity bound of the term of the amplitude linear in $s$ with just a few vector states, as done in \cite{perelstein}, cannot be considered a characteristic signature of the Higgsless models.

These 5D models fear no better than technicolor when confronted by EW precision measurements. There exists an order 1 mixing among the heavy vectors which contribute a tree level $W_\mu^3-B_\nu$ exchange and consequently a $S_{EWSB}\propto1/(gg')$. 
In 5D notation and for the simplest case of a flat metric, $S_{EWSB} = O(1)/g^2\simeq R/g^2_{(5)}$, in agreement with~\cite{BPR}.
This result can be ameliorated by the introduction of a warped 5D geometry, or boundary terms or even by a de-localization of the matter fields~\cite{cacciapaglia}.
In a certain sense these fine tuning can be seen as a 5D analog of the walking effect on a QCD-like Technicolor.

As it will become clear in the next section, our general analysis of the resonant spectrum relies on the presence of the linear term in $s$ and therefore any 5D Higgsless model is a priori excluded. Nevertheless, since we already know what is the spectrum, we can give some indicative result of what an Higgsless model implies for the coefficients $a_4$ and $a_5$. 

\begin{figure}[h]
\begin{center}
\includegraphics[width=5.5in]{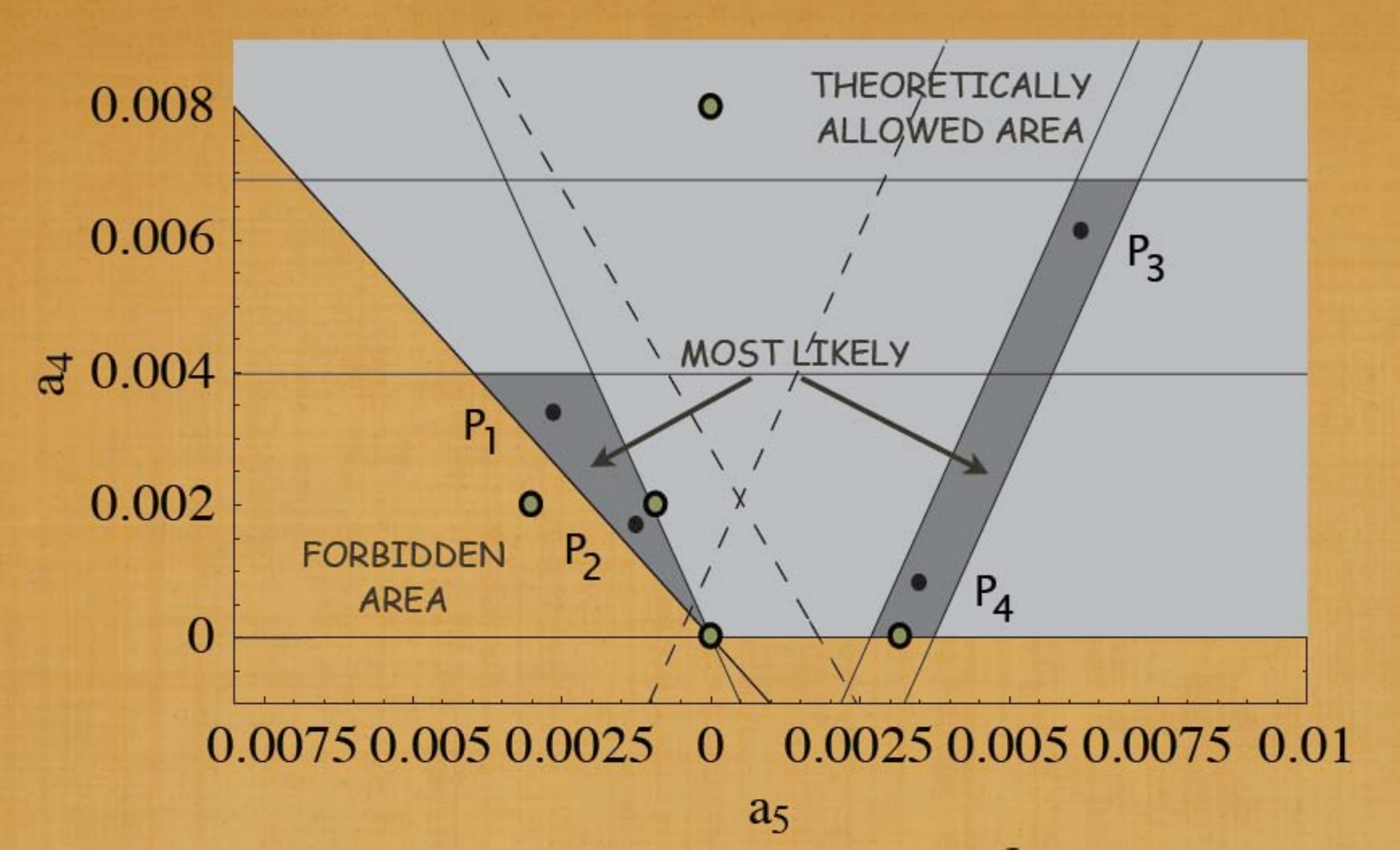}
\caption{\small  Model-dependent bounds for the coefficients. Horizontal lines mark the bounds from EW precision tests for the large-$N$ scenario (lower line) and heavy-Higgs scenario (higher line). Four representative points are indicated: $P_1$ and $P_2$ for the large-$N$ scenario and $P_3$ and $P_4$ for the heavy Higgs. The two oblique dashed lines represent, respectively, the region of vector resonances (left side of dashed line with positive angular coefficient) and of scalar resonances (right side  of dashed line with negative angular coefficient). Also indicated (large dots with dark circles) the points discussed in ref.~\cite{bcf}. Notice that the range of this figure is all within the black box of Fig.~\ref{fig1}. \label{fig2}}
\end{center}
\end{figure}

These models present the relation $a_4=-a_5$ which is characteristic of all models with vector resonances only. 
This line in the $a_4-a_5$ plane of Fig.~\ref{fig2} lies on the causality bound and coincides with the large-$N$ scenario in which the strong dynamical effect $\langle G^2\rangle$ is maximal or, equivalently, in the case in which the scalar resonances are excluded. If we content ourselves with an estimate in the 5D flat space approximation we can write some explicit result~\cite{simmons}.
For example, the asymptotic behavior of $t_{00}$ in the case of a flat 5D geometry is found to be 
\bea
t_{00}\sim\frac{M_1^2}{\pi^3v^2}\log\left(\frac{s}{M_1^2} \right) 
\label{higgslessunit}
\eea
and represents an upper bound on the mass $M_1$ of the lightest massive excitation of the $W^\pm,Z^0$.

The coefficient $a_4$ is related to $a_1$.  We find that
\be
a_4 = -\frac{1}{10} a_1 \, , 
\ee
and therefore, 
\be
a_4=-a_5 = \frac{\pi^2}{120}\frac{v^2}{M_1^2}=\frac{S_{EWSB}}{160 \pi}\, .
\ee
The constraints on $S$ of \eq{S}  lead to $M_1>$ 2.5 TeV which implies a violation of unitarity, and consequently the need of a UV completion for the 5D theory, at the scale $\sim M_1^2$. 

The parameters $a_4$ and $a_5$ are---as in the other scenarios considered---too small to be directly detected at the LHC. The large mass $M_1$ of the first vector state makes it hard for the LHC to find it. 

In case of a warped fifth dimension these relations are slightly changed but the tension existing between the unitarity bound~(\ref{higgslessunit}) (which requires a small $M_1^2$ to raise the cut off above 1.3 TeV) and the $S$ parameter (which requires a large $M_1^2$) remains a characteristic feature of these models.

\vskip1.5em
\section{Experimental signatures: resonances} 
\label{sec:res}

Even though the values of the coefficients may be too small for the LHC, the unitarity of the amplitudes is going to be violated at a scale around 1.3 TeV unless higher order contributions are included. Following the well-established tradition of unitarization in the strong interactions, we  consider the Pad\'e approximation, also known as the inverse amplitude method (IAM)~\cite{pade}. Other unitarization procedure have been used in the literature but we find them less compelling than IAM because they introduce further (unknown) parameters.

This procedure is carried out in the language of the partial waves introduced in (\ref{pertamplis}). In fact, using analytical arguments we  find that
\begin{equation}
t_{IJ}(s)=\frac{t^{(2)}_{IJ}}{1-t_{IJ}^{(4)}/t^{(2)}_{IJ}}+ O(s^3) \, .
\label{IAM}
\end{equation}
Equation (\ref{IAM}) is the IAM, which
 has given remarkable results describing meson interactions, 
having a symmetry breaking pattern almost identical to our present case. 
Note that this amplitude respects strict elastic unitarity, while keeping the
correct  low energy expansion. Furthermore, the extension of
(\ref{IAM}) to the complex plane can be justified using
dispersion theory. 
In particular, it
has the proper analytical structure and, eventually,
poles in the second Riemann sheet 
for certain $a_4$ and $a_5$ values, that can be interpreted as resonances.
Thus, IAM formalism can 
describe resonances without increasing 
the number of parameters and respecting chiral symmetry
and unitarity. 

By inspection of \eq{IAM}, the IAM yields the following masses and widths of the first resonances:
\be
\label{scalar0}
m_{S}^{2}=\frac{{4v^{2}}}{\frac{{16}}{3}\left[11a_{5}(\mu )+7a_{4}(\mu )\right]+\frac{{1}}{16\pi ^{2}}\left[ \frac{{101-50\log (m_{S}^{2}/\mu ^{2})}}{9}\right] } \, , \quad
\Gamma _{S}=\frac{{m_{S}^{3}}}{16\pi v^{2}}\, ,
\ee
for scalar resonances, and
\be
\label{vector1}
m_{V}^{2}=\frac{{v^{2}}}{4\left[a_{4}(\mu )-2a_{5}(\mu )\right]+\frac{{1}}{16\pi ^{2}}\frac{{1}}{9}} \, ,
\quad
\Gamma _{V}=\frac{{m_{V}^{3}}}{96\pi v^{2}} \, ,
\ee
for vector resonances.

A few words of caution about the IAM approach are in order.

The resonances thus obtained represent the lightest massive states we encounter (above the $Z$ pole) in each channel which are necessary in order for the amplitude to respect unitarity. These resonances are not  the only massive states produced by  the  non-perturbative sector but those with  higher masses give a contribution that is subdominant with respect to the IAM prediction and can safely be ignored.

Since we neglect $O(s^3)$ terms, the regime $s\sim m_{res}^2$ is not completely trustable. The larger the resonance peak, the larger the error and therefore we expect the IAM prediction to give  good results only in the case of very sharp resonances. This is the reason behind the  success of the IAM  for the  vector resonances in QCD  as opposed to  the more problematic very broad scalar $\sigma$. 

Similarly, if we integrate a Higgs boson at the tree level and substitute the $a_{4}$ and $a_5$ parameters we find in the IAM formula, we obtain a value for the scalar resonance mass given by \eq{scalar0} which is smaller, that is $m_S = 3m_H/4$. 

Nevertheless, we consider the IAM result a remarkable prediction, given the very small amount of information needed. 

One way to check the  reliability of this method consists in separating the $a_{4,5}$ plane  into three areas depending on the predicted lowest laying resonances being a vector, a scalar or even both of them. This partition follows the coefficients patterns one expects by studying the tree level values for $a_4$ and $a_5$ as given in section \ref{sec:db}. It is represented in Fig.~\ref{fig2} by the two oblique and dashed lines which mark the limit where $\Gamma/M$ is less or more than 1/4 for the case of scalar (oblique line with negative angular coefficient) and vector (oblique line with positive angular coefficient) resonances.

Another check on the consistency of the method is obtained by taking the unrealistic example in which $a_{4}=a_{5}=0$. In this case one finds a pole at an energy $s >(4\pi v)^2$---at which we already know unitarity is violated---thus indicating the unreliability of the input. More generally, a naive estimate---based on integrating out massive states like in the vector meson dominance of QCD---shows that for resonance masses $M$ between the range of hundreds GeV and a few TeV we should expect $a\simeq v^2/M^2$ from $10^{-2}$ to $10^{-3}$ which agrees with the IAM formula. 

Gauge boson scattering and the presence of resonances have previously been discussed in a number of papers~\cite{ww2,dobado}.

\subsection{Parton-level cross sections}

\begin{figure}[hbtp]
\includegraphics[width=4.5in]{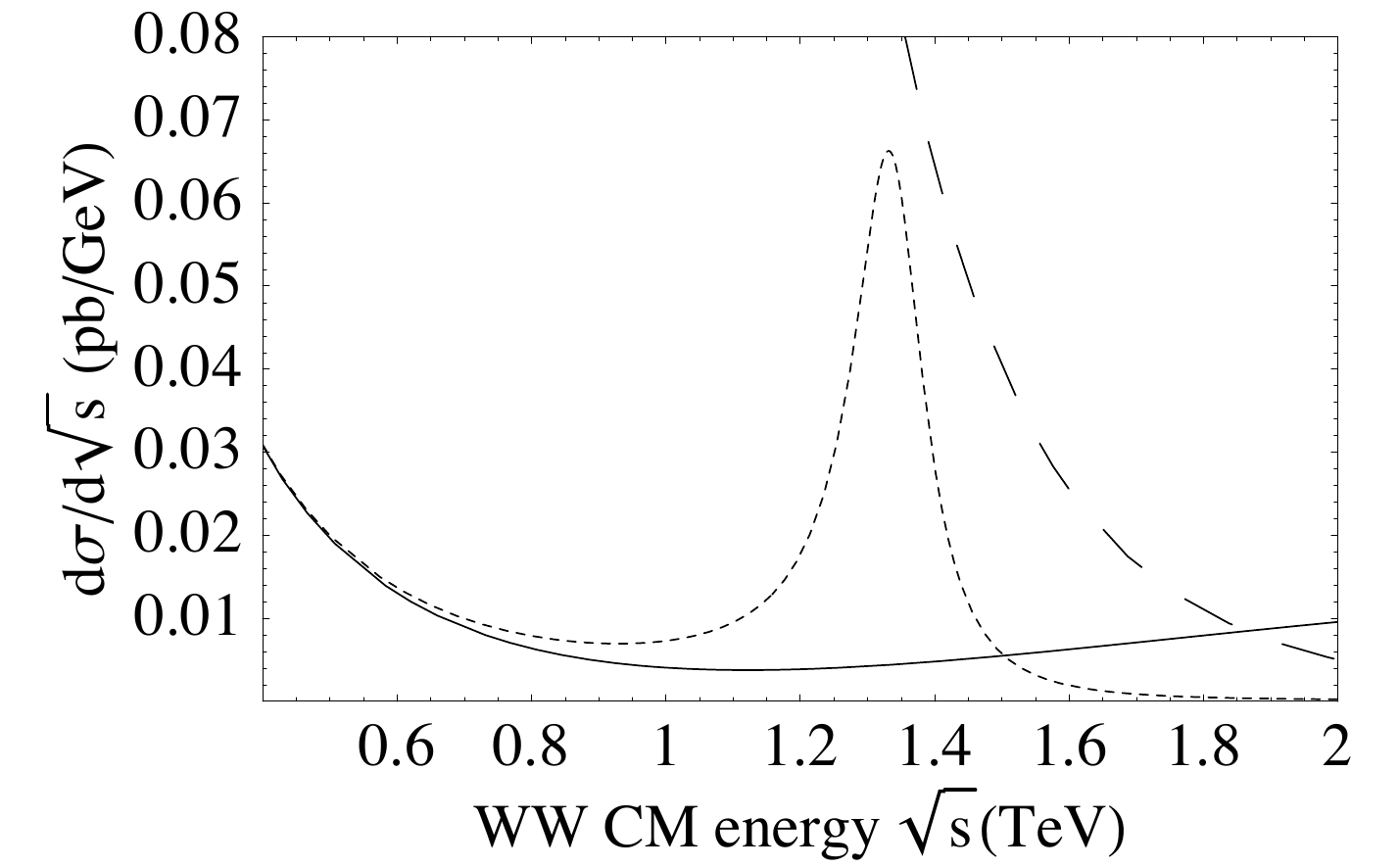} \\
 \includegraphics[width=4.5in]{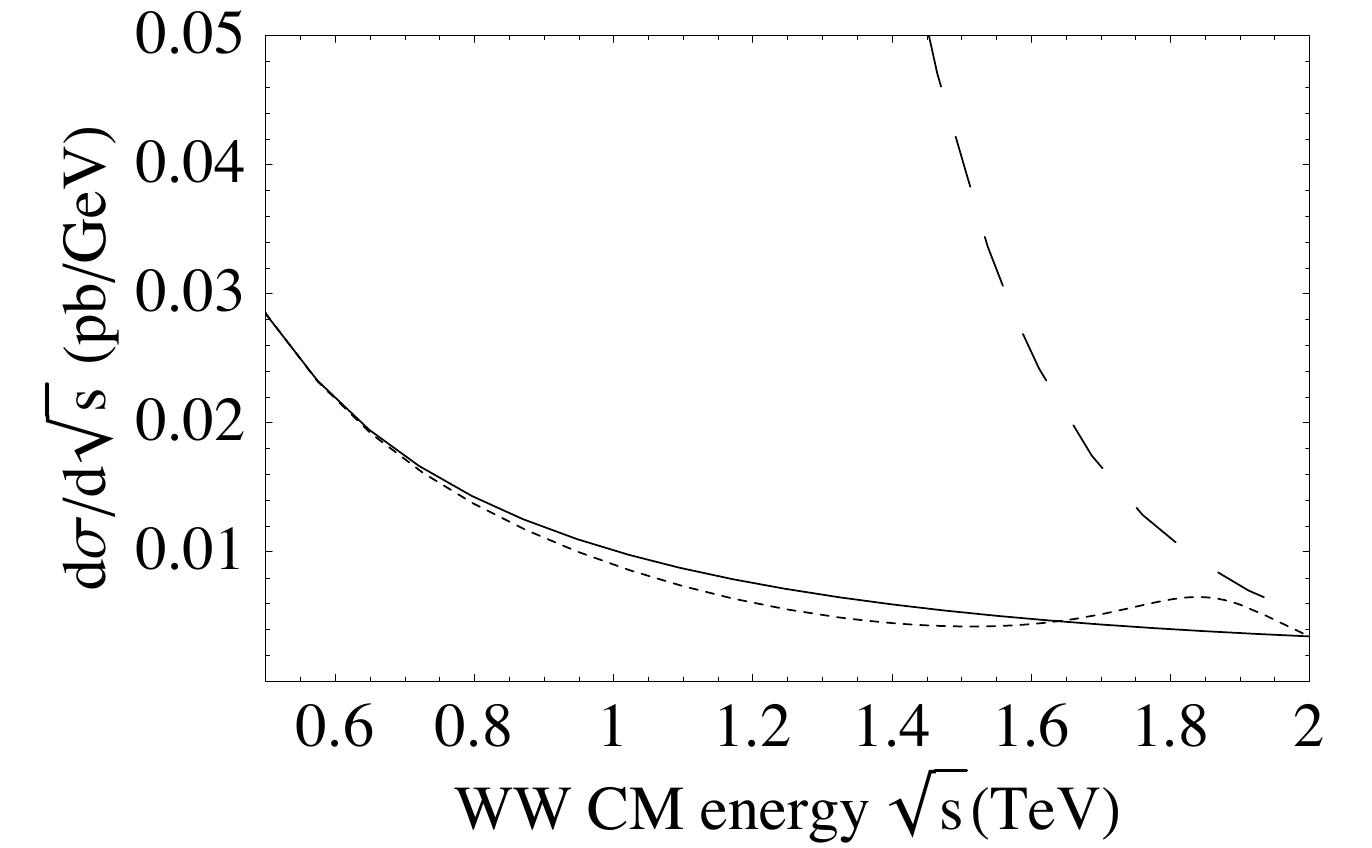} 
  \caption{\small  Parton-level cross sections for $WW$ scattering. In both figures, the continuous line is the result of the effective lagrangian. The long-dashed line is the  limit after which unitarity is lost. The dashed  line with a peak is the amplitude in presence of a vector resonance in the large-$N$ scenario.
  The two figures correspond to the  two representative points $P_1$ and $P_2$ discussed in the text. \label{fig3} }
 \end{figure}
\begin{figure}[hbdp]
\includegraphics[width=4.5in]{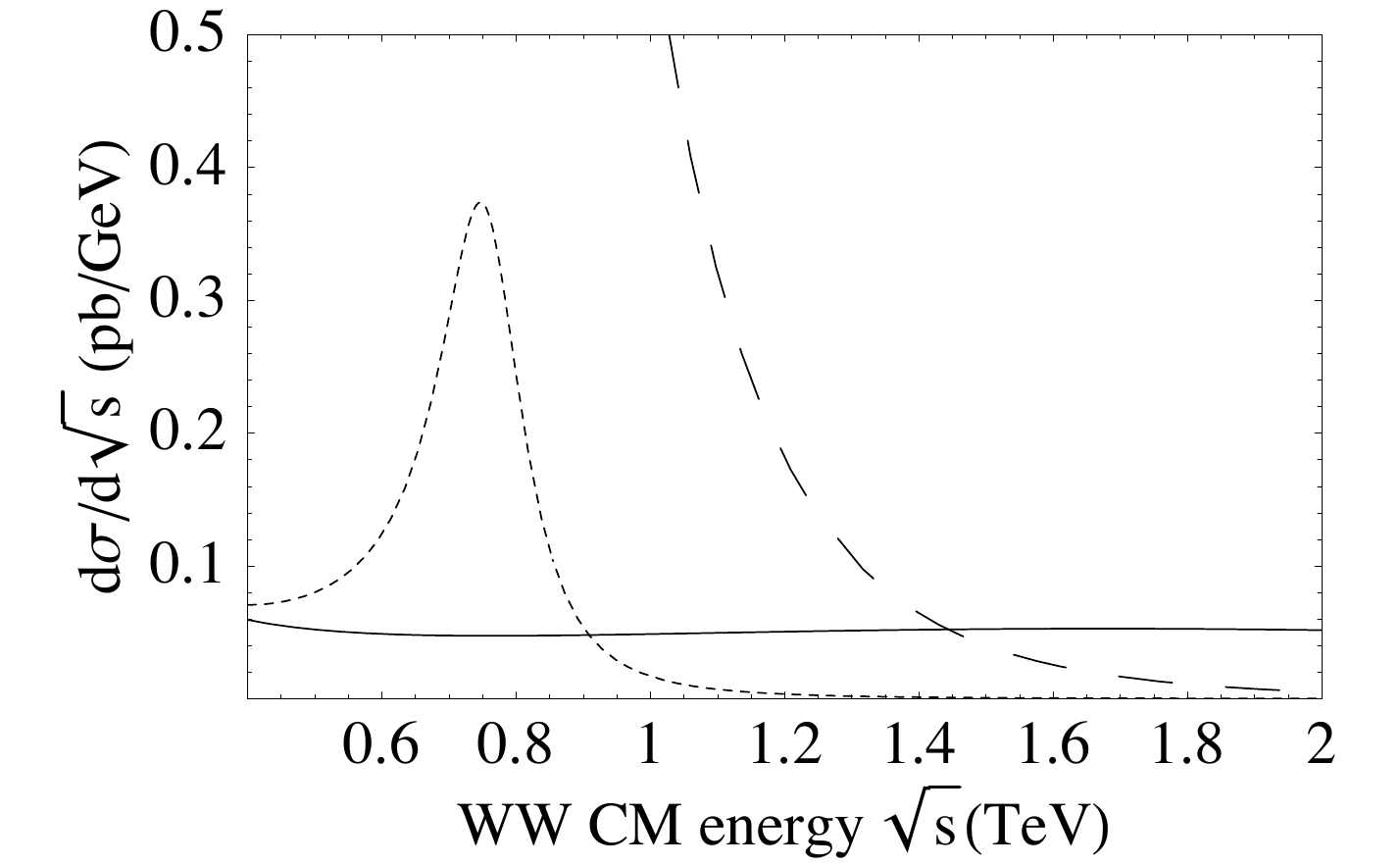}\\
\includegraphics[width=4.5in]{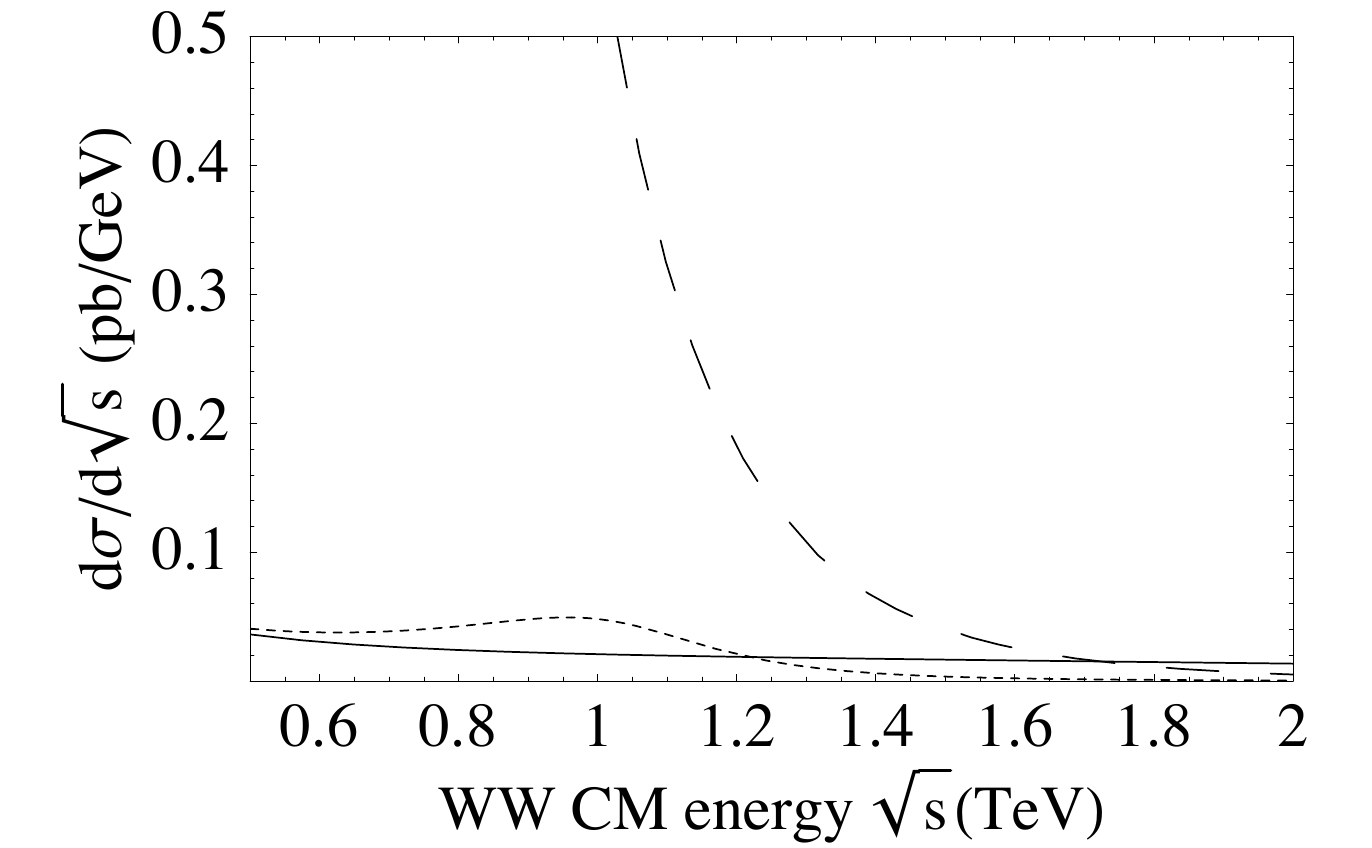}
\caption{\small  Parton-level cross sections for $WW$ scattering. The continuous line is the result of the effective lagrangian. The long-dashed line is the  limit after which unitarity is lost. The dashed line with a peak is the amplitude in presence of a scalar resonance in the heavy-Higgs scenario. The two figures correspond to the  two representative points $P_3$ and $P_4$ discussed in the text. Notice that the second plot has rescaled vertical axis because of the smallness of the resonant peak.\label{fig4}}
\end{figure}

Our plan is to choose two representative points for each of the considered scenarios  in the allowed $a_4-a_5$ region  and then find the first resonances appearing in the $W_LW_L$ elastic scattering using the IAM approximations. The  points are shown in Fig.~\ref{fig2}. We take
\be
P_1:\; \left\{ \begin{array}{ccc}  a_4 &=&  3.5  \times 10^{-3} \\
 a_5 &=& -2.5  \times 10^{-3} \end{array}
 \right.
 \quad \mbox{and} \quad 
P_2:\; \left\{ \begin{array}{ccc}  a_4 &=&  1.7  \times 10^{-3} \\
 a_5 &=& -1.3  \times 10^{-3} \end{array}
 \right.
\ee
for the large-$N$ scenario and
\be
P_3:\;\left\{ \begin{array}{ccc}  a_4 &=& 5.7  \times 10^{-3} \\
 a_5 &=& 6.0  \times 10^{-3} \end{array}
 \right.
 \quad \mbox{and} \quad 
P_4:\; \left\{ \begin{array}{ccc}  a_4 &=&  3.5  \times 10^{-3} \\
 a_5 &=& 0.7  \times 10^{-3} \end{array}
 \right.
\ee
for the heavy-Higgs scenario. 

The first pair corresponds to having  vector resonances at
\be
\left\{ \begin{array}{ccc}  m_V &=&  1340\; \mbox{GeV} \\
 \Gamma_V &=&  128\; \mbox{GeV}\end{array}
 \right.
 \quad \mbox{and} \quad 
 \left\{ \begin{array}{ccc}  m_V &=&  1870\; \mbox{GeV} \\
 \Gamma_V &=& 346\; \mbox{GeV} \end{array}
 \right.
\ee
together with  heavier (2 TeV) and very broad scalar states,
while the second pair to  scalar resonances at
\be
\left\{ \begin{array}{ccc}  m_S &=&  712\; \mbox{GeV} \\
 \Gamma_S &=&  78\; \mbox{GeV}\end{array}
 \right.
 \quad \mbox{and} \quad 
 \left\{ \begin{array}{ccc}  m_S &=&  1250\; \mbox{GeV} \\
 \Gamma_S &=& 237\; \mbox{GeV} \end{array}
 \right.
\ee
These points are representative of the possible values and span the allowed region.
The resonances become heavier, and therefore less visible at the LHC, for smaller values of the coefficients. Accordingly, whereas points $P_1$ and $P_3$ give what we may call an ideal scenario,  the other two show a situation that will be difficult to discriminate at the LHC.

We can now consider the physical process $pp\rightarrow W_LW_Ljj+X$ and plot its differential cross section in the $WW$ CM energy $\sqrt{s}$ for the values of the coefficients $a_4$ and $a_5$ we have identified. 
To simplify,  we will use the effective $W$ approximation~\cite{EWA}.

Once the amplitude $A(s,t,u)$ is given, the differential cross-section for the factorized $WW$ process is
\begin{equation}
\label{WWxsecn}
\frac{d\sigma_{WW} }{d\cos \theta }=\frac{|A(s,t,u)|^{2}}{32\pi \, s}.
\end{equation}
while the differential cross section for the considered physical transition $pp\rightarrow W_LW_Ljj+X$  reads:
\begin{equation}
\label{ppxsecn}
\frac{d\sigma }{ds}=\sum _{i,j}\int ^{1}_{s/s_{pp}}\int _{s/(x_{1}s_{pp})}^{1}\frac{dx_{1}\, dx_{2}}{x_{1}x_{2}s_{pp}}f_{i}(x_{1},s)\, f_{j}(x_{2},s)\frac{dL_{WW}}{d\tau }\int ^{1}_{-1}\frac{d\sigma_{WW}}{d\cos \theta }d\cos \theta 
\end{equation}
where \( \surd s_{pp} \) is the CM energy which we take to be 14
TeV, as appropriate for the LHC, and
\be
\frac{dL_{WW}}{d\tau }\approx \left( \frac{\alpha }{4\pi \sin ^{2}\theta _{W}}\right) ^{2}\frac{1}{\tau }\left[ (1+\tau )\ln (1/\tau )-2(1-\tau )\right] 
\ee
where \( \tau =s/(x_{1}x_{2}s_{pp}) \). For the structure functions $f_j$ we use those of ref.~\cite{lai}.

The high-energy regime will be very much suppressed by the partition functions so that the resonances found by (\ref{scalar0}) and (\ref{vector1}) turn out to be the only phenomenologically interesting ones. Because of this, we can safely make use of the approximation (\ref{IAM})  in the whole range from 400 GeV to 2 TeV and thus we  take $A(s,t,u)$ to be given by the IAM unitarization of (\ref{Wamp}). 

Figures~\ref{fig3} and \ref{fig4} give the cross section for the large-$N$ and heavy-Higgs scenario, respectively. The scalar resonance corresponding to $P_3$  is particularly high and narrow and a very good candidate for detection.
For a LHC luminosity of 100 fb$^{-1}$, it would yield  $10^4$ events after one year.  If it exists, it will appear as what we would have called the Higgs boson even though it is not a fundamental state and its mass is much heavier than that expected for the SM Higgs boson.

\section{Experimental analysis}

The actual signal at the LHC requires that the parton-level cross sections derived here  be included in a Montecarlo simulation (of the bremsstrahlung of the initial partons, QCD showers as well as of the final hadronization) and compared with the expected background and the physics of the detector. In the papers of ref.~\cite{dobado,bcf}  it has been argued that resonances in the range here considered can be effectively identified at the LHC. Similar signals have also been analyzed in~\cite{BM2}. 

Our plan is to do a preliminary
study choosing the scalar resonance corresponding to the rappresentative point $P_3$ because it is narrow and relatively light and therefore
good candidate for the detection.

We use PYTHIA~\cite{pythia} as Monte Carlo event generator to simulate a proton-proton collison events taking into account
intial state and final state radiation, QCD showers, final hadronization and decaying.
The fast detector simulator used in our study is PGS~\cite{pgs}.
The analysis of the PGS output has been done  using CHAMELEON~\cite{chameleon}, a MATHEMATICA package.

The PYTHIA Monte Carlo generator has been modified to include the EW effective approach using the IAM protocol~\cite{bcf}. Signal samples
containing the $W^\pm W^\pm$ final state (including all charge combinations) have been generated using PYTHIA 6.4 with
 the IAM unitarization scheme.

\begin{figure}[hbtp]
\begin{center}
\includegraphics[width=4in]{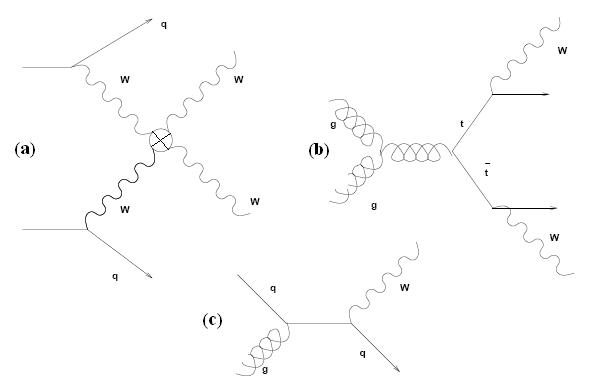}
\caption{\small Typical leading order Feynman diagrams for the signal (a) and backgrounds: $t\,\bar{t}$ (b)
e $W+jets$ (c).} \label{diagr}
\end{center}
\end{figure}

The relevant backgrounds are QCD $t\bar{t}$ production and QCD radiative $W+jets$ production, as illustrated in
Fig.~\ref{diagr}. These backgrounds have been generated using the standard version of PYTHIA 6.4.
The generated statistics for each process are described in Table~\ref{stat}.

\begin{center}
\begin{table}[h]
\begin{tabular}{|l|c|l|c|}
\hline
Process & N events generated & $\sigma (fb)$ & $L_{\mathrm{equiv}}(fb^{-1})$\\
\hline
$W_LW_L\to W_L W_L$ & $10^5$ & $\sim 10^2$ \small{(PYTHIA)} & $ 10^3$ \\ 
$t\bar{t}$ & $10^5$ & $\sim 10^6$ \small{(MCatNLO)}& $10^{-1}$\\
$W+jets$ & $10^5$ & $\sim 10^8$ \small{(PYTHIA)}& $10^{-3}$\\
\hline
\end{tabular}
\caption{\small Number of events generated for the signal and the backgrounds with the cross section (order of
magnitude) and integrated equivalent luminosity ($N =\sigma\cdot L_{\mathrm{equiv}}$).}\label{stat}
\end{table}
\end{center}

\subsection{Extracting the signal}

We focus on the selection of the semileptonic decay mode for the $WW$ system because this channel is cleaner with
respect to the statistics. To identify semileptonic decays and isolate the signal we select first  the leptonically
decaying $W$ (charge lepton and missing transverse energy),
then  the hadronically decaying $W$ (jet invariant mass) and finally we select  the event
enviroment (tagging jets, top veto).
We only keep events with 1 charged lepton with $p_t>40$ GeV and missing transverse energy (MET) $>40$ GeV in order to
eliminate leptons from non leptonically decayin $W$. The charged lepton+MET system is the leptonic $W$ candidate.
We next cut on the $p_T$ of the leptonic $W$ candidate selecting events in which this $W$ candidate has $p_T>250$ GeV.

To identify the hadronic $W$ candidate we select events in which the invariant mass of the system (hardest jet+second or
third hardest jet) reconstructs the $W$ mass. The range of this mass reconstruction is from $70$ GeV to $90$ GeV.

Finally, to further reduce the backgrounds, cuts related to the event enviroment must be applied:
\begin{itemize}
\item in the $WW$ scattering process the gauge bosons are radiated from quarks in the initial state
(see Fig.~\ref{diagr}). The quark from which the boson is radiated will give a jet at high pseudorapidity
(i.e. close to the direction of the hadron from which it emerged).
A forward (backward) tag jet is defined as the highest transverse energy jet in the forward (backward) region.
For an event to be included  it must have a tag jet with the forward and backward regions satisfying $p_T>20$ GeV
and $2<|\eta|<4$.
\item in the remaining $t\bar{t}$ events containing a genuine leptonic $W$, this $W$ will combine with a jet other
than the hadronic $W$ candidate to give a mass close to the top mass. Any event with a mass in the region $130$ GeV
$< M_{wj} < 240$ GeV is rejected.
\end{itemize}
The cut flow and the effect of each cut on signal and background are shown in Table~\ref{tabcut}.

\begin{table}[hbtp]
\begin{tabular}{|c|c|c|c|c|c|c|}
\hline
Cut & Signal & Efficiency & Bckg 1 & Bckg 2 & Signal/ & Signal/\\
&&Signal&($t\,\bar{t}$)& ($W+jets$)&$t\,\bar{t}$&$W+jets$\\
\hline
&&&&&&\\
Events& $10^5$ & 100\% & $10^5$ & $10^5$ & $10^{-4}$ & $10^{-6}$\\
generated&&&&&&\\
&&&&&&\\
\hline
&&&&&&\\
1 Lepton &&&&&&\\
e & 33400 & 33.4\% & 18723 & 10871 & $1.7\cdot 10^{-4}$&$3.1\cdot 10^{-6}$\\
MET &&&&&&\\
&&&&&&\\
\hline
&&&&&&\\
$p_T$(Lept.)&&&&&&\\
$>40$ GeV & 31342 & 31.3\% & 13521 & 2337& $2.3\cdot 10^{-4}$ & $1.3\cdot 10^{-5}$\\
&&&&&&\\
&&&&&&\\
\hline
&&&&&&\\
MET$>40$ GeV & 25189 & 25.1 \% & 6929 & 794 & $3.6\cdot 10^{-4}$ & $3.2\cdot 10^{-5}$\\
&&&&&&\\
\hline
&&&&&&\\
$p_T$($W_{\mathrm{Lept.}}$)& 13475 & 13.5\% & 588 & 9 & $2.3\cdot 10^{-3}$ & $1.5\cdot 10^{-3}$\\
$>250$ GeV &&&&&&\\
&&&&&&\\
\hline
&&&&&&\\
$70$ GeV $<$  & 5510 & 5.5\% & 96 &  0 & $5.7\cdot 10^{-3}$ & -\\
M($W_{\mathrm{Hadr.}}$)&&&&&&\\
$<$ $90$ GeV &&&&&&\\
&&&&&&\\
\hline
&&&&&&\\
``Tag jet'' & 1862 & 1.8 \% & 18 & 0 & $1.0\cdot 10^{-2}$& -\\
&&&&&&\\
&&&&&&\\
\hline
&&&&&&\\
``Top veto'' & 1338 & 1.3 \% & 0 & 0 & - &-\\
&&&&&&\\
\hline
\end{tabular}
\caption{\small Cut flow table. In the first column, the various  cuts are  described. The next four columns show the efficiency and the number
of remaining signal and background events after each cut. The last two columns show
the  signal over background ratio.}\label{tabcut}
\end{table}
\subsection{Results}

\begin{figure}[t]
\begin{center}
\includegraphics[width=0.9\textwidth]{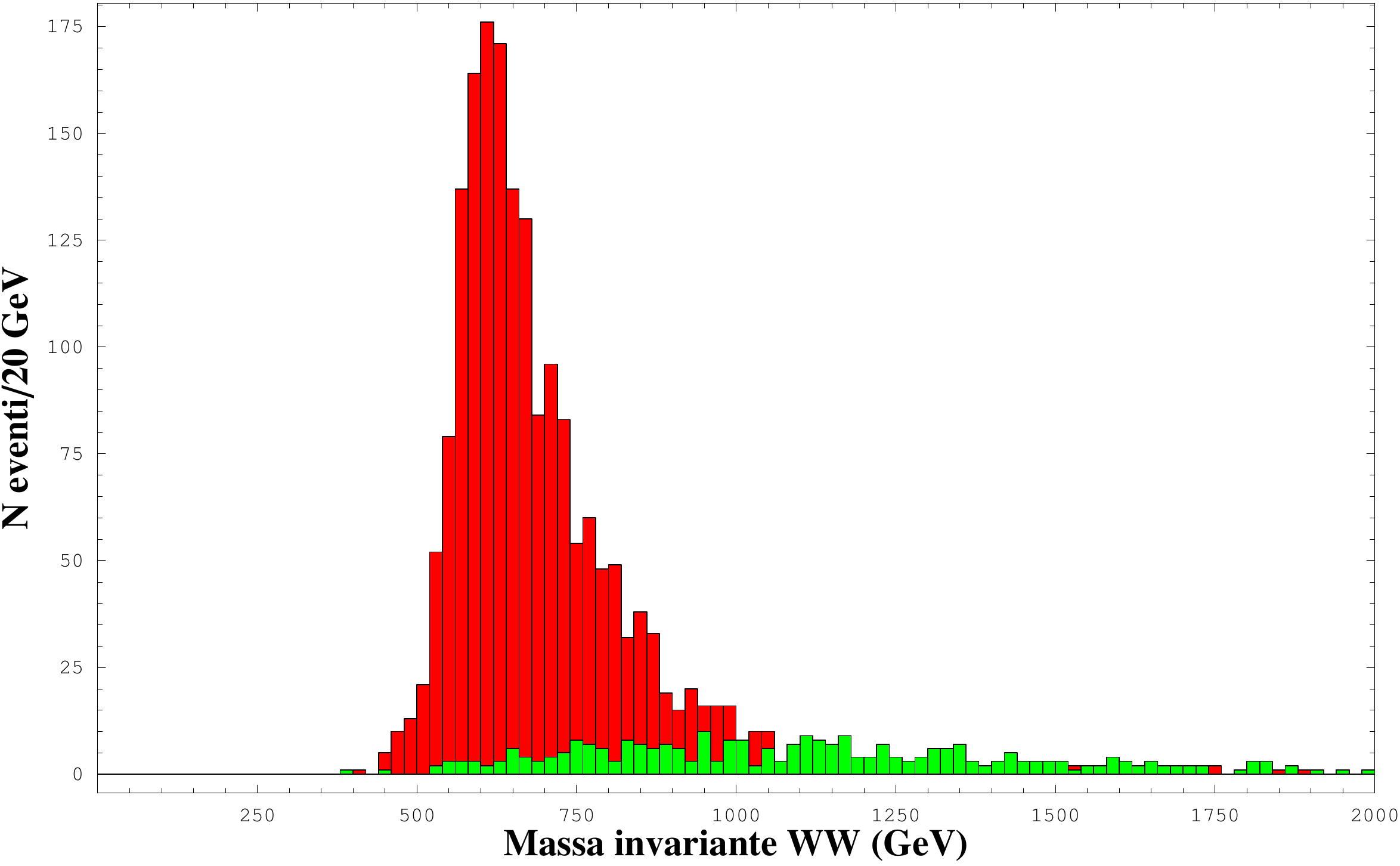}
\caption{\small Invariant $WW$ mass reconstruction for the signal after the cut flow described in Table~\ref{tabcut}. It can be compared with the parton cross section in Fig.~\ref{fig4}. In green, the continuum corresponding to the choice $a_4=a_5=0$.}
\label{resonS1}
\end{center}
\end{figure}

Even though our study is only preliminary, the results obtained from the analysis can be considered encouraging.
The selection of events imposing the cuts described in Table~\ref{tabcut} allows us to eliminate
completely the background with 1.3 \% of efficiency on the signal.
In Fig.~\ref{resonS1} is shown the reconstuction of the resonance corresponding to the rappresentative point $P_3$ and, for  comparison, the continuum corrisponding to the choice $a_4 = a_5=0$.


\addtocounter{chapter}{1}



\renewenvironment{2figures}[1]{\begin{figure}[#1]
  \begin{center}
    \begin{tabular}{p{.47\textwidth}p{.47\textwidth}} }
 {  \end{tabular}
  \end{center}
 \end{figure}
}

\def\taujet{\tau\rm{-jet}}%
\def\taujets{\tau\rm{-jets}}%

\renewcommand{\met}{\ensuremath{E^{miss}_T}}
\newcommand{\vmet}{\ensuremath{\vec{E}^{miss}_T}}
\newcommand{\Ztau}{\ensuremath{Z \rightarrow \tau \tau}}
\newcommand{\Atau}{\ensuremath{A \rightarrow \tau \tau}}
\newcommand{\Wen}{\ensuremath{W \rightarrow l \nu}}
\renewcommand{\ttbar}{\ensuremath{t \bar{t}}}
\newcommand{\sumet}{\ensuremath{\sum E_T}}

\newcommand{\fixme}{{\bf FIXME~}}


\mchapter{The experimental world}
{ Authors: Giuseppe Bagliesi, Leonardo Carminati, Andrea Giammanco,
Chiara Mariotti, Ernesto Migliore, Aleandro Nisati, Andrea Perrotta,
Andrea Rizzi, Stefano Rosati,
Francesco Tartarelli, Iacopo Vivarelli}\label{detectors}
\vskip 0.3cm\noindent
{\it Revisors: Paolo Nason}
\vskip 1cm
This chapter is meant to provide an introduction of the
actual implementation in ATLAS and in CMS of the 
experimental techniques used for the detection of the physics objects
introduced in Chapter~\ref{chap:Intro}. A detailed description of the individual
subdetectors can be found in the Technical Design Reports from the two
collaborations~\cite{ATLAStdr,CMStdr}.

\section{Muons}
{\large {\sl S. Rosati}}
\vspace{0.5cm}

Final state with muons will be amongst the most promising and robust physics signatures
at the LHC. Because of their crucial role in
the trigger of the experiment, the description of the muon system
of ATLAS and CMS should include both the online and
the offline identification and reconstruction of the muons.\\
Two different approaches have been chosen for the muon systems of ATLAS and CMS:
\begin{itemize}
\item[-] in ATLAS the system is an air-core spectrometer of three
  toroids, one for the barrel ($r_{IN}$=4.25 m $r_{OUT}$=10 m), 
  two for the endcaps, with an average magnetic field of 0.6 T. The inner tracking 
  detectors are instead placed in the central solenoid ($r$=1.2 m), in a 2 T magnetic field. The 
  bending planes are thus different for the two systems, respectively the $r-\phi$ and 
  the $r-z$ for the Inner Detector and the Muon Spectrometer.  
\item[-] in CMS the muon detectors are placed in the return yoke of
  the 4 T solenoid ($r$=3 m). 
\end{itemize}
The detectors composing the ATLAS muon system are the Muon Drift
Tubes (MDT), the  Cathode Strip Chambers (CSC), the Thin Gas Chambers
(TGC) and the Resistive Plate Chambers (RPC).  
In CMS they are the Drift Tubes (DT), the Cathode Strip Chambers (CSC) and the
Resistive Plate Chambers (RPC).
All these detectors are based on the collection of the ionization
produced by the passage of the muon in a gas filled volume.
The different choice of the detector mode of operation (drift,
proportional, streamer) depends on the value of the magnetic field and
the rate of charge particles expected in the region where the detector
is actually placed. 
In the case of CMS the resolution is dominated by the contribution of the multiple scattering
while for ATLAS by the calibration and alignment of the tracking
detectors. 
In both cases the combination of the track reconstructed by the muon detector, with the one 
reconstructed by the inner tracking detectors is necessary to obtain optimal resolution.\\

At LHC experiments, the trigger system has the task of reducing the event rate from the 
40 MHz bunch-crossing rate to the rate of about 100 Hz, which can be afforded by the event
storage system. The muon trigger is designed to accept events with one or more 
muons with $p_T$ above a given threshold; the trigger decision of the muon system can
then be combined with the one of the other subdetector systems to give the final 
experiment's trigger decision.
The trigger is organized over more than one level: the first one (L1) has to operate a 
fast choice (to be taken in less than $\simeq$10 ns) on the $p_T$ and also identify
the region of the detector, the so-called Region of Interest, 
that has to be taken into account by the following trigger 
levels. These levels reconstruct the muons with higher resolution and detail, refining 
the initial choice operated by the L1. The final trigger level uses algorithms very
close to those used for the offline reconstruction.\par   
The offline reconstruction has the task of providing optimal muon identification and
momentum resolution over the $p_T$ range $\simeq$ 5 GeV/$c$-1TeV/$c$. The reconstruction
in the muon spectrometer standalone can exploit the cleaner environment of the muon 
system, while the combination with the inner tracking detector is performed 
in order to improve the performance.\par
\begin{figure}[hbt]
  \includegraphics[width=0.45\textwidth]{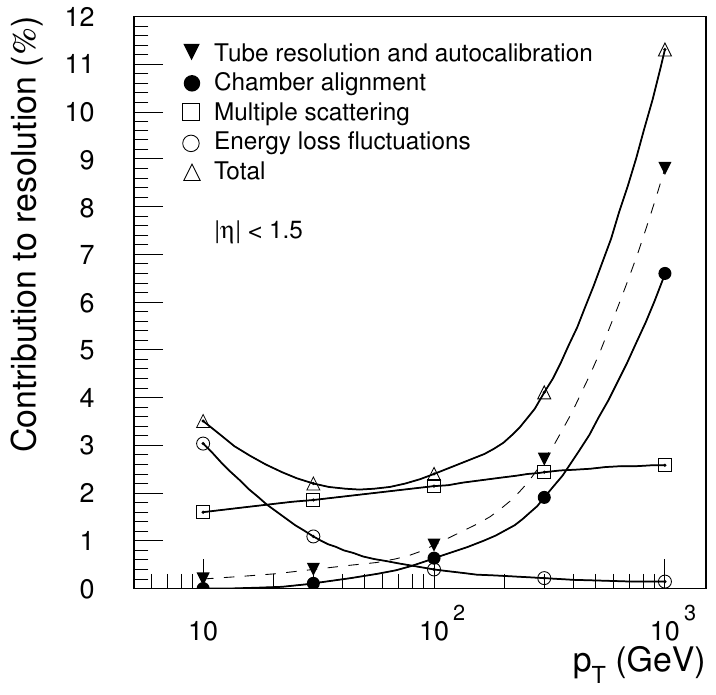} 
  \includegraphics[width=0.45\textwidth]{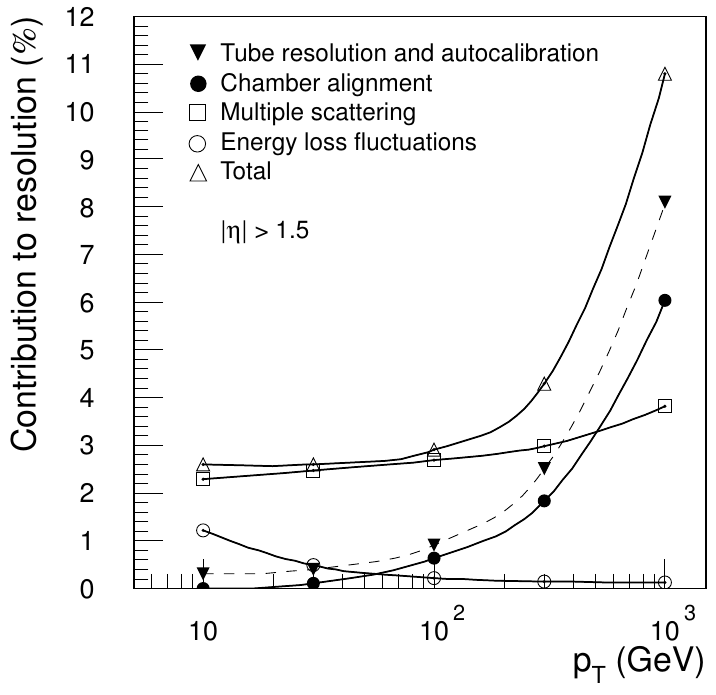} 
  \caption{Contributions to the $p_T$ resolution of the ATLAS Muon
  Spectrometer:  for $|\eta|<$ 1.5 (left) and for $|\eta|>$ 1.5 (right).}
  \label{fig:atlas_muon_resolution}
\end{figure}
\begin{figure}[hbt]
  \includegraphics[width=0.45\textwidth]{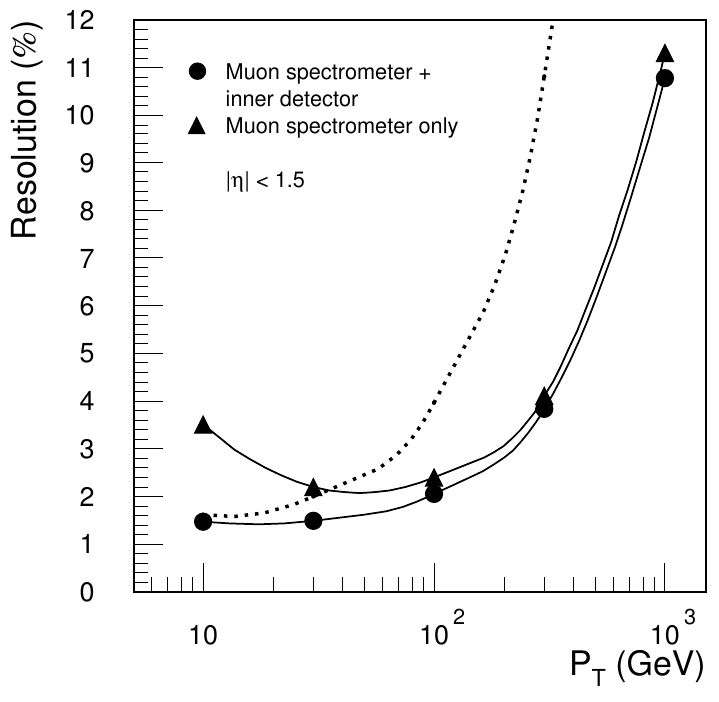}
  \includegraphics[width=0.45\textwidth]{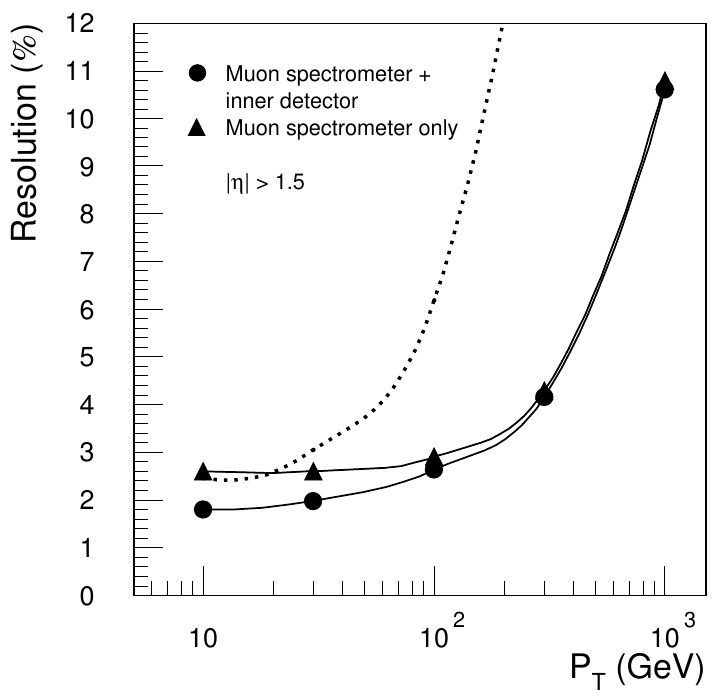}
  \caption{Resolution on $p_T$ as a function of $p_T$ 
    for standalone and combined muon reconstruction in ATLAS: 
    for $|\eta<$1.5 (left) and  for $|\eta|>$1.5 (right).
    The dashed line is the resolution obtained using only the Inner
    Detector.}  
    \label{fig:atlas_combined_resolution}
\end{figure}

The contributions to the momentum resolution for the standalone reconstruction 
in the ATLAS Muon Spectrometer are shown in
Figure~\ref{fig:atlas_muon_resolution}~\cite{ATLAStdr}. 
At low momenta, for $p_T<$20 GeV/$c$, the main contribution  
comes from the fluctuations of the energy loss in the calorimeters. The spatial resolution 
of the muon spectrometer tracking detectors and of their calibration and alignment becomes
relevant for $p_T>$200 GeV/$c$.
In Figure~\ref{fig:atlas_combined_resolution} 
the $p_T$ resolution using the Muon Spectrometer and the Inner Detector
is shown.\\

\begin{figure}[hbt]
  \includegraphics[width=0.45\textwidth]{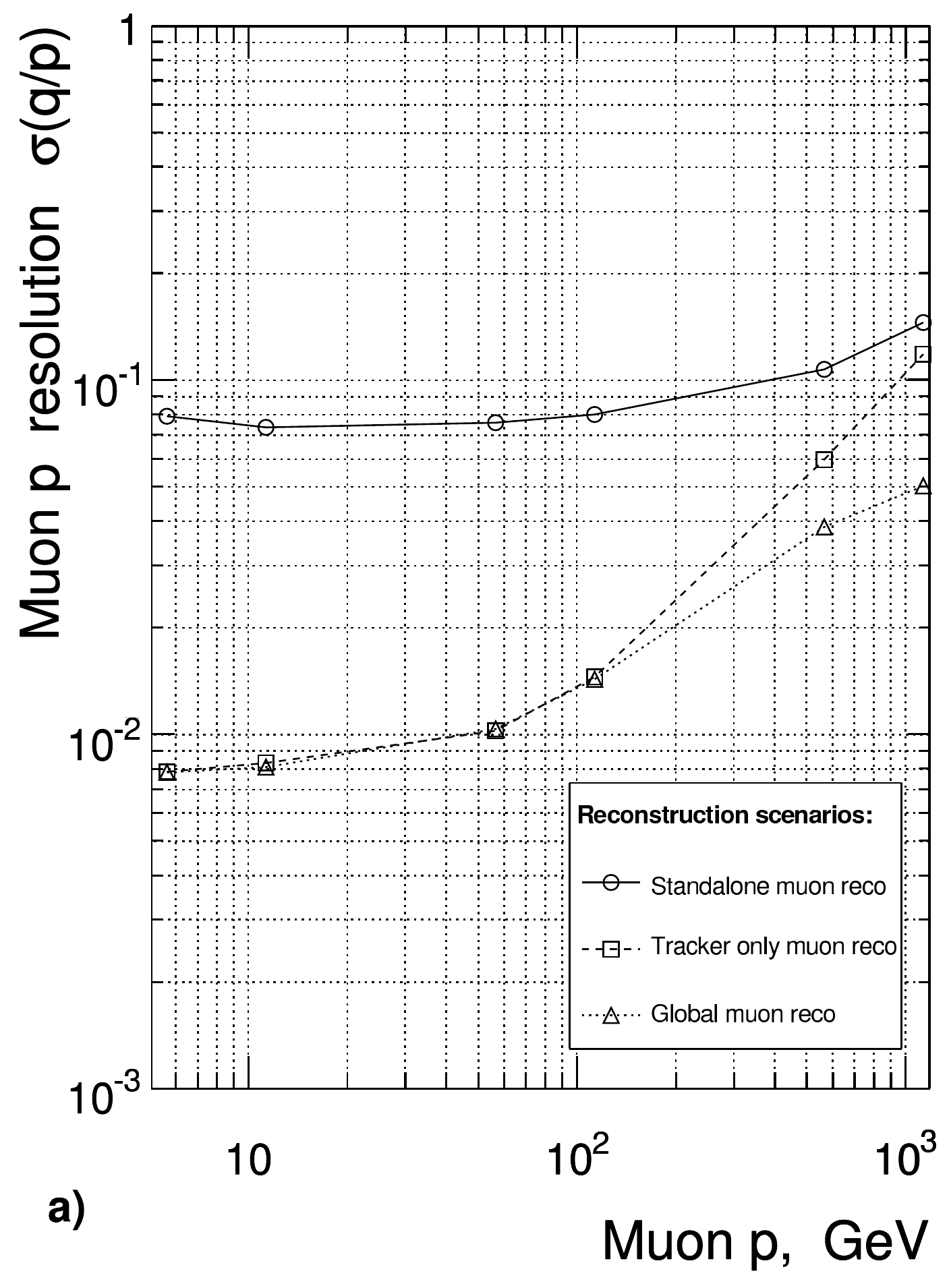}
  \includegraphics[width=0.45\textwidth]{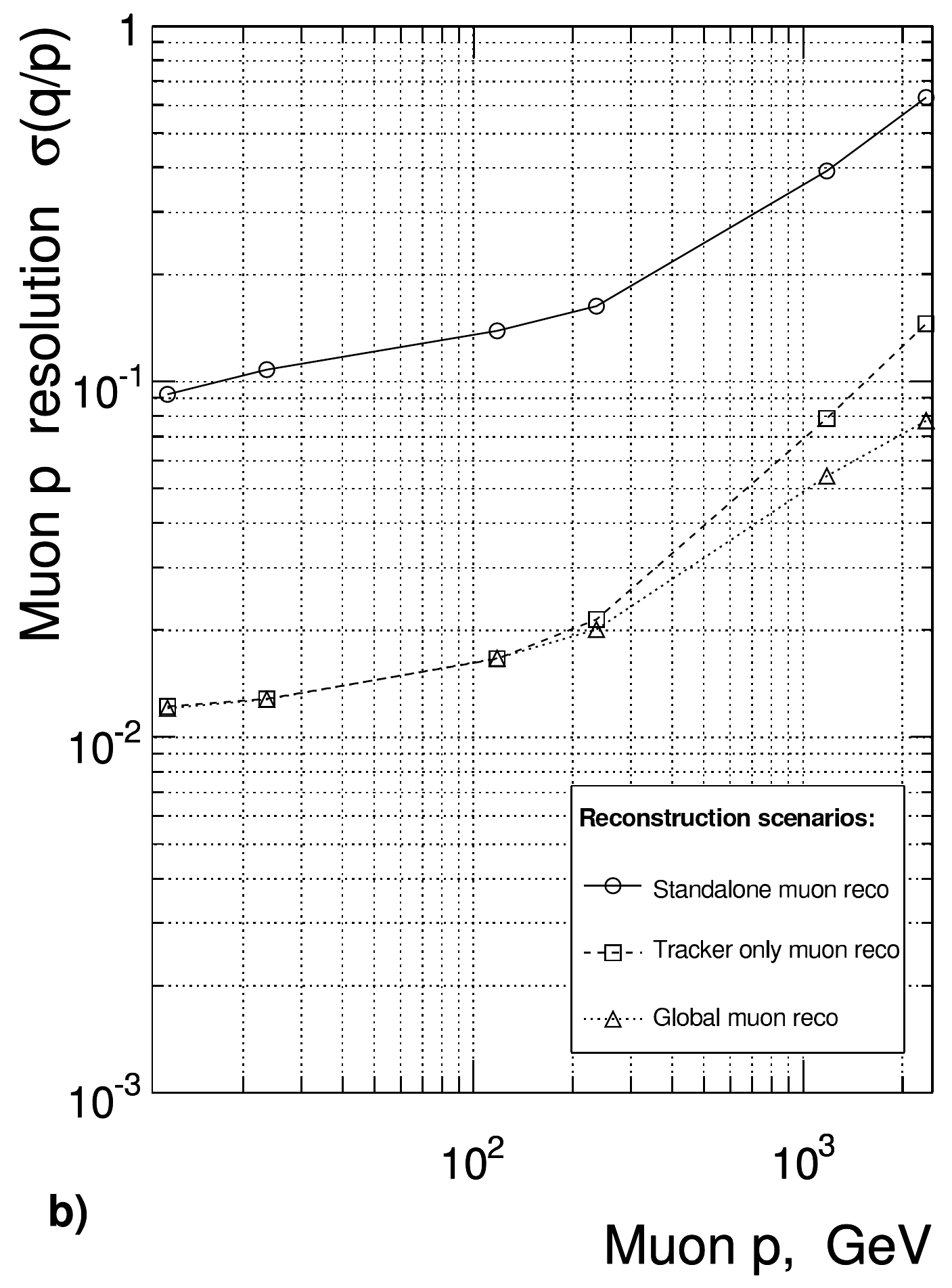}
  \caption{Momentum resolution of the muon reconstruction in the CMS experiment, 
    in the barrel at $\eta=$0.5 (left) and in the endcap at $\eta=$1.5 (right).}
    \label{fig:cms_muon_resolution}
\end{figure}

The $p$ resolution of the muon reconstruction in the CMS experiment is shown in 
Figure~\ref{fig:cms_muon_resolution} at two $\eta$ values
representative of the barrel and of the endcap regions, respectively
$\eta$=0.5 and $\eta$=1.5~\cite{CMStdr}. 
To obtain optimal resolution, the combination with the inner tracking
system can be
performed. The combined resolution, compared to the standalone resolutions of each of the
two systems, is shown in the two figures.
The expected efficiency and the resolution for the reconstruction of di-muon masses are shown 
in Figures~\ref{fig:cms_dimuon_efficiency} and~\ref{fig:cms_dimuon_resolution}. 
Shown in the figures is also the expected effect on efficiency and resolution
of the detector misalignment remaining after the calibration with the
first data (few 100 pb$^{-1}$) and after long-term (few fb$^{-1}$).
\begin{2figures}{hbt}
  \resizebox{\linewidth}{!}{\includegraphics{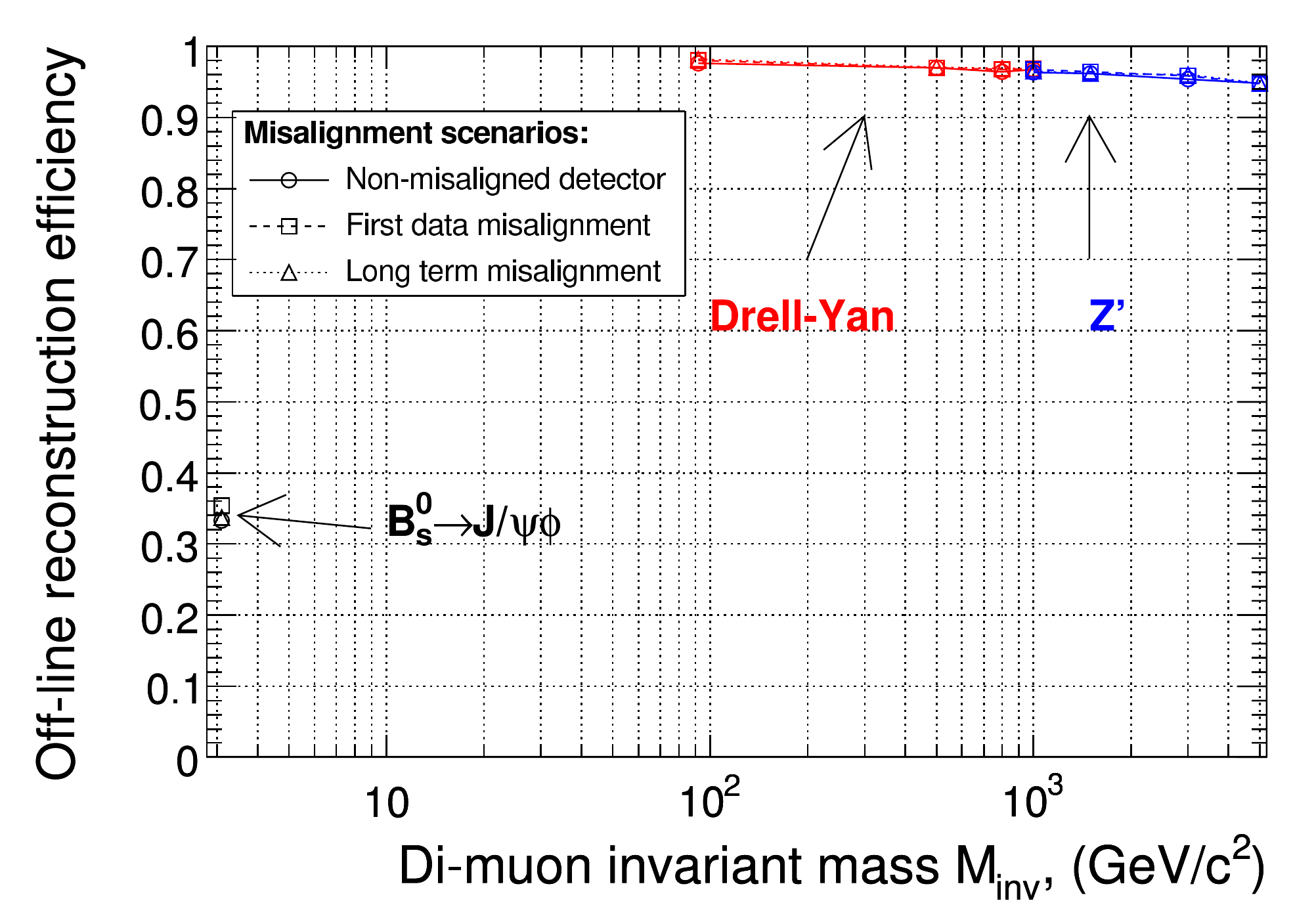}} &
  \resizebox{\linewidth}{!}{\includegraphics{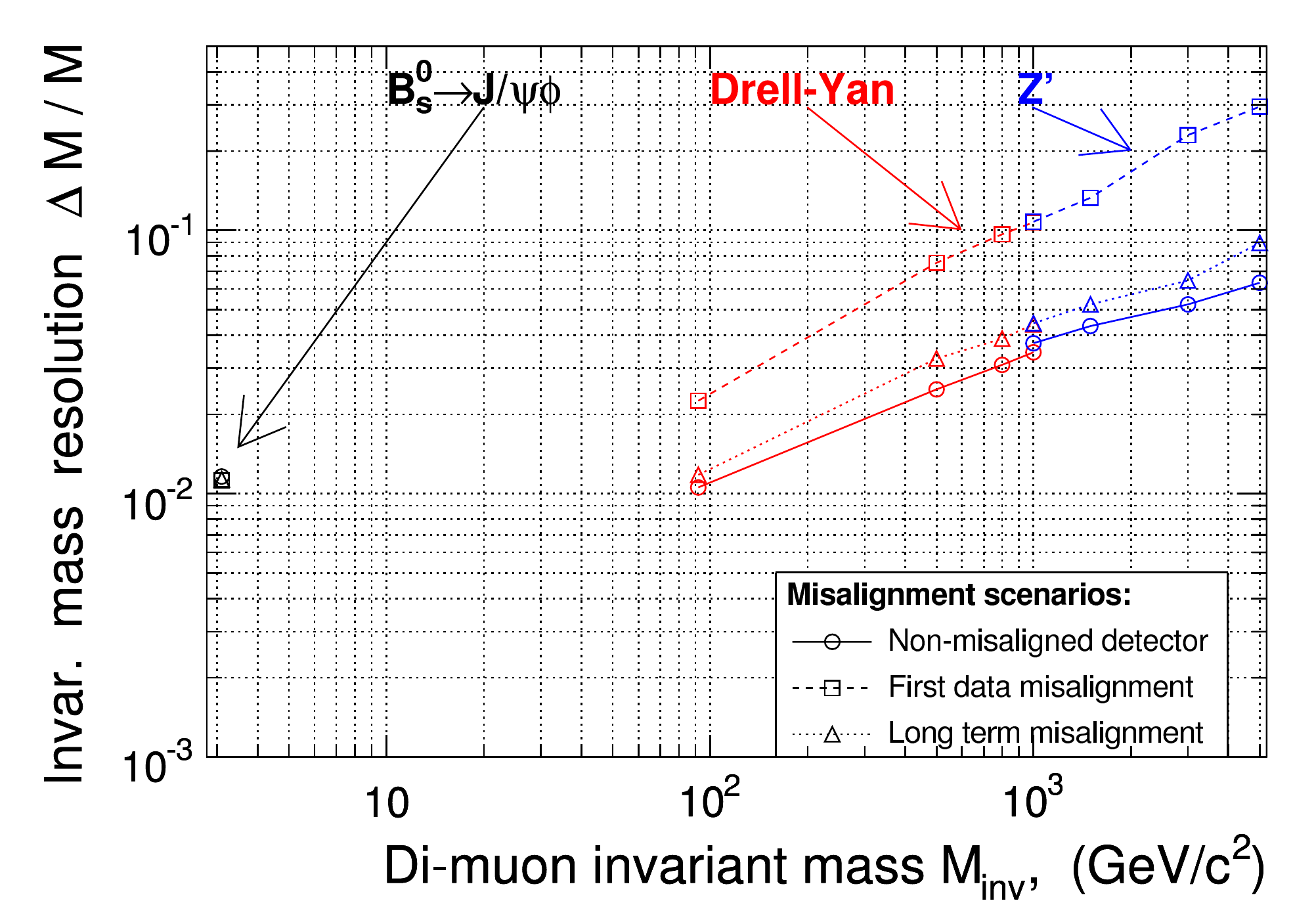}} \\
  \caption{Efficiencies for the reconstruction of di-muon masses
  in the CMS experiment.}
  \label{fig:cms_dimuon_efficiency}&
  \caption{Resolution  on the reconstructed di-muon masses
    in the CMS experiment.}
  \label{fig:cms_dimuon_resolution}\\
\end{2figures}
\section{Electrons and Photons}
{\large {\sl L.~Carminati, F.~Tartarelli}}
\vspace{0.5cm}

Electrons and photons (EM objects) are reconstructed using information
from the tracking detector and the calorimeters. 
The electromagnetic calorimeter occupies a cylindrical volume
located outside the tracking system (at smaller radii) and inside the hadronic 
calorimeter (at higher radii). An EM object looses its energy
in the calorimeter material so that an energy measurement can be performed. 
A high-energy EM object hitting the calorimeter will create lower energy 
electrons and photons (via bremssthralung and pair production), the so-called 
electromagnetic shower, in a process known as electromagnetic cascade. The lower
energy particles created in the cascade can then be detected using
different techniques:
\begin{itemize}
\item[-] in ATLAS the shower develops in several layers of lead
  plates. These are interleaved with 2 mm-thick layers of liquid Argon
  where the energy of the low energy electrons created in the cascade
  is deposited as ionization energy. The signal in the detector is
  generated by the drift of the ionization electrons in an electric
  field placed in the liquid Argon gap. The gap extends between
  the lead absorbers and copper-kapton electrodes where the signal is
  collected. To keep the Argon liquid at a temperature of about 90 K,
  the ATLAS electromagnetic calorimeter is kept into 3 cryostats (one
  for the barrel region and two for the endcaps). The alternance of
  layers of active and passive material makes it a so-called {\em
  sampling} calorimeter. 
\item[-] in CMS the same material, Lead Tungstate (PbWO$_4$) crystals,
  used to degrade the energy of the impinging EM object is also
  used to obtain a signal. The CMS calorimeter is a so-called
  {\em homogeneous} calorimeter. The low energy electrons created in the 
  cascade excite the crystal lattice which emits blue-green (420 nm)
  scintillation light.
\end{itemize}

The detection principle is different in the two experiments: ATLAS collects 
charge while CMS collects light. In ATLAS the electrical signal 
produced
in the liquid Argon gap in a purely ionization regime (no charge 
multiplication) is sent via transmission lines to the front-end electronics
located outside the cryostats where the signal is amplified, changed in
shape to optimize the signal-to-noise ratio and put in digital format.
In CMS, the relatively low light yield (30 photons/MeV) requires the use
of photodetectors (avalanche photodiodes in the barrel and vacuum 
phototriodes in the endcap) with intrinsic amplification. The produced
signal (about 4.5 photoelectrons/MeV) is sent to the front end 
electronic (located just outside the crystals) where the signal is
amplified, shaped and digitized.

The bulk calorimeter is subdivided in smaller units called towers or cells 
which project back to the interaction point. 
In CMS the crystals have a size (front face) of about 
22$\times$22 mm$^2$ (in the barrel), approximately the Moliere radius
in PbWO$_4$ and the towers cover regions of size 0.0175$\times$0.0175
in the $\Delta\eta\times\Delta\phi$ space; 
the CMS calorimeter has no longitudinal (along the radius) segmentation. 
In ATLAS the calorimeter
is subdivided in three sections called (from inward outward) strips, middle and
back. The middle section cells (which collects most of the energy) have a 
square size of 4$\times$4 cm$^2$. The strips have a rectangular size with a very 
small dimension along $\eta$, 4 mm. The longitudinal subdivision allows to 
sample the development of the shower in three points and helps in 
particle identification (see below).
In CMS, the gaps between each cell and the neighboring ones 
would produce inefficiency in the shower reconstruction in those regions. To
reduce this effect, the crystals are mounted in a quasi-projective 
configuration with the crystal axis making a 3$^\circ$ angle with a vector
coming from the nominal interaction point in both the $\eta$ and $\phi$
direction. In ATLAS, along $\eta$ these effects are much reduced as the cells
are not mechanical units but they are obtained by etching copper strips on
the readout electrodes and so the gaps are much smaller. Along $\phi$ the
geometry with accordion-shaped electrodes is such that there is 100\% coverage
with no gaps at all. 
The pseudorapidity coverage of the CMS calorimeter is $|\eta|$<1.479
in the barrel and 1.55<$|\eta|$<3 in the endcap. The ATLAS
calorimeter covers the regions $|\eta|$<1.475
in the barrel and 1.375<$|\eta|$<3.2 in the endcap.

\subsection{Effects of material}
Ideally one would like to have the EM object hitting the calorimeter,
to start the electromagnetic cascade just in the calorimeter material and
have all its energy lost in the calorimeter, in one cluster of its cells.
To accomplish these goals
it is necessary to keep the material in front of the calorimeter (the
so-called {\it material budget}) to a 
minimum so that the shower does not develop before the calorimeter
and to build a calorimeter that has enough thickness to accommodate the
development of the electromagnetic shower.

Concerning the latter point, one should note that luckily, when the length
needed to accommodate a certain shower is expressed in units of the radiation 
length ($X_0$) of the calorimeter material, it scales only with the
logarithm of the shower energy. Both the ATLAS and CMS calorimeters have a
longitudinal depth of 24-26 $X_0$ (it varies along $\eta$) which has been 
calculated to accommodate showers up to 500 GeV and keep to a minimum the
contribution to the energy resolution due to the energy fluctuation for
showers of higher energies.  

Incidentally, one should note that the electromagnetic calorimeter 
represent about one absorption length ($\lambda_I$) for charged hadrons. About
10 $\lambda_I$ are needed to contain hadronic showers and limit the background
in the muon system. This goal is accomplished by the hadronic calorimeter
which surrounds the electromagnetic calorimeter.

Every effort has been done in the design and construction phases of the 
ATLAS and CMS experiment to keep the material in front of the calorimeter 
to a minimum.
In both ATLAS and CMS the material in front of the calorimeter is represented 
by the beam pipe walls and by the inner tracking detectors 
which amounts to about 1 $X_0$ at small $\eta$'s and it increases in the endcap 
regions. 
The most of the material is due not just to the thin (usually
300 $\mu$m) active layers of silicon (strips or pixels) of the inner
tracking detectors but rather to the mechanical 
supports, electronics, cabling and services associated to the tracker
operation and readout. 

In ATLAS additional material in front of the calorimeter is due to the walls of
the cryostat used
to keep the liquid Argon at a temperature of about 90 K. To save material, this 
cryostat also integrates the cryostat for the superconducting coil that 
produces the 2 T magnetic field into the tracker volume. 
This last effect is not present in CMS as both calorimeters (the electromagnetic 
and most of the hadronic one) are placed inside a large solenoid, so 
that the material of the coil and of the cryostat does not enter in the 
EM calorimeter material budget. 
\begin{figure}[htb]
 \centering
   \includegraphics[width=0.3\textwidth,angle=-90]{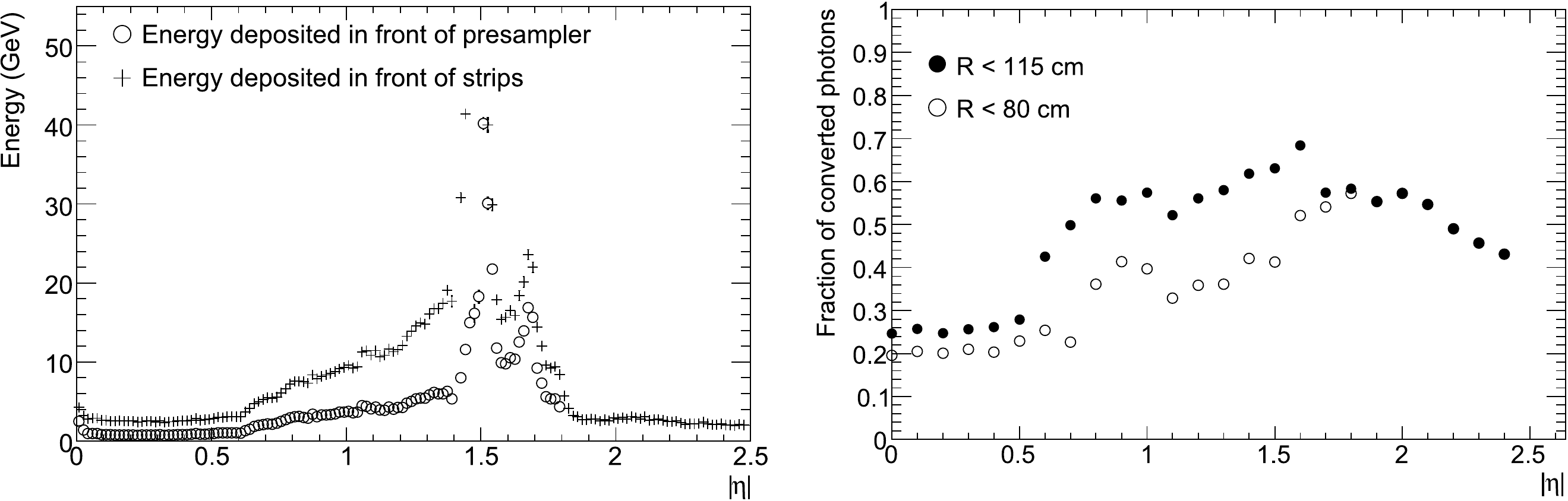}
  \caption{Left: average energy deposited by 100 GeV electrons in front of the
  presampler (open circles) and before the first compartment of the ATLAS 
  calorimeter (crosses) as a function of $\eta$. 
  Right: fractions of photons converted below a radius of 80 cm (open 
 circles) and 115 cm (full circles) as a function of $\eta$ in the ATLAS
 detector.}
 \label{fig:ATLAS_mateff}
\end{figure}

Electrons will undergo bremsstrahlung in the upstream material. Soft brems 
radiation will
increase the size of the cluster. The effect is larger along the $\phi$ 
direction
due to the effect of the magnetic field that bends the electron direction. 
As a consequence the cluster becomes larger and asymmetric.
If a hard bremsstrahlung photon is emitted along the electron direction, it is 
also
possible that the electron and the emitted photon reach the calorimeter into 
separate
clusters. Moreover, the electron trajectory is no more a helix
and this makes the electron track reconstruction in the tracker more difficult.
Fig.~\ref{fig:ATLAS_mateff} shows the average energy deposited by electrons
before the arrive to the ATLAS calorimeter and before the presampler. The curve
follows the material profile before the calorimeter and has a maximum 
around $\eta \sim$1.5: this corresponds to the gap between the barrel
and end-cap calorimeter (a region that cannot be used for precision physics).

Photons can convert in the tracker material and give origin to an
electron-positron pair. Fig.~\ref{fig:ATLAS_mateff} shows the fraction of
conversions as a function of $\eta$ for $H \rightarrow \gamma \gamma$
photons: the quantity is shown for two radii, corresponding approximately
to the end of the tracking detector (80 cm) and to the beginning of the
calorimeter (115 cm).    
With respect to an unconverted photon, a converted photon 
will deposit its energy in 
a larger and asymmetric cluster: the superposition of the two 
electron-positron clusters. 
Again the cluster is larger along the $\phi$ direction due to the bending of
the electrons along this direction.

In both cases, material at low radii is the most dangerous as these effects are
amplified by the longer electron(s) path into the magnetic field. Electrons 
from
early conversions might be reconstructed as two separate clusters into the 
calorimeter. Effects are also
larger in CMS where the magnetic field is 4 T (to be compared to the 2 T in 
ATLAS).
Material at high radius is anyway detrimental for the calorimeter 
performance due to the fluctuations in the energy lost before the 
calorimeter as the shower starts earlier. 
In ATLAS, where 
the effect is larger due to the presence of the coil and of the 
cryostat walls, a 
presampler detector
is placed just in front of the calorimeter. This consists of a 11 mm 
thick layer 
of liquid Argon that samples the early development of the cascade.

However, the tracker itself provides information useful to recover some of
the problems it creates. In case of hard electron bremsstrahlung one can
try to reconstruct the typical "kink" in the track trajectory (where the
photon is emitted). The calorimeter cluster also provides an additional
point that can be included in the track fit. Moreover, one can exploit the 
fact that the energy weighted barycenter of the electron and brems photon
clusters in the calorimeter provides the extrapolated trajectory of the
electron before the brems emission occurred.  

If a photon converts early (the most dangerous situation), the two electrons 
can leave enough hits in the silicon layers so that their tracks can be 
reconstructed: the converted photon energy and direction is then obtained from
the four-momenta of the two electrons. The situation is more difficult for late 
conversions as the number of hits left by the two electrons cannot be enough 
for 
them to be reconstructed with satisfying efficiency. In any case, an ad-hoc
tracking in which the electron track is reconstructed from the outer 
layers of the tracker inwards is usually needed. With respect to a track 
coming from the primary
vertex, a reduced number of hits in the detector is also allowed (at the 
expense
of an increased number of reconstructed fake tracks). There are cases in which 
a conversion cannot be reconstructed as one of the two electrons is not 
reconstructed:
this might happen in case of asymmetric conversions with one of the two 
electrons 
having a low transverse momentum. ATLAS studies have shown that a track matched
to a calorimeter cluster that does not have a hit in the innermost pixel layer 
are coming from conversions, if a non-negligible fake rate (around 8\%, 
from charged
pions) can be accepted. Of course this strategy strongly depends on pile-up 
and on the inefficiencies in the pixel layer.  

\subsection{Clustering}
The energy of the EM object is deduced by summing together the energy of
contiguous 
towers using an appropriate algorithm, usually seeded by energy deposits in
the calorimeter itself.  A pattern recognition algorithm is
needed to locate the EM clusters and discriminate between noise clusters,
by searching for local maxima (above a certain threshold). Once the cluster
is found, its energy and position are reconstructed by using all cells
included in a window (usually rectangular) centered around the direction
provided by the previous algorithm. The size of the window
is a compromise between the need to recover as much as possible of the 
particle energy
(which would favor a larger window) and the need to limit the noise (electronic
and pile-up noise). The
size can also depend on the particle type (electron or photon).

In the absence of material effects a square cluster would provide the optimal
energy resolution. As this is not the case, ATLAS uses 
rectangular clusters with the longer dimension along the $\phi$ direction: in
the barrel part of the calorimeter a 
3$\times$5 cell cluster (in term of middle cells, see above)  is usually used for 
unconverted photons and a 3$\times$7 cell cluster for electrons and converted photons.
In the endcap a 5$\times$5 clusters is used for both electrons and photons.
The longer dimension along the azimuthal angle, the direction of magnetic 
bending of the electrons, is used to recover energy lost in bremsstrahlung
emission or to correct for undetected converted photons.

In CMS a 5$\times$5 cluster provides best results for unconverted photons and 
electrons that have not radiated. In the other cases best energy 
resolution performance is obtained by algorithms that cluster together
cells dynamically according to a certain algorithm instead of using 
fixed-size arrays. These clusters, which are called {\em superclusters}
in CMS, have no predefined size nor a fixed number of cells. Two algorithms
have been developed: the {\em Hybrid} and the {\em Island} algorithms.
The Hybrid algorithm, as its name says, uses a standard fixed size approach
along the $\eta$ direction, while searching dynamically for separated
energy deposits along the $\phi$ direction. This algorithm gives the best
results for electrons in the barrel. In the endcap the Island algorithm 
is used. This algorithm starts from cells above a certain threshold (the
seed) and adds neighbouring cells (scanning first in $\phi$ then in
$\eta$) until there is a rise in the energy (or the crystal energy is
below threshold). Then clusters found like this can in turn be clustered
together into a supercluster, associating to a seed cluster nearby 
clusters in a narrow window along $\eta$ and in a wider $\phi$-window.

Dynamic clustering algorithms have also been studied in ATLAS (where they
are called {\em topoclusters}): however, at the moment, they are not used
in the studies of electromagnetic clusters but rather for hadronic ones.

Whatever the clustering method and the cluster size are, reconstructed clusters
undergo a series of offline corrections which are described in the following. 



\subsection{Calibration}\label{Sec:calib}
Calibration is the set of procedures needed to go from the energy deposited
into the calorimeter to the best estimate of the electron or photon energy
(and direction) produced in the interaction point. A set of corrections
to the raw energy is applied in various stages to correct for various effects.
These corrections have been studied using very detailed simulation of the
detector (not only of the calorimeter but also of the tracker and material 
in front of it) and using data collected exposing modules of the calorimeter
to test beams (mainly electrons of known energies).

There are various operation involved that may also be performed at the same
time  according to the calibration strategy adopted:
\begin{itemize}
\item[-] electronic calibration;
\item[-] cluster correction for containment and material effects;
\item[-] estimate of the material in front of the calorimeter;
\item[-] intercalibration of different regions/cell of the calorimeter
to ensure the uniformity of the response;
\item[-] absolute calibration of the response.
\end{itemize}

The calibration strategy adopted by the experiments,especially at the 
beginning of data taking, will evolve with time depending on the
performance available on the detectors involved (tracker and calorimeters) 
and on the availability of data samples. Cross-checks of various calibration
techniques and iterative procedures will be needed.
For example, one of the first information
needed is an accurate estimate of the material in front of the calorimeter.
Although various methods are planned and are briefly described below,
these will already require that an energy reconstruction scheme for electrons
and photons is in place. Then, once a new estimate of the material is 
available, the procedure will be iterated up to the desired precision.  

\subsubsection{Estimation of the material in front of the calorimeter}
An accurate modelling of the material in front of the calorimeter is needed
to reach the best performance in electron and photon reconstruction.
During construction of the detectors, components of the trackers have been
weighted and also data from beam tests are available. However, even if
these data will provide a reasonable starting point, the precision in the
estimation of the material which is needed (about 1\% of radiation 
length) can only be achieved using LHC data.

A very accurate estimation of the material in front of the calorimeter
will be performed using the radii of identified photon conversions. Another
method is the study of the $E/p$-ratio distributions, where $E$ is the 
energy of an electromagnetic cluster and $p$ is the momentum of the 
matching charged track. Other variables that have been studied and that
are sensible to the material are shower shape variables along $\eta$ and
$\phi$ and variables connected to the quality of the reconstructed 
associated track.

\subsubsection{ATLAS}
In ATLAS, the raw ADC values coming from the electronics are corrected for:
\begin{itemize}
\item[-] an optimal filtering technique to reconstruct the signal from 
a certain number of samples (usually 5 samples, taken every 25 ns);
\item[-] a factor that corrects for the different gain of the front-end 
electronics. An electronic calibration system generates and sends to the 
 preamplifiers a well known electrical signal; 
\item[-] a factor (studied on test beams) that translated the ADC counts 
 of the electronics signal into an energy value (GeV);
\item[-] a raw sampling fraction factor. The sampling fraction, typical of 
sampling calorimeter (like the ATLAS LAr calorimeter), gives the ratio 
between the active material and the total material. So that if a certain 
energy is deposited into the calorimeter one has to divide by the sampling 
fraction to have recover the energy of the incident particle. 
\end{itemize}

Once clusters are formed in each section in which the ATLAS calorimeter is 
divided (four sections, including the presampler) the energy of the EM object 
is the sum of four clusters located along the same direction. However, a 
better estimate of the produced EM objects is obtained by using appropriate 
weights in the sum. The weights correct for the energy lost in inactive
layers (like the absorbers) of the calorimeters and dead regions like the
solenoid, the cryostat and in particular the material of the tracking detector
in front of the calorimeters. These coefficients are determined using 
detailed Monte Carlo simulations of the detector. The energy lost in front of
the calorimeter is recovered by an appropriate weight of the energy 
deposited into the presampler. The energy lost in the dead material of
the calorimeter is corrected by weighting the energy deposited into the
calorimeter with a factor that depends on the longitudinal barycenter of
the shower.

The chosen technology for the EM calorimeter is such that enough uniformity
is guaranteed by design so that there is no need to intercalibrate at cell
level. Indeed, extensive test-beam studies on production modules have 
successfully verified that the response is uniform at better than 0.5\% on
regions of dimension of $\Delta \eta \times \Delta \phi = 0.2 \times 0.4$,
that's to say the dimension of an electronic readout board (128 middle
cells). Then it will be necessary to intercalibrate these regions at the
design level of 0.7\% (at the beginning of the data taking they are 
expected to be miscalibrated at the level of 2\%). These can be done using
the legs of $Z \rightarrow ee$ decays. However, as electrons are involved,
the procedure assumes an excellent knowledge of the material in front of
the calorimeter.  

\subsubsection{CMS}
In CMS, the reconstructed energy of a EM object can be written as:
\begin{equation}
E = G \times F \times \sum_i c_i \times A_i
\label{CMS-cal-eq}
\end{equation}

where G is the global absolute scale, F is a correction factor and $c_i$'s
are coefficients that intercalibrate the cell $i$ of amplitude (in ADC counts)
$A_i$ entering in the sum of the cells for this cluster. 

The largest source of channel-to-channel variation in the CMS barrel 
calorimeter is the spread in the scintillation light yield. Notwithstanding
it is possible to obtain a first estimate of these coefficients from lab and
test-beam measurements, the final calculation will be done {\em in situ}
using LHC data. Due to the large number of crystals involved, various methods 
have been studied to achieve this goal, depending on the available integrated
luminosity. Both electrons and photons can be used. Often the selected
calibration sample and method can be used for more than one calibration
task. 
\begin{enumerate}
\item[-] Assuming to virtually divide the calorimeter in $\eta$ rings, it is
possible to intercalibrate crystals within these rings by comparing the
total energy deposited in each crystal with the mean of total energy for
all crystals in the ring. Minimum bias events can be used to perform such
procedure. Of course, then all rings have to be intercalibrate each other:
this can be done using one of the methods below. This method has the advantage 
that can be used at the beginning of data taking with simple triggers. However
it is sensible to inhomogeneity in the tracker material and to asymmetries
coming from the geometry of the detector (boundaries between modules, 
off-pointing crystal angles, \ldots).
\item[-] Use single electrons to intercalibrate crystals. However as 
electrons are involved, this method requires that the tracking detector is
operational and aligned. In order to cope with detector effects, a series
of cluster variable sensitive to the amount of brems is studied in order to
control the quality of the electron.
\item[-] Use $Z \rightarrow ee$ decays. This method can be used, for example,
to intercalibrate rings in first method described here. Moreover it can be
used to determine the correction factor F of equation~\ref{CMS-cal-eq}.;
\item[-] Use $\pi^0/\eta \rightarrow \gamma \gamma$ events to intercalibrate 
crystals. This method has the advantage that photons are less sensitive
to material than electrons as long as a sample of unconverted photons 
can be selected by cutting on cluster variable such as the shower shape. 
\item[-] Use radiative decays of the Z boson to muons $Z \rightarrow \mu \mu
\gamma$. This sample would have the advantage of creating a sample of
photons with small background and known energy. Concerning this last point,
however, it is clear that this method is correlating the calibration of
photons to that of muons, assumed to be already calibrated at the right
scale. With such a sample it would be possible to intercalibrate crystals,
estimate the correction factor F and set the global energy scale. Once
the global energy scale has been set for photons, it can be transferred 
to electrons using conversions.
\end{enumerate}

\subsubsection{Absolute energy scale}
All the corrections described above, however, do not guarantee that the 
absolute scale is 
correct. To do this, a well known mass resonance is used. As ATLAS and CMS
are mainly interested in high $p_T$ physics above the 100 GeV mass where e.g 
the Higgs boson is expected, the closest resonance is the $Z\rightarrow ee$ 
decay. At much lower masses $J/\psi, \Upsilon \rightarrow ee$  can be used.

\subsection{Energy resolution and linearity}

The goal of both the ATLAS and CMS detectors is to keep the energy
resolution at the level of 1\% of better. This is motivated by the 
required mass resolution on important physics channels with electrons
and photons, like $H \rightarrow \gamma \gamma$ and $H \rightarrow 4e$.

The energy resolution can be parametrized as a function of the energy E as:
\begin{equation}
\frac{\sigma}{E} = \frac{a}{\sqrt{E}} \oplus \frac{b}{E} \oplus c
\label{eres}
\end{equation}

where:
\begin{itemize}
\item[$a$:] is called stochastic term. In ATLAS it accounts for the energy
fluctuations due to the presence of alternating layers of lead and liquid
Argon and is about 10\%. In CMS it includes contributions from photostatistics
as well as fluctuations in the shower containment: it amounts to $\sim$3 \%.
\item[$b$:] is the noise term. The noise in a calorimeter is coming both from 
the electronics (electronic noise) and  from the
energy deposited in the calorimeter by the underlying event and from additional
collisions in the same bunch crossing or from previous or successive ones 
(pile-up noise). The electronics noise amounts to 100-200 MeV in a typical
test-beam cluster (which is usually smaller than the cluster size that will
be used during LHC running). The pile-up noise depends on the LHC luminosity. 
\item[$c$:] is called the constant term. As both the stochastic and noise parts
of the energy resolution decrease with energy, this is term that becomes
dominant at increasing energy. In ATLAS contributions to this term come from
LAr impurities and temperature variation, high-voltage variations, mechanical
deformations, material inhomogeneities in front of the calorimeter. In CMS
the list of contributions include residual intercrystal miscalibration,
temperature stability, supply voltage stability, disuniformities in the 
light collection and damages due to radiation. The goal is to keep this 
term at the level of 0.5-0.7\%.
\end{itemize}

\begin{figure}
 \centering
    \includegraphics[width=0.45\textwidth,angle=-90]{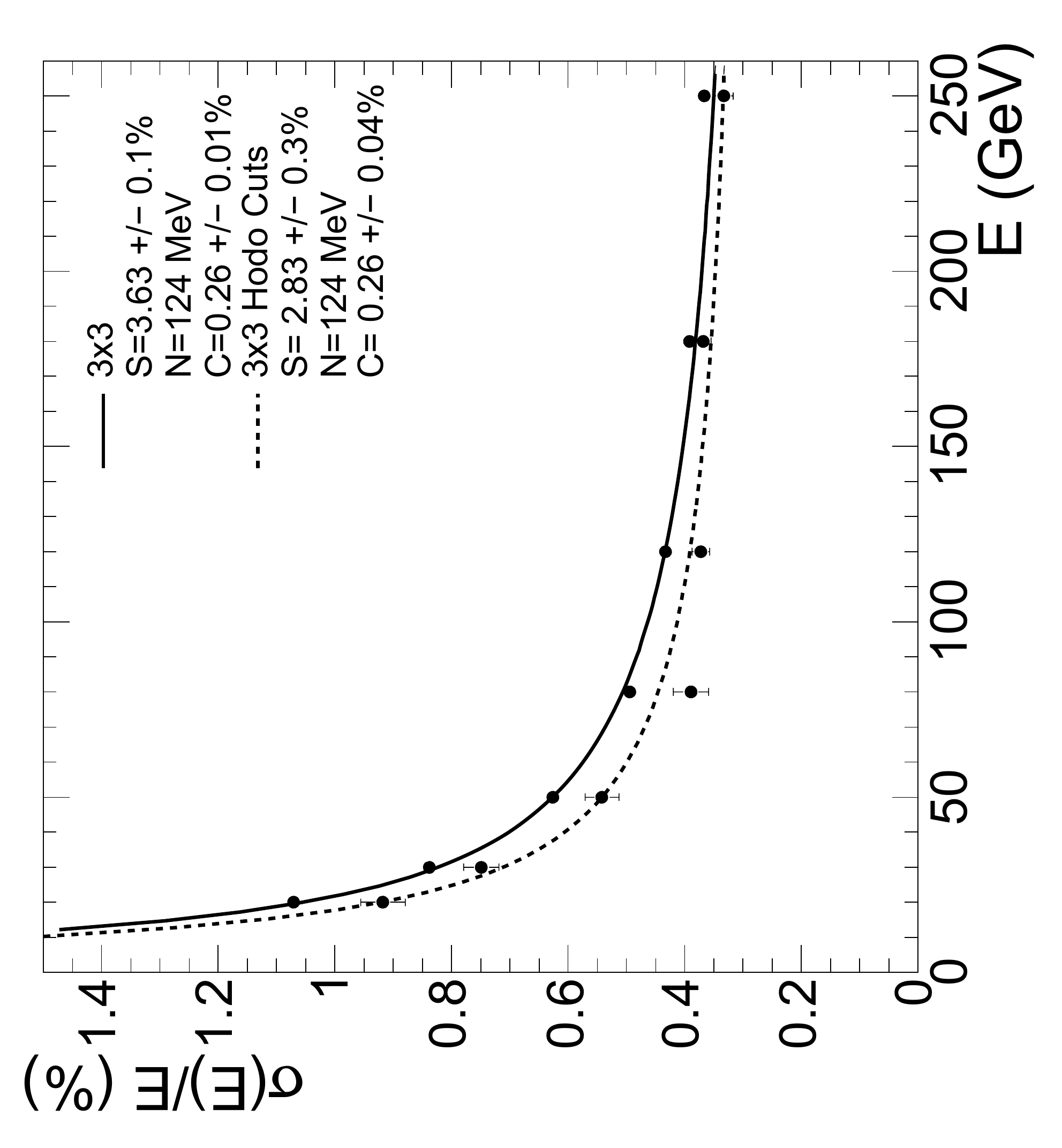}
    \includegraphics[width=0.45\textwidth,angle=-90]{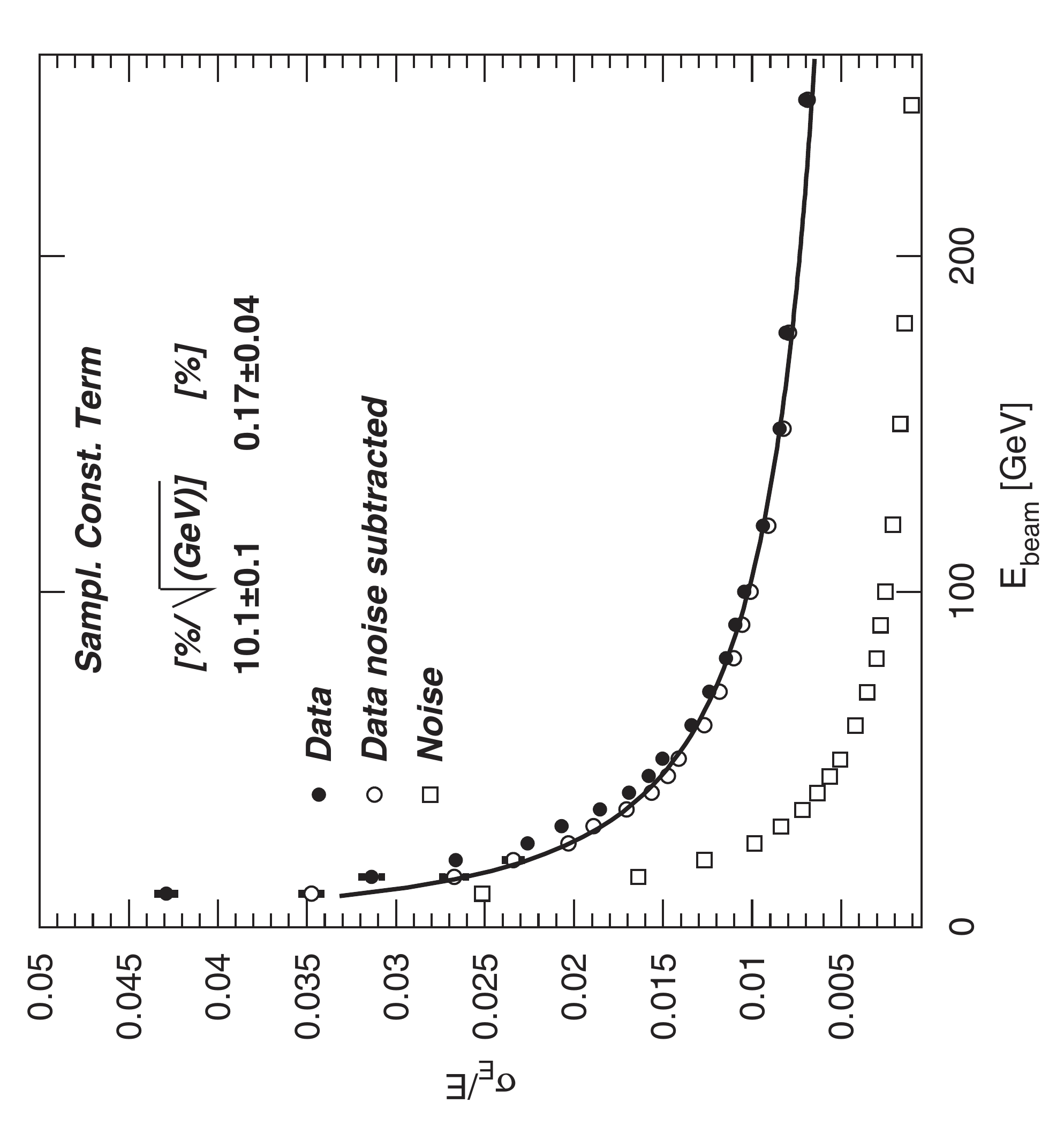}
  \caption{Relative energy resolution as a function of energy as measured
  for fixed-energy electrons in the CMS (left) and ATLAS (right) test-beams.
For CMS, the fit parameters S, N and C are the stochastic, noise and constant
term, respectively. For ATLAS, only the sampling and constant term are fitted.}
 \label{fig:CMS_ereso}
\end{figure}

The expected performances have been extensively tested using electron test
beams by both the ATLAS~\cite{carli} and the CMS collaborations. 
The measured energy
resolutions are shown in Fig.~\ref{fig:CMS_ereso}. The data have been obtained
at a fixed point in the calorimeter (corresponding to $\eta = 0.687$ and
$\phi = 0.28$ for ATLAS). The spectrum has been
fitted with the functional form in Eq.~\ref{eres}. In ATLAS, since the
electronic noise depends on the electronic gain (which may vary at different
energies), the noise is not included in the fit but rather measured independently
and subtracted by each energy point. It amounts to about 250 MeV (slightly
larger at high energies). The results are within the expectations. The
value extracted for the constant term, however, only accounts for local
disuniformities (restricted to the cluster cells, since the name {\em local 
constant term}) and not of all calorimeter ({\em global constant term}: see
also the discussion in Section~\ref{Sec:calib}). 

\begin{figure}
 \centering
    \includegraphics[width=0.45\textwidth,angle=-90]{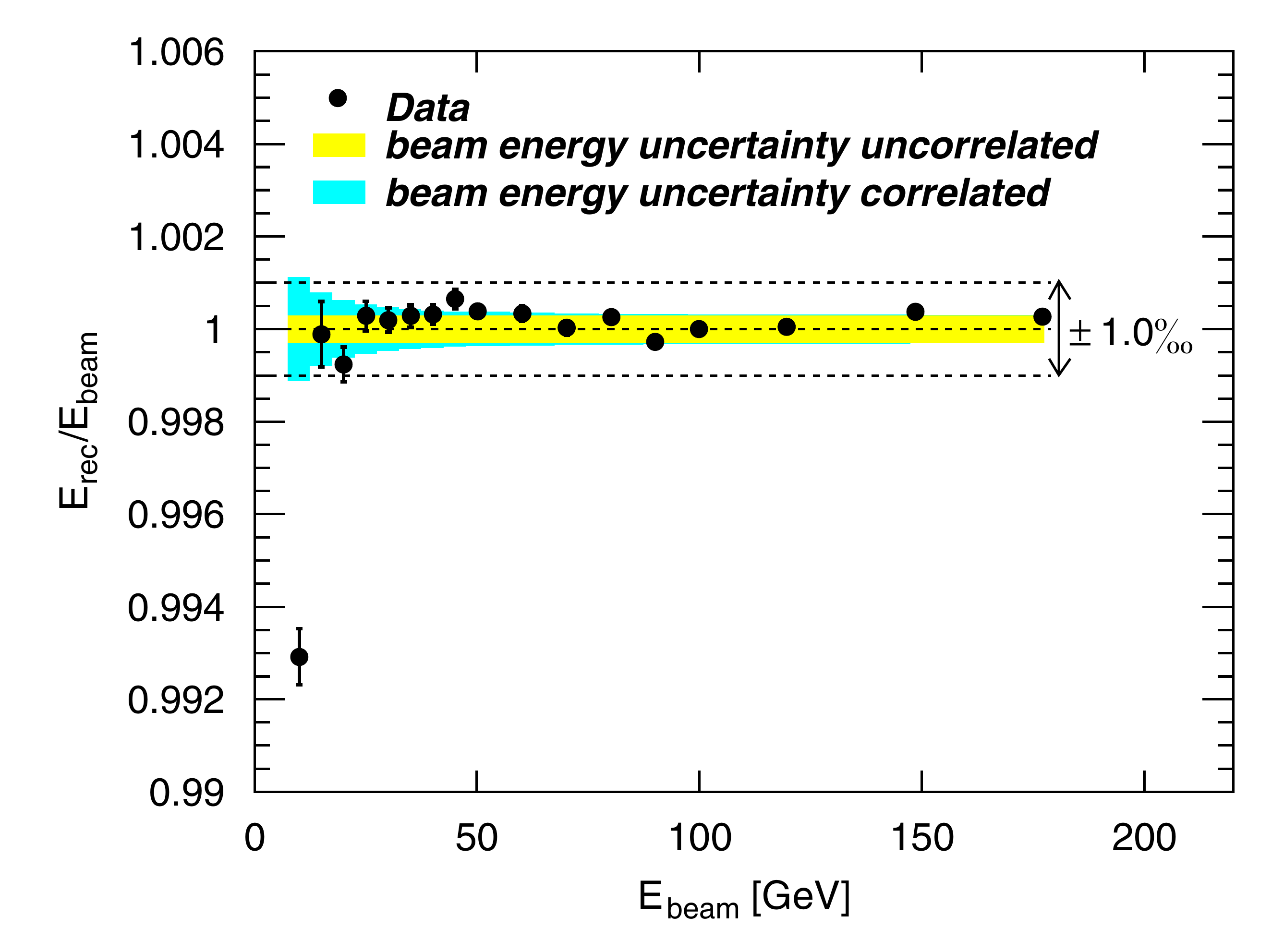}
  \caption{Ratio of reconstructed electron energy to electron beam energy as
     a function of the beam energy (ATLAS). The measured points are normalized to
    the 100 GeV point. The inner error band includes the uncorrelated uncertainty
on the beam energy measurement; the outer band adds in quadrature the correlated
 uncertainty.}
 \label{fig:ATLAS_lin}
\end{figure}

The linearity of the response of the ATLAS calorimeter is shown in
Fig.\ref{fig:ATLAS_lin}. For energies $E>10$ GeV, all measured points are
within $\pm 0.1$\%.

\subsection{Position measurements}
The shower direction is reconstructed by an 
energy weighted average of the coordinates of the cells of the
clusters. In ATLAS, the middle 
compartment only is used along the $\phi$ direction (as this is the most 
precise in this direction) while middle and strips 
are used along the $\eta$ direction. In CMS the average runs on the
position of the crystals in the clusters (or the position of the clusters 
inside a supercluster). The simple weighted 
average has to be corrected to take into account a few effects due to the
detector geometry. In ATLAS the $\eta$ positions as measured in the
middle and strips compartments can be combined to determine the direction
of the shower axis along $\eta$ (or $\theta$). Typical resolutions for
photons reconstructed in the ATLAS detectors are 4--6 mrad/$\sqrt(E)$ along
$\phi$ direction and 50--75 mrad/$\sqrt(E)$ along the $\theta$ direction.
For electrons, once a charged track has been successfully associated to the
electron cluster, the reconstructed electron position is better measured
from the track parameters measured in the tracking detectors.

The possibility of a stand-alone reconstruction of the photon direction
along $\eta$ plays an important role in $H \rightarrow \gamma \gamma$
events to identify the position of the Higgs vertex. The typical resolution
obtainable on the vertex position along the beam axis is about 16 mm. 
This method can be used
either alone or in combination with other methods which will be used to
determine the interaction vertex in these events. These methods (planned
by both ATLAS and CMS) rely on the determination of the event vertex 
using charged tracks produced in association with the Higgs boson or in
exploiting the fact that a good fraction of photons will convert in
the detector material. If the conversion is identified, additional 
direction information is provided from the converted electron tracks.
The possibility of combining several methods is particularly important
at high luminosity where one has to select the right vertex among
the additional interaction vertices due to minimum bias interactions.

\subsection{Particle identification}
Clusters reconstructed in the EM calorimeters are mainly due to  energy 
deposited by jets. At the LHC the electron-to-jet ratio would be very high, 
$\sim 10^-5$ is expected for electron around a transverse momentum of 40 GeV,
so an excellent rejection is needed in order to select an electron sample.

Some of the rejection is already done at trigger already. At offline level a 
series of additional cuts is applied to bring the rejection at the desired level.

First of all a track is loosely associated to the electromagnetic clusters. If 
such a track is found the cluster is classified as an electron, otherwise as a
photon. However this simple picture is spoiled by converted photons. 

Conversions have to be found using a dedicated algorithm based on tracking. 
Depending on the conversion radius, electrons from conversion might have less
pixel/strip hits than a primary track. That's why a dedicated tracking algorithm 
that builds tracks starting from the outer tracking layers is needed.

Once conversions have been found, an electron is defined as such if the EM has 
a track pointing to it but no associated conversion and a photon cluster has
no track pointing to it or an associated conversion.  

In CMS the track is searched for by starting from the EM supercluster itself.
The energy weighted position of the supercluster is propagated back to the 
interaction point to look for hits in two pixel layers.  This 2-hit track 
provides the seed for the electron track search into the outer tracking layers. 
Careful reconstruction strategies are applied as electrons are affected,
in addition to multiple scattering fluctuations, to non-gaussian fluctuations
due to the emission of brems photons along the electron path.

\begin{figure}[hbt]
  \includegraphics[width=0.3\textwidth,angle=-90]{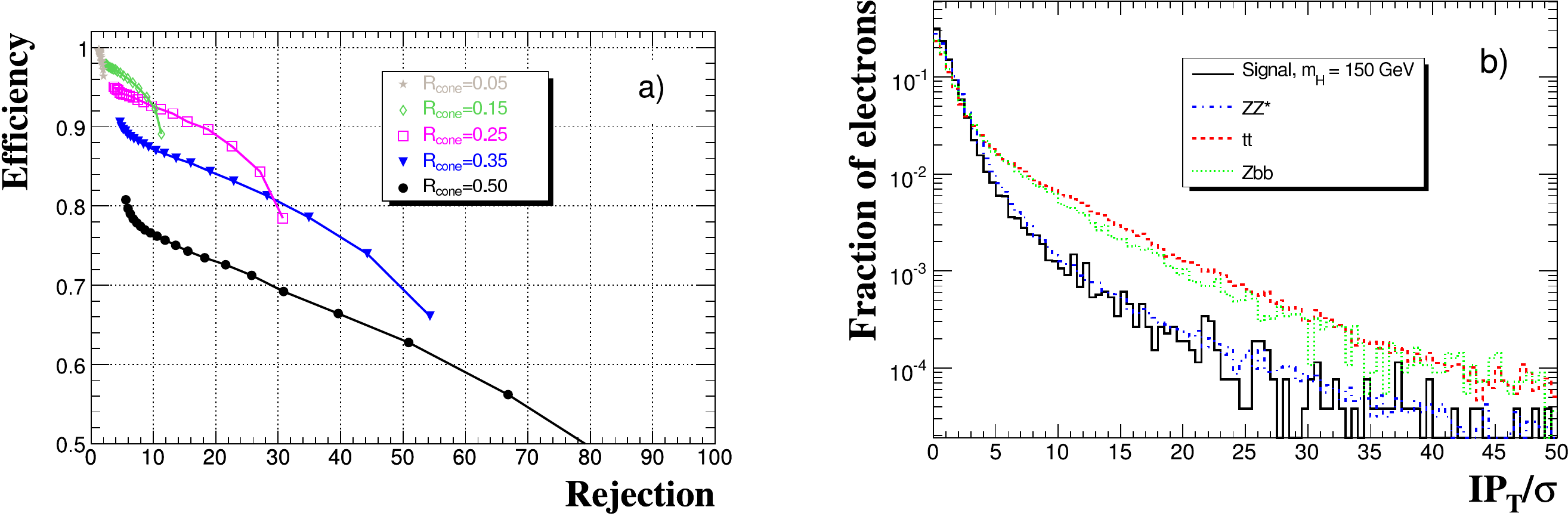}
  \caption{CMS: signal efficiency ($H \rightarrow 4e$) vs. background rejection
(\ttbar\ ) for various isolation cone widths around the electron direction. 
 Impact parameter significance for electrons in $H \rightarrow 4e$ events 
  and in three background samples.}
 \label{fig:CMS_egr}
\end{figure}

Other cuts are based on:
\begin{enumerate}
\item[-] the fraction of energy recorded in the hadronic calorimeter region
just beyond the EM cluster: this should be below a certain threshold to
reduce the jet contamination;
\item[-] variables sensible to the shower shape (lateral and longitudinal
shower shape profile) of the EM cluster in order to select narrow jets
compatible with the showering of a single particle 
\item[-] (for electrons only) variables that combine calorimeter and tracking 
information. Like E/p, the ratio of cluster energy to the track momentum and 
the quality of the $\eta$ and $\phi$ matching of the track with the cluster
barycenter. Additionally, in ATLAS, a cut on the ratio of high threshold hits 
(due to transition radiation generated by the electron track) to low threshold
hits is applied.
\item[-] (for electron only) cuts on the impact parameter,and/or its 
significance~\footnote{The impact parameter (or 3D impact parameter) is defined as the distance
  of closest approach of the track to a given point. The primary
  vertex of the event is considered as the reference point if not
  specified otherwise. The transverse impact parameter is the distance
  of the closest approach to the primary vertex in the plane perpendicular to the beam
  line.}, of the electron track with respect to the
event primary 
vertex (see Fig.~\ref{fig:CMS_egr}).
\item[-] isolation cuts.  
\end{enumerate}

Isolation cuts are one of the strongest way to reduce the jet background. 
Additional particles, and so hadronic energy, near the shower can be detected
either by looking for additional tracks above a certain $p_T$  in a cone (or
annulus) around the reconstructed EM cluster or by looking for additional 
energy into the calorimeter around the EM cluster. Of course, one can also use
a combination of the two. The first method is sensible to charged pions or 
kaons provided they can be efficiently reconstructed down to 1-2 GeV. The
second method is also sensible to neutrals. In both cases the size of 
the isolation cone needs to be carefully tuned. Fig.~\ref{fig:CMS_egr}
shows the rejection power of the isolation cut in $H \rightarrow 4e$
events from the \ttbar\ background. The cut requires no track with
$p_T > 1.5$ GeV/$c$ in an $\eta - \phi$ cone around the electron.  

This set of cuts can be used either in a traditional cut-based way or using 
more sophisticated estimators (multivariate techniques, neural nets,\ldots).
In both cases, as the required efficiency and jet rejections depends on
the physics channel under study, both experiments define electrons of
various classes of {\em quality} according to the tightness of the cuts
applied. 

\begin{figure}[htb]
 \centering
  \includegraphics[width=0.33\textwidth,angle=-90]{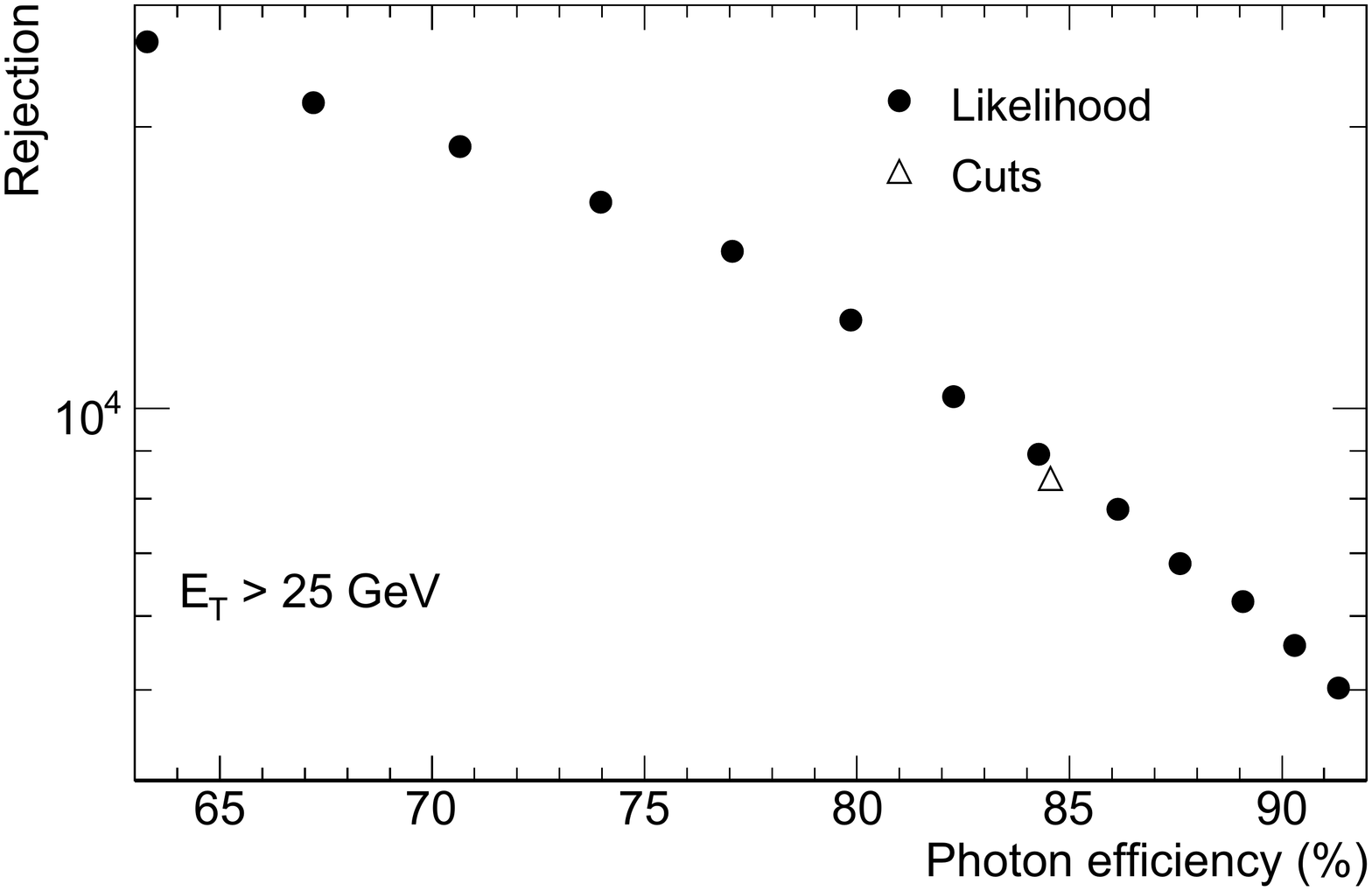}
  \includegraphics[width=0.33\textwidth,angle=-90]{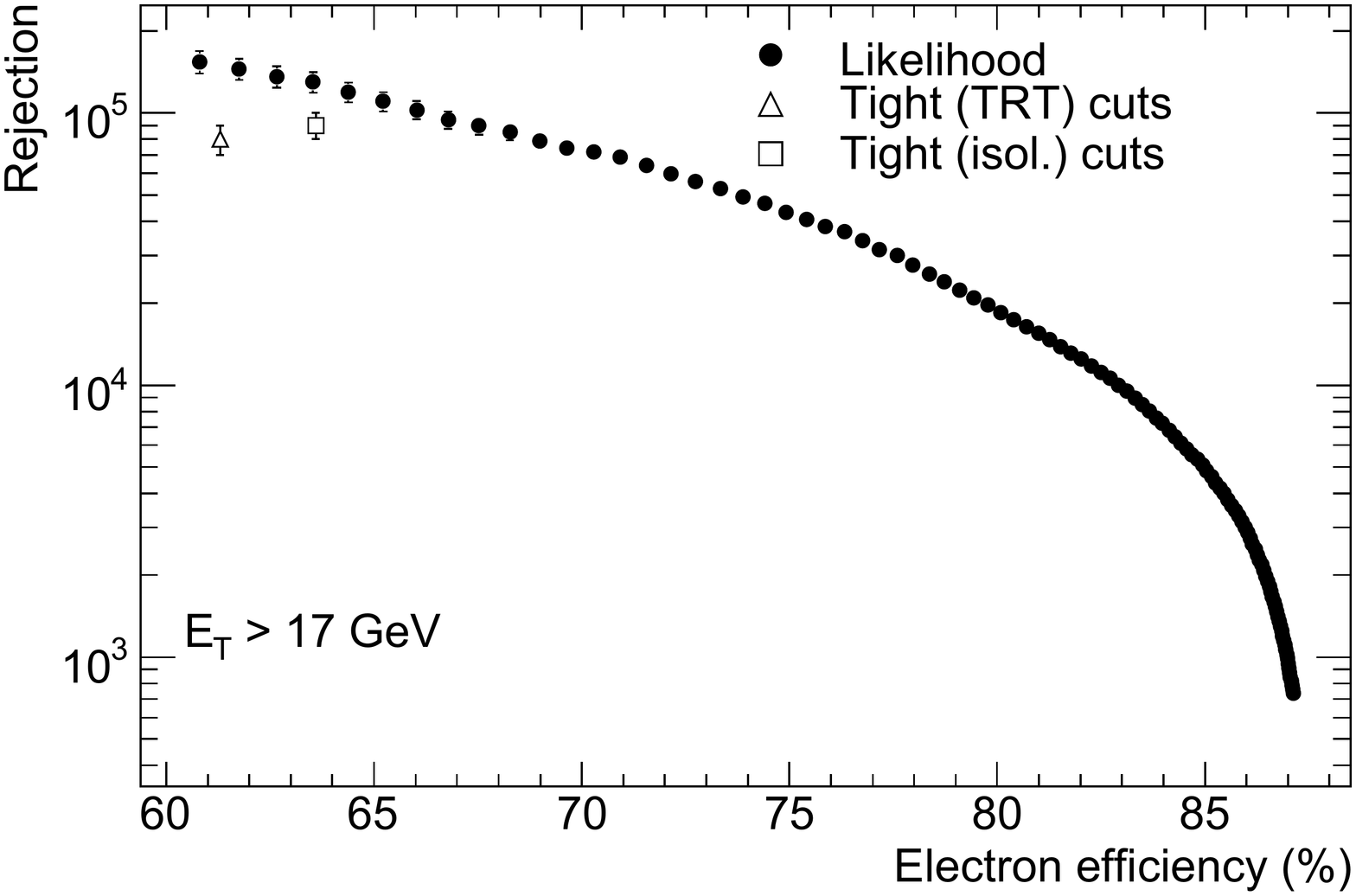}
  \caption{ATLAS: jet rejections versus photons (left) and electrons (right) 
   efficiency using a likelihood method (full points) and a cut-based
   method (open symbols).}
 \label{fig:ATLAS_egr}
\end{figure}

Fig.~\ref{fig:ATLAS_egr} shows the jet rejection vs. photon and electron
efficiency obtained using a likelihood using several of the identification
variable quickly introduced above. For comparison also a few points obtained
using a more traditional cut-based method are shown. For the electrons, two cut
based results are shown depending if a cut based on the transition radiation
detector has been applied to provide a 90\% efficiency for electrons
(``tight (TRT)'') or a 95\% one (``tight (isol.)''). The jet rejections are
computed with respect to truth-particle jets reconstructed with 0.4 wide
cone size on a generic di-jet sample. The jet rejection for photons (around  
9000) is an average over the sample content of quark jets (where the rejection
is about 3000) and gluon jets (rejection 28000). The difference is due to
the softer fragmentation and therefore broader later extent of gluon
jets that facilitate the rejection against photons.

\section{Tau leptons}
{\large {\sl G. Bagliesi}}
\vspace{0.5cm}

The $\tau$ leptons, which are the most difficult leptons to identify,
 are expected to be produced by the decay of several interesting physics channels,
 like Higgs (h/H/A$\to\tau\tau$ and $\rm{H}^\pm\to\tau\nu$), SUSY and other exotic 
particles decays. It has been shown \cite{tau:CMS_PTDR1} that in a large range of
the parameter space, $\tau$ identification is very effective in discarding 
the background, which is mainly due to QCD jets, keeping a good efficiency 
for signal. The most interesting and distinct decays are fully hadronic $\tau$
decays (called $\taujets$). Leptonic $\tau$ decays are usually identified 
through the muon or electron produced. In the following we will concentrate
on the methods developed by ATLAS and CMS to identify $\taujets$ and to 
trigger on them.  
\subsection{Properties of tau leptons at LHC}
\begin{figure}
\centering
\includegraphics[scale=0.40]{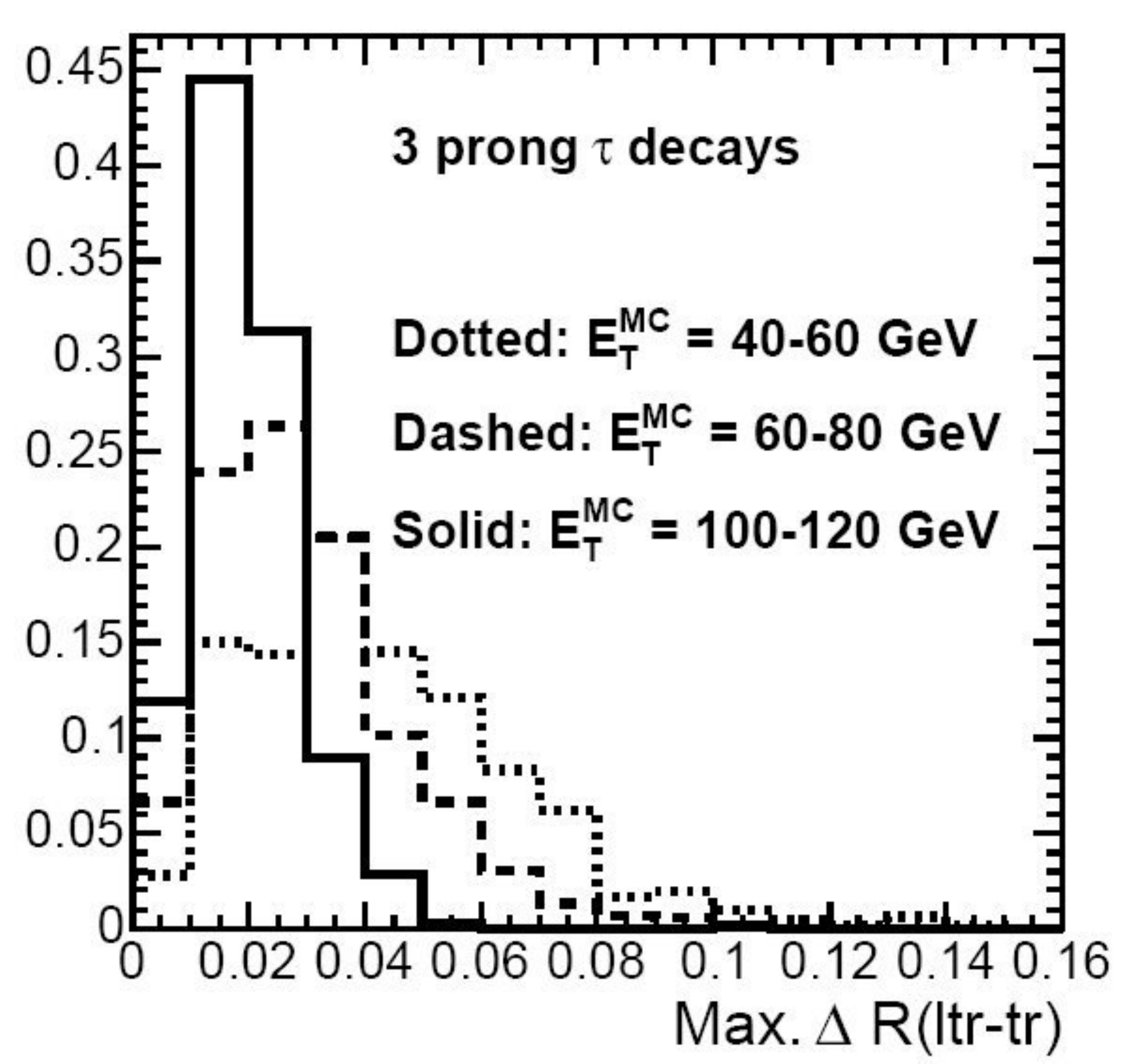}
\caption{Maximal distance $\Delta$R in $\eta-\phi$ space between the 
leading $p_T$ charged particle and the other two charged particles in
the three-prong $\tau$ decay for three intervals of the $\taujet$ 
transverse energy $\rm{E}_T^{MC}$}
\label{tau:fig-dr}
\end{figure}

The $\tau$ lepton decays hadronically with a probability of 65\%, producing a
$\taujet$, containing a small number of charged and neutral hadrons and a 
$\tau$-neutrino. When the momentum of the $\tau$ is large compared to the mass a 
very collimated jet is produced (see Figure~\ref{tau:fig-dr}).

\begin{figure}
\centering
\includegraphics[scale=0.40]{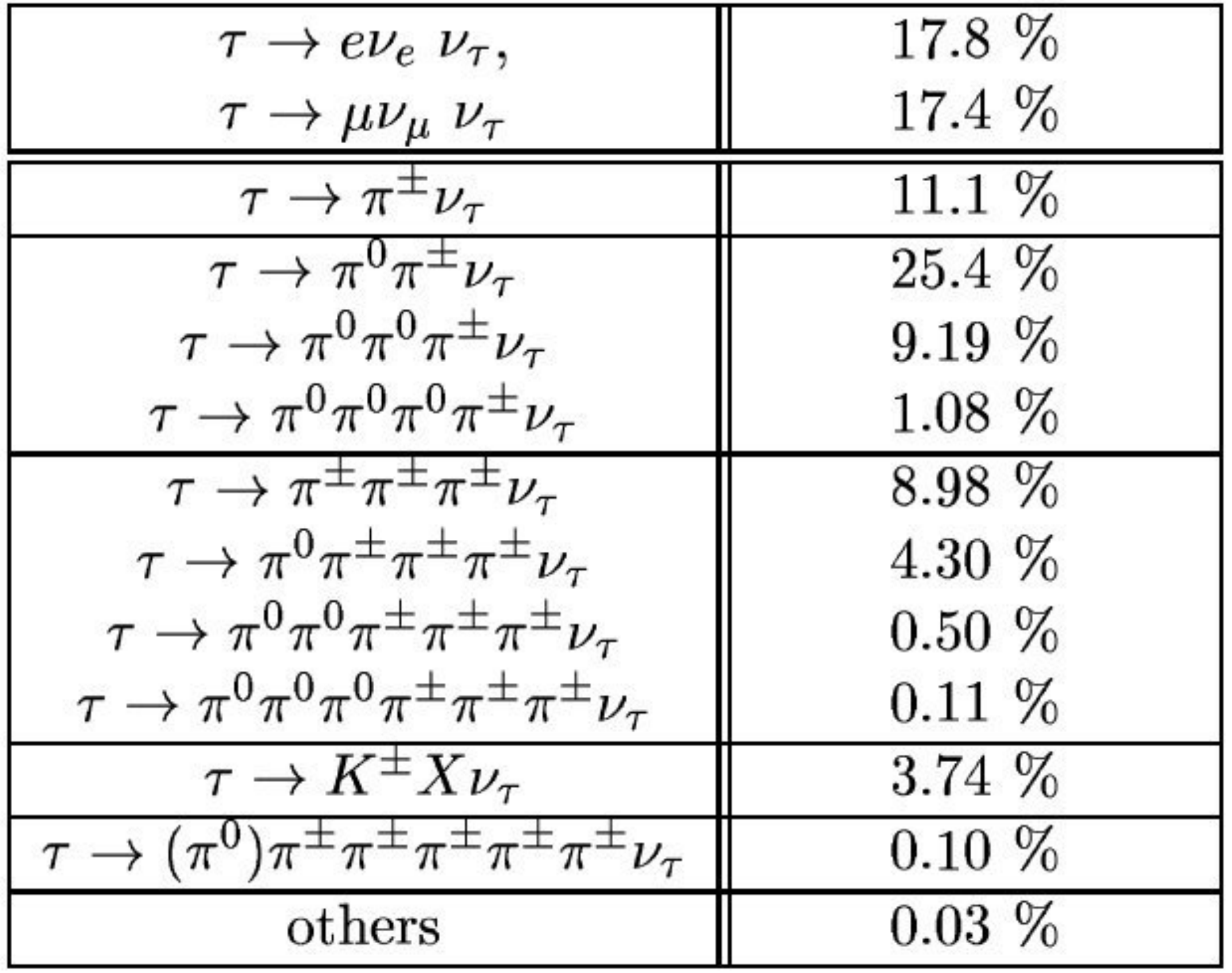}
\caption{Most relevant $\tau$ decay branching ratios}
\label{tau:br}
\end{figure}

For example for a transverse momentum $p_T>$ 50 GeV/$c$, 90\% of the energy 
is contained in a cone of radius $\Delta R=\sqrt{\Delta\eta^2+\Delta\phi^2}=0.2$.
Hadronic $\tau$ decays have low charged track multiplicity (one or three prongs)
and a relevant fraction of electromagnetic energy deposition in the calorimeters
due to photons coming from the decay of neutral pions. In Table~\ref{tau:br} the main $\tau$-decay branching ratios are shown. Quite often taus are produced
in pairs (like the decay h/H/A$\to\tau\tau$): in this case 42\% of the final states
will contain two $\taujets$.
ATLAS and CMS have developed dedicated algorithms for the identification of 
$\taujets$. 

\subsection{Identification of hadronic tau decays: methods based on isolation}

\subsubsection{Calorimetric isolation and shape variables} 
\begin{figure}
\centering
\includegraphics[scale=0.30]{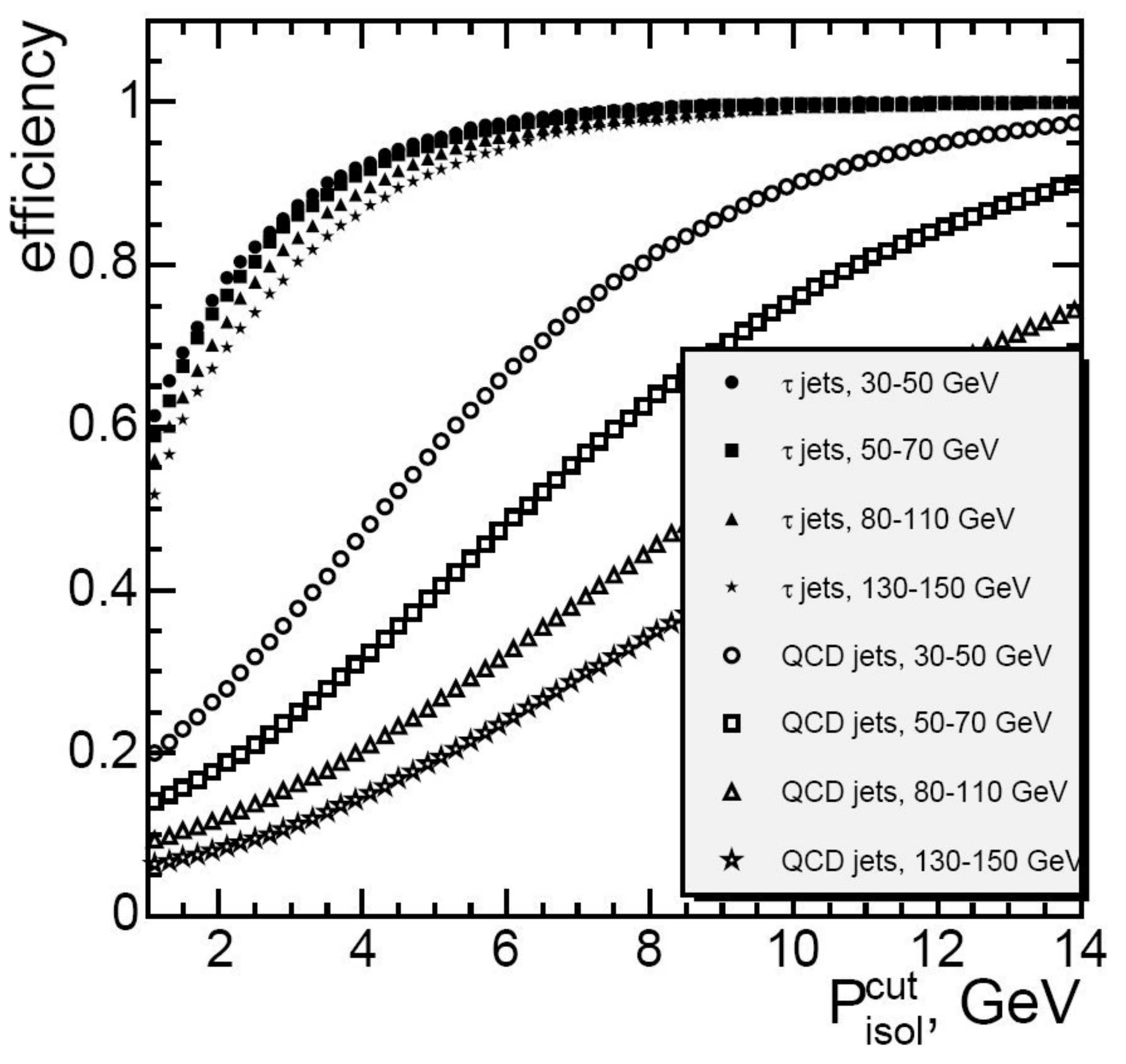}
\caption{The efficiency of the electromagnetic isolation for $\tau$ jets
and QCD jets in several intervals of the true transverse energy
when the value of the cut $P_{isol}$ is varied.}
\label{tau:pisol}
\end{figure}

Hadronic $\tau$ decays 
produce localized energy deposits in the electromagnetic and
hadronic calorimeters. To exploit this characteristic several isolation 
parameters which give a measurement of the energy in a ring around the core of
the jet have been defined: real taus are expected to release only a small
fraction of energy in this ring.
ATLAS defines a variable 
$\Delta{E_T^{12}}=\sum_{j=1}^{n'}E_{Tj}/\sum_{i=1}^nE_{Ti}$
where the sum in the numerator runs over all the calorimeter cells in the 
cluster with $0.1<\Delta{R}<{0.2}$ respect to the jet direction and the sum 
in the denominator runs over all the cells with $\Delta{R}<{0.4}$.
A similar variable is defined by CMS: 
$P_{isol}= \sum_{\Delta{R}<0.40}E_T-\sum_{\Delta{R}<0.13}E_T$. 
In Figure~\ref{tau:pisol} the performance of the $P_{isol}$ cut are shown for
tau and QCD events.

Calorimetric
variables are also very important for $\tau$-ID, in particular the EM
radius (see Ref.~\cite{tau:cavalli}) which exploits the smaller 
transverse profile of
$\tau$ jets. The electromagnetic radius is defined as follows:
\begin{equation}
\label{EQUEMRADIUS}
R_{em} = \frac{\sum_{i=1}^{n} E_{Ti} \sqrt{ \left( \eta_i - \eta_{cluster}
\right)^2 + \left( \phi_i - \phi_{cluster} \right)^2 } }
{\sum_{i=1}^{n}E_{Ti}}
\end{equation}

Where i runs over all electromagnetic calorimeters cells in the cluster
with $\Delta \rm{R} < 0.4$, n denotes their number and  $E_{Ti}$ is the
transverse energy in cell $i$.


\subsubsection{Charged track isolation} 
\begin{figure}
\centering
\includegraphics[scale=0.50]{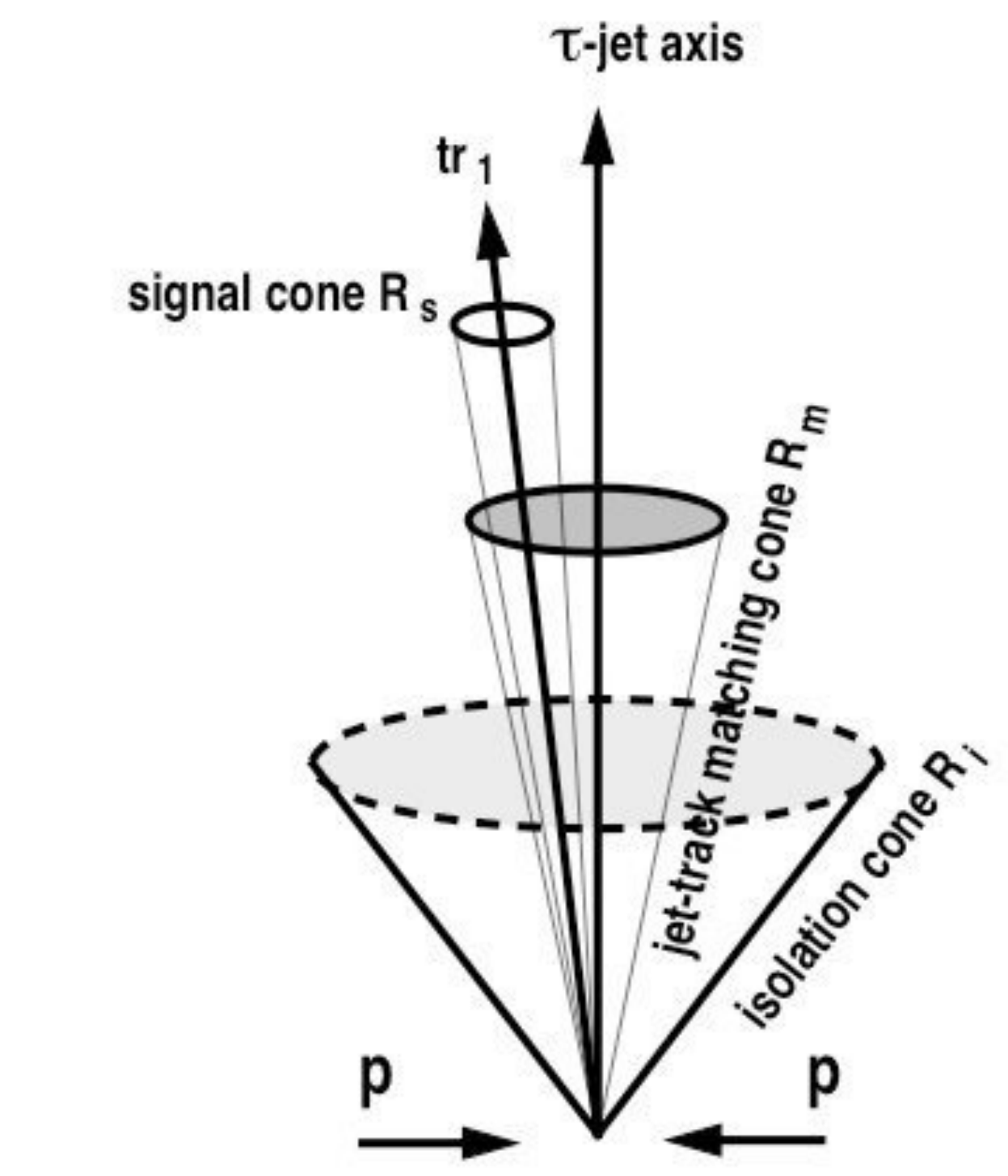}
\caption{Sketch of the basic principle of $\taujet$ identification using
the tracker isolation}
\label{tau:cones}
\end{figure}

The few and collimated charged
tracks contained in a $\taujet$ are the basic 
ingredients of a powerful selection algorithm based on isolation.
The principle is shown in Figure~\ref{tau:cones}. The direction of the $\taujet$
is defined by the axis of the calorimeter jet. The tracks above a threshold of
$p_T^{min}$ and in a matching cone of radius $\rm{R_m}$ around the
calorimeter jet direction are considered in the search for signal
tracks. 
The leading track ($\rm{tr_1}$) is defined as the track with the highest 
$p_T$. Any other track in the narrow signal cone of radius $\rm{R_S}$
around  $\rm{tr_1}$ and with $z$-impact parameter $z_{\rm{tr}}$ close to the
$z$-impact parameter of the leading track is assumed to come from the $\tau$ decay.
Tracks with $\Delta z_{\rm{tr}}$ (impact parameter distance from the leading 
track) smaller than a given cut-off and transverse momentum above a threshold
of $p_{Ti}$ are then reconstructed inside a larger cone of the size 
$\rm{R_i}$. If no tracks are found in the $\rm{R_i}$ cone, except the ones which
 are already in the $\rm{R_S}$ cone, the isolation criterion is fulfilled.

\subsection{Identification of hadronic tau decays: other methods}
In addition to the calorimetric and tracker isolation the $\tau$ lepton has 
other peculiarities which can be used successfully for tagging. 

\subsubsection{Number of tracks} 
Taus decay hadronically 
into one (49.5\%), three (15.2\%) and very seldom 
five (0.1\%) charged particles, often plus several $\pi^o\rm{s}$.
A tagging criterion is to define a track association algorithm which identifies
the tracks belonging to the $\taujet$ (like a cut on impact parameter
along the beam axis $\Delta{z_{imp}}$ or on the transverse 
plane $\Delta r$) and then require that precisely one 
or three tracks are associated. This cut can be either enforced by adding the
condition that the {\it total charge} has to be $\pm{1}$, or loosened by 
asking for a maximum of three tracks (in order to take into account possible
track reconstruction inefficiencies).

\subsubsection{Lifetime} 
The $\tau$ lifetime ($\rm{c\tau=87\mu{m}}$) and relatively low mass 
($m_\tau$=1.78 GeV/$c^2$) produce a sizeable decay length at the energies
of interest for LHC analyses.
 However since the tracks are very collimated
the reconstruction of the decay vertex poses a challenge: a big number of hits
are shared in the vertex detectors, which can lead to a reconstruction of
fake vertices. In the plane transverse to the $\taujet$ axis the resolution
of the reconstructed decay vertex is $\approx$ 20-30 $\mu$m.
In the direction parallel to the $\taujet$ axis the resolution depends on the
jet energy and is comprised in the range 0.5-1.5 mm. A somewhat more 
effective selection 
method is based on the {\it transverse} or {\it 3D impact parameter}
which does not depend at first order on the momentum of the decaying $\tau$ 
(Ref.~\cite{tau:bagliesi}).

\subsubsection{Invariant mass} 
The $\taujet$ mass  is reconstructed from the momentum of the 
tracks in the signal cone and the energy of the clusters in the calorimeter
within a certain cone $\rm{\Delta{R}_{jet}}$ around the calorimeter jet axis.
It is important to avoid double counting of particles by rejecting the 
calorimeter clusters which are matched to a given track. A possible 
un-matching condition could be that the cluster, taken for the mass 
calculation, must be separated from the track impact point on the 
calorimeter surface by a given distance $\rm{\Delta{R}_{tracks}}$.
Typical cuts used by CMS are $\rm{\Delta{R}_{jet}<0.4}$ and 
$\rm{\Delta{R}_{track}>0.08}$. More sophisticated algorithms based on the 
{\it particle flow} reconstruction are under study.
  
\subsection{ATLAS specific selection}
ATLAS has developed two independent algorithms for $\taujet$ selection: 
TauRec (Ref.~\cite{tau:cavalli}) and Tau1P3P (Ref.~\cite{tau:richter}).
The former is a general purpose algorithm based on calorimeters and inner detector
information, the latter is intended for studies of low mass Higgs decays.

\subsubsection{tauRec} This algorithm uses calorimeter clusters as $\taujet$ candidates.
These are provided by a sliding window cluster algorithm which runs on 
``CaloTowers'' which are the sum of all calorimeter layers summed up on a grid
of $\rm{\Delta{\eta}\times\Delta{\phi}=0.1\times{2\pi/64}}$. 
The $\taujets$ are identified
by looking at the following quantities: isolation, number of associated 
charged 
tracks with $p_T>$2 GeV/$c$ and with a distance $\rm{\Delta{R}<0.3}$ from
the barycenter of the cluster, charge, 2D signed impact parameter significance,
and other cluster shaping cuts in the electromagnetic calorimeter. 
A likelihood function is built with all the previous variables
including the ratio $\rm{E_T}/p_T$ of the leading track. A rejection against
QCD jets of $\rm{10^3-10^4}$ is obtained (depending on the jet energy) with an 
efficiency of about 40\% for $\taujets$.

\subsubsection{Tau1P3P} The Tau1P3P algorithm is specialized for low mass
Higgs ($m_H\approx$ 120 GeV/$c^2$), with visible energy from hadronic tau decays
in the range 20-50~GeV. One or three charged tracks are required plus associated
energy deposit in the calorimeter (from $\rm{\pi^{\pm}}$) and additional
electromagnetic energy from the accompanying $\rm{\pi^0s}$. The search for
calorimetric energy is seeded by the direction of the leading track. 
The main steps of the Tau1P3P algorithm are: look for a ``good'' hadronic 
leading track ($p_T>$9 GeV/$c$), zero or two nearby tracks with 
$p_T>$2 GeV/$c$, $\rm{\Delta{R}}$ (track-direction,jet-direction)$<$0.2,
 isolation around the $\taujet$ direction. Calorimeter cluster are
classified in {\it neutral electromagnetic, charged electromagnetic} and
{\it others} type with a simplified energy flow method. Several additional
discriminant variables are calculated by making use of the tracks and of the
clusters belonging to the $\taujet$. 
After
optimizing the cuts a selection efficiency of about 40 \% for $\tau$ jets
with a jet rejection of $102-103$ for $p_T < 50$ GeV is observed.
Alternative selections
have been developed based on multivariate analysis, which give 
somehow better results.

\subsection{CMS specific selection}
CMS selection is based on the calorimeter and tracker isolation described
previously. Referring to Figure~\ref{tau:cones} the optimization of the working
point of the tracker isolation algorithm is done by making a scan on the value 
of the isolation cone $\rm{R_i}$, with the value of $\rm{R_S}$ 
and $\rm{R_m}$ kept fixed. It is possible to reach good values of 
background rejection ($\rm{\epsilon(QCD jets)\approx{4-6\%}}$) with an 
efficiency for $\taujets$ of  $\rm{\approx{70\%}}$. The actual signal 
efficiency depends on the particular physics process considered.
A number of other selection methods (impact parameter, flight path, 
mass reconstruction) which can be applied after the isolation have been 
studied by the CMS collaboration. Most of these additional cuts have been
already described previously. Depending on the specific channel studied, the
application of these additional cuts can improve the overall signal/background
ratio. 

The CMS collaboration is optimizing all tau identification algorithms by making use of the particle flow approach. 
Better overall performance are expected 
since a particle flow algorithm can improve the reconstruction of the 
charged tracks and of the calorimetric deposit associated to a $\taujet$.

\subsection{Tau identification at trigger level}
The First Level Trigger ({\em L1}) for LHC experiments is implemented 
on custom hardware which performs a rapid decision based on calorimeters and muon chambers 
information.  
ATLAS has an intermediate level of trigger ({\em L2}) which is applied to the
region of interest pointed by the L1, followed by a High Level Trigger (HLT) 
selection. CMS instead implements a one-step HLT selection just after the 
L1 trigger.
The L1 selection for $\taujets$ starts by looking for 
collimated and isolated calorimetric jets. 
Given the huge QCD background cross-section, the goal of the HLT
is to reduce the rate of QCD events of a factor 
$\rm{\approx{10^{-3}}}$ after the L1 trigger in order 
to select a final rate of O(10~Hz) events containing one (or two) 
$\taujets$ candidates (Ref.~\cite{tau:CMS_PTDR1}). 

ATLAS and CMS HLT selection
is generally based on algorithms very similar to those applied for the off-line
selection. See for example Ref.~\cite{tau:cms_trigger} for a detailed and recent
study of CMS HLT trigger performance.

\section{The jets}
{\large {\sl I. Vivarelli}}
\vspace{0.5cm}

Many different requirements have to be satisfied in order to perform jet
measurements. The comparison with the theoretical predictions in a given
channel forces the experimentalists to use a reconstruction algorithm free
from collinear and infrared unsafetiness. The jet energy measurement is a
delicate issue as well. The first step consists in the removal of the detector
effects, i.e. nonlinearities in the measurement due to the
non-compensation~\footnote{A calorimeter is said to be {\it
    compensating} if it gives the same signal response for the
  for the electromagnetic component ($e$) and non electromagnetic ($h$) and  of a hadronic
  shower.} of the calorimeters, calorimeter cracks, etc. The
second step aims to correct for hadronization and thus obtain a measurement
which is directly comparable to the theoretical predictions. 
Detailed studies show that, in the measurement of the top quark mass
and of the inclusive jet cross section, the systematic error related to
the knowledge of the jet energy scale is the dominant term if the jet energy
scale is not known at a level better than 1-2\%.
In the following we will review the main experimental issues related to jets,
starting from the cluster reconstruction in the calorimeters to arrive to the
parton jet measurement.

\subsection{Clusterization and Jet Reconstruction} 

Before running the jet reconstruction algorithm, the calorimeter
cells are clusterized together. A very simple and fast clusterization is obtained building
calorimeter towers: all the cells lying in a square in the $\eta-\phi$
space are summed together in a tower. The tower size in 
ATLAS is $\Delta \eta \times \Delta \phi = 0.1 \times 0.1$ while CMS has $0.087 \times
0.087$. ATLAS is planning to make use of a more sophisticated, 3D,
clusterization algorithm, which takes into account both the lateral and
longitudinal development of the shower.

Once the clusters are built, the jet reconstruction algorithm groups them
together following its prescriptions. The most used reconstruction algorithm
so far is the cone algorithm~\cite{cone}:

\begin{itemize} 
\item[-]  A cone of radius $R$ (in the $\eta-\phi$ space) is built around the seed (trial seed, in the case of the seedless algorithm). 
\item[-] For each cluster (tower) $k$, with center ($\eta^k,\phi^k$),
  the center of the cone $\vec C_k = (\eta^{Ck}=\eta^k,\phi^{Ck} =
  \phi^k)$ is defined. A cluster (tower) $i$ is included in the cone
  if $\sqrt{(\eta^i - \eta^{Ck})^2 + (\phi^i - \phi^{Ck})^2} \le R$.  
\item[-] Then, the $E_T$--weighted centroid is evaluated
  $\vec{\overline C^k}=(\overline \eta^{Ck}, \overline \phi^{Ck})$ with:
\[
\overline \eta^{Ck} = \frac{\sum_{i \subset C^k} E_{Ti}\eta^i}{E_T^{Ck}} \qquad \overline \phi^{Ck} = \frac{\sum_{i \subset C^k} E_{Ti}\phi^i}{E_T^{Ck}} 
\]
where ${E_T^{Ck}} = \sum_{i \subset C^k} E_{Ti}$
\item[-] In general the centroid $\vec {\overline C^k}$ is not
  identical to the geometric center $\vec C_k$ and the cone is not
  stable. Therefore, an iterating procedure is needed until the cone
  found is stable. 
\item[-] The described procedure can lead to a final jet list where
  some of the jets overlap. A {\em split and merge} procedure has to
  be used to merge or separate jets which overlaps, in order to avoid
  the assignment of some particles to two jets. The way to deal with this,
  is to merge two jets if the overlapping energy percentage is above
  some threshold.  
\end{itemize}

The $K_T$ algorithm is implemented also~\cite{kt}:
\begin{itemize}
\item[-] For each cluster (tower) compute $d_i = E^2_{Ti}$. For each pair $i$,$j$ define 
\begin{equation}
d_{ij} = {\rm min}(E^2_{Ti},E^2_{Tj}) \frac{(\eta_i - \eta_j)^2 + (\phi_i - \phi_j)^2}{D^2}
\end{equation}
where $D$ is a resolution parameter (the current choice in ATLAS is $D=1$).
\item[-] Find $d_{min}=\min(d_i,d_{ij})$.
\item[-] If $d_{min} = d_{ij}$ for some $j$, merge tower $i$ and $j$ to a new tower $k$ with momentum $p^{\mu}_k = p^{\mu}_j + p^{\mu}_j$.
\item[-] If $d_{min} = d_{i}$ then a jet is found. 
\item[-] Iterate until the list of tower is empty.
\end{itemize}

\begin{figure}[hbtp]
  \begin{center}
    \includegraphics[width=0.55\textwidth]{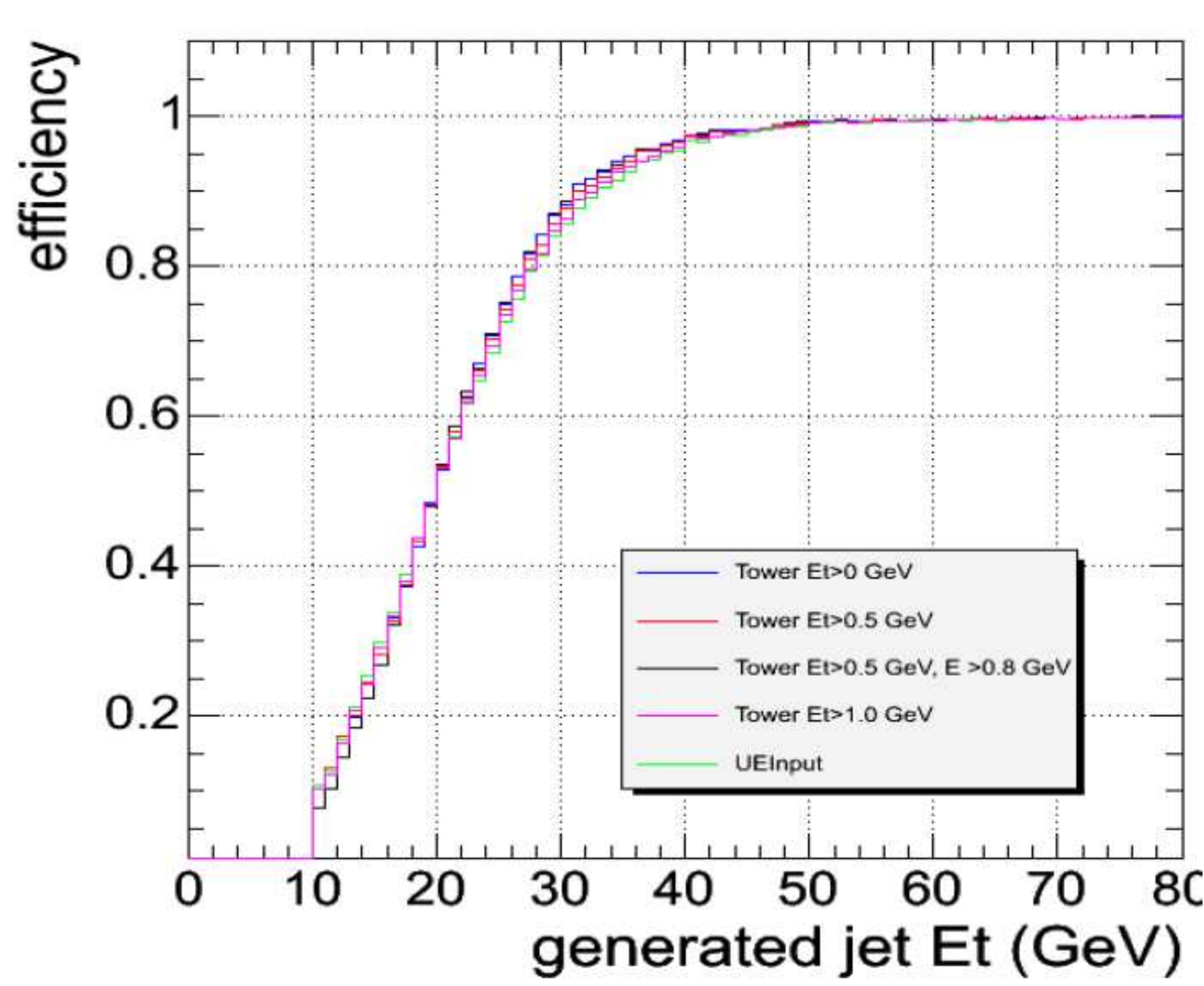}
    \caption{Jet reconstruction efficiency for the cone algorithm for
    different energy thresholds on the tower seeds (CMS).}
    \label{fig:rec_eff}       
  \end{center}
\end{figure}
The reconstruction efficiency of the different reconstruction algorithms is
estimated from the simulation. Typically, a matching procedure is defined to
associate jets reconstructed from the calorimeters to jets reconstructed from
the Monte Carlo final state particles (particle jets). A particle jet is
defined as reconstructed if there is a calorimeter jet within a given angular
distance. Figure~\ref{fig:rec_eff} shows the reconstruction efficiency as a
function of the particle jet $E_T$ for the seeded cone algorithm (different
colors are used for different energy thresholds on the tower seeds) in
CMS~\cite{CMS_jet}. A 90\% efficiency is obtained for $E_T \sim 30$ GeV.

\subsection{Calibration at the Particle Jet}

Jet energy measurements need to be corrected for non-uniformities and
non-linearities introduced by the detector itself. As well
known~\cite{wigmans}, if a calorimeter is non--compensating (as the
ATLAS and CMS 
ones are) the response to hadrons is lower with respect to electrons and
photons of the same energy. Moreover, its dependence on the impinging particle
energy is
non--linear. Finally, the structure of the calorimeters (gaps, cracks,
different technologies in different pseudorapidity regions) makes the response
also pseudorapidity dependent.

The present jet calibration in ATLAS is obtained from full simulated QCD
events. Calibration coefficients ${w_i}$ depending on the cell energy density are
extracted comparing the reconstructed energy of the jet with the energy of the
particle jet. For the same cell energy density, a different weight
is applied for different longitudinal samples and in the different sections of
the ATLAS calorimeters. Each reconstructed jet is associated with the
closest (in $\Delta R = \sqrt{\Delta\eta^2 + \Delta \phi^2}$) particle
jet. Once this association is done, the calibration coefficients can
be extracted minimizing a $\chi^2$: 
\begin{equation}
\chi^2 = \sum_e \frac{(E^{rec}_e - E^{true}_e)^2}{(E_e^{true})^2}
\end{equation}
\noindent The index $e$ runs on all the jets of all the considered events and $E^{rec}_e$ is defined as:
\begin{equation}
E^{rec}_e=\sum_i w_i\left(\frac{E_{ie}}{V_i}\right) E_{ie}
\end{equation}
 \noindent where $i$ is running on all the cells belonging to the jet,
 $E_{ie}$ is the energy deposit in the $i$-th cell for the jet $e$ and
 $V_i$ is the volume of the $i$-th cell. 

\begin{figure}[hbtp]
  \begin{center}
    \includegraphics[width=0.7\textwidth]{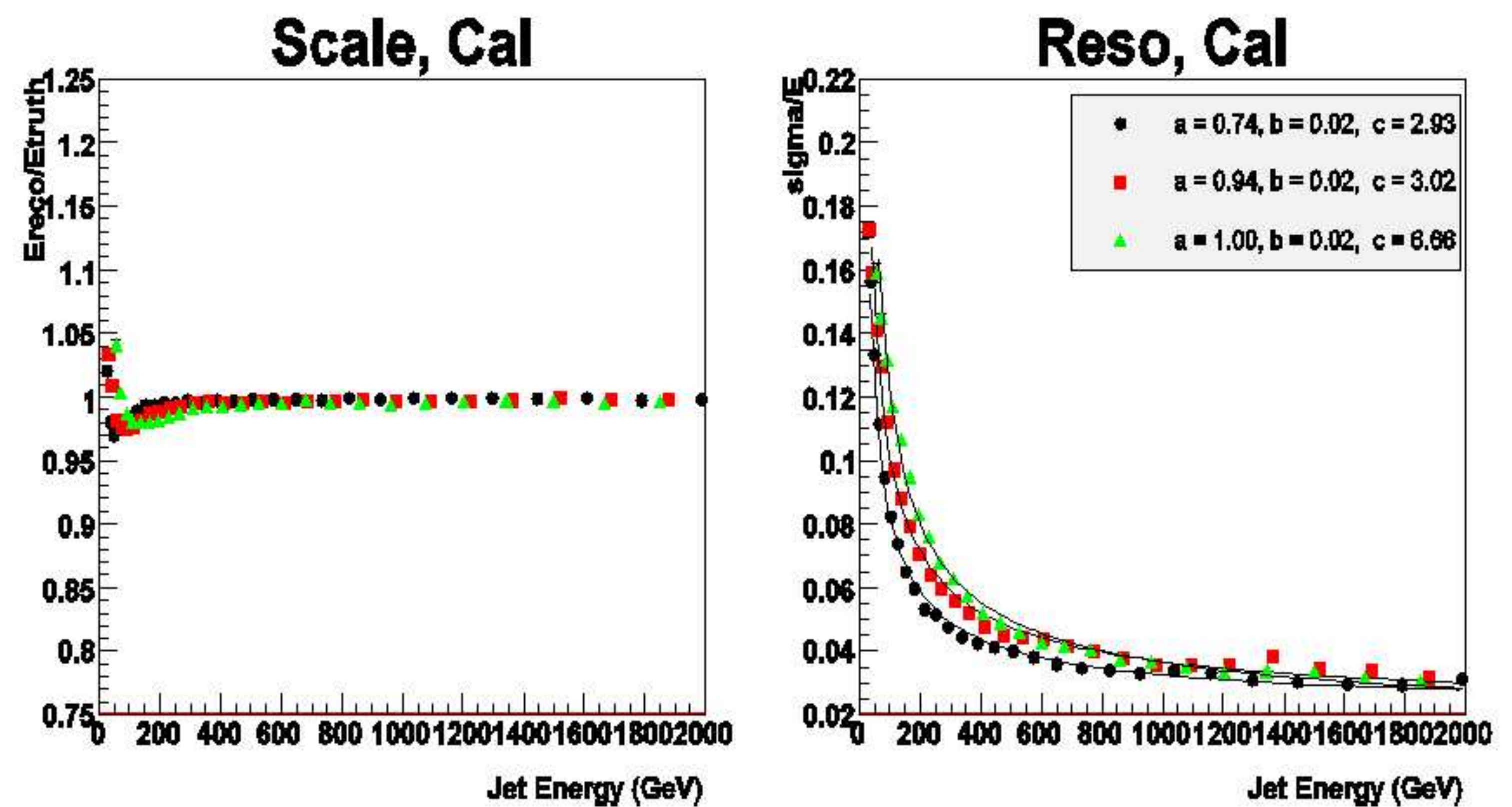}
    \caption{Linearity (left) and resolution (right) of jets, with respect to the particle jets,
      in ATLAS. The three curves refer to the central ($|\eta| < 0.7$,
      in black circles),
      intermediate ($0.7< |\eta| < 2.5$, in red square) and forward ($2.5< |\eta| < 3.5$,
      in green triangles) regions. 
      The energy resolution in the right plot is parametrized as
      $\frac{\sigma_E}{E}=\frac{a}{\sqrt{E}} \oplus b \oplus
      \frac{c}{E}$ with $E$ being the energy of the jet in GeV.}
    \label{fig:lin_res}       
  \end{center}
\end{figure}

In order to reduce the number of calibration coefficients to
calculate, the dependence of $w_i$ on the cell energy density is
parameterized with a polynomial function of $\log(E_i/V)$: 
\begin{equation}
w_i = a+b \log\frac{E_i}{V} +c \large( \log\frac{E_i}{V} \large)^2+d \large( \log\frac{E_i}{V} \large)^3
\end{equation}

Figure~\ref{fig:lin_res} shows the obtained linearity and resolution after the calibration procedure.

The CMS correction is instead obtained considering the ratio $r_{jet}$ between
the reconstructed jet transverse energy and that of the associated particle
jet ($r_{jet} = E_T^{rec}/E_T^{true}$). The corrections are computed
as a function of $\eta$ and $E_T$. Then, the reconstructed energy
is corrected by a factor $1/r_{jet}$. 

Both the approaches assume that the calorimeter response to jets is well
reproduced by the simulation of the detector. This has been verified in many
years of test beams. The agreement of the GEANT4 simulation with the test beam
data is within 2\% for both the experiments. 

\subsection{In Situ Calibration}

The {\em in situ} calibration will be performed using both the $E_T$ balance between a jet
and a vector boson (either $\gamma$ or $Z$) recoiling against that and the $W$
mass reconstruction in top decays. Focusing on the former, it will be used for two main purposes:
\begin{itemize}
\item[-]
  Verify that the unbalance in the data is well reproduce by the simulation
  (a validation of the particle jet calibration).
\item[-]
  Perform the so-called  {\it Parton Level Calibration} of the jet
  energy (see Section~\ref{mcws:jets} for a more detailed discussion).
\end{itemize}

As an example, we consider $\gamma + jet$ events in CMS~\cite{CMS_gamma}. Taking into account
trigger efficiencies and considering an integrated luminosity of $10$
fb$^{-1}$, CMS expects to have a statistical accuracy below 1\% up to about
100 GeV of photon transverse energy. A factor $K_{jet}(E_T^{\gamma})$ is defined as the
ratio of the mean reconstructed jet transverse energy with the mean photon
transverse energy in a given bin of the photon spectrum. This can be compared
with $K_{jet}^{true}$, defined as the ratio between reconstructed jet
transverse energy and the parton jet transverse energy:
\begin{eqnarray}
  \delta K = \frac{K_{jet} - K_{jet}^{true}}{K_{jet}^{true}}
\end{eqnarray}

\begin{figure}
\begin{center}
\resizebox{0.9\textwidth}{!}{%
  \includegraphics{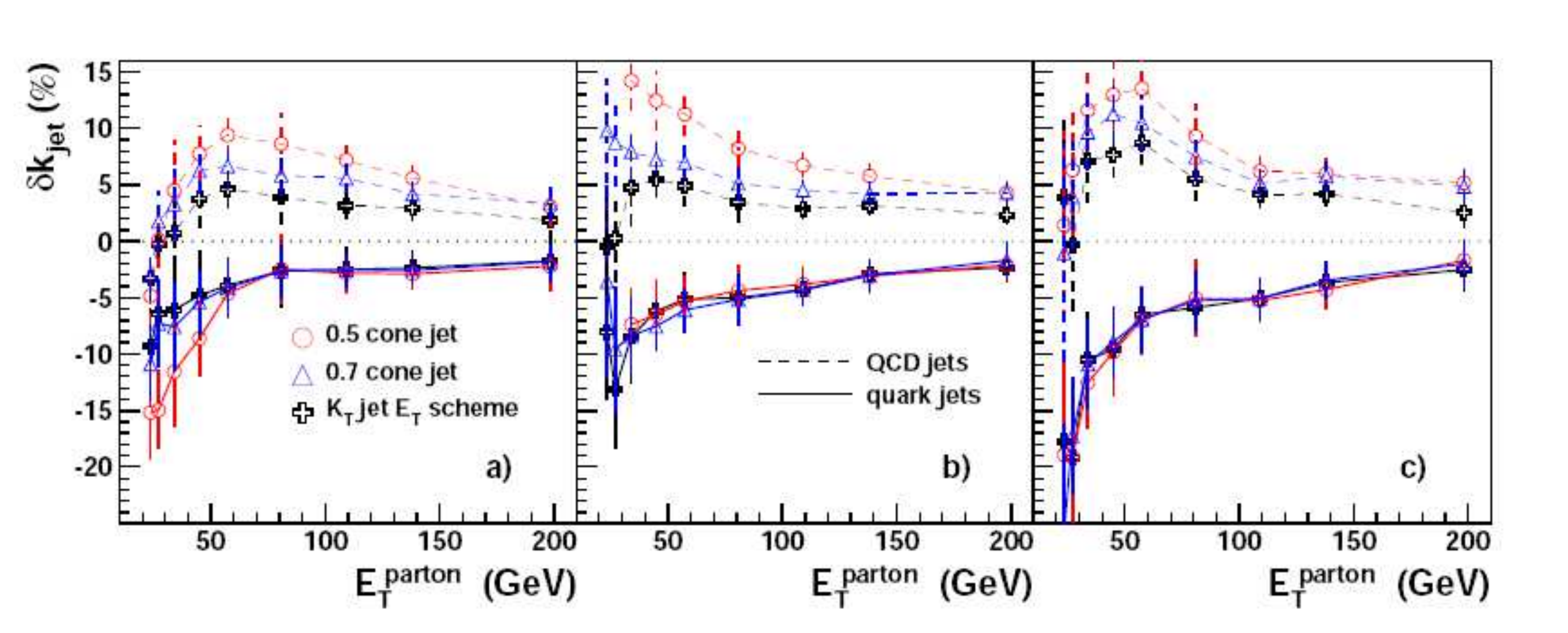}
}
\end{center}
\caption{$\delta K$ (described in the text) for the cone (two cone sizes) and
  the $K_T$ algorithms, for gluon and quark jets. From the left to the right,
  the plots are done for 0.5, 1 and 1.5 GeV tower seeds.}
\label{fig:CMS_bias}       
\end{figure}

Figure~\ref{fig:CMS_bias} shows $\delta K$ as a function of the parton jet
energy for different reconstruction algorithms and different originating
partons for three different energy thresholds on the tower seeds. As
can be seen, there are differences up to 
$\sim 10\%$ due to biases of the event selection and
non-leading radiation effects.    

%

\section{The missing transverse energy}
{\large{\sl F. Tartarelli}}
\vspace{0.5cm}

The presence of one or more energetic neutrinos or other weakly-interacting 
stable particles
is an important signature for several standard model and beyond the standard 
model physics
processes. Neutrinos appear in the leptonic decays of W's, decays of
Z's, in the semileptonic decays of heavy quarks and in the decays of
$\tau$'s. 
Weakly-interacting stable particles appear in SUSY
models and (if massive) can be candidate for the dark matter.

Multi-purpose collider experiments like ATLAS and CMS cannot detect directly these kind of 
particles. Due to the importance of these studies, since long time a technique has 
been used in order to infer indirectly the presence of these particles. 
If such a particle is produced in the collision, it will give an apparent imbalance in the 
total energy and momentum (so-called {\em missing energy} or {\em missing momentum}). 
In order to measure such imbalance, the detector need to be able
to measure the energy of {\em all} particles produced in the collisions. 
There are several 
instrumental effects that in practice limit this possibility: we will see a few examples in the
following. The main limitation, however, is due to fact that while detectors have usually a 
full azimuthal coverage (around the beam direction), the pseudorapidity coverage (along the
beam directions) is limited by the impossibility to instrument the region close to the beam
line. In ATLAS (CMS) the calorimetry coverage extends up to 4.9 (5.0) using dedicated "forward
calorimeters" that cover the higher pseudorapidity region, $3.1 < |\eta| < 4.9$ ($3 <|\eta| < 5$).
These are challenging detectors that have to operate in the extremely hostile radiation 
environment of such high pseudorapidities where the calorimeter performance 
deteriorates quickly. Indeed, the main physics motivation for forward calorimeters is just the 
extension of the detector coverage for missing energy measurement (together with the tagging of 
forward jets).

Nevertheless some particles produced in the collisions will escape undetected down the beam pipe so
that the momentum balance along the direction of the beam cannot be evaluated. However the 
transverse momentum of particles produced in the collision can be measured with enough precision
(the transverse momentum of particles escaping along the beam direction is small) so that the transverse 
energy balance (since the name {\em missing transverse energy} or {\em missing transverse momentum})
can be measured with an accuracy good enough to help establish the presence of 
one or more non-interacting particles. 

The missing transverse energy is defined as the vector sum of the 
energy deposits in the calorimeter towers (or cells):
\begin{equation}
 \vec{E}_T^{miss} = 
 \sum_n (E_n \sin{\theta_n} \cos{\phi_n} \hat{\imath} + 
          E_n \sin{\theta_n} \sin{\phi_n}\hat{\jmath})
        = (E_x^{miss}\hat{\imath} + 
           E_y^{miss}\hat{\jmath})  
  \label{etmiss}
\end{equation}

where $E_n$ is the energy in the calorimeter tower and $\theta_n$ 
and $\phi_n$ are the tower polar and 
azimuthal angle, respectively. In ATLAS, for example, best results have
been obtained always working at the level of calorimeter cells rather
than towers.

If for example a neutrino is present in the event, since the \vmet\ is a vector, it provides both the  
energy and the transverse direction of the escaping neutrinos. If two neutrinos are expected in the event, in
several cases the missing energy can still provide information on the kinematic of the event.

\subsection{Experimental issues}

Several experimental issues are involved in the calculation of the $E_T^{miss}$. We will briefly discuss noise suppression, muon
correction and cell energy calibration. 

\subsubsection{Noise suppression}
The sum in Eq.~\ref{etmiss} is extended to cells above a certain threshold. As the LHC calorimeters
have a large number of cells (i.e. ATLAS has about 2$\times$10$^5$ calorimeter cells) the 
contribution of noise can
quickly become significant if it is not kept under control. A good calorimeter noise suppression 
algorithm is needed.
The goal would be to
include in the sum the clustered (jets, photons, electrons) and unclustered energy deposits in the
calorimeter avoiding contributions due to noise. Electronic noise and pile-up noise are the sources of
noise in the calorimeter. Concerning the first contribution, the so-called {\em coherent} noise is
particularly dangerous and should be avoided by 
careful design (grounding and shielding) as it produces correlated
noise in a large number of calorimeters cells.

Several techniques are possible. Most of them in the end require the knowledge for each cell of $\sigma_{noise}$,
the quadratic sum of the estimated electronic and pile-up noises, and
allow cells with energy $|E_i| > n \sigma_{noise}$, where $n$ is an
appropriate cut.

\subsubsection{Muon correction}
If  muons are identified in the event a correction has to be implemented. In both experiments
to reduce the fake muon background, a muon is generally identified when there is a match between a muon
stub identified in the muon chambers and a track identified in the tracking system. The energy deposit
in the calorimeter cells crossed by the extrapolated muon track should be compatible with that of a MIP.

CMS adds to the sum in Eq.~\ref{etmiss} the muon transverse momentum as measured by the tracker and,
to avoid double counting, removes from the sum the tower crossed by the muon track. 

The ATLAS muon system
can provide a stand-alone (i.e., without using the tracker) measurement of the muon momentum: that's why 
the current ATLAS strategy is to leave in the sum of 
Eq.~\ref{etmiss} the muon energy deposit and
add the muon momentum as measured by the muon spectrometer only (provided
it is matched with a track reconstructed in the tracking detector). 
With this recipe, no double counting is  done.    

\subsubsection{Cell energy calibration}
Towers entering the missing transverse energy need to be calibrated. There 
are several issues connected to
the $E_T^{miss}$ calibration. Cell calibration means to recover the optimal 
calibration for physics object measured in a calorimeter (electrons,
photons, jets). It is obtained by correcting the cell energy for factors 
that depend on the particle type 
and that have been obtained by simulation or beam tests. As the $E_T^{miss}$ 
is an inclusive quantity also
the energy belonging to unclustered towers (not belonging to any identified 
physics object) need to be
calibrated. The cells can belong either to the electromagnetic or to
the hadron calorimeter and one 
has to take into account that in both experiments these are non compensating.

In ATLAS, the calibration procedure follows a multi-step strategy:
\begin{enumerate}
\item[-] Cells are calibrated using weights determined using a technique similar
to the one developed for the calorimeter of the H1 experiment at the HERA
collider. In this method each cell is corrected with a weight that depends
on the cell energy density ($\rho = E/V$ where E is the cell energy and V is 
the volume of the cell), on the cell pseudorapidity position and on the 
calorimeter module and compartment. The weights have been obtained using 
jets from QCD dijet events covering the whole kinematic range expected at 
the LHC, calibrating the reconstructed energy to the {\em truth} particle
energy. This procedure corrects for detector effects like: missing signals
from charged particle bent away from the calorimeter due to the tracking
magnetic field, energy losses in inactive materials, noise, non-compensation 
of the calorimeters, etc.).

\item[-] A dedicated correction is applied to recover the energy lost in the
inactive material (cryostat walls) between the electromagnetic and the
hadronic calorimeters

\item[-] As described elsewhere in this paper, electrons, muons and other physics
objects are accurately calibrated using dedicated procedures based on 
simulations and test beam data. It is possible to benefit from this work also
to improve the generic calibration procedure described above. 
To do this, cells belonging to reconstructed physics object (as there could 
be ambiguities, a well defined order has been chosen: electrons, photons,
hadronically decaying $\tau$ leptons, $b$-jets, light jets and muons) are
removed from the sum in Eq.~\ref{etmiss} and their total contribution 
replaced by the contribution from the calibrated physics object itself.     
\end{enumerate}

All ATLAS results presented in next section have been obtained using the 
calibration procedure just described.

CMS results have been obtained in a much less sophisticated way using 
photon calibration for cells belonging to the electromagnetic calorimeter
and hadron calibration for cells belonging to the hadronic calorimeter.
Studies are ongoing to improve the \met\ calibration using charged track
corrections and energy flow techniques. In the first case, for example, 
tracks (and their momenta) reconstructed in the tracker are used to correct
for tracks swept out by the magnetic field and to replace calorimeter
deposits by the more accurate tracker momentum measurement.

\subsection{Performance}
The \met\ performance of the detector is evaluated in term of: resolution,
linearity of response, direction resolution (in the transverse x and y
coordinates) and tails. A review of the main ATLAS and CMS results is presented
in this section. All the results are taken 
from~\cite{ATLAStdr} and~\cite{CMStdr}. 

It has since long observed that the \met\ resolution depends on the overall 
event activity that can be
characterized by the scalar sum of the transverse energy in all calorimeter 
cells (\sumet).  The \met\ resolution 
(in x or y) follows the simple stochastic 
law $\sigma(\met\ _{x,y}) = k \sqrt{\sumet}$. Deviations are observed at
both ends of the \met\ spectrum. For low \sumet\ values noise
becomes an important contribution while at very high \sumet\ values the
jet energy resolution constant term dominates.

\begin{figure}[hbtp]
 \centering
    \includegraphics[width=0.45\textwidth]{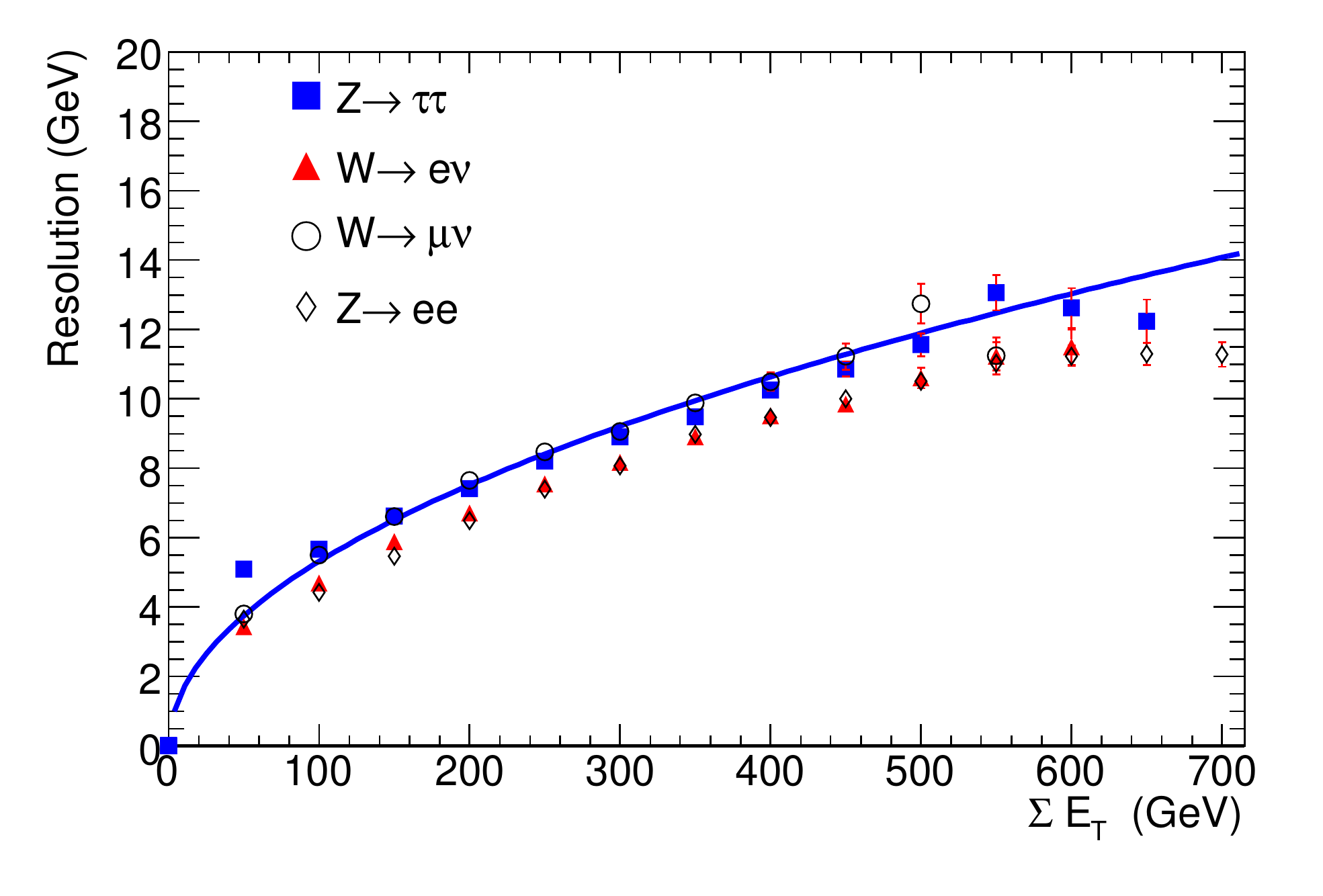}
    \includegraphics[width=0.45\textwidth]{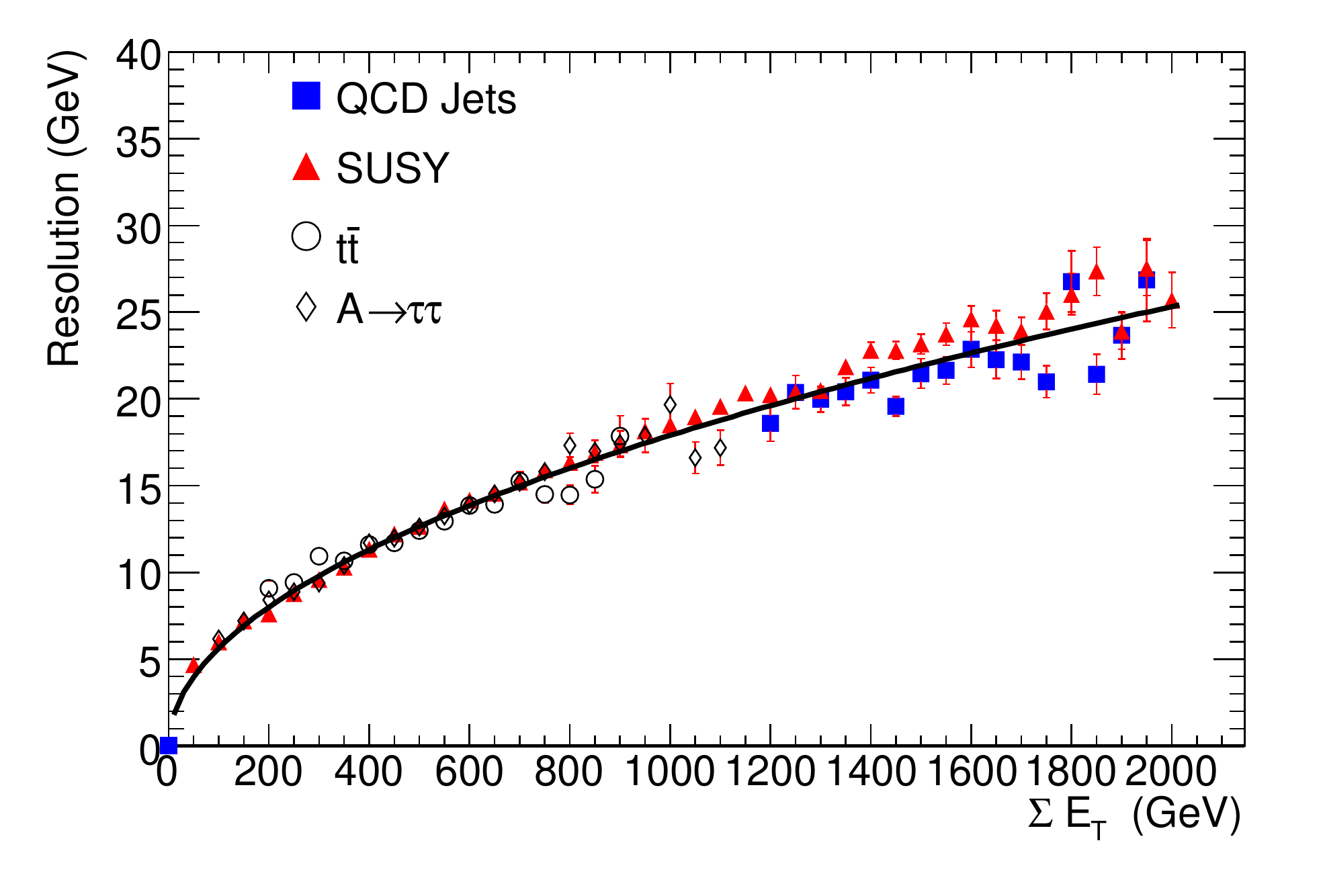}
  \caption{Resolution $\sigma$ of the two components of the \met\ vector,
   as a function of the total transverse energy,~$\Sigma E_T$, measured 
   in the ATLAS calorimeters for different physics processes corresponding to 
   low to medium values of~$\Sigma E_T$~(left) and to higher values 
   of~$\Sigma E_T$~(right). The 
   QCD~jets correspond to di-jet events with $560 < p_T < 1120$~GeV.}
 \label{fig:resol_1206_jsu3tt}
\end{figure}

The ATLAS \met\ transverse
resolution is shown in Figure~\ref{fig:resol_1206_jsu3tt} for various samples 
of interest (corresponding to different \sumet\ regions). The fit provides
$k=0.53$ in the low \sumet\ region (\Ztau\ events) and $k=0.57$ in the
hight \sumet\ region (\Atau\ events for masses $m_A$ ranging from~150 to 
800~GeV).

\begin{figure}[hbtp]
 \centering
    \includegraphics[width=0.45\textwidth,angle=-90]{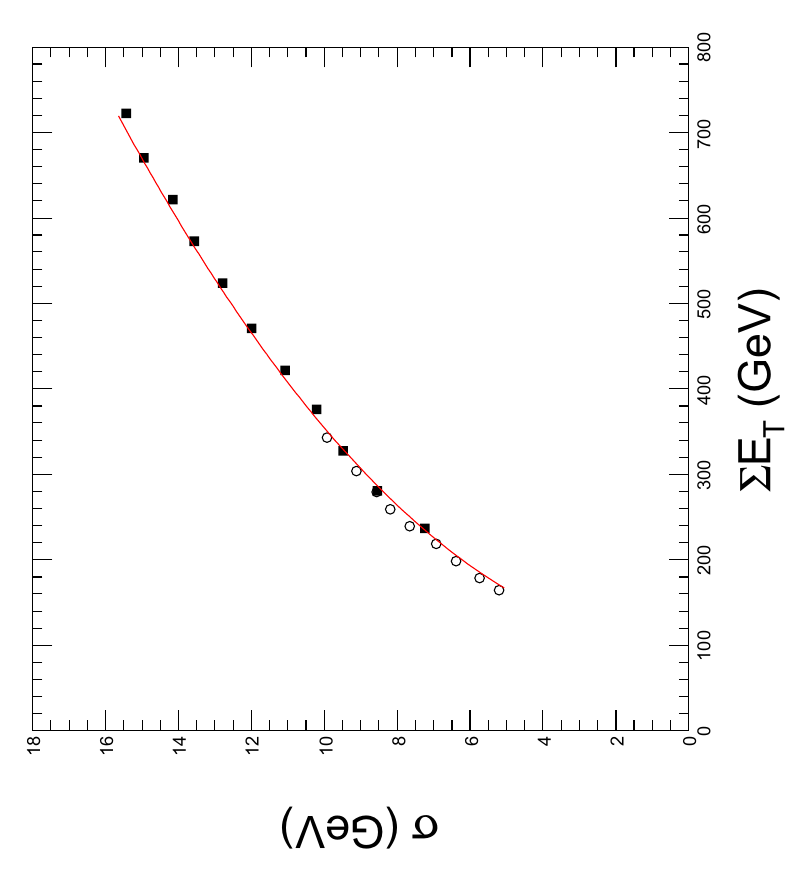}
    \includegraphics[width=0.45\textwidth,angle=-90]{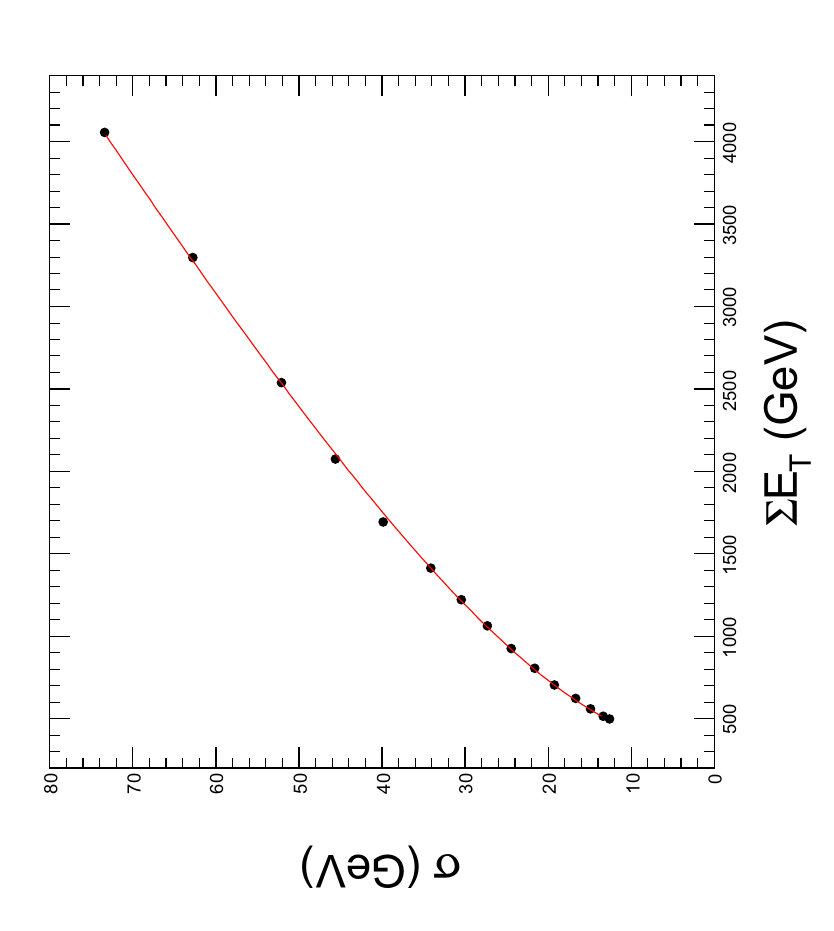}
  \caption{Resolution $\sigma$ of the two components of the \met\ vector,
   as a function of the total transverse energy,~$\Sigma E_T$, measured 
   in the CMS calorimeters for processes corresponding to 
   low to medium values of~$\Sigma E_T$~(left) and to higher values 
   of~$\Sigma E_T$~(right). The left plot is obtained with QCD soft events, 
   $0 < \hat{p}_T < 15$ GeV/C (squares) and minimum bias
    events (open circles). The right plot is for QCD events up to 
   $\hat{p}_T = 4000 $.}
 \label{fig:cms_2}
\end{figure}

The CMS \met\ resolution is shown in Figure~\ref{fig:cms_2} for minimum bias
events and QCD events in a wide range of parton transverse momentum values.
Low luminosity pile-up is included in all cases. The fit to the
distributions provides $k=0.65$ for minimum bias events. For hard QCD events
the resolution deteriorates to $k=1.23$. With the present calibration
scheme, for very active events the \met\ resolution degrades faster than 
what it is expected from, for example, minimum bias events.  

\begin{figure}[hbtp]
 \centering
    \includegraphics[width=0.45\textwidth]{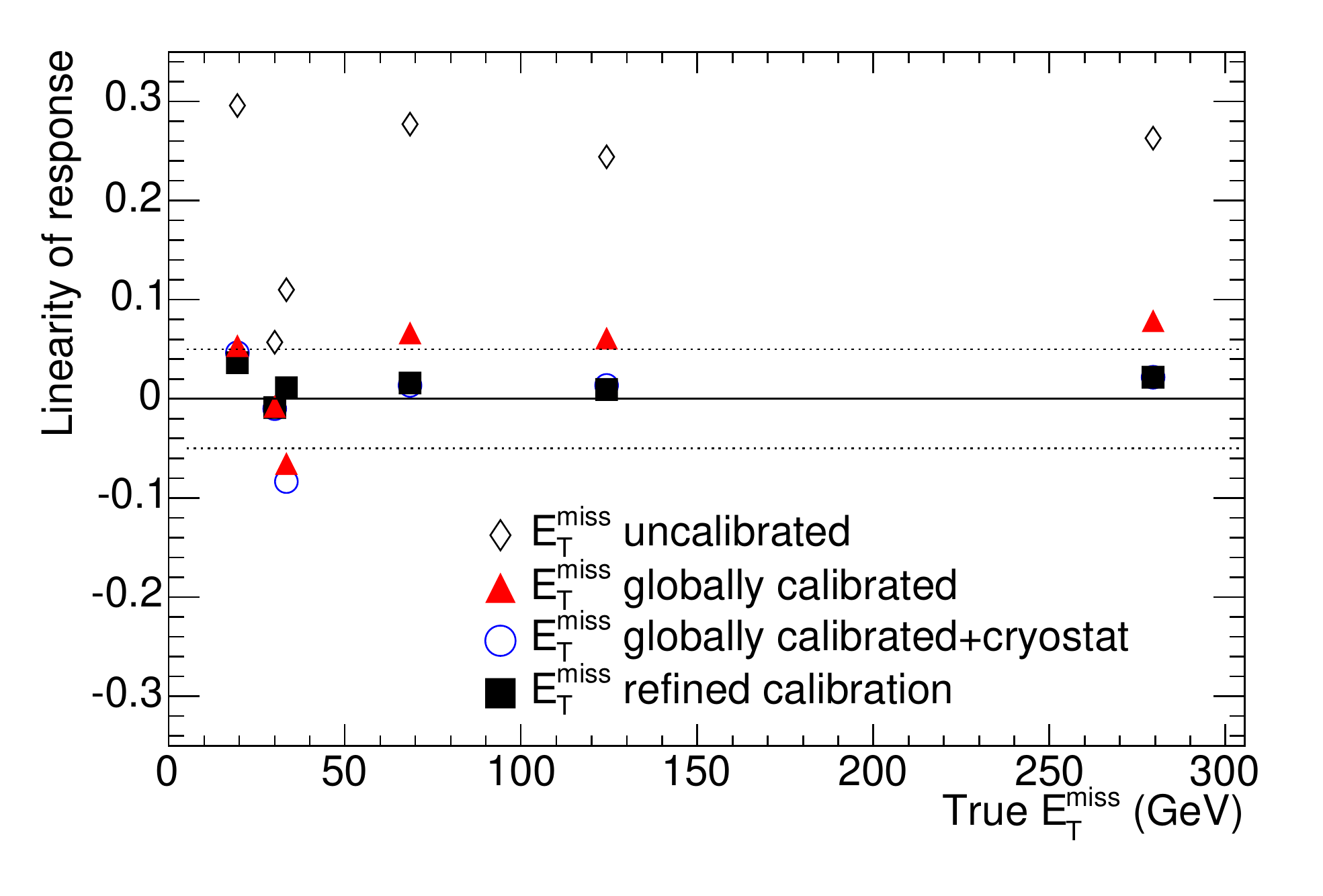}
    \includegraphics[width=0.45\textwidth]{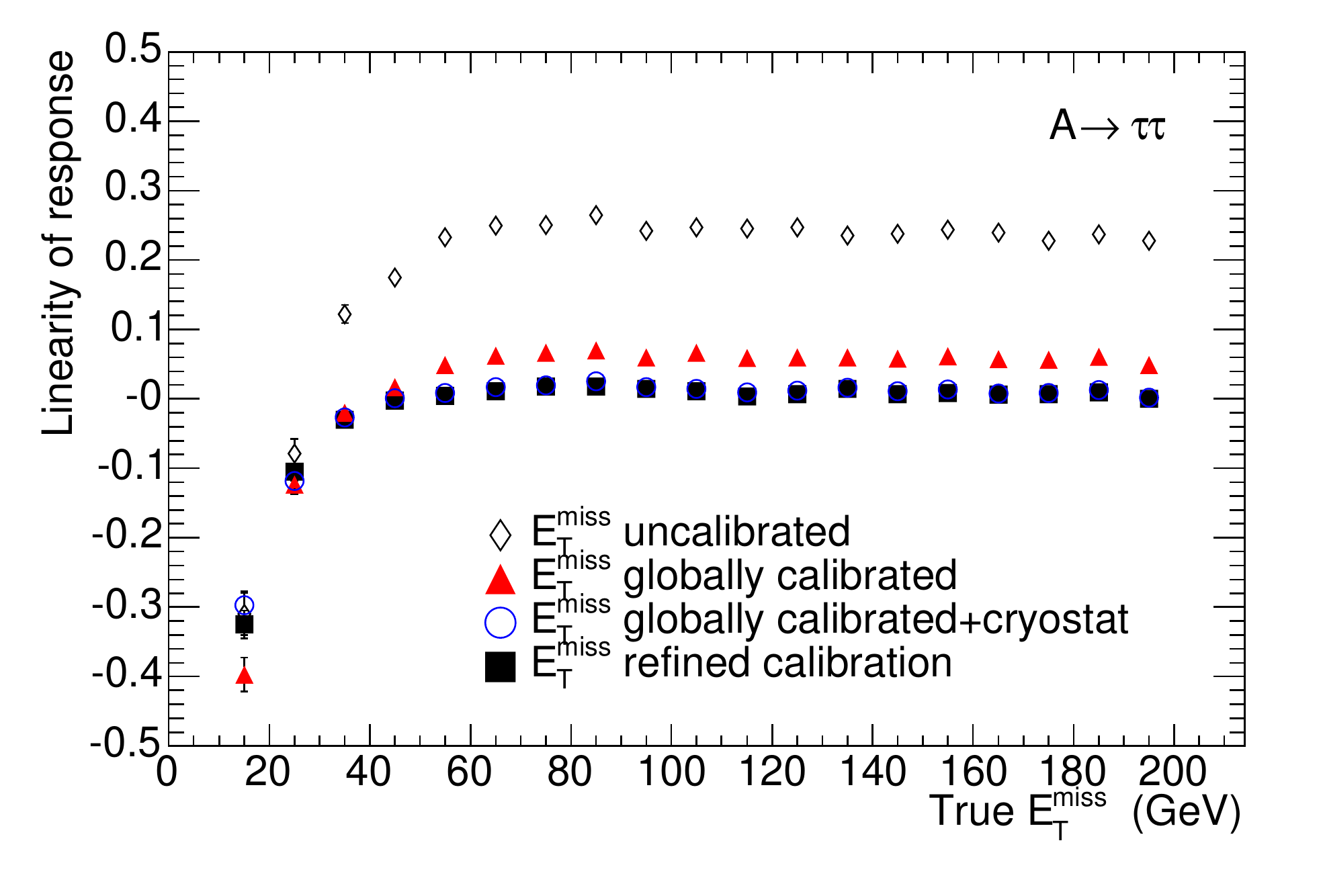}
 \caption{ATLAS linearity of response for reconstructed \met\ as a function of
   the average true \met\ for different physics processes covering a wide
   range of true~\met . In the left plot 
   the points at average true~\met\ of~20~GeV are from \Ztau~events, those
   at~35~GeV are from \Wen~events, those at~68~GeV are from semi-leptonic
   \ttbar~events, those at~124~GeV are from \Atau~events with
   $m_A$ ~=~800~GeV, and those at~280~GeV are from events containing
   supersymmetric particles at a mass scale of~1~TeV.
   The right plot has been obtained for \Atau~events with 
   $m_A$~=~800~GeV (right). The linearity is shown at three different steps
   of the calibration procedure.}
\label{fig:shiftresol_1206}
\end{figure}

The linearity for the \met\ as reconstructed and calibrated in the ATLAS 
detector is shown in Figure~\ref{fig:shiftresol_1206}. 
Except for $\met\ < 20-30$ GeV the linearity is better than about 5\%. At 
low \met\ there is a bias in the linearity due to the finite resolution
of the \met\ measurement (it is not a bias in the \met\ reconstruction
itself).

A good accuracy of the measurement of the \met\ direction is needed when
the \met\ vector is used to reconstruct the kinematic of the final state.
Moreover it is often necessary to apply a cut on the distance between the
reconstructed \met\ and the high $p_T$ jets in the event; this cut will
reject fake \met\ due to fluctuations of the reconstructed jet energy
due to problems in the jet reconstruction, like for example cracks or
dead regions inside the calorimeter acceptance.

Figures~\ref{fig:atlas_direction} and~\ref{fig:cms_direction} show the resolution on the measurement of the
\met\ azimuthal angle in ATLAS and CMS respectively. A resolution of about 100 mrad
(or better) can be obtained for high \sumet\ values. The resolution is 
better for samples with moderate hadronic activity.
\begin{2figures}{hbt}
  \resizebox{\linewidth}{!}{\includegraphics{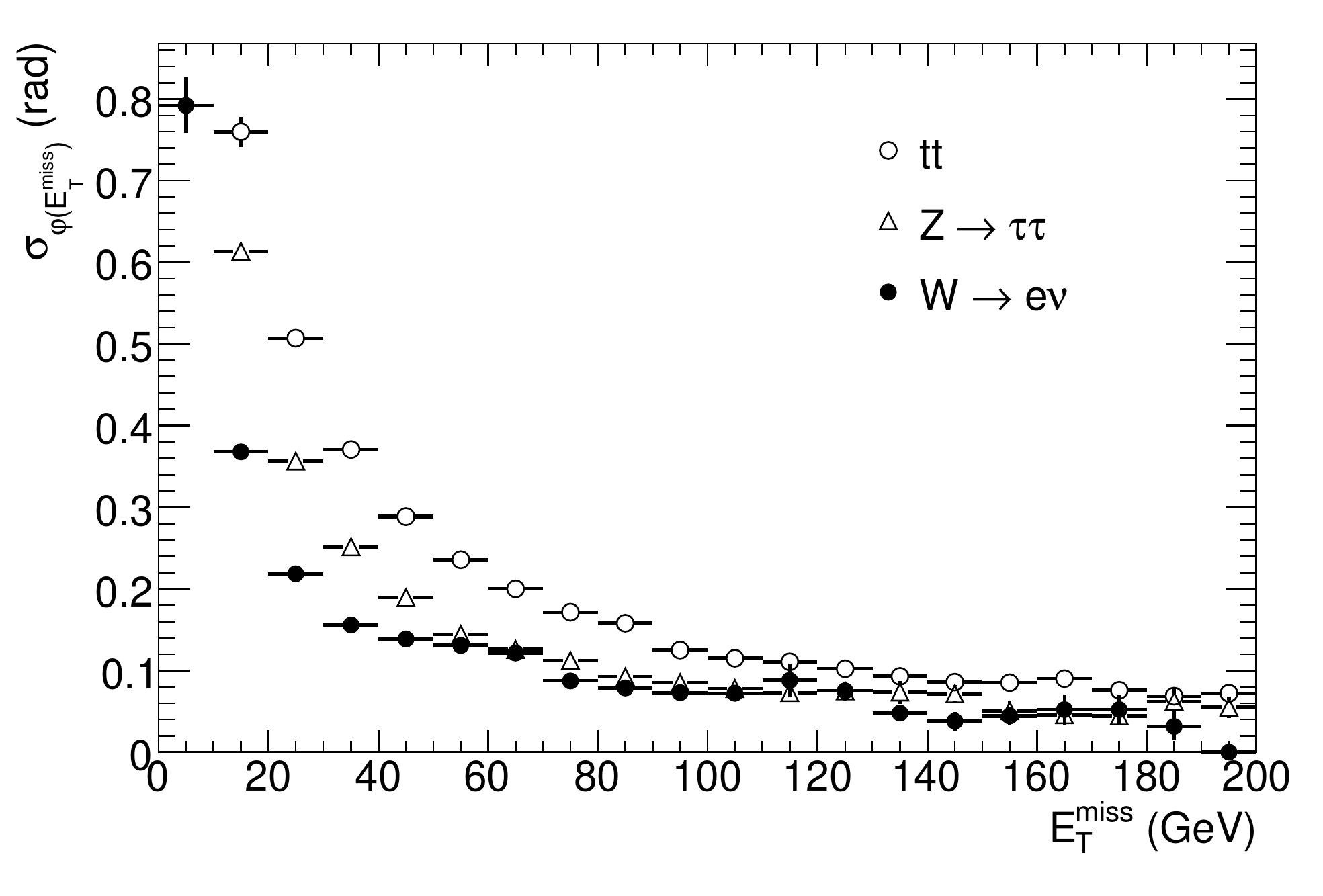}}&
  \resizebox{\linewidth}{!}{\includegraphics{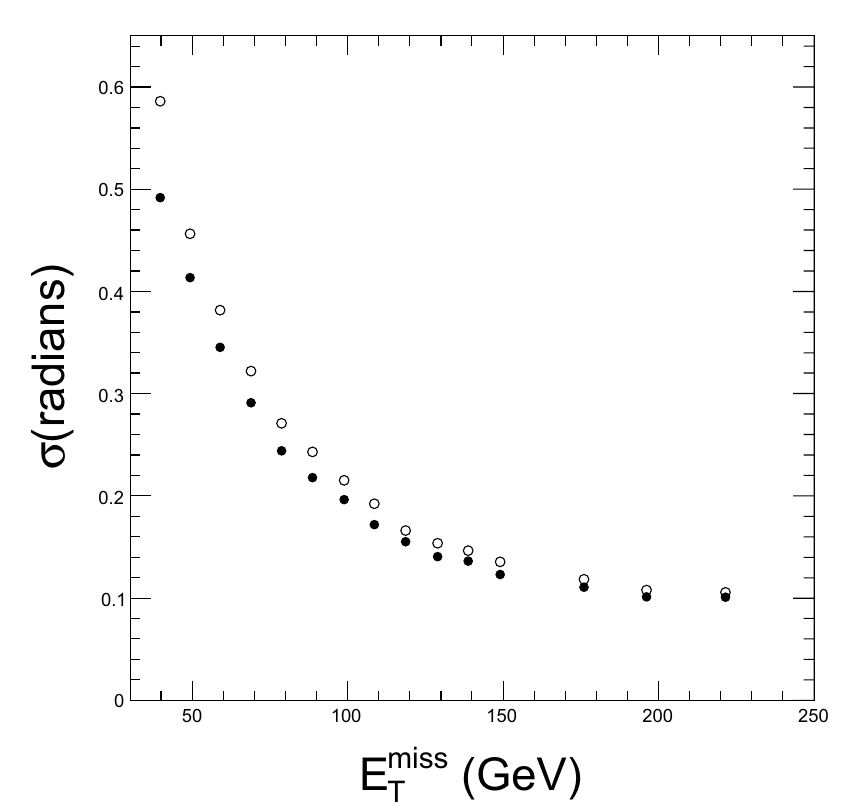}}\\
  \caption{Accuracy of the measurement of the azimuth of the \vmet\ vector as 
    a function of the true~\met\ for three different physics processes: 
    semi-leptonic \ttbar~events, \Ztau~ and \Wen~events for ATLAS.}  \label{fig:atlas_direction} &
  \caption{\met\ azimuthal angle accuracy as a function of the reconstructed 
    \met\ for \ttbar\ events in CMS (right) before (open circles) and after 
    (solid circles) a jet energy
    correction used to linearize the \met\ scale.}                  \label{fig:cms_direction}\\
\end{2figures}


Fake \met\ can come from various sources like dead or noisy cells or towers
in the calorimeter, energy losses in cracks and inactive materials, problems
in muon reconstruction (undetected or poorly reconstructed muons, fake muons).

\begin{figure}[hbtp]
 \centering
    \includegraphics[width=0.45\textwidth,clip]{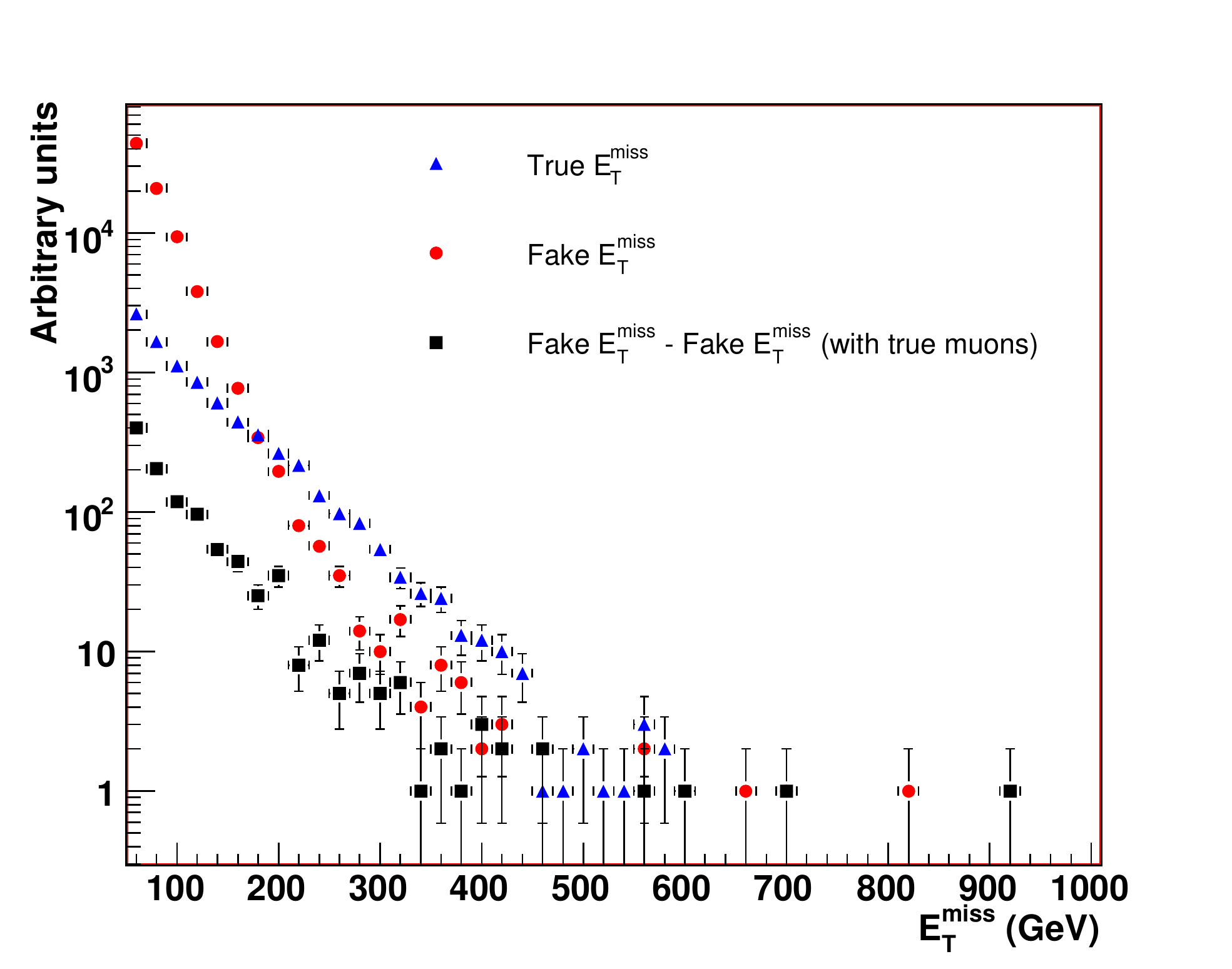}
    \includegraphics[width=0.45\textwidth,clip]{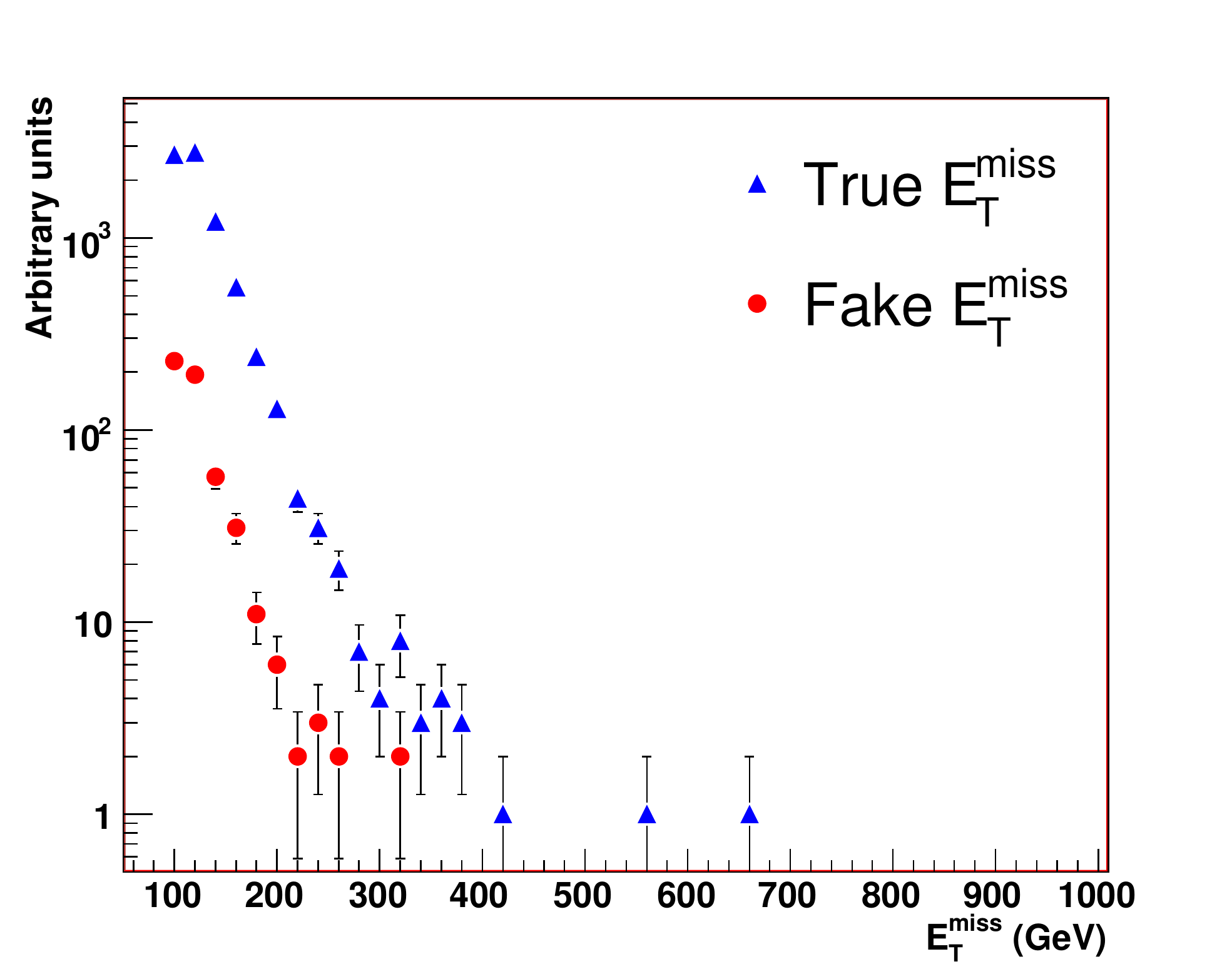}
 \caption{Fake (circles) and true (triangles) \met\  expected in a di-jet
 sample before (left) and after (right) a jet isolation cut (see text). }
\label{fig:fake}
\end{figure}

Figure~\ref{fig:fake} shows the fake and true \met\ reconstructed in a di-jet
sample with at least one jet with $560 < E_T < 1120$ GeV. The fake \met\
dominates the spectrum up to about 200 GeV. When the \met\ vector is required
to be more than 17$^\circ$ in azimuth from all reconstructed jets in the
event the \met\ spectrum is seen to be dominated by true \met . This confirm
the above statement that mis-measurement of the jet energies is the
main cause of  fake \met .

\section{b-tagging}
{\large{\sl A. Rizzi}}
\vspace{0.5cm}


The identification of the jets containing $b$ quarks ($b$-jets) relies
on the properties of B hadrons decays.
The B hadrons have a lifetime $\tau_B\sim$ 1.6 ps, which corresponds
to a $c\tau_B \sim$ 500 $\mu$m, and they produce, on average, 5 charged
particles per decay. One of the charged tracks is often a lepton, with
a branching ratio of $B\rightarrow l + X$ of $\sim$ 10\% for each
lepton family.\\ 
The two major features of the B hadrons decay, that can be exploited
to identify jets originating from $b$ quarks, are the lifetime and the
presence of a lepton. Obviously the efficiency of the second technique
is limited by the branching ratio of B hadrons to leptons.\\ 
Different algorithms can be implemented to identify $b$-jets using
these two properties. The goal of an algorithm is to have an high
efficiency in identifying the $b$-jets and a low probability of
mis-identifying a jet originated from a light flavour quark as a $b$-jet.\\ 
The algorithms act on the input data from the detectors which is
typically already processed with the so-called reconstruction
algorithms. 
The actual inputs are the reconstructed tracks, jets and
vertices. The tracks are represented as a momentum vector plus the
spatial information given by the impact parameter on the transverse
plane and on the coordinate along the longitudinal direction,
i.e. along the beam axis.  
The reconstructed jets
usually provide information on the jet energy and direction. The
vertices are points in three dimensional space where several tracks
cross; the point in which the LHC protons interacts is defined as
the \emph{primary} vertex.\\ 
To each quantity its uncertainty is assigned as computed by the
reconstruction algorithm. The uncertainties take into account the
precision of the measurements and the effects of trajectory
extrapolations: for example the parameters of a track are measured in
tracking detectors up to a distance of few cm from the interaction
point, the position near the interaction point is then obtained
extrapolating to the beam line the trajectory of the particle in the
detectors magnetic field.\\ 

In the next sections a brief description of how the two types of
algorithms work is given. Then the usage of b-tagging algorithm in the
trigger is discussed and finally the issues of calibration of the
algorithm are presented. 

\subsection{Lifetime based algorithms}
The lifetime information can be exploited in different ways. A first
class of methods is based on the observation of tracks with large
impact parameters.\\ 
\begin{figure}
  \begin{center}
    \includegraphics[width=9cm]{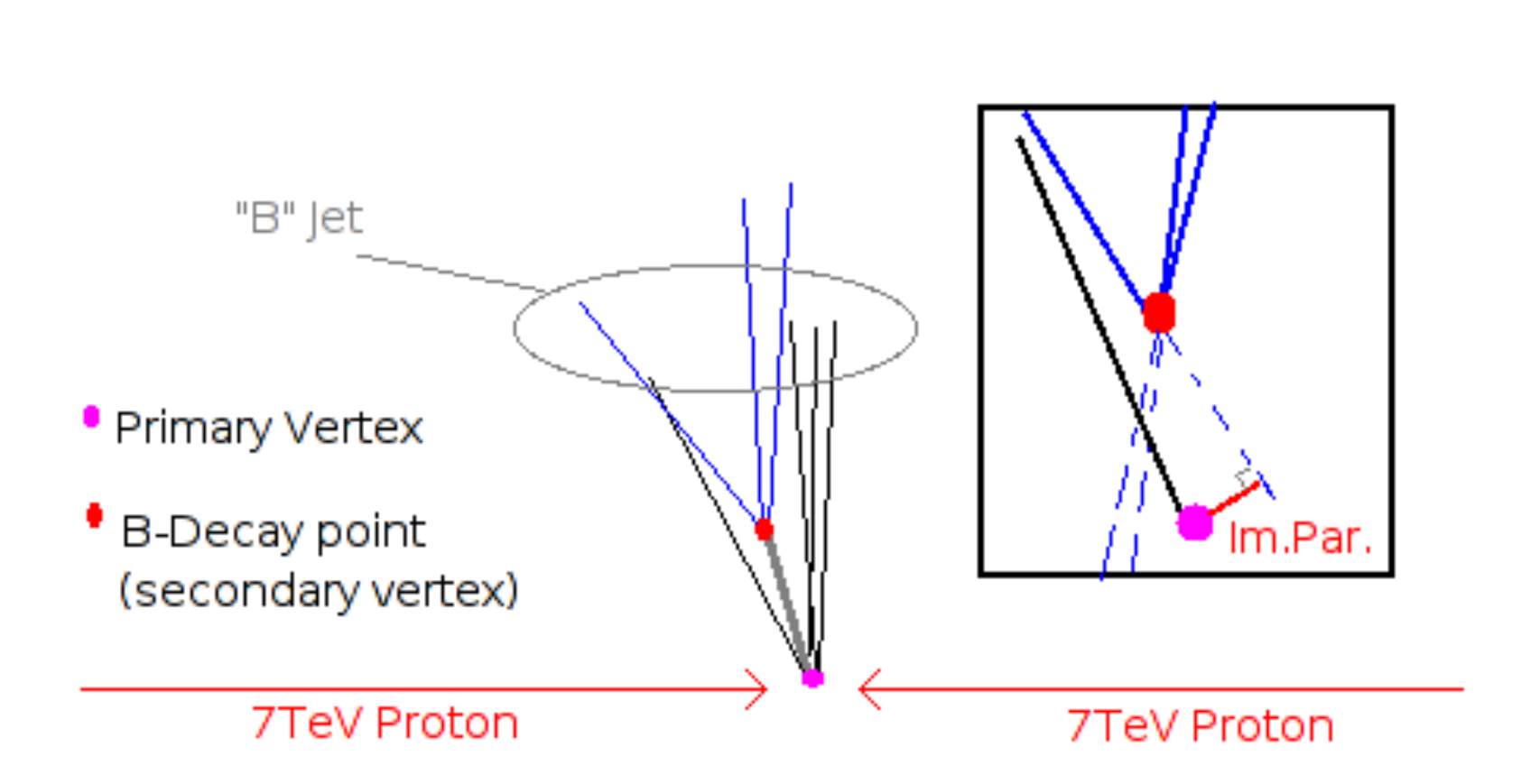}
    \caption{B hadron decay in a $b$-jet. The tracks originating from
    the B hadron decay have a large impact parameter with respect to
    the interaction point.} 
    \label{fig:jet}
  \end{center}
\end{figure}

As  shown in Figure~\ref{fig:jet}, tracks originating from B decays
have large impact parameters with respect to the primary vertex, since
they come from a displaced vertex. The impact parameters of particles
originated from the primary vertex 
are null and so the measured value is expected to be compatible with
the tracking resolution. The mean value of the impact parameters of
the tracks originating from the B hadron decay is not much affected
by the energy scale (i.e. B hadron momentum) as at higher energy the
tracks are more collimated because of the boost but the decay length
is higher.\\
A complementary approach is based on the reconstruction of secondary
vertices. In this case the information obtained from different tracks
is correlated: if several trajectories cross in a point a secondary
vertex is identified. The identification of a secondary vertex alone
is already a very discriminating quantity, i.e. a high fraction of 
$b$-jets produce secondary vertices during reconstruction while only few
light flavour jets produce this type of topology. Nevertheless by
computing quantities specific of the secondary vertex it is possible
to improve the discriminating power.\\ 

\begin{figure}
  \begin{center}
    \includegraphics[width=6cm]{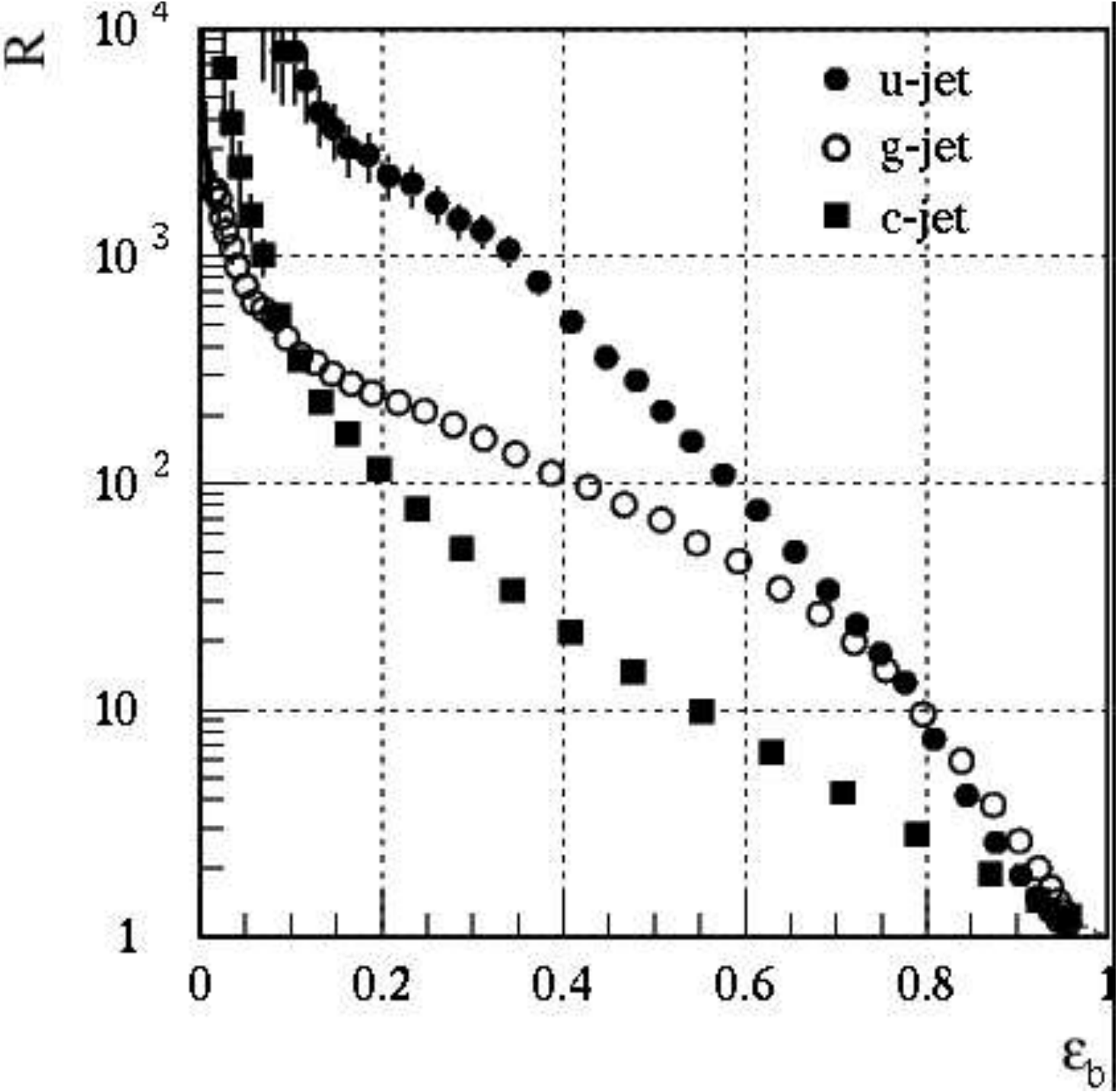}
    \includegraphics[width=6cm]{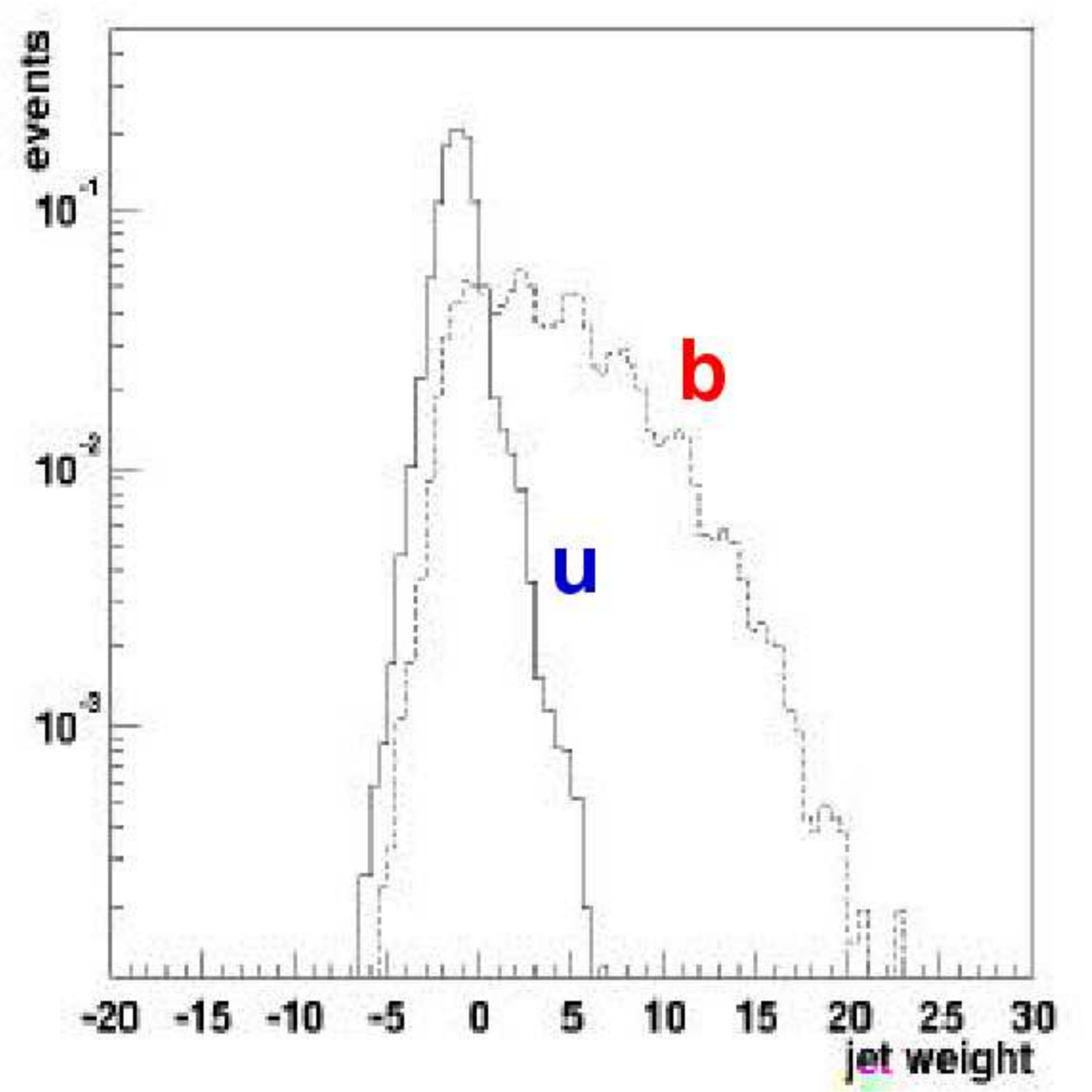}
    \caption{Left plot shows the b-tagging efficiency versus purity
    (inverse of the probability of tagging a light quark jet). On the
    right plot the distribution of the \emph{jet weight} is shown for
    light quark jets and for $b$-jets as computed by ATLAS simulation.} 
    \label{fig:eff}
  \end{center}
\end{figure}

A very simple algorithm to tag $b$-jets is the so called \emph{Track
  Counting} algorithm. This algorithm consists of counting the number
of tracks in a jet with the impact parameter $IP$ higher than a given
threshold. The impact parameter resolution $\sigma_{IP}$ can be
  computed from primary 
vertex and track parameters uncertainties and, because of
  extrapolation errors, can be different for
different tracks. Therefore the significance of the impact parameter
  $S=\frac{IP_{value}}{\sigma_{IP}}$ is used instead of its value. 
Jets with at least two or three tracks with $S$ higher then $\sim
  2\div3$ are likely to be $b$-jets.\\ 
More complex algorithms  need calibration either based on real data or
  Monte Carlo simulation.  This algorithms work by using the
  Probability Distribution Function (PDF) of
  impact parameters of tracks originating from light quarks jet and/or
  the one of heavy flavour quarks tracks. With the given PDFs it is
  possible to compute the probability that a track, with a given
  impact parameter, originates from a light or b quark jet. The ratio
  of the two probabilities can be computed and then a global weight for
  a jet, combining the ratios of individual tracks, is obtained.\\ 
In order to estimate the discriminating power of an algorithm, its
  performances are studied looking at the efficiency of tagging a
  $b$-jet versus the probability of wrongly tagging as a $b$-jet a
  light quark jet or a $c$-jet as shown in Figure~\ref{fig:eff}.\\ 
The \emph{working point} on the curves shown in Figure~\ref{fig:eff}
  is set by a cut on a continuous variable, as the \emph{jet weight}
  of the algorithm described above, which is the final result of any
  b-tagging algorithm. 

\begin{figure}
  \includegraphics[width=6cm]{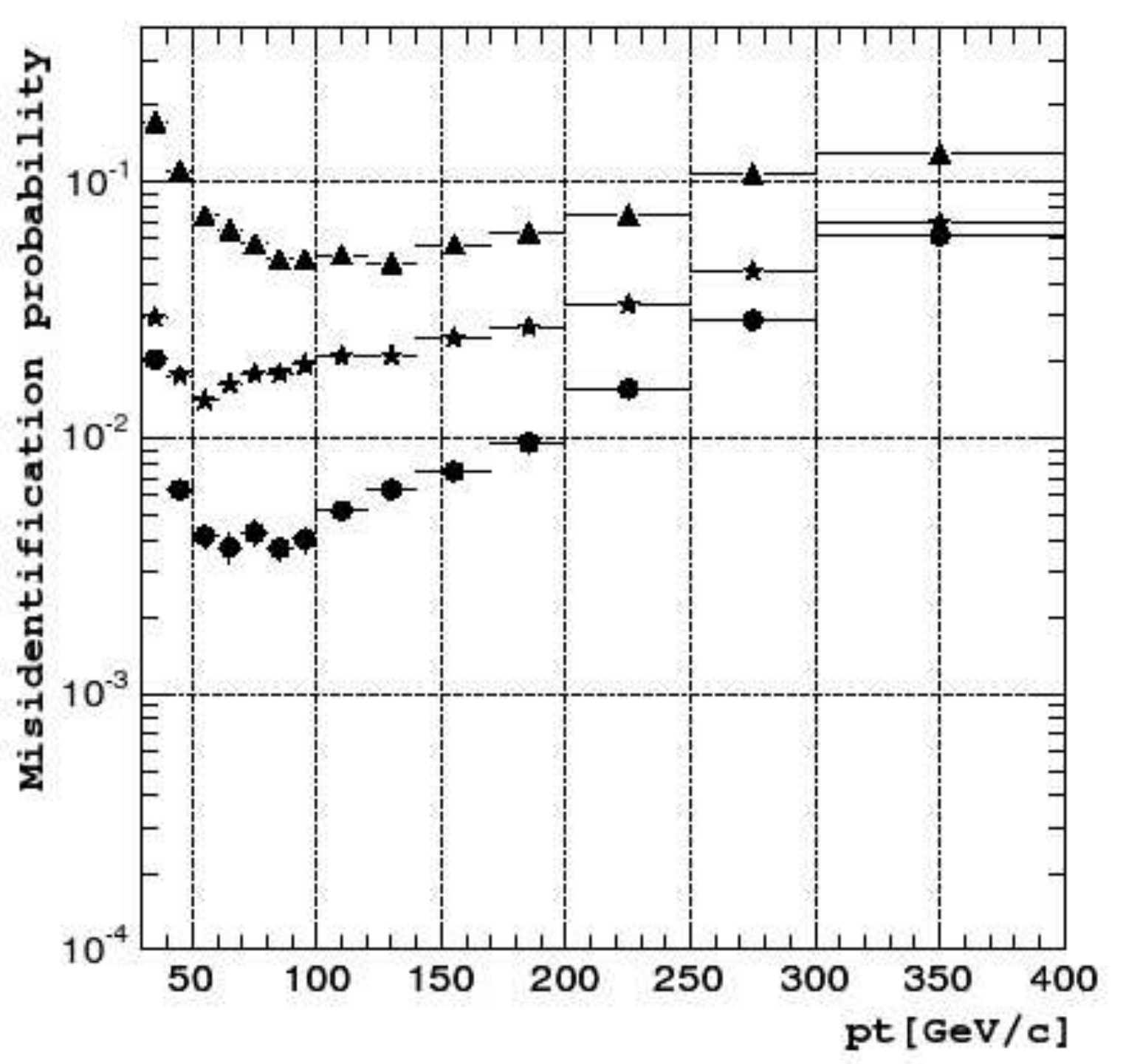}
  \includegraphics[width=6cm]{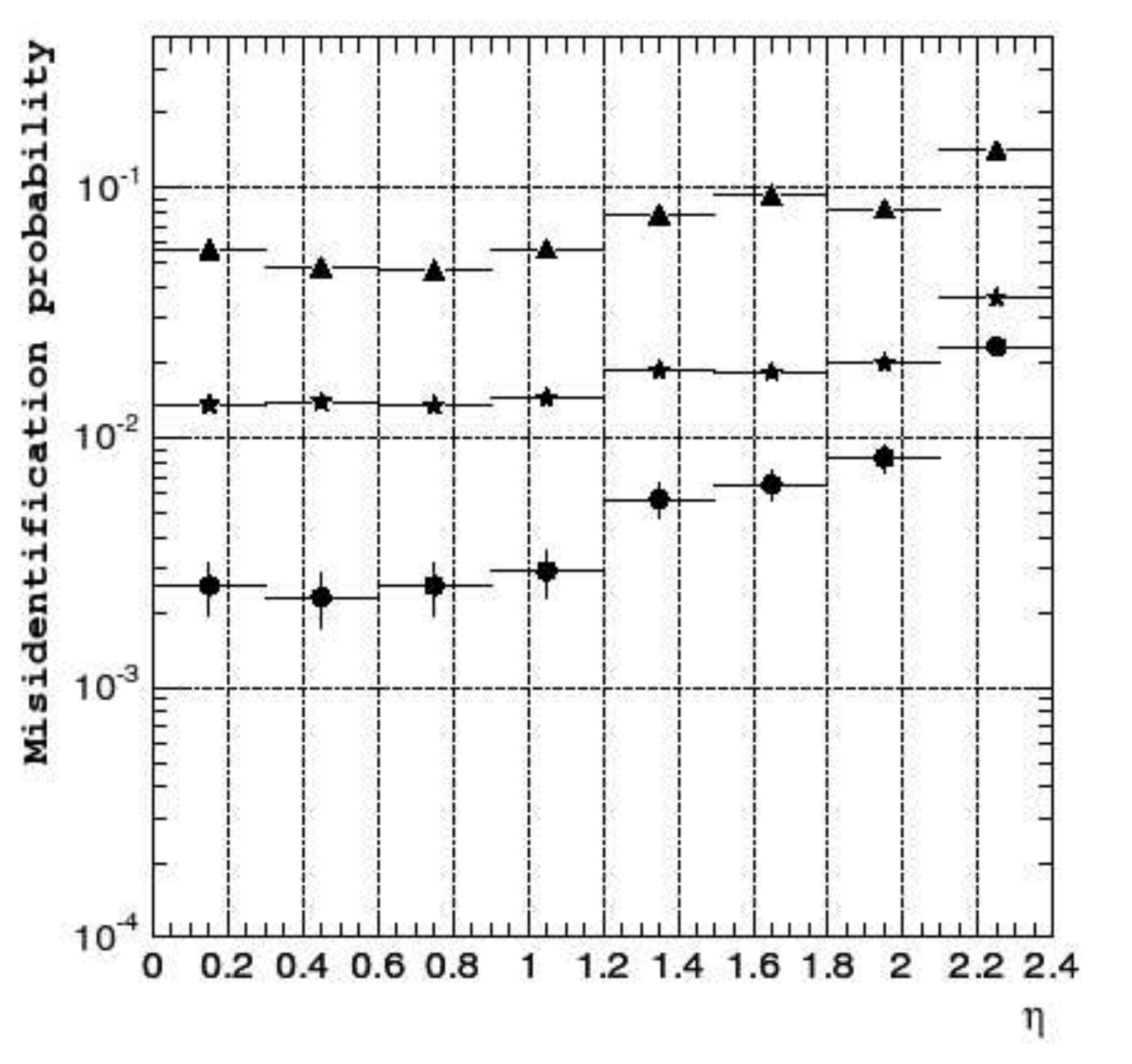}
  \caption{Probability of tagging uds (bottom points), gluon (middle
  points) or charm (top points) jets for a fixed b-tag efficiency
  ($50\%$) as a function of jet energy (left) and direction $\eta$
  (right), as computed with CMS simulation for a secondary vertex
  based algorithm.} 
  \label{fig:effvs}
\end{figure}

The performances can then be studied as a function of the jet energy,
or as function of the $\eta$ of the jet by looking at the mistagging
probability, at a fixed b efficiency, for different values of $p_T$ or
$\eta$ (Figure~\ref{fig:effvs}). 
\subsection{Soft lepton algorithms}
The soft lepton algorithms exploit the presence of muons or electrons in the $b$-jets.
The muons and the electrons can be easily identified using the muon
systems of the LHC experiments and the electromagnetic calorimeters. The main
drawback of these algorithms is that they are limited by the
$B\rightarrow l+X$ branching ratio. On the other hand they can work
even without a perfectly aligned tracker or in absence of the inner pixel
detectors which are instead crucial for lifetime based algorithms.\\ 
Information such as the component of the momentum of the lepton on the
plane orthogonal to jet direction and the $\eta-\phi$ distance of the
lepton from the jet axis, are used to improve  the
discrimination. Neural network algorithms can be used to obtain the
best performance. 

\subsection{b-tagging at trigger level}
The b-tagging can be exploited also at trigger level to improve
selection efficiency of events where $b$-jets are expected. In this
context, since for timing reasons it is not possible to perform the
full event reconstruction, the algorithms should be applied to a
subset of the event. 
A possible implementation is to apply it only
to the most energetic jets, performing track reconstruction only in a
small geometrical region containing the jet. In this way it is possible
to lower the jet energy threshold, for jets identified as $b$-jets,
without increasing the total trigger rate. 

\subsection{Calibration}
Two important issues of b-tagging algorithm are how we can tune the
algorithms and how we can measure their efficiency. In both cases the
usage of simulated data leads to high uncertainties, so
reliable methods based on real data should be implemented.\\ 
The tuning of some algorithms needs the knowledge of the impact parameter
PDF for light quarks and $b$-jets. While those are easily computed in
simulated data, it can be quite difficult to do the same on real data,
where no Monte Carlo truth is available.\\ 
For the impact parameter PDF of track originating from primary vertex
(such as most of those present in light quark jets) an efficient way
to measure it is by using the negative part of the impact parameter
distribution. The impact parameter is indeed \emph{lifetime signed} by
looking at its projection on the jet axis direction. If the track
originates from the decay of a long-lived particle the projection is on
the same side of the jet direction and the track is positively
signed. If the projection falls on the opposite side the sign is
assigned to be negative. The tracks originating from primary vertex
should have a symmetrical distribution while displaced vertices tracks
populate only the positive part of the distribution. The negative half
of the distribution is so with good approximation an estimate of the
PDF.\\ 

The second problem is trying to calibrate $b$-jets properties or to
measure b-tagging efficiency. In order to do that an independent way
of tagging a jet is needed. This can be obtained by fully
reconstructing events where two $b$-jets are expected, such as
$t\bar{t}$ events, identifying only one of the two top quarks with b-tagging
algorithms and considering the remaining jet as a $b$-jet. It has been
proved that in this way a quite pure sample of $b$-jets can be selected, so
that $b$-jets properties can be studied on real data and the
efficiency of the algorithm can be measured. 

\section{The Trigger System}
\label{sec:trigger}
{\large {\sl A. Nisati}}
\vspace{0.5cm}

\subsection{The requirements to the Trigger System}
\begin{figure}[hbtp]
  \begin{center}
    \includegraphics[width=0.55\linewidth]{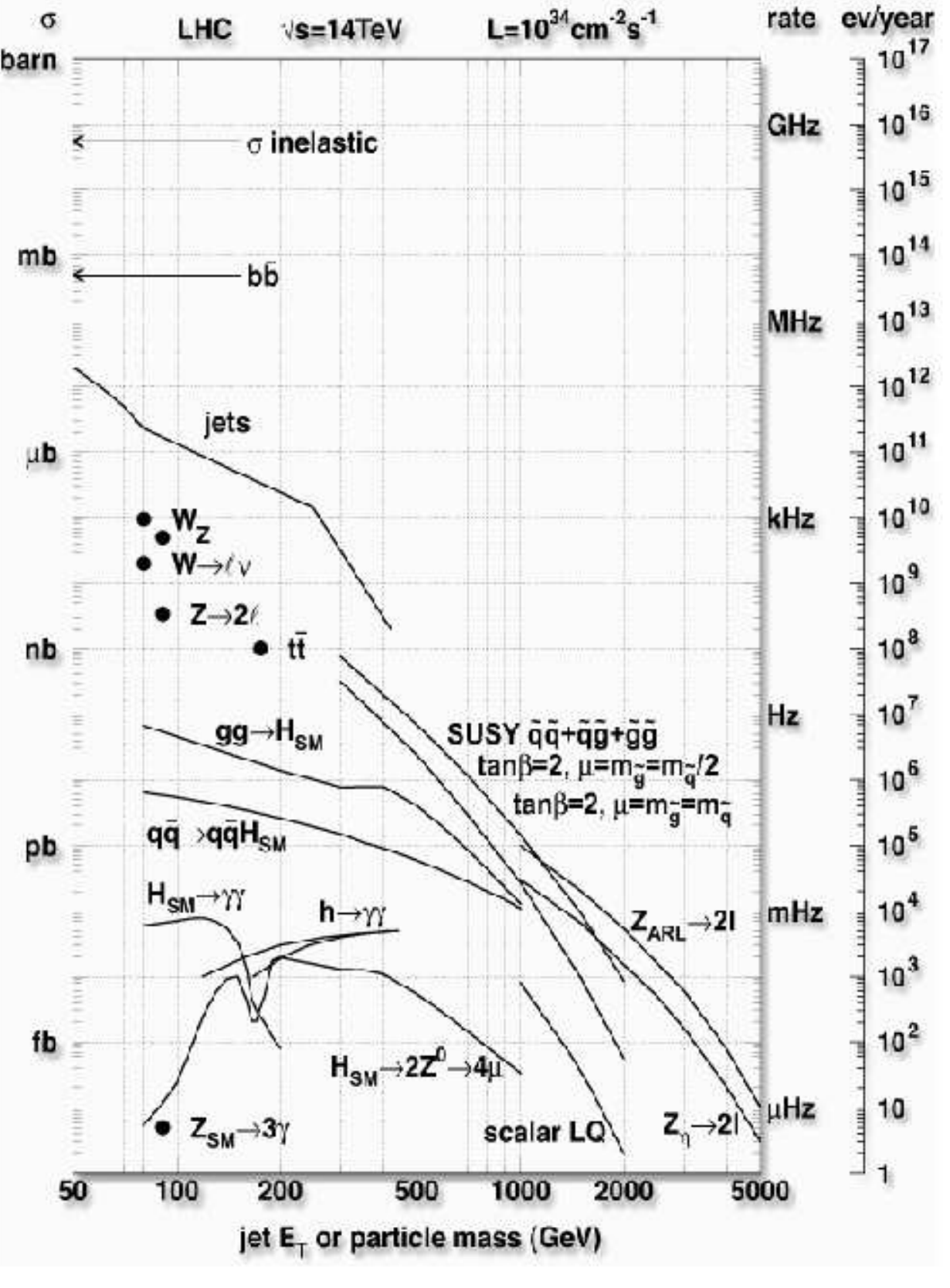}
    \caption{The proton-proton inelastic cross-section 
    at the LHC center-of-mass energy is compared
    to the cross-section of more specific physics processes such as
    the SM Higgs boson production and decay to photons or four-leptons
    in the final state, or SUSY particles. A rejection factor  
    of the order of 10$^{13}$ is
    needed to separate the physics of interest from the bulk of the inelastic
    p-p interaction. }
    \label{fig:xsect}
  \end{center}
\end{figure}

As mentioned in Chapter~\ref{cap:intro}, the intensity of the
signal of new physics at LHC is usually very weak (about 10$^{-4}$ Hz,
cf. Figure~\ref{fig:xsect}). An online  
selection of events of physics interest is mandatory to store
on permanent memory only a small fraction of the p-p events 
produced (about 10$^9$ Hz), for detailed offline reconstruction and analysis.
The role of the trigger system is to make an online selection of particle
collisions potentially containing physics of interest reducing 
at the same time the large rate of uninteresting
physics processes.
The event selection needed to isolate the physics of interest from the bulk
of minimum bias events requires typically a rejection of a factor
10$^{13}$, most of which (about 10$^{7}$) has to be performed
online. 
The physics process of interest can be tagged by looking to particles in the 
final state such as:
\begin {itemize}
 \item[-] electrons (Higgs, new gauge bosons, extra dimensions, 
                  SUSY, W, top);
 \item[-] photons (Higgs, extra dimensions, SUSY);
 \item[-] muons  (Higgs, new gauge bosons, extra dimensions, 
                  SUSY, W, top, B-physics);
 \item[-] jets  (SUSY, compositeness, resonances);
 \item[-] jets + missing E$_T$ (SUSY, leptoquarks);
 \item[-] tau + missing E$_T$ (MSSM Higgs, SUSY);
\end {itemize}
The selection efficiencies must be precisely known for the
different physics processes in order to evaluate correctly the
production cross-sections and the branching ratios without introducing
biases. 
Furthermore, the trigger systems for experiments at the LHC must be robust against
the physics background that is present in the experimental halls.
This is particularly the case for the muon trigger, exposed to the large rate
of low energy particles produced by the interaction of primary particles with
the forward detectors of the apparatus and the machine elements such as the
beam-pipe and the collimators.\\
Given the complexity of the events to be analysed,
the experimental apparatus will provide lots of precision measurements;
online systems with large bandwidth capabilities are therefore required
to move this amount of information 
from the on-detector electronics and from the readout buffers to the event builder
to compose the event fragment to be stored.\\
Last but not least, the system flexibility is important to optimize the trigger
selection for new possible physics signals that today are fully unexpected.
In the folowing the main aspectes of the trigger and
\subsection{The Trigger of ATLAS and CMS}
The online event rejection is performed with two or more trigger levels.
Multi-level triggers provide a rapid rejection of high-rate backgrounds
without incurring much deadtime. 
The First Level Trigger (\textit{L1})
is based on custom fast electronics that processes the signal coming from
detectors with reduced granularity. The event accept/reject decision is
produced with fixed latency with respect to the corresponding bunch
crossing time. A short latency is essential since information from all
detectors needs to be buffered, waiting for the L1 decision on
whether accept or reject the event for further online analysis performed
by the High Level Trigger system (HLT).

HLT selections are based on fast computer algorithms 
running on commercial PC farms at affordable cost,
and perform the final selection before the event storage on
memory mass for offline analysis. The resulting accepted event rate
has to be matched to the amount of data that can be reconstructed in the
offline computing farms.

\subsubsection{The First Level Trigger}
The First Level Trigger of ATLAS and CMS is based on very fast
reconstruction of muons, electrons/photons, jets, taus 
total transverse energy and missing transverse energy
performed with dedicated electronics.
The L1 system forms a trigger decision for each bunch-crossing
based on combinations of above. 
Both in ATLAS and CMS the measurements from the inner tracking detector are not
used to perform the L1 trigger selection.
The trigger accept signal is distributed
to the front-end electronics.
The latency is fixed and it is about 2.5 $\mu$s which implies that 100
events should be buffered in pipeline memories waiting for the trigger decision.

The maximum event rate accepted by the L1 trigger systems of
ATLAS and CMS is 100 kHz, and it is limited mainly by the 
input bandwidth of the HLT system can afford.
The representative event size of selected events is 1~MB
both for ATLAS and CMS.
Deadtime is artificially introduced in order to avoid data
loss or buffer overflow in front-end electronics. ATLAS plans
to introduce a deadtime of 4 bunch-crossings (100 ns), which
corresponds to a fraction of about 1\% for a 100 kHz L1 rate.

The L1 trigger selections at LHC are based on the presence of
inclusive muons with p$_T>20$ GeV/$c$ (corresponding rate at nominal
LHC luminosity: about 10 kHz), electromagnetic clusters (e/$\gamma$)
with E$_T>30$ GeV (rate: 10 $\div$ 20 kHz), jets with E$_T>300$ 
GeV (rate: 200 Hz).
 
\begin{figure}[hbtp]
  \begin{center}
    \includegraphics[width=0.55\linewidth]{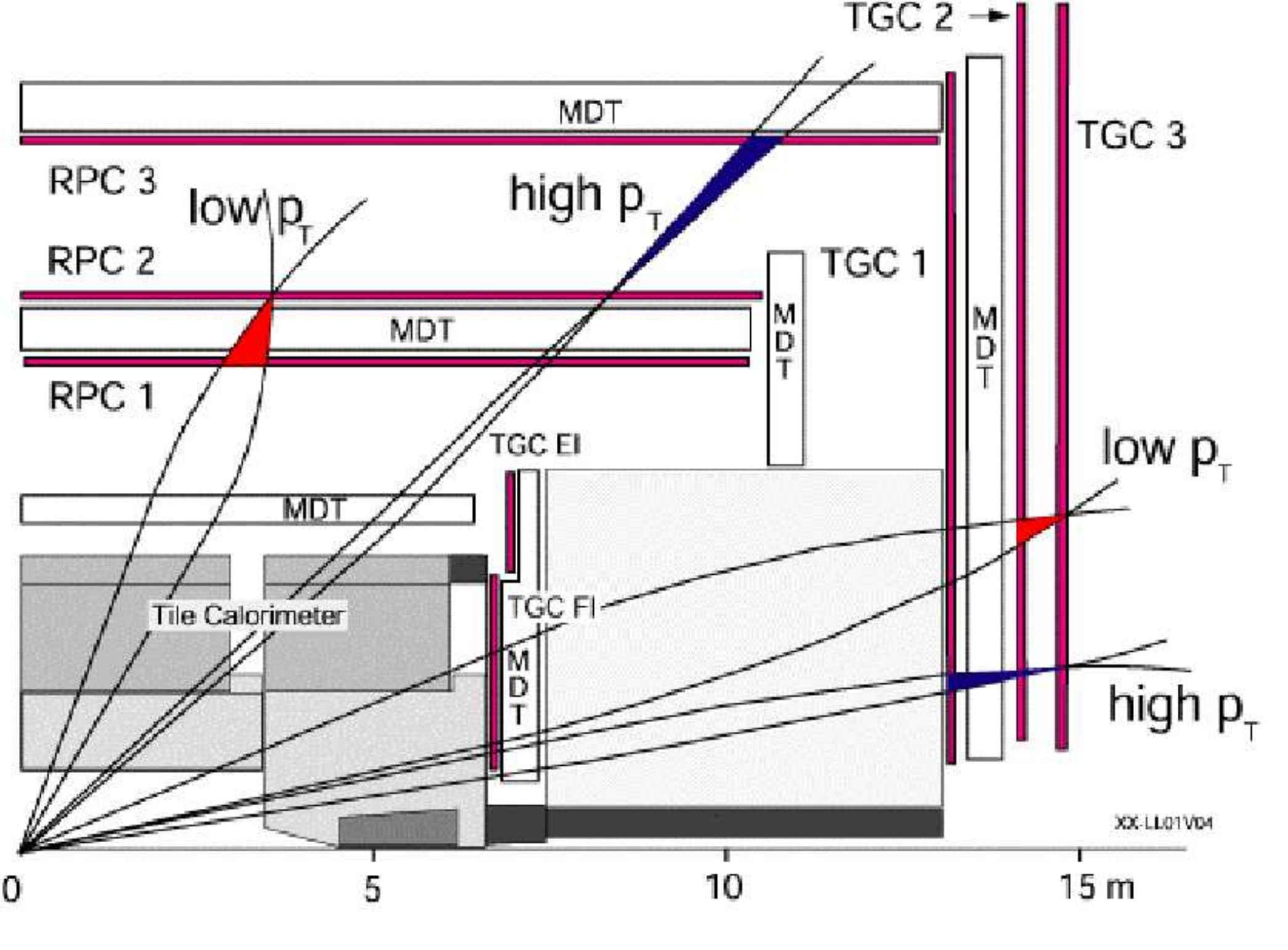}
    \caption{The L1 muon trigger system for LHC experiments must
    select with high efficiency genuine high-p$_T$ muons originating 
    from the p-p interaction point. Fake muons simulated by random
    coincidences produced by the physics background present in the muon
    system can raise the trigger rate to unacceptable values. The ATLAS
    muon trigger system sketched in this figure is based on a multiple
    detector tight coincidences with short time gates in both $\eta$ and
    $\phi$ views. In the barrel, the coincidence in both views of 
    RPC2 with RPC1 trigger
    stations provides the so-called low-p$_T$ trigger when running at
    ${\cal L}=2 \times 10^{33}$ cm$^{-2}$s$^{-1}$. For the nominal
    luminosity run 
    the high-p$_T$ trigger requires the low-p$_T$ selections $and$ the
    coincidence of RPC2 with RPC3. A similar scheme is adopted in the
    endcap where TGC stations replace the RPCs.
    Each trigger station is made by two gas gaps each read in two
    orthogonal projections.}
    \label{fig:LVL1mu}
  \end{center}
\end{figure}

In addition to the rejection of events with low-p$_T$ particles, the
L1 has to cope with physics background originating from the hadron
showers of primary particles from p-p collisions that can fake high-p$_T$
muon signatures in the muon spectrometer. To limit the rate from this
source, muon trigger systems must be fast and redundant, space 
and time coincidences must be taken as small as possible. 
Figure~\ref{fig:LVL1mu}
shows a simplified view of the ATLAS L1 Muon Trigger scheme.

Last but not least, the L1 trigger systems must assign the 
bunch-crossing to the event that has produced the trigger accept,
for correct event building by tha Data-Acquisition System (DAQ).

\subsubsection {The High-Level Trigger}

The events accepted by the L1 trigger need a further selection
in order to reduce the amount of data to be stored for the offline
reconstruction and physics analysis.
Both ATLAS and CMS plan to record about 100 events/s, that corresponds
to 10$^6$ TB data in a year of data taking. Since the L1 selects
events with a rate of 100 kHz, this implies that the HLT
system must provide another rejection factor 10$^3$ while keeping high
efficiency for the physics processes of interest.
The solutions adopted by ATLAS and CMS for the HLT are rather different:
the CMS Collaboration has decided to read all the detector data 
accepted by the L1, thus
performing the full event reconstruction with a PCs farm in one single
level of the HLT. On the contrary the ATLAS Collaboration adopted a 
different strategy, based on the so-called ''Region-of-Interest" (ROI)
approach,
that implies the movement only of a small fraction of the detector data 
(again, accepted by the L1 trigger) available in the readout buffers
of each subdetector.  

\subsubsection {The ATLAS High-Level Trigger}

\begin{figure}[hbtp]
  \begin{center}
    \includegraphics[width=0.75\linewidth]{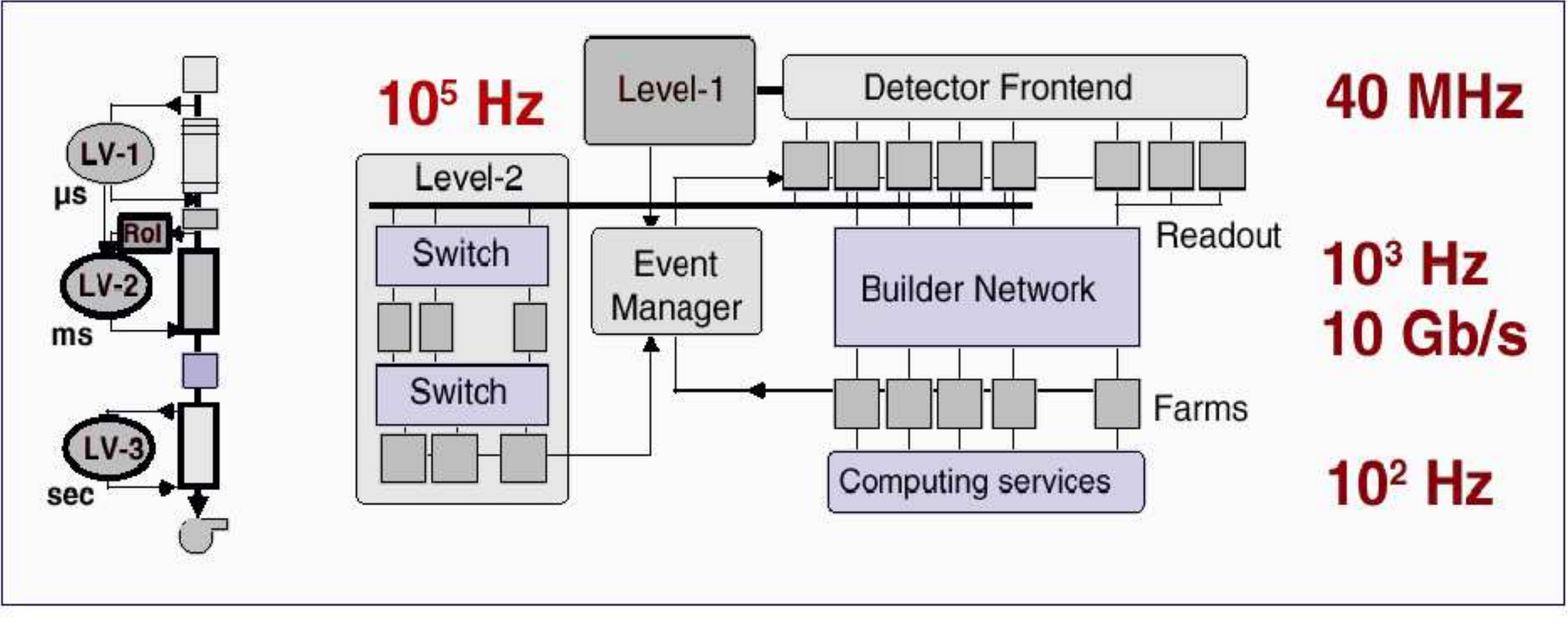}
    \caption{Simplified scheme of the ATLAS High-Level Trigger. Once the
    Level-1 trigger (LV1 in the
    figure) has accepted an event, the detector data are transferred
    from the front-end electronics to the Readout System (ROS) buffers
    (ROBs). Then fast reconstruction algorithms are executed by the PCs farm
    of the Level-2 Trigger System (LV2 in this figure). 
    Full granularity detector data belonging
    only small regions around the region(s) where the Level-1 trigger occurred 
    (ROI(s), see the text)
    are transferred and used by these programs to validate the Level-1
    selection. If the Level-2 system accepts the processed event, then the
    Builder Network transfers the whole amount of detector data to build
    the event fragment. Finally, the event
    can be fully reconstructed by the Event Filter farm (LV3 in the
    figure) to operate
    the last online selection before its recording on tape (or HDD).}
    \label{fig:ThreeLevels}
  \end{center}
\end{figure}

\begin{figure}[hbtp]
  \begin{center}
    \includegraphics[width=0.5\linewidth]{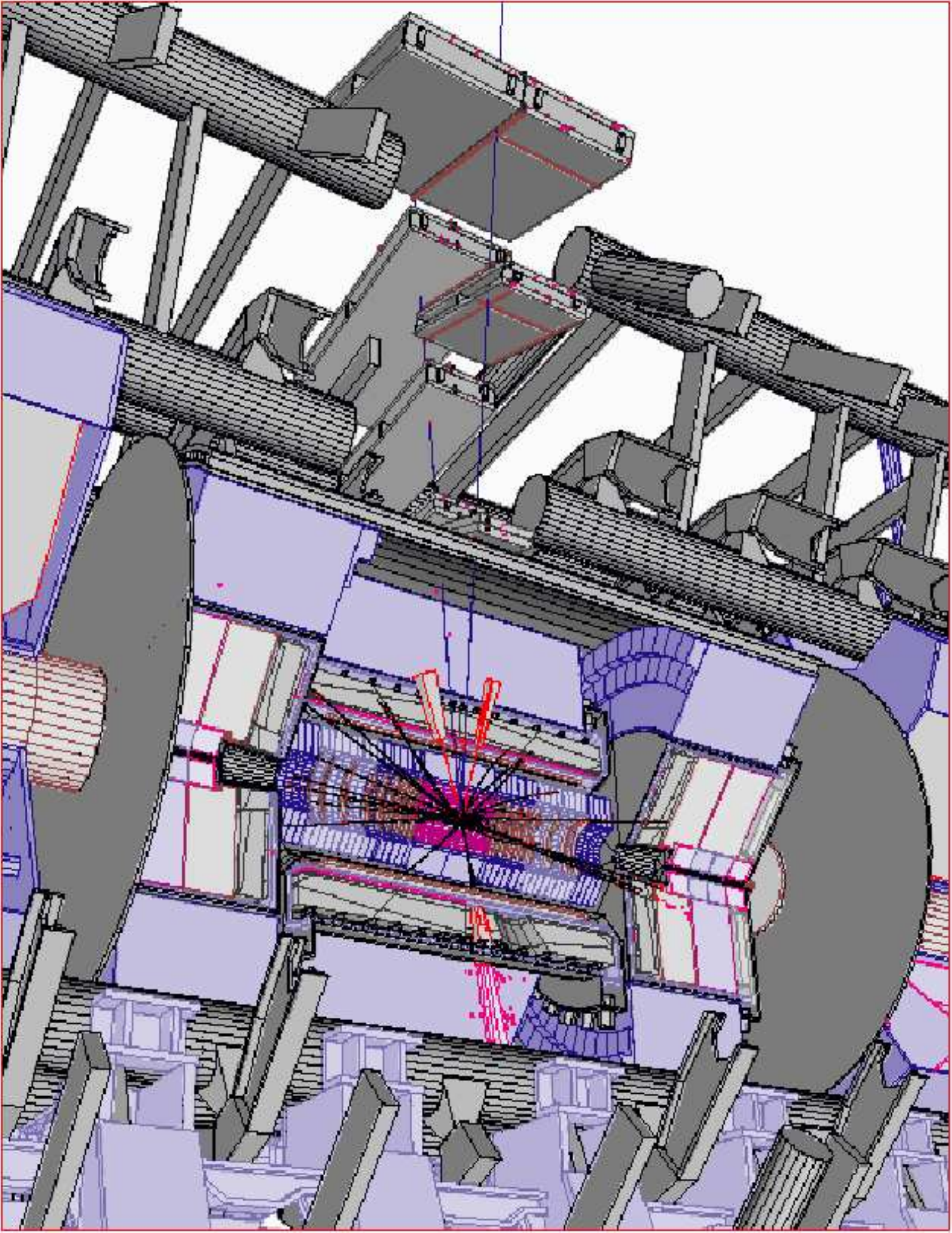}
    \caption{Illustration of the ROI approach in ATLAS: a muon is triggered by the
     Level-1 system (the two outermost RPC chambers visible in this figure);
      the $\eta-\phi$ addresses of a small region around the muon track 
      (the ROI) are transferred to the Level-2 Supervisor; the full granularity
       data set corresponding to a small number of muon chambers placed
       around the muon ROI are then transferred to the Level-2 farm
       to allow a precise and fast muon track reconstruction. On demand,
       other subdetectors data can be transferred to allow the overall muon
       measurement, including the track hits collected by the Inner Detector.  
      }
    \label{fig:ROI}
  \end{center}
\end{figure}

Assuming an event size of 1 MB, the L1 trigger (called Level-1 in ATLAS) accept rate of 100 kHz does
require a readout bandwidth of 1000 Gb/s in case the whole detector
data was accessed. This is possible with today's technology; however
ATLAS has decided to reduce the data readout volume by the implementation
of a Level-2 Trigger step that reduces by a factor 100 the event rate
to be passed to the Event Filter selection, where the events are fully
reconstructed and selected with offline-like algorithms.
In other words, the HLT of ATLAS is made by two 
sequential steps:
The Level-2 and the Event Filter~\cite{ATL_TDR}; see also Figure~\ref{fig:ThreeLevels}.
The Level-2 trigger is based on computer algorithms running on a reduced
set of full granularity detector data. In fact, in ATLAS the Level-1 
provides, in addition to the event selection and of the 
bunch-crossing, also the "Region-of-Interest" (ROI) identifier. The
ROI is small region in the $\eta-\phi$ space of a given detector (muon
system, electromagnetic calorimeter, hadron calorimeter), where the
Level-1 system has produced the trigger accept. This means that in this
region a high energy object (muon, electron/gamma, jet, tau, etc...)
has been found and satisfied the trigger menu conditions. The size
of this region is typically $\Delta \eta \times \Delta \phi = 0.1\times 0.1$.
When a Level-1 trigger is generated, the Level-2 algorithm refines
the measurements of the particle that has originated the Level-1 accept
using the full measurements available in a small region containing the
ROI. This is shown in Figure~\ref{fig:ROI}.

As an example, in the case of the muon trigger, the Level-1 uses
track measurements provided only by the RPCs; at Level-2, the
drift time measurements of the precision tracking system made by the
Muon Drift Tubes (MDT) are also used to improve the quality of the muon momentum estimate. 
To do this only the MDT and RPC data (or MDT/CSC plus the TGC data, in the
endcap) around the Level-1 muon ROI are read from the whole ReadOut System. 
The same approach is applied to electron/photon, taus, and energy
triggers.
Similarly, the Level-2 electron selection is based on a detailed shower analysis,
not performed at Level-1. In addition, the measurements from the Inner
Detector (not available at Level-1) are used to validate the reconstruction
 of this lepton:
a high-p$_T$ track is searched around the electromagnetic cluster and 
the matching
between the center-of-gravity of the cluster and the track is required.
After this reconstruction and selection, the electron rate is
reduced by a factor larger than 60, with an efficiency of about 85\%. 

Following this strategy, it is possible to precisely measure with a short 
average latency, of the order of 10 ms, the low-energy particles that have been
triggered as high-energy objects by the Level-1 trigger. An overall 
event rejection factor 100, relative to the Level-1 accepts, can be achieved.
At the same time, the data traffic is approximatively 
reduced accordingly to the fraction of
detector read out. Moreover, taking into account also the contribution 
given by the message passing among the several Level-2 Processors, 
it is estimated that
the readout bandwidth can be kept at the level of 100 Gbit/s. 

If an event is accepted by the Level-2 selection, the Trigger/DAQ system
allows the full detector data readout and the event building.
At this point the last event selection step is performed, the Event
Filter.
The event Builder Switch looks for the first PC ready for data processing
in the Event Filter Farm, transfers
the full event fragment in its memory and the PC starts the event 
reconstruction in the full detector (or in a fraction of it).
Further selection cuts are applied;
the event rate reduction is estimated a factor 10 with respect to the 
Level-2, with a latency of a few s.
If the event is accepted (''filtered"), it is recorded permanently in
the mass storage supports (tapes or HDDs); the rate of event recording
is about 100 Hz.

The ATLAS Trigger/DAQ system architecture allows the staging/deferral
scenarios of this system for the first years of data taking.
 In fact,
during the initial LHC operation, the machine luminosity will be well
below the nominal one, and in this condition the full Trigger/DAQ
potentiality is not needed. The HLT bandwidth can be staged, implementing
the two HLT processing farms with a reduced number of CPUs, to allow
for example, a 23 kHz Level-1 output rate. With increasing luminosity,
and financial resources, the HLT farms can be completed to match the
nominal system. 

More details on the ATLAS High-Level Trigger (rates and acceptances
for various physics channels) are reported in~\cite{parodi}.

\subsubsection {The CMS High-Level Trigger}

\begin{figure}[hbtp]
  \begin{center}
    \includegraphics[width=0.75\linewidth]{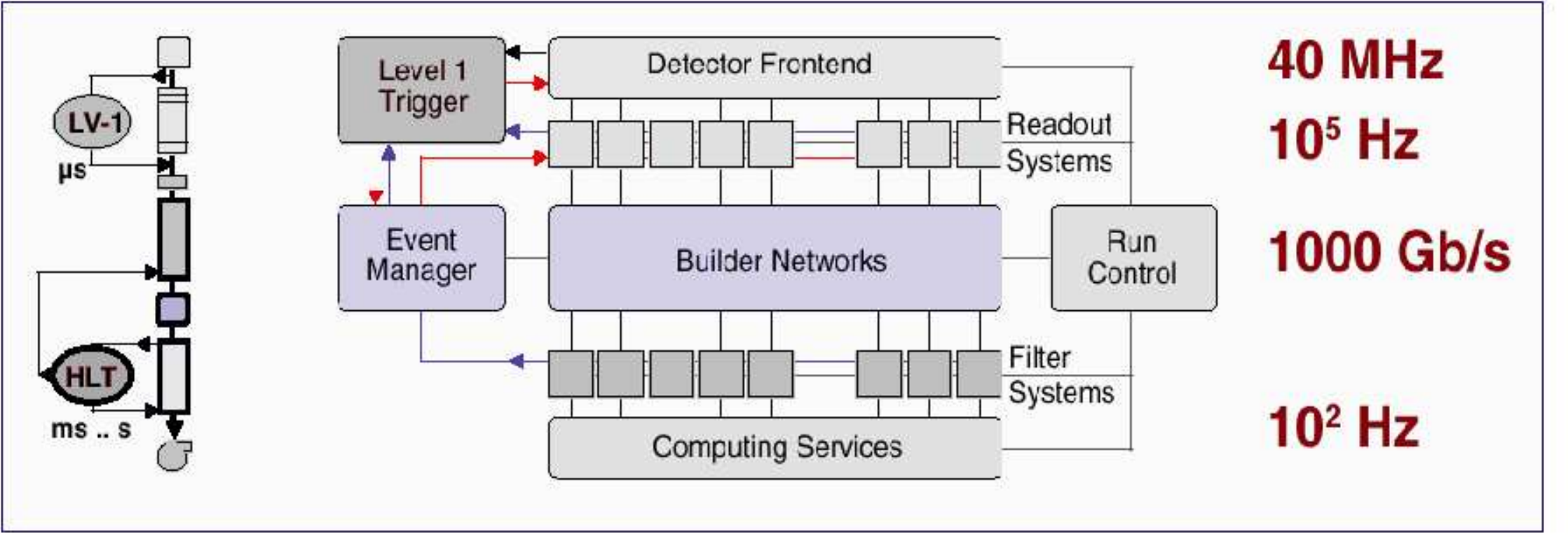}
    \caption{Simplified scheme of the CMS High-Level trigger. Once the
    Level-1 trigger has accepted an event, the detector data are transferred
    from the front-end electronics to the Readout System buffers.
    Then the whole amount of data are organized to build the event
    fragment. The event is then processed by the CPU farm of the
    HLT system to validate (or reject) the selection made by the
    first level trigger. The selected events are then stored on
    external memory supports.}
    \label{fig:TwoLevels}
  \end{center}
\end{figure}

The strategy adopted by the CMS Collaboration for the HLT System is simpler than
the one of ATLAS: when a L1 trigger accept is produced, the whole amount of
full granularity detector data are moved from the on-detector buffers to the DAQ
memories, to allow the complete event fragment building~\cite{CMS_TDR}.
 Once available, the
event is reconstructed by computer programs of different complexity and accuracy
to reduce the 100 kHz L1 rate to the about 100 Hz rate of
event recording. These algorithms run in large CPU farms, built with something
like 1000 dual-processor PCs. Similarly to the
ATLAS Event Filter trigger, the CMS HLT latency is of the
order of a few s. This approach, that differently from  
ATLAS consists in
one HLT step only, is very challenging from the point of view of the bandwidth
size required for the full detector data movement, 
that is estimated of be of the order of
1000 Gb/s; however, given the present available technologies and the extrapolation
of these to the next three years, this scheme can be realistically implemented. 
This CMS Trigger scheme is sketched in Figure~\ref{fig:TwoLevels}.

The architecture of the Trigger/DAQ system of CMS is made in a modular structure based on 
8 basic slices that can be inserted to match the output of the L1 trigger system. 
Each slice can process up to 12.5 kHz of events read from the Level-1
trigger. The system can run with one slice only up to the full 8 slices
that allow the processing of the nominal L1 throughput.
This approach is 
particular suitable during the initial LHC luminosity 
since the full potentiality of the Trigger/DAQ system is not needed.
As for ATLAS, this scheme 
allows also the deferral scenarios of the Trigger DAQ system to 
allocate financial resources needed for others CMS subdetectors on a 
critical path. 
More details on the CMS High-Level Trigger (rates and acceptances
for various physics channels) are reported in~\cite{amapane}.

\section{The simulation of events from p-p collisions}
{\large {\sl A. Giammanco, A. Perrotta}}
\vspace{0.5cm}

\begin{figure}[htb]
\centering
\includegraphics[height=60mm]{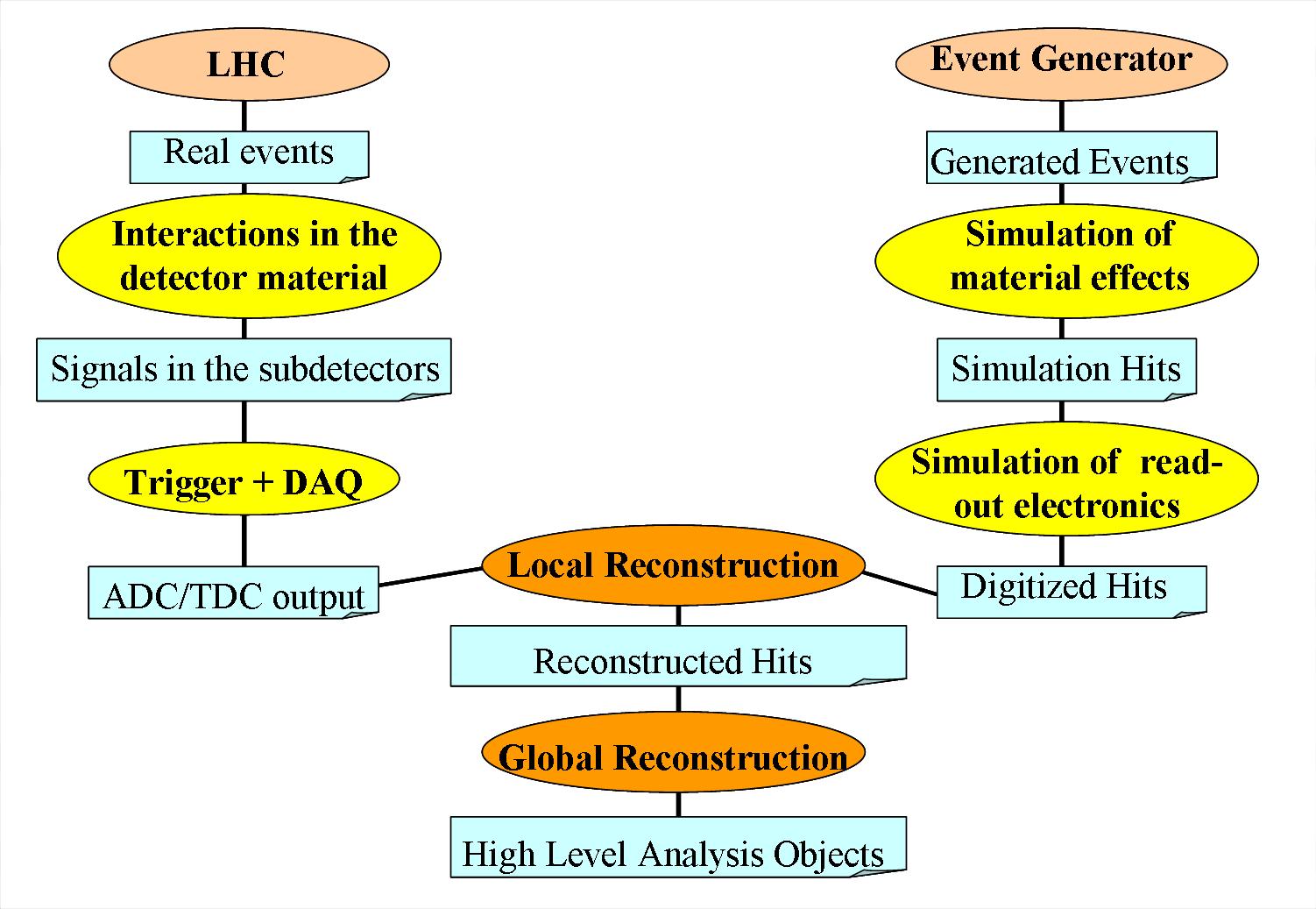}
\caption{Sketch of the parallel physics processes and simulated ones leading to the raw data,
first, and to the final high level analysis objects later on.}
\label{fig:fullsimu}
\end{figure}

\begin{figure}[htb]
\centering
\includegraphics[height=42mm]{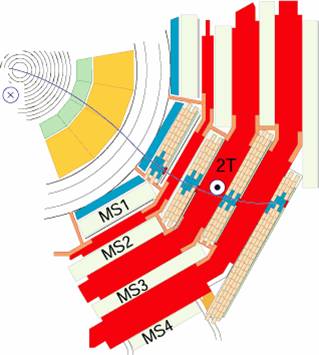}
\caption{Trajectory of a muon in a slice of the CMS detector. Outside the coil, the magnetic field
changes versus, as can be seen by the change of sign in the curvature radius of the trajectory.}
\label{fig:amuon}
\end{figure}

\begin{figure}[htb]
\centering
\includegraphics[height=60mm]{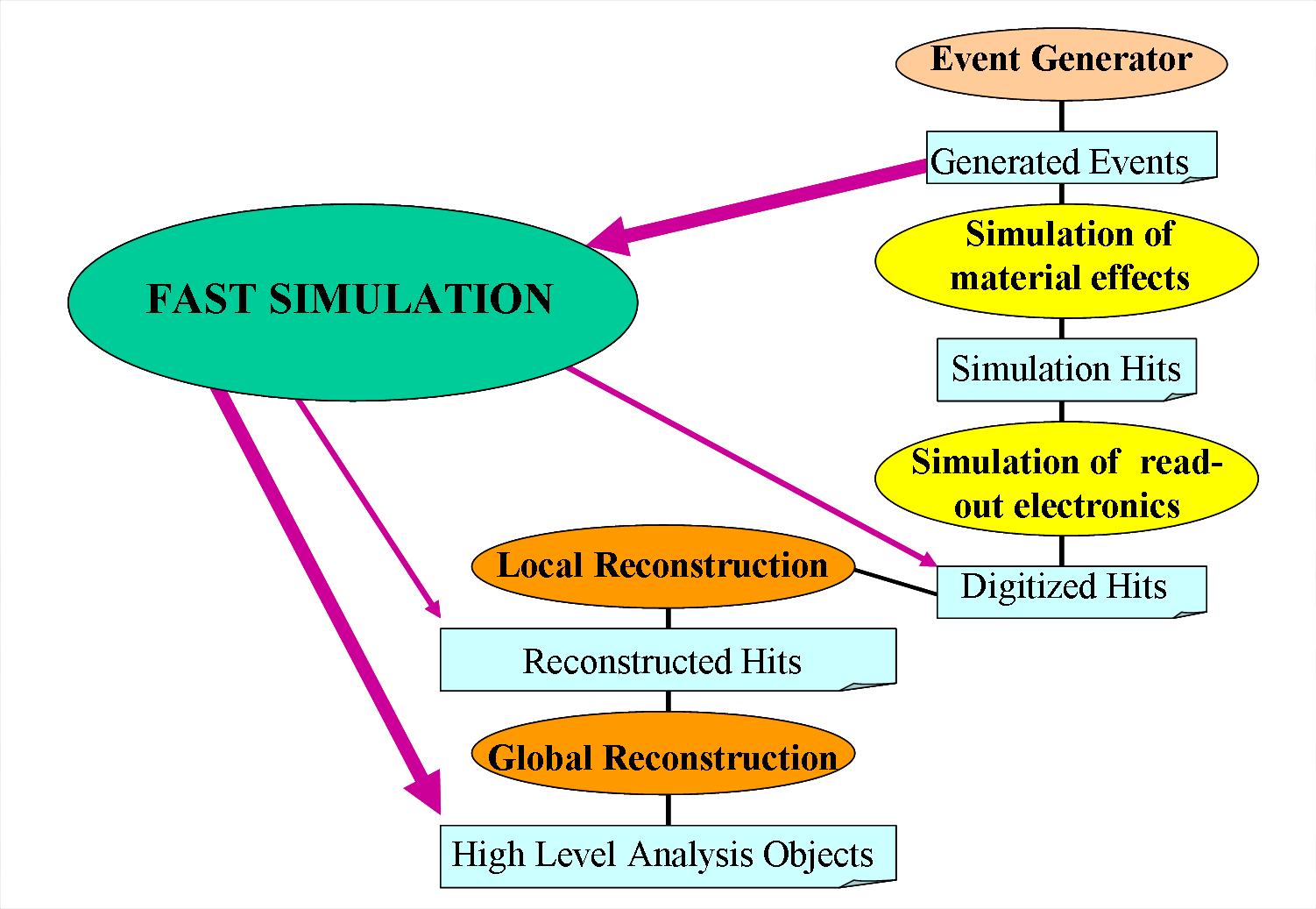}
\caption{Block diagram of a full and a fast simulation in a typical LHC experiment. They all start
from the same Monte Carlo generated events and aim to produce as similar as possible final analysis objects.}
\label{fig:fastsimu}
\end{figure}

Samples of simulated events are heavily used for the interpretation of
the data collected by high energy experiments to determine the
expected distributions of the particles in the final states accounting
for experimental effects like the resolution of the detectors or the
efficiency of selection cuts used to isolate a particular final state.
The simulation of an ``event'' requires:
\begin{itemize}
\item[-] the generation of the four-momenta of the particles in the final
  state;
\item[-] the simulation of the interaction of the generated particles with the detector;
\item[-] the simulation of the digitization phase, i.e. the process by
  which analogic electronic signals resulting from the particle-material
  interaction, get shaped, discriminated and read out by dedicated electronic devices
\item[-] running on the simulated digitized signals the programs which
  mimics the different levels of the trigger;
\item[-] running on the simulated digitized signals the programs for the
  reconstruction of the event both at the local subdetector level and
  to build the higher level analysis objects used for the final physics analysis.
\end{itemize}
If the output of the simulation after the digitization phase has the same format as the really
collected raw data, the same reconstruction software as used on the real data can be applied to
simulated ones. 
Effects as electronic noise in the detectors, event overlapping (``pile-up''),
instrumental dead-times, etc., must be properly taken into account to provide realistic reconstructed
analysis objects.
Figure~\ref{fig:fullsimu} summarizes the various steps leading to the final high level analysis 
objects of a typical LHC general purpose experiment, starting either from a real collider
interaction or from a Monte Carlo generated event.
A particle
(a muon, for example, that crosses the whole detector as shown in Figure~\ref{fig:amuon})
passes
through several layers of different subdetectors, built with different
materials; it passes through
passive material like the cables, the magnet, the mechanical support structure; it enters also
regions with different values of the magnetic field. All those effects and materials must be
properly taken into account for a precise detector simulation\footnote{Quite often, the very
final arrangement for auxiliary equipments, like cables, shieldings, etc., is not finalized until
the detector is fully built and closed, thus leading to some new ``final'' simulation samples to
be produced only when the correct account of the crossed material is known.}.
High level of details and precision can be achieved with an accurate full simulation. Detector
responses can further be validated and tuned with: test beam data;
\emph{in situ} calibration data
(e.g. cosmics, halo muons); calibration data from LHC collisions
($Z\to\mu^+\mu^- $, $Z\to e^+e^-$, $\pi^0\to\gamma\gamma$, etc.).
As experiments get more complex, also their simulations become more
complex and  CPU-time consuming. Therefore, while for several tasks the
most possibly detailed simulation is advised, there are many where the
required level of precision makes more suitable a less detailed but
much quicker simulation, the so-called {\it fast simulation}. 
Domains where a fast simulation is more suitable than a full one are:
\begin{itemize}
\item[-] quick and approximate estimates of signal and background rates;
\item[-] fast development of analysis methods and algorithms;
\item[-] test of new generators or new theoretical ideas in a realistic environment;
\item[-] scan of complex, multi-parameter spaces (like e.g. SUSY);
\end{itemize}
Emulation of intermediate quantities, as digitized or reconstructed detector hits,
could also be provided. Figure~\ref{fig:fastsimu} compares the job done by a fast simulation with what
done by a full simulation.
Fast simulation emulates the combined result of detector simulation and reconstruction, and it is
therefore generally tuned and validated with the full simulation results (while full simulation is
tuned and validated with the real data).
\subsection{Full simulation of the main detectors components}
In the following paragraphs the way how the main components of the
ATLAS and CMS detectors are accounted for in the full simulation will be
outlined and then compared with the methods used in the fast simulation. 
Details on the two detectors and their simulation and reconstruction
software can be found in~\cite{atlasdet,atlascptdr} for ATLAS
and~\cite{cmsdet} for CMS. 
\subsubsection{Simulation of the inner tracker systems}
\begin{figure}[htb]
\centering
\includegraphics[height=84mm]{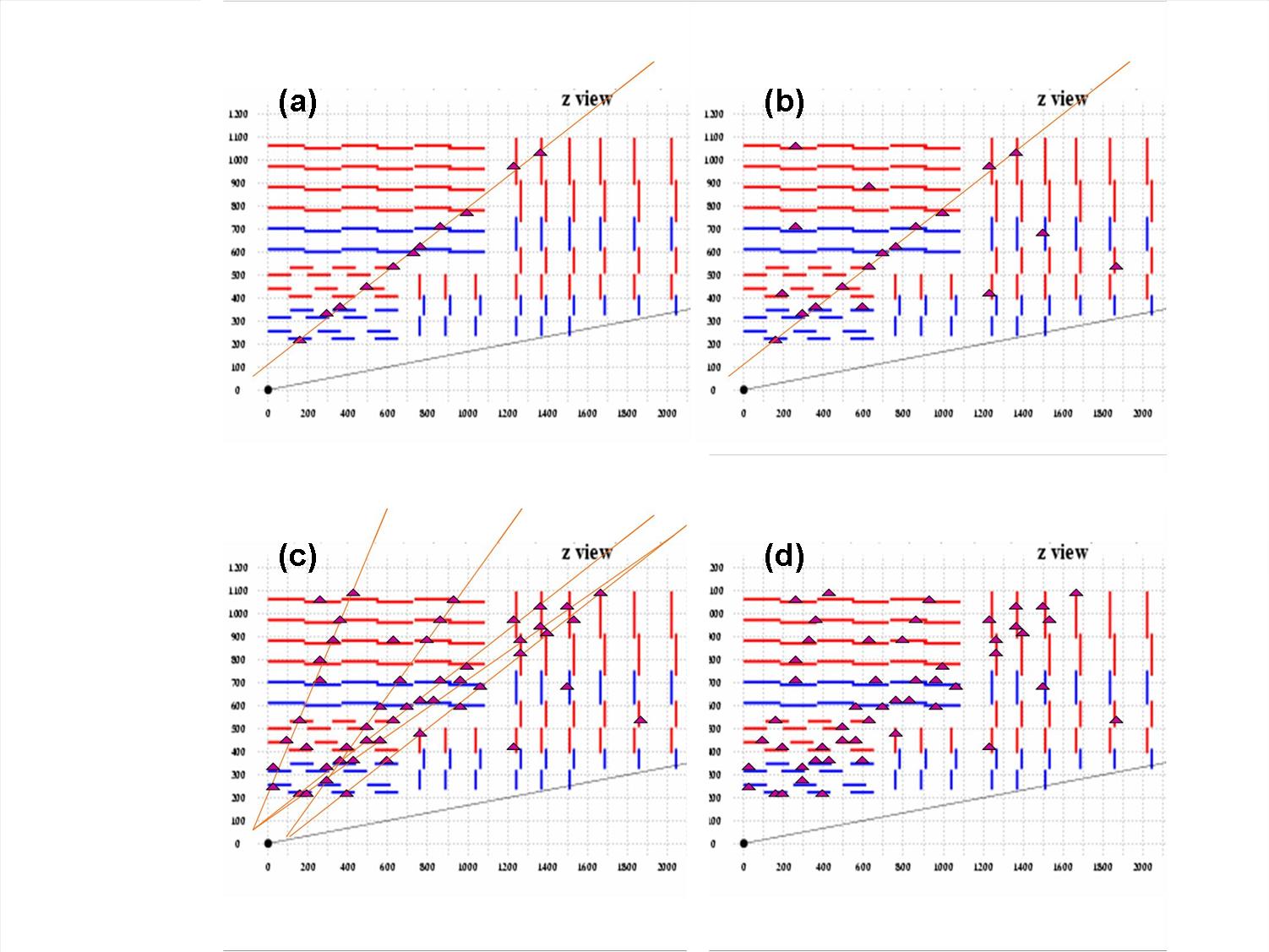}
\caption{Steps performed in the simulation of charged particles crossing the inner tracking devices
(see text).}
\label{fig:trk}
\end{figure}

A charged particle crosses the active layers of the inner tracking detectors (silicon strips and
pixels in CMS; silicon strips, pixels and an outer transition radiation detector in ATLAS).
Propagation is affected by multiple scattering in the detector and surrounding material.
Within each detector layer, the particle looses energy along the path between its entry and exit
point. The produced charges are collected causing a signal in the dedicated
electronics (Figure~\ref{fig:trk}a). Gaussian noise is added on top of those signals, and also to the
other channels not interested by the trajectory of any particle (Figure~\ref{fig:trk}b). In the same event
other particles add up, coming from the very same generated event, multiple interactions, 
in-time or out-of-time pile-up
(Figure~\ref{fig:trk}c). All charges are linearly added up in case of overlap, then discriminated
and digitized, ending up with the raw data of the tracking detector layers. Those raw data, separated
from the information of the generated particles, are the input for the reconstruction phase
(Figure~\ref{fig:trk}d).
Tracking algorithms apply pattern recognition and track fit; magnetic field, multiple scattering,
material effects are also taken into account. Different use cases can be considered:
low/high $p_T$, searches for displaced vertices, etc.
At the end of the reconstruction (as for the real data) the exact 1-to-1 correspondence between
generated charged particles and reconstructed tracks is generally lost, and it can only be
restored on a probabilistic basis.
\subsubsection{Simulation of the calorimeters}
\begin{figure}[htb]
\centering
\includegraphics[height=48mm]{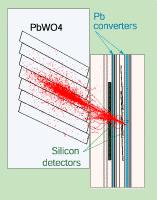}
\caption{Side view of the ECAL of CMS, with an electromagnetic shower
  that starts in the preshower and fully develops in the
  electromagnetic calorimeter.} 
\label{fig:ecal}
\end{figure}

Electrons and photons in the electromagnetic (ECAL) and hadrons in the
hadronic (HCAL) calorimeters generate large showers,
respectively via pair production and bremsstrahlung processes, see Figure~\ref{fig:ecal}, and via
hadronic interactions.
On the other hand  electromagnetic and hadronic 
calorimeters are coarser grained detectors if compared to the tracking  devices. 
To perform a realistic simulation, several effects must be taken into
account, i.e. for electromagnetic calorimeters:
variation of the light collection along the length of the crystal
(homogeneous ECAL) or of the fibers (sampling ECAL); modified crystal
transparency with large integrated doses (homogeneous ECAL); noise;
electronic thresholds. Simulation parameters must be tuned to reproduce the results of the test beams.
The whole charge collected in one, or even more than one, crystal or tile is read out together.
Therefore, in the reconstruction, exact 1-to-1 correspondence between generated and reconstructed
particles is lost and cannot be restored\footnote{The exception being isolated electrons,
photons or hadrons at low luminosity}. \\
Clusters of energy deposits in the HCAL represent the jets,
which are the high level analysis objects obtainable starting from the calorimetric showers;
different clusterization algorithms and recombination schemes are available, depending on the
needs of the specific analysis.
\subsubsection{Simulation of the muon detectors}
Muon detectors are tracking devices placed in the outer part of the detector and exploiting the large 
penetrating power of muons. Passing muons produce ionization charge in the drift cells; charges drift
towards the sense wires with a drift velocity which is in general dependent on the impact position, 
muon direction, residual magnetic field. Contributions from electronic noise, neutron background, halo
muons, muons from pile-up events (in-time or from a different beam crossing), punch-through hadrons,
must be taken into account.
Local reconstruction starts in a single layer
and continues by correlating track segments in the different substructures. Global
reconstruction matches these local segments with those of the inner
tracking system (plus possibly signals from the calorimeters that must
be compatible with the particle being a minimum ionizing particle). 
Exact 1-to-1 correspondence between generated and reconstructed muons is formally lost although,
given the lower track density, there is the matching probability is
higher than in the inner tracker.
\subsubsection{Simulation of the trigger}
As mentioned in Section~\ref{sec:trigger}, ATLAS and CMS achieve
rate reduction by means of their L1 and HLT trigger systems: events
rejected by the trigger are lost forever. 
The simulation must reproduce the trigger decision: it is not necessary to actually drop all events
that do not pass the trigger, but it must be made clear which can be used for the analysis, and which
cannot. Since the HLT reconstruction algorithms are similar but not
generally the same as those used in the off-line analysis, as for
example they cannot access the whole calibration data-base, 
to obtain realistic performance in the
simulation code specialized trigger modules must be developed.
\subsubsection{Timing}
\begin{figure}[htb]
\centering
\includegraphics[height=46mm]{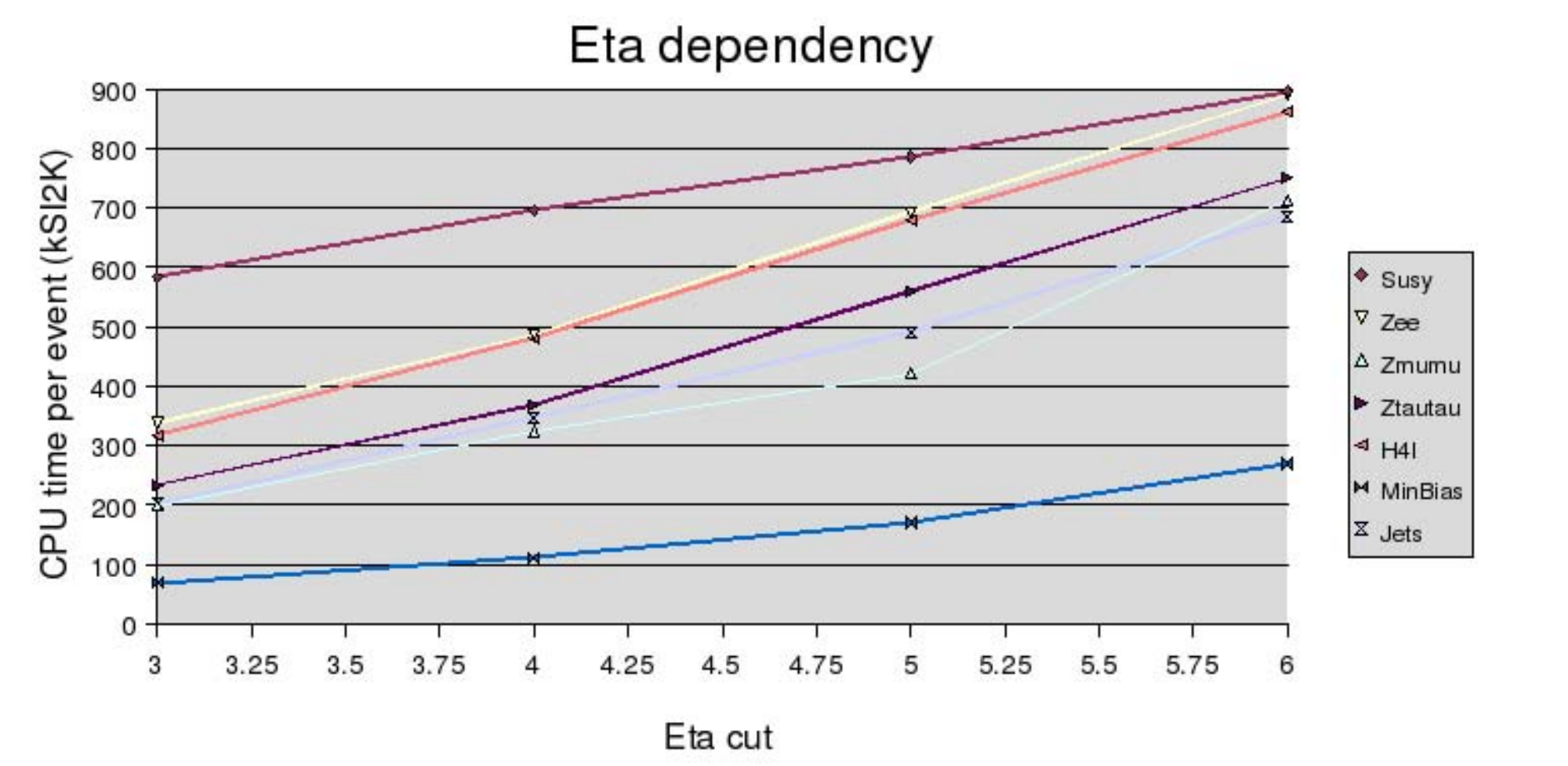}
\caption{Average CPU time, in kSI2k, needed to fully simulate different kind of events in ATLAS,
as a function of the upper limit of the interval of pseudorapidity in which particles are propagated
and their interaction with the detector simulated.}
\label{fig:atlascpu}
\end{figure}

To obtain the high level of details and precision of the full simulations a considerable amount of CPU
time is required. As an example, for CMS it was estimated~\cite{cmstiming} that for a typical LHC
high-$p_T$ p-p collision in a 1~GHz Pentium~III\footnote{To obtain the corresponding values in kSI2k-sec,
the standard CPU speed normalization between machines based 
on the SPECint\textregistered2000 benchmark for integer calculations,
those times obtained with a 1~GHz machine must be multiplied by a factor 0.46.} the required processing
times were:
\begin{itemize}
\item[-] less than 100~ms/evt for the Monte Carlo event generation;
\item[-] 100-200~s/evt for the simulation of the material effects;
\item[-] 1-10~s/evt for the digitization (simulation of the read-out electronics);
\item[-] 10-100~s/evt for the reconstruction.
\end{itemize}
Therefore, the total CPU-time spent before the analysis can start ranges from 3 to 5 minutes per 
event. Those estimates were done with the previous framework and event
data model of CMS: it is expected, however, that timings will not
change that much with the new CMS simulation code.
The CPU time needed for the event simulation in the present release of the ATLAS software can be derived
from Figure~\ref{fig:atlascpu}~\cite{rimoldi}, for different types of events and as function of the
largest absolute value of the pseudorapidity simulated (in a p-p collider the track density, and
therefore the CPU time needed to simulate the complete event, increases strongly with pseudorapidity).
\subsection{Fast simulation in ATLAS}
\label{par:atlfast}
ATLFAST~\cite{atlfast} is the package for fast simulation developed and used in ATLAS. It includes
most crucial detector aspects, as jet reconstruction in the calorimeter, momentum and energy smearing
for electrons and photons, effect of the magnetic field, and missing energy. It provides, starting
from the generated particles, the list of reconstructed jets, isolated leptons, photons, muons, and
missing transverse energy. It provides also (optionally) the list of reconstructed charged tracks.
No particle propagation, nor interaction with the detector material is simulated; a coarse detector
geometry is considered to define the acceptances. Fast simulation in ATLAS is therefore obtained by
smearing directly the Monte Carlo truth informations with efficiencies and resolutions as obtained from the
full simulation.

\subsubsection{Tracking}

Emulation of track reconstruction is provided (only optionally) for charged particles inside the inner
detector. It is obtained by smearing three-momenta and impact parameters, as indicated in the full 
simulation studies, with different parameterizations of the smearing and of the reconstruction
efficiency for muons, pions and electrons.

\subsubsection{Calorimetric clusters}

In the present implementation, all electron or photon energy is deposited in one single ECAL cell,
and all hadrons energy in one single HCAL cell. A new parameterization has been
studied~\cite{newatlascal} and is ready to be implemented. In this new parameterization, the
transverse energy of all undecayed particles is summed up in cells having the same granularity as
the calorimetric L1 trigger ($\Delta\phi \times \Delta\eta = 0.1 \times 0.1$), which is coarser
than the granularity of the full simulation; the longitudinal segmentation is limited to the
separation between ECAL and HCAL. The effect of the 2~T magnetic field is taken into account.
Generic calorimetric cluster reconstruction is started from those cells, and an appropriate energy
smearing and reconstruction efficiency is applied after cluster
identification from Monte Carlo truth as
electron, photon or hadron.

\subsubsection{Jets}

Calorimetric clusters non associated with isolated $e$ or $\gamma$ are associated into jets and
further smeared, with a parameterization which depends on the presence of quarks of a given flavour
in the generated particles that originated the calorimetric clusters. Different parameterizations
are also applied for different luminosity scenarios, reflecting the different amount of pile-up.
Reconstruction and tagging efficiencies are not included in ATLFAST, but they can be applied
``by hand'' at a later stage.

\subsubsection{Muons}

Three possibilities are foreseen for the parameterization of the momentum resolution, depending on
the subdetectors used for the muon reconstruction: muon system stand-alone, inner detector stand-alone,
or the two combined. Muons can be flagged as isolated or non-isolated. Muon tagging efficiency is
not included in ATLFAST, but it can be applied  at a later stage.

\subsubsection{Trigger}

Only primitive trigger routines are considered, not meant to cover all ATLAS triggers and levels.
They are aimed essentially at eliminating events which have no chance of passing ATLAS L1 and L2
triggers.

\subsubsection{Pile-up}
\begin{figure}[htb]
\centering
\includegraphics[height=82mm]{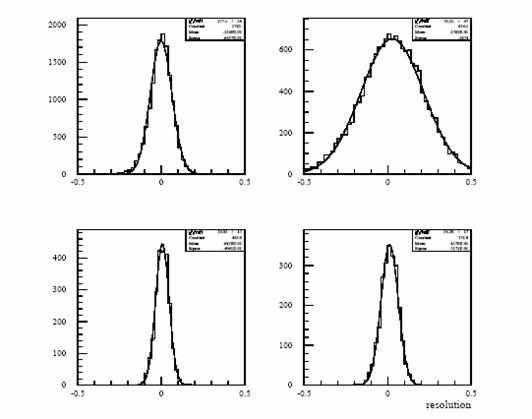}
\caption{The $p_T^{\rm jet}$ resolution for reconstructed jets with $40 < p_T^{\rm jet} < 50$~GeV$/c$
(top) and $200 < p_T^{\rm jet} < 250$~GeV$/c$ (bottom), obtained in ATLFAST with the default cone
algorithm for low (left) and high (right) luminosity.}
\label{fig:atlfastpileup}
\end{figure}

Pile-up events are not simulated in ATLFAST, but a different smearing of jets due to pile-up is provided
as a function of the luminosity, see Figure~\ref{fig:atlfastpileup}. Also the parameterization of the
trigger selection allows for the low and high luminosity options ($2\times 10^{33}$~cm$^{-2}$s$^{-1}$ and
$10^{34}$~cm$^{-2}$s$^{-1}$ respectively).

\subsubsection{Timing}

A very fast processing is obtained thanks to the approach chosen in ATLFAST of relying on
parameterizations of the properties of the final analysis objects, without simulating interactions
of particles with the detector material, nor attempting any reconstruction. A gain of about four
orders of magnitude is claimed with respect to fully simulated similar events, which corresponds to a
computation time of just a few hundred milliseconds per event.
\subsection{Fast simulation in CMS}
\label{par:cmsfast}
CMS software~\cite{cmsdet} has recently completed the migration from the previously adopted framework
to the present one. We describe here the package FAMOS for the fast simulation of particle
interactions in the CMS detector, based on the old framework; its main features will however remain
basically unchanged in the new framework.

The output of FAMOS is designed to be as close as possible to the output of the full simulation and
reconstruction of CMS. It delivers the same physics objects (calorimetric hits and clusters, tracker
hits, and reconstructed tracks and muons), with identical interface: they can be used as inputs of
the same higher-level analysis algorithms ($b$-tagging, electron, muon and tau candidates, jet
clustering, lepton isolation, etc.) as the real or fully simulated data.

Particles in FAMOS are propagated in the nominal magnetic field through the inner tracker and up to
the entrance in the calorimeters. The following interactions are simulated in the tracker material:
\begin{itemize}
\item[-] electron bremsstrahlung;
\item[-] photon conversion;
\item[-] energy loss by ionization for charged particles;
\item[-] multiple scattering for charged particles.
\end{itemize}
Electron, photon and hadron showering is allowed in the ECAL and HCAL. Nuclear interactions are not
simulated in FAMOS\footnote{Their implementation is indeed foreseen in the new fast simulation.},
which implies that hadronic showers never initiate before the calorimeters, and there is a lower
number of secondary vertices. As will be described in section~\ref{sec:btagging}, this implies in
turn a different $b$-tagging significance with respect
to the full simulation which needs therefore a separate tuning.

\subsubsection{Tracking}

\begin{figure}[htb]
\centering
\includegraphics[height=55mm]{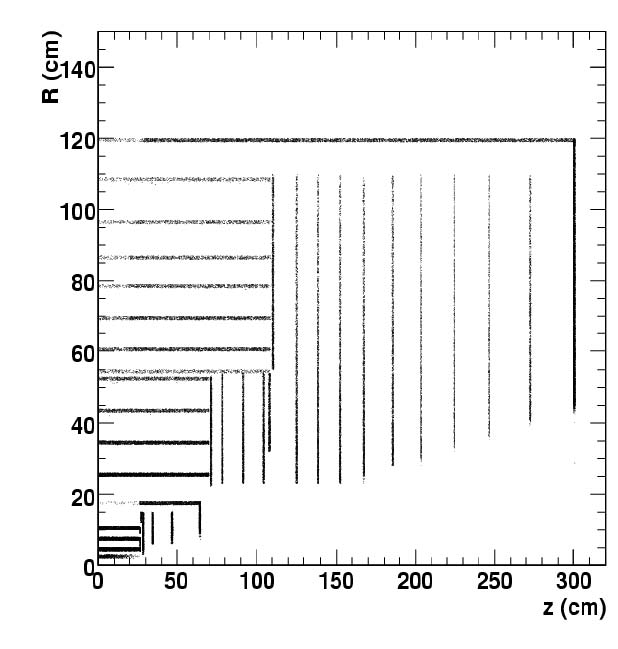}
\includegraphics[height=55mm]{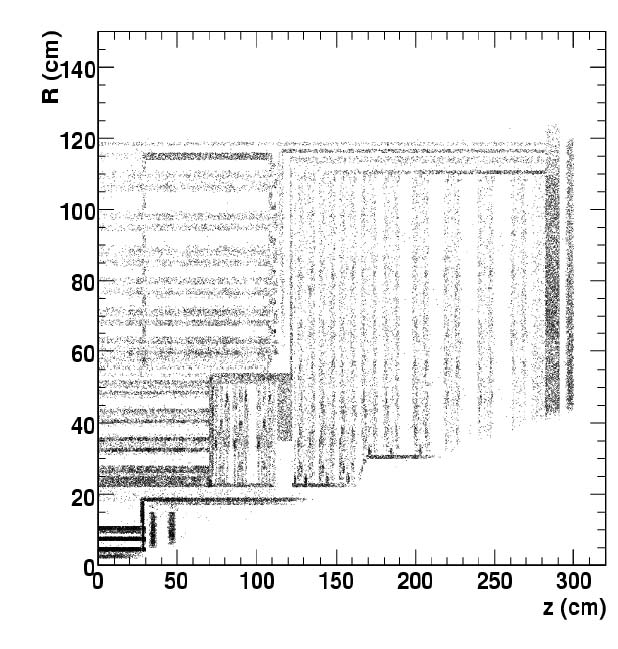}
\caption{A radiography of the inner part of the CMS detector, were are the tracker layers, obtained
by recording the points where a photon converted in the fast (left) and full (right) simulations.}
\label{fig:cmsradiography}
\end{figure}

Charged particles in FAMOS are traced through a simplified detector geometry. The inner part of CMS
is treated as composed by thin cylindrical layers of pure silicon, whose thickness is tuned on
the number of bremsstrahlung photons with $E_\gamma > 500$~MeV radiated by energetic electrons
traversing any such layer. A comparison of the material content of the inner CMS in FAMOS and in the
full simulation is shown in Figure~\ref{fig:cmsradiography}, where the photon conversion points in the
plane $R$-$z$ are recorded.

Charged particles in FAMOS propagate in the magnetic field through the tracker layers; multiple
scattering and energy loss by ionization are taken into account. Intersections between simulated
trajectories and tracker layers give the ``simulated hits''; they are then smeared and turned, with
a given probability, into ``reconstructed hits''. An emulation of seeding and pattern recognition is
performed with the reconstructed hits originating from a given propagated particle, followed by a fit
of the track done with the same fitting algorithms used the 
reconstruction of full simulated events.

\subsubsection{Calorimeter response to $e$ and $\gamma$}

In FAMOS, the simulation of an electron shower makes use of the Grindhammer 
parameterization~\cite{grindhammer}, implemented in the GFLASH code~\cite{geant}. The photon case
goes back to the electron case after the first $\gamma \to e^+e^-$ splitting.
Shower develops as if the whole ECAL were a homogeneous medium.
The energy deposits are sliced longitudinally; in each slice energy spots (calorimeter hits) are
distributed in space according to the radial profile and placed in the actual crystal geometry.
The following effects are simulated: leakage (which is propagated to the HCAL), gaps between
ECAL modules, shower enlargement due to the B-field, electronic noise and zero suppression.
Starting from the calorimeter hits, clustering is obtained as in the reconstruction of full simulated events.

\subsubsection{Calorimeter response to hadrons}

Charged and neutral hadrons propagate to the ECAL and HCAL entrances. The energy response is derived
from a full simulation of single pions generated at fixed $p_T$ values between 2 and 300~GeV$/c$.
Smeared energy distributes in the calorimeter cells using parameterized longitudinal and lateral
shower profiles. Other hadrons are treated as pions of the same $p_T$.

\subsubsection{Muons}

Muons in FAMOS are not propagated until the CMS muon chambers. Their calorimetric response is
tabulated in a similar way as for hadrons. The response of the muon chambers is parameterized on
samples of fully simulated single muons (with $2<p_T<1000$~GeV$/c$) to reproduce efficiencies and
resolutions, assuming a gaussian distribution for the final quantities. Different parameterizations
are provided for L1 trigger muons, HLT muons, and global muons. HLT and global muons may require a
correlation with the reconstructed track.

\subsubsection{Trigger}

L1 and HLT trigger signals and primitives are obtained as a ``by-product'' of the fast simulation of
the corresponding subdetectors. Decision functions are then reconstructed starting from those trigger
primitives with the very same logic as in the real data. 

\subsubsection{Pile-up}

In-time pile-up minimum bias generated events are superimposed to the signal events, and their
particles treated as all other particles in the event. No out-of-time pile-up is considered.

\subsubsection{Timing}

A complete event takes a couple of seconds to be simulated and reconstructed with FAMOS (about 1~s
in FAMOS itself, the rest in the analysis and framework overhead); it is slightly more with the
pile-up superimposed. It consists of more than two orders of magnitude gain with respect to the full
simulation and reconstruction.
\subsection{A case study: full vs. fast simulation in CMS}
\label{par:cmsdetails}
A few comparisons between the former fast and full simulations of CMS (respectively FAMOS and OSCAR,
based on GEANT4\cite{geant}) are shown here. Although the agreement between the results of the two
simulations is good for most of the relevant observables, emphasis will be given to the remaining
discrepancies, with a discussion of the possible causes.

\subsubsection{Electrons and photons}

In the fast simulation ECAL is represented as a homogeneous medium. This allows by itself such a saving
of CPU time, that a relatively high degree of realism can be afforded on other aspects:
\begin{itemize}
\item[-] a lot of details are allowed (after optimization, about 1500 hits are calculated per shower of
35~GeV);
\item[-] the front and rear leakage, the fraction of signal lost in the inter-module voids, and the shower
spread due to the magnetic field are simulated;
\item[-] the calorimetric noise is added to the signals;
\item[-] for very high energy electrons, the punch-through into HCAL is also parameterized;
\item[-] fake electrons can show up when an ECAL cluster is associated with a simulated seed originating
from hits produced by the tracks of the event.
\end{itemize}

\begin{figure}[htb]
\centering
\includegraphics[height=48mm]{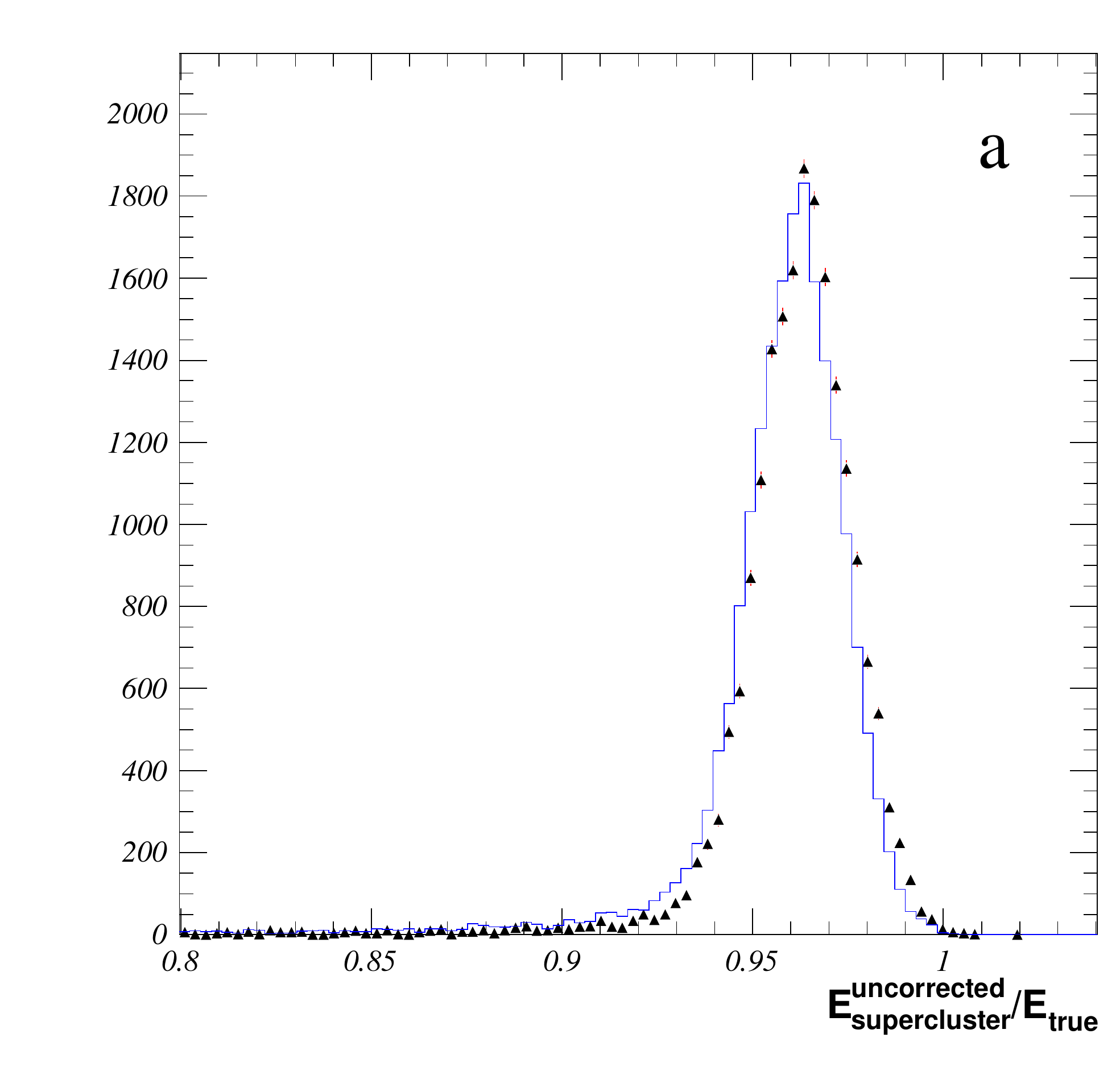}
\includegraphics[height=48mm]{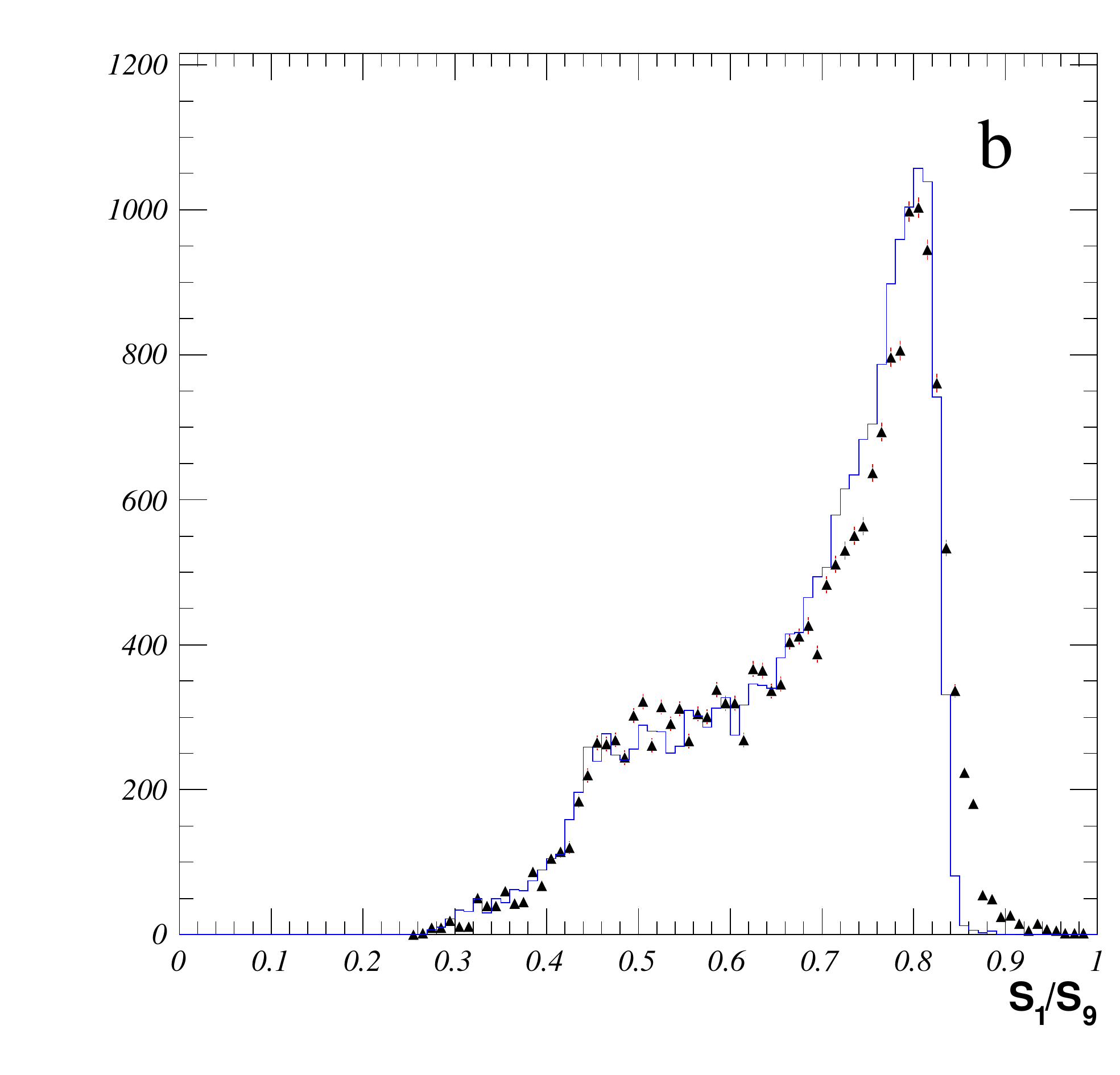}
\caption{Energy deposited in an ECAL supercluster over true energy (left) and the ratio of the energy
in the most energetic crystal to that in the surrounding $3\times 3$ crystals windows (right) for
isolated electrons in the CMS fast (dots) and full (histogram) simulations.}
\label{fig:electrons}
\end{figure}

The effect of all these details can be seen in Figure.~\ref{fig:electrons}: in general, the reconstructed
energies in FAMOS reproduce the corresponding ones from the full simulation with an accuracy at the
per mill level in the calorimeter barrel, and at the per cent level in the endcaps.

\subsubsection{Muons}

\begin{figure}[htb]
\centering
\includegraphics[height=48mm]{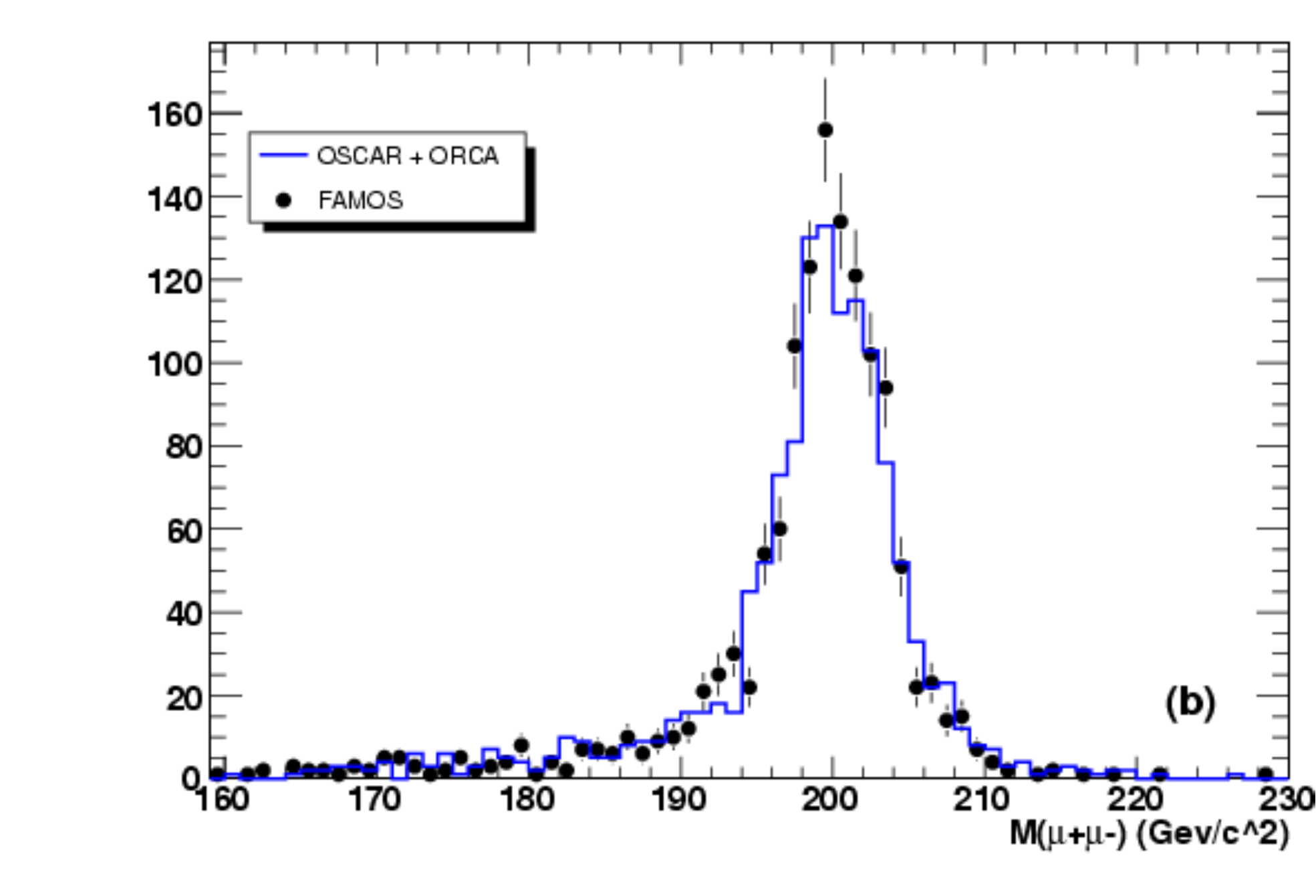}
\caption{Invariant mass peak of di-muons coming from the decay of a heavy Higgs, in the CMS fast (dots)
and full (histogram) simulations.}
\label{fig:h200mumu}
\end{figure}

As seen in~\ref{par:cmsfast}, the simulation of muons in FAMOS is not
very refined. 
In spite of that, the higher-level variables show a remarkable agreement with the full simulation, one
example being the invariant mass of a di-muon resonance, shown in Figure~\ref{fig:h200mumu}.

\subsubsection{Fake tracks}
\label{sec:faketracks}

As explained above, the tracks in FAMOS are not currently obtained from a pattern
recognition procedure, but from a fit of the hits associated to a ``true'' charged particle.
Because of this use of the Monte Carlo truth during the reconstruction step, no fake tracks 
(i.e., random combination of hits from more than one track, with or without the contribution of fake
hits coming from detector noise) can contaminate the final sample of reconstructed tracks.

Studies in full simulation show that 0.5\% of the tracks in the ``low luminosity'' scenario 
($2\times 10^{33}$~cm$^{-2}$s$^{-1}$) are fakes. At that level, the incidence of fake tracks is
irrelevant for most of the LHC studies, and a realistic simulation of this combinatorial background
starting from the hits would require a pattern recognition, which would result in a significant increase
of CPU time. 

\subsubsection{Impact parameter and $b$-tagging}
\label{sec:btagging}

\begin{figure}[htb]
\centering
\includegraphics[height=42mm]{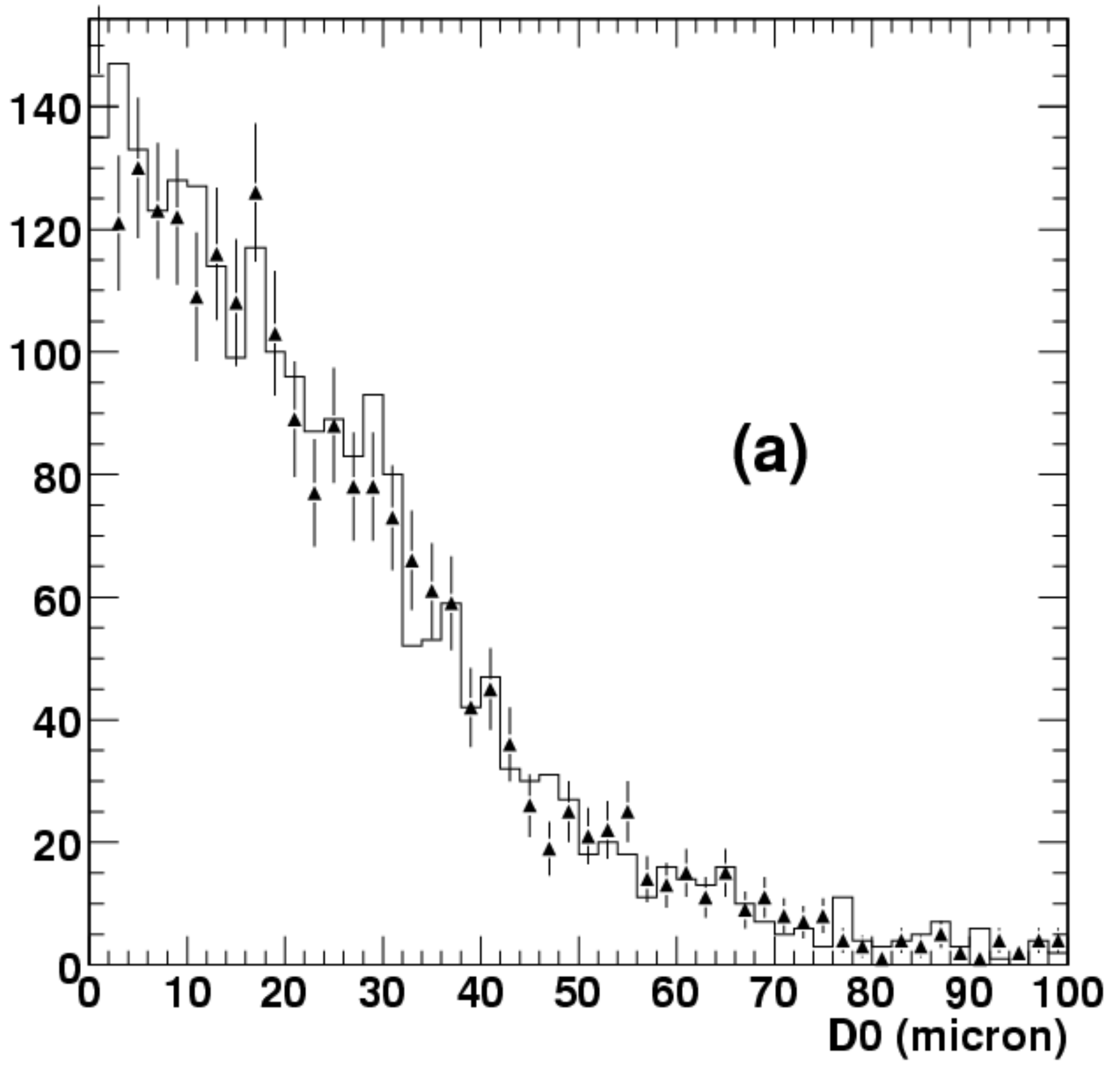}
\includegraphics[height=42mm]{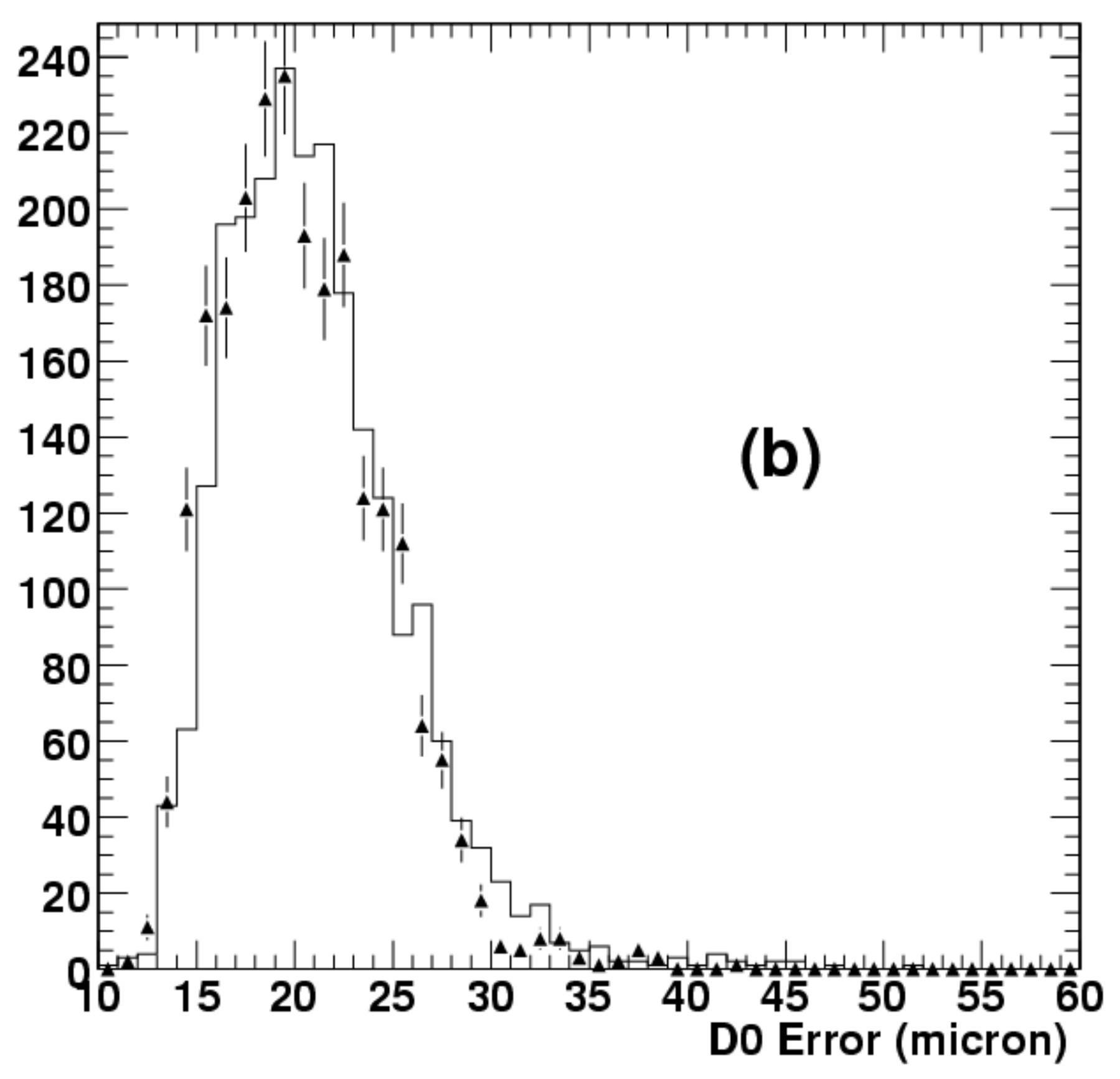}
\caption{Impact parameter (left) and its error (right) for isolated muons in the CMS fast (dots) and full (histogram) simulations.}
\label{fig:d0}
\end{figure}

FAMOS applies to the tracks the same $b$-tagging algorithms applied on data and full simulation.
Since the impact parameter is the key ingredient of some of the best performing $b$-tagging algorithms,
the validation of this variable (shown in Figure~\ref{fig:d0} for single muons) is of paramount importance.
It has to be remarked that the impact parameter was not directly tuned to reproduce the full simulation
shape, thus making this full/fast simulation agreement a particularly significant test.

\begin{figure}[htb]
\centering
\includegraphics[height=82mm]{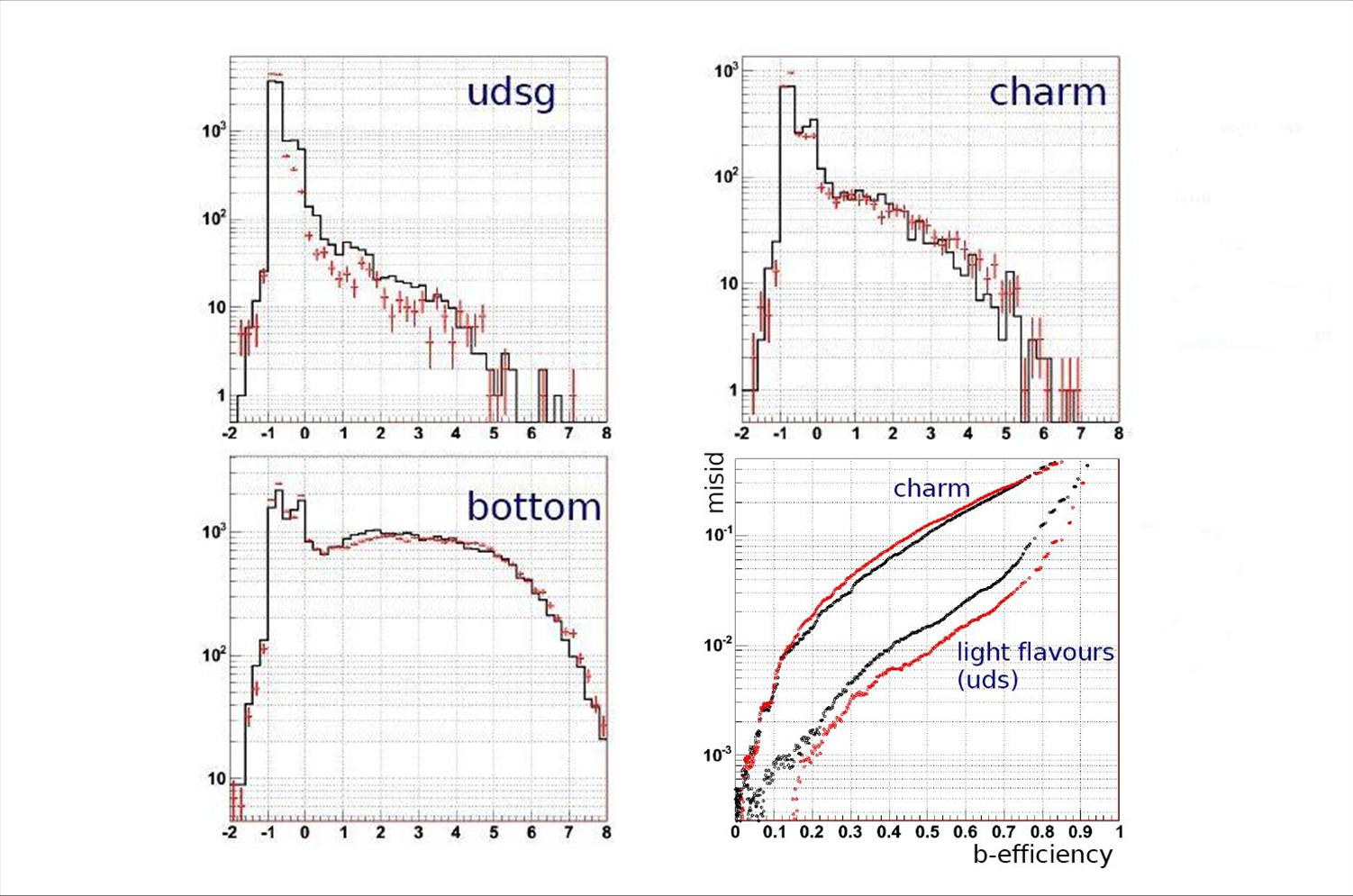}
\caption{
In the first three plots: output of the standard CMS $b$-tagging algorithm, in the CMS fast (red points) and full
(black histogram) simulation. In the last plot: probability of misidentification for non $b$-jets versus efficiency
of identification of true $b$-jets, in the CMS fast (red) and full (black) simulation.}
\label{fig:btag_disc}
\end{figure}

Unfortunately, the agreement observed in Figure~\ref{fig:d0} is not
enough to guarantee the same $b$-tagging performance on the fast
simulated events, as evidenced by the first three plots in
Figures~\ref{fig:btag_disc}, which show the output of one high-level
$b$-tagging algorithm mainly based on the impact parameters of the
charged tracks, for $b$, $c$ and $udsg$-initiated jets: the output of
this algorithm in the fast simulation reproduces quite well the
behavior of the full simulation 
for $b$ and $c$-jets, while the same is not true for jets originating from lighter partons.
This will affect all analyses in which significant sources of
background come from processes where some light jets are mistagged as
coming from heavy quarks. A common way to describe the performance of
a $b$-tagging algorithm is by showing its misidentification
probability as function of the efficiency. 
Such a representation is shown in the last plot in
Figure~\ref{fig:btag_disc}, for the fast and the full simulations of 
CMS: one can see that over a wide range of cuts, chosen such to fix
the rejection factor for the light-flavours related background, the
$b$-tag efficiency in the fast simulation is systematically
overestimated by some 5-10\%. 

\begin{figure}[htb]
\centering
\includegraphics[height=42mm]{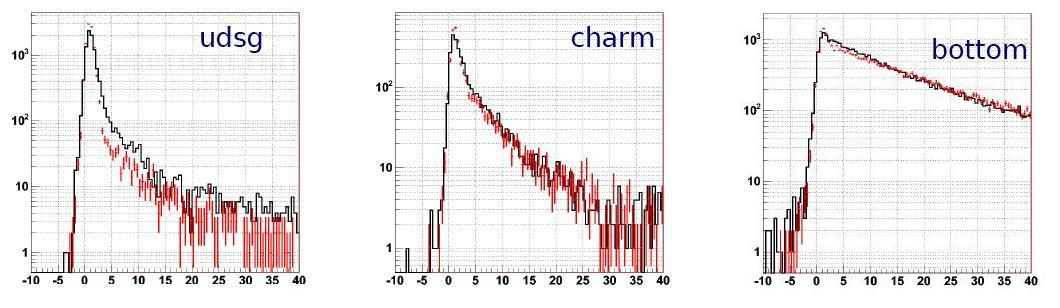}
\caption{Largest impact parameter for charged particles inside jets, in the CMS fast (red points) and full (black histogram) simulations.}
\label{fig:1st_ip_in_jets}
\end{figure}

\begin{figure}[htb]
\centering
\includegraphics[height=42mm]{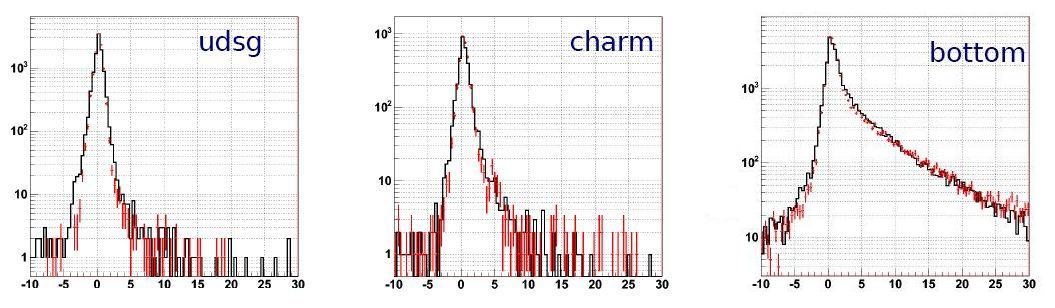}
\caption{Third largest impact parameter for charged particles inside jets, in the CMS fast (red points) and full (black histogram) simulations.}
\label{fig:3rd_ip_in_jets}
\end{figure}

In order to understand which, of the many simplifications meant to make FAMOS fast, is the culprit of this situation,
a closer look to the variables that enter in the definition of the $b$-tagging is needed.
In Figure~\ref{fig:d0} we had shown how well the impact parameter was reproduced in FAMOS, in the relatively easy case of isolated muons.
Figure~\ref{fig:1st_ip_in_jets} shows instead the largest impact
parameter among all the charged tracks (mostly hadrons) in each
jet. The comparison with the corresponding full simulation is not
satisfactory for jets from $udsg$ partons. 
The situation improves if one does not consider the tracks with the largest impact parameter: for instance,
in Figure~\ref{fig:3rd_ip_in_jets} the third largest impact parameter in each jet is shown.

All this suggests that FAMOS lacks the description of some process
which seldom produces a small number of tracks with
significant impact parameter. 
At first, it was thought that the lack of fake tracks (see previous
subsection) could have been the responsible 
of the discrepancy, but at a closer look they were found not sufficient to explain it.
Instead, the difference can be attributed to the nuclear interactions
of the hadrons with the tracker active and passive materials: they
were not simulated in FAMOS, but their implementation is planned for
the next release of the fast simulation of CMS.

\subsubsection{Hadrons and jets energy}

\begin{figure}[htb]
\centering
\includegraphics[height=40mm]{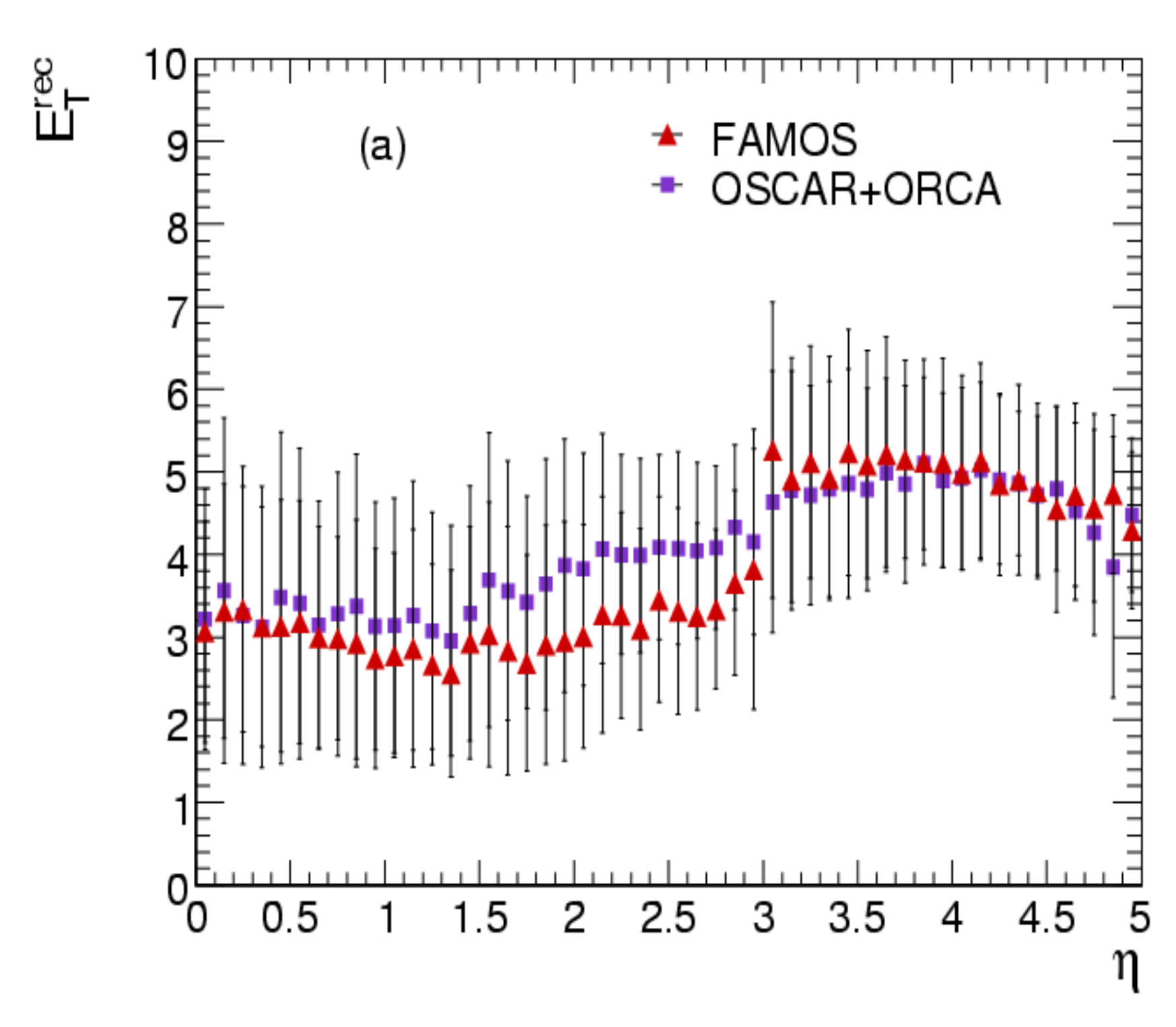}
\includegraphics[height=40mm]{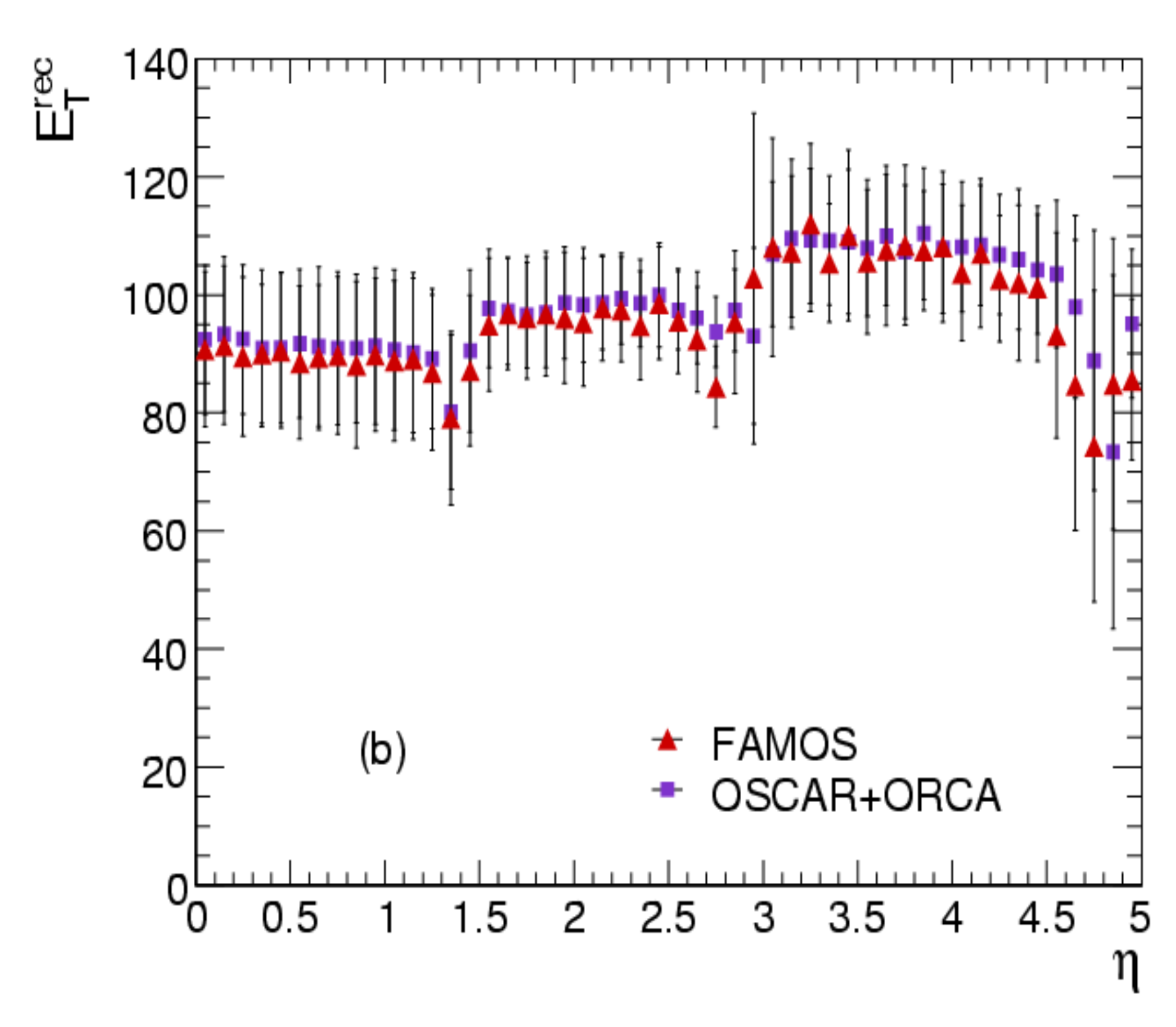}
\includegraphics[height=40mm]{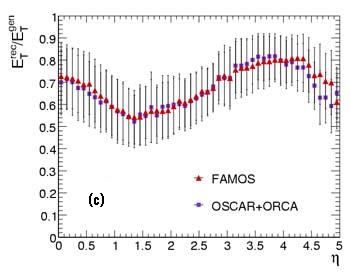}
\caption{Calorimetric response, as a function of $\eta$, to single pions of 5~GeV (a) and 100~GeV (b),
and to jets, reconstructed with the iterative cone algorithm, of $p_T$ between 80 and 120~GeV/$c$ (c) in the CMS
fast (triangles) and full (squares) simulations.}
\label{fig:pions}
\end{figure}

The calorimetric response (ECAL+HCAL) to single pions in FAMOS and in the CMS full simulation is shown in Figures~\ref{fig:pions}a and b.
In order to simplify the simulation, all the long-lived hadrons in FAMOS are treated as charged pions.
This proves to be enough to obtain a remarkable agreement with the full simulation, as shown in Figure~\ref{fig:pions}c
for jets between 80 and 120 GeV/$c$ in $p_T$. There are plans, however, to further improve the realism, by treating differently:
the long-lived neutral hadrons, since they don't release any signal before the first nuclear interaction;
protons and neutrons, whose kinematic is different due to the high mass;
anti-protons and anti-neutrons, which in addition can annihilate.

\end{document}